\documentclass[12pt,twoside]{report}
\pdfoutput=1
\usepackage{amsmath,amsfonts,physics,amssymb,amsthm,amssymb,bbm,bm}
\usepackage{pdfsync}
\usepackage{graphicx,import,wrapfig}
\usepackage[utf8]{inputenc}
\usepackage{empheq}
\usepackage[all]{xy}
\usepackage{stmaryrd}
\usepackage{rotating}
\usepackage[dvipsnames]{xcolor}  
\usepackage{slashed}
\usepackage{cancel}
\usepackage{array, makecell} %
\usepackage{comment}
\usepackage{tikz-cd}
\usepackage{hhline}
\usepackage{cancel}
\usepackage{pifont}
\usepackage{dsfont}
\usepackage{stackengine}
\usepackage[breakable]{tcolorbox}
\usepackage{mathrsfs}
\usepackage{booktabs}
\usepackage{float}
\usepackage{changepage}
\usepackage[normalem]{ulem}
\usepackage{abraces}
\usepackage{multicol}
\usepackage{pgfornament-han}
\usepackage[12pt]{moresize}
\usetikzlibrary{shapes.geometric,calc}

\usepackage[pagebackref]{hyperref}
\renewcommand*{\backref}[1]{}
\renewcommand*{\backrefalt}[4]{\small{
    \ifcase #1%
          \or (Cited on page~#2.)%
          \else (Cited on pages~#2.)%
    \fi%
    }}
\hypersetup{colorlinks,linkcolor={RedViolet},citecolor={Blue},urlcolor={Blue},
pdflang={en-CA},
pdftitle={Asymptotic Higher Spin Symmetries: Noether Realization and Algebraic Structure in Einstein-Yang-Mills Theory},
pdfauthor={Nicolas Cresto},
pdfsubject={Physics},
}

\usepackage{orcidlink}

\allowdisplaybreaks %long equations break on two pages
\numberwithin{equation}{section}

\DeclareMathAlphabet\mathrsfso{U}{rsfso}{m}{n}

\usepackage[scr=boondoxo,scrscaled=1.1]{mathalpha}

\makeatletter

\DeclareFontFamily{U}{matha}{\hyphenchar\font45}
\DeclareFontShape{U}{matha}{m}{n}{
      <5> <6> <7> <8> <9> <10> gen * matha
      <10.95> matha10 <12> <14.4> <17.28> <20.74> <24.88> matha12
      }{}
\DeclareSymbolFont{matha}{U}{matha}{m}{n}
\DeclareMathSymbol{\oright}       {2}{matha}{"69}

\usepackage[top=2.54cm, bottom=2.54cm, left=2.54cm, right=2.54cm]{geometry}

\usepackage{fancyhdr}

\newcommand{\doublehat}[1]{%
\begingroup%
  \let\macc@kerna\z@%
  \let\macc@kernb\z@%
  \let\macc@nucleus\@empty%
  \hat{\raisebox{.55ex}{\vphantom{\ensuremath{#1}}}\smash{\hat{#1}}}%
\endgroup%
}

\renewcommand{\ni}{\noindent}

\newcommand{\bit}{\begin{itemize}}
\newcommand{\eit}{\end{itemize}}
\newcommand{\bd}{\begin{description}}
\newcommand{\ed}{\end{description}}

\newcommand{\bc}{\begin{center}}
\newcommand{\ec}{\end{center}}
\newcommand{\C}{{\mathbb C}}
\newcommand{\N}{{\mathbb N}}
\newcommand{\R}{{\mathbb R}}
\newcommand{\cP}{{\cal P}}
\newcommand{\cD}{\mathscr{D}}
\newcommand{\E}{\mathsf{E}}

\newcommand{\gbms}{\mathfrak{gbms}}
\newcommand{\ebms}{\mathfrak{ebms}}

\newcommand{\cF}{{\mathcal F}}
\newcommand{\cM}{{\mathcal M}}

\newcommand{\cT}{{\mathcal T}}
\newcommand{\cA}{{\mathcal A}}

\def\be#1\ee{\begin{align}#1\end{align}}
\newcommand{\bea}{\begin{eqnarray}}
\newcommand{\eea}{\end{eqnarray}}
\newcommand{\bs}{\begin{subequations}}
\newcommand{\es}{\end{subequations}}
\newcommand{\nn}{\nonumber}

\newcommand{\mat}[4]{\left(\begin{matrix}{#1}&{#2}\\{#3}&{#4}\end{matrix}\right)}

\newcommand{\sds}{\,\tikz[baseline=1]{
		\draw[line width=.6pt] (0,.13) circle (.8ex);
		\draw[line width=.6pt] (-.05ex,.27) -- (-.05ex,0);
		\draw[line width=.6pt] (0,.13) -- (.8ex,.13);}\,}
\usepackage[numbers,sort&compress]{natbib}
\makeatletter
\newcommand{\oset}[3][0ex]{%
  \mathrel{\mathop{#3}\limits^{
    \vbox to#1{\kern-2\ex@
    \hbox{$\scriptstyle#2$}}}}}
\makeatother

\def\rd{\mathrm{d}}
\def\pa{\partial}
\newcommand{\Dcal}{\mathcal{D}}
\newcommand{\bDcal}{\overbar{\mathcal{D}}}
\newcommand{\DGR}{\overset{\mkern-1mu\sgr\mkern1mu}{\mathcal{D}}}
\newcommand{\DYM}{D_{\mkern-3mu A}}
\newcommand{\DEYM}{\overset{\mkern-0mu\scriptscriptstyle{\textsc{eym}}\mkern0mu}{\mathcal{D}}}
\newcommand{\DTOT}{\overset{\mkern-1mu\sym\mkern1mu}{\mathcal{D}}}
\newcommand{\DYMc}{D_{\Ac0}}

\newcommand{\nablaYM}{\nabla_{\mkern-4mu A}}
\newcommand{\nablaGR}{\overset{\sgr}{\nabla}}
\newcommand{\nablaEYM}{\overset{\scriptscriptstyle{\textsc{eym}}}{\nabla}}

\newcommand{\pui}[1][1]{\pa_u^{-#1}}

\newcommand{\dt}{\delta_\tau}
\newcommand{\dtp}{\delta_{\tau'}}
\newcommand{\dtpp}{\delta_{\tau''}}

\newcommand{\da}{\delta_\alpha}
\newcommand{\dap}{\delta_{\alpha'}}
\newcommand{\dapp}{\delta_{\alpha''}}
\newcommand{\dah}{\delta_{\fa}}

\newcommand{\dta}{\delta_{(\tau,\alpha)}}
\newcommand{\dtap}{\delta_{(\tau'\!,\alpha')}}

\newcommand{\tdt}{\tilde{\delta}_\tau}
\newcommand{\tdtp}{\tilde{\delta}_{\tau'}}

\newcommand{\hdt}{\hat{\delta}_\tau}
\newcommand{\hdtp}{\hat{\delta}_{\tau'}}

\newcommand{\dbt}{\delta_{\bar{\tau}}}
\newcommand{\dbtp}{\delta_{\bar{\tau}'}}
\newcommand{\dbtpp}{\delta_{\bar{\tau}''}}

\newcommand{\odt}{\delta_{\text{\scalebox{1.1}{$\ensurestackMath{\stackon[0.9pt]{\tau}{\mkern 1mu\scriptscriptstyle{\circ}}}$}}}}

\newcommand{\odtp}{\delta_{\text{\scalebox{1.1}{$\ensurestackMath{\stackon[-1.9pt]{\tau'}{\mkern-4mu\scriptscriptstyle{\circ}}}$}}}}

\newcommand{\odtpp}{\delta_{\text{\scalebox{1.1}{$\ensurestackMath{\stackon[-2pt]{\tau''}{\mkern-9mu\scriptscriptstyle{\circ}}}$}}}}

\newcommand{\hda}{\hat{\delta}_\alpha}
\newcommand{\hdap}{\hat{\delta}_{\alpha'}}
\newcommand{\hdapp}{\hat{\delta}_{\alpha''}}

\newcommand{\dT}{\delta_T}
\newcommand{\dTp}{\delta_{T'}}

\newcommand{\dPta}{\delta_\gamma}
\newcommand{\dPtap}{\delta_{\gamma'}}
\newcommand{\dPtapp}{\delta_{\gamma''}}

\newcommand{\dbPta}{\delta_{\bar{\gamma}}}
\newcommand{\dbPtap}{\delta_{\bar{\gamma}'}}
\newcommand{\dbPtapp}{\delta_{\bar{\gamma}''}}

\newcommand{\dA}{\delta_{\text{\scalebox{1.1}{$\mathsf{a}$}}}}
\newcommand{\dAp}{\delta_{\text{\scalebox{1.1}{$\mathsf{a}$}$'$}}}

\newcommand{\hdT}{\hat{\delta}_T}
\newcommand{\hdTp}{\hat{\delta}_{T'}}

\renewcommand{\d}[2]{\delta^{\scriptscriptstyle{[#1]}}_{#2}}
\newcommand{\td}[2]{\tilde{\delta}^{\scriptscriptstyle{[#1]}}_{#2}}

\newcommand{\Hd}[2]{\delta^{\scriptscriptstyle{[#1]}\mathsf{H}}_{#2}}
\newcommand{\Htd}[2]{\tilde{\delta}^{\scriptscriptstyle{[#1]}\mathsf{H}}_{#2}}

\newcommand{\Sd}[2]{\delta^{\scriptscriptstyle{[#1]}\mathsf{S}}_{#2}}
\newcommand{\Std}[2]{\tilde{\delta}^{\scriptscriptstyle{[#1]}\mathsf{S}}_{#2}}

\newcommand{\LL}{\mathscr{L}}
\newcommand{\sE}{\mathsf{E}}
\newcommand{\sI}{\mathsf{I}}

\newcommand{\Ll}{\mathbb{L}}
\newcommand{\nkor}{\rho}

\newcommand{\overbar}[1]{\mkern 3mu\overline{\mkern-3.5mu#1\mkern-2mu}\mkern 2mu}

\newcommand{\overbarT}[1]{\mkern 3mu\overline{\mkern-2.5mu#1\mkern-1.5mu}\mkern 2mu}

\newcommand{\ooverline}[1]{\mkern 3mu\overline{\overline{#1}}\mkern 2mu}

\newcommand{\overbarF}[1]{\mkern 2mu\overline{\mkern-4mu#1\mkern-1mu}\mkern 2mu}

\newcommand{\bnews}{\overbarF{F}}
\newcommand{\bA}{\bar{A}}

\newcommand{\bC}{\overbar{C}}
\newcommand{\bN}{\overbar{N}}
\newcommand{\bz}{\bar{z}}
\renewcommand{\bm}{\overbar{m}}
\newcommand{\bY}{\overbar{Y}}
\newcommand{\bD}{\overbar{D}}
\newcommand{\bP}{\overbar{P}}
\newcommand{\Vs}{\overline{\mathsf{V}}(S)}
\newcommand{\Vss}{\ooverline{\mathsf{V}}(S)}
\newcommand{\Ws}{\overline{\mathsf{W}}(S)}
\newcommand{\Wo}{\overline{\mathsf{W}}}

\newcommand{\bWcal}{\overline{\Wcal}}

\newcommand{\bgN}{\bar{g}_\textsc{n}}
\newcommand{\bbVcal}{\ooverline{\Vcal}}

\newcommand{\bJ}{\bar{J}}
\newcommand{\bcT}{\overbarT{\mathcal{T}}}
\newcommand{\bsfT}{\overbarT{\mathsf{T}}}

\newcommand{\bt}{\bar{\tau}}
\newcommand{\btp}{\bar{\tau}'}
\newcommand{\btpp}{\bar{\tau}''}

\newcommand{\bPta}{\bar{\gamma}}
\newcommand{\bPtap}{\bar{\gamma}'}
\newcommand{\bPtapp}{\bar{\gamma}''}

\newcommand{\hs}{\hat{s}}
\newcommand{\shs}{\hat{s}^*}
\newcommand{\hS}{\hat{S}}
\newcommand{\shS}{\hat{S}^*}
\newcommand{\hA}{\widehat{A}}
\newcommand{\hD}{\widehat{\Delta}}
\newcommand{\hV}{\widehat{\V}}

\newcommand{\overhat}[1]{\mkern 5mu\widehat{\mkern-4.5mu#1\mkern-1mu}\mkern 2mu}

\newcommand{\hh}{\overhat{\cH}}
\newcommand{\hq}{\widehat{\cq}}
\newcommand{\ft}{\hat{\tau}}
\newcommand{\fa}{\hat{\alpha}}
\newcommand{\hG}{\widehat{G}}
\newcommand{\fPta}{\hat{\Pta}}

\makeatletter
\newcommand{\hhat}[1]{% 
\begingroup%
  \let\macc@kerna\z@%
  \let\macc@kernb\z@%
  \let\macc@nucleus\@empty%
  \widehat{\mathchoice%
    {\raisebox{.2ex}{\vphantom{\ensuremath{\displaystyle #1}}}}%
    {\raisebox{.4ex}{\vphantom{\ensuremath{\textstyle #1}}}}%
    {\raisebox{.16ex}{\vphantom{\ensuremath{\scriptstyle #1}}}}%
    {\raisebox{.14ex}{\vphantom{\ensuremath{\scriptscriptstyle #1}}}}%
    \smash{\widehat{#1}}}%
\endgroup%
}
\makeatother

\makeatletter
\DeclareRobustCommand\widecheck[1]{{\mathpalette\@widecheck{#1}}}
\def\@widecheck#1#2{%
    \setbox\z@\hbox{\m@th$#1#2$}%
    \setbox\tw@\hbox{\m@th$#1%
       \widehat{%
          \vrule\@width\z@\@height\ht\z@
          \vrule\@height\z@\@width\wd\z@}$}%
    \dp\tw@-\ht\z@
    \@tempdima\ht\z@ \advance\@tempdima2\ht\tw@ \divide\@tempdima\thr@@
    \setbox\tw@\hbox{%
       \raise\@tempdima\hbox{\scalebox{1}[-1]{\lower\@tempdima\box
\tw@}}}%
    {\ooalign{\box\tw@ \cr \box\z@}}}
\makeatother

\newcommand{\hhG}{\widecheck{G}}
\newcommand{\overtilde}[1]{\mkern 5mu\widetilde{\mkern-4.5mu#1\mkern-1mu}\mkern 2mu}

\newcommand{\tQ}{\widetilde{Q}}
\newcommand{\tbQ}{\ensurestackMath{\stackon[0.8pt]{\overbarT{Q}}{\mkern1mu\text{\scalebox{1.25}{\texttildelow}}}}}
\newcommand{\tq}{\mkern0.7mu\widetilde{\cq}}
\renewcommand{\th}{\overtilde{\cH}}
\newcommand{\tT}{\widetilde{T}}
\newcommand{\tQcal}{\widetilde{\Qcal}}

\renewcommand{\tt}[1]{\tilde{\tau}_{#1}}
\newcommand{\ttp}[1]{\tilde{\tau}'_{#1}}

\newcommand{\tJ}{\overtilde{J}}

\newcommand{\cR}{\mathcal{R}}
\newcommand{\np}[2]{\phi_{#1}^{(#2)}}
\newcommand{\bnp}[2]{\bar{\phi}_{#1}^{(#2)}}

\newcommand{\news}{F}
\newcommand{\gN}{g_\textsc{n}}
\newcommand{\gYM}{g_\textsc{ym}}
\newcommand{\GN}{G_{\mkern-2mu N}}
\newcommand{\X}{\mathfrak{X}}
\newcommand{\g}{\mathfrak{g}}

\newcommand{\M}{\mathcal{M}}
\newcommand{\Ocal}{\mathcal{O}}
\newcommand{\A}{\mathcal{T}}
\newcommand{\Ah}{\widehat{\mathcal{T}}}
\newcommand{\Ao}{\mathcal{T}^+}

\newcommand{\cTo}{\ocirc{\mathcal{T}}}

\newcommand{\Cc}[1]{\sigma_{#1}}
\newcommand{\Ac}[1]{\beta_{#1}}

\newcommand{\G}[1]{\mathcal{G}\big(#1\big)}
\newcommand{\sn}{\mathsf{SN}}
\newcommand{\PS}{\mathcal{P}}

\newcommand{\cS}{\mathcal{S}}
\newcommand{\WS}[1]{\mathcal{W}^{#1}_{\bfAc}(S)}

\newcommand{\CP}[1]{\C\mathbb{P}^{#1}}
\newcommand{\sfY}{\mathsf{Y}}

\newcommand{\ym}{\textsc{ym}}
\newcommand{\gr}{\textsc{gr}}
\newcommand{\sym}{\scriptscriptstyle{\textsc{ym}}}
\newcommand{\sgr}{\scriptscriptstyle{\textsc{gr}}}
\newcommand{\QYM}[1]{\overset{\mkern-4mu\sym\mkern4mu}{Q^u}_{\mkern-11mu#1}}
\newcommand{\QGR}[1]{\overset{\mkern-4mu\sgr\mkern4mu}{Q^u}_{\mkern-11mu#1}}

\newcommand{\sfST}{\mathsf{ST}}
\newcommand{\sfSK}{\mathsf{S}_{\mkern-1mu K}}
\newcommand{\sfTK}{\mathsf{T}_{\mkern-3mu K}}
\newcommand{\sfSTK}{\mathsf{ST}_{\mkern-3mu K}}
\newcommand{\sfK}{\mathsf{K}}

\newcommand{\cST}{\mathcal{ST}}
\newcommand{\cSK}{\mathcal{S}_K}
\newcommand{\cTK}{\mathcal{T}_K}
\newcommand{\cSTK}{\mathcal{ST}_{\mkern-5mu K}}

\newcommand{\V}{\mathsf{V}}
\newcommand{\tV}{\widetilde{\mathsf{V}}}
\newcommand{\Vcal}{\mathcal{V}}

\newcommand{\W}{\mathsf{W}}
\newcommand{\Wcal}{\mathcal{W}}

\newcommand{\sfT}{{\mathsf{T}}}
\newcommand{\sfS}{\mathsf{S}}
\newcommand{\sfTo}{\mathsf{T}^+}
\newcommand{\sfTh}{\widehat{\mathsf{T}}}
\newcommand{\TN}{\mathscr{N}}
\newcommand{\ocT}{\ocirc{\mathcal{T}}}
\newcommand{\osfT}{\ocirc{\mathsf{T}}}

\newcommand{\Ccel}[1]{\mathcal{C}^\textsf{cel}_{(#1)}(S)}
\newcommand{\Ccar}[1]{\mathcal{C}^\textsf{car}_{(#1)}(\scri)}
\newcommand{\Ccelg}[1]{\mathcal{C}^\textsf{cel}_{(#1)}(S,\g)}
\newcommand{\Ccarg}[1]{\mathcal{C}^\textsf{car}_{(#1)}(\scri,\g)}
\newcommand{\dCar}{\delta}

\newcommand{\Wt}{\bt\vartriangleright}
\newcommand{\Wtp}{\btp\vartriangleright}
\newcommand{\Wtpp}{\btpp\vartriangleright}

\newcommand{\dotvartriangleright}{\ensurestackMath{\stackon[-6.6pt]{\vartriangleright}{\mkern-3mu\cdot}}}

\newcommand{\dotblacktriangleright}{\ensurestackMath{\stackon[-6.6pt]{\blacktriangleright}{\mkern-3mu\textcolor{white}{\cdot}}}}

\newcommand{\scriptdotblacktriangleright}{\text{\scalebox{1.1}{$\ensurestackMath{\stackon[-4.4pt]{\blacktriangleright}{\mkern-3mu\textcolor{white}{\cdot}}}$}}}

\newcommand{\dotblacktriangleleft}{\ensurestackMath{\stackon[-6.6pt]{\blacktriangleleft}{\mkern3mu\textcolor{white}{\cdot}}}}

\newcommand{\scriptdotblacktriangleleft}{\text{\scalebox{1.1}{$\ensurestackMath{\stackon[-4.3pt]{\blacktriangleleft}{\mkern3mu\textcolor{white}{\cdot}}}$}}}

\newcommand{\dWt}{\bt\mkern5mu\dotvartriangleright\mkern5mu}
\newcommand{\dWtp}{\btp\mkern5mu\dotvartriangleright\mkern5mu}

\newcommand{\blackwhitetriangle}{\ensurestackMath{\stackon[-1pt]{\,}{\mkern5mu\begin{tikzpicture}
\draw[fill=black] (0,-0.04) -- (0,0.075) -- (0.26,0.075) -- (0,-0.04);
\draw (0,0.075) -- (0,0.19) -- (0.26,0.075);
\end{tikzpicture}\mkern5mu}}}

\newcommand{\BWt}{\tau\blackwhitetriangle}
\newcommand{\BWtp}{\tau'\blackwhitetriangle}

\newcommand{\barWhiteTriangle}{\ensurestackMath{\mkern1mu\stackon[-1pt]{\,}{\mkern4mu\begin{tikzpicture}
\draw (0,0) -- (0,0.25) -- (0.28,0.125) -- (0,0);
\draw (0.05,0.03) -- (0.05,0.23);
\end{tikzpicture}}\mkern4mu}}

\newcommand{\bWt}{\tau\barWhiteTriangle}
\newcommand{\bWtp}{\tau'\barWhiteTriangle}

\newcommand{\ocirc}[1]{\ensurestackMath{\stackon[1.5pt]{#1}{\mkern 1mu\scriptscriptstyle{\circ}}}}

\newcommand{\ocircp}[1]{\ensurestackMath{\stackon[-2pt]{#1}{\mkern-3mu\scriptscriptstyle{\circ}}}}

\newcommand{\ocircpp}[1]{\ensurestackMath{\stackon[-2pt]{#1}{\mkern-6mu\scriptscriptstyle{\circ}}}}

\newcommand{\Bt}{\ocirc{\tau}\blacktriangleright}
\newcommand{\Btp}{\ocircp{\tau'}\blacktriangleright}
\newcommand{\Btpp}{\ocircpp{\tau''}\blacktriangleright}

\newcommand{\dBt}{\ocirc{\tau}\mkern5mu\dotblacktriangleright\mkern5mu}
\newcommand{\dBtp}{\ocircp{\tau'} \mkern5mu\dotblacktriangleright\mkern5mu}
\newcommand{\dBtpp}{\ocircpp{\tau''}\mkern5mu \dotblacktriangleright\mkern5mu}

\newcommand{\rBt}{\ocirc{\tau}\blacktriangleleft}
\newcommand{\rBtp}{\ocircp{\tau'}\!\blacktriangleleft}
\newcommand{\rBtpp}{\ocircpp{\tau''}\!\blacktriangleleft}

\newcommand{\drBt}{\ocirc{\tau}\mkern5mu \dotblacktriangleleft\mkern5mu}
\newcommand{\drBtp}{\ocircp{\tau'}\mkern3mu \dotblacktriangleleft\mkern5mu}
\newcommand{\drBtpp}{\ocircpp{\tau''}\mkern2mu \dotblacktriangleleft\mkern5mu}

\newcommand{\Pairg}[1]{\big\langle #1\big\rangle_\g}

\newcommand{\Bprod}{\mkern5mu\begin{tikzpicture}
\draw[fill=black] (0,0) -- (0,0.23) -- (0.175,0.115) -- (0,0);
\draw[fill=black] (0.175,0.115) -- (0.35,0.23) -- (0.35,0) -- (0.175,0.115);
\end{tikzpicture}\mkern5mu}

\newcommand{\Wprod}{\mkern5mu\begin{tikzpicture}
\draw (0,0) -- (0,0.23) -- (0.35,0);
\draw (0,0) -- (0.35,0.23);
\end{tikzpicture}\mkern5mu}

\newcommand{\bWprod}{\mkern5mu\begin{tikzpicture}
\draw (0,0) -- (0,0.25) -- (0.40,0);
\draw (0,0) -- (0.40,0.25);
\draw (0.05,0.03) -- (0.05,0.22);
\end{tikzpicture}\mkern6mu}

\newcommand{\BWprod}{\mkern5mu\begin{tikzpicture}
\draw (0,0) -- (0,0.23) -- (0.35,0);
\draw (0,0) -- (0.35,0.23);
\draw[fill=black] (0,0) -- (0,0.115) -- (0.175,0.115) -- (0,0);
\end{tikzpicture}\mkern5mu}

\newcommand{\ot}{\ocirc{\tau}}
\newcommand{\otp}{\ocircp{\tau'}}
\newcommand{\otpp}{\ocircpp{\tau''}}

\newcommand{\Pta}{\gamma}
\newcommand{\Ptap}{\gamma'}
\newcommand{\Ptapp}{\gamma''}

\renewcommand{\l}{\ell}
\newcommand{\sshp}{Y^s_{\ell,m}}

\newcommand{\adC}{\textrm{ad}_{\scriptscriptstyle{C}}}
\newcommand{\adA}{\mathrm{ad}\,A}
\newcommand{\ad}{{\textrm{ad}}_\sigma}

\newcommand{\adtG}{{\textrm{ad}}_{\sigma_t} \hat{G}_t}
\newcommand{\adaG}{{\textrm{ad}}_{\sigma_\alpha} \hat{G}_\alpha}
\newcommand{\ada}{\textrm{ad}_{\sigma_\alpha}}

\newcommand{\mTG}{\mathcal{T}^G}

\newcommand{\mTaG}{\mathcal{T}^{G_\alpha}}

\renewcommand{\t}[1]{\tau_{#1}}
\newcommand{\T}[1]{T_{#1}}
\newcommand{\tp}[1]{\tau'_{#1}}
\newcommand{\Tp}[1]{T'_{#1}}
\newcommand{\tpp}[1]{\tau''_{#1}}
\newcommand{\Tpp}[1]{T''_{#1}}

\renewcommand{\a}[1]{\alpha_{#1}}
\newcommand{\Aa}[1]{\mathsf{a}_{#1}}
\newcommand{\ap}[1]{\alpha'_{#1}}
\newcommand{\Aap}[1]{\mathsf{a}'_{#1}}
\newcommand{\app}[1]{\alpha''_{#1}}

\newcommand{\scri}{\mathrsfso{I}}
\newcommand{\scrip}{\scri^+}
\newcommand{\scrim}{\scri^-}

\newcommand{\Qt}{Q_\tau}
\newcommand{\Qa}{Q_\alpha}
\newcommand{\Qta}{Q_{(\tau,\alpha)}}
\newcommand{\QPta}{Q_\Pta}
\newcommand{\Ht}{H_\tau}
\newcommand{\cH}{\mathrsfso{H}}
\newcommand{\Qtp}{Q_{\tau'}}
\newcommand{\Qap}{Q_{\alpha'}}

\newcommand{\Qad}{\dot{Q}_\alpha}

\newcommand{\Qcal}{\mathcal{Q}}
\newcommand{\cq}{\mathscr{Q}}
\newcommand{\cqAas}{\mathscr{Q}_{\text{\scalebox{1.24}{$\Aa{s}$}}}}
\newcommand{\q}[2]{\hq_{#1}^{\scriptscriptstyle{\!(#2)}}}
\newcommand{\cQ}[2]{Q^u_{#1}\left[#2\right]}

\newcommand{\soft}[1]{#1^{\mathsf{S}}}
\newcommand{\hard}[1]{#1^{\mathsf{H}}}

\newcommand{\lbr}{\llbracket}
\newcommand{\rbr}{\rrbracket}
\newcommand{\poisson}[1]{\big\{#1\big\}}
\newcommand{\paren}[1]{\,\Lbag #1\Rbag\,}
\newcommand{\cyc}{\overset{\circlearrowleft}{=}}

\newcommand{\Ham}{\eta}

\newcommand{\hHam}{\eta_{\scriptscriptstyle{G}}}

\newcommand{\hHama}{\eta_{\scriptscriptstyle{G}_\alpha}}

\newcommand{\bfm}[1]{\boldsymbol{#1}}
\newcommand{\bft}{\bfm{\tau}}
\newcommand{\bfa}{\bfm{\alpha}}
\newcommand{\bfAc}{\bfm{\beta}}
\newcommand{\bfCc}{\bfm{\sigma}}
\newcommand{\bep}{\bfm{\epsilon}}

\usepackage{tikz}
\newcommand*\circled[1]{\tikz[baseline=(char.base)]{
            \node[shape=circle,draw,inner sep=2pt] (char) {#1};}}

\newcounter{dingstart}
\newcommand{\dn}[1]{\textrm{{\small\ding{\numexpr\value{dingstart}+ #1\relax}}}}

\definecolor{myorange}{RGB}{223, 109, 20}

\definecolor{argile}{RGB}{239, 239, 239}
\definecolor{beige}{RGB}{254, 253, 240}

\begin{document}

\newgeometry{margin=0.5cm,top=0.5cm,bottom=0.5cm}

\begin{newfamily}[pgfhan]
\begin{center}
\begin{tikzpicture}
\tikzset{every node/.append style={%
inner sep=0pt,
color= black}}
\node[minimum width=20.58cm,minimum height=26.9cm] (chframe){};

\node[anchor=north west] (nw) at (chframe.north west)
{\pgfornament[scale=0.3]{10}};

\node[anchor=north east] at (chframe.north east)
{\pgfornament[symmetry=v,scale=0.3]{10}};

\node[anchor=south west] (sw) at (chframe.south west)
{\pgfornament[symmetry=h,scale=0.3]{10}};

\node[anchor=south east] (se) at (chframe.south east)
{\pgfornament[symmetry=c,scale=0.3]{10}};

\node[anchor=south west,xscale=8] at (sw.south east) {\pgfornament[scale=0.3]{32}};

\node[shift={(4.41mm,0cm)},anchor=south west,yscale=10.8,rotate=90] at (sw.north west) {\pgfornament[scale=0.3]{32}};

\node[shift={(-4.41mm,0cm)},anchor=south east,yscale=10.8,rotate=-90] at (se.north east) {\pgfornament[scale=0.3]{32}};

\node[anchor=north west,xscale=8] at (nw.north east) {\pgfornament[scale=0.3]{32}};

\node[text width=20cm,align=center](Text){
\vspace{3.5cm}

\textbf{\huge{Asymptotic ~Higher~ Spin ~Symmetries:}}
\vspace{0.9cm}

\textbf{\LARGE{Noether ~Realization}}
\vspace{0.4cm}

\textbf{\Large{\&}}
\vspace{0.4cm}

\textbf{\LARGE{Algebraic ~Structure}}
\vspace{0.4cm}

\textbf{\Large{in}}
\vspace{0.4cm}

\textbf{\LARGE{Einstein-Yang-Mills~ Theory}}
\vspace{2cm}

\large{\textit{PhD Thesis in Physics}}
\bigskip

\Large{Nicolas Cresto\,\orcidlink{0009-0006-7263-8777}}
\vspace{2cm}

\large{\textit{Supervised by}}
\bigskip

\Large{Laurent Freidel}
\vspace{2.5cm}

\normalsize{September 2025}
\medskip

Waterloo, Ontario, Canada
\vspace{1.5cm}

\includegraphics[width=7cm]{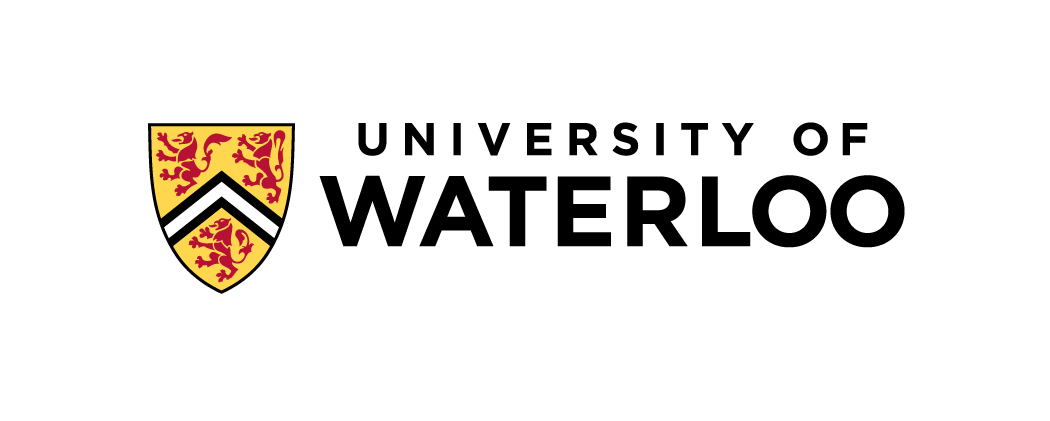}
\hspace{8.4cm}
\includegraphics[width=3.5cm]{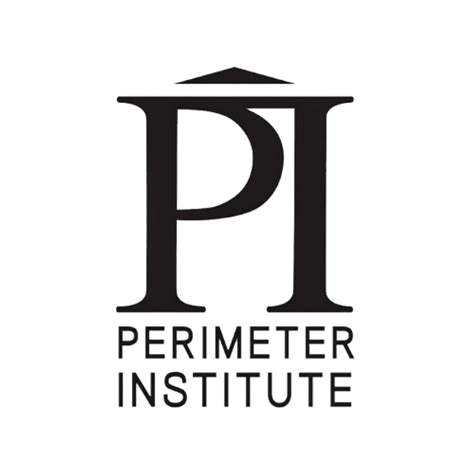}
};

\end{tikzpicture}
\end{center}
\end{newfamily}

\restoregeometry
\thispagestyle{empty}
\vspace*{2cm}

\begin{center}
\Large{\textit{Committee Members}}
\bigskip
\end{center}

\ni \normalsize{The following served on the Examining Committee for this thesis. 
Their decision is by majority vote.
In alphabetic order, hereafter the Internal Member, the Other Member, the External Examiner, the Supervisor, the Internal-External Member and the co-Supervisor.}

\vspace{1cm}

\begin{center}
\large{Niayesh Afshordi$^{1,2}$}
\bigskip

\large{Freddy Cachazo$^{2}$}
\bigskip

\large{Miguel Campiglia$^3$}
\bigskip

\large{Laurent Freidel$^2$}
\bigskip

\large{Florian Girelli$^{2,4}$}
\bigskip

\large{Robert Myers$^{1,2}$}
\vspace{2cm}

\normalsize{\textit{
$^1$Department of Physics \& Astronomy, University of Waterloo,\\200 University
Avenue West, Waterloo, Ontario, N2L 3G1, Canada\\\bigskip
$^2$Perimeter Institute for Theoretical Physics,\\ 31 Caroline Street North, Waterloo, Ontario, N2L 2Y5, Canada\\ \bigskip
$^3$Faculty of Sciences, Universidad de la Rep\'ublica,\\ Igu\'a 4225, Montevideo, Uruguay\\ \bigskip
$^4$Department of Applied Mathematics, University of Waterloo,\\200 University
Avenue West, Waterloo, Ontario, N2L 3G1, Canada}}
\end{center}

% \newpage
% \newgeometry{margin=2.5cm,top=3cm,bottom=3.5cm}
% \setcounter{page}{1}
% \pagenumbering{roman}

% \begin{center}
% \huge{\text{$\mathscr{Dedications}$}}\\
% \pgfornament[color=black,width=0.7\textwidth]{88}
% \end{center}
% \vspace{2.5cm}

% \begin{flushright}
% \large{\emph{To my parents, Hélène \& Franck, \\
% for their unconditional support throughout \\
% 28 years of pretty intense life.\!\!\!}}
% \vspace{1.5cm}

% \large{\emph{To Kree, my Mutu,\\
% whose love entangled my whole being\\
% and accompanies each one of my breaths \&\\
% each one of my steps.\!\!}}
% \vspace{1.5cm}

% \large{\emph{To Laurent.\\
% More than a supervisor,\\
% A friend \& mentor.\!\!}}
% \vspace{1.5cm}

% \large{\emph{To my dear friend Andrea,\\
% who makes this planet a better place.}}
% \end{flushright}

\newgeometry{margin=0.5cm,top=0.5cm,bottom=0.5cm}

\setcounter{page}{1}
\pagenumbering{roman}
\addcontentsline{toc}{chapter}{Dedications}
\phantomsection 

\begin{newfamily}[pgfhan]
\begin{center}
\begin{tikzpicture}

\tikzset{every node/.append style={%
inner sep=0pt,
color= black}}
\node[minimum width=20.58cm,minimum height=26.9cm] (chframe){};

\node[text width=20.57cm,align=center](Text){
\includegraphics[width=20.57cm]{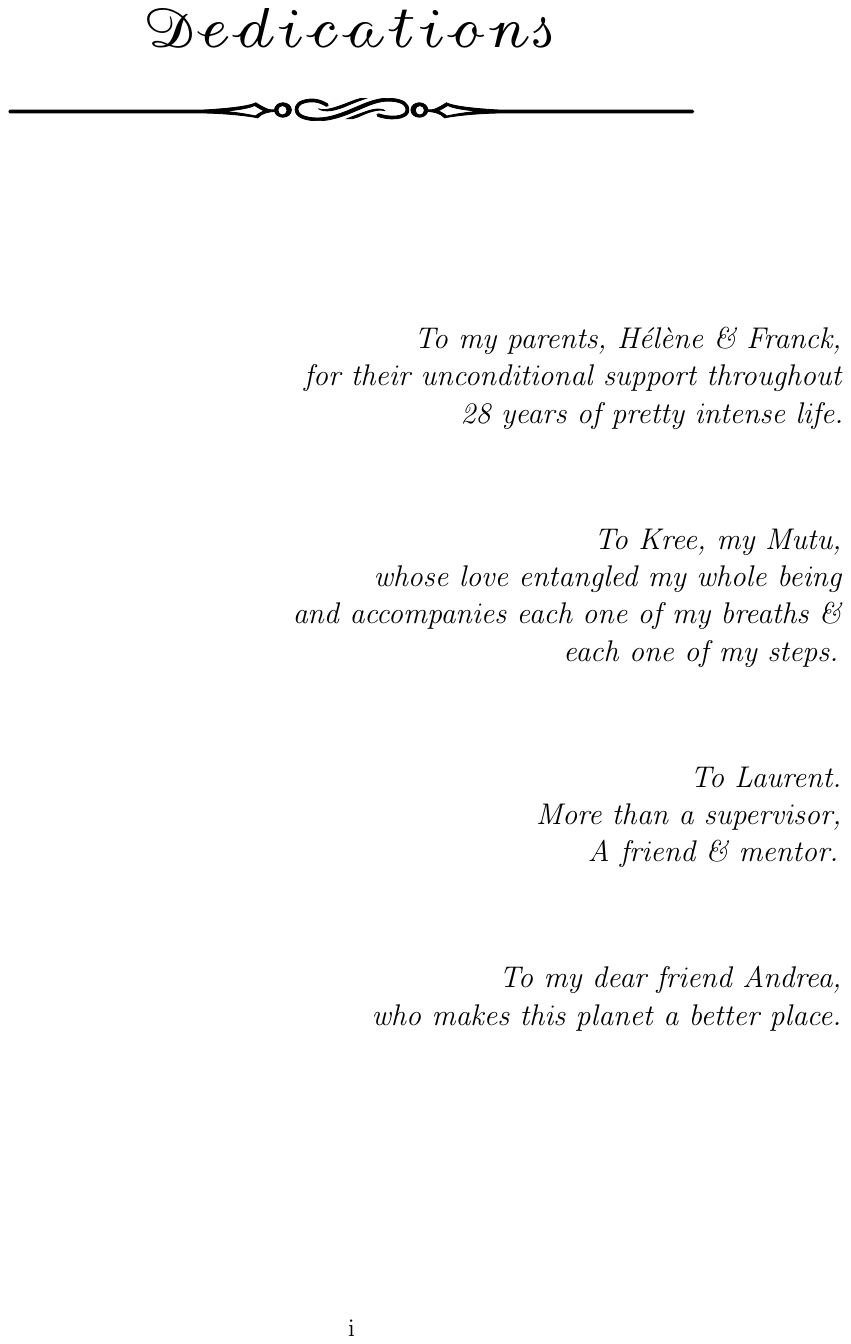}
};

\node[anchor=north west] (nw) at (chframe.north west)
{\pgfornament[scale=0.3]{10}};

\node[anchor=north east] at (chframe.north east)
{\pgfornament[symmetry=v,scale=0.3]{10}};

\node[anchor=south west] (sw) at (chframe.south west)
{\pgfornament[symmetry=h,scale=0.3]{10}};

\node[anchor=south east] (se) at (chframe.south east)
{\pgfornament[symmetry=c,scale=0.3]{10}};

\node[anchor=south west,xscale=8] at (sw.south east) {\pgfornament[scale=0.3]{32}};

\node[shift={(4.41mm,0cm)},anchor=south west,yscale=10.8,rotate=90] at (sw.north west) {\pgfornament[scale=0.3]{32}};

\node[shift={(-4.41mm,0cm)},anchor=south east,yscale=10.8,rotate=-90] at (se.north east) {\pgfornament[scale=0.3]{32}};

\node[anchor=north west,xscale=8] at (nw.north east) {\pgfornament[scale=0.3]{32}};

\end{tikzpicture}
\end{center}
\end{newfamily}

\restoregeometry
\chapter*{Abstract}
\addcontentsline{toc}{chapter}{Abstract}

This thesis deals with the phase space realization of asymptotic higher spin symmetries, in 4-dimensional asymptotically flat spacetimes.
These symmetries live on the conformal null boundary, namely null infinity $\scri$, and were first revealed few years ago via the study of conformally soft gluons and gravitons operator product expansions, in the context of celestial holography.
Connections with twistor theory and phase space realization followed soon after.
In the gravitational case  for instance, these symmetries generalize the BMS algebra to include an infinite tower of symmetry generators constructed from tensors on the sphere of arbitrary high rank $s$.
The bulk interpretation of these transformation parameters is still largely under investigation.

Building on a first series of results on their canonical representation, we develop the necessary framework to define Noether charges for all degrees $s$.
Importantly, we construct these charges out of a `holomorphic' asymptotic symplectic potential, such that we obtain an infinite collection of charges conserved in the absence of radiation.
The classical symmetry is then realized non-perturbatively and non-linearly in the so-called holomorphic coupling constant, generalizing the perturbative linear and quadratic approach known so far.
The infinitesimal action defines a symmetry algebroid which reduces to a symmetry algebra at non-radiative cuts of $\scri$.
The key ingredient for our construction is to consider field and time dependent symmetry parameters constrained to evolve according to equations of motion dual to (a truncation of) the asymptotic equations of motion in vacuum.
We expose our results for Yang-Mills, General Relativity, and Einstein-Yang-Mills theories.

This canonical analysis comes hand in hand with an in-depth study of the algebraic structure underlying the symmetry.
We reveal several Lie algebroid and Lie algebra brackets, which connect the Carrollian, celestial and twistorial realizations.
We show that these brackets are a deformation of the soft $sw_{1+\infty}$ celestial algebra, where the deformation parameters are the radiative asymptotic data.
On the one hand they allow us to define the symmetry algebra at non-radiative cuts of $\scri$, for arbitrary values of the asymptotic shear and gauge potential.
On the other hand, we can accommodate for radiation using the algebroid framework.

For the specific case of non-abelian gauge theory, we also investigate how the asymptotic expansion around null infinity of the full Yang-Mills equations of motion in vacuum can be recast in terms of the higher spin charge aspects.
Since the analysis of higher spin symmetries is for now inherent to a truncation of the latter equations of motion, this paves the way towards the understanding of the relevance of these symmetries in the full theory.

\paragraph{Keywords:}
Gravity, Yang-Mills, Non-Abelian Gauge Theory, Einstein-Yang-Mills, Asymptotic Symmetries, Asymptotically Flat Spacetimes, Flat Space Holography, Celestial Holography, Wedge Algebra, Covariant Wedge Algebra, $w_{1+\infty}$ algebra, $s$-algebra, $sw_{1+\infty}$ algebra, Lie Algebroid, Bicrossed Product, Semi-direct Product, Noether Charge, Charge Aspect, Canonical Representation, Poisson Bracket, Symplectic Potential, Radiation, Twistor, Good Cut Equation.

\chapter*{Acknowledgments}
\addcontentsline{toc}{chapter}{Acknowledgments}

From the bottom of my heart, I thank my parents. 
Without them, none of this would be possible.
To Kree, I can only say Dhanyabad.
Thank you Andrea for a wildly unexpected friendship which makes my days brighter.
I am immensely indebted to Laurent, whose enthusiasm, passion, mentorship, friendship, advices, stories, sense of humor, philosophy, insights, intuition, vision, sense of sharing, kindness, care, honesty, humility and wisdom made me grow into a better researcher, but more importantly into a better human being.
\medskip

I want to thank Lionel Mason, Romain Ruzziconi and Adam Kmec for the ten fruitful days of discussion while they were visiting the Perimeter Institute for the Celestial Summer School 2024, couple of weeks after the publication of their related work.
I thank Andrew Strominger for his remarks during the celestial school.
I am grateful to Atul Sharma and Roland Bittleston for insightful comments about twistor space.
Thanks Simon Pekar for pointing out a reference on chiral Poincaré.
I thank again Romain Ruzziconi, Lionel Mason and Adam Kmec, together with Daniele Pranzetti, Shreyansh Agrawal, Laura Donnay and Panagiotis Charalambous, for discussions during my visit respectively at the Mathematical Institute in Oxford and at SISSA in Trieste in December 2024.
\textit{Merci} to Marc Geiller together with Tom Wetzstein and Laurent Baulieu for the fruitful interactions while visiting ENS Lyon and LPTHE Paris respectively.
I thank Temple He and Jacob McNamara for our discussions while visiting Caltech and when Temple visited PI few months prior.
I am grateful to Sruthi Narayanan for her comments during the various stages of this thesis.
Besides, I would like to thank Céline Zwikel and Florian Girelli for their support during my PhD, as well as Rob Myers, Sabrina Pasterski and Kevin Costello for our various interactions.
\medskip

I give a special thank you to Hank Chen for countless hours of discussion and the numerous comments all along the development of this project.
And simply for being a great friend.
Thanks to my fellow students and friends Adri\'an Khalil Lopez, José Padua Argüelles, Christopher Andrew James Pollack and Cole Coughlin for always being open to chat.
\medskip

Last but not least, a big shout out to the Perimeter Institute Black Hole Bistro staff, without whom I would have withered long ago.
\medskip

Research at Perimeter Institute is supported by the Government of Canada through
the Department of Innovation, Science and Economic Development and by the Province of
Ontario through the Ministry of Colleges and Universities. 
This work was supported by the
Simons Collaboration on Celestial Holography.

\chapter*{Statement of contribution}
\addcontentsline{toc}{chapter}{Statement of contribution}

This thesis is based on \cite{Cresto:2024fhd, Cresto:2024mne, Cresto:2025bfo, Cresto:2025ubl} plus some unpublished notes from a former work of Laurent and I.
\cite{Cresto:2024fhd} has been published in \textit{Letter of Mathematical Physics} by Springer Nature.
The paper \cite{Cresto:2025bfo} is the result of own work.
The main body of the other three papers went through so many back and forth between Laurent and I that it is impossible to distinguish who wrote what.
The main point is that I contributed to the entirety of the content.
Concerning the calculations, my contribution goes approximately as follows: 50\% in \cite{Cresto:2024fhd}, 60\% of \cite{Cresto:2024mne} and 90\% in \cite{Cresto:2025ubl}.

\chapter*{List of Figures \& Tables}
\addcontentsline{toc}{chapter}{List of Figures \& Tables}

This thesis contains three figures:
\bigskip
\bigskip

\ni \ref{Fig:scri}: Penrose compactification for
asymptotically flat spacetimes.
\bigskip

\begin{wrapfigure}[10]{r}{5cm}
\vspace{-0.6cm}

\includegraphics[width=5cm]{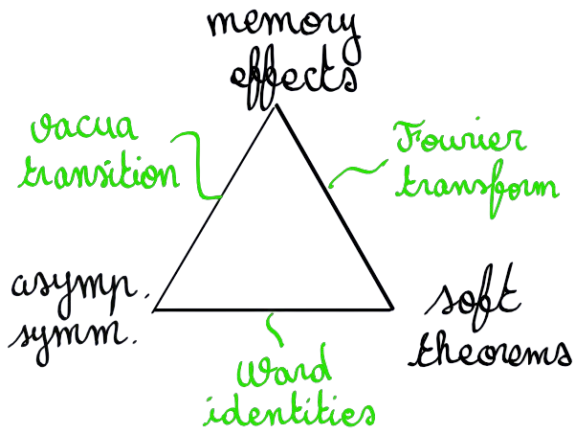}
\vspace{-0.6cm}

\caption{\textit{IR triangle \cite{Strominger:2017zoo}.} \label{Fig:IRtriangle}}
\end{wrapfigure}

\ni \ref{Fig:IRtriangle}: IR triangle.
\bigskip

\ni \ref{fig:softTheorem}: Soft theorem.
\bigskip

%\ni \ref{fig:algebroid}: Algebroid, algebra and dressing map.
\bigskip
\bigskip
\bigskip

\ni This thesis contains one table:
\bigskip
\bigskip

\ni \ref{table1}: Number of solutions to the wedge condition depending on the topology.

\newpage
\tableofcontents
\newpage

\pagestyle{fancy}
\pagenumbering{arabic}

\chapter{Introduction \label{secIntro}}

\section{Overview}

\textit{Symmetries...}
Symmetries are the \textit{raison d'être} of this thesis.
It is pretty humbling to realize that after more than a 100 years, we still do not fully understand what are the symmetries of General Relativity (GR).
Even the framework to ask that question in full generality is not well defined.
And even more surprisingly, we learned only about a decade ago that Maxwell's theory was more subtle than just global charge conservation and $U(1)$ redundancy.
The fact that a gauge transformation does not have to die off or be constant at infinity---contrary to what I had heard and read during my master in any field theory lecture---which implies that there exists a non-zero associated Noether charge and a \textit{local}\footnote{Local in the sense per direction on the sphere.} notion of electric charge conservation came as a shock.
Modern physics formulated as a gauge theory is thus not a split between local (synonym of redundancy) and global (synonym of conservation laws) symmetry parameters, but between zero and non-zero Noether charges.
And in order to be able to distinguish the two paradigms, the presence of \textit{corners}, namely co-dimension 2 surfaces, is essential, since this is where this new set of symmetry charges is supported.
If the Noether charge is non-zero---thus acting non-trivially on the phase space---for a subset of the local transformations,\footnote{This subset typically retains a sphere dependence.} the dimensionality of the resulting symmetry group becomes \textit{infinite}.
I had heard about the predictive power of 2 dimensional Conformal Field Theory (CFT) due to the enhancement of the usual conformal group to the infinite dimensional Virasoro, but this curiosity seemed disconnected from 4 dimensional spacetime.
The relevance of the corner, which is 2 dimensional in 4d, plus the fact that one also gets an infinite dimensional enhancement of the traditional global symmetry group are concepts I found intriguing.
Tie them together with the overarching idea of \textit{holography}, and you get the missing link to a new form of understanding of Nature.

Here I am referring to Asymptotically Flat Space (AFS) holography, which has several features I am particularly fond of.

Firstly, we are most likely living in an asymptotically de Sitter (dS) space, with a small positive cosmological constant. 
Thus flat space holography seems a natural step from the well-established AdS/CFT correspondence towards a potential dS counter-part.

Secondly, the goal is to understand to which extent the combination of \textit{corner data} with symmetries determine bulk dynamics. 
This means understanding how certain corner degrees of freedom allow to reconstruct bulk information via the constraints of the corner symmetries (recall: infinite dimensional).
This is an exciting task.

Thirdly, the endeavor is carried out from a bottom-up approach, a way of thinking I relate to in my daily life.
I find it humbling and appealing because moving forward is done out of necessity.
We are guided by Nature (or say, well tested theories), rather than our expectations and educated guesses.
Hence we infer \textit{universal} statements agnostic of any fundamental theory of Quantum Gravity (QG).

I purposefully mention the word `quantum' only now to emphasize that before all, we need a rework of GR at the classical level; and this comes with acknowledging that we do not have the classical theory under control yet.
The quantum cart shall thus not be put before the classical horse.
But yes, it goes without saying that pondering on the quantum gravitational path is a major part of the thrill.
This is therefore the fourth point.
Knowing the importance of representation theory to organize the Hilbert space, exploring the power of working with infinite dimensional symmetry groups is worth pursuing for this reason as well.

Fifthly, null physics is at the heart of the conversation.
This is especially due to the causal structure of flat space, which admits null boundaries at infinity (where past and future light rays end up), but it also more simply relates to the one to one correspondence between a corner and its causal domain of influence (or causal diamond).
Null lines are a profound\footnote{(and insane as soon as one thinks about it for too long)} feature of our world and should not be treated merely applying the motto ``Do the computation in Euclidean and Wick rotate'' (no denial of the power of analytic continuation here, just the thought that it can only lead us so far).

Sixthly, flat space holography addresses the elephant in the room of Quantum Field Theory (QFT), namely how to define the scattering $S$-matrix in the presence of long range interactions?
One has to deal with infrared (IR) physics and the QFT soft theorems have been recognized as the conservation laws associated to this infinite set of new symmetries. 
Remarkable.
Moreover, there are growing evidences that QG and the study of causal diamonds lead to a mix between ultraviolet (UV) and IR physics, with potentially observable consequences. 
This would signal a departure from the Effective Field Theory framework and would be a smoking gun that QG has the potential to cure QFT pathologies.

Seventhly, asymptotic symmetries (i.e. corner symmetries at infinity) manifest themselves through \textit{observable} memory effects.\footnote{A wave made of massless particles generically carries memory due to the presence and the scattering of soft (i.e. zero energy) modes.}
To be honest, I was very far from imagining (but for sure pleasantly astonished) that my research in theoretical physics could be related to an effect measurable by the Laser Interferometer Space Antenna (LISA) in a couple of decades.

Finally, the topic brings together aspects and communities of GR, scattering amplitudes, 2d CFT, twistor theory, Carrollian physics and experimental physics.
People try to build bridges and this feels wholesome and fruitful.
Besides, the encompassing and open-minded vision of Laurent is also a motivation by itself.

Hence, the main questions of interest that the framework of asymptotic symmetries can help to tackle are:
What are the fundamental symmetries of Nature?
What are the fundamental degrees of freedom?
How to define the gravitational $S$-matrix?
What are the observable signatures of Quantum Gravity?
\medskip

We now give a proper introduction to some of the aforementioned elements.

\section{Historical review \label{sec:IntroHis}}

\ni \uline{\textsc{Flat Space Holography:}}
\begin{wrapfigure}{r}{8cm}
\vspace{-0.3cm}

\includegraphics[width=8cm]{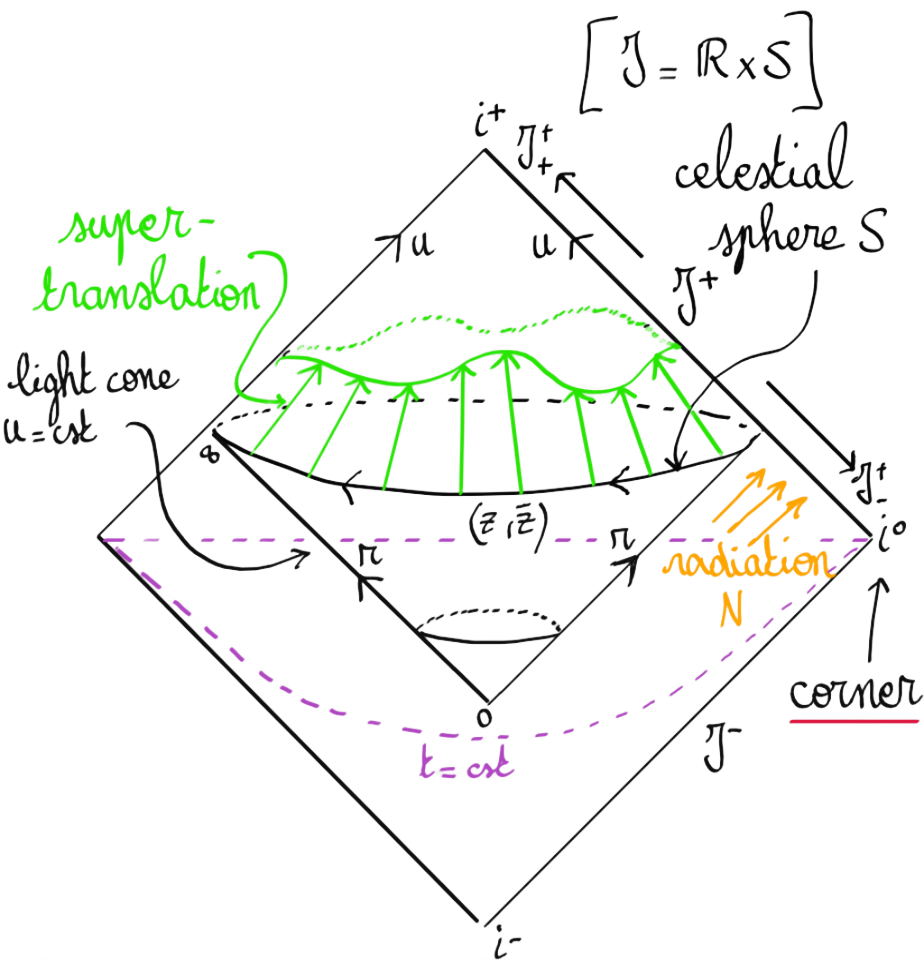}
\vspace{-0.5cm}

\caption{\textit{Penrose compactification for asymptotically flat spacetimes. $(u,r,z,\bz)$ are the Bondi coordinates.} \label{Fig:scri}}
\end{wrapfigure}
\ni For more than a century, Emmy Noether has guided us thanks to her theorems relating symmetries and conserved quantities \cite{noether1971invariant}.
Since the seminal work of Bondi, van der Burg, Metzner and Sachs \cite{Bondi:1962px, Sachs:1962wk}, we know that Asymptotically Flat Spacetimes (AFS) do not reduce to Minkowski space at infinity.
In particular, the symmetry group preserving the asymptotic structure was originally recognized to be the BMS group, namely an infinite dimensional enhancement of the Poincaré group where (null) time translations different at every point onto the sphere---called \emph{super-translations}---are physical symmetries on the null boundary $\scri$, see Fig.\,\ref{Fig:scri}.
Recently it was shown that it is necessary to consider extensions such as the extended BMS (eBMS) algebra \cite{Barnich:2010eb, Barnich:2011mi, Barnich:2016lyg}---where Lorentz rotations\footnote{SL$(2,\C)$ is also the group of global conformal transformations of the 2d sphere.} are replaced by Virasoro super-rotations---, the Generalized BMS (GBMS) algebra \cite{Campiglia:2014yka, Compere:2018ylh, Campiglia:2020qvc, Grant:2021sxk}---where Lorentz is further enhanced to diffeomorphisms of the sphere---or the Weyl BMS (BMSW) algebra \cite{Freidel:2021fxf, Freidel:2021qpz}---where GBMS is supplemented by Weyl rescaling of the sphere---and view BMS as a residual unbroken symmetry group \cite{Freidel:2024jyf}.  
These symmetry groups are infinite-dimensional, and the associated Noether charges depend on the point of the celestial sphere.
This led to the notion of charge \textit{aspect} which represents, through Noether's theorem, the charge density on the celestial sphere. 
It is now understood that asymptotic symmetries can be directly related to the broader concept of \emph{corner symmetries} \cite{Ciambelli:2022vot, Ciambelli:2025mex}, which appear as global symmetries associated with gauge symmetries \cite{Donnelly:2016auv, Freidel:2020xyx, Ciambelli:2021nmv}. 
These corner symmetries are associated with \textit{codimension 2 surfaces}, on which causal diamonds can be attached and which represent entangling surfaces \cite{Caminiti:2025hjq}.

To appreciate this paradigm of corner symmetries, recall that in any gauge theory, Noether's theorems not only tell us that there exists a conserved current\footnote{$\xi$ is a symmetry parameter and we use form notation for simplicity. $j_\xi$ is thus a co-dimension 1 form.} $j_\xi$ such that\footnote{Here we use $\approx$ to emphasize an equality valid on-shell of the equations of motion.} $\rd j_\xi\approx 0$, but also that this current reduces to a corner term (i.e. a co-dimension 2 form $q_\xi$) on-shell, namely that $j_\xi\approx\rd q_\xi$.
From there, we readily see that if the Cauchy surface $\Sigma$ on which we integrate the Noether current is boundaryless, i.e. $\pa\Sigma=\emptyset$, then the Noether charge $Q_\xi=\int_\Sigma j_\xi\approx 0$ vanishes on-shell, for any symmetry parameter $\xi$.
If however the boundary of $\Sigma$ is not empty, in other words $\pa\Sigma=S$, then $Q_\xi\approx \int_S q_\xi$.
Depending on the nature of $\xi$, this charge can either vanish or not.
If $Q_\xi=0$, $\xi$ is a so-called \emph{small} gauge transformation and corresponds to a genuine redundancy of the theory (since its action on physical states is trivial).
If $Q_\xi\neq 0$, $\xi$ is a so-called \emph{large} gauge transformation and corresponds to a genuine physical symmetry.
A super-translation $\xi\overset{\scri}{=}\T0(z,\bz)\pa_u$ is an example of this last category.
\textit{The set of asymptotic corner symmetries is then defined as the quotient of all gauge transformations by the small ones.}\footnote{One often treats this question by studying \textit{residual} gauge symmetries, i.e. imposing a partial gauge fixing first. 
For a gauge independent statement, see the BRST approach of \cite{Baulieu:2023wqb, Baulieu:2024oql, Baulieu:2024rfp}.}
In the context of AFS, we can take the limit of a $t=cst$ Cauchy slice, as depicted on figure \ref{Fig:scri}, to late times.
Assuming that we do not have any matter,\footnote{This is always true in this thesis.} especially massive matter, then the Cauchy slice reduces to\footnote{In the following we are only interested in quantities on $\scri^+$ and we shall usually drop the `$+$'.} $\scri^+$ (in other words there is no contribution from time-like future infinity $i^+$).
The boundary of the latter is made of two corners, $\scri^+_+$ and $\scri^+_-$, and we assume the boundary condition $\int_{\scri^+_+}q_\xi=0$ such that the Noether charge reduces to a corner integral at $\scri^+_-$.

Determining to which extent one can encode bulk information using corner data is the defining quest of \textit{flat space holography}, also branded \textit{celestial holography} \cite{Raclariu:2021zjz}.
In particular, of major interest over the past decade is the reformulation of the gravitational $S$-matrix in terms of correlators leaving on the celestial sphere \cite{Pasterski:2016qvg}. 
The dual theory to 4d AFS has been putatively called a \textit{Celestial Conformal Field Theory} and understanding its nature is the subject of tremendous investigations (we shall give specific references later in the introduction).
\textit{This thesis partakes in this endeavor, by studying the canonical representation of asymptotic higher spin symmetries and their algebraic structure.}\\

\ni \uline{\textsc{Infrared Triangle:}}

\begin{figure}[H]
    \centering
    \includegraphics[width=0.6\linewidth]{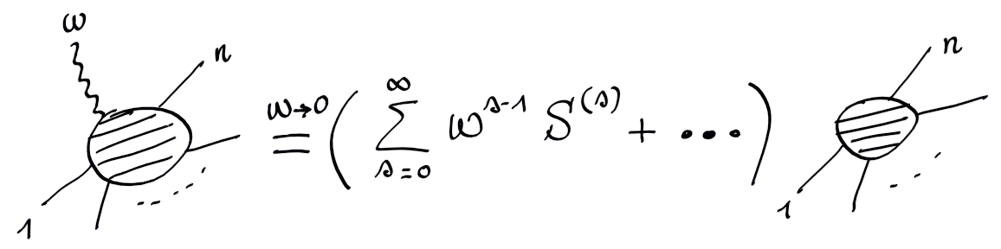}
    \caption{\textit{The soft theorem is a factorization property of scattering amplitudes containing at least one massless particle of energy $\omega$, when $\omega\to 0$. The leading soft theorem has the Weinberg pole \cite{PhysRev.140.B516} form $\cA_{n+1}\overset{\omega \to 0}{=}\big(\frac{1}{\omega}S^{(0)}+\ldots \big)\cA_n$. The leading soft factor $S^{(0)}$ only depends on the momenta of the $n$ external particles. A Taylor series in the energy can be extracted to all orders. The `$\ldots$' contain logarithmic, loop and theory dependent corrections. See the rest of the introduction for multiple references.}}
    \label{fig:softTheorem}
\end{figure}

\ni One achievement from a decade ago was the realization that the leading \textit{soft graviton theorem} \cite{PhysRev.140.B516}, cf.~Fig.\,\ref{fig:softTheorem}, could be reinterpreted as a Ward identity for the supertranslation charge (namely the Bondi mass aspect) \cite{Strominger:2013jfa, He:2014laa}. 
This relation between soft theorems and asymptotic charges is \textit{generic} and does not rely on the peculiarities of the gravitational analysis.
On top of that, the position space version of the leading soft graviton theorem was proven to be equivalent to the leading displacement memory effect\footnote{The fact that after the passage of a gravitational wave, two free falling masses end up at a different proper distance than the one they started at.} \cite{Strominger:2014pwa} (see \cite{braginskii1985kinematic, braginsky_gravitational-wave_1987, PhysRevLett.67.1486, PhysRevD.45.520, PhysRevD.46.4304} for the original work on displacement memory effect).
The main point is that an \textit{equivalence} between three seemingly disconnected topics related to infrared physics was unraveled for General Relativity. 
Nowadays known as \textit{infrared triangle}, it was then natural to look for a similar set of relationships in the case of gauge theories as well.

Besides, the Ward identity associated to an asymptotic symmetry can be phrased as the invariance of the $S$-matrix under an infinitesimal transformation generated by a charge which contains a soft ($\mathsf{S}$) and a hard ($\mathsf{H}$) part. 
In equations, this schematically gives 
\begin{equation}
    \bra{\mathrm{out}}Q_+ S-SQ_-\ket{\mathrm{in}}=0\quad\Leftrightarrow\quad  \bra{\mathrm{out}}\soft{Q}_+S-S\soft{Q}_- \ket{\mathrm{in}}=- \bra{\mathrm{out}} \hard{Q}_+S-S\hard{Q}_- \ket{\mathrm{in}}, \label{SmatrixInv}
\end{equation}
where $Q_{\pm}=\soft{Q}_{\pm}+\hard{Q}_{\pm}$.
$Q_+$ and $Q_-$ are respectively constructed on $\scrip$ and $\scrim$ and \eqref{SmatrixInv} thus states the conservation of the charge between past and future.\footnote{One assumes in particular the (potentially subtle) antipodal matching between $\scrim_+$ and $\scrip_-$ through spatial infinity $i^0$.}
The soft part, linear in the gauge field, only exists for local (i.e. not constant) large gauge transformations. 
It is related to the insertion of the soft mode while the hard part, quadratic in the gauge field, acts on the in- and out-states to create the soft factors appearing in soft theorems \cite{Strominger:2017zoo, Raclariu:2021zjz, Hamada:2018vrw}.

\ni \uline{\textsc{Asymptotic Symmetries versus Leading Soft Theorems:}}\\
The study of asymptotic symmetries for abelian and non-abelian gauge theories is much more recent that its gravitational counterpart which dates from the 1960' \cite{Bondi:1962px, Sachs:1962wk}.
This illustrates one of these situations where the quest for Quantum Gravity led to new discoveries and perspectives in non-gravitational theories.
Historically, in 2014, \cite{He:2014cra} established the equivalence between leading soft \textit{photon} theorem (see \cite{PhysRev.96.1428, PhysRev.96.1433, low1958bremsstrahlung, kazes_generalized_1959, yennie_infrared_1961, burnett1968extension} for the original references on the leading and sub-leading soft photon theorems) and the Ward identity for super-phase rotation $\Aa0(z,\bz)$---the so-called large gauge transformation of \cite{He:2014cra}.\footnote{See also \cite{Kapec:2015ena} for the generalization to massive charged particles.} 
Later that year, \cite{Lysov:2014csa} studied the implications of the subleading soft photon theorem and how it could be formulated as the Ward identity for a new set of charges associated to a vector onto the celestial sphere (see \cite{Laddha:2017vfh} for the generalization to massless particles and \cite{Hirai:2018ijc} for the extension to massive scalar QED and a discussion of the associated memory effects).
In our analysis, we shall denote this vector by $\Aa1(z,\bz)$.
The generalization to non-abelian gauge theories was developed in parallel:
\cite{Strominger:2013lka, He:2015zea} proved that one can re-interpret the leading soft \textit{gluon} theorem as the Ward identity for the super-phase rotation (in color space) symmetry.
The associated color memory effect was introduced in \cite{Pate:2017vwa} and the IR $S$-matrix was then further constrained by the discovery of the subleading soft gluon theorem \cite{Casali:2014xpa, Broedel:2014fsa, SabioVera:2014mkb}.
Exactly as in the abelian case, the latter can be recast as the Ward identity for a charge parametrized by the vector $\Aa1$ (now $\g$-valued, with $\g$ the gauge Lie algebra) \cite{Adamo:2015fwa}.

On the gravity side, on top of the original link between conservation of the super-translation charge and leading soft graviton theorem \cite{Strominger:2013jfa, He:2014laa}, Cachazo and Strominger uncovered a sub-leading and a sub-sub-leading soft theorem \cite{Cachazo:2014fwa}.\footnote{See \cite{Kalousios:2014uva, Zlotnikov:2014sva} for a generalization to higher dimensions.} 
\cite{Kapec:2014opa} then related the sub-leading soft factor to the Ward identity of super-rotations, i.e. of a meromorphic diffeomorphism of the sphere generating the Virasoro algebra.
In parallel, \cite{Campiglia:2014yka} showed the equivalence between the sub-leading soft theorem and the Ward identity this time for smooth diffeomorphisms of the sphere.
These two cases respectively correspond to the eBMS and GBMS asymptotic symmetry groups mentioned as the beginning of the introduction and their associated charge is the angular momentum aspect. 
We refer the reader to \cite{Compere:2018ylh} for a good summary of the subtle differences between the two cases.
For the analysis performed in this thesis, where we primarily focus on the generalization of (e/G)BMS to higher spin at the \textit{algebraic} level, these subtleties will not show up.
A vector onto the sphere will be then denoted by $\T1(z,\bz)$.
Nonetheless, the questions related to phase space extensions and Goldstone modes are definitely important. 
They are simply out of the scope of the present work.
Let us also point out that \cite{Pasterski:2015tva} worked out the spin memory effect related to the sub-leading soft theorem.

All the preceding results hold for the tree-level $S$-matrix. 
See \cite{He:2014bga} for one-loop correction to the subleading soft theorems and \cite{Feige:2014wja} for factorization properties of scattering amplitudes to all orders. 
See also \cite{Sahoo:2018lxl, Campiglia:2019wxe, AtulBhatkar:2020hqz, Sahoo:2020ryf} for the loop and logarithmic corrections to the soft theorems.

Despite the aforementioned relations between the most leading soft theorems and asymptotic charges, how to relate the charges associated to the sub-leading soft theorem in gauge theories, or the sub-sub-leading in GR, to asymptotic symmetries was not immediately transparent.
In 2016, Campiglia and Laddha \cite{Campiglia:2016hvg} proposed to interpret the sub-leading soft theorem as coming from the Ward identity associated to a charge parametrized by an over-leading, linear in $r$, gauge transformation---with $r$ the radial coordinate which tends to infinity on $\scri$.
In that context, the vector on the sphere previously considered by Strominger et al. then becomes the derivative onto the sphere of this over-leading gauge transformation.
Even if such a divergent gauge transformation does not preserve the asymptotic fall-offs of the gauge field and leads to divergent charges, there is still a natural finite part of the charge that one can relate to the sub-leading soft theorem.
The prescription is to neglect the divergent piece if the latter corresponds to the smearing of the \textit{leading} charge aspect with the divergent parameter.
This is an important point, also emphasized in \cite{He:2019pll}---where the analysis of \cite{Campiglia:2016hvg} is extended to the non-abelian case.
Indeed, this is because we know that the leading soft photon/gluon theorem holds and relates to the conservation of the leading charge that we can neglect the divergent contribution.
The fact that the divergent gauge transformation enters in a charge associated to a ``more leading'' soft theorem than the one associated to the finite part of the charge is key to get a finite Ward identity.\\

\ni \uline{\textsc{Tower of Asymptotic Symmetries and Tower of Soft Theorems:}}\\
The present work deals with asymptotic \textit{higher} spin symmetries, which in the context of Yang-Mills (YM) theory means that the symmetry parameters $\Aa{s}$ and the associated charges are labeled by their spin-weight $s\in\N$, hence generalizing the aforementioned $\Aa0$ and $\Aa1$.
In GR, $\T{s}$ generalizes $\T0$ and $\T1$ in a similar way. 
Their relevance became manifest in 2018, when \cite{Hamada:2018vrw} unraveled a set of Ward identities constraining the sub${}^s$-leading infrared structure of gauge theory and gravity.
This was understood \cite{Li:2018gnc} as the universal\footnote{On top of the loop and log corrections already mentioned, the sub${}^s$-leading soft theorems---$s\geq 2$ in gauge theory and $s\geq 3$ in gravity---are no longer universal and depend upon the details of the bulk interactions \cite{Li:2018gnc}.} tree-level part of sub${}^s$-leading soft theorems, thus giving information on the higher order soft behavior of gauge bosons and gravitons.
Thanks to the IR triangle, one knows that an interpretation of this result in terms of asymptotic symmetries should be possible.
In particular, \cite{Campiglia:2018dyi} conjectured\footnote{And showed for the sub-sub-leading case.} for massless QED how to relate these newly found sub${}^s$-leading soft theorems to the conservation of asymptotic charges associated to over-leading $\mathcal{O}(r^s)$ gauge parameters.
We expect the latter to be related to our parameters $\Aa{s}$ but this link is not the point of the present thesis.

We also refer the reader to \cite{Campiglia:2021oqz, Nagy:2022xxs, Nagy:2024dme, Nagy:2024jua} for an analysis of over-leading gauge transformations in YM theory and gravity and the definition of an extending phase space compatible with such transformations.\\

\ni \uline{\textsc{Celestial OPEs and Soft Algebras:}}\\
Importantly, asymptotic higher spin symmetries also appeared in the celestial holography context (see \cite{Pasterski:2021raf} and references therein for a review). 
There, the main objects of interest are celestial amplitudes/operators, obtained after a Mellin transform in energy of the usual momentum space amplitudes/operators. 
This transformation trades the energy dependence of the Fourier transform $\widetilde{O}(\omega,z,\bz)$ of the field $O(u,z,\bz)$ for a dependence in the conformal dimension\footnote{$\Delta$ together with the spin of the operator label SL$(2,\C)$ representations \cite{gelfand1966generalized}.} $\Delta$ as follows:
\begin{equation}
    O_\Delta(z,\bz)=\int_0^\infty\rd\omega \,\omega^{ \Delta-1}\widetilde{O}(\omega,z,\bz).
\end{equation}
The \textit{energetically} soft theorems described so far then morph into \textit{conformally} soft theorems, where the soft energy limit of the boson is replaced by a limit in the conformal weight of the associated conformal operator \cite{Pate:2019mfs}.
Equivalently, one can study conformally soft theorems and their Ward identities by considering 2-operators celestial operator product expansions (OPEs) where one operator is the soft current \cite{Fan:2019emx}.
The simple poles on the sphere appearing in celestial amplitudes and OPEs come from the collinear singularities in momentum space scattering amplitudes.
These collinear singularities can be studied on their own via Feynman diagrams.
However, one can bypass the diagrammatic approach and show that the 2-points celestial OPE coefficients are fully constrained by the asymptotic symmetries associated to the conformally leading and sub-leading soft gluon theorems (and similarly for GR including the sub-sub-leading order) \cite{Pate:2019lpp}.
In the same way \cite{Hamada:2018vrw, Li:2018gnc} discovered a whole tower of sub${}^s$-leading energetically soft theorems, \cite{Guevara:2021abz, Strominger:2021mtt, Himwich:2021dau} studied a tower of sub${}^s$-leading conformally soft currents, characterized by definite conformal weights,\footnote{The relation with the order $s$ of the soft theorems is $\Delta=1-s$.} whose OPEs satisfy respectively the $s$-algebra, $w_{1+\infty}$ algebra\footnote{Historically, $W_N$ algebras have been studied in the context of two-dimensional conformal field theories as extensions of the Virasoro algebra to higher conformal spin operators  \cite{zamolodchikov1995infinite,PseudoDiffBakas}.
The contraction (or classical limit) of $W_\infty$ to $w_{\infty}$ \cite{BAKAS198957}, together with the enlargement to $w_{1+\infty}$ can be interpreted in terms of area preserving diffeomorphisms \cite{Boyer_1985}. 
See \cite{Shen:1992dd} for a review of the mathematical and historical developments of $W_N$ algebras.} and $sw_{1+\infty}$ algebra\footnote{Technically the wedge sub-algebra of the loop algebra of these.} for pure YM, pure GR and Einstein-Yang-Mills (EYM) theory.
See also \cite{Himwich:2023njb} for an excellent account of the equivalence between momentum-space soft limits and emissions of particles of definite conformal weight; and how the $w_{1+\infty}$ soft algebra organizes the universal part of the infinite tower of soft theorems, both with massless and massive particles.\footnote{See \cite{Elvang:2016qvq, Melton:2022fsf} for the deformation of the soft gluon theorems and the associated soft algebra in the presence of a massless scalar field.}
Furthermore, the $w_{1+\infty}$ symmetry was found to be exact in self-dual gravity \cite{Ball:2021tmb}.\\

\ni \uline{\textsc{Asymptotic Higher Spin Symmetries:}}\\
Such symmetries have been referred to as higher spin symmetries because the symmetry generators—originally the Bondi mass aspect and the angular momentum aspect, generating respectively super-translations (spin 0) and sphere diffeomorphisms (spin 1 transformation)—included a spin 2 charge aspect. 
Allowing for such a spin 2 symmetry transformation then forces the inclusion of the whole infinite tower of positive spin in order to close the algebra.
This thesis focuses on the canonical asymptotic representation of the higher spin symmetry algebra at $\scri$. 
We follow the work of \cite{Freidel:2021dfs, Freidel:2021ytz, Freidel:2023gue, Geiller:2024bgf}, which formalized that there exist charge aspects of spin $s$ that one can build from the asymptotic gravitational data. These results generalized to higher spin the classic construction of \cite{Strominger:2013jfa, Strominger:2014pwa, Campiglia:2014yka, Campiglia:2015yka, Kapec:2014opa} for spin $0$ and $1$.
These charges are related to the sub$^s$-leading soft theorems and they are known to represent corner data that incorporate the necessary information to reconstruct the bulk spacetime \cite{Freidel:2021ytz, Geiller:2024bgf}.
They are also related to gravitational multipoles \cite{Compere:2022zdz}.
Besides, they furnish an operational description of General Relativity, where the metric components are recast as a set of non-commutative  observables satisfying a well-defined Poisson algebra. 
Moreover, they are associated to a higher spin generalization of the memory effect \cite{Pasterski:2015tva, Compere:2016gwf, Compere:2018ylh, Compere:2019odm, Nichols:2017rqr, Nichols:2018qac, Flanagan:2018yzh, Flanagan:2019ezo, Grant:2021hga, Grant:2023ged, Seraj:2021rxd, Seraj:2022qyt, Seraj:2022qqj, Faye:2024utu} (although the whole tower is not understood). 
This suggests that a proper non-perturbative understanding of the higher spin charges would then be invaluable for quantization.
Finally, the soft components of these charges and their dual can be used to construct a discrete asymptotic basis of states \cite{Freidel:2022skz, Cotler:2023qwh}.\\

\ni \uline{\textsc{Twistor:}}\\
The loop extension of $w_{1+\infty}$ denoted $Lw_{1+\infty}$ appears naturally in the twistorial approach of gravity as canonical transformations of the 2-dimensional fibers of the twistor fibration over the celestial sphere \cite{Dunajski:2000iq, Adamo:2021lrv, Mason:2022hly}. 
In the original Penrose's construction, self-dual spacetimes \cite{Penrose:1976js} are recast in terms of complex deformations of twistor space. 
Moreover, these deformations were obtained by applying a canonical transformation of the fiber on the overlap between the two coordinate patches. 
The fact that the topology of the base is $\C^*$ leads to a parametrization of the twistor deformations in terms of $Lw_{1+\infty}$. 
In this perspective, the latter algebra therefore parametrizes self-dual spacetimes. 
Recently, the connection between twistor and celestial OPE \cite{Adamo:2021zpw} was achieved through the study of a twistor sigma model for self-dual gravity developed in \cite{Adamo:2021bej,Adamo:2021lrv, Mason:2022hly}.
More recent developments involving $Lw_{1+\infty}$ on the twistor side involve Moyal deformations \cite{Bu:2022iak}, the connection with the Carrollian perspective \cite{Mason:2023mti} and,  quite remarkably, the relevance of $Lw_{1+\infty}$ as a symmetry algebra for de-Sitter amplitudes 
\cite{Taylor:2023ajd, Bittleston:2024rqe}.

Note that the direct proof of the equivalence between the twistorial and canonical realization of the $Lw_{1+\infty}$ symmetry was first given in \cite{Donnay:2024qwq} at the quadratic order in $\GN$. 
Recently in \cite{Kmec:2024nmu}, this connection was established further to all order.\\

\ni \uline{\textsc{The Freidel-Pranzeti-Raclariu (FPR) Construction and Results (in brief):}}\\
To understand the origin of this thesis, we first give a brief summary of \cite{Freidel:2021dfs, Freidel:2021ytz, Freidel:2023gue}:
FPR identified a particular truncation in the asymptotic Einstein's equations and YM equations of motion (EOM) in which the Weyl tensor, respectively field strength, was recast as a collection of higher spin fields $\tQ_s$.
The latter were shown to satisfy a simple set of evolution equations, which represents the input of the analysis of the current work, cf.\,\eqref{intro1} and \eqref{intro1YM}.
The conservation of these higher spin `charges' (i.e. the fact that they commute with the $S$-matrix) truncated to quadratic order was then equivalent to the infinite tower of conformally soft theorems.
In this approach, the label `charges' was justified as the (renormalized version of) $\tQ_s$ reproduced the celestial $w_{1+\infty}$ and $s$-algebras generated by the increasingly subleading soft modes of \cite{Guevara:2021abz, Strominger:2021mtt}, thus furnishing a phase space representation of this tower of symmetries.
Nevertheless, a derivation à la Noether for the entire tower of charges was missing, as well as a non-perturbative treatment.
This thesis remedies to this in the case of YM theory, General Relativity and finally EYM theory.

\section{The FPR construction and results \label{sec:FPR}}

We now review the procedure of \cite{Freidel:2021dfs, Freidel:2021ytz} in details.
The starting point was to consider a set of spin-weighted functionals $\tQ_s(u,z,\bz)$, $s\geqslant -2$, on $\scri$.\footnote{We use the $~\widetilde{~}~$ as a reminder of the $(u,z,\bz)$ dependence.}
It had been proven in \cite{Freidel:2021qpz} that a certain combination of them transformed homogeneously under the asymptotic BMSW algebra \cite{Freidel:2021fxf}, the symmetry algebra of null infinity, at least till spin 2.
These combinations took the form of evolution equations, with the simple recursive pattern
\begin{equation}
    \pa_u\tQ_s=D\tQ_{s-1}+(s+1)C\tQ_{s-2}, \qquad s\geqslant -1,\label{intro1}
\end{equation}
where $C$ is the asymptotic shear, $\tQ_{-2}=\pa_u\bN$, $\tQ_{-1}=D\bN$ and $\bN=\pa_u \bC$ is the news tensor.
$\tQ_{-2}$ and $\tQ_{-1}$ thus encode the radiation.
For $s\leq 3$, these are the \textit{exact} asymptotic EE \cite{Newman:1968uj, Freidel:2021qpz, Freidel:2021ytz}. 
For instance, the $s=0$ equation can  be recast into the Bondi mass loss formula.
The first five $\tQ_s$, i.e. $-2\leq s\leq 2$, are actually the leading order components in the asymptotic expansion\footnote{Namely the expansion in $1/r$: $\displaystyle{\Psi_{2-n}(u,r,z,\bz)=\sum_{s=0}^\infty \frac{\Psi^{(s)}_{2-n}}{r^{3+n+s}}}$.} of the Weyl scalars.
In other words, in the Newman-Penrose (NP) notation, $\Psi_{2-s}\approx\tQ_s/r^{3+s}+ \cdots$.
The rest of the tower $\tQ_s$ relates to $\Psi_0^{(s)}/r^{s+5}$; see \cite{Newman:1968uj} for the original NP work but also \cite{Barnich:2019vzx, Freidel:2021ytz} and recently \cite{Geiller:2024bgf} for the most up-to-date account. 
We also discuss the equivalent construction for Yang-Mills in chapter \ref{secNAGTI}.

Notice that the charge aspect $\tQ_s$ depends linearly on $\bN$ and polynomially\footnote{In a general sense since the coefficients of the polynomial can be differential operators.} on $C$. 
This means that we can decompose $\tQ_s$ for $s\geqslant -2$ as\footnote{This condition is valid for $s\geqslant -2$, with $\lfloor n\rfloor$ the floor function.
To see it, just notice that the initial data $\tQ_{-2}$ and $\tQ_{-1}$ are of degree 0 in $C$ and afterwards $\deg[\tQ_s]=\deg[\tQ_{s-2}]+1$.}
\begin{equation} \label{tQsk}
\tQ_s =\sum_{k=0}^{\lfloor s/2\rfloor+1} \tQ_s^{\scriptscriptstyle{(k)}},
\end{equation}
where $\tQ_s^{\scriptscriptstyle{(k)}}$ is homogeneous of degree $k$ in $C$ and linear in $\bN$ (due to the initial condition $\tQ_{-2}$).
The charge aspect $\tQ_s^{\scriptscriptstyle{(0)}}\equiv\soft{\tQ_s}$ is the soft charge while $\tQ_s^{\scriptscriptstyle{(1)}}\equiv\hard{\tQ_s}$ is the hard charge and $\sum_{k=2}^{\lfloor s/2\rfloor+1} \tQ_s^{\scriptscriptstyle{(k)}}$ is the super-hard contribution.
Independently rescaling $C\to\gN C$ and $\bC\to\bgN\bC$ by dimensionless `holomorphic' and `anti-holomorphic' couplings $\gN$ and $\bgN$, we see that $\tQ_s^{\scriptscriptstyle{(k)}}= \mathcal{O}(\bgN\gN^k)$.
As we are about to explain, the treatment of FPR was perturbative in $\gN$, since only involved $\tQ_s^{\scriptscriptstyle{(0)}},\tQ_s^{\scriptscriptstyle{(1)}}$ and $\tQ_s^{\scriptscriptstyle{(2)}}$.
\textit{One achievement of the present thesis is to be able to deal with the `non-perturbative' (in $\gN$) object $\tQ_s$ rather than its perturbative expansion $\tQ_s^{\scriptscriptstyle{(k)}}$.}

For spins  higher than $3$, \eqref{intro1} represents a truncation of EE on $\scri$.\footnote{The conjecture is that it corresponds to the part of Einstein's equations linear in $\bgN$.
The statement is however more subtle than it may seem, cf. \eqref{Qnevol}.
From a twistor perspective though, \eqref{intro1} arises from self-dual gravity.}
Nevertheless, studying this fundamental pattern, FPR were able to construct, after renormalization, a notion of higher spin \textit{charges} $q_s(\T{s})$,\footnote{$q_s(\T{s}):=\frac{1}{4\pi\GN}\int_{\scrip_-}\tilde{q}_s\T{s}$.} where the $\T{s}(z,\bz)$ are smearing (spin-weighted) functions on the sphere that play the role of symmetry parameters (recall that $\T0$ and $\T1$ corresponding to a super-translation and a sphere diffeomorphism respectively).
The reason why $q_s(\T{s})$ was called a charge is that on one hand it is conserved in the absence of radiation (at the linear level in \cite{Freidel:2021ytz} and then corrected at the quadratic order $\mathcal{O}(\bgN\gN)$ in \cite{Geiller:2024bgf}\footnote{See their section 6.4.2 for subtle points of the discussion.
\cite{Geiller:2024bgf} also makes very transparent that the procedure to build $\tilde{q}_s$ order by order is particularly cumbersome.
In chapter \ref{secGRII}, we prove a systematic approach to construct it.}) while on the other hand it satisfies the algebra 
\begin{equation}
    \poisson{q_s^{\scriptscriptstyle{(0)}}(T_s),q_{s'}^{\scriptscriptstyle{(1)}}(\Tp{s'})}+ \poisson{q_s^{\scriptscriptstyle{(1)}}(T_s),q_{s'}^{\scriptscriptstyle{(0)}}(\Tp{s'})} =q^{\scriptscriptstyle{(0)}}_{s+s'-1}\big((s+1)\T{s}D\Tp{s'}-(s'+1)\Tp{s'}D\T{s}\big). \label{intro2}
\end{equation}
Note that the left hand side (LHS) can also be written as\footnote{Using the Ashtekar-Streubel \cite{ashtekar1981symplectic} Poisson bracket on null infinity, namely $\poisson{\bN(x),C(x')}=4\pi\GN\delta^3(x-x')$, where $x=(u,z,\bz)$.} 
$\poisson{q_s(T_s),q_{s'}(\Tp{s'})}^{\scriptscriptstyle{(0)}}$ where the superscript `(0)' emphasizes the fact that the computation is valid for the soft part of the Poisson bracket.

On top of satisfying the soft algebra \eqref{intro2}, the second main result of FPR was to prove that the action of the tower of renormalized hard charge aspects $\tilde{q}_s^{\scriptscriptstyle{(1)}}$ on the asymptotic gravitational phase space was related to celestial OPEs with conformally soft modes \cite{Guevara:2021abz, Strominger:2021mtt, Himwich:2021dau}.
To say it differently, by assuming the invariance of the $S$-matrix under the asymptotic charges (truncated at quadratic order), and their conservation at spacelike infinity via antipodal matching (see \cite{Strominger:2013jfa, ashtekar_angular_1979, Prabhu:2019fsp, Prabhu:2021cgk} for a similar discussion about the BMS algebra), FPR could relate the Ward identities for $\tilde{q}_s$ with the tower of sub$^s$-leading conformally soft theorems.

Importantly, the combination of $\T{s}$ and $\Tp{s'}$ on the RHS of \eqref{intro2} is a shifted Schouten-Nijenhuis (SN) bracket for symmetric contravariant tensors on a 2-dimensional manifold $S$ \cite{Schouten, SNbracket}. 
We shall come back on this property in great details in chapter \ref{secGRI}.
When $S=\C^*$, one recovers the celestial holography $Lw_{1+\infty}$ algebra constraining the structure of the soft gravitons OPEs.
Although at leading order in perturbation around flat space, the result \eqref{intro2} was already a first step towards the understanding of the canonical realization of these higher spin symmetries.

The analysis and the results are similar for YM, treated in \cite{Freidel:2023gue}.
The counter-part of \eqref{intro1} takes the form
\begin{equation} \label{intro1YM}
    \pa_u\tQ_s=D\tQ_{s-1}+\big[A,\tQ_{s-1}\big]_\g,
\end{equation}
where $A$ is the asymptotic gauge field onto the sphere (playing a similar role as the shear) and the $\tQ_s$ are now related to the asymptotic expansion of the Newman-Penrose scalars built from the field strength.
Since the chapter \ref{secNAGTI} is dedicated to an in-depth review, demonstration and generalization of \eqref{intro1YM}, let us simply mention that FPR built YM charges satisfying the following algebra
\begin{equation}
    \poisson{q_s(\Aa{s}),q_{s'}(\Aap{s'})}^{\scriptscriptstyle{(0)}}=q^{\scriptscriptstyle{(0)}}_{s+s'}\big([\Aa{s},\Aap{s'}]_\g\big). \label{intro2YM}
\end{equation}
The $(z,\bz)$ mode expansion of the symmetry parameters $\Aa{s}$ then allows to recover the celestial $s$-algebra of conformally soft gluons OPE \cite{Guevara:2021abz}.
Note that we can use the same convention as in gravity with the superscript `$(k)$', where now $\tQ_s=\sum_{k=0}^{s+1}\tQ_s^{\scriptscriptstyle{(k)}}$.
This means that one can rescale the gauge field $A\to g A$ and $\bA\to\bar g A$ in order to get a `holomorphic' expansion in $g$ of the charges (which are still linear in $\bar g$).

Another point of discussion when dealing with higher spin symmetries concerns the potential restrictions imposed on the symmetry parameters $\T{s}$ and $\Aa{s}$. 
Most of the analysis done prior to the canonical one imposed a restriction called the  \textit{wedge} condition \cite{Guevara:2021abz}.
From the celestial CFT point of view, super-rotations are described by the $\ebms$ algebra, i.e. by two copies\footnote{Respectively for the holomorphic and anti-holomorphic diffeomorphisms.} of the Virasoro algebra, $\mathrm{Vir}\times\overline{\mathrm{Vir}}$.
Classifying fields and generators in terms of representations of $\mathrm{Vir}\times\overline{\mathrm{Vir}}$ is still a challenge.
The approach has been to reduce one of the two factors to its global sub-algebra, namely $\mathfrak{sl}(2,\R)$ when working in split (2,2) signature \cite{Atanasov:2021oyu, Guevara:2021abz}.
This truncation induces a finite dimensional multiplet structure of the modes of the soft currents, which can be recast (via the natural pairing of currents with symmetry parameters) as a condition on the parameters $\T{s}$ and $\Aa{s}$.
In our language, it reads $D^{s+2}\T{s}=0$ for GR and $D^{s+1}\Aa{s}=0$ for YM.
The wedge restriction is also a natural consequence of the twistorial treatment of $Lw_{1+\infty}$ \cite{Dunajski:2000iq, Adamo:2021lrv, Mason:2022hly}.
The wedge then amounts to a (anti-)holomorphicity condition on the transformation parameters in twistor space.
This restriction did not seem mandatory from the canonical perspective and thus required clarification.
\textit{A proper algebraic understanding of the wedge is a key aspect that we develop in this thesis.}

The next major step in the canonical approach to asymptotic higher spin symmetries was the computation of the algebras \eqref{intro2} and \eqref{intro2YM} at quadratic order for GR and Yang-Mills theory. 
For GR, the computation was done inside the wedge for arbitrary spins $s,s'$ and outside the wedge for $s=s'=2$.
For YM, no wedge condition was imposed and the calculation was valid for any spin.
Showing the consistency of the algebra beyond the wedge demanded a tremendous amount of calculation carried out in \cite{Freidel:2023gue}, and based on computational techniques developed in \cite{Freidel:2022skz}.
The fact that the gravitational algebra closed without any modifications of the aforementioned shifted SN bracket was so non-trivial that it suggested that a \textit{non-perturbative} description should exist.
Likewise, for YM, the Lie algebra bracket $[\cdot\,,\cdot]_\g$ of \eqref{intro2YM} did not receive any correction at quadratic order, even without imposing the wedge condition $D^{s+1}\Aa{s}=0$.
Besides, notice that these computations necessitated the cubic charge $q_s^{\scriptscriptstyle{(2)}}(T_s)$ (and $q_{s'}^{\scriptscriptstyle{(2)}}(\Tp{s'})$) to conspire with the soft charge $q_{s'}^{\scriptscriptstyle{(0)}}(\Tp{s'})$ (and $q_s^{\scriptscriptstyle{(0)}}(T_s)$) so that the Poisson brackets of those would precisely recombine with the Poisson bracket $\poisson{q_s^{\scriptscriptstyle{(1)}}(T_s),q_{s'}^{\scriptscriptstyle{(1)}}(\Tp{s'})}$ to close the algebra (see also \cite{Hu:2022txx, Hu:2023geb}).
\textit{The essential role of the cubic charge suggested that the higher spin symmetry should admit a \textit{non-linear realization} onto the phase space, a result that we shall indeed formalize at arbitrary high order in the holomorphic coupling constants $\gN$ and $g$.}

To conclude this summary, we want to give credit to Geiller \cite{Geiller:2024bgf}, who disentangled some nasty aspects of the computations mentioned throughout this exposition.
Let us add that the analysis of \cite{Geiller:2024bgf} was then reproduce for Einstein-Yang-Mills in \cite{Agrawal:2024sju}.
The reader who is at least familiar with the FPR-G series of papers will see a change of paradigm in the way we treat the asymptotic higher spin algebras.
This change of paradigm should bring a sense of clarity on some of the subtleties of the topic, especially at the algebraic level and in the canonical representation, but also in the relation to twistor theory.
The implications and the relation of our work to celestial OPEs are beyond the scope of this thesis. 
See however \cite{Freidel:2021dfs, Hu:2022txx, Hu:2023geb, Ball:2023sdz, Guevara:2024ixn, Jorstad:2025qxu} for multi-particles and multi-collinear contributions.

\section{Summary of results}

\ni \uline{\textsc{Original Motivation of this Work (Part I):}}\\
The first motivation for this thesis was to better understand the origin of the fundamental pattern \eqref{intro1YM} and the truncation of the equations of motion it corresponds to.
In other words, we wanted to analyze the corrections to \eqref{intro1YM} and how they impact celestial OPEs.
While we do have some promising preliminary results, the story has been more subtle than expected.
To avoid too many digressions, we limit ourselves to the presentation (without any analysis) of the full YM EOM in terms of higher spin charge aspects.
This is a result in itself, which we have not yet published elsewhere.\\

\ni \uline{\textsc{Original Motivation of this Work (Part II):}}\\
Although there is evidence, for spin $2$ only in the gravitational case, that the higher spin charges could be understood as overleading diffeomorphism changing the boundary condition \cite{Campiglia:2016jdj, Campiglia:2016efb, Horn:2022acq}, a fully consistent and non-perturbative Noether analysis is missing. 
In other words, the charges have \textit{not} been built as Noether charges derived from a symmetry action on the gravitational phase space, but from consistency with the OPEs and soft theorems. 
Indeed, the derivation of \cite{Freidel:2021ytz} related the Ward identities for the spin $s$ charge aspects $\tQ_s$ (or their renormalization)  with the tower of sub$^s$-leading conformally soft theorems (see \eqref{SmatrixInv} and \cite{Strominger:2017zoo, Hamada:2018vrw, Raclariu:2021zjz}---plus references therein---for the interpretation of soft theorems as conservation laws).

The goal of this thesis is to remedy this situation and give a first principle phase space derivation of the tower of higher spin charges from the Noether's theorem applied to the gravitational phase space.\footnote{Understand the logic of the motivation as: if we can figure it out for GR, then YM and EYM will follow.}
Understanding the tower of higher spin charges in a systematic, non-perturbative way, where one can leverage Noether's theorems, is primordial to determine the full set of symmetries of GR (and YM) and understand to which extent they constraint the gravitational dynamics and the quantum $S$-matrix \cite{Strominger:2013jfa, ashtekar_angular_1979, Prabhu:2019fsp, Prabhu:2021cgk, Strominger:2017zoo, Hamada:2018vrw, Raclariu:2021zjz}.

The purpose of this project is therefore to show that the higher spin symmetries can be realized canonically and non-perturbatively (in $\gN$) via Noether charges. 
One of the many challenges one faces is that the action of the symmetry generators on the gravitational (resp. YM) phase space is \emph{non linear} starting at spin 2 (resp. spin 1) and this non-linearity grows with the spin, cf.\,\eqref{tQsk}.  
The fact that this non-linear action closes into an algebra up to quadratic order was shown in \cite{Freidel:2021ytz, Freidel:2023gue}. 
These promising results looked rather miraculous and were due to the combination of highly non-trivial hypergeometric identities. 
Quite remarkably, the closure of the algebra  beyond the wedge required to include the commutator with the cubic order charges, beyond the quadratic hard charges. This suggests that the validity of the higher spin symmetry algebra controls some of the non-linearity of Einstein's equations---see also \cite{Hu:2022txx, Hu:2023geb}.
The validity of the algebra is so non-trivial that it suggested that a \textit{non-perturbative} description should exist.
This was the second original motivation for the present work.\\

\ni \uline{\textsc{Present Work:}}\\
We give a summary of the main results which are conceptually common to the YM, GR, and Einstein-Yang-Mills analyses.
Each subsequent chapter also includes a more specific introduction.
To avoid repetition, we present few major equations written directly in the EYM case.
Of course in practice, the latter is the most complicated in terms of computations.
Yet, conceptually, all three analyses have a common rationale.

\begin{itemize}
    \item[$\star$] We identify a higher spin symmetry algebroid\footnote{$\cS$, $\A$ and $\cST$ are respectively the YM, GR and EYM algebroids.} $\cST$.
    The ``structure constants'' of the associated algebroid $\sfST$-bracket depend explicitly on the shear $C$ and the asymptotic gauge potential $A$ (as well as its time derivative $F=\pa_u A$).
    This bracket is proven to satisfy the Jacobi identity. 

    \item[$\star$] We show that the non-linearity and non-locality of the action of higher spin symmetries on the EYM phase space can be recast into a simple set of equations of motion for the time dependent transformation parameters $\Pta\equiv(\t{},\a{})$, while the action itself involves only spin $0$, spin $1$ and spin $2$ generators.

    \item[$\star$] This radical simplification allows to construct, \textit{non-perturbatively} in $g$ and $\gN$, the canonical representation of the higher spin generators. 
    The fact that these generators satisfy the algebra non-perturbatively then follows from the power of Noether's theorem and the covariant phase space formalism \cite{kijowski1976canonical, Wald:1999wa, crnkovic1987covariant, ashtekar1991covariant, lee1990local, barnich1991covariant, Freidel:2020xyx, Freidel:2021cjp}.

     \item[$\star$] We establish that the renormalization procedure developed (perturbatively) in \cite{Freidel:2021ytz, Geiller:2024bgf} for the charge aspects simply amounts to evaluating the transformation parameters in terms of their initial conditions $(\T{},\Aa{})$.

     \item[$\star$] We relate all the preceding elements to twistor space.

     \item[$\star$] From the general $\cST$-algebroid, we define a symmetry algebra which preserves the asymptotic data at a non-radiative cut of null infinity.
\end{itemize}

Let us now flesh out some of these items with equations.
One of the key ingredients of the new paradigm is to introduce Carrollian smearing parameters $\t{s}(u,z,\bz)$ and $\a{s}(u,z,\bz)$ on $\scri$, of spin $-s$. 
These parameters are dual to the GR and YM spin $s$ charge aspects.
For instance in gravity, the pairing $\QGR{s}[\t{s}]$ defines a scalar charge via integration on the sphere (denoted with $[\,\cdot\,]$ in that context). 
One can then form the object of central interest, called \textit{master charge}, and defined as follows:
\begin{equation}
    \QPta^u:=\sum_{s=-1}^\infty\QGR{s}[\t{s}]+ \sum_{s=0}^\infty\QYM{s}[\a{s}].
\end{equation}
We constraint the time evolution of $\Pta$ to follow the \textit{dual equations of motion}\footnote{$[\cdot\,,\cdot]_\g$ denotes the YM Lie algebra commutator.}
\bs \label{introDualEOM}
\begin{align} 
    \pa_u\t{s}&=\big(\DGR\tau\big)_s :=D\t{s+1}-(s+3)C\t{s+2}, \\
    \pa_u\a{s}&=\big(\DTOT\Pta\big)_s:=D\a{s+1}+[A,\a{s+1}]_\g-(s+2)C\a{s+2}-(s+2)F\t{s+1},
\end{align}
\es
for $s\geq 0$. 
The RHS of the evolution equations \eqref{introDualEOM} reveals a generalization of the covariant derivative on the sphere, which depends explicitly on the asymptotic data and plays an essential role in our construction. 
The derivatives $\DGR$ and $\DTOT$ encompass the algebraic non-linearity due to the presence of a non-vanishing shear and of the YM self-interaction.

These evolution equations ensure that the master charge is \textit{conserved in time} in the absence of left-handed radiation $\bN=0=\bnews$. 
This is essential in our reasoning.
Indeed, by turning off the flux of radiation through $\scri$, one should be able to construct a conserved quantity.
We precisely define the dual EOM such that the master charge is this conserved quantity.
We can then show that the limit of the master charge to the corner $\scrip_-$, i.e. $\QPta\equiv\QPta^{-\infty}$ is a Noether charge built from the \textit{holomorphic} Ashtekar-Streubel asymptotic symplectic potential.
Denoting $\Omega$ the symplectic form, then $\QPta$ satisfies the Hamiltonian flow equation 
\begin{equation}
    I_{\dPta}\Omega=-\delta\QPta
\end{equation}
and generates the higher spin symmetries action
\begin{equation}
    \poisson{Q_{(\t{},\a{})},O}= \delta_{(\t{},\a{})}O,
\end{equation}
where $O$ is a functional of $(C,\bC,A,\bA)$.
The infinitesimal symmetry transformations of the asymptotic phase space variables are given by\footnote{$N=\pa_uC$ is the news.}
\bs \label{dtauCintro}
\begin{align}
    \dta C &\equiv\dt C=N\t0-D^2\t0+2DC\t1+3CD\t1-3C^2\t2, \\
    \dta A &=F\t0-D\a0-[A,\a0]_\g+C\a1.
\end{align}
\es

This canonical action, supplemented by the dual EOM, reproduces the linear and quadratic actions of higher spin symmetries derived in \cite{Freidel:2021ytz, Freidel:2023gue}.
Notice two important features about \eqref{dtauCintro}: 
Firstly, when written in terms of $\tau$, the symmetry is realized non-linearly due to the presence of the $C^2$ term associated with the spin $2$ symmetry.
Secondly, the symmetry transformation of the shear contains only the parameters $\t0,\t1$ and $\t2$.
However, they themselves contain arbitrary high spin symmetry generators when expressed in terms of the celestial symmetry parameters given by the initial value $\T{s}:=\t{s}(u=0)$.
When written in terms of the latter, the associated symmetry transformation $\delta_{\tau(\T{s})}C$ becomes polynomial in $C$ with a degree growing linearly with $s$ and also non-local along the null time direction $u$.
The reasoning is the same for $\dPta A$.

Remarkably, we also prove that the commutator of $\dPta$ closes thanks to the validity of the dual EOM. 
This means that there exists a bracket $\lbr\Pta,\Ptap\rbr$ such that
\begin{equation}
    \big[\dta,\dtap\big]\cdot=-\delta_{\lbr (\tau,\a{}),(\tau'\mkern-3mu,\ap{})\rbr}\cdot.
\end{equation}
This is the aforementioned $\sfST$-bracket which encodes the commutator of Noether charges and symmetry transformations on the EYM phase space.
It includes a pure GR and a pure YM part, while incorporating the effect of the coupling in terms of a bicrossed product structure.

The $\sfST$-bracket can be cast in many equivalent forms.
We give one of them:
\begin{equation}
    \lbr\Pta,\Ptap\rbr=[\Pta,\Ptap]+\dPtap\Pta-\dPta\Ptap,
\end{equation}
with 
\begin{align}
\big[\Pta,\Ptap\big]^\gr_s &=\big[\tau,\tp{}\big]^C_s= \sum_{n=0}^{s+1}(n+1)\big(\t{s}(\DGR\tau')_{s-n}-\tp{s}(\DGR \tau)_{s-n}\big), \\
\big[\Pta,\Ptap\big]^\ym_s &= \big[\a{},\ap{}\big]^\g_s+\bigg\{\sum_{n=0}^s(n+1) \Big(\t{n}(\DTOT\Ptap)_{s-n}-\ap{n+1}(\DGR\t{})_{s-n-1}\Big)-\Pta\leftrightarrow\Ptap\bigg\}.
\end{align}
The GR projection, called the $C$-bracket, reveals a shear-dependent bracket given in terms of the covariant derivative $\DGR$, as a deformation of a shifted Schouten-Nijenhuis bracket for symmetric tensors on 2d manifolds \cite{Schouten, SNbracket}.
In chapter \ref{secGRI}, we show how it relates to the $w_{1+\infty}$ bracket.
The YM-bracket consists in the pure YM Lie bracket $[\cdot\,,\cdot]^\g$, whose modes expansion forms the $s$-algebra, plus cross GR/YM terms.
In chapter \ref{secEYM}, we show how this last part reduces to the $sw_{1+\infty}$ algebra as a particular case.
\medskip

We also discuss in details when does the algebroid reduce to an algebra, the so-called \textit{covariant wedge}, which defines the symmetry algebra of a non-radiative cut $S$ of $\scri$.
In gravity for instance, we denote it $\Wcal_\sigma(S)$, where $\sigma\equiv C\big|_S$ is the shear at the non-radiative cut.
The defining characteristic of $\Wcal_\sigma(S)$ is a constraint on the set of symmetry parameters $\T{s}$.
In terms of the covariant derivative $\DGR$---evaluated at $S$ as well---it takes the form $\big(\DGR{}^{s+2}T\big)_{-2}=0$.
Notice that in the special case where $\sigma=0$, then $0=\big(\DGR{}^{s+2}T\big)_{-2} \overset{\sigma=0}{=}D^{s+2}\T{s}$ is the `usual' wedge condition described previously in \ref{sec:FPR}.
The \textit{covariant} wedge condition ensures that $\big(\dt C\big)\big|_S=0$ and that the symmetry transformation does not induce any radiation.\footnote{For the scope of this thesis, the assumption that no radiation should be collected at $S$ is crucial.}
A transformation in the covariant wedge thus \textit{leaves invariant} the defining properties of the cut under consideration.  
The general theorem for the Noether charge $\Qt$ also ensures that the latter\footnote{More precisely $\cq_{\T{s}}\equiv Q_{\tau(\T{s})}$, where $\tau(\T{s})$ is the solution of the dual EOM determined by the initial condition $\t{s}|_{u=0}=\T{s}$ and $\t{n}|_{u=0}=0$ for $n\neq s$.} forms a representation of the algebra $\Wcal_\sigma(S)$ at non-radiative cuts.
This result is \emph{non-perturbative} and guarantees that the symmetry is realized \emph{non-linearly} on the gravitational phase space to all order in $\gN$.\footnote{This discussion holds for EYM too.}

Let us emphasize that the algebroid framework we introduce in this work is essential to carry over the calculations efficiently.
In addition to uncovering a clear algebraic structure that gives access to a non-perturbative description, we think that this framework naturally accommodates the idea of radiation as a transition between two non-radiative states. 
Indeed, if $\Wcal_\sigma(S)$ and $\Wcal_{\sigma'}(S)$ represent the gravitational symmetry algebras at two different non-radiative cuts of $\scri$, and that radiation was registered on $\scri$ in the interval, then $\sigma\neq\sigma'$ due to the memory component carried by this radiation.
A Lie algebroid describes how these algebras relate to one another and provides a notion of path in that space that we can physically interpret (at least classically) as a transition between the two non-radiative spacetimes.
%For a visual, see the drawing \ref{fig:algebroid}.

\chapter{Notation and Preliminaries \label{sec:Preliminaries}}

We present the basic elements of notation which shall appear all throughout the manuscript.
When necessary, we give extra details at the beginning of each chapter.

\section{Notation \label{sec:Notation}}

We use the Bondi coordinates $(u,r,z,\bz)$ together with the null dyad $(m,\bm)$ on the sphere $S$. 
We choose the normalization as follows:\footnote{For instance, on a round sphere $m=\frac{1+z\bz}{\sqrt2}\pa_z\equiv P\pa_z$.} 
\begin{equation}
    m=P\pa_z,\qquad\bm=\bP\pa_{\bz},\qquad m^A\bm^B\gamma_{AB}=1, \label{mdef}
\end{equation}
for a prefactor $P=P(z,\bz)$ and where $\gamma_{AB}=m_A\bm_B+\bm_A m_B$ is the asymptotic sphere metric.
The measure on the sphere takes the form
\begin{equation}
\bep_S=\frac12\epsilon_{AB}\, \rd\sigma^A\wedge\rd\sigma^B=i\sqrt{\gamma}\,\rd z\wedge\rd\bz\qquad\mathrm{with}\qquad\epsilon_{AB}=i(\bm_A m_B-m_A\bm_B).
\end{equation}
The latter is induced by pull back to $S$ from the 4-dimensional volume form\footnote{This orientation amounts to $\bep=r^2\sin\theta\,\rd t\wedge\rd r\wedge\rd\theta\wedge\rd\varphi$ in usual spherical coordinates with $z=e^{i\varphi}\cot\frac{\theta}{2}$.}
\begin{equation}
    \bep=ir^2\sqrt{\gamma}\,\rd r\wedge\rd u\wedge\rd z\wedge\rd \bz,
\end{equation}
where $\sqrt{\gamma}\equiv\sqrt{|\det\gamma|}=|P|^{-2}$.
Explicitly,
\begin{equation}
    \bep_{ru}\equiv i_{\pa_u}i_{\pa_r}\bep\equiv r^2\bep_S. \label{spheremeasure}
\end{equation}

Rather than using the covariant derivative on the sphere $D_A$ with a free index $A$, one can use projected quantities and work with spin-weighted operators \cite{ethop, Eastwood_Tod_1982, Chandrasekhar}. 
Namely, an operator $O_s$ of spin-weight, or helicity, $s$ is defined by
\begin{equation}
O_s\equiv O_{\left\langle A_1\ldots A_s\right\rangle}m^{A_1}\ldots m^{A_s}.
\end{equation}
Similarly, the operator $O_{-s}$ is defined by the same contraction with $m\rightarrow\bm$. Angle brackets stand for symmetric and traceless. The tracelessness condition allows us to map any symmetric traceless tensor to a Lorentz scalar of helicity $s$ and a Lorentz scalar of helicity $-s$. To make things more explicit, one can write\footnote{The isomorphism is justified by a counting argument: A symmetric tensor with $s$ indices in dimension 2 has $\binom{2+s-1}{s}=s+1$ independent components. 
Imposing it to be traceless adds $\binom{s-1}{s-2}=s-1$ constraints, so that the number of independent components reduces to two, encoded in the two helicities.}
\begin{equation}
O_{\left\langle A_1\ldots A_s\right\rangle}=O_s\,\bm_{A_1}\ldots\bm_{A_s}+O_{-s}m_{A_1}\ldots m_{A_s}.
\end{equation}
For instance, $C\equiv C_{AB}m^Am^B$ and its complex conjugate $\bC$ will denote the asymptotic shear. Similarly, the news tensor will be denoted by $N=\pa_u C$.

We introduce the intrinsic-Cartan derivatives on the sphere\footnote{Historically the \textit{edth} differential operator has been defined with a different normalization \cite{ethop}.} $\sqrt{2}D=\eth$ and $\sqrt{2}\bD=\bar{\eth}$ defined as the sphere covariant derivative projected along the dyad. For instance, on an operator of spin-weight $s$,
\begin{equation}
DO_s\equiv m^{A_1}\ldots m^{A_s}m^A D_AO_{A_1\ldots A_s}.
\end{equation}
Changing $m^A\rightarrow\bm^A$, we get a similar definition for $\bD$. 
Using \eqref{mdef}, we have equivalently
\begin{subequations}
\label{intrinsic1}
\begin{align}
DO_s&=P\bP^{-s}\pa_z\big(\bP^{s}O_s\big),\\
\bD O_s&=\bP P^s\pa_{\bz}\big(P^{-s}O_s\big).
\end{align}
\end{subequations}
See the appendix \ref{AppCNCD} for extra details.

Like a function on the sphere which admits an expansion in terms of spherical harmonics $Y^0_{\l,m}$, a spin-weighed function of helicity $s$ admits an expansion in terms of spin-spherical harmonics $\sshp$ \cite{ethop}, where $\ell\geq |s|$,
\begin{equation}
    O_s=\sum_{\l=|s|}^\infty\sum_{m=-\l}^\l O_s^{\l,m}\sshp,\qquad O_s^{\l,m}\in\C. \label{sshExpansion}
\end{equation}
$D$ and $\bD$ act as raising and lowering operators for the helicity. 
Their action on the basis element $\sshp$ is given by
\begin{align} 
    D\sshp &=\sqrt{\frac{(\l-s)(\l+s+1)}{2}}Y^{s+1}_{\l,m}, \label{raisingD} \\ 
    \bD \sshp &=-\sqrt{\frac{(\l+s)(\l-s+1)}{2}}Y^{s-1}_{\l,m}, \label{loweringbD}
\end{align}
from which we also get 
\be 
\bD D \sshp=-\frac{(\l-s)(\l+s+1)}{2}\sshp,
\qquad 
D \bD \sshp= -\frac{(\l+s)(\l-s+1)}{2}\sshp. \label{casimirDD}
\ee

\section{Definition of conformal weight and helicity \label{Sec:Celest}}

We define $S_0$ to denote the 2d-sphere and $S_n$ to be the 2d-sphere minus $n$-points.
In the following we denote $S$ any one of the $S_n$ and we define $\mathcal{C}(S)$ to be the space of smooth functions on $S$.
The most studied cases are $S_0$ in the gravitational literature and the cylinder $S_2$ in the celestial and twistor literature.

We start by defining the notion of celestial primary fields $\Ccel{\Delta,s}$ on the 2-sphere (or the punctured 2-sphere, on which $D\to\pa_z$), where $\Delta$ is the celestial conformal dimension and $s$ the helicity (or spin-weight) as before.
By definition a $\phi(z,\bz) \in \Ccel{\Delta,s}$ transforms under sphere diffeomorphisms $\mathcal Y\in \mathrm{Diff}(S)$ as
\be 
\delta_{\mathcal Y}  \phi
&=\big(YD+\tfrac12(\Delta + s) DY\big)\phi+\big(\bY\bD + \tfrac12(\Delta - s)\bD\bY\big) \phi. \label{sDeltaDef}
\ee
This reduces to the usual definition of a conformal primary field in 2d \cite{francesco2012conformal, Barnich:2021dta, Donnay:2021wrk} when $Y$ ($\bY$) is (anti-)holomorphic.
Such fields commonly appear in the celestial holography context when transforming momentum space scattering amplitudes to the conformal (boost) basis by a Mellin transform, see \cite{Raclariu:2021zjz} and references therein.
Here we extend the definition of conformal primary fields to arbitrary sphere diffeomorphisms \cite{Freidel:2021dfs}.
What we call celestial fields also corresponds to the restriction of Carrollian fields at the cut $u=0$ of null infinity, with $u$ the Bondi time \cite{Freidel:2021dfs}, see the following section.

$Y$ and $\bY$ are the projections of $\mathcal Y$ along $\bm$ and $m$ respectively.
When writing $\mathcal Y\in\mathrm{Diff}(S)$, we mean the full vector on the sphere, namely
\begin{equation}
    \mathcal Y=\mathcal Y^A\pa_A=(m^A\mathcal Y_A)\bD+(\bm^A\mathcal Y_A)D\equiv \bY\bD+YD.
\end{equation}
Notice that  the parameter $Y$ is the component of a holomorphic vector field $Y\pa_z\in \Gamma(T^{1,0}S)$ and comes with a negative helicity degree, i.e.  $Y$ is the contraction of $\mathcal Y$ with $\bm$.
Hence $Y$ has weight $(-1,-1)$\footnote{This is clear from the fact that \eqref{sDeltaDef}  matches with the Lie derivative $\LL$ when acting on a vector field. Indeed, if we take $\Wcal\in \mathrm{Diff}(S)$, then 
\begin{equation}
    \delta_{\Wcal}Y= \bm_C\big(\LL_{\mathcal W}Y\big)^C=\bm_C\big(\Wcal^AD_AY^C-Y^A D_AW^C\big)=\big(WD+\overbar W\bD\big)Y-YDW.
\end{equation}} while $\bY$ has weight $(-1,1)$.\footnote{Remark that in general the multivector $\t{s}=\tau^{\langle A_1\cdots A_s\rangle}\bm_{A_1}\cdots\bm_{A_s}$, which shall play the role of dual variable to the higher-spin charges, has spin $-s$.}

We can restrict the conformal transformation labeled by the vector $\mathcal Y$ to preserve the round sphere metric. 
$Y$ is then holomorphic, i.e. $\bD Y=0$.
In this case, the operators $D$ and $\bD$ are respectively operators of celestial weight $(1,1)$ and $(1,-1)$.\footnote{To extend this property to Diff$(S)$ we need to introduce super-rotation Goldstone field
$\Phi$ and a boost connection ${\mathcal D}O_{(\Delta,s)} 
= (D+(s+\Delta)  \Phi)O_{(\Delta,s)}$,
see \cite{Campiglia:2020qvc, Donnay:2021wrk, Freidel:2021dfs}.
}

Once a conformal structure is chosen one  introduces homogeneous coordinates $\lambda_\alpha$ on the sphere such that  $z= \lambda_1/\lambda_0$. 
One can understand the conformal primaries as homogeneous section $O(n,\bar{n})$ over $\mathbb{CP}_1$ \cite{Eastwood_Tod_1982, gelfand1966generalized}, where the homogeneity degrees are given by the celestial weights as $n=-(\Delta+s)$ and $\bar{n}= -(\Delta -s) $.   This means that 
\begin{equation} \label{defCelHomogeneous}
    \phi(a \lambda_\alpha,\bar{a} \bar{\lambda}_{\dot{\alpha}})=a^{-(\Delta+s)}\bar a^{-(\Delta-s)}\phi(\lambda_\alpha, \bar{\lambda}_{\dot{\alpha}}),\qquad a\in\C^*.
\end{equation}
In this formulation the SL$(2,\C)$ action on the sphere is linear on $\lambda_\alpha$.
Importantly, $(\Delta,s)$ label $\mathrm{SL}(2,\C)$ representations, naturally realized on the space of complex homogeneous functions as above.

\section{Carrollian fields}
\label{Sec:Carroll}

On $\scri=\R\times S$, the symmetry generators include the supertranslations $T(z,\bz) \pa_u$. Moreover, the action of sphere diffeomorphisms on Carrollian fields is given by the vector field  $\mathcal Y \equiv \frac{1}{2}\big(DY+\bD\bY\big)u \pa_u+\bY\bD+YD$, which includes a boost component  determined by the components $Y$ and $\bY$ of an actual diffeomorphism of $S$.
The combination of supertranslations and sphere diffeomorphisms form the generalized BMS algebra $\gbms$ \cite{Barnich:2011mi,  Campiglia:2016efb, 
Compere:2018ylh,Freidel:2021qpz}.
If we demand in addition that the vector field is holomorphic, i.e. $\bD Y=0$ then we get the algebra $\mathfrak{ebms}(S)$. We have that $\mathfrak{ebms}(S_0)=\mathfrak{sl}(2,\C)$ and $\mathfrak{ebms}(S_2)= \mathrm{Vir}\times \overline{\mathrm{Vir}}$.

The Carrollian field $\Phi(u,z,\bz) \in \Ccar{\dCar,s}$ of Carrollian weight $\delta$ and spin-weight $s$ admits the action of supertranslations $\delta_T\Phi= T\pa_u \Phi$. 
It also transforms under sphere diffeomorphisms as
\be
\delta_{\mathcal Y}  \Phi 
= \big(YD + \tfrac12 (\delta  + u\pa_u + s) DY   \big)\Phi
+ \left(\bY\bD+\tfrac12 (\delta + u\pa_u  - s)\bD\bY \right) \Phi .
\label{sDeltaDefCarroll}
\ee
We see that the connection between the Carrollian and celestial weights \eqref{sDeltaDef} is simply that 
\be 
\hat\Delta = \delta + u\pa_u.
\ee 
$\hat\Delta$ is the operator which once diagonalized in the celestial basis, has eigenvalues $\Delta$. 
For instance, if $\Phi_\delta\in \Ccar{\dCar,s}$, we can construct a celestial field $\phi_{\Delta}\in \Ccel{\Delta,s}$ using the time Mellin transform \cite{Freidel:2021ytz,Freidel:2022skz}:
\begin{equation}
    \phi^\pm_\Delta (z,\bz) = 
    (\mp i)^{(\Delta+1-\delta)} \Gamma(\Delta+1-\delta)
    \int_{-\infty}^\infty\!\rd u\,\frac{\Phi_\delta(u,z,\bz)}{( u \pm i \epsilon)^{(\Delta-\dCar+1)}}. \label{CelCarMap}
\end{equation}
This means that a Carrollian field can be expressed as a direct sum (or direct integral) of celestial fields after diagonalization of the boost operator. This decomposition is either done in terms of the principal \cite{Pasterski:2017kqt} or discrete series of representation \cite{Freidel:2022skz, Cotler:2023qwh, Mitra:2024ugt}.  

Note that from our definition \eqref{sDeltaDefCarroll} we easily infer that $\pa_u$ is an operator of weight $(1,0)$.\footnote{Use that $ \delta_{\mathcal Y}(\pa_u\Phi)= \pa_u\big(\delta_{\mathcal Y}\Phi\big)$.}

As before, introducing homogeneous coordinates $\lambda_\alpha$ on the sphere such that  $z= \lambda_1/\lambda_0$, the Carrollian primaries in $\Ccar{\dCar,s}$ is equivalently given by the homogeneous sections over $\scri$ of degree $\big(-(\delta+s),-(\delta-s)\big)$, cf. \cite{Eastwood_Tod_1982}. 
They transform as 
\begin{equation} \label{defCarrollHomogeneous}
    \Phi(|a|^2 u, a \lambda_\alpha,\bar{a} \bar{\lambda}_{\dot{\alpha}})=a^{-(\delta+s)}\bar a^{-(\delta-s)}\Phi(u, \lambda_\alpha, \bar{\lambda}_{\dot{\alpha}}),\qquad a\in\C^*.
\end{equation}
In the following we will use the natural projection 
\be 
\Ccar{\dCar,s} & \to \Ccel{\dCar,s} \cr 
\Phi &\mapsto \phi(z,\bz) = 
\Phi(u=0,z,\bz),
\ee 
which follows from a choice of embedding of $S$ into $\scri$.
This projection preserves the conformal dimension $\Delta=\dCar $ since the boost operator action vanishes $(u \pa_u \Phi)|_{u=0} = 0$.\footnote{This projection can also be written in terms of \eqref{CelCarMap} as a limit $\lim_{\Delta \to \delta}(\phi^+_\Delta) $. }

\section{Various vector spaces \label{sec:Vspace}} 

\paragraph{The $\V$-space:}
We introduce the following space of celestial fields:
\be 
\V_s\equiv  \Ccel{-1,-s}, \qquad 
\V(S)\equiv \bigoplus_{s=-1}^\infty\V_s.
\ee 
$\V(S)$ is a graded vector space where the spin  $s$ denotes the grade. Note that $\V_s\equiv \V_s(S)$ depends on the topology of $S$. 
To avoid notational cluttering we keep this dependence implicit in the notation $\V_s$. 
The gradation of $\V(S)$ induces a filtration denoted 
\be \label{filtration}
\V^s(S) := \bigoplus_{n=-1}^s \V_n(S).
\ee

In the following we denote by $T$ the elements of $\V(S)$ and by $T_{s}$ its grade $s$ element.
In other words, $T$ corresponds to the series of spin weighted field  $(\T{s})_{s+1\in\N}$ and $\T{s}$ is the result of the evaluation map $\pi_s$ to the subspace of degree $s$:\footnote{The associated \textit{projection} operator is $\iota\circ\pi_s$.}
\begin{align}
    \pi_s :\,&\V(S)\to\V_s \nn\\
         &T\mapsto \T{s}.
\end{align}
Conversely, the inclusion map $\iota$,
\begin{align}
    \iota :\,&\V_s\to\V(S) \nn\\
         &\T{s}\mapsto T=\iota(\T{s}), \label{inclusionMap}
\end{align}
takes the element $\T{s}$ and sends it to the series $\iota(\T{s})=(0,\ldots,0,\T{s},0,\ldots)$ for which $\pi_p(T)=0$ for $p\neq s$ and $\pi_s(T)=\T{s}$.
Finally, for later convenience, we also define 
\bs \label{VstarstarDef}
\begin{align}
    \Vs &:=\bigoplus_{s=0}^\infty \V_s \\
    \Vss &:=\V(S)\backslash \{\V_{-1} \oplus\V_0\}=\bigoplus_{s=1}^\infty\V_s.
\end{align}
\es

\paragraph{The $\tV$-space:}
We introduce the corresponding spaces of Carrollian  fields:
\be \label{VCcardef}
\tV_s\equiv  \Ccar{-1,-s}, \qquad 
\tV(\scri)\equiv \bigoplus_{s=-1}^\infty\tV_s,\qquad \tV^s\equiv \tV^s(\scri) := \bigoplus_{n=-1}^s \tV_n.
\ee 
We denote by $\t{}$ the elements of $\tV(\scri)$ and by $(\tau)_s\equiv \t{s}$ a grade $s$ element.
The evaluation and inclusion maps are defined as for $\V(S)$.

\paragraph{For $\g$-valued fields:}
When dealing with Lie algebra $\g$-valued fields, we use the following notation:\footnote{To avoid having even more symbols, we use $\tV_s$ and $\V_s$ both in the GR and the YM cases. 
The context shall make it clear which space we are talking about.}
\begin{equation}
    \a{}\in\tV(\scri,\g), \qquad \tV(\scri,\g):=\bigoplus_{s=0}^\infty\tV_s\quad\textrm{with}\quad \tV_s\equiv\Ccarg{0,-s}.
\end{equation}
Notice that the Carrollian weight of $\a{s}$ is 0 while it was $-1$ for $\t{s}$.
The reason of the difference will become obvious once these fields, playing the role of symmetry generators, are paired with charge aspects.
The filtration associated to the gradation of $\tV(\scri,\g)$ is denoted by
\begin{equation}
    \tV^s\equiv\tV^s(\scri,\g):= \bigoplus_{n=0}^s\tV_n.
\end{equation}
We also introduce the celestial counterpart of $\a{s}$, denoted $\Aa{s}$, where
\begin{equation}
    \Aa{}\in\V(S,\g), \qquad \V(S,\g):=\bigoplus_{s=0}^\infty\V_s\quad\textrm{with}\quad \V_s\equiv\Ccelg{0,-s}.
\end{equation}

\chapter{Non-Abelian Gauge Theory I \label{secNAGTI}}

This chapter is based on unpublished notes.
It serves as a first exposure to the higher spin charge aspects $\tQ_s$, even if at that point they only correspond to a field redefinition.
Their interpretation as charge aspects will become clear in the next chapter (see also the introduction \ref{sec:IntroHis}).
We start in section \ref{secNAGTI:Pre} by giving few complementary elements of notation with respect to chapter \ref{sec:Preliminaries} and we detail some properties of the covariant derivative.
We then recast Maxwell's equations and the Yang-Mills' EOM in the Newman-Penrose formalism in section \ref{secNAGTI:NP}.
This is essential to understand how the $\tQ_s$ enter the picture.
Next, section \ref{secNAGTI:Asymp} describes the fall-off conditions in the radial direction $r$ that we use to study the asymptotic solutions of the EOM.
We also analyze the asymptotic phase space at null infinity.
Finally section \ref{Sec:AsympSolutionYM} deals with the YM EOM written for the asymptotic expansion of the NP scalars and we introduce the field redefinition in terms of higher spin charge aspects.
The exact time evolution of $\tQ_s$ \eqref{Qevolve}, without any truncation, is the main result of this chapter.
In the subsequent chapters, we restrict our analysis to a part of these evolution equations, from which we build a \textit{Noether symmetry preserving the symplectic potential}.
Certain details of computation are gathered in appendix \ref{AppNAGTI}.

\section{Preliminaries \label{secNAGTI:Pre}}

In order to keep track of the conformal dimension of the various equations, it is useful to introduce a curvature parameter $\cR\in\Ccel{2,0}$ such that the flat space metric is  
\be 
\label{ret-metric2}
ds^2 = -\cR \rd u^2 -2 \rd u \rd r +  r^2 \gamma_{AB} \rd\sigma^A \rd\sigma^B,
\ee 
where
\begin{equation}
\gamma_{AB}\rd\sigma^A \rd\sigma^B=\frac{2 \rd z \rd\bz}{|P(z,\bz)|^2}\qquad\textrm{and}\qquad
|P(z,\bz)|= \frac{1+ \cR z\bz}{\sqrt{2}}. \label{defP}
\end{equation}
The exact phase of $P$ is irrelevant for our discussion and we shall consider formulas (especially for the covariant derivative (\ref{intrinsic1})) valid for any $P$.
When $\cR=1$ this is the usual parametrization of $\scri$ while $\cR=0$ corresponds to the flattened space description \cite{Donnay:2022wvx}.
The inverse metric components are 
\begin{equation}
g^{uu}=0,\quad g^{ur}=-1,\quad g^{rr}=\cR. \label{metriccomp}
\end{equation}
The null vectors transverse to the sphere $S$ located at a constant value of $(r,u)$ are  $k= \pa_r$ and $\l=\pa_u -\frac{\cR}{2}\pa_r$. They satisfy
\begin{equation}
    k\cdot k=0,\qquad \l\cdot\l=0,\qquad\l\cdot k=-1. \label{defkl}
\end{equation}
Supplemented by the dyad $(m,\bm)$, these vectors form the Newman-Penrose (NP) null tetrad and appear in the definition of the NP scalars \cite{Newman:1968uj}.

Furthermore, the \textit{edth} derivatives satisfy the fundamental commutation relation 
\begin{equation}
    \boxed{[\bD,D] O_s=\cR s O_s}. \label{ethcommut}
\end{equation}
We refer the reader to appendix \ref{AppCNCD} for extra properties of these derivatives, especially about their commutator.

\section{EOM à la Newman-Penrose \label{secNAGTI:NP}}
\subsection{Electrodynamics}

We start with Maxwell's equations in four dimensional Minkowski spacetime
\begin{equation}
\label{eq:Maxwell}
\nabla^{\mu}F_{\mu\nu} = J_{\nu}.
\end{equation} 
Using the formula for the divergence of an antisymmetric tensor $\nabla_{\mu} F^{\mu\nu} = \frac{1}{\sqrt{|g|}}\pa_{\mu}\left(\sqrt{|g|} F^{\mu\nu}\right)$, \eqref{eq:Maxwell} reduces to
\bs
\label{Maxwell}
\begin{align}
-r (2 +r \pa_r) F_{ur}  + \gamma^{AB} D_A F_{Br} &= r^2J_r, \label{M1}\\
 r^2 \pa_u F_{ur} + \gamma^{AB} D_A \big(F_{Bu}-\cR F_{Br}\big) &= r^2\big(J_u-\cR J_r\big),\label{M2} \\
r^2 \pa_u F_{Ar} + r^2\pa_r\big(F_{Au} - \cR F_{Ar}\big) +  \gamma^{BC} D_B F_{CA}  &= r^2J_A. \label{M3}
\end{align}
\es
The Bianchi's identity $\nabla_\mu (\star F)^{\mu\nu}=\tJ^\nu$ gives a similar set of equations. In Maxwell's theory, only the electric current $J$ is non-zero while the magnetic current vanishes $\tJ\equiv 0$. 
It is convenient to keep it alive for when dealing with YM theory.
Indeed, we can understand the asymptotic expansion of  Yang-Mills' theory in terms of a collection of QED like quantities where both electric and magnetic currents are turned on. 
The dual tensor $\star F$ is defined by
\begin{equation}
(\star F)_{\mu\nu}=\frac{1}{2}\sqrt{-g}\,\epsilon_{\alpha\beta\mu\nu}g^{\alpha\rho}g^{\beta\lambda}F_{\rho\lambda}.
\end{equation}
For instance
\begin{equation}
    r^2\star\!F_{ur}=\frac12 r^4\sqrt{\gamma}\epsilon_{urAB}g^{AC}g^{BD}F_{CD}=\frac12\sqrt{\gamma}\epsilon_{urAB}\gamma^{AC}\gamma^{BD}F_{CD}=-\frac12\epsilon^{AB}F_{AB}.
\end{equation}
Moreover
\begin{equation}
    \star F_{rA}=\epsilon_A^{~\,B}F_{rB}\qquad\textrm{and}\qquad\star\! F_{uA}=\epsilon_A^{~\,B}(\cR F_{rB}-F_{uB}),
\end{equation}
meaning also that
\begin{equation}
   r^2\star\! F_{ur}=iF_{m\bm},\qquad\star F_{rm}=iF_{rm} \qquad\textrm{and}\qquad\star\! F_{um}=-i(F_{um}-\cR F_{rm}),
\end{equation}
or equivalently
\be 
\boxed{r^2 (\star F)_{\ell k } = i  F_{m\bm}, \qquad
(\star F)_{k m } = i F_{k m},
\qquad 
(\star F)_{\ell m } = -i F_{\ell m}}.
\ee
It becomes then convenient to form the self-dual part of the field-strength, namely $\bfm{F}^+\equiv\frac12(\bfm{F}-i\star\! \bfm{F})$ which we choose to repackage as follows:
\begin{subequations}
\label{NPscalars}
\begin{align}
    \phi_{\bm} &\equiv \frac{1}{r}F_{\bm u}-\frac{\cR}{2r}F_{\bm r}=\frac{1}{r}F^+_{\bm u}, \\
    \phi &\equiv \frac12\left( F_{ru}+\frac{1}{r^2}F_{\bm m}\right)=F_{ru}^+=\frac{1}{r^2}F^+_{\bm m}, \\
    \phi_m &\equiv\frac{1}{r} F_{rm}=\frac{1}{r}F_{rm}^+.
\end{align}
\end{subequations}
We also have that  that $F^+_{r\bm}=0$ and $F^+_{um}=\frac{\cR}2 F_{rm}^+$. Since the dyad $(m,\bm)$ is defined for the unit sphere, the factors of $1/r$ are introduced for the sake of normalization. The scalars $\phi_{\bm}$, $\phi$ and $\phi_m$ correspond to the Newman-Penrose scalars $\phi_2, \phi_1$, and $\phi_0$ respectively \cite{Newman:1968uj}. Indeed, in terms of the tetrad,
\begin{equation}
    \boxed{\phi_{\bm}=\frac1r F^+_{\bm\l},\qquad \phi=F^+_{k\l}\qquad\textrm{and}\qquad\phi_m=\frac1r F^+_{k m}}.
\end{equation}
According to subsec.\ref{Sec:Carroll}, they respectively have dimension/helicity $(1,-1)$, $(0,0)$ and $(-1,1)$. 

We then introduce the combination of currents \begin{equation}
    J^+_\mu\equiv \frac12\big(J_\mu-i\tJ_\mu\big),\qquad\textrm{so that}\qquad \nabla^\mu F_{\mu\nu}^+=J^+_\nu,
\end{equation} 
and therefore recast the Maxwell's equations (\ref{M1}) and the projection of (\ref{M3}) along $\bm$, combined with the corresponding Bianchi's identities, respectively as
\bs
\label{MaxwellNPradial}
\begin{empheq}[box=\fbox]{align}
\big(2+r\pa_r\big)\phi &=\bD\phi_m+r J^+_r,\label{NPr}\\
\big(1+r\partial_r\big)\phi_{\bm} &=\bD\phi+J^+_{\bm}.\label{NPbm}
\end{empheq}
\es
These two radial evolution equations determine---almost entirely---the data of $\phi$ and $\phi_{\bm}$ in terms of $\phi_m$ and the currents. We also get two time evolution equations starting from (\ref{M2}),\footnote{For convenience, we actually write the linear combination \eqref{M2}$-\frac{\cR}{2}$\eqref{NPr}=\eqref{NPu}.} the projection of  (\ref{M3}) along $m$, and the associated Bianchi's identities:
\begin{subequations}
\label{MaxwellNPtime}
\begin{empheq}[box=\fbox]{align}
\left(r\pa_u- \frac{\cR}{2} (2+r\partial_r) \right)\phi &= D\phi_{\bm} - r\left(J^+_u-\frac{\cR}2 J^+_r\right)\label{NPu}\\
\left(r\pa_u- \frac{\cR}{2} (1+r\partial_r) \right)\phi_m &= D\phi-J^+_m. \label{NPm}
\end{empheq}
\end{subequations}

One can also get a closed form of these equations. For instance, if we multiply (\ref{NPm}) by the operator $(2+r\pa_r)$, recalling that $\partial_r$ and $\partial_u$ commute with $D$ and $\bD$, and  then use expression (\ref{NPr}), we get
\begin{align}
(2+r\partial_r)\left(r\pa_u- \frac{\cR(1+r\partial_r)}{2}\right)\phi_m=rD J^+_r+D\bD\phi_m-(2+r\pa_r)J^+_m. \nn
\end{align}
Simplifying and repeating similar steps, we find that the three NP scalars satisfy the following closed equations:\footnote{These equations can also be obtained  by starting from the Bianchi identity \begin{equation}
    \nabla_\gamma F_{\alpha \beta} = \nabla_\alpha F_{\gamma \beta}- \nabla_\beta F_{\gamma \alpha}
\end{equation} 
and contracting it  with $\nabla^\gamma$ which gives 
\be 
\square F_{\alpha \beta}= \nabla_\alpha J_\beta - \nabla_\beta J_\alpha.
\ee }
\begin{subequations}
\label{phiclose}
\begin{empheq}[box=\fbox]{align}
\left(\frac{\cR}{2}(1+r\pa_r)(2+r\pa_r) -r\pa_u(3+r\pa_r)\right)\phi_m+D\bD\phi_m &=(2+r\pa_r)J^+_m-r D J^+_r, \label{phiclose0}\\
\left(\frac{\cR}{2}(1+r\pa_r)(2+r\pa_r)-r\pa_u(2+r\pa_r)\right)\phi+D\bD\phi &= \nn\\
&\hspace{-2cm}=r(2+r\pa_r)\left(J^+_u-\frac{\cR}2{J^+_r}\right)- D J^+_{\bm}, \label{phiclose1}\\
\left(\frac{\cR}{2}(1+r\pa_r)(2+r\pa_r)-r\pa_u(1+r\partial_r)\right)\phi_{\bm}+\bD D\phi_{\bm} &= \nn\\
&\hspace{-4cm}=\left(\frac{\cR}2(2+{r\pa_r})-r\pa_u\right)J^+_{\bm}+r\bD\left(J^+_u-\frac{\cR}2 {J^+_r}\right). \label{phiclose2}
\end{empheq}
\end{subequations}

\subsection{YM theory}

We generalize the previous equations of motion (\ref{MaxwellNPradial}-\ref{MaxwellNPtime}) to the non-abelian case before studying the solution around null-infinity. 
The linear component is essentially the same as before. 
The difference comes from the gauge covariant derivative $\nabla_\mu+A_\mu$, inducing a self-interacting contribution. 
We work with Lie algebra valued quantities to lighten the notation, i.e. $A_\mu=A_\mu^b Z_b$ with $\{Z_b\}$ a basis of anti-hermitian Lie algebra generators. 
Besides, we adopt the convention that the complex conjugation acts only on the spacetime components, i.e. relates to the frame fields, and not on the algebra generators (in case the latter are complex).\footnote{Note that we used the notation $\phi_m$ and $\phi_{\bm}$ as a mere reminder of the helicity of these scalars, the latter are thus \textit{not} complex conjugate of one another.}
Even if we shall write expressions such as $\bar{\phi}_m$, it must then be understood as $r\bar{\phi}_m=r\bar{\phi}_m^bZ_b= \overbar{F_{rm}}^bZ_b=F_{r\bm}^bZ_b$.

To keep the same structure as Maxwell's equations, the non-linear piece of YM equations can effectively be written as a current $J_\mu^{\mathsf{QCD}}$ purely involving gauge fields. 
Hence YM equations in vacuum and the Bianchi's identities take the form
\begin{equation}
\boxed{\nabla_\mu F^{+\mu}_{~~~\nu}+\big[A_\mu,F^{+\mu}_{~~~\nu}\big]_\g=0\qquad\Leftrightarrow\qquad \nabla_\mu F^{+\mu}_{~~~\nu}=J_\nu^{\mathsf{QCD}}}.
\end{equation}
where $J^{\mathsf{QCD}}_\nu = [F^{+}_{\mu\nu}, A^\mu]_\g$. $J^{\mathsf{QCD}}$ thus plays the role of $J^+$ in equations (\ref{MaxwellNPradial}-\ref{MaxwellNPtime}). 
In other words, without matter, a non-abelian gauge theory can be viewed as an abelian gauge theory sourced by self-interacting gauge fields that include both electric and magnetic charges---contrary to ordinary matter which does not source magnetic monopoles. 
Therefrom we just have to specify
\begin{subequations}
\label{currentYM}
\begin{empheq}[box=\fbox]{align}
J_r^{\mathsf{QCD}} &=-\big[A_r,\phi\big]_\g+\frac{1}{r}\big[A_{\bm},\phi_m\big]_\g,\label{currentr}\\
J_u^{\mathsf{QCD}} &= +\big[A_u,\phi\big]_\g-\frac{1}{r}\big[A_m,\phi_{\bm}\big]_\g+\frac{\cR}{2r}\big[A_{\bm},\phi_m\big]_\g- \cR\big[A_r,\phi\big]_\g,\label{currentu}\\
J_m^{\mathsf{QCD}} &=-\big[A_m,\phi\big]_\g +r\big[A_u,\phi_m\big]_\g-\frac{r\cR}{2}\big[A_r,\phi_m\big]_\g,\label{currentm}\\
J_{\bm}^{\mathsf{QCD}} &=+\big[A_{\bm},\phi\big]_\g -r\big[A_r,\phi_{\bm}\big]_\g .\label{currentbm} 
\end{empheq}
\end{subequations}

Using these currents we see that the radial evolution equations are, in the radial gauge $A_r=0$, simply given by\footnote{$\mathrm{ad}\,A_{\bm}(\cdot)=[A_{\bm},\cdot]_\g$.}
\bs
\label{YMNPradial}
\begin{empheq}[box=\fbox]{align}
\big(2+r\pa_r\big)\phi &=\big(\bD+\mathrm{ad}\, A_{\bm}\big)\phi_m,\label{YMr}\\
\big(1+r\pa_r\big)\phi_{\bm} &=\big(\bD+\mathrm{ad}\,A_{\bm}\big)\phi.\label{YMbm}
\end{empheq}
\es
The time evolution equations for Yang-Mills are in turn given by
\begin{subequations}
\label{YMNPtime}
\begin{empheq}[box=\fbox]{align}
\left(r\pa_u- \frac{\cR}{2} (2+r\pa_r) \right)\phi &= \big(D +\mathrm{ad}\,A_m\big)\phi_{\bm} - r\big[A_u,\phi\big]_\g,\label{YMu}\\
\left(r\pa_u- \frac{\cR}{2} (1+r\pa_r) \right)\phi_m &= \big(D+\mathrm{ad}\,A_m\big)\phi-r\big[A_u,\phi_m\big]_\g. \label{YMm}
\end{empheq}
\end{subequations}
This also implies that the closed form for $\phi_m$ can be written like
\bs \label{YMphiclose}
\begin{equation}
\left(\frac{\cR}{2} (1+r\partial_r)(2+r\pa_r)-r(3+r\pa_r)\pa_u \right)\phi_m +\big(D+\mathrm{ad}\,A_m\big)\big(\bD +\mathrm{ad}\,A_{\bm}\big)\phi_{m}=J,
\end{equation} 
where 
\be 
J&:=  r(3+r\pa_r)\big[A_u,\phi_m\big]_\g+\big[\phi, r\pa_r A_m\big]_\g.
\ee 
\es

\section{Asymptotic behavior \label{secNAGTI:Asymp}}

\subsection{Asymptotic expansion}

We look for a solution of (\ref{phiclose}) analytic around $r=\infty$. 
See \cite{Strominger:2017zoo} for a review.
In particular we assume that the Peeling theorem holds, meaning that we discard any logarithmic behavior.
We thus consider the large $r$ expansions employed by \cite{Newman:1968uj} and consistent with the boundary conditions suggested by the soft theorems. The components of the field strength take the form
\begin{subequations}
\label{radialexp}
\begin{equation}
F_{ur} = \frac{1}{r^2}\sum_{n = 0}^{\infty} \frac{F_{ur}^{(n)}}{r^n}, \quad
F_{uz} = \sum_{n = 0}^{\infty} \frac{F_{uz}^{(n)}}{r^n}, \quad
F_{rz} = \frac{1}{r^2}\sum_{n = 0}^{\infty} \frac{F_{rz}^{(n)}}{r^n}, \quad
F_{z\bz} = \sum_{n= 0}^{\infty} \frac{F_{z\bz}^{(n)}}{r^n}. \label{radialexpF}
\end{equation}
In the following, we shall impose the radial gauge $A_r=0$, so that for the gauge field, the fall-off conditions (\ref{radialexpF}) correspond to
\begin{equation}
A_u=\sum_{n=0}^\infty\frac{A_u^{(n)}}{r^n},\qquad A_{m}=P(z)A_z=\sum_{n=0}^\infty\frac{A_{m}^{(n)}}{r^n}. \label{ansatzAs}
\end{equation}
Concerning the Newman-Penrose scalars, this amounts to
\begin{equation}
\phi_{\bm}=\frac{1}{r}\sum_{n=0}\frac{\np{\bm}{n}}{r^n},\qquad \phi=\frac{1}{r^2}\sum_{n=0}\frac{\np{}{n}}{r^n}, \qquad
\phi_m=\frac{1}{r^3}\sum_{n=0}\frac{\np{m}{n}}{r^n}. \label{ansatzphis}
\end{equation}
\end{subequations}
Physically, $\phi$ contains the Coulomb's field while $\phi_{\bm}$ contains the radiative component, which is part of the asymptotic phase space. Notice that once we assume that $\phi_m\sim\np{m}{0}/r^3+\cdots$, the behavior of $\phi$ and $\phi_{\bm}$ is fixed. 
Indeed, say $\phi\sim \np{}{0}/r^p+\cdots$, and integrate (\ref{NPr}). If $p=0$, we get a logarithm, that we explicitly disregard. 
If $p=1$, then $\np{}{0}$ would be 0. If $p=2$, then $\np{}{0}$ is unconstrained by (\ref{NPr}).
In other words, the equations of motion indicate that we cannot impose a stronger fall-off condition without extra assumptions. 
A similar reasoning leads to $\phi_{\bm}\sim\np{\bm}{0}/r$ once we integrate (\ref{NPbm}). 
The relation between fall-off behavior and helicity (dimension) is thus $\phi_\bullet\sim 1/r^{2+s[\bullet]}$ ($\phi_\bullet\sim 1/r^{2-\dCar[\bullet]}$).

\subsection{Asymptotic Phase Space \label{secNAGTI:PS}}

In order to analyze the asymptotic solution of YM equations, we have to identify which variables constitute the phase space at null infinity. 
On $\scri$, the radiative phase space consists of $\big(A_z^{(0)},A_{\bz}^{(0)}\big)$ and their conjugate variables $\big(\pa_u A_{\bz}^{(0)},\pa_u A_z^{(0)}\big)$. 
Indeed, we start from the symplectic potential current\footnote{$\Pairg{Z_a,Z_b}=-\Tr(Z_aZ_b)$, where $\Tr$ denotes the trace, i.e. the Cartan–Killing form for $\g$, normalized such that $\Tr(Z_aZ_b)=-\delta_{ab}$ (recall that we use anti-hermitian generators).} $\bfm{\theta}$
\begin{align}
    \bfm{\theta}=-\frac{1}{\gYM^2}\bep_\mu \Pairg{F^{\mu\nu},\delta A_\nu},
\end{align}
where $\bep_\mu$ is the 3-form obtained by contraction of the 4d volume form, i.e. $i_{\pa_\mu}\bep=\bep_\mu$. Pulling back on $\scri$, and using the metric components \eqref{defP}-\eqref{metriccomp} together with the expansion (\ref{radialexp}), we get
\begin{align}
    \gYM^2\bfm{\theta}&=\lim_{r\rightarrow\infty} \bep_r\Tr\big( F^{r\nu}\delta A_\nu\big) =\lim_{r\rightarrow\infty}\bep_r\Tr\big(F^{ru}\delta A_u +F^{rA} \delta A_A\big) \nn\\
    &= \lim_{r\rightarrow\infty}\bep_r\Tr\big(F_{ur}\delta A_u  -r^{-2}(F_{um}-\cR F_{rm}) \delta A_{\bm}
    -r^{-2}(F_{u \bm}-\cR F_{r \bm}) \delta A_{m} \big)\nn\\
    &=\left( \lim_{r\rightarrow\infty}r^{-2}\bep_r\right) \Tr\big( F_{ur}^{(0)} \delta A_u^{(0)}-F_{um}^{(0)} \delta A_{\bm}^{(0)}-F_{u\bm}^{(0)} \delta A_m^{(0)}\big).
     \label{AsympPS}
\end{align}
The limit of the measure is simply $\lim_{r\rightarrow\infty}r^{-2}\bep_r =\rd u\wedge\bep_S$ where $\bep_S$ is the sphere measure \eqref{spheremeasure}.
To continue we introduce a phase field $\varphi$ valued in the group such that $A_u^{(0)} =\varphi^{-1}\pa_u \varphi$ and we denote 
\be 
A:= \varphi\big(A^{(0)}_{m}+D\big)\varphi^{-1}, \qquad \bA:= \varphi\big(A^{(0)}_{\bm}+\bD\big)\varphi^{-1}.
\ee
These fields are gauge invariant combinations\footnote{We use the right group multiplication law.} such that 
\be 
F_{um}^{(0)} =\varphi^{-1}(\pa_u A) \varphi
\qquad\textrm{and}\qquad  F_{u\bm}^{(0)}=\varphi^{-1}(\pa_u\bA) \varphi.
\ee 
The variables $A\in\Ccarg{1,1}$ and $\news \equiv\pa_u A\in\Ccarg{2,1}$ represent the radiative degree of freedom.
$A$ is the analog of the shear in GR and $\news$ of the news.
To evaluate the symplectic structure we use that 
\be
\delta A_m^{(0)} =\varphi^{-1} \left( \Dcal[\delta\varphi \varphi^{-1}] +\delta A \right)\varphi\qquad\textrm{and}\qquad \delta A_u^{(0)}=\varphi^{-1}\big(\pa_u[\delta\varphi \varphi^{-1}]\big)\varphi,
\ee
where we introduced the asymptotic gauge covariant derivative
\begin{equation}
    \boxed{\Dcal:=D+\mathrm{ad}\,A}, \label{gaugecovd}
\end{equation}
that incorporates only the radiative phase space variable. 
Any sub-leading quantity in the asymptotic expansion is determined by the latter.
The symplectic potential current finally simplifies to\footnote{Here we used the equation of motion $\nabla_\mu F^\mu_{~\,\nu}+[A_\mu,F^\mu_{~\,\nu}]=0$. 
Indeed, if we select $\nu=u$, use the asymptotic expansion \eqref{radialexp}, focus on the leading $(1/r)^2$ contribution, and multiply by $\varphi\cdot\varphi^{-1}$, we find that
\begin{equation}
    \varphi\big(\nabla_\mu F^\mu_{~\,u}+[A_\mu,F^\mu_{~\,u}]\big)\varphi^{-1}=0\quad\Rightarrow\quad \pa_u\big(\varphi F_{ur}^{(0)}\varphi^{-1}\big)-\Dcal(\pa_u\bA)-\bDcal(\pa_u A)=0.
\end{equation}}
\begin{align}
    \gYM^2\bfm{\theta}=\rd u\wedge\bep_S\,\Tr\Big(\pa_u\big(F_{ur}^{(0)}\varphi^{-1}\delta\varphi\big) +\bD\big(\pa_u A\varphi\delta\varphi^{-1}\big) &+D\big(\pa_u\bA\varphi\delta\varphi^{-1}\big)-\news\delta\bA-\bnews\delta A\Big).
\end{align}
After integration over $\scri$, the total divergence on the sphere vanishes and the symplectic structure simply reduces to
\begin{equation}
    \Theta=\int_{\scri}\bfm{\theta}= -\frac{1}{\gYM^2}\int_{\scri^+_-}^{\scri^+_+} \bep_S\, \Pairg{F_{ur}^{(0)},\varphi^{-1}\delta\varphi} +\frac{1}{\gYM^2} \int_{\scri} \rd u\wedge\bep_S\Big(\Pairg{\news,\delta\bA}+ \Pairg{\bnews,\delta A}\Big).
\end{equation}
The first expression on the RHS is a corner term that provides the canonical pair conjugated to the memory effect.
The integral over $\scri$ corresponds to the Ashtekar-Streubel symplectic potential \cite{ashtekar1981symplectic}.

\section{Higher spin charge aspects \label{Sec:AsympSolutionYM}}

\subsection{Time evolution of $\np{m}{n}$}

We are now ready to analyze the equations of motion (\ref{YMNPradial})-(\ref{YMNPtime}). 
The first step consists in expressing the data in $\phi$ and $\phi_{\bm}$ as a function of those of $\phi_m$. 
We work in the radial gauge where $A_r=0$ and we also impose that $A^{(0)}_u=0$ on $\scri$.\footnote{Strictly speaking this is not a gauge condition as the soft goldstone mode $\int_{\scri}A_u^{(0)}$ enters the symplectic potential as a zero mode conjugated to the memory variable. In this work we ignore these subtleties associated with the presence of soft goldstone modes since we focus on the resolution of the bulk equations of motion.} 
In this gauge we have that $A= A_m^{(0)}$. 

Let us show that (almost) any sub-leading quantity in the asymptotic expansion is expressed in terms of $\phi_m$.
Once we plug the radial expansion (\ref{radialexp}) into the radial evolution equation \eqref{YMr}, we get $\np{}{1}=-\bDcal\np{m}{0}$ and
\begin{equation}
    -(n+1)\np{}{n+1}=\bDcal\np{m}{n}+\sum_{p=0}^{n-1}\big[A_{\bm}^{(n-p)},\np{m}{p}\big]_\g,\qquad n\geqslant 1. \label{phi10}
\end{equation}
Similarly, from \eqref{YMbm}, we infer that $\np{\bm}{1}=-\bDcal\np{}{0}$ and 
\begin{equation}
    -(n+1)\np{\bm}{n+1}=\bDcal\np{}{n}+\sum_{p=0}^{n-1}\big[A_{\bm}^{(n-p)},\np{}{p}\big]_\g, \qquad n\geqslant 1. \label{phi21}
\end{equation}
We thus get an expression for the NP scalars of spin 0 and $-1$ in terms of the spin 1 component, with the exception of two coefficients. 
For later purposes, we rename the part of $\phi$ and $\phi_{\bm}$ not determined by $\phi_m$ and the dominant part of $\phi_m$ respectively as 
\begin{eqnarray} 
 \tQ_{-1}&\equiv& \np{\bm}{0}= F_{\bm u}^{(0)} = -\pa_u \bA,\cr
 \tQ_{0}&\equiv&\np{}{0}= \frac{1}{2}\big(F^{(0)}_{\bm m}-A_u^{(1)}\big), \label{leadingQs}\\
 \tQ_{1} &\equiv& \np{m}{0}. \nn
\end{eqnarray}
The radial gauge has the advantage of having a simple dictionary between the Newman-Penrose scalars and the gauge potential (see Appendix \ref{APPphi0} for further details):
\begin{align}
-(n+1)A_u^{(n+1)} &=2\textrm{Re}\big(\np{}{n}\big),\qquad n\geqslant 0, \nn\\
-(n+1)A_{m}^{(n+1)} &=\np{m}{n},\qquad ~~\qquad n\geqslant 0. \label{AsToPhis}
\end{align}
Gathering the former results and definitions, we find
\bs
\label{phis0}
\begin{empheq}[box=\fbox]{align}
\phi_m &=\frac{\tQ_1}{r^3} +\sum_{n=1}^\infty\frac{\phi_m^{(n)}}{r^{n+3}}, \\
\phi &=\frac{\tQ_{0}}{r^2}
-\frac{\bDcal\tQ_{1}}{r^3}-\sum_{n=1}^\infty\frac{1}{n+1}\frac{1}{r^{n+3}}\left(\bDcal\np{m}{n}-\sum_{p=0}^{n-1}\frac{1}{p+1}\big[\bnp{m}{p},\np{m}{n-1-p}\big]_\g\right), \label{phi10final}\\
\phi_{\bm} &=\frac{\tQ_{-1}}{r}-\frac{\bDcal\tQ_{0}}{r^2}-\sum_{n=1}^\infty\frac{1}{n+1}\frac{1}{r^{n+2}}\left(\bDcal\np{}{n}-\sum_{p=0}^{n-1}\frac{1}{p+1}\big[\bnp{m}{p},\np{}{n-1-p}\big]_\g\right). 
\end{empheq}
\es

We now move our attention to the time evolution equations. 
We have to distinguish the ones for $\tQ_{0}$ and $\tQ_{1}$ from the one for $\np{m}{n},n>1$. 
For the former, we first look at the order $1/r$ of the evolution equation for $\phi$  (\ref{YMu}). At this order the only source term contributing is $-rJ^{\mathsf{QCD}}_u \approx [A,\np{\bm}{0}]_\g/r$, and we simply get that 
\begin{equation}
\pa_u\np{}{0}=\bD\np{\bm}{0} + \big[A,\np{\bm}{0}\big]_\g
\qquad
\Leftrightarrow\qquad\boxed{\pa_u\tQ_{0}=\Dcal\tQ_{-1}}. \label{Q0evol}
\end{equation}

Similarly, we then consider the time evolution equation  (\ref{YMm}).
Using the expansions (\ref{radialexp}) we can easily extract the lowest order $(1/r)^2$ term. 
It is given by 
\begin{equation}
\pa_u\np{m}{0}= D\np{}{0} +\big[A, \np{}{0}\big]_\g \qquad\Leftrightarrow\qquad\boxed{\pa_u\tQ_{1}=\Dcal\tQ_{0}}, \label{Q1evol}
\end{equation}
where we identified $\np{m}{0}\equiv\tQ_{1}$.\footnote{The general formula is given by (\ref{ansatzcharge}).} 
We have checked that the whole tower of time evolution equations for $\phi$ associated to higher powers than $(1/r)$ is trivially satisfied---as it should since all the rest of the information is already contained into $\phi_m$ and thus (\ref{phi0n}) below---once we use the solution (\ref{phi0n}) to replace the appearance of $\pa_u\np{m}{n+1}$.

The next equation to analyze concerns the sub-subleading charge of spin $2$. For this we need an evolution equation for $\np{m}{1}$. We present here the gist of the computation and refer the reader to Appendix \ref{APPphi0} for the general demonstration of the evolution equation satisfied by $\np{m}{n}$. 

The starting point is the closed form \eqref{YMphiclose}.
If we plug in the leading orders in $1/r$ of each quantity using \eqref{leadingQs}, \eqref{AsToPhis} and \eqref{phi10final}, namely
\begin{equation}
     \phi_m\approx\frac{\tQ_{1}}{r^3}+\frac{\np{m}{1}}{r^4},\quad \phi\approx \frac{\tQ_{0}}{r^2}-\frac{\bDcal\tQ_{1}}{r^3},\quad A_u\approx -\frac{\tQ_{0}+\tbQ_{0}}{r}\quad\textrm{and}\quad A_m\approx A-\frac{\tQ_{1}}{r},
\end{equation}
we find that the leading contribution is proportional to $(1/r)^3$ and takes the form
\begin{equation}
     \pa_u\np{m}{1}+\cR\tQ_{1}+\Dcal\bDcal\tQ_{1} -2\big[\tQ_{0},\tQ_{1}\big]_\g-\big[\tbQ_{0},\tQ_{1}\big]_\g=0.
\end{equation}
Finally, we commute the derivatives thanks to \eqref{ethcommutnYM}\footnote{In our case it simplifies to $\Dcal\bDcal\tQ_{1}=\bDcal\Dcal\tQ_{1}-\cR\tQ_{1}-\big[F_{\bm m}^{(0)},\tQ_{1}\big]_\g$.} and make use of \eqref{leadingQs} one more time to replace $F_{\bm m}^{(0)}\to \tQ_{0}-\tbQ_{0}$. Hence the evolution equation satisfied by $\np{m}{1}$ is
 \begin{equation}
     \boxed{\pa_u\np{m}{1}+\bDcal\Dcal\tQ_{1}-3\big[\tQ_{0},\tQ_{1}\big]_\g=0}. \label{phim1evol}
 \end{equation}

Generalizing the previous calculation to any $s>0$, we find (cf. App.\,\ref{APPphi0}) the evolution equation for $\np{m}{s}$:
\begin{adjustwidth}{-0.3cm}{-0.1cm}
\begin{empheq}[box=\fbox]{align}
    &\qquad (s+1)\pa_u\np{m}{s+1}+\cR\frac{s(s+3)}{2}\np{m}{s}+\bDcal\Dcal\np{m}{s} \nn\\
    &\quad -(s+3)\big[\tQ_{0},\np{m}{s}\big]_\g+\sum_{p=0}^{s-1}\frac{(s+1)(s+2)}{(p+1)(p+2)(s-p)}\big[\bDcal\np{m}{p},\np{m}{s-p-1}\big]_\g \label{phi0n}\\
    &=s\big[\tbQ_{0},\np{m}{s}\big]_\g+\sum_{p=0}^{s-1}\frac{1}{p+1}\Dcal\big[\bnp{m}{p},\np{m}{s-1-p}\big]_\g-\sum_{p=0}^{s-1}\frac{s+1}{(p+1)(p+2)}\big[\Dcal\bnp{m}{p},\np{m}{s-1-p}\big]_\g \nn\\
    &\quad+\sum_{p=0}^{s-1}\sum_{k=0}^{p-1}\frac{(s+1)\big[(s+2)(p-k)-(k+1)(s-p)\big]}{(p+1)(p+2)(p-k)(k+1)(s-p)}\big[[\bnp{m}{k},\np{m}{p-1-k}]_\g,\np{m}{s-1-p}\big]_\g.\nn 
\end{empheq}
\end{adjustwidth}
\textit{This is the full non-linear evolution equation for the NP scalar $\phi_m$ in non-abelian gauge theory, under the assumption that the solution is analytic in $1/r$.}

\subsection{Time evolution of $\tQ_s$}

Similarly to the gravitational case treated in \cite{Freidel:2021ytz}, we introduce the set of spin $s$ objects $\tQ_s$, called \textit{higher spin charge aspects}, defined via
\begin{equation}
\boxed{\np{m}{s}\equiv \np{m,G}{s}+\np{m,L}{s}=\np{m,G}{s}+\frac{(-1)^s}{s!}\bDcal^s\tQ_{s+1}}. \label{ansatzcharge}
\end{equation}
The split between $\np{m,G}{s}$ and $\np{m,L}{s}$, called \textit{global} and \textit{local} part respectively, is determined by SL$(2,\C)$ representation theory \cite{gelfand1966generalized} (see also the celestial diamond discussion of \cite{Freidel:2021ytz} for a concise summary).
The statement goes as follows.
As mentioned around \eqref{defCelHomogeneous} and \eqref{defCarrollHomogeneous}, the conformal and spin weights are labels for SL$(2,\C)$ representations.
In the case of $\np{m}{s}\in\Ccarg{s+2,1}$,\footnote{We keep the notation $\Ccarg{\delta,s}$ but what really matters for the representation theory is the transformation property described in \eqref{defCelHomogeneous}.} the representation is reducible:
\begin{equation}
    \Ccarg{s+2,1}=G_{(s+2,1)}\oplus L_{(s+2,1)},
\end{equation}
where one can prove that 
\begin{equation}
    L_{(2+s,1)}=\Ccarg{s+2,1}/G_{(s+2,1)}\simeq \Ccarg{2,s+1}. 
\end{equation}
We precisely take $\tQ_{s+1}$ as an element of the latter, namely $\tQ_{s}\in\Ccarg{2,s}$.
Next, notice that $\Ccar{s+2,1}$ is the image of $\Ccarg{2,s+1}$ by $\bD^s$.
Trading the \textit{edth} operator for its gauge covariant version,\footnote{According to the pattern that started to appear in (\ref{Q0evol}) and (\ref{Q1evol}), it is natural to extend the definition of \cite{Freidel:2021ytz, Freidel:2023gue} given in terms of $\bD$, to the one given in \eqref{ansatzcharge} where $\bD\to\bDcal$.} i.e. $\bD\to\bDcal$, we deduce that $\np{m}{s}$ is partially made of the irreducible subspace $L_{(s+2,1)}$ generated by $\bDcal^s\tQ_{s+1}$.
We can now infer that $G_{(s+2,1)}$ is the co-kernel of $\bD^s$, namely the part of the NP coefficient $\np{m}{s}$ \textit{not} captured by $\bD^s\tQ_{s+1}$.
Denoting the vector space of definite helicity $s$ and angular momentum $\l$ (and conformal dimension $s+2$) by $\sfY^s_\l$, we have that $\np{m}{s}\in\bigoplus_{\l=1}^\infty \sfY_\l^1$ while $\bDcal^s\tQ_{s+1}\in \bigoplus_{\l=s+1}^\infty \sfY_\l^1$.
Therefore $G_{(s+2,1)}\simeq \bigoplus_{\l=1}^s \sfY_\l^1$ and is captured in $\np{m,G}{s}$.

If we were in Maxwell's theory or in linearized gravity, then the EOM for the global and local parts would split and become independent of one another (see \cite{Freidel:2021ytz} for a demonstration). 
In the following, for simplicity we set $\np{m,G}{s}\equiv 0$ and compute the evolution equation for $\tQ_s$.
Under this restriction, plugging \eqref{ansatzcharge} in \eqref{phi0n} and using (\ref{ethcommutnYM}) to commute the derivatives on the sphere, we get an evolution equation for $\tQ_{s}$, $s\geq 2$. 
The general result is reported in appendix \ref{APPphi0Q} and takes the form
\begin{empheq}[box=\fbox]{align}
    &\quad \bDcal^{s+1}\big(\pa_u\tQ_{s+2}-\Dcal\tQ_{s+1}\big)-\sum_{p=0}^s\bDcal^p\big[ \tQ_{-1},\bDcal^{s-p}\tQ_{s+2}\big]_\g +(2s+3)\big[\tQ_0,\bDcal^s\tQ_{s+1}\big]_\g \nn\\
    &+\sum_{p=0}^{s-1}\frac{(s+1)!}{(p+2)!(s-p-1)!}\big[\bDcal^{p+1} \tQ_{0},\bDcal^{s-p-1}\tQ_{s+1}\big]_\g \nn\\
    &+\sum_{p=0}^{s-1}\frac{(s+2)!}{(p+2)!(s-p)!}\big[\bDcal^{p+1}\tQ_{p+1},\bDcal^{s-1-p}\tQ_{s-p}\big]_\g \nn\\
    &\!\!=\sum_{p=0}^{s-1}\frac{(s+1)!}{(p+2)!(s-p-1)!}\big[\bDcal^{p+1}\tbQ_{0},\bDcal^{s-p-1}\tQ_{s+1}\big]_\g \label{Qnevol}\\
    &-\sum_{p=0}^{s-1}\frac{(s+1)!}{(p+2)!(s-1-p)!}\big[\Dcal^{p+1}\tbQ_{p+1},\bDcal^{s-1-p}\tQ_{s-p}\big]_\g \nn\\
    &+\sum_{p=0}^{s-1}\frac{s!}{(p+1)!(s-1-p)!}\Dcal\big[\Dcal^p\tbQ_{p+1},\bDcal^{s-1-p}\tQ_{s-p}\big]_\g\nn\\
    &+ \sum_{p=0}^{s-1}\sum_{k=0}^{p-1}\frac{(s+1)!\big((s+2)(p-k)-(k+1)(s-p)\big)}{(p+1)(p+2)(p-k)!(k+1)!(s-p)!}~\cdot \nn\\
    &\hspace{5cm}\cdot \big[\bDcal^{s-1-p}\tQ_{s-p},[\Dcal^k\tbQ_{k+1},\bDcal^{p-1-k}\tQ_{p-k}]_\g\big]_\g. \nn
\end{empheq}

As already computed in \eqref{phim1evol}, the evolution of $\tQ_2$ is given by
\begin{equation}
    \boxed{\bDcal\big(\pa_u\tQ_{2}-\Dcal\tQ_{1}\big)-[\tQ_{-1},\tQ_2]_\g +3[\tQ_{0},\tQ_{1}]_\g =0}. \label{Q2evol}
\end{equation}
The appearance of the commutators breaks the pattern initiated in (\ref{Q0evol})-(\ref{Q1evol}). 
This is not surprising in the sense that any commutator is at least made of two charges, hence has a minimal Carrollian weight of 4, so that such a term could not appear at a lower order. 
Note that if we were in the abelian theory, the commutators would not exist and $\Dcal,\bDcal$ would reduce to $D,\bD$ so that we would leverage the injectivity of $\bD$ and recover the pattern $\pa_u \tQ_{s}=D\tQ_{s-1}$, $\forall\,s\in\N$, which was satisfied for $s=0,1$. 

In the next chapter, we truncate this EOM and analyze the recursion relation
\begin{equation} \label{fundamentalPattern}
    \pa_u\tQ_s=\Dcal\tQ_{s-1}
\end{equation}
for $s\geq 0$. 
Notice that this amounts to neglecting all the commutators, both the $\big[\tQ,\tQ\big]_\g$ present on the LHS of \eqref{Qnevol} and the $\big[\tQ,\tbQ\big]_\g$ present on the RHS.
The latter naturally disappear if we impose the self-duality condition $\bar\phi_m=\bar\phi= \bar\phi_{\bm}=0$ (see also the RHS of \eqref{phi0n}).
Yet, the former are still present in this case.
According to the recent result \cite{Kmec:2025ftx}, the pattern \eqref{fundamentalPattern} can be recovered starting from the self-dual action of YM in twistor space.
This would suggests that there is a non-linear redefinition of the charges $\tQ_s$ which re-absorbs the $\big[\tQ,\tQ\big]_\g$ commutators into the recursion relation \eqref{fundamentalPattern}.
From a different (related?) viewpoint, one could truncate \eqref{Qnevol} at linear order in the anti-holomorphic coupling constant $\bar g$---cf. the discussion around \eqref{intro2YM}.
This allows us to discard the $\big[\tQ,\tQ\big]_\g$ terms, since they are at least $\Ocal(\bar g^2)$.
This truncation also turns $\bDcal$ into $\bD$.
However, the part of the $\big[\tQ,\tbQ\big]_\g$ commutators which involves the soft part of $\tbQ_s$, i.e. the $\Ocal(g)$ piece, still contributes to the evolution equation of $\tQ_s$.
This may suggest that there exists a better field redefinition that \eqref{ansatzcharge}, where $\tQ_s$ is dressed by $\tbQ{}^{\mathsf{S}}_s$.

Note that we can at least state that by imposing the field strength self-duality condition---namely setting the complex conjugate NP scalars to 0---and simultaneously taking the $\Ocal(\bar g)$ truncation, then \eqref{Qnevol} does reduce to \eqref{fundamentalPattern}.\footnote{More precisely, we are left with $\bD^{s+1}\big(\pa_u \tQ_{s+2}-\Dcal\tQ_{s+1}\big)=0$, which amounts to \eqref{fundamentalPattern} since $\bD$ is injective on positive spin-weight operators.}

\chapter{Non-Abelian Gauge Theory II \label{secNAGTII}}

This chapter corresponds to the material published in \cite{Cresto:2025bfo}.

\section{Introduction}

We start in section \ref{secYM:Preliminaries} with a reminder of the notation introduced in the previous chapter, together with the defining evolution equations for the YM higher spin charge aspects $\tQ_s$.

In section \ref{secYM:Master}, we then define the \textit{master charge}.
This is a pairing, via integration over the sphere, between $\tQ_s$ and a dual variable called $\a{s}$, such that one gets a scalar.
On top of that, the `$s$' dependence disappears by summing over all spin-weights.
In other words, the master charge $\Qa^u$ takes the form
\begin{equation}
    \Qa^u=\sum_s\cQ{s}{\a{s}}.
\end{equation}
If I had to pick one equation to memorize from this thesis, it would be this one.
It is extremely simple, and from it, one can deduce all the key aspects of our analysis.
Mathematically, it corresponds to a pairing between an algebra and its dual. 
Physically, it is a \textit{conserved quantity}.
More precisely, by letting $\a{s}(u,z,\bz)$ depend on time,\footnote{Contrast this with the traditional celestial picture where the smearing symmetry parameters are time independent, cf.\,\ref{secIntro}.} we can \emph{constraint} their time evolution such that the master charge is conserved in the absence of radiation (i.e. in the absence of flux through $\scri$).
This leads to the notion of dual equations of motion and allows us to construct the symmetry transformation of the asymptotic gauge potential $A$, respectively
\begin{equation}
    \pa_u\a{s}=\Dcal\a{s+1}\qquad\quad \textrm{and}\qquad\quad \da A=-\Dcal\a0.
\end{equation}
Hence, $\da A$ depends on the whole collection of $\a{s}$ implicitly through their recursive time evolution.

Next, we show that this transformation is the realization of the $\cS$-algebroid in section \ref{secYM:algebroid}.
The use of algebroids is essential all throughout this thesis (notably to describe radiation) and we will give ample explanations in due time.
For now, simply notice that the $\a{s}$ are field dependent (since $\Dcal=D+\adA$) and algebroids are therefore the adapted mathematical framework.

Furthermore, we relate the master charge to the Noether charge $\Qa\equiv\Qa^{-\infty}$ in section \ref{secYM:Noether}.
The essence of the demonstration consists in constructing $\da \bnews$ such that the asymptotic \textit{holomorphic} symplectic potential is invariant under the symmetry transformation.
\textit{We thus obtain a canonical representation of the symmetry algebroid for all spins and at all orders in $g$.}

We then give an explicit solution of the dual EOM in section \ref{secYM:solEOM}.
This allows to relate Carrollian and celestial symmetry parameters $\Aa{s}$ by defining the latter as the initial condition of the former, namely $\Aa{s}:=\a{s}\big|_{u=0}$.
Next, in section \ref{secYM:RenormCharge}, we show how to algorithmically and non-perturbatively build a renormalized charge aspect for each spin which is conserved in time in the absence of radiation.
This is a generalization of what we called $q_s(\Aa{s})$ in the introduction \ref{sec:FPR}.
We then dedicate section \ref{secYM:ActionOnA} to the computation of the quadratic action on $A$, recovering the results of \cite{Freidel:2023gue}.

In section \ref{secYM:wedge}, we introduce the \textit{covariant wedge algebra}, both on $\scri$ and on the sphere.
$\Wcal_A(\scri)$ is the \textit{Lie algebra} associated to the \textit{Lie algebroid} $\cS$.
In particular at a \textit{non-radiative cut} of $\scri$, it corresponds to a restriction on the symmetry parameters, of the form $\Dcal^{s+1}\Aa{s}=0$, $s\geq 0$.
This thus generalizes the wedge condition $D^{s+1}\Aa{s}=0$ mentioned in the FPR summary \ref{sec:FPR}.
The general Noether result for the algebroid guarantees that the covariant wedge algebra is canonically realized.

Finally, section \ref{secYM:twistor} shows how to relate the graded vector $\a{}$ to a function $\fa$ which naturally appears in twistor space.
This link with twistor space will be discussed in way more details in sections \ref{sec:twistor} and \ref{secEYM:twistor}.
The purpose here is just to give the gist of the change of perspective.
A collection of appendices \ref{AppNAGTII} complements the main text with details of demonstrations.

\section{Symplectic potential and equations of motion \label{secYM:Preliminaries}}

We study non-abelian gauge theory over flat space, with the associated Lie algebra $\g$, and denote by $A_\mu$ and $F_{\mu\nu}$ the gauge potential and field strength respectively.
We work with spin-weighted scalars rather than tensors over the sphere.
As an example, when performing a $1/r$ expansion around null infinity of the YM gauge field \cite{Strominger:2017zoo}, where for instance $A_z$ takes the form
\begin{equation}
    A_z=\sum_{n=0}^\infty\frac{A_z^{(n)}(u,z,\bz)}{r^n},
\end{equation}
then we can show that only $A_z^{(0)}$ is part of the asymptotic phase space (cf. section \ref{secNAGTI:PS}) and we thus denote its contraction with the frame field by $A\equiv A^{(0)}_B m^B$.
Similarly, the canonically conjugated variables will be denoted by $\news=\pa_u A$ and its complex conjugate $\bnews$.
Notice also that we work with Lie algebra valued quantities to lighten the notation, i.e. $A_\mu=A_\mu^b Z_b$ with $\{Z_b\}$ a basis of anti-hermitian Lie algebra generators. 
Besides, we adopt the convention that the complex conjugation acts only on the spacetime components, i.e. relates to the frame fields, and not on the algebra generators (in case the latter are complex).
Even if we shall write expressions such as $\bA$, the latter must then be understood as $\bA=\bA^bZ_b=A_B^{(0)b}\bm^B Z_b$.

$A$ and $\news$ parallel the role of the shear and news in asymptotically flat spacetimes.
This is transparent since the asymptotic YM symplectic potential is given by\footnote{$\Pairg{Z_a,Z_b}=-\Tr(Z_aZ_b)$, where $\Tr$ denotes the trace, i.e. the Cartan–Killing form for $\g$, normalized such that $\Tr(Z_aZ_b)=-\delta_{ab}$ (recall that we use anti-hermitian generators). \label{footPairg}}${}^,$\footnote{$\int_\scri\equiv\int_\scri\rd u\wedge\bep_S$.}
\begin{equation}
\boxed{\Theta=\frac{2}{\gYM^2}\int_{\scri} \Pairg{\bnews,\delta A}}, \label{SymPotYM}
\end{equation}
which is the holomorphic Ashtekar-Streubel asymptotic symplectic potential.
Recall that the Ashtekar-Streubel symplectic potential \cite{ashtekar1981symplectic} is real and takes the form 
\begin{equation}
\Theta^{\mathsf{AS}}=\frac1{\gYM^2} \int_{\scri}\Big(\Pairg{\bnews,\delta A}+\Pairg{F,\delta\bA}\Big).
\end{equation}
If we add a total variation together with a boundary term, namely 
\begin{equation}
    \frac{1}{\gYM^2}\left(\delta\left(\int_\scri \langle\bnews, A\rangle_\g\right)- \int_\scri \pa_u\langle A,\delta\bA\rangle_\g\right),
\end{equation}
we transform it into its holomorphic version.
Notice that we allow for \textit{complex canonical transformations}.
This is essential in the upcoming Noether analysis.

By performing an asymptotic expansion around $\scri$ and relating the field strength to the higher spin charges aspects $\tQ_s$, $s\geqslant -1$, the YM EOM in vacuum are approximated by \cite{Freidel:2023gue}
\begin{equation}
    \boxed{\pa_u\tQ_s=D\tQ_{s-1}+\big[A,\tQ_{s-1}\big]_\g},\qquad s\geqslant 0. \label{EOMYM}
\end{equation}
We define $\Dcal$ the asymptotic gauge covariant derivative on the sphere,
\begin{equation}
    \boxed{\Dcal:=D+\adA}, \label{defDcalYM}
\end{equation}
where $\adA(\cdot)=[A,\cdot]_\g$ is the adjoint action in the Lie algebra $\g$.
The EOM are thus expressed as $\pa_u\tQ_s=\big(\Dcal\tQ\big)_s\equiv\Dcal\tQ_{s-1}$.
The initial condition $\tQ_{-1}=-\bnews$ encodes the radiative content of the gauge potential.
Here we take \eqref{EOMYM} as a defining property for $\tQ_s$ and it is sufficient to know that the evolution equations are the exact asymptotic EOM for $s=0,1$ and then become an approximation.
We refer the reader to the previous chapter and in particular \eqref{Qnevol} for the non-approximate equation.
From \eqref{defDcalYM}, it is clear that $A\in\Ccarg{1,1}$.
Moreover, since the initial charge $\tQ_{-1}=-\bnews\in\Ccarg{2,-1}$, we deduce that $\tQ_s\in\Ccarg{2,s}$.

\section{Master charge and symmetry transformation \label{secYM:Master}}

The main insight from this thesis is to introduce $\g$-valued time dependent symmetry parameters $\a{s}\in\Ccarg{0,-s}$.
Besides, we denote the series $(\a{s})_{s\in\N}$ by $\a{}$.
In other words, $\a{}$ is a graded vector such that
\begin{equation}
    \a{}\in\tV(\scri,\g), \qquad \tV(\scri,\g):=\bigoplus_{s=0}^\infty\tV_s\quad\textrm{with}\quad \tV_s\equiv\Ccarg{0,-s}.
\end{equation}

We define the smeared charges\footnote{$\int_S\equiv\int_S\bep_S$ and the product $\Tr\big(\tQ_s\a{s}\big)\in\Ccar{2,0}$ is a scalar density.}
\begin{equation}
\cQ{s}{\a{s}}:=\frac{2}{\gYM^2}\int_S\Pairg{\tQ_s,\a{s}}(u,z,\bz). \label{Qsas}
\end{equation}
Since $\cQ{s}{\a{s}}$ involves an integral over the sphere (at $u=\mathrm{cst}$), we can freely integrate by parts.
Using \eqref{defDcalYM} to construct 
\begin{equation}
    (\Dcal\a{})_s=\Dcal\a{s+1}:=(D+\adA)\a{s+1},
\end{equation}
we get that\footnote{$\Tr\big(\tQ_s[A,\a{s+1} ]_\g\big)=-\Tr\big([A,\tQ_s]_\g\a{s+1}\big)$  and we assume that $\sum_s\tQ_s\a{s}$ is regular at the punctures if $S$ is not the regular sphere.}
\begin{equation}
    \cQ{s}{\Dcal\a{s+1}}=-\Dcal\cQ{s}{\a{s+1}}, \label{QDDQ}
\end{equation}
where
\begin{equation}
    \Dcal\cQ{s}{\a{s+1}}=\frac{2}{\gYM^2}\int_S\Pairg{(\Dcal\tQ)_{s+1},\a{s+1}}.
\end{equation}

We then construct the \textit{master charge} $\Qa^u$,
\begin{equation}
\boxed{\Qa^u := \sum_{s=0}^{\infty} \cQ{s}{\a{s}}}. \label{masterChargeYM}
\end{equation}
Using \eqref{EOMYM} and \eqref{QDDQ}, its time evolution takes the form
\begin{equation}
\pa_u\Qa^u = \sum_{s=0}^\infty \Big(\cQ{s}{\pa_u \a{s}} - \cQ{s-1}{\Dcal\a{s}}\Big) = -\cQ{-1}{\Dcal \a0}+\sum_{s=0}^\infty Q^u_{s}\big[\pa_u \a{s} - \Dcal\a{s+1}\big].
\end{equation}
By imposing what we shall refer to as the dual EOM, 
\begin{equation}
\boxed{\pa_u\a{s}\equiv\dot\alpha_s= \Dcal\a{s+1}}, \quad s\geqslant 0, \label{dualEOMYM}
\end{equation}
and remembering that $\tQ_{-1}=-\bnews$, we obtain that 
\begin{equation}
\Qad^u = \bnews\big[\Dcal\a0\big]=-\bnews\big[\da A\big], \label{Qadot}
\end{equation}
where we have defined the transformation\footnote{This choice of sign implies that the anchor $\delta$ \cite{algebroid} is an anti-homomorphism of Lie algebroids.
We also take that sign convention in GR. 
Sign consistency is especially important when coupling GR and YM.}
\begin{equation}
\boxed{\da A:= -\Dcal\a0}.  \label{deltaA}
\end{equation}

The master charge is thus conserved in the absence of left-handed gluons, namely $\Qad^u=0$ when $\bnews=0$.
Moreover, by imposing the fall-off condition $\Qa^{+\infty}=0$,\footnote{Namely $\lim_{u\to\infty}\tQ_s\a{s}=0$.} it takes the form of an integral over a portion of $\scri$,
\begin{equation}
    \Qa^u=\int_u^{\infty}\bnews\big[\da A\big]. \label{Qau}
\end{equation}
Notice that $\Qa^{-\infty}\propto\int_{\scri}\Pairg{\bnews,\da A}$. 
In other words $\Qa^{-\infty}$ can be constructed from the symplectic potential \eqref{SymPotYM}.
Hence, the dual EOM are key in order for $\Qa^{-\infty}$ to be a Noether charge, see Sec.\,\ref{secYM:Noether}.

\section{$\cS$-algebroid \label{secYM:algebroid}}

In this section, we prove that the $\alpha$-transformation \eqref{deltaA} is a realization on the asymptotic Yang-Mills phase space $\PS$ of an algebroid extension of the $s$-algebra \cite{Guevara:2021abz}.

\subsection{Comments about Lie algebroids  \label{Sec:Liealg}}

For the reader's convenience, we provide a reminder about Lie algebroids \cite{algebroid, Barnich:2010xq, Ciambelli:2021ujl}, since  this notion will be essential in our understanding of the nature of the time dependent symmetry. 

\begin{tcolorbox}[colback=beige, colframe=argile] \label{ReminderAlgebroid}
\textbf{Reminder [Lie algebroid -- Definition]}\\
A Lie algebroid $\big(\mathcal{A},\lbr\cdot\,,\cdot\rbr,\hat{\nkor} \big)$ is a vector bundle $\mathcal{A}\to\M$ with a Lie bracket  $\lbr\cdot\,,\cdot\rbr$ on its space of sections $\Gamma(\mathcal{A})$ and a vector bundle morphism $\hat\nkor:\mathcal{A}\to T\M$, called the anchor.
The anchor gives rise to an anchor map $\nkor:\Gamma(\mathcal{A})\to\X(\M)$ with two fundamental properties, namely the compatibility between Lie algebras and the Leibniz rule:
\begin{subequations}
\begin{align}
    i)\hspace{-2cm}& &\nkor\big(\lbr\alpha,\alpha'\rbr\big) &=\big[\nkor(\alpha),\nkor(\alpha')\big]_{T\M}, \label{anchorRep}\\
    ii) \hspace{-2cm}& &\big\lbr\alpha,f\alpha'\big\rbr &=f\big\lbr\alpha,\alpha'\big\rbr+\big(\nkor(\alpha)\triangleright f\big)\alpha', \label{anchorLeib}
\end{align}
\end{subequations}
for $\alpha,\alpha'\in\Gamma(\mathcal{A})$ and $f\in \mathcal{C}^\infty(\M)$. The RHS of $i)$ is the Lie bracket of vector fields on $\M$ while in $ii)$, $\nkor(\alpha)\triangleright f$  is the derivative of $f$ along the vector field $\nkor(\alpha)$.
\end{tcolorbox}

A canonical way to construct algebroids arises when the action of a Lie algebra on a given manifold exists. In this case we can construct a \emph{symmetry algebroid} as follows.

\begin{tcolorbox}[colback=beige, colframe=argile, breakable] \label{ReminderAlgebroidbis}
\textbf{Reminder [Symmetry algebroid]}\\
Let us assume that we have a manifold $\M$ equipped with the action of a Lie algebra $(\g,[\cdot\,,\cdot]_\g)$. 
In practice this means that there exists an
infinitesimal left action  $\hat\nkor : \g\to\X(\M)$, such that 
 \be \label{symcomp}
 \big[\hat\nkor(a), \hat\nkor(b)\big]_{T\M} 
= -\hat\nkor\big([a,b]_\g\big), \qquad a,b 
\in \g.
 \ee 
One can build the associated Lie algebroid $\mathcal A$ as the trivial bundle  $\M\times\g\to  \M$ with anchor map $\nkor:\Gamma(\mathcal{A})\to\X(\M)$ such that $\nkor(\alpha)(x)=\hat\nkor(\alpha(x))$ for each $x\in \M$ and $\alpha\in\Gamma(\cA)$. 
We identify sections of $\mathcal{A}$ with Lie algebra functions on $\M$. This naturally extends the vector fields action to $\Gamma(\mathcal{A})$.
The Lie algebroid bracket is then given by
\begin{equation}
    \lbr\alpha,\alpha'\rbr :=[\alpha,\alpha'] +\nkor(\alpha')\triangleright\alpha-\nkor(\alpha)\triangleright\alpha', \label{assocLieoidBracket}
\end{equation}
where we have defined $[\alpha,\alpha'](x):= [\alpha(x),\alpha'(x)]_\g$
for $\alpha,\alpha'\in\Gamma(\mathcal{A})$. 
In other words, the bracket $[\cdot\,,\cdot]$ over $\Gamma(\mathcal A)$ is the  fiberwise lift of $[\cdot\,,\cdot]_\g$ when we identify the sections $\alpha,\alpha'\in\Gamma(\mathcal A)$ with $\g$-valued functions over $\M$.
\end{tcolorbox}

In the physics literature the notion of symmetry algebroid appears naturally in field theory (see e.g. \cite{Gomes:2019xto} and references therein). 
It is used when we have the action of a symmetry algebra on the space of fields $\cF$ defined as the space of sections over a spacetime manifold $\cM$.
In this case the algebroid is a bundle over field space $\cF$ and the algebroid bracket is a generalization of the Lie bracket to field dependent transformations. It is exactly the modified bracket introduced in \cite{Barnich:2010eb}.

\subsection{$\sfS$-bracket}

Let us first show that one can build a bracket $\lbr\alpha,\alpha'\rbr^\g$ such that\footnote{$\big[\da,\dap\big]$ is the Lie bracket of vector fields over fields space \cite{Freidel:2020xyx, saunders1989geometry} and \eqref{actiondelta}.}
\begin{equation}
\boxed{\big[\da,\dap\big]A=-\delta_{\lbr\alpha,\alpha'\rbr^\g}A}. \label{SalgebraMorphism}
\end{equation}
From the transformation \eqref{deltaA}, we deduce that
\begin{align}
\da\dap A &=-\da\big(D\ap0+[A,\ap0]_\g\big) \nn\\
&=-D\da\ap0-\big[A,\da\ap0\big]_\g+\big[\Dcal\a0,\ap0\big]_\g \\
&=-\Dcal(\da\ap0)+\big[\Dcal\a0,\ap0\big]_\g. \nn
\end{align}
The action of the commutator $\big[\da,\dap\big]$ on $A$ is thus
\begin{equation}
    \big[\da,\dap\big]A =-\Dcal\big(\da\ap0-\dap\a0\big) +\big[\Dcal\a0,\ap0\big]_\g- \big[\Dcal\ap0,\a0\big]_\g.
\end{equation}
Merely using the Leibniz rule on the third term, we get
\begin{equation}
    \big[\da,\dap\big]A=\Dcal \big([\a0, \ap0]_\g+\dap\a0-\da\ap0\big)= \Dcal\lbr\alpha,\alpha'\rbr^\g_0,
\end{equation}
for 
\begin{equation}
    \lbr\alpha,\alpha'\rbr^\g_0=[\a0,\ap0]_\g+\dap\a0-\da\ap0. \label{Sbracket0}
\end{equation}
Notice that the Leibniz rule for the derivative operator $\Dcal$ on the Lie algebra bracket $[\cdot\,,\cdot]_\g$ is nothing else than the Jacobi identity for this bracket.
Indeed,
\begin{equation}
    \Dcal[\a{s},\ap{n}]_\g=[D\a{s},\ap{n}]_\g+[\a{s},D\ap{n}]_\g+\big[A,[\a{s},\ap{n}]_\g\big]_\g=[\Dcal\a{s},\ap{n}]_\g+[\a{s},\Dcal\ap{n}]_\g, \label{DcalLeibniz}
\end{equation}
after using Jacobi.
Since the evolution in time of $\a0$ is related to $\a1$ via \eqref{dualEOMYM}, we have to define a bracket of degree 1 which respects this property as well.
By induction, we thus need to construct $\lbr\a{},\ap{}\rbr^\g_s$, for all $s\in\N$.

Hence we generalize the bracket \eqref{Sbracket0} for $s\geqslant 0$ by writing
\begin{equation} \label{Sbracket}
    \boxed{\lbr\alpha,\alpha'\rbr^\g= [\alpha,\alpha']^\g+\dap\alpha-\da\alpha'},
\end{equation}
where the evaluation at degree $s$ satisfies
\begin{equation}
    [\alpha,\alpha']^\g_s=\sum_{n=0}^s\big[\a{n} ,\ap{s-n}\big]_\g\qquad\textrm{and}\qquad (\da\alpha')_s=\da\alpha'_s.
\end{equation}
Notice that $[\alpha,\alpha']^\g$ is the natural extension of the Lie algebra bracket $[\cdot\,,\cdot]_\g$ to the full graded vector space $\tV(\scri,\g)$.
Indeed, using the inclusion map $\iota:\tV_s\to\tV(\scri,\g)$, such that $\a{s}\mapsto (0,\ldots,0,\a{s},0,\ldots)$, we see that
\begin{equation}
    \big[\iota(\a{s}),\iota(\ap{s'})\big]^\g_n= \left\{\begin{array}{ll}
        [\a{s},\ap{s'}]_\g & \textrm{if }n=s+s', \\
        0 & \textrm{otherwise}.
    \end{array}\right.
\end{equation}
This also means that $[\cdot\,,\cdot]^\g$ is a bracket of degree 0.

\subsection{$\cS$-algebroid construction}

Since the Carrollian symmetry parameters $\alpha$ are restricted to follow the dual EOM \eqref{dualEOMYM}, we define the $\sfS$-space accordingly.

\begin{tcolorbox}[colback=beige, colframe=argile]
\textbf{Definition [$\sfS$-space]}
\be 
\sfS :=\Big\{\,\alpha\in\tV(\scri,\g)~\big|~\pa_u\a{s}=\Dcal\a{s+1},~  s\geqslant 0~ \Big\}.
\ee 
\end{tcolorbox}

\ni The fact that the parameters $\a{s}$ are constrained to follow an evolution equation involving $A$ implies that they are field dependent.
It is thus natural and necessary that the $\sfS$-bracket \eqref{Sbracket} involves the variations $\da\ap{}$ and $\dap\a{}$.
This is a well-known fact in the algebroid literature \cite{algebroid}.
The following theorem then represents one of the main result of this chapter.

\begin{tcolorbox}[colback=beige, colframe=argile] \label{theoremSAlgebroid}
\textbf{Theorem [$\cS$-algebroid]}\\
The space $\cS\equiv\big(\sfS,\lbr\cdot\,,\cdot\rbr^\g,\delta\big)$ equipped with the $\sfS$-bracket \eqref{Sbracket} and the anchor map $\delta$,
\begin{align}
    \delta : \,\,&\cS\to \X(\PS) \nn\\
     &\alpha\mapsto\da, \label{defdeltaYM}
\end{align}
is a Lie algebroid over $\PS$.
This means that the anchor map is a morphism of Lie algebroid, namely $\big[\da,\dap\big]A=-\delta_{\lbr\alpha,\alpha'\rbr^\g}A$, cf. \eqref{SalgebraMorphism} and also that the $\sfS$-bracket closes, i.e. $\pa_u\lbr\a{},\ap{}\rbr^\g= \Dcal\lbr\a{},\ap{}\rbr^\g$.
\end{tcolorbox}

\paragraph{Proof:}
First of all, the proof that the $\sfS$-bracket satisfies the Jacobi identity follows from a general result about symmetry algebroids, see \cite{Cresto:2024mne,algebroid}.
In words, since $[\cdot\,,\cdot]^\g$ is built from the Lie bracket $[\cdot\,,\cdot]_\g$---and thus satisfies Jacobi on its own---and that the anchor $\da$ acts as a derivative operator on $[\cdot\,,\cdot]^\g$, i.e. $\da[\cdot\,,\cdot]^\g=[\da\cdot\,,\cdot]^\g+[\cdot\,,\da\cdot]^\g$, then the Jacobi identity for $\lbr\cdot\,,\cdot\rbr^\g$ is guaranteed to hold.
For the sake of completeness and to illustrate this general fact, we give the explicit demonstration in App.\,\ref{App:JacobiYM} (this result even holds over the whole space $\tV(\scri,\g)$).

Second of all, we already proved that $\delta$ is a Lie algebroid anti-homomorphism, cf. \eqref{SalgebraMorphism}, and thus an anchor map.

Third of all, in App.\,\ref{App:dualEOMYM}, we report the technical details to show that $\lbr\cdot\,,\cdot\rbr^\g:\sfS\times\sfS\to\sfS$ is a well-defined bracket over $\sfS$, i.e. that it satisfies the dual EOM,
\begin{equation}
    \boxed{\pa_u\lbr\a{},\ap{}\rbr^\g= \Dcal\lbr\a{},\ap{}\rbr^\g}. \label{dualEOMYMbracket}
\end{equation}
The essence of the proof goes as follows:
We mentioned already that $\Dcal$ is the asymptotic gauge covariant derivative onto $S$.
Its appearance in the current analysis is thus natural.
However, $\Dcal$ may or may not be a \textit{derivative operator} depending on which bracket is acts on.
Hence, we compute the potential Leibniz rule anomaly of $\Dcal$ (and similarly for $\pa_u$) on the various brackets.
This then allows us to straightforwardly ensure that the bracket $\lbr\cdot\,,\cdot\rbr^\g$ closes in $\sfS$.

Concretely, we define the Leibniz rule anomaly for any bracket $[\cdot\,,\cdot]$ and differential operator $\mathscr{D}$ as
\begin{equation}
    \cA\big([\cdot\,,\cdot],\mathscr{D} \big)=\mathscr{D}[\cdot\,,\cdot]-[\mathscr{D}\cdot\,,\cdot]-[\cdot\,,\mathscr{D}\cdot].
\end{equation}
We find that
\begin{equation} \label{DcalLeibniz1}
    \cA_s\big([\a{},\ap{}]^\g,\Dcal\big)= \big[\dap A,\a{s+1}\big]_\g-\big[\da A,\ap{s+1}\big]_\g
\end{equation}
and
\begin{equation}\label{DcalLeibniz2}
    \cA\big(\lbr\a{},\ap{}\rbr^\g, \Dcal\big)=\delta_{\Dcal \a{}}\ap{}-\delta_{\Dcal \ap{}}\a{}.
\end{equation}
We can also compute
\begin{equation}\label{DcalLeibniz3}
    \cA\big(\lbr\a{},\ap{}\rbr^\g, \pa_u\big)=\delta_{\pa_u\a{}}\ap{}-\delta_{\pa_u\ap{}}\a{}.
\end{equation}
Combining \eqref{DcalLeibniz2} with \eqref{DcalLeibniz3}, we obtain
\begin{equation}
    \cA\big(\lbr\a{},\ap{}\rbr^\g, \pa_u-\Dcal\big)=\delta_{(\pa_u-\Dcal)\a{}}\ap{}-\delta_{(\pa_u-\Dcal)\ap{}}\a{}.
\end{equation}
If $\a{},\ap{}\in\sfS$, then the RHS of this last equation vanishes while its LHS reduces to $(\pa_u-\Dcal)\lbr\a{},\ap{}\rbr^\g$.
Therefore
\begin{equation}
    \pa_u\lbr\a{},\ap{}\rbr^\g=\Dcal \lbr\a{},\ap{}\rbr^\g,
\end{equation}
so that $\lbr\a{},\ap{}\rbr^\g\in\sfS$ if $\a{},\ap{}\in\sfS$, which concludes the proof of \eqref{dualEOMYMbracket} and of the theorem.

\section{Noether charge \label{secYM:Noether}}

In quantum mechanics, the presence of a symmetry associated with the Lie algebra $\g$ acting canonically on the phase space $\PS$, means that there exists a moment map\footnote{In the physics literature, it is customary to work with the co-moment map instead 
 $Q: \mathfrak{g} \to  \mathcal{C}(\PS)$, which is simply the dual map.} $Q^*: {\cal{P}} \to \mathfrak{g}^*$, \cite{da2001lectures, weinstein1981symplectic}. This is the essence of Noether's theorem \cite{noether1971invariant}.

In practice the Noether's theorem means that for any element $X\in \g$ we have a Noether charge $Q[X]$ which is a functional on phase space $\PS$ such that 
\be 
\poisson{Q[X], Q[Y]}= Q\big[[X,Y]_\g\big], \qquad 
\poisson{Q[X], O} =\delta_X O.
\ee 
where $O\in \mathcal{C}(\cP)$ is a phase space functional and  $\delta_X$ represents the action of $\g$ on $\cP$.
In our case, we have a symmetry algebroid and a generalization of the moment map theorem for the \hyperref[theoremSAlgebroid]{$\cS$-algebroid}.
\medskip

The goal of this section is thus to prove that there exists a canonical representation of the higher spin algebroid bracket \eqref{Sbracket} on the holomorphic Ashtekar-Streubel phase space \eqref{SymPotYM}. 
This is the central result of this chapter, which is valid provided we impose proper boundary conditions. 
In this work, we choose the fields $(A,\bA)$ to belong to the Schwartz space.\footnote{
The Schwartz space $\cS^{ch}$ is defined \cite{schwartz2008mathematics} as the set of continuous functions $f(u)$ that decay faster than any positive inverse power of $u$ as $|u| \rightarrow \infty$. Formally,
\be 
\cS^{ch} = \left\{f \in C^{\infty}\,|\, \forall \,\alpha, \beta \in \mathbb{N}, || f ||_{\alpha, \beta} < \infty \right\},
\ee
where $
||f||_{\alpha, \beta} = \underset{u \in \mathbb{R}}{\rm sup} \left|u^{\alpha} \pa_u^{\beta} f(u) \right|.$ 
\label{foot1}}
This hypothesis about the fall-off in $u$ is important for two reasons.
On one hand, at this stage of the analysis, we discard any boundary terms that arise when integrating by parts $\pa_u$ derivatives.
Therefore, the fields must tend to 0 when $|u|\to\infty$.\footnote{We let for future work a potential relaxation of this condition.
See \cite{Nagy:2024dme, Nagy:2024jua} for some recent developments.}
On the other hand, in section \ref{secYM:solEOM}, we construct solutions of the dual EOM \eqref{dualEOMYM} which are polynomials in $u$.
Hence, in order for $\tQ_s\a{s}$ to tend to 0 when $|u|\to\infty$, the gauge potential must decay faster than any positive inverse power of $u$, namely be part of the Schwartz space.
\medskip

We now show that the action of $\cS$ on the YM phase space is canonical.
So far, we have only described how $\cS$ acts on $A$. 
We also need to describe its action on $\bnews$.
For this, we use that
$\da A$ is a functional of $A$ only. 
This action involves local terms $\pa_u^n D^m A$, but it also involves non local terms 
$\pa_u^{-n} D^m A$ where we choose\footnote{The definition of $\pui$ is ambiguous and depends on a base point $\beta$. We could also choose  $\pui O =\int_\beta^u \rd u' O(u')\rd u'$ and more generally 
\begin{equation}
    \big(\pui[n]O\big)(u)=\int_\beta^u \rd u_1\int_\beta^{u_1}\rd u_2\ldots\int_\beta^{u_{n-1}}\rd u_n O(u_n).
\end{equation} 
In this section we take $\beta=+\infty$. Another choice will be used in the next section. The context makes it clear which inversion we use.}
$\pui O := \int_{+\infty}^u O(u') \rd u'$.
Moreover, if $\a{} \in \cS^s$ (cf.\,\eqref{defFiltrationYM}), the non-locality is bounded since the most non-local term involves $\pui[s]A$ at most.\footnote{As we shall see explicit in \eqref{solalpha210}, $\da A$ contains terms of the type $\pui(A\pui A)$, thence $\delta(\da A)$ involves $\pui(\delta A\pui A)$ and $\pui\big(A\pui(\delta A)\big)$.
Using the generalized Leibniz rule \eqref{LeibnizRuleGeneral} for $\pui$, the first variation contributes to terms of the type $\big(\pui[(2+n)]A\big)(\pa_u^n\delta A)$ while the second variation involves terms like $\big(\pui[(1+n)]A\big)(\pa_u^{n-1}\delta A)$, for $n\in\N$.}
The general variation rule \eqref{actiondelta} then implies that we can write\footnote{$\delta$ here is the fields space exterior derivative, not to be confused with the anchor. 
By definition, both are related according to $I_{\da}(\delta A)=\da C$, for $I$ the fields space interior product.}${}^,$\footnote{Notice that by construction, the operator $\cD_{\a{}}$ incorporates the variation of the field dependent symmetry parameters $\a{}$.} 
\be 
\delta (\da A)(y) =
\int_{\scri} \rd^3 x \, \cD_\alpha(y,x)  \delta A(x),
\qquad
\cD_\alpha(y,x) := \sum_{n=-s}^\infty \sum_{m=0}^{\infty}
\frac{ \delta (\da A(y))}{\delta( \pa_u^nD^m A(x))}
\pa_u^nD^m,
\ee 
where $x=(u,z,\bz)$.
To write down the action of $\cS$ on $\bnews$ we need the dual operator $\cD_\alpha^*(x,y)$, where the duality is defined by 
\be 
\int_{\scri} \rd^3 x \rd^3 y  \, \big(\cD_\alpha^*(x,y) A(y)\big)   B(x):= \int_{\scri} \rd^3 x \rd^3 y \, A(x) \big(\cD_\alpha(x,y) B(y)\big).
\ee 
In order to construct $\cD^*_\alpha$ we therefore need to construct $(\pa_u^n D^m)^*$. 
For $D$, we have $D^*=-D$ since we are integrating over the sphere. 
Moreover, we also have $\pa_u^* =-\pa_u$ since we are assuming that the fields are Schwartzian, thence we can safely integrate by parts.
To construct $(\pa_u^{-1})^*$ we use that that
\begin{align}
    \int_{-\infty}^\infty\rd u\, A(u)[\pui B](u) &=\int_{-\infty}^\infty\rd u\, \pa_u\left(\int_{-\infty}^u \rd u'A(u')\right)[\pui B](u) \nn\\
    &=\left.\left(\int_{-\infty}^u\rd u' A(u')[\pui B](u)\right)\right|_{u=-\infty}^{u=+\infty}-\int_{-\infty}^\infty\rd u\left[\int_{-\infty}^u\rd u' A(u')\right]B(u) \nn\\
    &=\int_{-\infty}^\infty\rd u\left[\int^{-\infty}_u\rd u' A(u')\right]B(u),
\end{align}
where the evaluation at the boundary drops since we picked $\pui=\int_{\infty}^u$.
Hence in that case, $(\pui)^*=\int_u^{-\infty}$.
We give the general expression of $(\pui)^*$ for arbitrary base point $\beta$ in App.\,\ref{App:pui}.

The knowledge of $\big(D^*,\pa_u^*, (\pa_u^{-1})^* \big)$ defines $\cD_\alpha^*$ and we find that the canonical action of $\cS$ on $\bnews$ is simply given by 
\be 
\da \bnews(x) = -\int_\scri \rd^3 y\, \cD^*_\alpha (x,y) \bnews(y).
\ee 
It is now straightforward to see that\footnote{We use that $[L_{\da},\delta]=0$.}
\begin{align}
    \frac{\gYM^2}{2}\big(L_{\da}\Theta\big) &=L_{\da} \int_\scri\Pairg{\bnews,\delta A} =\int_\scri\Big(\Pairg{\da\bnews,\delta A}+\Pairg{\bnews,\delta(\da A)}\Big) \\
    &=\int_\scri \Big(-\big\langle\cD^*_\alpha\bnews,\delta A\big\rangle_{\includegraphics[width=1.4mm]{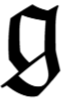}} +\big\langle\bnews,\cD_\alpha\delta A\big\rangle_{\includegraphics[width=1.4mm]{test2}}\Big) =0. \nn
\end{align}
From there we infer that\footnote{Using $L_{\da}=I_{\da}\delta+\delta I_{\da}$.} 
\begin{equation}
    \boxed{I_{\da}\Omega=-\delta\Qa}, \label{canonicalRepYM}
\end{equation}
where $\Omega\equiv\delta\Theta$ is the symplectic form\footnote{The fact that the symmetry action is integrable is non-trivial and due to the fact that all the field dependency of $\a{s}$ is taken into account in the definition of the dual action operator $\cD^*_\alpha$.} and 
\begin{equation}
    \boxed{\Qa:=\Qa^{-\infty}=I_{\da}\Theta} \label{NoetherChargeYM}
\end{equation}
is the Noether charge. 
This charge can be written as an integral over $\scri$ as a direct consequence of \eqref{Qau}, namely
\begin{equation}
    \boxed{\,\Qa=-\frac{2}{\gYM^2}\int_{\scri}\Pairg{\bnews,\Dcal\a0}\,}. \label{QaNoetherYM}
\end{equation}
The dependence on the higher spin $\a{s}$ is implicit via the dual EOM \eqref{dualEOMYM}.

The equations \eqref{canonicalRepYM} and \eqref{NoetherChargeYM} prove that $\Qa$ is the Noether charge for the $\cS$-algebroid action. 
Using the morphism \eqref{SalgebraMorphism} and by definition of the Poisson bracket,\footnote{According to the symplectic form $\Omega$ (or the symplectic potential \eqref{SymPotYM}), we have that $\poisson{\bnews(x),A(x')}=\frac{\gYM^2}{2}\delta^3(x-x')$, where $x=(u,z,\bz)$.} we get that\footnote{We used that the morphism property $\big[\dap,\da\big]O =\delta_{\lbr\a{},\ap{}\rbr^\g}O$ of the symmetry action implies 
\begin{equation}
 \poisson{\Qap,\{\Qa,O\}}-\poisson{\Qa,\{\Qap,O\}}=\poisson{\{\Qap,\Qa\},O} =\poisson{Q_{\lbr\a{},\ap{}\rbr^\g},O}.
\end{equation}
\vspace{-0.4cm}}
\begin{equation}
    \boxed{\poisson{\Qa,O}=\da O} \qquad \textrm{and}\qquad \boxed{\poisson{\Qa, Q_{\ap{}}}=-Q_{\lbr\a{},\ap{}\rbr^\g}},
\end{equation}
where $O$ is an arbitrary functional of $(A,\bA)$.
Notice that we find an equivariant moment map,\footnote{Sign convention: $I_{\dap}I_{\da}\Omega=\Omega\big(\da,\dap\big)=\poisson{\Qa,\Qap}=L_{\da}\Qap=\big(\mathrm{ad}^*_\alpha (Q^*)\big)(\ap{})=-Q(\mathrm{ad}_\alpha\ap{})=-Q\big(\lbr\a{},\ap{}\rbr^\g\big)\equiv-Q_{\lbr\a{},\ap{}\rbr^\g}$, with $\mathrm{ad}^*$ the coadjoint action \cite{Hall2013}.} namely a representation without any 2-cocycle.

\section{Solution of the dual EOM \label{secYM:solEOM}}

In this section, we study the polynomial class of solutions to the dual EOM and give explicit expressions to all order in the $\gYM$ expansion.
We especially introduce the Lie algebroid map $\bfa$ which defines a unique $\bfa(\Aa{})\in\cS$ given an element $\Aa{}\in\V(S,\g)$.
This is essential to define a \textit{corner} symmetry charge with an associated \textit{celestial} symmetry parameter $\Aa{}$, namely a parameter living on the sphere.
Indeed, for now $\Qa$ is either defined over $\scri$ \eqref{QaNoetherYM}, using the $u$-dependent Carrollian symmetry parameter $\a0$, or as a corner integral \eqref{masterChargeYM} by summing over multiple $\a{s}$. 
The solution $\bfa(\Aa{})$ will allow us to write the Noether charge as a corner integral for a \textit{single} $\Aa{s}$ with definite spin-weight.

In order to define such a spin-$s$ Noether charge $\cqAas$, we need to understand the space of solutions of \eqref{dualEOMYM}.
Therefore, we now describe the filtration and the associated gradation of the $\cS$-algebroid.

\subsection{Filtration and gradation \label{secYM:filtration}}

$\cS$ has a natural filtration,
\begin{equation}\label{defFiltrationYM}
    \{0\} \subset\cS^0\subset \cS^1\subset \ldots\subset\cS^s\subset \ldots\subset \cS \quad\textrm{such that}\quad \cS=\bigcup_{n\in\N}\cS^n,
\end{equation}
where the $\cS^s=\cS\cap\tV^s$ are the subspaces for which $\a{n}=0$ for $n>s$.
This filtration is compatible with the bracket since
 \begin{equation}
    \lbr\cS^s,\cS^{s'}\rbr^\g\subseteq \cS^{s+s'},\qquad s,s'\geqslant 0. \label{filtrationCompatibleYM}
 \end{equation}
The associated graded algebroid $\G{\cS}$ is defined as
\begin{equation}
\G{\cS}=\bigoplus_{s=0}^\infty\sfS_s,
\end{equation}
with 
\begin{equation}
    \sfS_0=\cS^0\qquad \textrm{and}\qquad \sfS_s=\cS^s/\cS^{s-1}\quad\textrm{for }s>0.  \label{defgradationYM}
\end{equation}
Each $\sfS_s$ is an equivalence class and we can write $\cS^s= \bigoplus_{n=0}^s\sfS_n$.
We can easily show that the following isomorphism holds:
\vspace{-0.2cm}
\be 
\sfS_s \simeq  \V_s. \label{IsoYM}
\ee 
To prove this, we just have to pick a representative element of $\sfS_s$ and show that it is fully determined by an element $\Aa{s}\in\V_s$.
The mapping goes as follows: 
For $\a{}\in\cS^s$, $\pa_u\a{s}=0$ so that $\a{s}$ is equal to its value at any cut of $\scri$.
For definiteness, we take the $u=0$ cut.
Therefore,
\be 
\a{s}=\Aa{s},\quad  \mathrm{where}\quad  \Aa{s}:= \a{s}|_{u=0}\in \V_s.
\ee
The other values, $\a{s-n}\neq 0$ for $1\leq n\leq s$ are determined recursively from $\a{s}$ by the dual equations of motion. 
We thus have constructed the unique section $\a{}$
\begin{align}
    \a{} :\,&\V_s\to \cS^s \nn\\
     &\, \Aa{s}\mapsto \a{}(\Aa{s}), \label{mapalpha}
\end{align} 
which is a solution of the dual EOM with initial condition given by
\be 
\a{s}(\Aa{s})\big|_{u=0}=\Aa{s}, \qquad \a{s-n}(\Aa{s})\big|_{u=0}=0,\quad \mathrm{for} \quad 1 \leq n \leq s.
\ee 
By construction, such $\a{}$ is precisely a representative element of $\cS^s/\cS^{s-1}$ and is fully given in terms of $\Aa{s}\in\V_s$, which proves \eqref{IsoYM}.
As we mentioned already, these solutions are polynomial in $u$ and thus diverge at infinity, which forces us to take the charge aspects $\tQ_s$ as part of the Schwartz space.

Since $\a{}:\V_s\to \cS^s$ is a linear map, we can extend it by linearity to the map $\bfa: \V(S) \to \cS$ on the full space,\footnote{We use the notation $\bfa(\Aa{})\equiv\a{}(\Aa{})$ interchangeably.}
\begin{align} \label{mapalphabis}
    \bfa :\,& \,
    \V(S,\g) \to \cS \nn\\
     & \,\, \Aa{} \mapsto \bfa(\Aa{}) := \sum_{s=0}^{\infty} \a{}(\Aa{s}).
\end{align}

\subsection{Explicit construction of the section $\alpha$}

We recast the dual evolution equation \eqref{dualEOMYM} with initial condition $\bfa(\Aa{})\big|_{u=0}=\Aa{}$ as follows:
\begin{equation}
    \boxed{\bfa{}(\Aa{})=\pui\Dcal\bfa{}(\Aa{}) +\Aa{}}. \label{solEOMYMcompact}
\end{equation}
More explicitly,
\begin{equation}
    \a{s}(\Aa{})=\big(\pui D+\pui[1]
    [A,\cdot]_\g\big)\a{s+1}(\Aa{})+\Aa{s},
\end{equation}
where it is understood that $\pui=\int_0^u$ acts on all products on its right.\footnote{Here we choose the base point $\beta=0$, namely the cut of $\scri$ where $\Aa{}$ is defined.}
Therefore, the solution for $\a{s-1}$, $\a{s-2},\ldots$ and so on, is built by taking one more power of the operator $\pui\Dcal$ at each step.
Concretely, picking $\a{}\in\cS^s$, we can write
\begin{equation}
    \a{n}(\Aa{})=\sum_{k=0}^{s-n}\big(\pui \Dcal\big)^k\Aa{n+k}.
\end{equation}
To find a representative element of $\sfS_s$, we take only $\Aa{s}\neq 0$, cf. \eqref{IsoYM},
\begin{equation}
    \boxed{\a{n}(\Aa{s})=\big(\pui \Dcal\big)^{s-n}\Aa{s}}, \qquad 0\leq n\leq s.\label{solAlphanAs}
\end{equation}
This is a non-perturbative solution.
To illustrate this, let us see how it expands in power of $A$.
Using the formula
\begin{equation}
    (B+A)^k=\sum_{\l=0}^k\sum_{P=k-\l}B^{p_0}AB^{p_1}AB^{p_2}\ldots AB^{p_\l},\qquad\textrm{with}\quad P=\sum_{i=0}^\l p_i,
\end{equation}
which holds for arbitrary operators $A$ and $B$, we get that
\begin{align}
    \big(\pui\Dcal\big)^k=\left(\pui D+\pui \adA\right)^k &=\sum_{\l=0}^k\sum_{P=k-\l}\pui[p_0]D^{p_0}\bigg(\pui\Big[A,\pui[p_1]D^{p_1}\Big(\pui\Big[A,\pui[p_2]D^{p_2}\Big(\ldots \nn\\
    &\hspace{-1.2cm}\ldots \pui\Big[A,\pui[p_{\l-1}]D^{p_{\l-1}}\Big(\pui \Big[A,\pui[p_\l]D^{p_\l}\Big]_\g\Big)\Big]_\g\ldots \Big)\Big]_\g\Big)\Big]_\g\bigg). \label{puiDcalkYM}
\end{align}
Notice that each term in the sum over $\l$ contains $\l$ gauge fields $A$.
We shall use this way \eqref{solAlphanAs} and \eqref{puiDcalkYM} of writing the solution of \eqref{dualEOMYM} in the discussion about the renormalized charges in section \ref{secYM:RenormCharge}.
As an example, we give the explicit expressions of \eqref{solAlphanAs} when $\a{}\in\cS^2$, which implies that $\Aa{}=(\Aa0,\Aa1,\Aa2,0,\ldots)$:
\bs \label{solalpha210}
\begin{align}
    \a2(\Aa{}) &=\Aa2, \\
    \a1(\Aa{}) &=uD\Aa2+\Aa1+\pui[1][A,\Aa2]_\g, \\
    \a0(\Aa{}) &=\frac{u^2}{2}D^2\Aa2+uD\Aa1+ \Aa0+\pui[2]D[A,\Aa2]_\g \\
    &+\pui\big[A,uD\Aa2+ \Aa1\big]_\g+\pui\big[A,\pui[1][A,\Aa2]_\g\big]_\g. \nn
\end{align}
\es

Let us emphasize that the construction of $\bfa(\Aa{})$ is algorithmic. 
In particular, the symmetry transformation $\da A=-\Dcal\a0$ is simply expressed in terms of $\a0$ but the latter incorporates a great deal of non-linearity and non-locality via the solution $\a0(\Aa{s})$.
From a perturbative point of view, the number of gauge fields appearing in the transformation $\d{s}{\Aa{s}}A\equiv\delta_{\bfa(\Aa{s})}A$ grows linearly with the helicity $s$.
However, from the perspective of the variable $\a{}$, the infinitesimal transformation is still $-\Dcal\a0$.
This is the key reason why our approach is \textit{non-perturbative}.
Moreover, since the proof that the action $\da A$ is canonically represented on the phase space holds for generic $\a{}$, we get for free that the \textit{non-linear} action $\d{s}{\Aa{s}}A$ in terms of the celestial symmetry parameters $\Aa{s}$ is canonically represented too.
The Noether charge associated to this transformation is denoted $\cqAas$, i.e. $\d{s}{\Aa{s}}\cdot=\poisson{\cqAas,\cdot}$, and we give its explicit expression in the next section.

To be more accurate about the notion of `non-perturbative' used here, if we formally introduce $g$ and $\bar g$ as small dimensionless parameters, then by rescaling $A\rightarrow gA$ and $\bA\rightarrow\bar g\bA$, our result is non perturbative in $g$ and at leading order in $\bar g$,\footnote{Recall that $\tQ_{-1}=-\bnews\propto\bar g$ while the EOM \eqref{EOMYM} or their dual \eqref{dualEOMYM} only introduce extra $A$'s.} see also section \ref{sec:FPR}.
$g$ and $\bar g$ are a bookkeeping device and play the role of a complexification of the gauge coupling constant once we impose $g\bar g=\gYM^2$.

\section{Renormalized charge \label{secYM:RenormCharge}}

In \cite{Freidel:2023gue}, the charge renormalization procedure was necessary to get a finite action of the charge aspect in the limit $u\to-\infty$.
In \cite{Geiller:2024bgf}, it was then discussed at length (in the case of General Relativity, but YM would be the same) that even if the final formula for the action on phase space was correct, the actual expression of the renormalized charge aspect $\hq_s$ in terms of the original charges aspects $\tQ_s$ was subtler than the ansatz of \cite{Freidel:2023gue}.

Here we prove (see also the work of \cite{Kmec:2024nmu} for a twistorial perspective) that the charge renormalization procedure amounts to trading the Carrollian symmetry parameter $\a{s}$ for its celestial counterpart $\Aa{s}$.
The renormalization is thus no longer an iterative procedure to remove divergences.
It rather reflects the change of perspective between the Carrollian and the celestial descriptions of the symmetry.
The complicated field dependency necessary to construct the corner/celestial charge aspect $\hq_s(z,\bz)$ is then fully determined by the dual EOM for $\a{}$.

The systematic usage of the dual EOM furnishes an \textit{algorithmic, non-ambiguous and non-perturbative} (in $g$) way of defining the renormalized charges for any helicity.

On top of that, as it was pointed out in \cite{Freidel:2021dfs, Freidel:2021ytz, Geiller:2024bgf}, the finiteness of the action generated by $\hq_s$ comes hand in hand with the answer to the question ``In the absence of radiation, what is the conserved charge associated to $\tQ_s$?''.

Let us now see how the knowledge of the explicit solution $\bfa(\Aa{})\in\cS$ allows to construct the Noether charge $\cqAas$ associated to the celestial symmetry parameter $\Aa{s}$.

Turning words into equations, we define (cf. \eqref{masterChargeYM})
\begin{equation}
   \cqAas^u:= Q^u_{\bfa(\Aa{s})}=\sum_{n=0}^sQ^u_n\big[\a{n}(\Aa{s})\big]. \label{defcqasu}
\end{equation}
The following theorem then summarizes the aforementioned points.

\begin{tcolorbox}[colback=beige, colframe=argile]
\textbf{Theorem [Noether charge of spin $s$]}\\
The Noether charge $\cqAas$ associated to the celestial symmetry parameter $\Aa{s}$ of helicity $s$ is written as the following corner integral:
\begin{equation} \label{NoetherhatqsYM}
    \cqAas\equiv\cqAas^{-\infty} =\frac{2}{\gYM^2}\int_{\scri^+_-}\big\langle\hq_s,\Aa{s} \big\rangle_{\includegraphics[width=1.4mm]{test2}}= \lim_{u\to-\infty}\cq_s^{u}[\Aa{s}], \qquad s\geqslant 0, 
\end{equation}
with $\cqAas^u$ given by \eqref{defcqasu} and
\begin{equation}
    \hq_s(z,\bz)=\lim_{u\to -\infty}\tq_s(u,z,\bz).
\end{equation}
The renormalized charge aspect $\tq_s$ satisfies $\pa_u\tq_s=0$ in the absence of left-handed radiation $\bnews=0$.
\end{tcolorbox}

\paragraph{Proof:}
We construct $\tq_s$ systematically in the appendix \ref{AppYM:renormcharge}, cf. \eqref{deftqsYM}, where we also give explicit expressions for $s=0,1,2$ in \eqref{renormchargeYM}.
To show that $\tq_s$ is conserved in the absence of left-handed radiation, notice that (cf. \eqref{Qadot})
\begin{equation}
    \pa_u \cqAas^u =\frac{2}{\gYM^2}\int_S\big\langle\bnews,\delta_{\bfa{}(\Aa{s})}A\big\rangle_{\includegraphics[width=1.4mm]{test2}}=0 \qquad\textrm{if}\qquad\bnews=0,
\end{equation}
while we also have that
\begin{equation}
    \pa_u \cqAas^u =\frac{2}{\gYM^2}\pa_u\left(\int_S\big\langle\tq_s,\Aa{s}\big\rangle_{\includegraphics[width=1.4mm]{test2}}\right)= \frac{2}{\gYM^2}\int_S \big\langle\pa_u\tq_s,\Aa{s}\big\rangle_{\includegraphics[width=1.4mm]{test2}}.
\end{equation}
This concludes the proof.

\paragraph{Remark:}
In a non-radiative strip of $\scri$, where $\news=0$ for $u\in[0,u_0]$, we get that
\begin{equation}\label{deftqs0radYM}
    \boxed{\tq_s=\sum_{k=0}^s\frac{(-u)^k}{k!}\Dcal^k\tQ_{s-k}}.
\end{equation}
The proof is straightforward since for $u\in[0,u_0]$,
\begin{align}
    \cq^u_s[\Aa{s}]\overset{\eqref{defcqasu}}{=}\sum_{n=0}^sQ^u_n\big[\a{n}(\Aa{s})\big] &\overset{\eqref{solAlphanAs}}{=}\sum_{n=0}^sQ^u_n\big[\big(\pui\Dcal\big)^{s-n}\Aa{s}\big] \\
    &\overset{F=0}{=} \sum_{n=0}^s\frac{u^{s-n}}{(s-n)!}Q^u_n\big[\Dcal^{s-n}\Aa{s}\big]\overset{\eqref{QDDQ}}{=} \sum_{n=0}^s\frac{(-u)^{s-n}}{(s-n)!}\Dcal^{s-n}Q^u_n\big[\Aa{s}\big], \nn
\end{align}
where to go from the first to the second line, we used that $F=0$ so that $\big(\pui\Dcal\big)^{s-n}\Aa{s} =\Dcal^{s-n}\pui[(s-n)]\Aa{s}=\frac{u^{s-n}}{(s-n)!}\Dcal^{s-n}\Aa{s}$.
We then extract \eqref{deftqs0radYM} upon changing $s-n\to k$.
The reader can check that this renormalized charge aspect is conserved in the strip if we also assume that $\bnews=0$.

\section{Action on the gauge potential \label{secYM:ActionOnA}}

As an application of the general formalism presented so far, we compute the soft and hard actions of the higher spin charges $\cqAas$ on the asymptotic gauge potential $A$.

The Noether charge $\cqAas$, or equivalently the charge aspect $\hq_s$, depends linearly on $\bnews$ and has a polynomial dependence\footnote{The coefficients of the polynomial are in general differential operators.} on $A$. 
More precisely, for $s\geqslant -1$, we can decompose $\hq_s$ as
\be 
\hq_s =\sum_{k=0}^{s+1} \q{s}{k},
\ee 
where $\q{s}{k}$ is homogeneous of degree $k$ in $A$ and linear in $\bnews$.\footnote{Recall that this is equivalent to the expansion in terms of the coupling constant $g$ (while all charges are at leading order in $\bar g$), cf.\,\ref{sec:FPR}.}
The charge aspect $\q{s}{0}\equiv\soft{\hq_s}$ is the soft charge while $\q{s}{1}\equiv\hard{\hq_s}$ is the hard charge and $\sum_{k=2}^{s+1} \q{s}{k}$ is the super-hard contribution.
Here we compute the soft and hard actions for any spin. 
We let for future work the study of the super-hard contribution.
However, we underline that the algebra generated by $\Qa$ does close \textit{without} relying on any soft or hard truncation; this is the main point of this work.

We classify the action as a function of the helicity by defining 
\begin{equation}
    \d{s}{\Aa{s}}A :=\delta_{\bfa(\Aa{s})}A= \poisson{\cqAas,A}, \label{defdeltaAs}
\end{equation}
where $\bfa(\Aa{s})$ is the solution \eqref{solAlphanAs}.
The superscript $\d{s}{}$ refers to the helicity of the charge the transformation is associated with.
This notation will come handy in the following.
Moreover, since the fields do not diverge at $|u|\to\infty$ due to the Schwartz fall-off condition, the operation $\pa_u\pui$ is the identity and we can conveniently write
\begin{equation}
    -\d{s}{\Aa{s}}A=\Dcal\a0(\Aa{s})=\Dcal \big(\pui\Dcal\big)^s\Aa{s}=\pa_u\big(\pui\Dcal\big)^{s+1}\Aa{s}.
\end{equation}
We also denote\footnote{$\soft{\cq_{\Aa{s}}} =\frac{2}{\gYM^2}\int_S\big\langle\soft{\hq_s}, \Aa{s}\big\rangle_{\includegraphics[width=1.4mm]{test2}}$ and similarly for $\hard{\cq_{\Aa{s}}}$.}
\begin{equation}
    \poisson{\soft{\cqAas},A} =\Sd{s}{\Aa{s}} A\qquad\textrm{and}\qquad \poisson{\hard{\cqAas},A}=  \Hd{s}{\Aa{s}}A.
\end{equation}
Before stating the results, we define yet another piece of notation, namely 
\begin{equation}
    \td{s}{\Aa{s}}A\equiv\deg^u_0\big(\d{s}{\Aa{s}}A\big),
\end{equation}
where $\deg^u_0(O)$ stands for the coefficient of the term $u^0$ in the expression $O$ (when $O$ is a polynomial in $u$).

\begin{tcolorbox}[colback=beige, colframe=argile]
\textbf{Lemma [Soft action]}\\
The soft action for arbitrary spin $s\in\N$ is given by
\begin{equation}
    \Sd{s}{\Aa{s}}A=-\frac{u^s}{s!} D^{s+1}\Aa{s}. \label{SoftActionYM}
\end{equation}
\end{tcolorbox}

\paragraph{Proof:} 
It is direct since the soft part of $\Dcal^{s+1}$ is $D^{s+1}$, so that
\begin{equation}
    \Sd{s}{\Aa{s}} A=-\soft{\Big(\pa_u\big( \pui\Dcal\big)^{s+1}\Aa{s}\Big)}=-\pa_u\big(\pui D\big)^{s+1}\Aa{s}=-\frac{u^s}{s!}D^{s+1}\Aa{s}.
\end{equation}
Notice that $\Sd{s}{\Aa{s}}A=\frac{u^s}{s!}\Std{0}{D^s\Aa{s}}A$, where $\Std{0}{\Aa0}A=-D\Aa0$.
This means that the soft transformation of spin $s$, parametrized by the tensor $\Aa{s}$, is simply the soft part of a super-phase rotation parametrized by the tensor $D^s\Aa{s}$ (times the $u$ dependence $\frac{u^s}{s!}$).

\begin{tcolorbox}[colback=beige, colframe=argile, breakable]
\textbf{Lemma [Hard action]}\\
The hard action for arbitrary spin $s\in\N$ takes the form
\begin{subequations}
\label{YMHardAction}
\begin{equation}
    \Hd{s}{\Aa{s}}A=\sum_{p=0}^s\frac{u^{s-p}}{(s-p)!}\Htd{p}{D^{s-p}\Aa{s}}A,
\end{equation}
where
\begin{equation}
    \Htd{0}{\Aa0}A=-[A,\Aa0]_\g \quad~\textrm{and }\quad \Htd{p}{\Aa{p}}A=-D\big[\pui[p]D^{p-1}A,\Aa{p}\big]_\g,\quad p\geq 1.
\end{equation}
\end{subequations}
\end{tcolorbox}

\paragraph{Proof:}
We have to consider 
\begin{equation}\label{puiDcalHard}
    \Hd{s}{\Aa{s}} A=-\hard{\Big(\pa_u\big( \pui\Dcal\big)^{s+1}\Aa{s}\Big)},
\end{equation}
where $\big( \pui\Dcal\big)^{s+1}$ is given by \eqref{puiDcalkYM}.
In \eqref{puiDcalkYM}, the sum over $\l$ precisely counts the number of gauge potential, i.e. the expansion in $g^\l$.
The hard part, namely the part proportional to $g$, is thus the term $\l=1$.
We present the proof of how to relate \eqref{puiDcalHard} to \eqref{YMHardAction} in appendix \ref{App:HardActionYM}.

A similar computation was carried out in \cite{Freidel:2023gue} using a discrete basis of modes for $A$ (see also \cite{Freidel:2022skz}).
Since it was not obvious from \cite{Freidel:2023gue}, we find remarkable that the hard action specific to the degree $p$, namely $\Htd{p}{\Aa{p}}A$, reduces to a total derivative (for $p\geq 1$).
More precisely, our formula \eqref{YMHardAction} is the same as equation (50) in \cite{Freidel:2023gue}.
The proof necessitates some manipulations so we report it in appendix \ref{AppYM:comparison}.
The point is that \eqref{YMHardAction} now appears as a consequence of the general, non-perturbative action of the Noether charge $\Qa$, in a much simpler way than the perturbative treatment known so far.

\section{Wedge sub-algebras of $\cS$ \label{secYM:wedge}}

In this section, we discuss (wedge) sub-algebras present in $\cS$ or its projection at a cut of $\scri$. 

\subsection{$\Wcal_A(\scri)$-algebra}

The Lie algebroid $\sfS$-bracket $\lbr\cdot\,,\cdot\rbr^\g$ reduces to the Lie algebra bracket $[\cdot\,,\cdot]^\g$ when $\da\ap{}\propto\da A\equiv 0$.
This imposes a set of constraints on the symmetry parameters $\a{}$.
Indeed, by definition $\da A=-\Dcal\a0$ is set to 0.
Then, since $\da A=0$ implies $\pa_u(\da A)=\da F=0$, and that the commutator of $\pa_u$ with $\Dcal$ is just
\begin{equation}
    [\pa_u,\Dcal]\,\cdot=[F,\cdot\,]_\g,
\end{equation}
we also have that
\begin{equation}\label{wedgeStart}
    0=-\da F=\pa_u(\Dcal\a0)=\Dcal (\pa_u\a0)+[F,\a0]_\g=\Dcal^2\a1+[F,\a0]_\g.
\end{equation}
Iterating this reasoning, the $s^{\rm{th}}$-time derivative of $\da A$ then generates a condition on $\Dcal^{s+1}\a{s}$.
To summarize, the first 3 conditions are
\bs
\begin{align}
    \Dcal\a0 &=0, \\
    \Dcal^2\a1 &=-[F,\a0]_\g, \\
    \Dcal^3\a2 &=-[\Dcal F,\a1]_\g -3[F,\Dcal\a1]_\g -[\pa_u F,\a0]_\g.
\end{align}
\es

We thus define the covariant wedge space as follows.
\begin{tcolorbox}[colback=beige, colframe=argile]
\textbf{Definition [Covariant wedge space on $\scri$]}
\begin{equation}
    \W_A(\scri):=\Big\{\,\a{}\in\sfS~ \big|~\pa_u^s\big(\Dcal\a0\big)=0,\, \forall \,s\in\N\,\Big\}.
\end{equation}
\end{tcolorbox}

\begin{tcolorbox}[colback=beige, colframe=argile]
\textbf{Theorem [Covariant wedge algebra on $\scri$]}\\
$\Wcal_A(\scri)\equiv\big(\W_A(\scri),[\cdot\,,\cdot]^\g\big)$ is a Lie algebra over $\scri$.
\end{tcolorbox}

\paragraph{Proof:}
Since $\da A=0=\dap A$ implies that $\pa_u[\a{},\ap{}]^\g=\Dcal[\a{},\ap{}]^\g$, cf. App.\,\ref{App:dualEOMYM}, we are allowed to evaluate $\pa_u^s(\da A)$ for $\a{}\to[\a{},\ap{}]^\g$. 
The $s^{\rm{th}}$-time derivative then generates a condition on $\Dcal^{s+1}\a{s}$ which is automatically preserved by the bracket.
In brief,
\begin{equation}
    \da A=0\quad\Longrightarrow\quad \Big(\da\big(\pa_u^s A\big)=0\quad \&\quad\delta_{[\a{},\ap{}]^\g}\big(\pa_u^s A\big)=0\Big),\quad\forall \,s\in\N.
\end{equation}

Even if we let a detailed analysis of $\Wcal_A(\scri)$ for another work, we can say that it encompasses the symmetry transformations that leave the asymptotic gauge field $A$ invariant.
It is thus the symmetry algebra of such an asymptotic solution.

\subsection{$\Wcal_{\beta}(S)$-algebra \label{subsecYM:wedgeAlgebraOnS}}

We now focus on the particular case of a non-radiative cut of $\scri$, for which the analysis simplifies greatly.

First of all, notice that the $\cS$-algebroid over $\scri$ can be projected at any cut. 
For definiteness and to match with the solution of the dual EOM discussed in section \ref{secYM:solEOM}, we always consider the cut $S=\{u=0\}\subset\scri$.
Therefore, there exists a Lie algebroid bracket $\lbr\cdot\,,\cdot\rbr^\g$ over $S$ which is naturally given by the evaluation of its Carrollian version at $u=0$, namely\footnote{Of course $[\Aa{},\Aap{}]^\g_s=\sum_{n=0}^s[\Aa{n} ,\Aap{s-n}]_\g$ and we use the fact that $\pui(\ldots)\big|_{u=0}=0$.
\vspace*{-0.8cm}

}
\begin{equation}
    \lbr\Aa{},\Aap{}\rbr^\g:=\lbr\a{},\ap{}\rbr^\g \big|_{u=0}=[\Aa{},\Aap{}]^\g+\dAp\Aa{}-\dA\Aap{}.
\end{equation}

Second of all, if $\Dcal\Aa0=0=\Dcal\Aap0$, then the action of the anchor map vanishes and $\lbr\Aa{},\Aap{}\rbr^\g$ reduces to the Lie algebra bracket $[\Aa{},\Aap{}]^\g$.
When writing $\Dcal\Aa0$, it is understood that the gauge covariant derivative is evaluated at the cut too. 
In other words $\Dcal\Aa{s}:=D\Aa{s}+[\Ac0,\Aa{s}]_\g$, where $\Ac0:=A\big|_{u=0}$.
With that notation, $\dA \Ac0=\big(\da A\big)\big|_{u=0}=-\Dcal\Aa0$.

Third of all, if we consider a fully non-radiative cut, by which we mean that $\Ac{n}:=\big(\pa_u^n A\big)\big|_{u=0} \equiv 0$ for $n\geq 1$, then the symmetry of the cut has to preserve these conditions.
Using the calculation \eqref{wedgeStart}, we get
\begin{align}
    -\dA\Ac1=-\big(\da F\big)\big|_{u=0} =\Dcal^2\Aa1+\big[\Ac1,\Aa0 \big]_\g.
\end{align}
We thus see that in order to preserve the condition $\Ac1=0$, we need to take $\Dcal^2\Aa1=0$.
Iterating this computation, we deduce that
\begin{equation}
    -\dA\Ac{n}=\big(\pa_u^n(\Dcal\a0)\big) \big|_{u=0}=\Dcal^{n+1}\Aa{n},\qquad\textrm{if}\quad \Ac{n}=0,\,n\geqslant 1.
\end{equation}
Hence preserving the non-radiative nature of the cut under consideration---namely $\bfAc=(\Ac0,0,\ldots)$ where $\bfAc$ denotes the collection $\{\Ac{n}\}_{n=0}^\infty$---amounts to imposing the \textit{covariant wedge} condition $\Dcal^{s+1}\Aa{s}=0$, $s\in\N$.

To summarize our findings, we define the $\W_{\bfAc}(S)$-space (in the special case $\bfAc=(\Ac0,0,\ldots)$) and state the associated theorem.

\begin{tcolorbox}[colback=beige, colframe=argile]
\textbf{Definition [Covariant wedge space on $S$]}
\begin{equation}
    \W_{\bfAc}(S):=\Big\{\,\Aa{}\in\V(S,\g)~ \big|~\Dcal^{s+1}\Aa{s}=0,\, \forall\, s\in\N\,\Big\}.
\end{equation}
\end{tcolorbox}

\begin{tcolorbox}[colback=beige, colframe=argile]
\textbf{Theorem [Covariant wedge algebra on $S$]}\\
$\Wcal_{\bfAc}(S)\equiv\big(\W_{\bfAc}(S),[\cdot\,,\cdot]^\g\big)$ is a Lie algebra over $S$.
\end{tcolorbox}

\paragraph{Proof:}
From the intrinsic perspective of the sphere, i.e. not using the projection from $\scri$ of $\Wcal_A(\scri)$, the reason why the bracket preserves the covariant wedge condition is that 
\begin{equation}
    \Dcal^{s+1}[\Aa{},\Aap{}]^\g_s=\sum_{n=0}^s \sum_{k=0}^{s+1}\binom{s+1}{k}\big[ \Dcal^k\Aa{n},\Dcal^{s+1-k}\Aap{s-n}\big]_\g=0 \label{proofCovWedgeYM}
\end{equation}
since $\Dcal^k\Aa{n}\equiv 0$ for $k\geqslant n+1$ and $\Dcal^{s+1-k}\Aap{s-n}\equiv 0$ for $k\leqslant n$.
Hence $[\Aa{},\Aap{}]^\g\in\Wcal_{\bfAc}(S)$ if $\Aa{},\Aap{}\in\Wcal_{\bfAc}(S)$.

\paragraph{Remark:}
As we can see from the proof \eqref{proofCovWedgeYM}, the $\cS$-algebroid admits a collection of sub-algebras $\WS{n}$ which are organized into a filtration
\begin{equation}
    \WS{}\subset\ldots\subset\WS{n+1} \subset\WS{n} \subset\ldots\subset\WS0, 
\end{equation}
where\footnote{We use the same letter $\cS$ to designate both the algebroid over $\scri$ and the one over $S$.
Since we use $\a{}$ and $\Aa{}$ for the respective variables, the context always makes it clear which algebroid we are referring to.}${}^,$\footnote{The condition $\Dcal\Aa0=0$ is always satisfied, so that the anchor $\dA$ always vanishes and $\WS{n}$ is a Lie algebra for any $n\in\N$.}
\begin{equation}
    \WS{n}:=\Big\{\,\Aa{}\in\cS~ \big|~\Dcal^{s+1}\Aa{s}=0,\, 0\leqslant s\leqslant n\,\Big\}\subset\cS
\end{equation}
and $\WS{}=\lim_{n\to\infty}\WS{n}$.

\paragraph{Remark:}
From the general result about the Noether representation in Sec.\,\ref{secYM:Noether}, we infer that the covariant wedge algebra is represented canonically, so that
\begin{equation}
    \boxed{\,\poisson{\cq_{ \text{\scalebox{1.24}{$\Aa{}$}}},\cq_{ \text{\scalebox{1.24}{$\Aap{}$}}}}=- \cq_{\text{\scalebox{1.24}{$[\Aa{},\Aap{}]$}}{}^{\includegraphics[width=1.4mm]{test2}}}\,}.
\end{equation}

\section{Relation to twistor theory \label{secYM:twistor}}

We present an alternative description of the Carrollian symmetry parameters $\a{}$ by trading the spin degree for a continuous dimension labeled by a spin 1 variable $q \in \Ccar{0,1}$.
We shall give much more explanations about the origin of $q$ in section \ref{sec:twistor}.
For now, let us notice that it allows to promote the series $\a{}=(\a{s})_{s\in \N}$ to a $\g$-valued holomorphic function $\fa$ of $q$\footnote{The Bondi coordinates dependence is implicit.} 
\be
\fa(q)=\sum_{s=0}^{\infty} \a{s}q^s \in \tV_0.
\ee 
The graded vector $\a{}$ and the function $\fa$ are two representations of the same abstract vector in $\tV(\scri,\g)$.
The dual EOM are represented on these functions as the functional
\begin{equation}
    \sE_{\fa}(q):= \sum_{s=0}^\infty \big(\pa_u\a{s}-(\Dcal\a{})_s\big)q^s.
\end{equation}
Moreover we see that the function representation of the bracket $[\a{},\ap{}]^\g$ is the Lie algebra $\g$-bracket of the function representation of $\a{},\ap{}$.
Indeed,
\begin{align}
    [\fa, \fa'](q):=\sum_{s=0}^\infty[\a{}, \ap{}]^\g_sq^s =
    \sum_{s=0}^\infty\sum_{n=0}^s[\a{n},\ap{s-n}]_\g q^s &=\sum_{n=0}^\infty \sum_{s=n}^\infty\big[\a{n}q^n,\ap{s-n}q^{s-n}\big]_\g \cr &=\big[\fa(q),\fa'(q)\big]_\g.
\end{align}
Finally, introducing the covariant derivative 
\be 
\nabla := q\pa_u -D-\adA,
\ee
we obtain that 
\begin{align}
    \nabla \fa(q) =\sum_{s=0}^\infty \big(\pa_u\a{s}- 
    \Dcal\a{s+1}\big) q^{s+1}
    - \Dcal\a0= q\E_{\fa}(q)+\da A.\label{bNIYM}
\end{align}
The condition $\a{}\in\sfS$, which imposes $\E_{\fa}=0$, simply reads $\pa_q \nabla \fa=0$.
It shows that when $\a{}\in \cS$, then $\nabla\fa$ is independent of $q$ and
\begin{equation}
\da A\equiv\dah A=\nabla \fa.
\end{equation}
Here we see that the dual EOM appear when requiring that the infinitesimal field transformation $\dah A$ be independent of the coordinate $q$.
This is natural since the asymptotic gauge field $A$ is by construction only a function of $(u,z,\bz)$.

It is customary to get field dependent symmetry parameters after a partial gauge fixing procedure.
In our case, we can tie this $q$-independence of $\dah A$ to the analogous twistor gravitational study of \cite{Kmec:2024nmu}, where $q$ is precisely the twistor coordinate in the fiber above $\scri$ \cite{Adamo:2021lrv}.\footnote{See also \ref{sec:twistor} for a discussion about the interpretation of $q$ in the context of Newman's good cut equation \cite{newman_heaven_1976, Adamo:2010ey} and as a parametrization of an Ehresmann connection in the Carrollian context \cite{Mars:1993mj, Freidel:2024emv}.}
Indeed, assuming that the twistor gauge potential comes from the lift of a $\scri$ gauge potential, the former must be independent of the fiber coordinate $q$.
From a twistor space perspective, this can be seen as a gauge fixing.
With that insight, $\dah A$ should result from a twistor gauge transformation when studying self-dual YM theory in twistor space.
This was confirmed in \cite{Kmec:2025ftx}.

\chapter{General Relativity I \label{secGRI}}

This chapter contains the material of \cite{Cresto:2024fhd} and of the first half of \cite{Cresto:2024mne}.

\section{Introduction}

In the quest for understanding the symmetries of asymptotically flat spacetimes, the study of conformally soft gravitons symmetries led to the discovery of a $Lw_{1+\infty}$ algebra structure that organizes the OPEs of the soft modes \cite{Guevara:2021abz, Strominger:2021mtt, Himwich:2021dau, Ball:2021tmb}.
Understanding this algebraic structure is particularly relevant for the study of the gravitational $S$-matrix, infrared physics and the relationship with asymptotic symmetries.

On top of this celestial treatment of $Lw_{1+\infty}$ (see the introduction \ref{sec:IntroHis}), FPR identified a collection of higher spin charge aspects that generalize the Bondi mass aspect (spin $0$) and the angular momenta aspect (spin $1$) to higher spins \cite{Freidel:2021dfs, Freidel:2021ytz, Geiller:2024bgf} (see Sec.\,\ref{sec:FPR} for a detailed summary).
The soft theorems are then understood as a matching condition for this set of higher spin charges at spacelike infinity \cite{Strominger:2017zoo, Hamada:2018vrw, Raclariu:2021zjz}.
 
This point of view is closer to a Noetherian perspective where the charge aspects are symmetry generators which act on the gravitational phase space and are dual to a symmetry algebra.
The algebra that arises from this analysis is not directly $Lw_{1+\infty}$: it is a shifted version of the Schouten-Nijenhuis algebra for \textit{symmetric} contravariant tensor fields \cite{Schouten, SNbracket}.
The advantage of this description is that the symmetry generators do not have to be holomorphic or to have specific singularity properties on the complex plane, as it is the case for $Lw_{1+\infty}$.
In fact, one can define the symmetry algebra on any Riemann surface $S$.
This algebra is encoded into a set of spin $s$ fields on $S$ denoted $T_s(z,\bz)$ for $s\geq -1$.\footnote{See section \ref{sec:Preliminaries} for our conventions of notation and a reminder about conformal and spin weights.}
$\T0$ and $\T1$ correspond respectively to the super-translations and super-Lorentz-rotations parameters.  
The parameters $T_{-1}$ labels the center.
The shifted Schouten-Nijenhuis algebra bracket is given by 
\begin{equation} 
    [T,T']^\V_s= \sum_{n=0}^{s+1} (n+1)\big(\T{n} D \Tp{s+1-n}-\Tp{n} D \T{s+1-n}\big),
\end{equation}
where $D$ is the holomorphic covariant derivative on $S$.
If we truncate this algebra to spin $0$ and spin $1$ fields we recover the  $\gbms$ algebra which includes 2d diffeomorphisms \cite{Barnich:2016lyg, Campiglia:2014yka, Compere:2018ylh, Campiglia:2020qvc, Freidel:2021fxf}, cf. Sec.\,\ref{sec:DeformingWedge}.

In section \ref{sec:WedgeAlgebra}, we establish that the consistent inclusion of the supertranslations into the \textit{higher spin}---i.e. not just super-rotations---algebra forces the parameters to satisfy the condition $D^{s+2} T_s=0$, which defines the \emph{wedge} sub-algebra denoted $\Wcal(S)$. 
As we will see, this condition is \emph{necessary} in order for the Jacobi identity to be valid.
The connection between $\Wcal(S)$ and $Lw_{1+\infty}$ is revealed if one chooses $S$ to be the cylinder. 
Indeed, it was established in \cite{Donnay:2024qwq} that $\Wcal(\C^*) = Lw_{1+\infty}$. 
The correspondence between the twistor and celestial sphere actions is such that the wedge condition on $S=\C^*$ follows from the demand for holomorphicity of the twistor transformation parameters.
On the other hand, if we choose $S=S^2$ as the global 2-sphere then we find that $\Wcal(S^2)$ is a finite-dimensional algebra given by a 3-dimensional central extension of the chiral Poincaré algebra $\mathfrak{sl}_-(2,\C) \sds \C^4$ (see sections \ref{sec:Wspace} and \ref{Sec:Walgebra}), a fact first noted in \cite{Krasnov:2021cva}.
This means that the wedge algebra is a global symmetry algebra that depends on the 2d surface topology.

Moreover, the higher spin fields $s\geq 1$ form a Lie sub-algebra that extends $\mathrm{Diff}(S)$ to higher spins.
The wedge condition is \textit{not} required in this case.
This explains why \cite{Freidel:2023gue} had been able to relax the wedge restriction at spin 2 when checking the closure of the algebra at quadratic order, cf.\,\ref{sec:FPR}.
This property thus clarifies why their computation is not in contradiction with the aforementioned Jacobi identity anomaly.
The latter vanishes in the absence of time (namely spin 0 parameter).

In the present work, we investigate the possibility of deforming the wedge algebra.
There are several reasons to expect such a deformation. 
The main one is physical: While the wedge algebra is a perturbative symmetry of amplitudes near the flat space vacuum, we expect the symmetry to exist at any non-radiative cut of $\scri$. 
Due to the memory effect, we know that such cuts will carry a non-trivial value of the shear $\sigma$. 
Therefore, promoting the wedge algebra as a symmetry algebra of gravity requires us to find what symmetry algebra is represented at cuts with non-zero shear.

Quite remarkably, a deformation of the wedge algebra exists, which respects Jacobi and is labeled by a shear parameter $\sigma(z,\bz)$.
The bracket for this algebra, denoted $\Wcal_\sigma(S)$, reads 
\begin{equation}
    [T,T']^\sigma_s = 
    [T,T']^\V_s-(s+3)\sigma\big(\T0\Tp{s+2}-\Tp0\T{s+2}\big). \label{sigmaBracketIntro}
\end{equation}
This bracket respects the Jacobi identity for a set of parameters that satisfy a deformed wedge condition (see sec. \ref{sec:NewFieldDepBracket}).
We also introduce the notion of a shear dependent covariant derivative that represents the adjoint action of the time translation element. 
Besides, we use this covariant derivative to present equivalent forms of the $\sigma$-bracket, which highlight some of its key properties, cf.\,\ref{sec:CovFormBracket}.

Next, in section \ref{sec:CovWedgeSol} we show that this deformed algebra is isomorphic to the undeformed one through a \emph{dressing} transformation, which allows us to describe symmetry parameters in $\Wcal_\sigma(S)$ in terms of dressed symmetry parameters in $\Wcal(S)$. 
This involves the definition of a dressing map $\mTG$ as a path ordered exponential along a flow in the space of Lie algebras, where the flow of the $\sigma$-bracket is determined by the adjoint action of the Goldstone field $G$.
The possibility of removing the shear through a redefinition of the symmetry parameter provides an algebraic equivalent of Newman's $\mathcal{H}$-theorem and good cut equation \cite{newman_heaven_1976, Adamo:2009vu}, which established that we can reabsorb the asymptotic shear into a redefinition of the cut (see also the next chapter).

\paragraph{Remark:}
Let us mention that one can also construct the $\sigma$-bracket \eqref{sigmaBracketIntro} following a logic similar to chapter \ref{secNAGTII}, namely by studying the flux of a master charge.
This will become clear in chapter \ref{secGRII}.
In chapter \ref{secEYM}, we even present a way to get \eqref{sigmaBracketIntro} from the on-shellness of an off-shell, kinematical bracket.
All of that to say that there are many equivalent ways of introducing \eqref{sigmaBracketIntro} and we shall expose some of them in due time.
Concerning the present chapter, we really want to emphasize the extent to which the discussion in independent of GR.\footnote{This is why we partitioned chapters \ref{secGRI} and \ref{secGRII} this way.}
Even if throughout the introduction we used words like `shear', `Goldstone', `dressing', `non-radiative' or `memory', that are loaded with physical meaning, this was mostly to guide the reader towards the final physical interpretation of the various objects.
These mathematical objects are however agnostic about any physical input and are defined to be consistent with the underlying mathematical structure.
Although in the main text we do not commit entirely to a ``maths first'' approach,\footnote{After all, this is a thesis in Physics, not in Mathematics.} let us point out that the deformation of the wedge bracket into the $\sigma$-bracket can also be deduced through the Jacobi identity anomaly \eqref{JacCF}.
The latter then allows the identification of an anchor map, which itself defines an algebroid bracket.
The latter naturally \textit{defines} a flow, that we can dub `temporal'.
Pairing the algebroid with its dual leads to a temporal flow on the dual variables,\footnote{To confuse the reader: what we just referred to as `dual' variables are what we call charge aspects and the temporal flow of the symmetry generators is called `dual' equations of motion. 
Again, it is a \textit{physics} thesis, so the duality is reversed...} which turns out to reproduce the asymptotic Einstein's equations till spin 3!
\textit{This means that a sector of the GR dynamics is determined by a Jacobi identity anomaly!}\\

Getting back to business, we give a one page summary of the main results of the previous sections in \ref{sec:Wedge}.
The next step consists in introducing time and generalizing the objects discussed so far (i.e. in \cite{Cresto:2024fhd}) to adapt them to $\scri$.
This is carried over in section \ref{sec:celestialToScri}.
We define the dual equations of motion in \ref{Sec:DualEOM} and show that they are consistent with the algebroid bracket $\lbr\cdot\,,\cdot\rbr$ in \ref{sec:TimeEvol}.
Section \ref{secAlgebroid} deals with the main object of interest in this work, namely the $\A$-algebroid.
The two main characteristics of $\A$ are on one hand the dual EOM satisfied by its elements,
\begin{equation}
    \pa_u\t{s}= D\t{s+1}- (s+3) C \t{s+2},
\end{equation}
and on the other hand the action of the anchor map onto the shear,
\begin{equation}
    \dt C=\t0\pa_u C-D^2\t0+2DC\t1+3CD\t1-3C^2\t2.
\end{equation}
This is the symmetry action on the asymptotic GR phase space discussed in chapter \ref{secGRII}.
As we just noticed in the remark, there are several ways of understanding the relevance of these equations.
Here we focus on an exposition which builds up from the construction onto $S$ laid down in the preceding sections.
The algebroid is essential to waive the covariant wedge restriction\footnote{The covariant wedge space precisely becomes the restriction of the anchor map $\delta$ to its kernel, which defines an algebra within an algebroid.} and describe radiation within an algebraic framework.
Then \ref{sec:InitialCondition} completes \ref{secAlgebroid} with a discussion about the sub-algebroids $\Ao$ and $\Ah$, which differ by their treatment of the central element.

Finally, section \ref{secGR:glossary} is a glossary of the notation, both for the current and the next chapter.
Besides the main text, we gather the technical proofs of our results in an extensive collection of appendices \ref{appGRI}.

\section{The wedge algebra \label{sec:WedgeAlgebra}}

There is by now ample evidence that the infinite dimensional Lie algebra $Lw_{1+\infty}$ is a symmetry algebra of perturbative gravity in the self dual sector \cite{Adamo:2021lrv,Donnay:2024qwq}. 
One direct way to see this is to use the twistor representation of self dual solutions and use the fact that $Lw_{1+\infty}= L(\mathrm{Ham}(\mathbb{C}^2))$ is a loop algebra over Hamiltonian vector field acting on the fiber of the twistor fibration $\mathbb{T}^* \to \mathbb{CP}_1$.

Another way to see this is to show that this algebra possesses a canonical action on the asymptotic data at $\scri$ \cite{Freidel:2021dfs, Freidel:2021ytz, Geiller:2024bgf, Freidel:2023gue}.
This chapter lays down all the necessary mathematical material in order to find such a canonical action from a physically motivated viewpoint in chapter \ref{secGRII}.

In this section, we remind the reader about the \textit{wedge} algebra, first studied in the twistor context by Penrose \cite{Penrose:1976js}, and in celestial holography context by Strominger and al. \cite{Himwich:2021dau,Strominger:2021mtt, Guevara:2021abz} and then in the canonical framework by Freidel, Pranzetti and Raclariu \cite{Freidel:2021dfs,Freidel:2021ytz, Freidel:2023gue}. 
The connection between the twistor and canonical action of $Lw_{1+\infty}$ has been studied in \cite{Donnay:2024qwq, Kmec:2024nmu}.

\subsection{The $\W$-space \label{sec:Wspace}}

An important subspace of $\V(S)$, cf.\,\ref{sec:Vspace}, corresponds to the \textit{wedge} vector space:
\begin{equation}
    \boxed{~\W(S):= \Big\{T\in \V(S)\,\big|\, D^{s+2}\T{s}=0,~s\geqslant -1\Big\} \subset \V(S)~}. \label{DefWedge}
\end{equation}
The $\W$-space inherits the grading of $\V(S)$, namely
\be 
\W(S)= \bigoplus_{s=-1}^\infty \W_s(S),
\ee 
where $\W_s(S):= \{ T_s\in \V_s\,|\, D^{s+2} T_s=0\}$ and $\V_s\equiv\Ccel{-1,-s}$.

Depending if we work on the sphere $S_0$, the complex plane $\C\equiv S_1$ or the cylinder $\C^*\equiv\C\backslash \{0\}\equiv S_2$,\footnote{Recall that $S_n$ generically denotes a 2-sphere with $n$ points removed.} the restriction $D^{s+2}\T{s}=0,~s\geqslant -1$, on the parameters $\T{s}$ takes different forms (see \cite{Barnich:2021dta} for a similar discussion about $\mathfrak{bms}$).

On the sphere the solution depends on the spin $s$. 
It is direct to establish that the space of solutions of $D^2 T_0=0$ is 4-dimensional,\footnote{All the vector spaces discussed in this section are complex vector spaces, over $\mathbb{C}.$} parametrized by a 4-vector $x^a\in\C^4$ and given by
\be \label{T0T}
T_0(x) = x_a  q^a (z,\bz), \qquad 
q(z,\bz) = \frac{1}{(1+|z|^2)} \big( 1+|z|^2, z+\bz, -i(z-\bz), 1-|z|^2\big).
\ee 
The vector $q$ is null and of the form $(1, n^i)$ where $n$ is normalized, $n^i n^j\delta_{ij}=1$, and represents a point on the sphere.  
The space of solutions of $D T_{-1}=0$ and $D^3 T_{1}=0$ are three  dimensional and respectively given by 
\be 
T_{-1}(y) = y^iD n_i, \qquad T_{1}(w)= w^i\bD n_i, \qquad y,w\in\C^3.\label{TDT}
\ee 
On the other hand, the space of solutions  of $D^{s+2} T_{s}=0$ for $s\geq 2$ is zero-dimensional. It only  contains the null solution $T_s=0$.
The reader can check the appendix \ref{AppWedgeOnSphere} for a  proof using spherical harmonics.

This result means that on the sphere the wedge algebra is 10 dimensional. 
We will see around equation \eqref{central-1} that the algebra reduces to a central extension of the  chiral Poincaré group. The chiral Poincaré algebra is obtained if we restrict to $s\geq 0$.

On the complex plane $\C\equiv S_1$, the situation is different.
The wedge condition $D^{s+2}\T{s}=0$ is merely the anti-holomorphicity of $D^{s+1}\T{s}$ (since $D\to\pa_z$).
$\T{s}$ is thus an arbitrary entire series in $\bz$ and a polynomial of degree $s+1$ in $z$:
\begin{equation}    
\T{s}=\sum_{m=0}^{s+1} T_{(s,m)}(\bz) z^m, \qquad 
T_{(s,m)}(\bz) = \sum_{k=0}^\infty T_{(s,m,k)} \,\bz^k.
\end{equation}
The algebra it generates is $L_+(w_{1+\infty})$, a sub-algebra of $Lw_{1+\infty}$.

Finally, on $S_2\equiv \C^*$, we are allowed to take a Laurent series in $\bz$.\footnote{Recall that since $D(1/\bz)=2\pi\delta^{(2)}(z)$, we had to discard these solutions on $S_1$.}
Therefore, the general solution of $D^{s+2}\T{s}=0$ is
\begin{equation}    
\T{s}=\sum_{m=0}^{s+1} T_{(s,m)}(\bz) z^m, \qquad 
T_{(s,m)}(\bz) = \sum_{k=-\infty}^\infty T_{(s,m,k)} \,\bz^k. \label{TsOnS2}
\end{equation}
This generates $Lw_{1+\infty}$.
Notice that while the vector space $W(S_0)$ is finite dimensional, $W(S_1)$ and $W(S_2)$ are infinite dimensional.

The relationship between  $T_s$ and the modes commonly used in the celestial literature (see e.g. \cite{Strominger:2021mtt}) arises if we take $s=2p-3$ and write \eqref{TsOnS2} as
\begin{equation}
    \T{s}= \sum_{m=1-p}^{p-1}\sum_{k=-\infty}^\infty T_{(2p-3,p-1+m,k)}  z^{p-1+m}\bz^k. \label{Tsvsw}
\end{equation}

The following table summarizes the previous discussion about the space of solutions of the wedge condition together with the algebra structure that we discuss in subsection \ref{Sec:Walgebra}.

\begin{table}[H]
\centering
\begin{tabular}{c|cccccc|l}
\toprule
manifold & $\T{-1}$ & $\T0$ & $\T1$ & $\T2$ & $\cdots$ & $\T{s}$ & \qquad algebra\\
\midrule
$S_0$   &  3  &  4  &  3  & 0 & $\ldots$ & 0   & $ \Wcal(S_0)=  \mathfrak{iso}_-(3,1)\,\oplus\, \C^3$ \\
$S_1$   &  1  &  2  &  3  & 4 & $\ldots$ & s+2 & $\Wcal(S_1)= L_+(w_{1+\infty})$\\
$S_2$   &  1  &  2  &  3  & 4 & $\ldots$ & s+2 & $\Wcal(S_2)= Lw_{1+\infty}$ \\
\bottomrule
\end{tabular}
\caption{\textit{Number of independent complex solutions to $D^{s+2}\T{s}=0$ depending on the choice of manifold. For $S_1$ and $S_2$, each subspace of degree $s$ has complex dimension $(s+2)\cdot\infty$ due to the loop part.} \label{table1}}
\end{table}

\subsection{The $\W$-bracket \label{sec:Wbracket}}

We equip $\W(S)$ with the following $\W$-bracket:
\be 
[T,T']^\W_s := \sum_{n=0}^{s+1} (n+1)\left( \T{n} D \Tp{s+1-n}-\Tp{n} D \T{s+1-n}\right). \label{Wbracket}
\ee
Notice that the bracket acts on the family $T= (T_s)_{s+1\in \N}$ rather than on a single object $\T{s}$ with definite helicity $-s$.
This notation will turn out to be very convenient and allow us to derive statements valid for any spin-weights. Quite remarkably, this bracket is defined on $\W(S)$ independently of the topology of $S$.

This bracket is usually written in term of its components. 
This means that  $\big[\iota(\T{p}),\iota(\Tp{q})\big]_{s}$  vanishes unless $s=p+q-1$. 
Hence one could also write \eqref{Wbracket} as
\begin{equation}
    \big[\iota(\T{p}), \iota(\Tp{q})\big]_{p+q-1}^\W=(p+1)\T{p} D\Tp{q}-(q+1)\Tp{q} D\T{p}, \label{Wbracketbis}
\end{equation}
which shows that $[\cdot\,,\cdot]^\W$ is a bracket of degree $-1$, namely
\begin{equation}
    [\cdot\,,\cdot]^\W: \W_p\times\W_q\to\V_{p+q-1}.
\end{equation}
In our case, we choose to define the natural extension of the latter to the full graded space $\W(S)$, namely
\begin{equation}
    [\cdot\,,\cdot]^\W: \W(S)\times\W(S)\to\V(S),
\end{equation}
such that we take the sum over all combinations of $p$ and $q$ with $p+q-1=s$ to recover \eqref{Wbracket}:
\begin{equation}
    \boxed{\big[T,T'\big]^\W_s= \sum_{p+q=s+1}\big[\iota(\T{p}),\iota(\Tp{q})\big]_{p+q-1}^\W}.
\end{equation}

An important property of this bracket (which is the reason why we use the superscript $\W$) is that it closes on the wedge. 
In other words, $D^{s+2}[T,T']_s=0$ when both $T$ and $T'$ are in $\W(S)$. 
Hence we get (see the proof in Appendix \ref{AppProofWedge})
\begin{equation}
[\cdot\,,\cdot]^\W:\W(S)\times\W(S)\to\W(S). \label{Wbracketclosure}
\end{equation}

\ni \textbf{Relation between $T_s$ and the celestial modes $w^p_m$:}\\
Defining the modes $w^p_{m,k}=-\iota\left(\frac12 z^{p-1+m}\bz^k \right)$, cf.\,\eqref{Tsvsw}, we find that the $\W$-bracket of the latter reduces to 
\begin{equation}
    \big[ w^p_{m,k}, w^q_{n,\l}\big]^\W_{2(p+q-2)-3}=\big(m(q-1)-n(p-1)\big)w^{p+q-2}_{m+n,k+\l}.
\end{equation}
This is the usual form of the $Lw_{1+\infty}$ Lie bracket.

\subsection{The $\Wcal$-algebra \label{Sec:Walgebra}} 

We are now ready to show the following lemma.

\begin{tcolorbox}[colback=beige, colframe=argile] \label{LemmaWedgeAlg}
\textbf{Lemma [Wedge algebra]}\\
The wedge algebra $\Wcal(S)\equiv\big(\W(S),[\cdot\,,\cdot]^\W\big)$ equipped with the bracket \eqref{Wbracket} forms a Lie algebra. 
\end{tcolorbox}

\ni As we already emphasized in the table \ref{table1}, $\Wcal(S)$ is a \emph{generalization} of $Lw_{1+\infty}$ since it is defined for any 2d surface $S$ of arbitrary topology, while  $Lw_{1+\infty}$ is equal to $\Wcal(S_2)$.

\paragraph{Proof:}  
The linearity and anti-symmetry of \eqref{Wbracket} are obvious. 
Besides, we already discussed the closure of the bracket in \eqref{Wbracketclosure}.
In Appendix \ref{AppJacobiVbracket} we also show that 
\be\label{Jac1}
\big[T,[T',T'']^\W\big]_s^\W \cyc (s+3) D^2\T0\big(\Tp0 \Tpp{s+2} - \Tp{s+2} \Tpp0\big).
\ee
The symbol $\cyc$ means that we add the cyclic permutation of $(T,T',T'')$ on both sides of the equal sign, i.e.
\begin{align}
    & f(T,T',T'')\cyc g(T,T',T'') \\
    \Leftrightarrow \quad & f(T,T',T'')+f(T',T'',T)+f(T'',T,T')=g(T,T',T'')+g(T',T'',T)+g(T'',T,T'), \nn
\end{align}
for some arbitrary functionals $f$ and $g$.
It means in particular that
\begin{equation}
    f(T,T',T'')\cyc f(T',T'',T)\cyc f(T'',T,T').
\end{equation}
We see that the RHS of \eqref{Jac1} vanishes when $D^2T_0=0$ which is implied  by the wedge condition.\footnote{The closure of the algebra requires the wedge condition $D^{s+2}\T{s}=0$ for all $s\geqslant -1$ (and not only for $s=0$).} And therefore this shows that the Jacobi identity is satisfied when $T,T',T''\in \W(S)$. 

\paragraph{Twistorial proof:}
We present for completeness an alternative proof of the \hyperref[LemmaWedgeAlg]{Lemma} which highlights the relationship with twistor space when $S=S_2$.
Indeed, the $\W$-bracket satisfies the Jacobi identity since it originates from the image of $L(\mathrm{Ham}(\mathbb{C}^2))$ when projected from twistor space to $\scri$. 
We refer the reader to \cite{Donnay:2024qwq}, whose notation we follow, for extra details.
Projective twistor space can be written as an $O(1)\oplus O(1)$ bundle over $\mathbb{CP}_1$. 
The anti-holomorphic fibers coordinates of this bundle are represented as coordinate $\bar\mu^{\alpha}$ on $\C^2$.
A hamiltonian generator of area preserving diffeomorphism  (i.e. with unit Jacobian\footnote{Which amounts to require the tracelessness of the symmetry parameters. 
This is trivial in twistor space since using \eqref{twistorbracket}, we know that $\poisson{g,\cdot}=n\,g_{\alpha\alpha_2\ldots\alpha_n}(\bz)\bar\mu^{\dot\alpha_2}\ldots\bar\mu^{\alpha_n}\epsilon^{\alpha\beta}\frac{\pa}{\pa\bar\mu^\beta}$. $g_{\alpha(n)}$ being symmetric, we immediately get that the symmetry parameter is trace-free, i.e. $n\,g_{\alpha\beta\alpha_3\ldots\alpha_n}\epsilon^{\alpha\beta}=0$.}) of degree $s+1$ is given by 
\begin{equation}
    g_s=g_{\alpha(s+1)}(\bz)\, \bar\mu^{\alpha(s+1)}, \label{HamGen}
\end{equation}
where $\alpha(n)$ is a multi-index notation that stands for\footnote{$\bar\mu^{\alpha(n)}=\bar\mu^{\alpha_1}\ldots \bar\mu^{\alpha_n}$.} $\alpha_1\cdots\alpha_n$ and  where $g_{\alpha(s+1)}(\bz)$ is  an anti-holomorphic function on $S_2$, 
\begin{equation}
    g_{\alpha(s+1)}(\bz)=\sum_{k=-\infty}^\infty g_{\alpha(s+1)}^{(k)}\bz^k.
\end{equation}
We consider another hamiltonian generator $g'_{s'}$ of degree $s'+1$,
\begin{equation}
    g'_{s'}=g'_{\alpha(s'+1)}(\bz)\, \bar\mu^{\alpha(s'+1)}.
\end{equation}
The algebra $Lw_{1+\infty}$ is then realized through the Poisson bracket
\begin{equation}
    \{\cdot\,,\cdot\}= \epsilon^{\alpha\beta}\frac{\pa}{\pa\bar\mu^\alpha}\frac{\pa}{\pa\bar\mu^\beta}, \label{twistorbracket}
\end{equation}
so that
\begin{equation}
    \{g_s,g'_{s'}\}=g''_{s+s'-1}=g''_{\alpha(s+s')}(\bz)\,\bar\mu^{\alpha(s+s')},
\end{equation}
where
\begin{equation}
    g''_{\alpha(s+s')}(\bz)=(s+1)(s'+1)\epsilon^{\alpha\beta} g_{\alpha\alpha_1\ldots\alpha_s}(\bz)g'_{\beta\alpha_{s+1}\ldots\alpha_{s+s'}}(\bz).
\end{equation}
Still following the notation of \cite{Donnay:2024qwq}, we project the algebra down to $\scri$.
The hamiltonian generator $g_s$ is mapped to the function $\tT_s(\lambda^\alpha,\bz)$ and then to the function $\T{s}(z,\bz)$ given by  
\begin{equation}
    g_s \quad\longrightarrow\quad \tT_s(\lambda^\alpha,\bz)=g_{\alpha(s+1)}(\bz)\lambda^{\alpha(s+1)}=(-\lambda^1)^{s+1}\T{s}(z,\bz), \label{HamGenScri}
\end{equation}
where $z\equiv-\lambda^0/\lambda^1=\lambda_1/\lambda_0$. The function $T_s$ is such that 
\begin{equation}
    \T{s}(z,\bz):=g_{\alpha(s+1)}(\bz)\frac{\lambda^{\alpha(s+1)}}{(-\lambda^1)^{s+1}}=\sum_{k=0}^{s+1}\binom{s+1}{k}(-1)^{s+1-k}z^k\, g_{0(k)1(s+1-k)}(\bz). \label{Tstwistor}
\end{equation}
We define similarly $\tT'_{s'}(\lambda^\alpha,\bz)$ and $\Tp{s}$ for $g'_{s'}$.
Besides, the Poisson bracket \eqref{twistorbracket} is sent to
\begin{equation}
    \{\cdot\,,\cdot\}^{\scri}= \epsilon^{\alpha\beta}\frac{\pa}{\pa\lambda^\alpha}\frac{\pa}{\pa\lambda^\beta}. \label{twistorbracketscri}
\end{equation}
The easiest way to see the correspondence with the $\W$-bracket is to evaluate the bracket in terms of $\pa_z$ acting on $T_s$:
\begin{align}
    \poisson{\tT_s,\tT'_{s'}}^{\scri} &=\epsilon^{\alpha\beta}\left(
    (-\lambda^1)^{s+1}\frac{\pa z}{\pa\lambda^\alpha}\pa_z\T{s}+\T{s}\frac{\pa(-\lambda^1)^{s+1}}{\pa\lambda^\alpha}\right)\left((-\lambda^1)^{s'+1}\frac{\pa z}{\pa\lambda^\beta}\pa_z\Tp{s'}+\Tp{s'}\frac{\pa(-\lambda^1)^{s'+1}}{\pa\lambda^\beta}\right) \nn\\
    &=(-\lambda^1)^{s+s'}\epsilon^{\alpha\beta} \big(\pa_z\T{s}(\delta^0_\alpha+z\delta^1_\alpha)-(s+1)\T{s}\delta^1_\alpha\big)\big(\pa_z\Tp{s'}(\delta^0_\beta+z\delta^1_\beta)-(s'+1)\Tp{s'}\delta^1_\beta\big) \nn\\
    &=(-\lambda^1)^{s+s'}\big((s+1)\T{s}\pa_z \Tp{s'}-(s'+1)\Tp{s'}\pa_z\T{s}\big).
\end{align}
From there we identify\footnote{On $\mathbb{C}^*$ we have $D =\pa_z$.}
\begin{equation}
    \Tpp{s+s'-1}\equiv (s+1)\T{s}\pa_z \Tp{s'}-(s'+1)\Tp{s'}\pa_z\T{s} =[\iota(\T{s}),\iota(\Tp{s'})]_{s+s'-1}^\W.
\end{equation}
The $\W$-bracket thus descends from the canonical Poisson bracket on twistor space, which proves the isomorphism between the projection of $L(\mathrm{Ham}(\C^2))$ on $\scri$ and the $\Wcal$-algebra.
Notice in particular that since $D g_{\alpha(s)}=0$ we have that  $\T{s}$ given by \eqref{Tstwistor} satisfies  $D^{s+2}\T{s}=0$.
The fact that the Jacobi identity holds for $[\cdot\,,\cdot]^\W$ is thus merely a consequence of $\{\cdot\,,\cdot\}$ satisfying Jacobi. 

We bring the reader's attention to the shift in degree between the twistor perspective compare to the gradation in $\V(S)$.
Indeed, the element $\T{s}\in\V_s$ originates from a polynomial of degree $s+1$ in $\bar\mu^\alpha$ or $\lambda^\alpha$ in the twistor perspective (cf. \eqref{HamGen} and \eqref{HamGenScri}).
In particular, the spin 0 super-translations are polynomials of degree 1 in $\bar\mu^\alpha$, so that for $s=0=s'$, we get $\poisson{g_\alpha\bar\mu^\alpha,g'_\beta\bar\mu^\beta}=\epsilon^{\alpha\beta}g_\alpha g'_\beta\neq 0$.
In other words, the super-translations do not commute in twistor space.
From the $\Wcal$-algebra viewpoint, this amounts to the presence of central elements of degree $-1$, as we now discuss.

\paragraph{Central elements:}
Notice that the space  $\W_{-1}(S)$ is central for the wedge bracket. 
Given $c\in \W_{-1}$, we denote by $\hat{c}:=\iota(c) \equiv(c,0,\ldots)$ the series for which $\pi_{-1}(\hat c)=c$ and $\pi_{s}(\hat c)=0$ for $s\geqslant 0$.
It is easy to see that $\hat c$ commutes with any element of $\Wcal$,
\begin{equation}
    [\hat c,T]_s^\W =0,\qquad\forall \,T\in\W(S),\quad s\geqslant -1. \label{central-1}
\end{equation}
The fact that $\W_{-1}(S)$ is central and thus forms an ideal, implies that we can define the quotient algebra 
$\bWcal(S)\equiv\big(\Ws,[\cdot\,,\cdot]^{\overline\W}\big)$ where 
\be
\Ws = \W(S)/\W_{-1}(S) \simeq  \bigoplus_{s=0}^{\infty } \W_s(S). 
\ee 

To define the quotient bracket $[\cdot\,,\cdot]^{\overline\W}$ one recalls that, since the bracket is of degree $-1$, the only way to obtain a degree $-1$ element is through the commutation of two supertranslations, namely
\be 
[T,T' ]^{\W}_{-1}=\T0 D\Tp0-\Tp0 D\T0~ \in \W_{-1}(S).
\ee 
This means that the supertranslations do not commute in $\Wcal(S)$ while they do in $\overline\Wcal(S)$. Overall we have that 
\be 
[T,T' ]^{\Wo}_{-1}=0, 
\qquad
[T,T' ]^{\Wo}_{s}= [T,T' ]^{\W}_s, \quad \mathrm{for}\quad s\geq 0.
\ee 
Therefore all the brackets
in $\bWcal(S)$ agree with the ones in $\Wcal(S)$ except for  the bracket of two supertranslations.

\paragraph{Wedge and Poincaré algebra:}
It is interesting to understand how the central extension works in the case of $S=S_0$ where $\bWcal(S)$ is the self dual truncation of the Poincaré algebra and $\Wcal(S)$ is a three dimensional central extension of it.
As we discussed, we have that $\W(S_0) = \W_{-1}(S_0)\oplus \W_{0}(S_0)\oplus\W_{1}(S_0)$ is ten dimensional.
The basis elements are parametrized by
\be 
C_i := D n_i \in \W_{-1}(S_0), \qquad 
P_{a} := q_a \in \W_0(S_0), 
\qquad
\bJ_i  := \bD n_i \in \W_{1}(S_0),
\ee 
where $i=1,2,3$ and $a=0,1,2,3$, while $q(z,\bz)= (1, n(z,\bz))$ was defined in \eqref{T0T}.
From the definition of the wedge algebra bracket, we can compute the algebra between these generators.
We find that $C_i$ commutes with $(C_j,P_a,\bJ_j)$, while the non trivial commutators are 
\bs \label{PoincB}
\be 
[\bJ_i,\bJ_j] &= -2i\epsilon_{ij}{}^k \bJ_k,\\
[\bJ_i,P_b] &= \bar\eta_{abi}P^a,\\
[P_a ,P_b ] &= \eta_{a b}{}^i C_i,
\ee
\es
where $\epsilon_{ijk}$ is the Levi-Civita tensor, while 
$\eta_{ab}{}^i$ and $\bar\eta_{ab}{}^i$ are the Lorentzian versions of the t'Hooft symbols \cite{t1976computation}. 
They are antisymmetric in $(a,b)$ and their components are given by 
\be 
\eta_{ij}{}^k =\bar\eta_{ij}{}^k= i \epsilon_{ij}{}^k, \qquad 
\eta_{0i}{}^k =-\bar\eta_{0i}{}^k= \delta_i{}^k. \label{tHooftSymbol}
\ee
$\eta_{ab}{}^i$ and $\bar\eta_{ab}{}^i$ are respectively projectors onto the self-dual and anti-self-dual basis of 2-forms.\footnote{One has $(\star\,\eta)_{ab}{}^k=\frac12\epsilon_{ab}{}^{cd} \eta_{cd}{}^k= i \eta_{a b}{}^k$ and $(\star\,\bar\eta)_{ab}{}^k=\frac12\epsilon_{ab}{}^{cd} \bar\eta_{cd}{}^k= -i\bar\eta_{a b}{}^k$, where $\star$ is the Hodge star associated to the Minkowski metric.}
In other words, the three generators $\bJ^k$ are nothing else than the anti-self-dual part of the usual Lorentz generators $\mathcal{J}^{ab}$, i.e. $\bJ^k=\frac{i}{2}\bar\eta_{ab}{}^k\mathcal{J}^{ab}$.
The quotient algebra $\overline{\mathcal{W}}(S_0)$ is obtained by imposing $C_i=0$. In this case we get the anti-self-dual Poincaré algebra
\be 
\overline{\mathcal{W}}(S_0)
= \mathfrak{sl}_{-}(2,\C)\sds \C^4\equiv \mathfrak{iso}_-(4),
\ee 
where $ \mathfrak{sl}(2,\C)_{-}$ denotes the anti-self-dual projection of the complexified Lorentz algebra.
Although the Poincaré algebra $\big(\mathfrak{sl}_{-}(2,\C)\oplus\overline{\mathfrak{sl}}_{+}(2,\C)\big)\sds\C^4$ does not admit any central extension \cite{Nakayama:2023xzu}, it is remarkable that its (anti-)self-dual part does.
The proof of the brackets \eqref{PoincB} is given in appendix \ref{app:Poinc} and relies on the validity of the following identities:
\be \label{nids}
D^2 n_i &=0, & 2 D n_i \bD n_j + n_i n_j &= \delta_{ij} + i \epsilon_{ij}{}^k n_k,\cr
D\bD n_i &= - n_i, &
n_i D n_j - n_j D n_i &= i \epsilon_{ij}{}^k D n_k.
\ee
It is interesting to note that the centrality of $C_i$ is consistent with the Jacobi identity.\footnote{Requiring $[\bJ_i,[P_a,P_b]]=0$ demands that  
$\eta_{b}{}^{cj}\bar\eta_{cai} - \eta_{a}{}^{cj}\bar\eta_{cbi} =0$, which is satisfied due to the orthogonality of self-dual and anti-self-dual projections.}

The central extension of the chiral Poincaré algebra can also be written in spinor variables. In this case the different generators are $(\bJ_{\dot\alpha\dot\beta}, P_{\dot\alpha\alpha}, C_{\alpha \beta})$ which renders manifest the respective chirality of each generator.\footnote{$\bJ_{\dot\alpha\dot\beta}$ is symmetric in $(\dot\alpha,\dot\beta)$ and similarly $C_{\alpha \beta}$ is symmetric.} The translation commutator is given by 
\be 
[P_{\dot\alpha \alpha}, P_{\dot\beta \beta}]= \epsilon_{\dot\alpha\dot\beta} C_{\alpha \beta}.
\ee 
This expression makes it clear that while the anti-self-dual rotation $\bJ_{\dot\alpha\dot\beta}$ acts on $P_{\dot\alpha\alpha}$ on the left, it does not act on $C_{\alpha\beta}$, respecting its chirality. 
Note that the dual chiral algebra is given by generators 
$(J_{\alpha \beta}, \bP_{\alpha \dot\alpha}, \bC_{\dot\alpha\dot\beta})$, where $\bP_{\alpha \dot\alpha}$ denotes the opposite chirality translation generators that commute with $\bJ$ and transform as a vector under the action of $J$.

To recover the usual action of Poincaré one needs to impose the reality condition 
$\bP_{a}=P_{a}\equiv\mathcal{P}_a$. 
This condition means that $[J_i, \mathcal{P}_b]=\eta_{abi}\mathcal{P}^a$ and 
$[\bJ_i, \mathcal{P}_b]=\bar\eta_{abi}\mathcal{P}^a$. The Jacobi identity then implies that $[J_i,[\mathcal{P}_a,\mathcal{P}_b]]\neq 0$. 
Hence  the only way to respect the centrality of $C$'s is to impose $C_{i}=\bC_{i}=0$.\footnote{Another way to be consistent with Jacobi is to impose $C_i =\Lambda J_i$ and $\bC_i=\Lambda \bJ_i$. The $C$'s are no longer central and we recover the de Sitter or anti-de Sitter algebra depending on the sign of $\Lambda$.} 
With these 10 constraints we have that 
$\mathcal{P}_a, J_i, \bJ_i$ represent $4+3+3=10$ generators which form the Poincaré algebra.
We learned \textit{a posteriori} that this central extension of chiral Poincaré was already discussed in \cite{Krasnov:2021cva}.

\section{Deforming the wedge algebra \label{sec:DeformingWedge}}

In this section we show that the presence of a shear element $\sigma \in \Ccel{1,2}$ allows for the deformation of the wedge algebra.

\subsection{Can we extend the $\W$-bracket beyond the wedge?}

We know that the $\gbms$ algebra closes without any wedge restriction on $\T0$ and $\T1$.
Besides, it was proven in \cite{Freidel:2023gue} that the wedge condition on $\T2$ is not necessary for the algebra of higher-spin charges to close at the quadratic level. In the same work, it was also proven that, in Yang-Mills, the wedge restriction can be lifted at quadratic order for all spin $s$ charges.
These results strongly suggest that the symmetry can be extended beyond the wedge and we now investigate how we can waive the restriction $D^{s+2}\T{s}=0$ at the algebraic level. 

The first question that arises is whether we can simply consider the bracket $[\cdot\,,\cdot]^\W$ 
to be valid beyond the wedge, i.e. to be viewed as a map
$[\cdot\,,\cdot]^\V:\V(S)\times\V(S)\to\V(S)$. This means that
\begin{equation}
    [T,T']^\V_s= \sum_{n=0}^{s+1} (n+1)\big(\T{n} D \Tp{s+1-n}-\Tp{n} D \T{s+1-n}\big),  \label{Vbracket}
\end{equation}
for $T,T'\in\V(S)$. 
This $\V$-bracket is simply the naive extension of the $\W$-bracket. 
As we are about to see, it is ill-defined since fails to satisfy the Jacobi identity, once we go beyond the wedge.
This is surprising at first sight since the $\V$-bracket is very closely related to a well-known bracket: 
the Schouten-Nijenhuis (SN) bracket for \textit{symmetric} contravariant tensor fields.
Be mindful that the literature often deals with the SN bracket for \textit{antisymmetric} multivectors. 
For the symmetric version, check the original reference \cite{Schouten} and \cite{SNbracket} for a modern perspective.

\begin{tcolorbox}[colback=beige, colframe=argile, breakable] \label{SNreminder}
\textbf{Reminder [Schouten-Nijenhuis bracket]}\\
Consider $\X^\bullet(M)= \bigodot^\bullet\Gamma(TM)$, the algebra of symmetric contravariant tensor fields of any rank over a $d$ dimensional manifold $M$ with coordinates $\{x^i\}_{i=1}^d$. For $S_p\in\X^p(M)$ and $\tilde S_q\in\X^q(M)$, the SN bracket is a graded commutator of degree 1, i.e.
\begin{center}
\begin{math}
    [\cdot\,,\cdot]^\sn:\X^p(M)\times \X^q(M)\to\X^{p+q-1}(M),\quad \forall\,p,q\in\N,
\end{math}
\end{center}
with components given by
\begin{equation}
    \big[S_p,\tilde S_q\big]^{\sn, i_1\ldots i_{p+q-1}}=pS_p^{j(i_1\ldots i_{p-1}}\frac{\pa}{\pa x^j}\tilde S_q^{i_{p} \ldots i_{p+q-1})}-
    q\tilde S_q^{j(i_1\ldots i_{q-1}}\frac{\pa}{\pa x^j} S_p^{i_q\ldots i_{q+p-1})}. \label{SNbracketDef}
\end{equation}
This is a well-defined tensorial expression, so that the partial derivatives can be replaced by any  torsion-free covariant derivative.
In our case where the manifold is the 2-sphere/complex plane (such that $i_1,i_2\ldots\to A_1,A_2\ldots$ and $\{x^A\}\equiv(z,\bz)$) and we restrict to symmetric \textit{holomorphic traceless} tensors, the projection of \eqref{SNbracketDef} along $\bm_{A_1}\ldots\bm_{A_{p+q-1}}$ reads\footnote{We use that the sphere metric $\gamma_{AB}=m_A\bm_B+\bm_A m_B$.}${}^,$\footnote{Notice the natural abuse of notation where on the LHS, $S_p$ refers to the full tensor as in \eqref{SNbracketDef}, while on the RHS, it stands for the projected quantity along $\bm$'s.}
\begin{equation}
    \big[S_p,\tilde S_q\big]_{p+q-1}^\sn=pS_p D\tilde S_q-q\tilde S_q DS_p.
\end{equation}
This means that for generic $S,\tilde S\in\X^\bullet(M)$ we have 
\begin{equation}
    \boxed{\big[S,\tilde S\big]^\sn_s= \sum_{p+q=s+1}\big[S_p,\tilde S_q\big]_{p+q-1}^\sn=\sum_{n=0}^{s+1} n\big(S_nD\tilde S_{s+1-n} -\tilde S_nDS_{s+1-n}\big).} \label{SNbracketDefbis}
\end{equation}
\end{tcolorbox}

We thus see that what we are calling $\V$-bracket is actually a shifted SN bracket where the prefactor between \eqref{SNbracketDefbis} and \eqref{Vbracket} is shifted from $n$ to $n+1$.
Now comes the intriguing part, with far reaching consequences: Although the original SN bracket satisfies the Jacobi identity (see App.\,\ref{JacobiSN} for a reminder), this is not the case for \eqref{Vbracket}.

In the appendix \ref{AppJacobiVbracket}, we  demonstrate that
\begin{equation}
    \boxed{\left[T,[T',T'']^\V\right]^\V_s \cyc D^2\T0 \paren{T',T''}_s}, \label{Jacobianomaly}
\end{equation}
where we introduced the new quantity\footnote{I refer to $\paren{\cdot\,,\cdot}$ as the Dali-bracket, as a reference to the paintings with melting watches.}
\begin{equation}
\boxed{\paren{T,T'}_s:=(s+3) \left( \T0\Tp{s+2}-\Tp0\T{s+2}\right)} \quad\in ~{\Ccel{-2,-s-2}}. \label{Cbracket}
\end{equation}
Notice that the Jacobi anomaly is proportional to $D^2\T0$.
In other words, it disappears if we restrict the algebra to the wedge, as it should.

\paragraph{Relation to $\gbms$:}
If we consider solely transformations of helicity $s=0$ and $s=1$, for which $\T{-1}\equiv 0$ and $\T{s+2}\equiv 0$ for $s\geqslant 0$, then \eqref{Cbracket} vanishes. 
Formally, this amounts to $T,T'\in\V_0\oplus\V_1$.
The bracket \eqref{Vbracket} restricted to these symmetry parameters is explicitly given by
\bs
\label{bracketdeg01}
\be
[T,T']^\V_0 &= (\T0 D\Tp1 - \Tp0 D\T1)  + 2(\T1 D\Tp0 - \Tp1 D\T0), \label{bracketdeg0}\\
[T,T']^\V_1 &= (\T0 D\Tp2 - \Tp0 D\T2)  + 2(\T1 D\Tp1 - \Tp1 D\T1). \label{bracketdeg1}
\ee
\es
The quotient algebra obtained by imposing the vanishing of the central elements, i.e. $\T{-1}=0$ is the $\gbms$ algebra \cite{Campiglia:2014yka, Compere:2018ylh, Campiglia:2020qvc, Freidel:2021fxf}.
To recognize the latter, we denote $\T0\equiv T$, the super-translations parameter and $\T1\equiv Y/2$, the super-Lorentz-rotations parameter. 
We get that the super-translations commute,
\begin{subequations}
\begin{equation}
   [T,T']^{\gbms}=0. \label{gbmstranslation}
\end{equation}
Picking respectively $\T0=0,\T1=Y/2,\Tp0=T', \Tp1=0$ in \eqref{bracketdeg0} and $\T0=0,\T1=Y/2,\Tp0=0, \Tp1=Y'/2$ in \eqref{bracketdeg1}, we get the commutation relations with super-rotations:
\be 
    [Y, T']^{\gbms}= Y DT'-\frac12 T' DY,\qquad [Y,Y']^{\gbms} = YDY'-Y'D Y. \label{gbmsrotation}
\ee 
\end{subequations}
We employ the notation $[\cdot\,,\cdot]^{\gbms}$ which represents  $[\iota(\cdot)\,,\iota(\cdot)]^{\overline{\V}}$  restricted to spin $0$ and $1$, where $\Vs$ is the quotient $\V(S)/\V_{-1}$.
Importantly, $\gbms$ forms an algebra without any wedge restriction $D^2 T=0=D^3 Y$ on its parameters.\\

There is even a third case where \eqref{Cbracket} cancels out. 
Indeed, even if the anomalous quantity depends on arbitrary high spins $s$, it also always includes $\T0$.
By restricting the algebra to $\Vss$ (where precisely $\T0\equiv 0$, cf. \eqref{VstarstarDef}), we thus restore the Jacobi identity.
On top of that, notice that the $\V$-bracket closes on $\Vss$,\footnote{It is clear since for any $T,T'\in\Vss$, we have that $[T,T']_0^\V = 0$. Hence $[T,T']\in \Vss$ .}
\begin{equation}
[\cdot\,,\cdot]^\V:\Vss\times\Vss\to\Vss.
\end{equation}
We thus proved that 
\begin{tcolorbox}[colback=beige, colframe=argile]
\textbf{Lemma [$\bbVcal$ algebra]}\\
The space $\bbVcal(S)\equiv \big(\Vss,[\cdot\,,\cdot]^\V\big)$ forms a Lie algebra.
\end{tcolorbox}
This Lie algebra is a higher spin generalization of $\mathrm{Diff}(S)$ which is contained in $\bbVcal(S)$ as its spin $1$ component. 
This proves that there exist three algebras, $\Wcal(S),\,\gbms(S)$ and $\bbVcal(S)$, for which the $\V$-bracket is a genuine Lie bracket.\footnote{This includes any sub-algebras of these, such as the $\ebms(S)$ sub-algebra of $\gbms(S)$ which consists of holomorphic vector fields on $S$, i.e. vector fields satisfying $\bD Y=0$. }
The next step consists in building a new bracket, denoted $[\cdot\,,\cdot]^\sigma$, that reduces to the $\V$-bracket for $\Wcal$, $\gbms$ and $\bbVcal$.

\subsection{Covariant wedge algebra \label{sec:NewFieldDepBracket}}

In order to resolve the Jacobi identity issue \eqref{Jacobianomaly}, we introduce a new field dependent bracket 
\begin{empheq}[box=\fbox]{align}
    [T,T']^\sigma_s &=[T,T']^\V_s -\sigma\paren{T,T'}_s,\qquad s\geqslant -1\nn\\
    &=\sum_{n=0}^{s+1}(n+1)\big(\T{n}D\Tp{s+1-n}-\Tp{n}D\T{s+1-n}\big)-(s+3)\sigma\big(\T0\Tp{s+2}-\Tp0\T{s+2}\big), \label{CFbracket}
\end{empheq}
where $\sigma(z,\bz)$ is a spin-weighted function on $S$.
At that stage, it can be viewed as a parameter of the algebra. 
The physical motivation for this deformed bracket will appear in chapter \ref{secGRII}.
From a purely algebraic point of view, notice that the $\sigma$-bracket is built from the $\V$-bracket by subtracting the part of the Jacobi identity anomaly which looks like a bracket, namely what we called the Dali-bracket $\paren{\cdot\,,\cdot}$.
The latter is multiplied by the deformation parameter $\sigma$, whose properties we constraint in order to construct a Lie bracket.

The difference between $[\cdot\,,\cdot]^\sigma$ and $[\cdot\,,\cdot]^\V$ involves $\paren{T,T'}_s$, which vanishes for $\gbms$
and $\Vss$. Therefore, $[\cdot\,,\cdot]^\sigma$ reduces to the $\V$-bracket for these cases.
Note that $\paren{T,T'}_s$ has celestial weight $-2$ and helicity $-(s+2)$; therefore the additional field dependent term $\sigma\paren{T,T'}_s$ belongs to $\V_s$ provided we choose $\sigma$ of celestial/spin weights $(1,2)$.\footnote{This implies that $\pi_s (\sigma \paren{T,T'})= \sigma \paren{T,T'}_s $.}

Let us investigate the Jacobi identity for that newly proposed bracket. 
The double commutator expands as
\begin{align}
     \big[T,[T',T'']^\sigma\big]_s^\sigma
     &= \big[T,[T',T'']^\V\big]_s^\V
     - \sigma \paren{T,[T',T'']^\V}_s
     - \big[T, \sigma\paren{T',T''}\big]_s^\V
     + \sigma \paren{T, \sigma\paren{T',T''}}_s.
\end{align}
We then evaluate the cyclic permutation of each term entering this expression. 
We report the details of computation in appendix \ref{AppJacobiCF}.
Besides \eqref{Jacobianomaly}, we have 
\begin{equation}
    - \sigma \paren{T,[T',T'']^\V}_s -\big[T,\sigma\paren{T',T''}\big]_s^\V \cyc -\big(2D\sigma\T1+3\sigma D\T1\big)\paren{T',T''}_s \label{Jacobi1}
\end{equation}
and
\begin{equation}
    \sigma \paren{T, \sigma \paren{T',T''}}_s \cyc 3 \sigma^2\T2\paren{T',T''}_s. \label{Jacobi2}
\end{equation}
Therefore we conclude that
\begin{align}
     \boxed{\big[T,[T',T'']^\sigma\big]_s^\sigma\cyc -\hdT \sigma\paren{T',T''}_s}, \qquad s\geqslant -1,\label{JacCF}
\end{align}
where
\begin{equation}
    \boxed{\hdT \sigma :=-D^2\T0+2D\sigma\T1+3\sigma D\T1-3\sigma^2\T2} \quad \in\, \Ccel{1,2}. \label{dTCv1}
\end{equation}

In order for the Jacobi identity to hold, the simplest choice is to set $\hdT \sigma=0$, a condition under which $[\cdot\,,\cdot]^\sigma$ is a well-behaved Lie algebra bracket.
In that case, $\sigma$ is a  field whose field variation vanishes.\footnote{It is a functional constant in the sense that $\hdT \sigma=0$.}
It plays the role of a background structure that represents the symmetry algebra atop a coherent state of gravitons. 
As we will see in more details, $\sigma$ also represents a deformation parameter mapping the wedge algebra onto the generalized wedge algebra $\Wcal_{\sigma}(S)$.
Note that the condition $\hdT\sigma=0$ only gives a relation between $(\T0,\T1,\T2)$. 
However if we demand that such elements form an algebra, i.e. that $\hat\delta_{[T,T']^\sigma}\sigma=0$ when $\hdT\sigma=0=\hdTp\sigma$, then this leads to a generalization of the wedge condition on the parameters $T_s$.
To linear order in $\sigma$, this generalized wedge condition reads 
\be \label{wedge approximate}
0= D^{s+2} T_s -\sum_{k=0}^{s+1} (k+1) D^k\big(\sigma D^{s+1-k}T_{s+1}\big) +\cdots,\qquad s\geqslant -1,
\ee 
where the dots stand for terms of higher powers in $\sigma$ when $s>-1$.
\bigskip

We now demonstrate the various statements of the last paragraph. 
For this, it is essential to introduce a preferred element $\Ham \in \V_{-1} \oplus \V_0$.\footnote{Any element of $\V_{-1}$ is central for the $\sigma$-bracket, like $\W_{-1}$ was for the $\W$-bracket.}
In other words $\Ham :=(\Ham_{-1},\Ham_0,0,0\ldots)$ is such that $\pi_s(\Ham)=0,\,s>0$. 
Note that $\Ham_0\in \V_{0}$ is, by definition, of conformal dimension $-1$.\footnote{Like the inverse of the  square root of a density on a 2d manifold.}
It therefore transforms non trivially under sphere diffeomorphisms.
We assume that $\Ham_0\neq 0$.
Quite remarkably, this means by Moser's theorem \cite{moser1965volume} that there exists a diffeomorphism frame such that $\Ham_0$ is constant, i.e. $D\Ham_0=0=\bD\Ham_0$.\footnote{The condition $D\Ham_0=0$ is not preserved by diffeomorphisms: 
under an infinitesimal diffeomorphism generated by the element $\iota(Y)$ with $Y\in \V_1$, we have that  $[\Ham, \iota(Y)]^\sigma_0= \Ham_0 DY \neq 0$.
This is analogous to the statement that the energy is not boost invariant.} 
We choose the scale of $\Ham_0$ to be such that $\Ham_0=1$ in this frame. 
We are then left with the choice of $\Ham_{-1}$. 
This freedom can be parameterized by another spin 0 element $G \in \V_0$ through $\Ham_{-1} =\Ham_0 DG \in \V_{-1}$---we can check that the dimensions match as $DG\in \Ccel{0,1}$. 
In the following we fix $G$ to be the \emph{Goldstone} mode associated with $\sigma$. 
In other words, $G$ is such that
\be 
  D^2 G =\sigma. \label{Goldstone}
\ee 
To summarize we chose a frame of reference where $\Ham\equiv\hHam$ is given by 
\be 
  \Ham_0=1, \qquad \Ham_{-1}= DG, \qquad \Ham_s=0, \quad \mathrm{for}~ s\geq 1.
\ee

Plugging in \eqref{CFbracket}, we get that
\begin{equation}
    \boxed{[\hHam,T]^\sigma_s= \Ham_0\big(D\T{s+1} -(s+3)\sigma \T{s+2}\big)}~\in\,\V_s,\qquad s\geqslant -1. \label{HamTbracket}
\end{equation} 
We kept $\Ham_0$ explicit to avoid confusion with the weights counting (since $[\hHam,T]_s^\sigma$ and $DT_{s+1}$ have different weights). 
In the frame we work in $\Ham_0=1$, so we could leave it implicit, but it is useful to keep it to remember that $\Ham_0$ is of conformal weight $-1$. 
This reflects the general fact that any equality between expressions with the same spin can be written in an arbitrary frame using powers of $\Ham_0$ to ensure that the conformal weight matches.  

The quantity in parenthesis in \eqref{HamTbracket} is of fundamental importance in the following.
We thus define
\begin{equation}
    \boxed{(\Dcal T)_s:=D\T{s+1} -(s+3)\sigma \T{s+2}}~\in\,\Ccel{0,-s}. \label{CovDer}
\end{equation}
Depending on the context, either \eqref{HamTbracket} or \eqref{CovDer} will turn out to be the most convenient way to write or manipulate expressions.
Both equations eventually carry the same information and can be used interchangeably recalling that\footnote{Notice that $\ad\hHam=\ad\Ham_{\scriptscriptstyle{G'}}$ for two Goldstone fields $G$ and $G'$.}
\begin{equation}
    \boxed{\Ham_0\Dcal T=\ad\hHam(T)=[\hHam,T]^\sigma}.
\end{equation}
We introduced the natural notation 
$\ad: \V(S) \to \mathrm{End}(\V(S))$
for the adjoint map. By definition this map is such that 
$ \ad T( T')=[T,T']^\sigma.
$ It sends an element $T\in \V(S)$ to a linear map in $\V(S)$. 
We can restrict this map to $\W_\sigma$ and anticipating the upcoming result that $[\cdot\,,\cdot]^\sigma$ is a Lie algebra bracket, we find that 
$\ad: \Wcal_\sigma(S) \to \mathrm{Der}(\Wcal_\sigma(S))$, i.e.
$\ad T$, with $T \in \Wcal_\sigma(S)$ maps $\Wcal_\sigma$ onto itself and is a derivation of the $\sigma$-bracket. 

We also point out the difference of notation when we write \eqref{CovDer}.
$D\T{s+1}$ is the concise version of $(DT)_s\equiv\pi_s(DT)$.\footnote{Here the celestial weight is irrelevant, so we just assume that $\pi_s$ is defined the same way as the helicity evaluation at degree $s$, on fields with arbitrary conformal dimension.}
While this convention is not confusing when using $D$, it is not possible for $\Dcal$. It is important to keep the parenthesis when dealing with $(\Dcal T)_s$, because of the extra shift in degree for the part proportional to $\sigma$.\footnote{To say it in another way: the action of $\Dcal$ (and powers thereof) is well-defined only on a graded vector $T\in\V(S)$, and not on a single grade $s$ element $\T{s}$.}

The introduction of $\hHam$ and $\Dcal$ allows us to give a new flavor to the quantity $\hdT\sigma$.
To see this, first notice that \eqref{CovDer} makes sense for $s=-2$, even if the $\sigma$-bracket \eqref{HamTbracket} is only defined for $s\geqslant -1$.
In the following, we thus denote
\begin{equation}
    [\hHam,T]^\sigma_{-2}:=\Ham_0(\Dcal T)_{-2}\qquad\textrm{with}\qquad (\Dcal T)_{-2}=D\T{-1}-\sigma\T0. 
    \label{DefAdjoint-2}
\end{equation}
Using this notation we can conveniently write the variation of $\sigma$ as a double commutator:\footnote{Strictly speaking we have $\big[\hHam,[\hHam,T]^\sigma\big]^\sigma_{-2}=-\Ham_0^2 \hdT \sigma$.}
\begin{align}
    \big(\ad^2\hHam(T)\big)_{-2}\equiv \big[\hHam,[\hHam,T]^\sigma\big]^\sigma_{-2}
    &= D[\hHam,T]^\sigma_{-1}- \sigma [\hHam, T]_0^\sigma \cr
    &= D\big(D\T0 - 2 \sigma \T1\big)- \sigma \big(D\T1 -3\sigma \T2\big) \label{doubleB}\\
    &= - \hdT \sigma. \nn
\end{align}
Moreover, since we know the variation $\hdT\sigma$, we can compute $\big[\hdT,\hdTp\big]\sigma$ and $\hat\delta_{[T,T']^\sigma}\sigma$.
The sum of these two reduces to\footnote{The fact that the RHS is non-zero, in other words that $\dT$ is not an algebra action, is connected to the fact that  $[\cdot\,,\cdot]^\sigma$ is not a Lie bracket when $\dT\sigma\neq 0$.}
\begin{align}
    &\big[\hdT,\hdTp\big]\sigma +\hat\delta_{[T,T']^\sigma}\sigma =\T0\,\hat\delta_{[\hHam,T']^\sigma}\sigma -\Tp0\,\hat\delta_{[\hHam,T]^\sigma}\sigma, \label{algebraRepinterm}
\end{align}
where 
\begin{equation}
    -\hat\delta_{[\hHam,T]^\sigma} \sigma =\big(\ad^3\hHam(T)\big)_{-2}:= \left[\hHam,\big[\hHam,[\hHam,T]^\sigma\big]^\sigma\right]^\sigma_{-2}. \label{deltaHamTsigma}
\end{equation}
We report the brute force demonstration of \eqref{algebraRepinterm} in appendix \ref{AppAlgebraAction}.

From  these computations, we deduce that:
$\hdT\sigma$, and thus $(\ad^2\hHam(T))_{-2}$, has to vanish in order for the  Jacobi identity to hold for the $\sigma$-bracket. 
Now given two elements $T,T'\in \V(S)$ such that 
$\hat\delta_T\sigma=\hat\delta_{T'}\sigma=0$ we need to have that $\hat\delta_{[T,T']^\sigma}\sigma=0$ in order for the set of such elements  to form a Lie algebra.

From \eqref{algebraRepinterm} we conclude that this happens if  $ \hat\delta_{[\hHam, T]^\sigma}\sigma$ and $ \hat\delta_{[\hHam, T']^\sigma}\sigma$ vanish too.
To summarize, the property necessary for the Jacobi identity is preserved by the bracket  if $T$ is such that $\hdT\sigma=(\ad^2\hHam (T))_{-2}=0$ and $\hat\delta_{[\hHam, T]^\sigma}\sigma=(\ad^3\hHam(T))_{-2}=0$. Applying the latter to $T=\ad \hHam(T')$ means that we have to demand $(\ad^4\hHam(T))_{-2}=0$ and so on.
This leads us to define the parameter space $\W_\sigma(S)$ as follows.

\begin{tcolorbox}[colback=beige, colframe=argile]
\textbf{Definition [Covariant wedge space]} \label{Wdef}\\
The covariant wedge parameter space is denoted $\W_\sigma(S)$ and given by 
\begin{equation}
\W_\sigma(S):= \Big\{ T \in \V(S)\,~\big|~ \,D\big(\ad^n\hHam(T)\big)_{-1}=\sigma \big(\ad^n\hHam(T)\big)_{0}, \,\forall\,n\geqslant 0 \Big\},
\label{Wdefeq}
\end{equation}
where $\ad\hHam (T):= [\hHam,T]^\sigma$ is the adjoint action of the \textit{Hamiltonian} $\hHam$.
\end{tcolorbox}

\ni This definition is such that if $T\in \W_\sigma(S)$ then $[\hHam, T]^\sigma \in \W_\sigma(S)$.

\paragraph{Remark:}
Note that according to our definition \eqref{DefAdjoint-2} the wedge condition can also be written, more succinctly, as $\left(\ad^{n+1}\hHam(T)\right)_{-2}=0$. 
This suggests to identify  the graded vector space $\V(S)$ with the subspace of $V_{-2}\oplus \V(S)$ such that all degree $-2$ elements vanishes, i.e. such that $T_{-2}=0$. 
This is trivial addition but it turns out to be very convenient once endowed with an algebra structure. 
It simply requires to extend the $\sigma$-bracket to the degree $-2$ taking $[\cdot\,,\cdot]^\sigma_{-2}:=0$. 
This choice is consistent with the definition of the space $\W_\sigma(S)$ and  perfectly natural since this is precisely what one obtains by evaluating our original definition \eqref{CFbracket} at $s=-2$. 
Notice how on one hand, $[T,T]^\V_{-2}=0$ since the sum is empty, while on the other hand $\paren{T,T'}_{s}=0$ for the special case $s=-2$. \\

The definition of $\hHam$ and its repeated action $\ad^{n+1}\hHam (T)= \big[\hHam,\ad^n\hHam(T)\big]^\sigma$ implies that the action of $\Dcal$ is recursively given by 
\be 
(\Dcal^{n+1} T)_{s}:=D (\Dcal^{n}T)_{s+1}-(s+3)\sigma(\Dcal^{n}T)_{s+2}\,\, \in\,\,\Ccel{n,-s},
\ee 
and is such that $\Ham_0^n(\Dcal^n T)_s =\big(\ad^n\hHam(T)\big)_s$.
From there, one straightforwardly obtains that $\big(\ad^{s+2}\hHam (T)\big)_{-2}$ is given to first order in $\sigma$ by the RHS of \eqref{wedge approximate}.
When $\sigma =0$, this condition reduces to the wedge condition $D^{s+2} T_s=0$.
Moreover, since the condition on $T$ is linear, $\W_\sigma(S)$ is a linear space. 
We wrote down already the expression for $(\Dcal^n T)_{-2},\,n=2$, cf. \eqref{doubleB}. 
It is interesting to look at $n=1,3$ too.
In brief, $(\Dcal^n T)_{-2}=0$  for $n=1,2,3$ gives
\bs
\begin{align}
D T_{-1}&=\sigma T_0, \label{Initialcond}\\
D^2\T0 & =2D(\sigma\T1)+\sigma D\T1-3\sigma^2\T2, \\
D^{3} T_1 &= 3 D^2(\sigma T_2)+ 2D(\sigma D\T2) + \sigma D^2 T_2 \cr
& - 8 D(\sigma^2 T_3)- 4 \sigma D(\sigma \T3)- 3 \sigma^2 D T_3 +15 \sigma^3 T_4.
\end{align}
\es

\paragraph{Remark:} 
$\hHam\in\W_\sigma(S)$ because of its non zero central part, namely $\Ham_{-1}$, which has been defined precisely such that $D\Ham_{-1}=D^2G=\sigma=\sigma\Ham_0$ in our preferred frame, and because $\ad\hHam(\hHam)=0$.\\

The previous analysis leads to one of the main results of this paper. 

\begin{tcolorbox}[colback=beige, colframe=argile] \label{theoremWalgebra}
\textbf{Theorem [$\Wcal_\sigma$-algebra]}\\
The space $\Wcal_\sigma(S)\equiv\big(\W_\sigma(S),[\cdot\,,\cdot]^\sigma)$ equipped with the $\sigma$-bracket is a Lie algebra.  
\end{tcolorbox}

\paragraph{Proof:}
The fact that the $\sigma$-bracket satisfies Jacobi follows from \eqref{JacCF} and \eqref{doubleB}.
So far we have also seen from \eqref{algebraRepinterm}, that when  $T,T'\in \W_\sigma$ then $\hat{\delta}_{[T,T']^\sigma}\sigma=0$.
This is a necessary but not sufficient condition for $[T,T']^\sigma$ to belong to $\W_\sigma$.  
To show that the $\sigma$-bracket closes on $\W_\sigma$, i.e. $[T,T']^\sigma\in \W_\sigma$ when $T,T'\in  \W_\sigma$, we first use that when $T,T'\in  \W_\sigma$, then (see App.\,\ref{AppCovWed1})\footnote{Notice that \eqref{doubleB} is compatible with \eqref{com1}.}
\be \label{com1}
\big[\hHam, [T,T']^\sigma\big]^\sigma_{-2}=D[T,T']^\sigma_{-1}- \sigma[T,T']^\sigma_0 
= \Tp0 \hdT\sigma - \T0 \hdTp\sigma=0.
\ee
This in turn allows us to compute the quantity
\begin{align}
    \Big(\ad^2\hHam\big([T,T']^\sigma\big) \Big)_{-2} &=\Big(\ad\hHam\Big(\big[[\hHam,T]^\sigma,T'\big]^\sigma+\big[T,[\hHam,T']^\sigma\big]^\sigma\Big) \Big)_{-2} \cr
    &=\Big(\Tp0\,\hat\delta_{ [\hHam,T]^\sigma} \sigma-[\hHam,T]^\sigma_0\hdTp\sigma\Big)-T\leftrightarrow T' \label{adjointsquare}
    =0.
\end{align}
In the first step, we used that $\hdT\sigma=0=\hdTp\sigma$, i.e. $T,T'\in\Wcal_\sigma(S)$, so that we could leverage the Jacobi identity\footnote{Obviously $\hat\delta_{\hHam}\sigma=0$.} on the  internal elements $\left[\hHam,[T,T']^\sigma\right]^\sigma \cyc 0 $.
The second line follows from \eqref{com1} applied to $(T,T')\to \big([\hHam, T]^\sigma, T'\big)$, while in the last step, we  used again that $T,T'\in\Wcal_\sigma(S)$ together with \eqref{doubleB}.
We can then show recursively that $\big(\ad^n\hHam\big([T,T']^\sigma\big) \big)_{-2}=0$, $n\geqslant 1$, when $T,T'\in\Wcal_\sigma(S)$, which concludes the proof.

\subsection{Covariant form of the $\sigma$-bracket \label{sec:CovFormBracket}}

The fact that $\Wcal_\sigma$ is a Lie algebra suggests that it should be possible to write the bracket in a strict Lie algebra form, namely with field independent structure constants.  
This is what we describe in this subsection, using the  surprising appearance and importance of the degree $-2$ seen in the previous subsection.

Previously, we proved that for $[\cdot\,,\cdot]^\sigma$ to be a Lie bracket, $D\T{-1}$ has to equate $\sigma\T0$. 
Using this condition, we find that 
the $\sigma$-bracket can be simply written as 
\begin{equation}
    \boxed{[T,T']^\sigma_s =\sum_{n=0}^{s+2}(n+1)\big(\T{n}D\Tp{s+1-n}-\Tp{n}D\T{s+1-n}\big)}, \qquad s\geqslant -2. \label{CFbracketSphereVersion}
\end{equation}
The RHS of this expression is similar to the $\V$-bracket, but the sum now runs till $s+2$ and the bracket is naturally defined for all elements $\T{s},~s\geqslant -2$.
The boundary term $n= s+2$ precisely generates the piece $-(s+3)\sigma\paren{T,T'}_s$ once we use the initial constraint $DT_{-1}=\sigma T_0$. Under this constraint we also have that  $[T,T']^\sigma_{-2}= \T0 D\Tp{-1} - \Tp0 DT_{-1}=0$.
It is also worth emphasizing that the degree $-1$ elements are central, as they should. 
Indeed, denoting $\hat c=(c,0,0,\ldots)\in \iota(\V_{-1})\cap\W_\sigma(S)$ (which means that $Dc=0$), we readily get that $[T,\hat c]^\sigma_s=(s+3)\T{s+2}Dc=0$.

\paragraph{Remark:}
This new way of writing $[\cdot\,,\cdot]^\sigma$ sheds light on the fact that $\Wcal_\sigma$ is a Lie \textit{algebra}.
In other words, that \eqref{CFbracketSphereVersion} is a genuine Lie algebra bracket on the sphere, with field \textit{independent} structure constants.
Indeed, so far we said that $\sigma=\sigma(z,\bz)$ was a special field such that $\hdT\sigma=0$. 
Hence, rather than being a structure constant, it played the role of a \textit{functional} structure constant.\footnote{This suggests an algebroid framework to tackle the general case $\hdT\sigma\neq 0$.}
What we show by \eqref{CFbracketSphereVersion} is that the residual field dependency of the functional structure constants can be re-absorbed into the constraint between the generators $\T{-1}$ and $\T0$. 

\paragraph{Remark:}
In the appendix \ref{AppJacobi-1Version}, we give a proof of the Jacobi identity starting from the definition \eqref{CFbracketSphereVersion}, where we stay agnostic about the element $\T{-1}$, i.e. we do not impose $D\T{-1}=\sigma\T0$ a priori.
To avoid confusion, we remove the superscript $\sigma$ to denote the RHS of  \eqref{CFbracketSphereVersion}.
We find that
\begin{align}
    \big[T,[T',T'']\big]_s \cyc - \big[T',T''\big]_{-2} D\T{s+3}-(s+4)\T{s+3}D\big[T',T''\big]_{-2}, \qquad s\geqslant -2, \label{Jacobi-1Version}
\end{align}
which vanishes when $\big[T,T'\big]_{-2}\equiv 0$.
Since 
\begin{equation}
    \big[T,T'\big]_{-2}=\T0 D\Tp{-1}-\Tp0 D\T{-1},
\end{equation}
this happens if and only if the initial constraint \eqref{Initialcond} is satisfied.\footnote{The vanishing of $[T,T']_{-2}$ implies that $DT_{-1}$ is proportional to $T_0$, with a coefficient in $\Ccel{1,2}$. This coefficient is $\sigma$.}
Under this condition we have that $[\cdot\,,\cdot]\equiv[\cdot\,,\cdot]^\sigma$, and the analysis of subsec.\,\ref{sec:NewFieldDepBracket} carries over.
This result is interesting on its own since it shows how $[\cdot\,,\cdot]$ could also be a legitimate modification of the $\W$-bracket and another starting point for the analysis performed in subsec.\,\ref{sec:NewFieldDepBracket}.\\

Next, the bracket \eqref{CFbracketSphereVersion} can also be recast as
\begin{equation}
    \boxed{[T,T']^\sigma_s =\sum_{n=0}^{s+1}(n+1)\big(\T{n}(\Dcal T')_{s-n}-\Tp{n}(\Dcal T)_{s-n}\big)}, \qquad s\geqslant -2.\label{CFbracketCovariantSphere}
\end{equation}
To see this, note that the quantity
\begin{equation}
    \sum_{n=0}^{s+2}(n+1)(s+3-n)\sigma\big(\T{n}\Tp{s+2-n}-\Tp{n}\T{s+2-n}\big)
\end{equation}
is equal to its opposite (after a change of variable $n \to s+2-n$), and thus vanishes.
Adding it to \eqref{CFbracketSphereVersion}, it precisely changes $D\to\Dcal$, so that 
\begin{equation}
    [T,T']^\sigma_s =\sum_{n=0}^{s+2}(n+1)\big(\T{n}(\Dcal T')_{s-n}-\Tp{n}(\Dcal T)_{s-n}\big), \qquad s\geqslant -2.\nn
\end{equation}
The term $n\equiv s+2$ of this sum drops since it involves $(\Dcal T)_{-2}$ (and $(\Dcal T')_{-2}$), namely the initial condition \eqref{Initialcond} written in covariant form.

The viewpoint \eqref{CFbracketCovariantSphere} is particularly pleasing since it eventually shows that the $\sigma$-bracket on the covariant wedge space $\W_\sigma(S)$ really is the covariantized version of the $\W$-bracket on the wedge space $\W(S)$. 

Finally, we show that $\Dcal$ is indeed a \textit{derivative operator}, namely that it satisfies the Leibniz rule.

\begin{tcolorbox}[colback=beige, colframe=argile]
\textbf{Lemma [Leibniz rule]} \label{LemmaLeibniz}\\
$\Dcal$ is a covariant derivative and satisfies the Leibniz rule on the covariant wedge algebra bracket $[\cdot\,,\cdot]^\sigma$:
\begin{subequations}
\begin{align}
    & i)\qquad \Dcal\big(f T \big)=f(\Dcal T)+ T Df, \\
    & ii) \!\!\!\qquad \Dcal[T,T']^\sigma 
    =\big[\Dcal T,T'\big]^\sigma +\big[T,\Dcal T'\big]^\sigma, 
\end{align}
\end{subequations}
for $f\in\Ccel{0,0}$ a smooth function on $S$.\footnote{To evaluate the first identity at spin $s$ we use that $(T Df)_s= T_{s+1} Df$.}${}^,$\footnote{Strictly speaking, by $ii)$ we mean $\Ham_0\Dcal[T,T']^\sigma=\big[\Ham_0\Dcal T,T'\big]^\sigma+\big[T,\Ham_0\Dcal T'\big]^\sigma$.}
\end{tcolorbox}

\paragraph{Proof:}
$i)$ follows directly from the definition \eqref{CovDer} while $ii)$ follows from the Jacobi identity (that holds when $T,T'\in\Wcal_\sigma(S)$), which ensures that $\ad \hHam =\Ham_0\Dcal$ is a derivation.

\paragraph{Remark:}
For the sake of completeness and as a follow-up to the former remark, we give another demonstration of the property $ii)$ of the previous lemma in appendix \ref{AppLeibniz}.
In this version of the proof, we use the form \eqref{CFbracketCovariantSphere} of the $\sigma$-bracket.
The fact that the covariant wedge condition, now simply written with $\Dcal$ rather than $D$, is preserved by the bracket becomes trivial thanks to the Leibniz rule. 
Indeed,
\begin{equation}
    \big(\Dcal^{s+2}\big[T,T'\big]^\sigma \big)_{-2}=\sum_{k=0}^{s+2}\binom{s+2}{k}\big[\Dcal^k T,\Dcal^{s+2-k} T'\big]^\sigma_{-2}=0
\end{equation}
since the bracket at degree $-2$ is 0 by definition of \eqref{CFbracketCovariantSphere}.

\paragraph{Remark:}
Coming back to the Jacobi identity anomaly \eqref{JacCF} that motivated the whole analysis of the last two subsections, we notice that the most general solution of $\big[T,[T',T'']^\sigma\big]_s^\sigma\cyc 0$ is not $\hat\delta_T\sigma=0$ but 
\be \label{sigmanu}
\hat\delta_T \sigma =-\nu T_0,
\ee
where one has to take $\nu\in\Ccel{2,2}$ to get $\nu\T0\in\Ccel{1,2}$.
This is due to the fact that  the combination $\nu\T0\paren{T',T''}_s$ vanishes under cyclic permutation, namely
\begin{equation}
    \nu\T0\paren{T',T''}_s = \nu(s+3) \left( \T0\Tp0\Tpp{s+2}-\T0\Tpp0\Tp{s+2}\right)\cyc 0,
\end{equation}
for any $\nu$.
We denote by $\W_{\sigma,\nu}$ the set of elements of $\V(S)$ that satisfy \eqref{sigmanu}.\footnote{More precisely, $\W_{\sigma,\nu}:=\big\{T\in\V(S)~\big|~ (\Dcal T)_{-2}=0~\&~(\Dcal^{s+2}T)_{-2}= \nu(\Dcal^s T)_0,\,s\geqslant 0\big\}$.}
Remarkably, we have that if $T,T'\in \W_{\sigma,\nu}$ then
$[T,T']^\sigma\in \W_\sigma.$ 
This follows from the fact that identities \eqref{com1} and \eqref{adjointsquare} are still satisfied for $T,T' \in \W_{\sigma,\nu}$.
Therefore  the proof given there is still valid and we conclude that $\big(\Dcal^n\big([T,T']^\sigma\big)\big)_{-2} =0$ if $T,T'\in \W_{\sigma,\nu}$, for all $n\geq 0$.
This shows that $\Wcal_\sigma$ is the maximal set which forms a Lie algebra in the sense that $\Wcal_\sigma=[\Wcal_{\sigma,\nu},\Wcal_{\sigma,\nu}]^\sigma$ is the commutator sub-algebra of $\Wcal_{\sigma,\nu}$.

\subsection{Covariant wedge solution \label{sec:CovWedgeSol}}

In this section we show that the two notions of wedge algebra that we have constructed are in fact equivalent: they are related by a Lie algebra isomorphism. 
In practice this requires constructing an element of the covariant wedge  $ \Wcal_\sigma$  given an element of $\Wcal$ and vice-versa. 
This construct relies on the  Goldstone mode $G\in \V_{0}$ introduced in \eqref{Goldstone}.
We also make use of the filtration of $\W_\sigma$
\begin{equation}
    \{0\} \subset\W_\sigma^{-1}\subset \W_{\sigma}^0\subset\W_{\sigma}^1\subset \ldots\subset\W_{\sigma }^s\subset\ldots\subset \W_{\sigma} \quad\textrm{such that}\quad \W_{\sigma}=\bigcup_{n+1\in\mathds{N}}\W_{\sigma }^n,
\end{equation}
where $\W_{\sigma}^s = \W_{\sigma} \cap \V^s$ are the subspaces  of $\W_{\sigma}$ for which $\T{n} = 0$ when $n>s$.\footnote{Note that $\{0\}$ can equivalently be viewed as the trivial space $\W_\sigma^{-2}:= \W_\sigma\cap\V_{-2}$.}
Importantly, the $\sigma$-bracket is of degree $-1$ for this filtration
\be 
[\W_\sigma^s, \W_{\sigma}^{s'}]^\sigma=\W_\sigma^{s+s'-1}.
\ee

Let us start with the definition of the map $\mTG: \hV(S) \to \hV(S) $---where $\hV(S)=\V_{-2}\oplus\V(S)$---given by
\be 
  \boxed{\,\mTG(T):= \sum_{n=0}^{\infty} \frac{G^n}{n!} T^G\big(\ad^n \hHam(T)\big),\,} \label{mTGdef} 
\ee
where $T^G$ is the following filtration preserving map,\footnote{We use the notation $\mTG_s(T)\equiv\big(\mathcal{T}^{G}(T)\big)_s$ and similarly for $T^G$.} 
\be 
T_{s}^G(T):=
\sum_{k=0}^{\infty} 
\frac{(s+k+1)!}{k!(s+1)!} (-DG)^k T_{s+k}, \qquad s\geq -1 \label{TGdef}
\ee 
and is such that $T_{-2}^G(T):=\T{-2}$.\footnote{This can be seen as an extension of the last formula since 
\be
T_{s}^G(T):= T_s + (s+2)
\sum_{k=1}^{\infty} 
\frac{(s+k+1)!}{k!(s+2)!} (-DG)^k T_{s+k},
\ee 
from which we get that $T_{-2}^G(T):= T_{-2}$.} 
Note that if $T\in \hV^p =\bigoplus_{n=-2}^p\V_n$, then both series truncate to finite sums where $n+k\leq p-s$ since  $(\Dcal^nT)_{s+k}=0$ when $s+k+n > p$. In this case we have   $\mTG_p(T) = T_p$ and  the next two terms read
\bs
\begin{align}
\mTG_{p-1}(T) &=
T_{p-1} + G (\Dcal T)_{p-1} - (p+1) DG T_{p} , \\
\mTG_{p-2}(T) &=
T_{p-2} + G (\Dcal T)_{p-2} - p  DG T_{p-1} \cr
& + \frac{G^2}{2} (\Dcal^2 T)_{p-2} - p\, G DG (\Dcal T)_{p-1} + \frac{p(p+1)}{2}
(DG)^2 T_{p}.
\end{align}
\es

This map is linear on $\hV(S)$ and possesses three remarkable properties.

\begin{tcolorbox}[colback=beige, colframe=argile]
\textbf{Lemma [Intertwining properties]} \label{LemmaTG}\\
The map $\mTG$ preserves the filtration, i.e. $\mTG: \hV^s \to \hV^s$; it intertwines the regular and covariant derivative operators
\vspace{-0.1cm}
\begin{subequations}
\begin{align}\label{Dintertwin}
D\mTG(T)& = \mTG\big([\hHam,T]^\sigma\big); 
\end{align}
and  is equal when restricted to $\W_\sigma$ to  the path ordered exponential of the  adjoint action of the element $\hat{G}= (0, G, 0, 0, \cdots) \in \V^0$ along the path $G_t :=e^t G$. 
In other words we have   
\vspace{-0.1cm}
\begin{align}\label{adjointaction}
    \mTG(T) = \overrightarrow{\mathrm{Pexp}}\left(\int_{-\infty}^{0}   \adtG \, \rd t \right)(T),
\end{align}
where  $\sigma_t:=e^t \sigma$ and for $T\in \W_\sigma$.
\end{subequations}
\end{tcolorbox}

\ni The identity \eqref{Dintertwin}  means that $D\mTG_{s+1}(T) = \mTG_{s}\big([\hHam,T]^\sigma\big)$ for $s\geq -2$.
Applying this intertwining equality recursively, we get that 
\be 
D^{s+2} \mTG_s(T) = \mTG_{-2}( \Dcal^{s+2}T) =\sum_{n=0}^{\infty} \frac{G^n}{n!} \big(\Dcal^{s+2+n}T\big)_{-2}\overset{\W_\sigma}{=}0.
\ee
This shows that $\mTG(T) \in \Wcal$ when $T\in \Wcal_\sigma$.

To understand the nature of \eqref{adjointaction}, we recall that the path ordered exponential 
$
W_\alpha:= \overrightarrow{\mathrm{Pexp}}\left(\int_{-\infty}^{\alpha}   A_t \rd t \right)
$ is, by definition, the unique solution to the differential equation $\pa_\alpha W_\alpha = W_\alpha A_\alpha $ with initial condition $W_{-\infty}= 1$. 
From \eqref{adjointaction} we have that
\be \label{PexpT}
\mTaG =  \overrightarrow{\mathrm{Pexp}}\left(\int_{-\infty}^{\alpha}   \adtG \,  \rd t \right),\qquad \Rightarrow \qquad   \pa_{\alpha} \mTaG = \mTaG \circ \adaG.
\ee 
The expression \eqref{adjointaction} also shows that $\mTG$ is invertible with inverse  the opposite path ordered exponential $(\mTG)^{-1} =  \overleftarrow{\mathrm{Pexp}}\left(-\int_{-\infty}^{0}   \adtG \,  \rd t \right)$.
This inverse can be written explicitly as
\be \label{mTinv}
\boxed{\,(\mTG)^{-1}(T)
= 
\sum_{n=0}^{\infty}\frac{(-G)^n}{n!} T^{-G}(D^n T),\,}
\ee
where $T^{-G}$ is the map \eqref{TGdef} with $G\to-G$.
We check in the appendix \ref{app:IInt} that this expression satisfies the intertwining property
$\Dcal (\mTG)^{-1}(T)= (\mTG)^{-1}(DT)$.

The identity \eqref{adjointaction} for $\mTG$ also means that we have the following theorem.

\begin{tcolorbox}[colback=beige, colframe=argile]
\textbf{Theorem [Wedge isomorphism]}\\
$\mTG: \Wcal_\sigma \to \Wcal$,  provides a Lie algebra  isomorphism:
\vspace{-0.1cm}
\begin{align}\label{morphismP}
    \mTG \big([T,T']^\sigma\big)= \left[\mTG(T),\mTG(T')\right]^\W. 
\end{align}
\end{tcolorbox}

\ni 
The morphism property naturally follows from  \eqref{adjointaction}. It is surprising at first since we know that 
$e^{\ad T}$ is an \emph{internal} automorphism of $\Wcal_\sigma$ when $T\in \Wcal_\sigma$.\footnote{In other words $ 
e^{\ad T}\big([T',T'']^\sigma\big) = \big[e^{\ad T}(T'),e^{\ad T}(T'')\big]^{\sigma}$ when $T\in \Wcal_\sigma$.} 
The key point here is that $\hat{G}\notin \W_\sigma $ since it satisfies $\hat{\delta}_{\hat{G}} \sigma = -D^2G =-\sigma$. 
This means that $\ad \hat{G}$ is not a morphism of the $\sigma$-bracket. 
Instead, using the violation of Jacobi identity \eqref{JacCF} on $(G,T,T')$, where $T,T'\in \W_\sigma$, we have that 
\be \label{adG1}
\ad \hat{G} \big([T,T']^\sigma\big) -  \sigma\paren{T,T'}
= \big[\ad \hat{G}(T), T'\big]^\sigma + \big[T,\ad \hat{G}(T')\big]^\sigma. 
\ee 
This equality can be promoted to a differential identity. To do so we  define  
\be T_\alpha := \overleftarrow{\mathrm{Pexp}}\left(-\int_{-\infty}^{\alpha}   \adtG \,  \rd t \right)(T),
\ee 
where $T\in \W$ and similarly for  $T'_\alpha$. By construction 
$T_\alpha, T'_\alpha \in \W_{\sigma_\alpha}$. They are solution of the differential equation 
\be \label{Tader}
\pa_\alpha T_\alpha = - \adaG(T_\alpha),
\ee 
and similarly for $T'_{\alpha}$.
We are now ready to evaluate the following derivative
\begin{align}
& \,\pa_\alpha  
\left( 
\mTaG \left(\left[ T_\alpha, T'_{\alpha}\right]^{\sigma_\alpha}  \right)
\right) \cr
= &\,    
\mTaG \left(\adaG \left(\left[ T_\alpha, T'_{\alpha}\right]^{\sigma_\alpha}  \right)
+ \pa_\alpha \left[ T_\alpha, T'_{\alpha}\right]^{\sigma_\alpha}  
\right) \cr
= &\, 
\mTaG \left(\adaG \left(\left[ T_\alpha, T'_{\alpha}\right]^{\sigma_\alpha}  \right)
- \sigma_\alpha \paren{T_\alpha,T'_{\alpha}} + 
\left[ \pa_\alpha T_\alpha, T'_{\alpha}\right]^{\sigma_\alpha}
+ \left[  T_\alpha, \pa_\alpha T'_{\alpha}\right]^{\sigma_\alpha}
\right) \\
= &\, 
\mTaG \left(\adaG \left(\left[ T_\alpha, T'_{\alpha}\right]^{\sigma_\alpha}  \right)
- \sigma_\alpha \paren{T_\alpha,T'_{\alpha}} - 
\left[ \adaG(T_\alpha), T'_{\alpha}\right]^{\sigma_\alpha}
- \left[  T_\alpha, \adaG(T'_{\alpha})\right]^{\sigma_\alpha}
\right) \cr
=&\, 0. \nn
\end{align}
In the first equality we use \eqref{PexpT}, in the second equality  we use that $\pa_\sigma [T,T']^\sigma = - \paren{T,T'}$, in the third we use the evolution equation  \eqref{Tader} and in the last we leverage \eqref{adG1}.
This shows that $\mTaG \left(\left[ T_\alpha, T'_{\alpha}\right]^{\sigma_\alpha}  \right)$ is independent of $\alpha$. We can evaluate it at $\alpha=0$ and $\alpha=-\infty$. Using that $[\cdot\,,\cdot]^{\sigma=0}\equiv [\cdot\,,\cdot]^{\W}$ and $T_{-\infty}:=T \in  \W$ together with $\mTG(T_{\alpha=0})=T$, we get the morphism property \eqref{morphismP} of $\mTG$.

\paragraph{Proof of $\mTG$ being the path ordered exponential of the adjoint action \eqref{adjointaction}:}
Let us first evaluate the adjoint action of 
$\hat{G}=(0, G, 0,0,\cdots)$ on an arbitrary element $T\in \V(S)$. It is given by  
\vspace{-0.2cm}
\begin{align}
\big(\ad \hat{G} (T)\big)_s=[\hat{G}, T]^\sigma_s = G (\Dcal T)_s - (s+2) DG T_{s+1} \label{ad1G}.
\end{align}
Then we use again the auxiliary parameter $\alpha$ and evaluate in appendix \ref{app:Adj}  the derivative of $\mathcal{T}^{ G_\alpha}$. We use the fact that the $\alpha$  dependence of $\mathcal{T}^{G_\alpha }$  appears directly from the polynomial  dependence
on $G$ and $DG$ but also through the $\sigma$ dependence of $\Dcal\equiv \Dcal_{\sigma} $ since $\sigma= D^2G$.
In the following we denote by $T_\alpha \in \W_{\sigma_\alpha}$ a family of elements which depend smoothly on $\alpha$. We do not assume that $T_\alpha$ satisfies the identity \eqref{Tader}.
We find that
\begin{align}
    \pa_\alpha \big(\mathcal{T}^{ G_\alpha}_s (T_\alpha)\big)
&= \sum_{n,k =0}^{\infty} 
\frac{(s+k+1)!}{n! (s+1)! k!} G_\alpha^{n}(-DG_\alpha)^k
\left(\big(\pa_\alpha + \adaG \big)\big(\Dcal_{\sigma_\alpha}^n T_\alpha \big)  
\right)_{s+k}, \label{palT}
\end{align}
where we used \eqref{ad1G}. 
Next we recast \eqref{adG1} as a differential identity:
\be \label{adG2}
\big(\pa_\alpha+ \adaG \big) \big([T_\alpha,T'_\alpha]^{\sigma_\alpha} \big) 
= \big[(\pa_\alpha+ \adaG)(T_\alpha), T'_\alpha \big]^{\sigma_\alpha} + \big[T_\alpha,(\pa_\alpha+ \adaG)(T'_\alpha)\big]^{\sigma_\alpha}. 
\ee 
We can rewrite this equality 
in terms of the commutator of the adjoint action as 
\be \label{adG3}
\left[\big(\pa_\alpha + \adaG\big), \ada T_\alpha \right] 
= \ada\big(\pa_\alpha(T_\alpha)+ [ \hat{G}_\alpha,T_\alpha]^{\sigma_\alpha} \big),
\ee 
which is valid when $T_\alpha\in\W_{\sigma_\alpha}$. On the LHS the commutator is simply the commutator of linear operators acting on $\V(S)$.

We can apply this identity for $T_\alpha=\hHama$ and need to remember that  since $\hat{G}_\alpha,\hHama\in \V^0$, their commutator is in $\V_{-1}$ and is therefore a central element. 
Similarly $\pa_\alpha\hHama \in \V_{-1}$ and is central. This implies that
$\big[(\pa_\alpha+ \adaG), \ada \hHama \big] =0$. 
Therefore,
\begin{align}
\big(\pa_\alpha + \adaG \big)\big(\Dcal_{\sigma_\alpha}^n T_\alpha\big) 
& = \big(\pa_\alpha+ \adaG \big)
\big( \ada^n \hHama ( T_\alpha )\big)=  
\ada^n \hHama \big(
(\pa_\alpha + \adaG)(T_\alpha)\big) \cr
& =\Dcal_{ \sigma_\alpha}^n \big(\pa_\alpha  T_\alpha + [\hat{G}_\alpha, T_\alpha]^{\sigma_\alpha}\big). \nn
\end{align} 
From this and \eqref{palT} we conclude that 
$\pa_\alpha \big(\mTaG (T_\alpha)\big) = 
 \mTaG \big( \pa_\alpha T_\alpha +[\hat{G}_\alpha,T_\alpha]^{\sigma_\alpha}\big)$, hence 
\be \label{paltad}
 \pa_\alpha \mTaG  = 
 \mTaG  \circ\adaG.
\ee 
This equation supplemented by the initial condition $\mathcal{T}^0=\mathrm{Id}$ implies that
\be \mTaG=\overrightarrow{\mathrm{Pexp}}\left(\int_{-\infty}^{\alpha}   \adtG \,  \rd t \right).
\ee

\paragraph{Proof of the intertwining property \eqref{Dintertwin}:}
We first show in appendix \ref{app:Int} that the map $T^G$ satisfies the key relationship\footnote{The degree $s-1$ component of this equation reads
$DT_s^G(T) + DG T_{s}^G\big([\hHam, T]^\sigma\big)= T_{s-1}^G\big([\hHam, T]^\sigma\big)$ since $(DG T^G)_{s-1}= DG T_{s}^G$ and $(D T)_{s-1}= D T_s$.}
\be \label{DTrecursion}
DT^G(T) + DG T^G\big([\hHam,T]^\sigma\big)= T^G\big([\hHam, T]^\sigma\big).
\ee 
This in turn implies that 
\begin{align}
 D \mTG(T) 
 &= \sum_{n=0}^{\infty} \frac{G^n}{n!}\Big( D T^G\big(\ad^n \hHam(T)\big)+ DG T^G\big(\ad^{n+1} \hHam(T)\big)\Big) \cr
 &=\sum_{n=0}^{\infty} \frac{G^n}{n!} T^G\big(\ad^{n+1}\hHam(T)\big) = \mTG\big([\hHam, T]^\sigma\big).
\end{align}

We come back in sections \ref{sec:GoodCut} and \ref{sec:RelationDressingMap} to this dressing map and its interpretation in the context of Newman's good cut equation \cite{newman_heaven_1976, Adamo:2010ey}.
Since this chapter emphasizes the mathematical aspects of the thesis, we postpone the physical picture attached to these various theorems to the next chapter.

\section{Time and Algebroid}

In this section we construct an algebroid bracket on $\scri$ which generalizes the covariant wedge bracket from section \ref{sec:NewFieldDepBracket}.

\subsection{The Wedge bracket \label{sec:Wedge}}

So far we constructed a Lie algebra $\Wcal_\sigma(S)$ which generalizes the wedge algebra. 
It is labeled by an element $\sigma \in \Ccel{1,2}$ and this construction relies on the introduction of a bracket $[\cdot\,,\cdot]^\sigma$ acting on $\V(S)$ and defined by\footnote{This is a deformation of the $\W$-bracket, defined as $[\cdot\,,\cdot]^\W\equiv\big([\cdot\,,\cdot]^\sigma\big)\big|_{\sigma=0}$. \label{footWbracket}} 
\begin{equation}
[T,T']^\sigma_s:=\sum_{n=0}^{s+1}(n+1)\big(\T{n}D\Tp{s+1-n}-\Tp{n}D\T{s+1-n}\big)-(s+3)\sigma \big(\T0\Tp{s+2}-\Tp0\T{s+2}\big). \label{Wbracket2}
\end{equation}
We have shown in that this bracket satisfies the Jacobi identity when $T$ belongs to a subspace of $\V(S)$ called the covariant wedge space and denoted $\W_\sigma(S)$.
To describe the latter, it is necessary to introduce the key element $\hHam\in \V^0(S)$ given by $\hHam=(DG, 1, 0,0,\cdots)$, where $G \in \V_0$ is the Goldstone field connected to $\sigma$ by 
\be 
D^2 G =\sigma.
\ee 
In other words, $\hHam $ is such that $({\hHam})_{-1}=DG$, 
$({\hHam})_0=1$ and $({\hHam})_s=0$ for $s>0$.
As we are about to see, it is convenient to think of $\hHam$ as a Hamiltonian generator and define its adjoint action $\ad \hHam: \V(S) \to \V(S)$ to be the linear operator $\ad \hHam(T)=[\hHam, T]^\sigma$ with grade $s$ element given by 
\be
\big(\ad \hHam(T)\big)_s=[\hHam, T]_s^\sigma = D T_{s+1} -\sigma (s+3) T_{s+2}\equiv(\Dcal T)_s.
\ee 
Furthermore, we define $(\Dcal T)_{-2}:=D\T{-1}-\sigma\T0$. 
We also showed that the $\sigma$-bracket has a convenient expression in terms of the covariant derivative $\Dcal$, namely
\begin{equation}
[T,T']^\sigma_s:=\sum_{n=0}^{s+1}(n+1)\big(\T{n}(\Dcal\Tp{})_{s-n}-\Tp{n}(\Dcal T)_{s-n}\big). \label{WbracketDcal}
\end{equation}
The covariant wedge is then defined as  
\be 
\W_\sigma(S) = \Big\{ T\in \V(S) ~\big|~  D\big(\ad^n\hHam (T)\big)_{-1} =  \sigma \big(\ad^n\hHam (T)\big)_{0},\, n \in \mathbb{N}\Big\}.
\ee
Note that the covariant wedge condition is also compactly written as $(\Dcal^{s+2}T)_{-2}=0,\,s\geq -1$.
One of the main result so far (i.e. of \cite{Cresto:2024fhd}) is that 
$\Wcal_\sigma(S)=\big(\W_\sigma(S),[\cdot\,,\cdot]^\sigma \big)$ is a Lie algebra. 
The Jacobi identity for the $\sigma$-bracket follows from the fact that $\hdT \sigma =0$, where 
\begin{align}
-\hdT \sigma =(\Dcal^2 T)_{-2}= D [ \hHam, T ]^\sigma_{-1} -\sigma [\hHam, T]_0^\sigma 
= D^2 T_0 -2 D(\sigma T_1) - \sigma DT_1 + 3\sigma^2 T_2. 
\end{align}

When $\sigma=0$ we recover the usual wedge algebra with bracket $[T,T']^0\equiv[T,T']^\W$ (the shifted \hyperref[SNreminder]{SN bracket} \eqref{Wbracket}) and the familiar wedge condition \cite{Strominger:2021mtt, Mason:2022hly} implying that $(D^{s+2}T)_{-2}= D^{s+2} T_{s}=0$ for $s\geq -1$.

Moreover, it was also shown that $\Wcal_\sigma$ is isomorphic as a Lie algebra to $\Wcal_0\equiv\Wcal$ with the isomorphism given as the path ordered exponential of an adjoint action,
\begin{align}
\mTG: \Wcal_\sigma &\to \Wcal_0 \cr
  T &\mapsto \overrightarrow{\mathrm{Pexp}}\left(\int_0^{1}\mathrm{ad}_{\lambda \sigma} \hat{G}_\lambda\,\rd \lambda \right)(T).
\end{align} 
where $\hat{G}_\lambda=\lambda\hat{G}$ and $\hat{G}\in \V^0$ is such that $ \hat{G}_n= 0$ for $n\neq 0$ and $\hat{G}_0 = G$.

\subsection{From celestial to sky  \label{sec:celestialToScri}}

The  Lie algebra $\Wcal_\sigma(S)=\big(\W_\sigma(S),[\cdot\,,\cdot]^\sigma \big)$ is a celestial symmetry algebra supported on the 2d surface $S$. It depends on the value of the shear $\sigma$ on $S$.  
In order to understand how this symmetry acts on the gravitational phase space one needs to include time into the picture and extend this symmetry from $S$ to $\scri$. 
One also needs to go beyond the wedge in order to allow for the shear to be time dependent. 
Quite remarkably the two issues are connected and there exists a choice of time dependence for the transformation parameters that allows to go beyond the wedge. This is what we now develop.

The first step consists in  promoting the transformation variables $T_s(z,\bz) \in \V(S)$ to time dependent variables $ \tau_s(u, z,\bz) \in \tV(\scri)$ where $\tV(\scri)$ is  introduced in \eqref{VCcardef}. This change of support is reflected in the change of notation $T\to \tau$.
As we have seen, $\tV(\scri)= \bigoplus_{s=-1}^\infty \tV_s$ is graded, and we denote by $\t{}$ the elements of $\tV(\scri)$ and by $\t{s}$ its grade $s$ elements.

Now that we introduced a time dependency in the transformation parameters, we need to allow $\sigma$ to also depend on time. 
For clarity, we shall henceforth use $C(u,z,\bz)$ to denote the time dependent version of the parameter $\sigma(z,\bz)$. 
We did not choose the letters $\sigma$ and $C$ without reason.
In section \ref{SecGR:Noether}, $C$ will play the role of the asymptotic shear on the gravitational phase space. In particular this means that we take $C\in\Ccar{1,2}$.

Given these data, we  can naturally extend the $\sigma$-bracket to $\scri$:
given $\tau,\tau'\in\tV(\scri)$  we define the bracket $[\cdot\,,\cdot]^C$ to be the same as the bracket $[\cdot\,,\cdot]^\sigma$ after replacing $\sigma \to C$, i.e.
\begin{equation}
[\tau,\tau']^C_s:=\sum_{n=0}^{s+1}(n+1)\big(\t{n}D\tp{s+1-n}-\tp{n}D\t{s+1-n}\big)-(s+3)C\big(\t0\tp{s+2}-\tp0\t{s+2}\big). \label{CFbrackettau}
\end{equation}
For later convenience, we denote the two pieces as
\bs
\begin{align}
    [\tau,\tau']_s^{\tV}&:=\sum_{n=0}^{s+1}(n+1)\big(\t{n}D\tp{s+1-n}-\tp{n}D\t{s+1-n}\big) & & \hspace{-1.5cm}\in\tV_s, \label{tVBracket}\\
    \paren{\tau,\tau'}_s &:= (s+3)(\t0 \tp{s+2}-\tp0\t{s+2}) & & \hspace{-1.5cm}\in \Ccar{-2,-s-2}, \label{DaliBracket}
\end{align}
\es
so that $[\tau,\tau']^C=[\tau,\tau']^{\tV}-C\paren{\tau,\tau'}$.
We refer to $\paren{\cdot\,,\cdot}$ as the Dali-bracket, in homage to Dali's paintings with melting watches.
Note that the term proportional to $C$ introduces a field dependency on $C$ in the structure constants of the bracket. 
This field dependency encodes the algebra deformation in the presence of shear and reflects the  non-linearity of the Einstein's theory. 
It will prove convenient to recast this non-linear dependence into a covariant derivative operator. 
Since the context is clear, we keep the symbol $\Dcal$ for the $C$-bracket adjoint action of $\Ham$, where the element $\Ham:=(0,1,0,0,\cdots)$ is such that $\eta_n=\delta_{n,0}$. 
This means that\footnote{We use $\Dcal$ and $\adC\Ham$ interchangeably, with appropriate multiplication by $\Ham_0\in\tV_0$ in order for the Carrollian boost weight to be consistent; see the corresponding discussion in section \ref{sec:NewFieldDepBracket}.}
\be\label{DcalC}
\Dcal\tau:=[\Ham,\tau]^C, \qquad
(\Dcal\tau)_s = D\tau_{s+1} - C(s+3) \t{s+2}.
\ee 
Although $\tau_s$ is defined only for $s\geq -1$, it will be convenient to use the previous definition of $\Dcal$ for $s=-2$, where by definition we denote $(\Dcal\tau)_{-2} := D\t{-1} - C\t0$.
Notice that the  value of $\Ham_{-1}$ is irrelevant for the definition of $\Dcal$, since the degree $-1$ elements are central for the $C$-bracket.\footnote{ This implies that the adjoint action of $\Ham$ and $\hHam$ are the same. The condition $D(\hHam)_{-1}=\sigma=\sigma(\hHam)_0$ was needed for the wedge but is not necessary here so we chose $\Ham_{-1}=0$ for simplicity.}

Now notice that $\sum_{n=0}^{s+2}(n+1)(s+3-n)\t{n}\tp{s+2-n}$ is symmetric under the exchange of $\tau$ and $\tau'$ (using the change of variable $n\leftrightarrow (s+2-n)$).
Isolating the term $n=s+2$ in the sum  implies the important identity:
\begin{equation}
    (s+3)\big(\t0\tp{s+2}-\tp0\t{s+2}\big)=\sum_{n=0}^{s+1}(n+1)(s+3-n)\big(\t{n}\tp{s+2-n}-\tp{n}\t{s+2-n}\big).
\end{equation}
This allows us to recast the  term $C\paren{\tau,\tau'}_s$ as a deformation of the sphere derivative $D\t{s+1-n}\to D\t{s+1-n}-(s+3-n)C\t{s+2-n}=(\Dcal \tau)_{s-n}$.
The $C$-bracket \eqref{CFbrackettau} then also takes the form
\begin{equation}
    \boxed{\,[\tau,\tau']^C_s= \sum_{n=0}^{s+1}(n+1)\big(\t{n}(\Dcal\tp{})_{s-n}-\tp{n}(\Dcal\t{})_{s-n}\big).\,} \label{CFbrackettauDcal}
\end{equation}
This shows that the field dependency of the $C$-bracket can remarkably entirely be recast in the deformation $D\to \Dcal$ of the holomorphic derivative.
The analysis of the Jacobi identity for the $C$-bracket is similar to the one done in \ref{sec:NewFieldDepBracket} for the $\sigma$-bracket since the time dependence plays no role.
For completeness, we also report another proof in appendix \ref{AppJacobiDcal}, that uses the form \eqref{CFbrackettauDcal} of the bracket.
We find that
\be \label{JacCFC}
\big[\tau,[\tau',\tau'']^C\big]_s^C \cyc -\hdt C \paren{\tau',\tau''}_s,
\ee
where $\cyc$ denotes the equality after summing over cyclic permutation of $(\tau, \tau', \tau'')$.
The hatted variation is given by
\be \label{hataction}
\hdt  C = -D^2 \tau_0 +2 D(C \tau_1) + C D\tau_1 - 3C^2 \tau_2=-(\Dcal^2\tau)_{-2}.
\ee 
Note that \eqref{JacCFC} is not the most general way to write the Jacobi identity violation. 
We  have the freedom to add a term proportional to $\t0$ in the expression of $\hdt C$ since the following cyclic combination vanishes identically:
\begin{equation}
    \t0 \paren{\tau',\tau''}_s\cyc 0, \quad \forall \,s.
\end{equation}
This justifies the addition of a term $\alpha \t0 N$, $\alpha\in\R$, to the initial transformation $\hdt$. 
This additional term is however constrained to have definite Carrollian weights. In order for $\tau_0 N$ to have Carrollian weights $(1,2)$ we need $N$ to be in $\Ccar{2,2}$.
Next, remember that $\pa_u$ is an operator of weight $(1,0)$.
Therefore, the only local functional with the right weight is
\begin{equation}
    \boxed{N:=\pa_u C}.
\end{equation}
This allows to conclude that we have $\big[\tau,[\tau',\tau'']^C\big]_s^C \cyc -\dt C \paren{\tau',\tau''}_s$ with 
\be \label{dtauCalpha}
\dt C = \alpha \t0 \pa_u C + \hdt C.
\ee 
This possibility is a crucial difference between working solely on $S$ and working on $\scri$: the field variations $\hdT\sigma$ and $\dt C$ are \textit{not} the same. 
In the following we assume that $\alpha\neq 0 $ since the case $\alpha=0$ was thoroughly investigated in the previous sections. 
When $\alpha\neq 0$  we can always rescale the time coordinate $u \to \alpha u$ to ensure that $\alpha=1$.
This means that the expression for $\dt C$ is
\begin{equation}
    \boxed{\dt C = \t0  N -D^2\t0+2D(C\t1)+C D\t1-3 C^2\t2}\quad\in\,\Ccar{1,2}, \label{dtauC}
\end{equation}

\subsection{Time as a canonical transformation \label{Sec:DualEOM}}

So far we have left arbitrary the time dependence of $\tau_s$. 
As we will see in the next chapter, in order to have a well defined action of  symmetries generated by $\tau$ on the gravitational phase space, we need to assume that $\tau_s$ satisfies a differential equation in time.

The key idea comes from realising that the aforementioned element $\Ham=(0,1,0,0,\cdots)$ is the generator of a constant supertranslation.\footnote{See remark \ref{remarkH} for an alternative time translation generator that generalizes $\Ham$.}${}^,$\footnote{Notice in particular that the action of $\Ham$ on the phase space is consistent with this fact since $\delta_\Ham C=\pa_u C$.}
Therefore, we naturally give $\Ham$ the status of Hamiltonian in the space of $\tau$ and identify its adjoint action with the flow of time. 
We thus consider, for $s\geq 0$, the \textit{dual equations of motion}
\be 
\boxed{\,\, \pa_u\tau= \Ham_0^{-1}[\Ham, \tau]^C \quad \Leftrightarrow \quad  
\pa_u\t{s}= (\Dcal\tau)_s=D\t{s+1}-(s+3) C\t{s+2}.\,\,} \label{dualEOM}
\ee 
In the following we refer to these dual equations as $\E_s(\tau)=0$, where $\E(\tau) :=\pa_u \tau -\Dcal\tau$ and denote the corresponding space of parameters by $\sfT$.
If the context is clear, we often write $\E_s$ instead of $\E_s(\tau)$.

\begin{tcolorbox}[colback=beige, colframe=argile]
\textbf{Definition [$\sfT$-space]}
\be \label{TspaceGR}
\sfT :=\Big\{\,\tau\in\tV(\scri)~\big|~\pa_u\t{s}=D\t{s+1}-(s+3) C\t{s+2},~  s\geqslant 0~ \Big\}.
\ee 
\end{tcolorbox}

\ni These evolution equations imply that elements of $\sfT$ are uniquely determined by their celestial data $T_s(z,\bz)\equiv \tau_s(u=0,z,\bz)$ which play the role of initial conditions for the $\tau_s$.

\subsection{Consistent time evolution \label{sec:TimeEvol}}

In order to impose the dual evolution equations \eqref{dualEOM} we need to allow for a field dependency of the transformation parameters. 
Therefore one defines an algebroid bracket\footnote{See \ref{Sec:Liealg} for a reminder about Lie algebroids.} $\lbr\cdot\,,\cdot\rbr$ associated with $[\cdot\,,\cdot]^C$ and which acts on $\sfT$.

\begin{tcolorbox}[colback=beige, colframe=argile]
\textbf{Definition [$\sfT$-bracket]} \label{Tbra}\\
The $\sfT$-bracket $\lbr\cdot\,,\cdot\rbr :\sfT\times\sfT\to\sfT$, between $\tau,\tau'\in\sfT$ takes the form\footnote{If there is no possible confusion, then we write $\big(\dt\tp{}\big)_s\equiv\dt\tp{s}$ for shortness.}
\begin{equation}
\lbr\tau,\tau'\rbr_s :=  [\tau,\tau']^C_s+\big(\dtp\tau\big)_{s}-\big(\dt\tau'\big)_{s},\qquad s\geqslant -1. \label{Tbracket}
\end{equation}
\end{tcolorbox}

\ni The first remarkable fact is that the bracket is well defined on $\sfT$. 
It satisfies the following closure property.

\begin{tcolorbox}[colback=beige, colframe=argile] \label{LemmaClosure}
\textbf{Lemma [$\sfT$-bracket closure]}\\
The $\sfT$-bracket closes, i.e. satisfies the dual EOM \eqref{dualEOM}.
If $\tau$ and $\tau'$ are in $\sfT$ then $\E_s\big(\lbr\tau,\tau'\rbr\big)=0,\, s\geqslant 0.$ 
Hence $\lbr\tau,\tau'\rbr \in \sfT$.
\end{tcolorbox}

\paragraph{Proof:}
For any bracket $[\cdot\,,\cdot]$ and differential operator $\mathscr{D}$, we define the Leibniz rule anomaly as
\begin{equation}
    \cA\big([\cdot\,,\cdot],\mathscr{D} \big)=\mathscr{D}[\cdot\,,\cdot]-[\mathscr{D}\cdot\,,\cdot]-[\cdot\,,\mathscr{D}\cdot].
\end{equation}
In appendix \ref{Apptppsevol}, we compute that
\begin{align}
    \cA_s\big([\tau,\tau']^C,\Dcal \big)=-(s+3)\big(\t{s+2}\,\hdtp C-\tp{s+2}\,\hdt C\big). \label{AnomalyLeibniz1}
\end{align}
This anomaly disappears for the algebroid bracket and is replaced by 
\begin{equation}
    \cA_s\big(\lbr\tau,\tau'\rbr,\Dcal \big)=-N\paren{\tau,\tau'}_s+ \delta_{\Dcal\tau}\tp{s}-\delta_{\Dcal\tau'}\t{s}. \label{AnomalyLeibniz2}
\end{equation}
Moreover, $\pa_u$ is also anomalous:
\begin{equation}
    \cA_s\big(\lbr\tau,\tau'\rbr,\pa_u \big)=-N\paren{\tau,\tau'}_s+ \delta_{\pa_u\tau}\tp{s}-\delta_{\pa_u\tau'}\t{s}. \label{AnomalyLeibniz3}
\end{equation}
Hence, taking the difference gives 
\begin{equation}
    \cA_s\big(\lbr\tau,\tau'\rbr,\pa_u-\Dcal \big)= \delta_{(\pa_u\tau-\Dcal\tau)}\tp{s} -\delta_{(\pa_u\tau'-\Dcal\tau')}\t{s},
\end{equation}
which amounts to
\begin{equation} \label{AnomalyLeibniz}
    (\pa_u-\Dcal)\lbr\tau,\tau'\rbr= \big\lbr(\pa_u-\Dcal)\tau,\tau'\big\rbr +\big\lbr\tau,(\pa_u-\Dcal)\tau'\big\rbr+\delta_{(\pa_u-\Dcal)\tau}\tau'-\delta_{(\pa_u-\Dcal)\tau'}\tau.
\end{equation}
The RHS vanishes on shell of the dual EOM \eqref{dualEOM}, i.e. if $\tau,\tau'\in\sfT$, which proves that the bracket $\lbr\cdot\,,\cdot\rbr$ closes on $\sfT$.

\subsection{A symmetry algebroid \label{secAlgebroid}}

The next step is to show that  $\dt$  is an algebroid action for the $\sfT$-bracket. 
$\dt$ is a vector field in the gravitational phase space denoted $\PS$, namely the space of functionals of $C$ and $\bC$, cf. \eqref{actiondelta}. 
Holomorphic functionals on $\PS$ are simply functionals of $C \in \Ccar{1,2}$.
When acting on 0-forms on fields space, $\dt$ is nothing else than the Lie derivative $L_{\dt}$ in $\PS$ along the vector field generated by the symmetry parameter $\tau$.

\begin{tcolorbox}[colback=beige, colframe=argile]
\textbf{Lemma [Algebroid action and dual EOM]} \label{LemmaDualEOM}\\
The action $\dt$ admits a representation of the bracket $\lbr\cdot\,,\cdot\rbr$ onto $\PS$ when $\tau\in\sfT$:
\begin{equation}
    \tau,\tau'\in \sfT\quad \Rightarrow \quad 
    \big[\dt,\dtp\big]C =-\delta_{\lbr \tau,\tau'\rbr}C. \label{RepdeltaPS}
\end{equation}
\end{tcolorbox}

\ni While we already gave a physical motivation for the form of the time evolution of $\tau$, the validity of this morphism property for the $\dt$ action is what formally justifies the introduction of the dual equations of motion \eqref{dualEOM}. 
Indeed, the proof shows that the violation of the morphism property 
is parametrized by   $\E_s$ (see \eqref{algebroidRepinterm}).

The next step requires to show that  if $\delta$ is an algebroid action,  then the $\sfT$-bracket $\lbr\cdot\,,\cdot\rbr$ is indeed a Lie (algebroid) bracket, so that the Jacobi identity holds.

\begin{tcolorbox}[colback=beige, colframe=argile]
\textbf{Lemma [Jacobi identity]} \label{PropJacId}\\
The Lie algebroid bracket $\lbr\cdot\,,\cdot\rbr$ satisfies Jacobi if and only if  $\dt$ is an algebroid action:
\begin{equation}
    \big[\dt,\dtp\big]+\delta_{\lbr \tau,\tau'\rbr}=0\qquad\Leftrightarrow \qquad \big\lbr \tau,\lbr \tau',\tau''\rbr\big\rbr_s\cyc 0.
\end{equation}
\end{tcolorbox}

Since the penultimate lemma means that $\tau \to \delta_\tau$ is a  Lie algebra anti-homomorphism when $\tau \in \A$, then piecing everything together, the following theorem represents one of the main results of this thesis. 

\begin{tcolorbox}[colback=beige, colframe=argile] \label{theoremAlgebroid}
\textbf{Theorem [$\A$-algebroid]}\\
The space $\A\equiv\big(\sfT,\lbr\cdot\,,\cdot\rbr,\delta\big)$ equipped with the $\sfT$-bracket \eqref{Tbracket} and the anchor map $\delta$,
\begin{align}
    \delta : \,&\A\to \X(\PS) \nn\\
     &\tau\mapsto\dt \label{defdelta}
\end{align}
is a Lie algebroid over $\PS$.
\end{tcolorbox}

\paragraph{Remark:} \label{remark1}
The  $\sfT$-bracket construction is similar to the notion of symmetry algebroid explained in \ref{ReminderAlgebroid}, where $\delta$ plays the role of the anchor map $\rho$, $\dt$ plays the role of  the vector field $\nkor(\alpha)$ and the $C$-bracket plays the role of the Lie bracket $[\cdot\,,\cdot]$ (or equivalently $[\cdot\,,\cdot]_\g$) while $\PS$ corresponds to the base manifold $\M$.
The difference in our case is that the $C$-bracket is field dependent and is \textit{not} a Lie bracket. 
What is remarkable is that the violation of its Jacobi identity is cancelled by the $\sfT$-bracket field dependency.
This result is therefore a \textit{generalization} of the symmetry algebroid construction (see also the next \hyperref[remarkBB]{remark}).

\paragraph{Proof of lemma [Jacobi identity]:}
We rewrite the property \eqref{JacCFC} for convenience,
\begin{equation}
    \big[\tau,[\tau',\tau'']^C\big]^C_s\cyc
    -\dt C \paren{\tau',\tau''}_s.
\end{equation}
We then establish that we have a Leibniz anomaly involving the $C$-bracket and the variation $\dt$,
\begin{equation} \label{vanomaly}
 \cA_s\big([\tau',\tau'']^C,\dt \big):= \dt \big[\tau',\tau''\big]^C_s-\big[\dt \tau',\tau''\big]^C_s-\big[\tau',\dt \tau''\big]^C_s = -\dt C\paren{\tau',\tau''}_s.
\end{equation}
We then turn to the quantity of interest: the expansion of the double commutator of the $\sfT$-bracket reads 
\begin{align}
    \big\lbr \tau,\lbr \tau',\tau''\rbr\big\rbr & = \big[\tau,[\tau',\tau'']^C\big]^C+\big[\tau,\dtpp \tau'-\dtp \tau''\big]^C +\delta_{\lbr \tau',\tau''\rbr}\tau-\dt\lbr \tau',\tau''\rbr\cr
    &\cyc \big[\tau,[\tau',\tau'']^C\big]^C 
    - \cA\big([\tau',\tau'']^C,\dt \big)
    +\delta_{\lbr \tau',\tau''\rbr}\t{} + [\dtp, \dtpp] \tau. 
\end{align}
In the second equality we have used the cyclic permutation to reconstruct the variational Leibniz anomaly \eqref{vanomaly}.
In the context we are in, the first two terms cancel each other and we are simply left with 
\vspace{-0.1cm}
\begin{equation}
    \big\lbr \tau,\lbr \tau',\tau''\rbr\big\rbr_s \cyc \delta_{\lbr \tau',\tau''\rbr}\t{s} +[\dtp,\dtpp]\t{s}, \qquad s\geqslant -1.
\end{equation}
Therefore the RHS vanishes iff $\big[\dtpp,\dtp\big]=\delta_{\lbr \tau',\tau''\rbr}$, on all functionals of $C$. 
This concludes the proof of this lemma.

\paragraph{Remark: \label{remarkBB}}
The fact that $\lbr\cdot\,,\cdot\rbr$ is a Lie algebroid bracket was not guaranteed since the $C$-bracket  fails to be a Lie bracket.
The validity of the former lemma follows from  the fact that if $[\cdot\,,\cdot]$ is a bracket with Jacobi anomaly  $\big[\cdot\,,[\cdot\,,\cdot]\big]\cyc\mathcal{B}$ 
and with Leibniz anomaly $\cA\big([\cdot\,,\cdot] ,\nkor(\cdot)\big) =\mathcal{B'}$ one can still build a Lie algebroid bracket $\lbr\cdot\,,\cdot\rbr$ via the \hyperref[ReminderAlgebroidbis]{`symmetry algebroid' procedure}, provided that both anomalies are equal, i.e. $\mathcal{B}=\mathcal{B'}$.
We are using  the notation of the reminder of section \ref{Sec:Liealg}.

\paragraph{Remark:}
As mentioned in section \ref{Sec:DualEOM}, we can introduce a flow of time by noticing that $\Ham$ is a constant supertranslation.
This supposes that one has noticed the relationship between the $\W$-bracket and the $\gbms$ or Poincaré algebras.
One thing we find remarkable is the extent to which one can be agnostic about any physical input.
Indeed, we can choose the flow generated by $\Ham$ to \textit{define} the flow of time $\pa_u$, i.e imposing that $\delta_\Ham C=:\pa_u C$.\footnote{This in turn tells us that $N$ has to be $\pa_u C$ (since what we call $N$ is after all just $\delta_\Ham C$).}
Quite remarkably this means that the dual time evolution equation can then be simply written as a constraint using the algebroid bracket.
Indeed, from what we just said around equation \eqref{dtauCalpha}, the $C$-bracket is only a Lie bracket in the associated covariant wedge where $\dt C=0$.
The introduction of time therefore comes hand to hand with the introduction of the algebroid bracket $\lbr\cdot\,,\cdot\rbr$, cf.\,\eqref{Tbracket}, for which $\dt C\neq 0$.
The adjoint action of $\Ham$ is then given by 
\begin{equation}
    \lbr\Ham,\tau\rbr=\Dcal\tau-\delta_\Ham\tau=:\Dcal\tau-\pa_u\tau.
\end{equation}
Time evolution (what we shall sometimes refer to as ``going on-shell of the dual EOM'') thus amounts to imposing the \textit{`dual Hamiltonian' constraint}
\begin{equation} \label{HamiltonianConstraint}
    \boxed{\lbr\Ham,\tau\rbr\overset{!}{=}0}.
\end{equation}
In section \ref{secGR:MasterCharge}, we will even show that \eqref{HamiltonianConstraint} is equivalent by duality to a truncation of Einstein's equations, so that \textit{the sole knowledge of the $C$-bracket} (from which the construction of the $\A$-algebroid is unique) \textit{determines part of the GR dynamics!}
Furthermore, if Jacobi holds, then we immediately infer that
\begin{equation}
    \big\lbr\Ham,\lbr\t{},\tp{}\rbr\big\rbr= \big\lbr\lbr\Ham,\t{}\rbr,\tp{}\big\rbr+ \big\lbr\t{},\lbr\Ham,\tp{}\rbr\big\rbr,
\end{equation}
which is another proof that the \hyperref[LemmaClosure]{$\sfT$-bracket closes on-shell}.\\

We now give the proof of the \hyperref[LemmaDualEOM]{algebroid action lemma}. This requires to investigate under which conditions $\dt$ is an algebroid action.

\begin{tcolorbox}[colback=beige, colframe=argile]
\textbf{Lemma [Pre-algebroid action]}\\
Let us assume that $\tau,\tau' \in \tV(\scri)$, hence that no evolution equations are imposed, then 
\begin{equation}
    \big[\dt,\dtp\big]C+\delta_{\lbr \tau,\tau'\rbr}C =\tp0\delta_{(\dot\tau-\Dcal\tau)}C-
    \t0\delta_{(\dot\tau'-\Dcal\tau')}C
    \label{algebroidRepinterm}.
\end{equation}
The RHS vanishes if the $\tau$'s satisfy the recursion relations 
\be \label{trecursion}
\dot\tau_s =(\Dcal\tau)_s,\qquad s\geq 0.
\ee
In particular, no condition is required on $\tau_{-1}$.
\end{tcolorbox}

\paragraph{Proof:}
Notice that $\big[\dt,\dtp\big]C$ splits into 3 contributions, namely
\bs
\label{ddC0}
\begin{align}
    \big[\dt,\dtp\big]C &=
    \dt\Big\{ N\tp0 -D^2\tp0+2DC\tp1+3C D\tp1-3 C^2\tp2\Big\}-\tau\leftrightarrow \tau' \nn\\
    &=\Big\{ N\dt\tp0 -D^2\big(\dt\tp0\big) +2DC\dt\tp1+3CD\big(\dt\tp1\big)-3 C^2\dt\tp2 \label{ddC01}\\
    &\qquad +2\tp1 D\big(\hdt C\big)+3D\tp1\big(\hdt C\big)-6C\tp2\big(\hdt C\big) \label{ddC02}\\
    &\qquad + \tp0\dt N+2\tp1 D(N\t0)+3 N\t0 D\tp1-6 CN\t0\tp2\Big\}-\tau\leftrightarrow \tau'. \label{ddC03}
\end{align}
\es
The first line is equal to $\delta_{\dt\tau'} C$, it contains all the terms of the type $\dt\tau'$; 
the second line all the terms that do not involve any news $N$ nor any variations of $\tau$ (it can be written using the hatted action \eqref{hataction}).
The third line collects all the terms that depend on $N$. 
Next, $\delta_{\lbr \tau,\tau'\rbr}C$ also splits into the 3 same contributions:
\begin{align} \label{dttC} 
    \delta_{\lbr \tau,\tau'\rbr}C =  
    \delta_{(\dtp\tau-\dt\tau')} C +\hat{\delta}_{[\tau,\tau']^C} C
     + N[\tau,\tau']^C_0 . 
\end{align}
Once we sum \eqref{ddC0} with \eqref{dttC}, the terms involving $\dt\tau'$ clearly cancel.
In appendix \ref{AppAlgebraAction2}, we evaluate the sum of the terms containing the variation $\hat\delta$. It reads
\begin{equation}
    \eqref{ddC02}+ \hat{\delta}_{[\tau,\tau']^C} C=\t0 \hat\delta_{\Dcal\tau'}C-\tp0 \hat\delta_{\Dcal\tau}C. \label{hatdTT}
\end{equation}
Finally, using the variation of $N$
\begin{equation}
    \dtp N= \pa_u(\dtp C)=\delta_{\dot \tau'}C+ \tp0 \dot N+2DN\tp1+3ND\tp1-6CN\tp2,
\end{equation}
and rewriting the spin $0$ commutator as
\begin{align}
[\tau,\tau']^C_0 &=\big(\t0 D\tp1+2\t1 D\tp0-3 C\t0\tp2\big) -\tau\leftrightarrow\tau' \cr &=\big(2\t1 D\tp0+\t0(\Dcal\tau')_0\big)-\tau\leftrightarrow\tau',
\end{align}
we obtain that
\begin{align}
    \eqref{ddC03}+N[\tau,\tau']^C_0 &=
     \Big(\big(\tp0\dt N+2\tp1 D(N\t0)+3 N\t0 D\tp1-6 CN\t0\tp2\big) -  \tau\leftrightarrow\tau'\Big) + N[\tau,\tau']^C_0 
    \nn \\ 
    &=  \t0\Big(-\dtp N +2 DN\tp1+3ND\tp1-6CN\tp2  + N(\Dcal\tau')_0\Big)  -\tau\leftrightarrow\tau'
    \nn\\
    &=  \t0\Big(-\delta_{\dot\tau'}C+ N(\Dcal\tau')_0  \Big) -\tau\leftrightarrow\tau'. \label{ddC5}
\end{align}
In the second equality we swapped $\tp0\dt N$ for $-\t0\dtp N$ and in the last equality we use that the term $-\t0\tp0 \dot{N}$ vanishes after anti-symmetrization. 
Adding \eqref{hatdTT} and \eqref{ddC5}, we finally obtain \eqref{algebroidRepinterm}:
\begin{equation}
     \big[\dt,\dtp\big]C+\delta_{\lbr \tau,\tau'\rbr}C =\tp0\delta_{(\dot\tau-\Dcal\tau)}C-
    \t0\delta_{(\dot\tau'-\Dcal\tau')}C.
\end{equation}

Note that the RHS vanishes when $\delta_{\dot\tau-\Dcal\tau}C  \in \Ccar{2,2}$ is proportional to $\t0$.  Since $\t0 \in \Ccar{-1,0}$ the proportionality coefficient is in $\Ccar{3,2}$. 
The only element of that weight that can be constructed in a local manner from $C$ is $\pa_u^2C$. This means that the RHS of 
\eqref{algebroidRepinterm} vanishes when 
$\delta_{(\dot\tau-\Dcal\tau)}C=\beta \t0 \pa_uN$, where $\beta$ is an arbitrary constant. 
In the following we choose the condition $\beta=0$ and therefore assume that
\be \delta_{(\dot\tau-\Dcal\tau)}C=0.
\ee
Since $\dt C$ involves only $\t0$, $\t1$ and $\t2$, this condition implies that the $\tau$'s satisfy the recursion relations\footnote{Here we precisely want to keep $\dt C\neq 0$ in general, so that we set the argument $\dot\tau-\Dcal\tau$ to 0, rather than restricting the map to its kernel.
This would lead us back to the covariant wedge, as explained around \eqref{WCIntersection} and in further details in Sec.\,\ref{Sec:algebroidTs}.}
\vspace{-0.2cm}
\be \label{trecursioninterm}
\dot\tau_s- (\Dcal\tau)_s= 0\qquad s=0,1,2.
\ee 
If we demand, in addition, that the condition \eqref{trecursioninterm} stays valid for all $\tau$'s in the image of the $\sfT$-bracket, i.e. parameters of the form $\tau= \lbr\tau',\tau''\rbr$, then this implies that
$\dot \tau_s=(\Dcal\tau)_s$ for all $s\geq 0$.
 
The proof goes as follows: since $\tau_s$ and $\tau'_s$ have to follow the differential equation \eqref{trecursioninterm} for $s=0,1,2$, the same has to hold for the bracket $\lbr\tau,\tau'\rbr_s$, $s=0,1,2$.
In particular, we find that in order to get
\begin{equation}
    \pa_u\lbr\tau,\tau'\rbr_1= D\lbr\tau,\tau'\rbr_2-4C\lbr\tau,\tau'\rbr_3, \label{tt1dot}
\end{equation}
we need to impose 
\begin{equation}
    \pa_u\t3=D\t4-6C\t5, \label{t3dot}
\end{equation}
and similarly for $\tau'$---which is nothing else than \eqref{trecursion} for $s=3$.
The general proof then goes on as such recursively, so that \eqref{trecursioninterm} does imply \eqref{trecursion}.
To give the gist of it, we report the demonstration of ``\eqref{tt1dot} implies \eqref{t3dot}'' in App.\,\ref{App:1implies3}.

This concludes the proof of this lemma and by extension the proof of the \hyperref[LemmaDualEOM]{lemma [Algebroid action and dual EOM]}.
The demonstration of the \hyperref[theoremAlgebroid]{theorem [$\A$-algebroid]} follows from combining the various lemmas.

\paragraph{Remark:}
The anchor map is a left anchor, which implies that it is  an anti-homomorphism\footnote{This is the same convention used for $\nkor$  in the \hyperref[ReminderAlgebroidbis]{reminder} ($\tau$ here plays the role of $\alpha$ there).}
of Lie algebras, i.e. satisfies\footnote{Since the notation shall never be confusing, we use $[\dt,\dtp]\equiv[\dt,\dtp]_{\mathrm{Lie}(J^\infty\bfm p)}$, cf. \eqref{actiondelta}.}
$[\dt,\dtp]\cdot=-\delta_{\lbr \tau,\tau'\rbr}\cdot\,$; the reason being that when acting on a functional $O$, the field space action $\dt O$ is realized as a differential operator $\LL_\tau O$ on $\scri$, namely $\dt O= \LL_\tau O$.
We purposefully chose the letter $\LL$ since $\LL_\tau$ can be viewed as a generalization of the spacetime Lie derivative.\footnote{After all, when $\tau$ is just a vector field $Y$, then $\LL_\tau$ \textit{is} the Lie derivative.}
Because the field space action $\dt$ commutes with the spacetime action $\LL_{\tau'}$ up to the derivative along the variation of $\tau'$,\footnote{In general we have $[\delta, \LL_V]= \LL_{\delta V}$ for an arbitrary vector field.} we have that $\dt\dtp O=\dt \LL_{\tau'} O=\LL_{\tau'}\dt O + \LL_{\delta_{\tau} \tau'}O=\LL_{\tau'} \LL_{\tau}O + \LL_{\dt\tau'}O$.
Therefore, $[\dt,\dtp]O=[\LL_{\tau'}, \LL_\tau]O+ \LL_{\dt\tau'}O- \LL_{\dtp \tau}O =\LL_{[\tau',\tau]}O+\LL_{\dt \tau'-\dtp\tau}O=\LL_{\lbr\tau',\tau\rbr}O =\delta_{\lbr\tau',\tau\rbr}O$, where we applied the usual commutation rule for our Lie derivative-like operator $\LL_\tau$.

\subsection{Initial condition \label{sec:InitialCondition}}

The $\A$-algebroid defined in the previous section only  uses the dual equation of motions $\E_s=0$ for $s\geq 0$. 
No condition on $\tau_{-1}$ was necessary. This is due to the fact that $ \tV_{-1}$ is central and acts trivially on the shear $C$.

This leaves free the possibility to impose additional conditions on $\tau_{-1}$ provided such conditions are compatible with the $\sfT$-bracket.
One option is simply to also demand that the dual equation of motion $\E_{-1}=0$ is satisfied at level $-1$.
Remarkably, there is another option, highly relevant for the canonical analysis, which is to impose the \textit{initial condition} $\sI_\tau=0$ where
\begin{equation}
    \boxed{\,\sI_\tau:=C\t0-D\t{-1}\,}. \label{initialCondition}
\end{equation}  
Depending on which condition we impose for $\tau_{-1}$ we have two different time dependent parameter spaces.

\begin{tcolorbox}[colback=beige, colframe=argile]
\textbf{Definition [$\sfTh$-space \& $\sfTo$-space]}
\bs \label{defTbarThat}
\begin{align}
    \sfTh &:=\Big\{\,\tau\in\tV(\scri)~\big|~\pa_u\t{s}=(\Dcal\tau)_s,~  s\geqslant -1\, \Big\}, \\
    \sfTo &:=\Big\{\,\tau\in\tV(\scri)~\big|~\pa_u\t{s}=(\Dcal\tau)_s,~  s\geqslant 0~\, \& ~ D\t{-1}= C\t0\,\Big\} . 
\end{align}
\es
\end{tcolorbox}

\ni Both $\sfTh$ and $\sfTo$ are subsets of $\sfT$.  $\sfTh$ and $\sfTo$ differ by the condition imposed on $\t{-1}$: the dual  equation of motion $\E_{-1}=0$ for $\sfTh$ or the initial condition $\sI_\tau=0$ for $\sfTo$.
Quite remarkably, the $\sfT$-bracket is compatible with both spaces.

\begin{tcolorbox}[colback=beige, colframe=argile] \label{LemmaClosure2}
\textbf{Lemma [$\sfT$-bracket closure---degree $-1$]}\\
The $\sfT$-bracket closes on $\sfTh$ and on $\sfTo$, i.e. it satisfies
\begin{align}
    \tau,\tau' \in \sfTh \Rightarrow \lbr\tau,\tau'\rbr\in \sfTh  \quad \mathrm{and}\quad 
    \tau,\tau' \in \sfTo \Rightarrow \lbr\tau,\tau'\rbr\in \sfTo .
\end{align}    
\end{tcolorbox}

\paragraph{Proof:}
In practice this means that 
$D\lbr\tau,\tau'\rbr_{-1}=C\,\lbr\tau,\tau'\rbr_{0}$ 
when $\tau,\tau' \in \sfTo$ and that 
 $\pa_u\lbr\tau,\tau'\rbr_{-1}= \big(\Dcal\lbr\tau,\tau'\rbr\big)_{-1}$ when $\tau,\tau' \in \sfTh$.
The former is proven in appendix \ref{App:InitialConstraint} while the latter follows from the demonstration of the \hyperref[LemmaClosure]{lemma [$\sfT$-bracket closure]}, which was based on the computation of the Leibniz anomalies in App.\,\ref{App:dualEOM}, a result which holds at degree $-1$ too.\\

This lemma implies the following theorem. 
\begin{tcolorbox}[colback=beige, colframe=argile] \label{theoremAlgebroids}
\textbf{Theorem [$\Ah$- and $\Ao$-algebroid]}\\
Both spaces $\Ah\equiv\big(\sfTh,\lbr\cdot\,,\cdot\rbr,\delta\big)$  and 
$\Ao \equiv\big(\sfTo,\lbr\cdot\,,\cdot\rbr,\delta\big)$ are Lie algebroids 
equipped with the $\sfT$-bracket \eqref{Tbracket} and the anchor map $\delta: \A \to \X(\PS)  $ \eqref{defdelta} restricted to $\Ah$ and $\Ao$.
They are distinct sub-algebroids of $\A$.
Moreover, the intersection between these two algebroids defines an algebra\footnote{An algebra can be characterized as an algebroid with trivial anchor map.}  
\be 
\Wcal_C:=\Ah\cap \Ao. \label{WCIntersection}
\ee
which is a generalized wedge algebra on $\scri$.
\end{tcolorbox}

\paragraph{Proof:}
To prove the last part of the theorem, let us assume that  both the initial condition and the initial EOM are satisfied for $\tau$. This means that 
\begin{align}
    D \pa_u \t{-1} &= D^2 \t0 - 2 DC \t1 -2 C D\tau_1, \cr
    \pa_u D \t{-1} &= N\t0 + C D\t1 -3 C^2 \t2.
\end{align}
Therefore, taking the difference we get that 
\be 
\dt C=0, \quad \mathrm{when}\quad \tau \in \Wcal_C= \Ah\cap\Ao,
\ee 
When this condition holds\footnote{ $\Wcal_C$ is the time dependent analog of $\Wcal_\sigma$, the condition $\dt C=0$ replacing  $\hdT\sigma=0$. 
Note that the condition $\dt C=0$ seems to only imply $\pa_u\sI_\tau=-D\sE_{-1}(\tau)$.
However, using \eqref{AnomalyLeibniz-2}, we know that it actually implies $\sI_{[\tau,\tp{}]^C}=0$ as well, which then imposes $\sI_\tau=0$ by consistency and thus $D\sE_{-1}=0$.} we have that the $C$-bracket $[\cdot\,,\cdot]^C$ satisfies the Jacobi identity, thanks to \eqref{JacCFC}, and thus,  there is  no need to use  field dependent transformation parameters and algebroid extension.\footnote{$\tau$  still  depends on $C$. This dependence is compatible with the condition $\dt\tau'=0$, since $\dt\tau'\propto \dt C=0$.}
In other words, the anchor map $\delta$ is identically $0$ on $\Wcal_C$, namely over its kernel, which means that $\Wcal_C$ is an algebra.

What needs to be established next is whether $\Wcal_C$ is non trivial.
The answer depends on the value of $C$: If $C$ is a non radiative shear, i.e. such that $\pa_u C=0$, then $\Wcal_C$ is isomorphic to the celestial wedge algebra $\Wcal_\sigma(S)$ for $\scri = \R \times S$ (see section \ref{sec:Wedge}) and we have shown that $\Wcal_\sigma(S)$ is a non trivial algebra isomorphic to $Lw_{1+\infty}$ when $S=S_2$.

In general when $N\neq 0$ but $\pa_u^{n+1} C=0$ for some $n>0$ we also expect $\Wcal_C\neq \{0\}$. 
For instance, we can find a two dimensional space of solutions such that $\tau_n=0$ for $n>0$. 
In this case, elements of $\Wcal_C$ are solutions of $D^2 \tau_0 - N \tau_0=0$. This is a second order differential equation on $S$ that admits a 2 dimensional set of solutions. 
If we denote by $ f := \frac{\tau_0}{\tau_0'}$ the ratio of these two solutions we have that 
\be 
-2 N=  \{ f, z\}:= D \left(\frac{D^2 f}{Df}\right)-\frac12 \left(\frac{D^2 f}{Df}\right)^2. 
\ee 
We recognize on the RHS  the  covariant Schwarzian derivative of $f$ along $z$. This shows that $\Wcal_C$ is non trivial.
Understanding the exact nature of $\Wcal_C$ is an interesting question which goes beyond the scope of this thesis.

\paragraph{Remark:} As we are about to see in the next section, the initial condition $D\t{-1}=C\t0$, which defines $\sfTo$, is primordial to recover the Bondi mass aspect  as the canonical generator of supertranslations; i.e. when we show that  the algebroid constructed here can be represented, through Noether's theorem, in terms of  higher spin charges living on the gravitational phase space $\PS$.
It is quite remarkable that the algebroid $\A$ has ``room'' for this condition, in the sense that we are able to accommodate for such a constraint without changing anything to any of the previous discussions.

\paragraph{Remark:}\label{remarkH} Note that $\Ham$ is in $\sfTh$, but it is not in $\sfTo$. The time translation generator in $\sfTo$ is denoted $\hHam$ and given by $\hHam= (DG, 1, 0, 0, \ldots)$ where $G$ is the Goldstone, i.e. the  solution of $D^2 G = C$. $\hHam$ differs from $\Ham$ by an element of $\tV_{-1}$, which is central. 
Therefore the adjoint action of $\hHam$ coincides with the adjoint action of $\Ham$.
This means that  we could have used $\hHam$ everywhere instead of $\Ham$ in the previous section.  
We will see  in section \ref{SecGR:Noether} that the shift from $\Ham$ to $\hHam$ corresponds to, at the level of charge aspect, shifting the mass generator from the covariant mass to the Bondi mass.

\newpage
\section{Glossary \label{secGR:glossary}}

\textbf{General notation, cf. section \ref{sec:Notation}:}
\begin{itemize}
    \item $S$: a 2d complex manifold.
    \item $S_n$: a 2d complex manifold with $n$ punctures.
    \item $(m,\bm)$: null dyad on $S$.
    \item $D=m^A D_A$: the covariant derivative on $S$ along $m$.
    \item $(u,r,z,\bz)$: Bondi coordinates.
    \item $C$ \& $N\equiv\pa_u C$: the shear and the news on $\scri$.
    \item $\sigma$: a deformation parameter of the $w_{1+\infty}$ algebra that plays the role of the shear at a cut $S$ of $\scri$.
    \item $\PS$ \& $\PS_h$: the Ashtekar-Streubel and the ``twistor/self-dual'' phase spaces.
    \item $\TN$: the Newman space \eqref{hhGdiff}.
\end{itemize}

\ni \textbf{Graded and filtered vector spaces:}
\begin{itemize}
    \item $\V(S)$: the space of celestial symmetry parameters $T$, cf.\,\ref{sec:Vspace}.
    \item $\W$: the wedge subspace of $\V(S)$ \eqref{DefWedge}.
    \item $\W_\sigma$: the covariant wedge subspace of $\V(S)$ \eqref{Wdefeq}.
    \item $\tV(\scri)$: the space of time dependent symmetry parameters $\tau$,  cf.\,\ref{sec:Vspace}.
    \item $\sfT$: a subspace of $\tV(\scri)$ where $\tau$ satisfies the dual EOM $\E_s=0$ for $s\in\N$, cf.\,\eqref{TspaceGR}.
    \item $\sfTh$: a subspace of $\tV(\scri)$ where $\tau$ satisfies $\E_s=0$ for $s+1\in\N$, cf.\,\eqref{defTbarThat}.
    \item $\sfTo$: a subspace of $\tV(\scri)$ where $\tau$ satisfies $\E_s=0$ for $s\in\N$ and $\sI_\tau=0$, cf.\,\eqref{defTbarThat}.
\end{itemize}

\ni \textbf{Graded vectors:}
\begin{itemize}
    \item $T$ \& $\tau$: graded vector in $\V(S)$ and $\tV(\scri)$ respectively. $\T{s}$ and $\t{s}$ are the degree $s$ elements,  cf.\,\ref{sec:Vspace}.
    \item $\ft$: a representation of a vector in $\tV$ in terms of a holomorphic function of a spin-1 variable $q$, cf.\,\eqref{ftauDef}.
\end{itemize}

\ni \textbf{Dual equations of motion (EOM):}
\begin{itemize}
    \item $\E_s(\tau)\equiv\E_s$: the dual EOM of degree $s$, \eqref{dualEOM} and \eqref{tausevolb}.
    \item $\sI_\tau$: the initial constraint \eqref{initialCondition} and \eqref{tauintial}.
    \item $\tau(T)$: a solution of $\E(\tau)=0$ for which $\tau(T)\big|_{u=0}=T$, cf.\,\eqref{maptaubis} and \eqref{solEOMcompact}.
\end{itemize}

\ni \textbf{Brackets:}
\begin{itemize}
    \item $[\cdot\,,\cdot]^\V$: the $\V$-bracket on $S$ \eqref{Vbracket}.
    \item $[\cdot\,,\cdot]^\W$: the $\W$-bracket \eqref{Wbracket}, a restriction of the $\V$-bracket to the wedge. It reduces to the $w_{1+\infty}$ bracket on $S_2$.
    \item $[\cdot\,,\cdot]^\sigma$: the $\sigma$-bracket \eqref{CFbracket}, a deformation of $[\cdot\,,\cdot]^\W$, which is a Lie bracket over $\W_\sigma$.
    \item $[\cdot\,,\cdot]^{\tV}$: the $\tV$-bracket \eqref{tVBracket}. Same as $[\cdot\,,\cdot]^\V$ but over $\scri$.
    \item $\paren{\cdot\,,\cdot}$: the Dali-bracket \eqref{DaliBracket}.
    \item $[\cdot\,,\cdot]^C$: the $C$-bracket \eqref{CFbrackettau}. Same as $[\cdot\,,\cdot]^\sigma$ but over $\scri$.
    \item $\lbr\cdot\,,\cdot\rbr^\sigma$: the algebroid bracket \eqref{TbracketSphere} over $S$.
    \item $\lbr\cdot\,,\cdot\rbr$: the algebroid $\sfT$-bracket \eqref{Tbracket} over $\scri$.
\end{itemize}

\ni \textbf{Covariant derivative:}
\begin{itemize}
    \item $\Dcal T=\frac{1}{\Ham_0}[\Ham,T]^\sigma$: the adjoint action \eqref{CovDer} of the particular super-translation $\Ham$.
    \item $\Dcal \tau=\frac{1}{\Ham_0}[\Ham,\tau]^C$: the adjoint action but for the $C$-bracket, cf.\,\eqref{DcalC}.
\end{itemize}

\ni \textbf{Algebra and algebroids:}
\begin{itemize}
    \item $\Wcal(S)\equiv\big(\W,[\cdot\,,\cdot]^\W\big)$: the \hyperref[LemmaWedgeAlg]{wedge algebra}.
    \item $\Wcal_\sigma(S)\equiv\big(\W_\sigma,[\cdot\,,\cdot]^\sigma\big)$: the \hyperref[theoremWalgebra]{covariant wedge algebra}.
    \item $\A\equiv\big(\sfT,\lbr\cdot\, ,\cdot\rbr,\delta\big)$: the \hyperref[theoremAlgebroid]{$\A$-algebroid}.
    \item $\Ao\equiv\big(\sfTo,\lbr\cdot\, ,\cdot\rbr,\delta\big)$: the \hyperref[theoremAlgebroids]{$\Ao$-sub-algebroid}, that admits a representation on $\PS$.
    \item $\Ah\equiv\big(\sfTh,\lbr\cdot\, ,\cdot\rbr,\delta\big)$: the \hyperref[theoremAlgebroids]{$\Ah$-sub-algebroid}.
    \item $\Ah\equiv\big(\sfTh,\lbr\cdot\, ,\cdot\rbr,\tilde\delta\big)$: an extension of the \hyperref[theoremAlgebroidh]{$\Ah$-sub-algebroid} that admits a representation on $\PS_h$.
\end{itemize}

\ni \textbf{Anchor map and infinitesimal variation:}
\begin{itemize}
    \item $\delta$ \& $\tilde\delta$: anchor maps of the various algebroids.
    \item $\hat\delta$: `non-radiative' anchor map \eqref{dTCv1} and \eqref{hataction}.
    \item $\dt$: fields space vector field that generates an infinitesimal symmetry transformation on functional over $\scri$, \eqref{dtauC}.
    \item $\dT$: same as $\dt$ but on functional over $S$.
    \item $\d{s}{\T{s}}\cdot=\delta_{\tau(\T{s})}\cdot$: the infinitesimal variation of spin-weight $s$ along the parameter $\T{s}$, cf.\,\eqref{defdeltaTs}.
    \item $\td{s}{\T{s}}\cdot$: the coefficient of $u^0$ when $\d{s}{\T{s}}\cdot$ is a polynomial in $u$, cf.\,\eqref{deftildedeltaTs}.
\end{itemize}

\ni \textbf{Charges:}
\begin{itemize}
    \item $\tQ_s(u,z,\bz)$: charge aspect of helicity $s$ that satisfies the truncation of EE \eqref{Qevolve}.
    \item $\tq_s(u,z,\bz)$: renormalized charge aspect \eqref{defcqTsu}. Bondi version.
    \item $\th_s(u,z,\bz)$: renormalized charge aspect \eqref{RenormChargeCov}. Covariant version.
    \item $\hq_s(z,\bz)=\lim_{u\to-\infty}\tq_s$: `Bondi' renormalized charge aspect at $\scri^+_-$.
    \item $\hh_s(z,\bz)=\lim_{u\to-\infty}\th_s$: `Covariant' renormalized charge aspect at $\scri^+_-$.
    \item $\q{s}{k}$: the part of $\hq_s$ homogeneous of degree $k$ in $C$, cf.\,\eqref{hqsk}.
    \item $Q^u_s[\t{s}]\propto\int_S \tQ_s\t{s}$: $\tQ_s$ smeared over the sphere.
    \item $\mathds{Q}_{\tau}^u$: master charge, the sum over all spins of $Q^u_s[\t{s}]$. 
    \item $\Qt^u$: `Bondi' master charge \eqref{masterChargeGR}. Conserved when $\bN=0$.
    \item $\Ht^u$: `Covariant' master charge. Conserved when $\dot{\bN}=0$, cf.\,\eqref{QH}.
    \item $\Qt=\Qt^{-\infty}$: `Bondi' Noether charge \eqref{defNoetherCharge}.
    \item $\Ht=\Ht^{-\infty}$: `Covariant' Noether charge \eqref{canonicalRepH}.
    \item $\cq_{\T{s}}=Q_{\tau(\T{s})} \propto\int_S\hq_s\T{s}$: Noether charge of the helicity $s$ parameter, cf.\,\eqref{Noetherhatqs}.
    \item $\mathsf{Q}$: charge aspect 2-form \eqref{Qaspect2form}.
    \item $\mathsf{Q}_\tau$: master charge 2-form \eqref{Qaspect2form}. `Bondi' version.
    \item $\mathsf{H}_\tau$: master charge 2-form \eqref{Haspect2form}. `Covariant' version.
    \item $\Qt^U$: master charge on an arbitrary cut $S(U)$ of $\scri$, cf.\,\eqref{QtArbitraryCut}.
\end{itemize}

\ni \textbf{Goldstone:}
\begin{itemize}
    \item $G(z,\bz)$ \& $G(u,z,\bz)$: the Goldstone field \eqref{Goldstone} \& \eqref{GoodCutEq}.
    \item $\hG$: the Goldstone diffeomorphism on $\scri$ \eqref{GoodCutEq}.
    \item $\hhG$: the Goldstone diffeomorphism on $\TN$, \eqref{hhGdiff}.
    \item $\mTG$: the dressing map \eqref{mTGdef}, see also \ref{sec:RelationDressingMap}. 
\end{itemize}

\chapter{General Relativity II \label{secGRII}}

This chapter gathers the material of the second half of \cite{Cresto:2024mne}.

\section{Introduction}

We now describe the canonical realization of $\Ao$ and $\Ah$ in section \ref{SecGR:Noether} and prove that the symmetry algebroids are realized as Noether charges on the gravitational phase space.
This chapter follows more closely the style of the \hyperref[secNAGTII]{YM discussion}.
We also give a definition of the surface charges for arbitrary cuts of $\scri$ in \ref{sec:ArbitraryCut}.
We discuss the filtration and the gradation of $\A$, together with a class of solutions of the dual EOM in section \ref{Sec:SolDualEOM}.
This allows us to define the renormalized charges in \ref{sec:RenormCharge} and we give some insights about our systematic, non-perturbative (in $\gN$, cf.\,\ref{sec:FPR}) procedure by comparing with the findings of \cite{Geiller:2024bgf}.
We then compute the action of the associated Noether charge onto the shear in \ref{SecComparison} and show that we recover previously known results as a particular application of our formalism.
In section \ref{Sec:algebroidTs}, we make the connection with the covariant wedge algebra from chapter \ref{secGRI}.

While this work was in completion, the Oxford group published a beautiful paper \cite{Kmec:2024nmu} which obtained independently results in agreement with ours. 
They showed, starting from a twistor space formulation of self-dual gravity, that the twistor symmetry charges satisfy, after integration over the fiber, the equation \eqref{Qevolve}. 
They also proved that the Noether's theorem ensures a realization of the twistor symmetry beyond the wedge on the self-dual phase space.
The dual equations of motion arise there from a gauge fixing condition that projects twistor functionals onto spacetime functionals. 
One apparent difference between their work and ours is that the symmetry bracket used in twistor theory is simply a Poisson bracket on the twistor fiber and is, therefore, independent of the shear. 
In our work, the symmetry bracket that stems from the spacetime analysis is shear-dependent. 
In section \ref{sec:twistor}, we show that, quite remarkably, the shear non-linearity can be reabsorbed through the introduction of an extra spin 1 variable $q$ and after using the dual EOM.
It should be clear to the reader that our work, which focuses on the spacetime and canonical formulation, was done independently of \cite{Kmec:2024nmu}. 
It shows that spinor and canonical approaches lead to similar results from very different perspectives. 
For instance, the canonical framework allows for two series of charges: the one conserved when $\dot{\bC}=0$, which includes the Bondi mass, and the covariant charges conserved when  $\ddot{\bC}=0$. 
The latter are the ones that appear naturally from the twistor analysis. 
Both are represented canonically and differ only by how one treats the spin $-1$ transformation.

Finally, we discuss the relationship to Newman's good cut equation \cite{newman_heaven_1976} in section \ref{sec:twistor}.
We are able to construct a field dependent frame\footnote{Recall that the definition of the higher spin charges always depends on a choice of Newman-Penrose tetrad, as we reviewed in details in chapter \ref{secNAGTI}.} along which the time evolution \eqref{dualEOM} of the transformation parameters now no longer involves any shear $C$.\footnote{By duality, this also means that $C$ disappears from the charge aspects time evolution \eqref{Qevolve}.}
In other words, we find a frame which brings the dynamics back to its flat space version.
Interestingly, this necessitates the definition of a diffeomorphism on what we call the Newman space $\TN$, namely the line bundle of $\scri$-complexified by a fiber parametrized by the variable $q$ mentioned above.
This new coordinate is related to Carrollian physics as well as work by Adamo and Newman \cite{Adamo:2009vu}.
When the diffeomorphism is time-independent, we recover the definition of the dressing map $\mTG$ from section \ref{sec:CovWedgeSol}.
The isomorphism between the wedge\footnote{Recall that the wedge is precisely the 0 shear limit of the covariant wedge.} and covariant wedge algebras therefore originates from the possibility of dressing the symmetry parameters using a shear (or more precisely its associated Goldstone) dependent diffeomorphism.
Geometrically, we interpret these findings as the possibility of constructing the wiggly cuts of $\scri$ which result from the light-cones emitted by the bulk points of an asymptotically flat spacetime.
The symmetry algebra associated to a (shear-determined) wiggly cut effectively does not see the shear.
The dressing map (or the generalized diffeomorphism in $\TN$) makes concrete and formal this last statement.
As usual, a collection of appendices \ref{appGRII} supplements the main text.

\section{Noether charge \label{SecGR:Noether}}

The goal of this section is to prove that there exists a canonical representation of the higher spin algebroid bracket \eqref{Tbracket} on the Ashtekar-Streubel phase space. 
This phase space possesses a symplectic potential which depends on the shear $C(u,z,\bar{z})$ and the news $\bN\equiv \pa_u \bC$. 
We use the convention where 
$C:=\tfrac12C_{mm}= \frac{P^2}{2}C_{zz}$,\footnote{This is the Newman-Penrose's shear $\sigma_{\textsc{np}}$.} 
instead of the one of \cite{Freidel:2021ytz}. 
The holomorphic Ashtekar-Streubel symplectic potential \cite{ashtekar1981symplectic} is given by an integral over $\scri$,\footnote{$\delta$ here is the fields space exterior derivative, not to be confused with the anchor. 
By definition, both are related according to $I_{\dt}(\delta C)=\dt C$, for $I$ the fields space interior product \cite{Donnelly:2016auv}.}
\be 
\Theta^{\mathsf{HAS}} = \frac1{4\pi \GN} \int_{\scri}  \bN \delta C.
\ee 
It is related to the usual, real Ashtekar-Streubel symplectic potential by a complex canonical transformation.
Indeed, 
\begin{equation}
    \Theta^{\mathsf{HAS}}=\Theta^{\mathsf{AS}} +\frac{1}{8\pi\GN}\left(\delta\left(\int_\scri \bN C\right)-\int_\scri \pa_u(C\delta\bC) \right),
\end{equation}
where
\be 
\Theta^{\mathsf{AS}}=\frac1{8\pi \GN} \int_{\scri}\big(\bN \delta C +N \delta \bC\big).
\ee 
This section aims to show that the \hyperref[theoremAlgebroids]{$\Ao$-algebroid} is represented canonically on the gravitational phase space.
This is a major result of this chapter, which is valid provided we choose the fields $(C,\bC)$ to belong to the \hyperref[foot1]{Schwartz space}.
\textit{Mutatis mutandis}, this result follows exactly the same reasoning as the demonstration we provided in section \ref{secYM:Noether}.
Consequently, in section \ref{Sec:CanAction}, we only emphasize the main arguments and refer the reader to \ref{secYM:Noether} for details.

\subsection{Master charge \label{secGR:MasterCharge}}

Let us start with the explicit construction of the algebroid Noether charge. 
For each $s\in \{-1,0,1,\cdots,\}$ we define a charge aspect $\tQ_s$ to be the spin $s$ charge aspect. 
This is an element of $\Ccar{3,s}$ and its evaluation at a point $(u,z,\bz)$ on $\scri$ is denoted $ \tQ_s(u,z,\bz)$. These charge aspects are defined recursively by imposing that 
\be 
\tQ_{-2}= \dot{\bN},\quad\qquad \tQ_{-1}=  D\bN,
\ee
and through the evolution equations
\be \label{Qevolve}
\pa_u \tQ_{s}= D\tQ_{s-1}+(s+1) C \tQ_{s-2}, \qquad s\geqslant -1.
\ee 
This set of evolution equations encodes the asymptotic evolution of the Weyl tensor \cite{Adamo:2009vu, Freidel:2021ytz, Geiller:2024bgf}.
Imposing the asymptotic condition that  $\tQ_s|_{u=+\infty}=0$ allows us to define $\tQ_s(u)$, after recursive $u$-integration, as a function of $(C,\bN)$ which is \emph{linear} in $\bN$ and polynomial of degree $\lfloor s/2\rfloor+1$ in $C$ and its derivatives\footnote{This condition is valid for $s\geqslant -2$, with $\lfloor n\rfloor$ the floor function.
To see it, just notice that the initial data $\tQ_{-2}$ and $\tQ_{-1}$ are of degree 0 in $C$ and afterwards $\deg[\tQ_s]=\deg[\tQ_{s-2}]+1$.} \cite{Freidel:2021ytz}. 

At any given $u=\mathrm{const}$ cut of $\scri$ we  define, for $s\geqslant -1$, the smeared charges\footnote{In the following we shall write $\int_S:=\int_S\bep_S$.}
\begin{equation}
Q_s^u[\tau_s]:=
\frac{1}{4\pi \GN}\int_S(\tQ_s\t{s})(u,z,\bz), \label{Qstaus}
\end{equation}
where $\tau_s$ satisfy the dual evolution equations  $\E_s(\tau)=0$ for $s\geq 0$ \eqref{dualEOM}, where
\be 
 \E_s(\tau) := \dot \tau_s - D \t{s+1} + (s+3) C \t{s+2},\label{tausevolb}
\ee  
and are subject to the initial condition \eqref{initialCondition} $\sI_\tau=0$, where
\be  \label{tauintial}
\sI_\tau:=C\t0-D \t{-1}.
\ee 
When the context is clear, we henceforth use $\E_s$ instead of $\E_s(\tau)$.
The fact that $\t{s} \in \Ccar{-1,-s}$ while $\tQ \in \Ccar{3,s}$ implies that the product $\tQ_s \tau_s \in \Ccar{2,0}$ is a scalar density on the sphere and 
\eqref{Qstaus} actually defines, after integration on $S$, a pairing between $\Ao$ and its dual $\big(\Ao\big)^*$.

It is important to remember that $\cQ{s}{\tau_s}$ involves an integral over the sphere, so that we can freely integrate by parts sphere derivatives.
We then construct the time-dependent \emph{master charge} $\Qt^u$,
\begin{equation} \label{masterChargeGR}
\boxed{\Qt^u := \sum_{s=-1}^{\infty} Q^u_s[\t{s}]}.
\end{equation}
The time evolution of the master charge is remarkably simple. 
One first uses the charge aspect evolution equation \eqref{Qevolve} to get 
\begin{align}
\pa_u \Qt^u  &= Q^u_{-1}[\dot\tau_{-1}]-Q^u_{-2}[D\t{-1}]+\sum_{s=0}^\infty \Big(Q^u_s[\dot\tau_{s}] - Q^u_{s-1}[D\t{s}] + (s+1) Q^u_{s-2}[C \t{s}]\Big) \label{timeEvolQt}\\
& = Q^u_{-2}\big[-D \t{-1} +C\t0\big] + Q^u_{-1}\big[\dot \tau_{-1} - D\t0 +2C \t1\big] +\sum_{s=0}^\infty Q^u_s\big[\dot\tau_s -D\t{s+1} + (s+3) C\t{s+2}\big]. \nn
\end{align}
Using the dual evolution equations \eqref{tausevolb} and initial condition \eqref{tauintial}, i.e. when $\tau \in \Ao$, only the second term survives.
The latter can be simplified using that $\tQ_{-1}=  D\bN$, so that 
\begin{align}
\dot{Q}^u_\tau &= - \bN \left[ D \dot\tau_{-1} - D^2 \t0 +  2 D(C \t1)\right] \nn \\
&= - \bN  \left[  \big(N\t0 + C(D \t1 - 3C \t2)\big) - D^2 \t0 + 2  D(C \t1)\right] \label{dotQtuinterm}\\
& =- {\bN} \left[ -\left( D^2-  N \right) \t0 + \big(2 DC+ 3 CD\big)\t1 - 3 C^2 \t2 \right]. \nn
\end{align}
Hence
\begin{equation}
\pa_u\Qt^u=- \frac{1}{4\pi \GN}\int_S \bN \dt C , \label{Qtdot}
\end{equation}
where we used the transformation \eqref{dtauC}
\begin{equation}\label{deltaC}
\dt C:= -D^2 \t0+ N\t0+ 2 DC\t1+ 3 CD \t1- 3 C^2\t2.
\end{equation}
On the RHS, all functions are evaluated at the same point $(u,z,\bz)$ where the shear $C$ is evaluated on the LHS. 
We have seen in sections \ref{secAlgebroid} and \ref{sec:InitialCondition} and equation \eqref{RepdeltaPS} that this transformation provides a representation of the $\Ao$-algebroid on phase space. 
Besides, notice that $\Qt^u$ is conserved in the absence of left-handed radiation, i.e. when $\bN\equiv 0$. 
From \eqref{Qtdot} and the final condition $\Qt^{+\infty}=0$,\footnote{Note that what matters for the definition of the charge is that $\lim_{u\to +\infty} \tQ_s\t{s}=0$, which is much less restrictive than imposing $\lim_{u\to +\infty} \tQ_s=0$ as it can be read as a condition on the transformation parameters $\t{s}$ instead. \label{foot2}}
we can express $\Qt^u$ as the integral 
\begin{equation}
\Qt^u = \int^{+\infty}_u  \bN[\dt C]. \label{defNoetherChargeinterm}
\end{equation}

\subsection{Covariant charge \label{Sec:CovCharges}}

It is interesting to note that the initial condition $D \t{-1} =  C \t0$  is crucial in order to obtain a series of charges which is conserved when $\bN=0$ and for which the total mass satisfies the Bondi  loss formula. 
We now show that the total mass is the charge associated to $\hHam$ (see e.g. remark \ref{remarkH}).

The initial condition ensures that $\t{-1}$ is not an independent parameter, so there is no independent spin $-1$ charge. Instead, it implies that the charge associated with $\tau_0$ is given by the combination,\footnote{Notice however that
\begin{equation}
    \poisson{\cQ{-1}{\t{-1}},C(u',z,\bz)}=- C\t0\,\delta(u-u').
\end{equation}
We thus see that the bracket with the $\tQ_{-1}$ charge contributes to a contact term which becomes 0 upon taking the limit $u\to -\infty$, cf. the definition of the Noether charge \eqref{defNoetherCharge}.}
\be
 Q^u_{-1}[\t{-1}] +Q^u_0[\t0]= \frac{1}{4\pi \GN}\int_S \left(  - \bN D\tau_{-1} + \tQ_0\t0 \right)=
 \frac{1}{4\pi \GN}\int_S \big(\tQ_0 - \bN C\big)\t0. \label{bondimass}
\ee 
$\tQ_0$ is the covariant mass aspect equal to the leading order in the $1/r$ expansion of the Weyl tensor component $\psi_2$. On the other hand the combination $\tQ_0 - \bN C$ is the Bondi mass aspect!\footnote{This was first noted by Moreschi in \cite{moreschi2004intrinsic}.} The total mass is the charge evaluated for $\tau=\hHam$ and is given by the Bondi flux formula
\be 
\pa_u Q^u_{\hHam}= -\frac{1}{4\pi \GN} \int_S \bN N \leq 0.
\ee

Note that a choice was made in \eqref{timeEvolQt} about the behavior of $\tau_{-1}$. 
If we only assume that the evolution equations $\E_s=0$ for $s\geq0$ are imposed, we obtain that 
\be 
\pa_u \mathds{Q}_{\tau}^u = Q^u_{-2}\big[-D\t{-1} +C\t0\big] + Q^u_{-1}\big[\dot\tau_{-1} - D\t0 +2C\t1\big],
\ee
where the notation $\mathds{Q}_{\tau}$ refers to the fact that $\tau_{-1}$ is left unconstrained, i.e. that $\tau\in \sfT$.
From this expression we see that two natural possibilities open: either we impose the initial condition which gives a charge $\mathds{Q}^u_{\tau}\to Q^u_\tau$ conserved when $\bN=0$; or we impose the condition $\E_{-1}(\tau)=0$, namely
\be 
\dot \tau_{-1} = D \t0 -  2 C \t1. \label{dualEOM-1}
\ee 
In this case we denote the charge by $\mathds{Q}^u_{\tau}\to H^u_\tau$, which is conserved when $\dot\bN=0$. This is what we call the covariant charge in \cite{Freidel:2021qpz}. In order to understand the relationship between $H_\tau$ and $Q_\tau$, we reveal the important identity 
 \be
 \pa_u (D \t{-1} -  C \t0) &=
 D \dot \tau_{-1} -  N \t0 
 - C \dot \tau_0 \nn\\
 &= D (\dot\tau_{-1}-D \t0 +  2 C \t1 )
 +(D^2-N) \t0 -  2 DC \t1 - 2C D\t1  -C( D\t1 -3 C\t2) \nn\\
&= D (\dot\tau_{-1}-D \t0 +  2 C \t1)
 +(D^2-N) \t0 -  2 DC \t1 - 3 C D\t1  + 3 C^2\t2 \nn\\
&= D \E_{-1} - \dt C.
 \ee 
In other words,
\begin{equation}\label{-1id}
     \boxed{\dt C=D\E_{-1}+\pa_u\sI_\tau },
\end{equation}
with $\dt C$  given in \eqref{deltaC}. 
From this identity we conclude that if we impose the initial condition, then we come back to the definition of $\dt C$ and the original computation \eqref{dotQtuinterm} for which $\pa_u Q_\tau^u =- \bN[\dt C]$.
On the other hand, if we impose,   $\E_{-1}=0$, then we conclude  that  we can interpret
$\sI_\tau$ as the transformation of the potential $h:= \pa_u^{-1} C$.\footnote{$h=\pa_u^{-1}C$ appears as the twistor potential evaluated at $\scri$ \cite{Adamo:2021lrv,Donnay:2024qwq,Kmec:2024nmu}.} We denote this transformation by $\tdt$ and define 
\be
\boxed{\tdt h := \t0 \pa_u h  - D\t{-1}=\sI_\tau=-(\Dcal\tau)_{-2}}. \label{dtauh}
\ee
From the condition $\sE_{-1}=0$ we get that the charge associated to $\tilde\delta_\tau$ satisfies
\be \label{QH}
\pa_u H_{\tau}^u = Q^u_{-2}[C\t0-D \t{-1}] = \dot\bN\big[\tdt h\big].
\ee
This evolution equation shows that the charge is conserved when $\dot \bN=0$.

A priori the symmetry transformation $\tdt$ associated with the charge $H_\tau$ is different than the one $\dt$ associated with
the charge $\Qt$ due to the different  conditions on $\t{-1}$.
However, the equation \eqref{-1id} means, under the condition
$\sE_{-1}=0$, that $\pa_u\tdt h = \dt C$. This signifies that although the charges $\Qt$ and $\Ht$ are different, their action on the shear $C$ coincides!
The difference only lies in the fact that the twistor potential $h$ is transformed by the $\Ht$ action since the $\Ht$ action of $\t{-1}$ is non trivial on $h$.

In \cite{Kmec:2024nmu} it was recently showed that the covariant charge $H_\tau$ can be naturally derived from the analysis of the twistorial formulation of self-dual gravity after a  gauge fixing that projects the twistor description onto spacetime.
Equation \eqref{QH} shows that the charge $H_\tau$ is naturally associated to the symplectic potential
\be 
\Theta_h = -\frac1{4\pi \GN} \int_{\scri}  \dot \bN \delta h.
\ee 
This potential is, after integration by part, equal to the holomorphic Ashtekar-Streubel symplectic potential provided we assume the boundary condition $\displaystyle{\lim_{u\to \pm \infty}  (\bN \delta h)(u,z,\bz) =0}$.
To avoid confusion, we also label the associated phase space with $h$, namely $\PS_h$.

In the appendix \ref{App:hE-1}, we show that we can readily adapt the $\Ah$-algebroid construction of Sec.\,\ref{sec:InitialCondition} to accommodate for this change of symplectic potential.
In particular, re-defining the algebroid bracket using $\tdt$ instead of $\dt$ for the anchor map, i.e.\footnote{$[\tau,\tau']^C_s$ is still the bracket \eqref{CFbrackettau}.}
\begin{equation}
    \lbr\tau,\tau'\rbr_s :=[\tau,\tau']^C_s+\tdtp\t{s}-\tdt\tp{s}, \label{hbracket}
\end{equation}
we prove that 
\begin{equation}
    \boxed{[\tilde{\delta}_\tau, \tilde{\delta}_{\tau'}] h= 
- \tilde{\delta}_{\lbr \tau, \tau'\rbr } h}. \label{algebroidReph}
\end{equation}
when $\tau,\tau'\in \sfTh$.

Moreover, if instead of requiring $\tau$ and $\tau'$ to satisfy the initial condition $\sI_\tau=0$ \eqref{tauintial}, we impose the dual EOM $\mathsf{E}_{-1}=0$ \eqref{dualEOM-1}, then the latter is preserved by $\lbr \tau, \tau'\rbr$:
\begin{equation}\label{bibi}
    \pa_u\lbr\tau,\tau'\rbr_{-1}
    =\big(\Dcal\lbr\tau,\tau'\rbr\big)_{-1}.
\end{equation}
The proof of \eqref{bibi} and that \eqref{hbracket} respects the EOM is the same as for the \hyperref[LemmaClosure]{lemma [$\sfT$-bracket closure]}.
In other words,
\begin{equation} \label{AnomalyLeibnizh}
    (\pa_u-\Dcal)\lbr\tau,\tau'\rbr= \big\lbr(\pa_u-\Dcal)\tau,\tau'\big\rbr +\big\lbr\tau,(\pa_u-\Dcal)\tau'\big\rbr+\tilde\delta_{(\pa_u-\Dcal)\tau}\tau'-\tilde\delta_{(\pa_u-\Dcal)\tau'}\tau,
\end{equation}
and the RHS vanishes for $\tau,\tau'\in\sfTh$.

We thus have two notions of $\A$-algebroid, the first one is $\Ao$, which is naturally adapted to the AS symplectic potential; the other denoted $\Ah$ is adapted to the twistor (or self-dual) symplectic potential.
Both share the same action on the shear.

We can then state the same theorem as in Sec.\,\ref{secAlgebroid}, but for the $\Ah$-algebroid.

\begin{tcolorbox}[colback=beige, colframe=argile] \label{theoremAlgebroidh}
\textbf{Theorem [$\Ah$-algebroid]}\\
The space $\Ah\equiv\big(\sfTh,\lbr\cdot\,,\cdot\rbr,\tilde\delta\big)$ equipped with the bracket \eqref{hbracket} and the anchor map,
\begin{align}
    \tilde\delta : \,&\Ah\to \X(\PS_h) \nn\\
     &\,\tau\mapsto\tilde\delta_\tau \label{defdeltah}
\end{align}
 forms a Lie algebroid over $\PS_h$.
\end{tcolorbox}

\ni Since the action $\tdt$ reduces to $\dt$ on $\PS$, we have that $\Ah$ here reduces to $\Ah$ from section \ref{theoremAlgebroids} if all the functionals $\tau$ depend only on $C$ and not $h$. 
To put it differently, $\big(\sfTh,\lbr\cdot\,,\cdot\rbr,\tilde\delta\big)$ is an extension of $\big(\sfTh,\lbr \cdot\,,\cdot\rbr,\delta\big)$ to $\PS_h$.

\subsection{Canonical action \label{Sec:CanAction}}

We now show that the action of $\Ao$ and $\Ah$ on the gravity phase space is canonical.
So far, we have only described how $\Ao$ and $\Ah$ act on $C$.
As in \ref{secYM:Noether}, the whole point of the demonstration is to show that one can build $\dt\bN$ such that 
\begin{equation}
    L_{\dt}\Theta^{\mathsf{HAS}}=0,
\end{equation}
where $L_{\dt}$ is the Lie derivative in fields space along the vector field $\dt$. 
Using Cartan calculus \cite{Freidel:2020xyx}, this amounts to $I_{\dt}\delta\Theta^{\mathsf{HAS}} +\delta I_{\dt}\Theta^{\mathsf{HAS}}=0$, which means that
\begin{equation} \label{defNoetherCharge}
    \boxed{I_{\dt}\Omega=-\delta\Qt}\qquad \textrm{for}\qquad\boxed{\Qt:=\Qt^{-\infty}=I_{\dt}\Theta^{\mathsf{HAS}}}, 
\end{equation}
with $\Omega=\delta\Theta^{\mathsf{HAS}}$ the symplectic form.
From \eqref{defNoetherChargeinterm}, the charge can be written as an integral over $\scri$,
\be 
\boxed{\Qt =\frac{1}{4\pi \GN}\int_\scri \bN 
\Big( ( N-D^2 ) \t0+  2 DC\t1+ 3 CD \t1- 3 C^2\t2\Big)}. 
\ee 
Notice that the dependence on the higher spin $\tau_s$ is implicit through the time evolution $\E_s=0$ \eqref{tausevolb}.
The equations \eqref{defNoetherCharge} show that $\Qt$ is the \textit{Noether charge} for the $\Ao$-algebroid action.
Using the morphism \eqref{RepdeltaPS} and by definition of the Poisson bracket,\footnote{According to the symplectic form $\Omega$, we have that $\poisson{\bN(x),C(x')}=4\pi\GN\delta^3(x-x')$, where $x=(u,z,\bz)$.} we get that
\begin{equation} \label{NoetherChargeRep}
    \boxed{\poisson{\Qt,O}=\dt O} \qquad \textrm{and}\qquad \boxed{\poisson{\Qt,\Qtp}=-Q_{\lbr\t{},\tp{}\rbr}},
\end{equation}
where $O$ is an arbitrary functional of $(C,\bC)$.

\paragraph{Remark:}
We expect the discussion to generalize straightforwardly when working with the covariant news tensor $\hat N=N-\rho$, where $\rho=\rho_{AB}m^Am^B$ is the Geroch tensor \cite{Geroch1977, Compere:2018ylh, Dray:1984rfa}. 
The latter is necessary when considering a conformal compactification which is not the round sphere.
It also ensures that the BMS charge algebra is 2-cocycle free \cite{Rignon-Bret:2024wlu, Rignon-Bret:2024gcx}.
Besides, notice that the spin 1 charge matches with the Dray-Streubel charge \cite{Dray:1984rfa} and was proven to define a $\gbms$ moment-map in \cite{Barnich:2021dta, Freidel:2024jyf}.\\

The same analysis goes through if we consider the covariant charge $\Ht$ instead. 
In that case,
\begin{equation}
    \Omega_h=\delta\Theta_h,\qquad I_{\tdt}\Omega_h=-\delta\Ht,\qquad \Ht:=\Ht^{-\infty}=I_{\tdt}\Theta_h, \label{canonicalRepH}
\end{equation}
where the Noether charge is written as the following integral over $\scri$:
\be 
\boxed{\Ht =-\frac{1}{4\pi \GN}\int_\scri \dot{\bN}\big(C\t0-D\t{-1}\big)}. 
\ee 
$\Ht$ is the Noether charge for the $\Ah$-algebroid action. 
It satisfies
\be 
\boxed{\{\Ht, O\} =\tdt O \qquad \textrm{and}\qquad
\poisson{\Ht, H_{\tau'}}= -H_{ \lbr \tau, \tau' \rbr }}. \label{NoetherChargeRepH}
\ee

\subsection{Arbitrary cut \label{sec:ArbitraryCut}}

So far the charges we have constructed are associated with constant $u$ cuts. 
We need to be able to construct Noether charges associated with any cut  $S(U)=\{ u= U(z,\bz)\}\subset\scri$.
 where $U \in \Ccel{-1,0}$. 
Here, we generalize the Noether charge construction to accommodate any such cuts of $\scri$.
To do so, we promote the charge aspect to a 2-form on $S$. 
One defines\footnote{Recall that $\bfm\epsilon_S\equiv i\frac{\rd z \wedge \rd \bz}{P\bP}$.}
\begin{empheq}[box=\fbox]{align}
    \mathsf{Q}_s : = \tQ_s\,\bfm\epsilon_S + \tQ_{s-1} \,\rd u \wedge \frac{i\rd \bz}{\bP}, \qquad
 \mathsf{Q}_\tau := \sum_{s=-1}^\infty \tau_s \mathsf{Q}_s. \label{Qaspect2form}
\end{empheq}
Denoting $\rd^3 x:= \rd u \wedge \bep_S$ the volume element on $\scri$ and $\rd = \rd u \,\pa_u + \frac{\rd z}{P} D + \frac{\rd \bz}{\bP} \bD$ its differential, one evaluates
 \be
 \rd \mathsf{Q}_\tau = &
 \sum_{s=-1}^\infty
 \left(\tau_s \big(\pa_u \tQ_s
 - D\tQ_{s-1}\big) +\pa_u\tau_s \tQ_s - D\tau_s \tQ_{s-1}\right)
 \rd^3 x \cr
 = &
 \left( \sum_{s=0}^\infty (s+1) C \tau_s \tQ_{s-2} - D\tau_{-1} \tQ_{-2}
 + (\pa_u\tau_{-1}- D\t0) 
 \tQ_{-1}  + 
 \sum_{s=0}^\infty ( \pa_u\tau_s- D\tau_{s+1}) \tQ_s \right)
 \rd^3 x \cr 
 = &
 \left( \big(C \t0- D\t{-1}\big) \tQ_{-2}
 + \big(\pa_u\t{-1}- D\t0 + 2 C\t1\big) 
 \tQ_{-1} \right)
 \rd^3 x,
\ee
where we used the EOM \eqref{Qevolve} in the second equality and the dual EOM \eqref{tausevolb} in the last one. 
Hence,
\begin{equation}
    \boxed{\rd \mathsf{Q}_\tau=\big(\sI_\tau \tQ_{-2} +\E_{-1}\tQ_{-1}\big)\rd^3 x}.
\end{equation}
The initial condition \eqref{tauintial} implies that the first term vanishes while we know that $D\E_{-1}=\dt C$, cf.\,\eqref{-1id}.
Therefore, since $\tQ_{-1}= D\bN$, we get that\footnote{$\mathsf{Q}_\tau=\mathsf{Q}_\tau'$ on a constant $u$ cut, as it should.}
\be \label{FluxBalanceLaw}
\rd \mathsf{Q}_\tau' 
=  - \big(\bN \dt C \big) \rd^3 x, \qquad
\mathsf{Q}_\tau'
:=
\mathsf{Q}_\tau + \E_{-1}(\tau) \bN \,\rd u \wedge \frac{i\rd \bz}{\bP}. 
\ee 
When the left-handed radiation vanishes, i.e. $\bN=0$, we have that 
$\mathsf{Q}_\tau'=\mathsf{Q}_\tau$ is covariantly conserved: $\rd \mathsf{Q}_\tau =0$.
One can use $\mathsf{Q}_\tau$ to define the Noether charge at any cut,
\be \label{QtArbitraryCut}
Q_\tau^{U} \equiv
\frac{1}{4\pi\GN}\int_{S(U)} \mathsf{Q}_\tau
= \frac{1}{4\pi\GN}\sum_{s=-1}^{\infty} \int_{S} \big(\tau_s \tQ_s + \tau_s \tQ_{s-1} DU \big)(U(z,\bz),z,\bz)\, \bfm\epsilon_S, 
\ee 
and similarly for $\Qt^{'U}$.

Following the discussion in Sec.\,\ref{Sec:CovCharges}, if we impose $\E_{-1}=0$ rather than the initial constraint $\sI_\tau=0$, then $\rd\mathsf{H}_\tau=(\dot{\bN}\tdt h)\rd^3x$, or equivalently (using $\pa_u\sI_\tau=\dt C$)
\begin{equation} \label{Haspect2form}
    \rd\mathsf{H}'_\tau=-(\bN\dt C)\rd^3x,\qquad \mathsf{H}'_\tau:=\mathsf{H}_\tau-\sI_\tau\bN\bep_S.
\end{equation}

\paragraph{Remark:}
The equation \eqref{FluxBalanceLaw} is the usual flux balance law satisfied by the Wald-Zoupas charge prescription \cite{Wald:1999wa}.
If there exists a spacetime realization in terms of bulk diffeomorphisms of the higher spin symmetries, it would be interesting to make contact with this prescription. 
See \cite{Rignon-Bret:2024wlu} and references therein for a recent review.

\section{Solution of the dual EOM \label{Sec:SolDualEOM}}

In this section, we study the solution to the dual EOM and give explicit solutions to all order in the $\gN$ expansion.
In particular, we introduce a Lie algebroid map $\bft$ which defines a unique $\bft(T)\in\Ao$ or $\bft(T)\in\Ah$ given an element $T\in\V(S)$.
As we will see in section \ref{sec:RenormCharge}, the knowledge of this explicit solution gives a geometrical and natural understanding of the charge renormalization procedure that was devised in \cite{Freidel:2021ytz, Geiller:2024bgf}.

We have proven that the symmetry algebroids $\Ao$ and $\Ah$, which are sub-algebroids of $\A$ are canonically represented on the gravitational phase space.
This algebroid $\A$ is filtered and it is natural to understand the associated gradation. 
This is what we now describe, in a similar fashion to section \ref{secYM:solEOM}.

\subsection{Filtration and gradation \label{sec:filtration}}

There exists a natural filtration of $\A$,
\begin{equation}\label{defFiltration}
    \{0\} \subset\A^{-1}\subset\A^0\subset \A^1\subset \ldots\subset\A^s\subset \ldots\subset \A \quad\textrm{such that}\quad \A=\bigcup_{n+1\in \mathbb{N}}\A^n,
\end{equation}
where the $\A^s\subset\A$ are the subspaces for which $\t{n} = 0$ for $n>s$.
This filtration is compatible with the bracket in the sense that
 \begin{equation}
    \lbr\A^s,\A^{s'}\rbr\subseteq \A^{s+s'-1},\quad \mathrm{when}\quad s,s'>0. \label{filtrationCompatible}
 \end{equation}
 The spin $0$ is an exception since in general we only have that  $\lbr\A^s,\A^0\rbr\subseteq \A^s$.\footnote{Indeed, if $\tau\in\A^s$ and $\tau'\in\A^0$, then $\lbr\tau,\tau'\rbr_s=\dtp\t{s}$ which does not generically vanish.}
There exists an associated graded algebroid $\G{\A}$ to the filtered algebroid $\A$, defined as
\begin{equation}
\G{\A}=\bigoplus_{s=-1}^\infty\sfT_s,
\end{equation}
with 
\begin{equation}
    \sfT_{-1}=\A^{-1}\qquad \textrm{and}\qquad \sfT_s=\A^s/\A^{s-1}\quad\textrm{for }s\geqslant 0.  \label{defgradationTh}
\end{equation}
Each $\sfT_s$ is an equivalence class and we can write $\A^s= \bigoplus_{n=-1}^s\sfT_n$.
It is easy to show that we have the isomorphism 
\vspace{-0.2cm}
\be \label{IsoQ}
\sfT_s \simeq  \V_s.
\ee 
To prove this we just look at the evolution equations for the equivalence class $[\tau]\in \sfT_s$. 
Let us assume that $\tau\in\A^s$ for $s\geq 0$.
Since $\pa_u [\tau]_s=0$,
this means that $[\tau]_s$ is constant in time and therefore equal to its value at any cut of $\scri$. 
For definiteness we choose the cut to be at $u=0$, hence we have that  
\be 
[\tau]_s=\T{s},\quad  \mathrm{where}\quad  T_s:= \tau_s|_{u=0}\in \V_s.
\ee
The other values, $[\tau]_{s-n}\neq 0$ for $1\leq n\leq s$  are  determined recursively from $[\tau]_s$ by the equation of motion. 
Describing this construction explicitly is the purpose of the next subsections.

In the following, rather than working with equivalent classes, it is more convenient to pick the natural representative element $\tau(\T{s})\in\A^s,\,s\geq 0,$ which is fully determined by $\T{s}\in\V_s$.
To do so we use that 
the isomorphism \eqref{IsoQ} means that we have a projection map $p: \A^s \to \V_s$ which maps $\tau \to T_s = \tau_s|_{u=0}$. 
This map admits an ``inverse'', i.e. a section $\tau: \V_s \to \A^s$,
\begin{align}
    \tau :\,&\V_s\to \A^s \nn\\
     &\, T_s\mapsto \tau(T_s), \label{maptau}
\end{align} 
such that  $(p \circ \tau)(T_s) = T_s$.
This section denoted $\tau(T_s)(u,z,\bz) \in \A^s$ is the unique solution of the dual EOM with initial condition given by\footnote{By definition, we also have that $\tau_{s+n}(u,z,\bz)\equiv 0$ for all $n>0$ since  $\tau(T_s)\in \A^s$.}
\be 
\tau_s(T_s)\big|_{u=0}=T_s, \qquad \tau_{s-n}(T_s)\big|_{u=0}=0,\quad \mathrm{for} \quad 1 \leq n \leq s.
\ee 
These solutions are polynomial in $u$ and thus diverge at infinity.
This is however not an issue since the charge aspects $\tQ_s\in\cS^{ch}$ are part of the \hyperref[foot1]{Schwartz space} (see also footnote \ref{foot2}).
For simplicity, we study the cut $S\equiv S(0)$, i.e. $T$ is the value of $\tau$ at the cut $u=0$ of $\scri$. The construction done here can be adapted to any other choice of cut.
In particular, a constant finite cut $S(u_0),\,u_0\in\R$ is equivalent to $S$ by translation, i.e. by replacing $u\rightarrow u-u_0$ in the upcoming analysis, for instance in \eqref{soltaup}.

Since $\tau:\V_s\to \A^s$ is a linear map, it can be extended by linearity to a map $\bft: \Vs \to \A$  on the full space using that $\Vs =\bigoplus_{s=0}^\infty \V_s$:
\begin{align} \label{maptaubis}
    \bfm{\tau} :\,& \,
    \Vs \to \A \nn\\
     & \, T \mapsto \bft(T) := \sum_{s=0}^{\infty} \tau(\T{s}).
\end{align}

The filtration for $\A$ extends to a filtration for $\Ah$ and $\Ao$ which treats the modes $\t{-1}$ differently. 
On one hand, the construction of this subsection goes along the same lines with $\Ah$, where $\t{-1}$ is treated on par with the positive degrees.\footnote{We have in particular the map $\bft:\V(S)\to\Ah$.}
On the other hand, $\t{-1}$ is determined by $\t0$ through the initial condition when dealing with $\Ao$.\footnote{The only quantity that matters to define the Noether charge is $D\t{-1}$, which precisely is purely given in terms of $\t0$.
Furthermore, the transformation of the shear $\dt C$ does not involve $\t{-1}$ at all.
Therefore, it is sufficient to consider elements $\bft(T)$ with $T\in\Vs$.}
In this case, since $(\Ao)^{-1}$ is central,\footnote{Indeed, if $\tau\in(\Ao)^s$ and $\tau'\in(\Ao)^{-1}$, then $\lbr\tau,\tau'\rbr_n=0,\,n\geqslant 0$ and $\lbr\tau,\tau'\rbr_{-1}=-\dt\tp{-1}$. 
Since $D\tp{-1}=0$, $\tp{-1}$ is field independent and $\dt\tp{-1}=0$.
Therefore $\lbr(\Ao)^s,(\Ao)^{-1}\rbr=\{0\}$.}
we can consider the quotient Lie algebroid $\Ao/(\Ao)^{-1}$ for which $\G{\Ao/(\Ao)^{-1}} =\bigoplus_{s=0}^\infty\sfT_s$.
By construction, the bracket at degree $-1$ vanishes on this quotient space (and the other degrees are unchanged). 
In particular, the super-translations commute.
This is thus the right object to consider if we want to recover the generalized BMS algebra as a sub-case of the higher spin symmetries analysis.

\subsection{Explicit construction of the section for $s\leq 4$}

To illustrate the construction of the map $\bfm{\tau}$, we start with an explicit example. 
Let us assume that $\t{n}\equiv 0,\,n\geq 5$. 
Since $\pa_u\t{s}=D\t{s+1}-(s+3)C\t{s+2}$, we deduce that $\t4$ is constant in time.
By taking\footnote{We anchor the inverse operator at the cut $0$ where the initial condition is set up.} $\pui=\int_0^u$, we can then write $\t{s},\,s=0,1,2,3,4$ as follows:
\bs
\label{soltau43210}
\begin{align}
    \t4(T) &=\T4, \\
    \t3(T) &=\T3+uD\T4, \\
    \t2(T) &=\T2+uD\T3+\frac{u^2}{2}D^2\T4-5(\pui C)\T4, \\
    \t1(T) &=\T1 +uD\T2+\frac{u^2}{2}D^2\T3+\frac{u^3}{3!}D^3\T4-5\pui D\big((\pui C)\T4\big)-4\pui \big(C(\T3+uD\T4)\big), \\
    \t0(T) &=\sum_{k=0}^4\frac{u^k}{k!}D^k\T{k}-5(\pui D)^2\big((\pui C)\T4\big)-4(\pui D)\pui\big(C(\T3+uD\T4)\big) \nn\\
    &-3\pui\left(C\left(\T2+ uD\T3+\frac{u^2}{2}D^2\T4 -5(\pui C)\T4\right)\right).
\end{align}
\es
Now that the pattern is clear, we write a general solution for $\tau\in \A^s$.

\subsection{General solution}

When $\tau\in\A^s$, then $\dot\tau_s\equiv 0$ and we can solve the dual EOM \eqref{tausevolb} recursively.
The solution which satisfies $\t{p}(u=0,z,\bz)= 0$ for $p > s$ and
\begin{equation}
    \t{p}(u=0,z,\bz)=\T{p}(z,\bz), \quad 0\leqslant p\leqslant s,
    \label{intconstant}
\end{equation}
is given by $\t{s}=T_s$, $\t{s-1}= \T{s-1} + uD T_s$ and in general by
\begin{equation}
    \boxed{\t{s-p}(T)=\sum_{k=0}^{p}\frac{u^{p-k}}{(p-k)!}D^{p-k}T_{s-k}- \sum_{k=0}^{p-2}(s-k+1)\big(\pui D\big)^{p-k-2} \pui(C\t{s-k})\in\A_s\,}, \label{soltaup}
\end{equation}
for $2\leqslant p\leqslant s$.
These relations show that $\tau_{s-p}$ is defined recursively in terms of $\tau_s,\tau_{s-1}, \ldots, \tau_{s-p +2 }$ and $T_s, \ldots, T_{s-p}$.
It also shows that $\tau_{s-p}$ is a polynomial in the shear $C$ and its derivatives of degree $\lfloor p/2\rfloor$.
Solving \eqref{soltaup} can then be done recursively starting from the expression for  $\t{s}$, then $\t{s-1}$, $\t{s-2}$ and so on, as we did in the warm-up.
Doing so, \eqref{soltaup} is an explicit solution of \eqref{tausevolb}.
The reader can readily check that  this recursive definition satisfies the 
dual equation of motion
\be 
\dot\tau_{s-p}(T) &=\sum_{k=0}^{p-1}\frac{u^{p-k-1}}{(p-k-1)!}D^{p-k}\T{s-k}-\sum_{k=0}^{p-2}(s-k+1)\big(\pui D\big)^{p-k-2} (C\t{s-k})\cr
& =D\t{s-p+1}(T)-(s-p+3)C\t{s-p+2}(T).
\ee

Because when pairing $\t{s}$ with $\tQ_s$ the initial condition \eqref{tauintial} amounts to a change of the spin-0 charge from $\tQ_0$ to $\tq_0:=\tQ_0-\bN C$---cf. \eqref{bondimass}---we do not need to solve for the parameter $\t{-1}$ when we restrict to $\tau(T)\in\Ao$. On the other hand, if we consider $\tau(T)\in\Ah$, then we just have to extend the range of $p$ to $2\leqslant p\leqslant s+1$ in the solution \eqref{soltaup}.

Next, let us choose a solution for which only $\T{s}\neq 0$.
We get that\footnote{Recall that $\bfm{\tau}(T)=\sum_s\tau(\T{s})$, where the sum runs over all the values of $s$ for which $\pi_s(T)\neq 0$. \label{foot:linearity}}
\begin{equation}
    \boxed{\t{s-p}(\T{s}) =\frac{u^{p}}{p!}D^{p}\T{s}-\sum_{k=0}^{p-2}(s-k+1)\big(\pui D\big)^{p-k-2}\pui (C\t{s-k})},\qquad p\geqslant 0. \label{soltaupTs}
\end{equation}
By construction, this $\tau(\T{s})$ is precisely a representative element of $\sfT_s$,\footnote{Notice however a subtlety in the fact that the gradation depends on the cut $S(U)$ of $\scri$, so that to be precise, we should have written $\mathcal{G}_{U}(\A)$ in section \ref{sec:filtration}.}
which makes explicit the aforementioned isomorphism with $\V_s$, since for a given shear $C$, $\tau(\T{s})$ is fully determined by $\T{s}\in\V_s$.

The form \eqref{soltaupTs} of the `polynomial class' of solutions of \eqref{tausevolb} is handy for section \ref{SecComparison}.
However, we also present a systematic approach in Sec.\,\ref{Sec:systApproach}, more convenient for Sec.\,\ref{sec:RenormCharge}.

\subsection{Systematic approach \label{Sec:systApproach}}

First of all, we introduce $\hs$ and $\hS$, respectively the spin and shift operators, acting on the series $\tau\in \tV(\scri)$ as\footnote{The same formulas are valid for any $\lambda_s\in \Ccar{\delta,-s}$ or $\lambda_s\in\Ccel{\Delta,-s}$.}
\begin{equation}
    \hs \tau_s =s\t{s}\qquad \textrm{and} \qquad 
    \hS\t{s}=\t{s+1}.
\end{equation}
Hence $\hs$ is an operator of degree $(0,0)$ and $\hS$ is an operator of degree $(0, 1)$. 
The usage of $\hS$ turns out to be convenient for any systematic approach.
Furthermore, $\hs$ and $\hS$ satisfy the following commutation relations with $D$ and $C$:\footnote{Notice that from this point of view, $C$ acts as an operator that shifts the degree by $-2$ and multiplies the graded vector by its ``eigenvalue'' $C$.}
\begin{equation}
    \hS D=D\hS,\quad\hS C=C\hS,\quad D\hs=(\hs+1)D,\quad\hs C=C(\hs-2),\quad\hs\hS=\hS(\hs+1).
\end{equation}
This formalism then allows us to write $\Dcal$ as follows:
\begin{equation}
    \Dcal \tau=\big(D-C (\hs+1)\big)\tau, \label{DcalOp}
\end{equation}
which in particular means that
\begin{align}
    (\Dcal \tau)_s
    =(D\tau)_s-(\hs+3 )(C\tau)_{s}=\big(D-C(\hs+1)\hS\big)\t{s+1}.
\end{align}
Hence $(\Dcal^n \tau)_s$ takes the compact form
\begin{equation}
    (\Dcal^n \tau)_s = \big(D-C(\hs+1)\hS\big)^n\t{s+n}.
\end{equation}

Next, we recast the dual evolution equation \eqref{tausevolb} as follows:
\begin{equation}
    \boxed{\tau=\pui\Dcal\tau +T}. \label{solEOMcompact}
\end{equation}
More concretely,
\begin{equation}
    \t{s}=\big(\pui D-\pui C(\hs+1)\hS\big)\t{s+1}+\T{s},
\end{equation}
where it is understood that $\pui$ acts on all products on its right.
Therefore, when looking for $\t{s-1}$, $\t{s-2},\ldots$ and so on, we see that formally, one does nothing more than taking an extra power of the operator $\pui\Dcal$ at each step.
Concretely, picking $\tau\in\A^s$, we can write
\begin{equation}
    \t{s-n}(T) =\sum_{k=0}^n
    \left(\pui D-\pui C(\hs+1)\hS
    \right)^k T_{s-n+k},
\end{equation}
or equivalently
\begin{equation}
    \t{n}(T)=\sum_{k=0}^{s-n}\left(\pui D-\pui C(\hs+1)\hS\right)^k\T{n+k}.
\end{equation}
At that stage, this solution is still very formal.
As we did in \eqref{soltaupTs}, the next step is to find a representative element of $\sfT_s$, i.e. we take only $\T{s}\neq 0$.
By inspection, we find that\footnote{The key to find the lower bound in the sum over $k$ is to realize that for each term in the sum, the shift operator acts between 0 and $k$ times.
Each term at a definite $k$ is thus a linear combination of $\T{n+k},\ldots,\T{n+2k}$.
$k$ needs to be sufficiently large, so that $n+2k\geqslant s$.
We therefore have to take the integer part of $\frac{s-n+1}{2}$.}
\begin{equation}
    \boxed{\t{n}(\T{s})=\sum_{k=\lfloor\frac{s+1-n}{2}\rfloor}^{s-n}\left(\pui D-\pui C(\hs+1)\hS\right)^k\T{n+k}}. \label{soltaunTs}
\end{equation}
Notice that we have to deal with an operator of the form $(A-B)^k$, where in our case $B$ is linear in the shear and in $\hS$.
A good way to organize the solution is thus to make patent this expansion in powers of $C$.
For this, we can use the following formula:
\begin{equation}
    (A-B)^k=\sum_{\l=0}^k(-1)^\l\sum_{P=k-\l}A^{p_0}BA^{p_1}BA^{p_2}\ldots BA^{p_\l},\qquad\textrm{with}\quad P=\sum_{i=0}^\l p_i.
\end{equation}
Therefore,
\begin{align}
    &\left(\pui D-\pui C(\hs+1)\hS\right)^k = \label{puiDcalk}\\
    &=\sum_{\l=0}^k(-1)^\l\sum_{P=k-\l}\pui[p_0]D^{p_0}\bigg(\pui\Big\{C(\hs+1)\pui[p_1]D^{p_1}\Big(\pui\Big\{C(\hs+1)\pui[p_2]D^{p_2}\Big(\ldots \nn\\
    &\qquad\qquad\qquad\ldots \pui\Big\{C(\hs+1)\pui[p_{\l-1}]D^{p_{\l-1}}\Big(\pui\Big\{C(\hs+1)\pui[p_\l]D^{p_\l}\hS^\l\Big\}\Big)\Big\}\ldots \Big)\Big\}\Big)\Big\}\bigg). \nn
\end{align}
For visual clarity, we used round and curly parentheses to distinguish (when necessary) between the action of $D$ versus $\pui$.

This way \eqref{soltaunTs} and \eqref{puiDcalk} of writing the solution of \eqref{tausevolb} is particularly convenient for the discussion about the renormalized charges in section \ref{sec:RenormCharge}.

This result is also an opportunity to show how algorithmic the construction is and how concise the results and the notation we have been using are.
Indeed, $\dt C$ is very simply written in terms of $\t0,\t1$ and $\t2$---cf. \eqref{deltaC}---but the amount of complexity hidden in the latter is tremendous.
Having proven that $\dt$ is a representation of the symmetry algebroid $\Ao$ on the phase space---and here we want to emphasize that $\tau$ contains all the symmetry parameters $\T{s}$ and arbitrarily high powers of $C$---is clearly a non-perturbative result.
To be more accurate, recall that if we formally introduce $\gN$ and $\bgN$ as small dimensionless parameters, then by rescaling $C\rightarrow\gN C$ and $\bC\rightarrow\bgN\bC$, our result is non perturbative in $\gN$ and at leading order in $\bgN$.
We can use $\gN$ and $\bgN$ as a bookkeeping device.

We now turn to the next section, where we exploit the machinery developed so far in order to clarify the notion of renormalized charge introduced in \cite{Freidel:2021dfs, Freidel:2021ytz} and partially corrected in \cite{Geiller:2024bgf,Kmec:2024nmu}.

\section{Renormalized charge and its action on the shear \label{sec:RenormAction}}

In this section we show that the charge renormalization procedure devised in \cite{Freidel:2021ytz, Geiller:2024bgf}, amounts to evaluate the smearing variable at the cut $u=0$.
We then compute the action of the renormalized charges at quadratic order onto the shear to show that we recover the results of the aforementioned papers.

\subsection{Renormalized charge aspect \label{sec:RenormCharge}}

The goal of this subsection is to recast the Noether charge $\Qt$, when $\tau=\bft(\T{s})$, as the integral over the sphere of a certain charge aspect $\hq_s(z,\bz)$ smeared against the symmetry parameter $\T{s}$ that defines the $\tau(T_s)$.

As we did in section \ref{Sec:SolDualEOM}, let us start with an example for the lowest spin-weights.
Consider successively $\tau=\bft(\T{s})$ for $s=0,1,2,3,4$.
We denote
\begin{equation}
    \cq^u_{\T{s}}\equiv Q^u_{\bft(\T{s})}=\sum_{n=-1}^sQ^u_n\big[\t{n}(\T{s})\big].
\end{equation}
Using the explicit computation \eqref{soltau43210}, we deduce that
\bs
\begin{align}
    (4\pi \GN)\cq^u_{\T0} &=\int_S\big(\tQ_0-\bN C\big)\T0\equiv \int_S\tq_0\T0, \\
    (4\pi \GN)\cq^u_{\T1} &=\int_S\left(\tq_0 uD\T1+\tQ_1\T1\right)=\int_S\big(-uD\tq_0+\tQ_1\big)\T1, \\
    (4\pi \GN)\cq^u_{\T2} &=\int_S\left(\tq_0\left(\frac{u^2}{2}D^2\T2-3\T2\pui C\right)+\tQ_1 uD\T2+\tQ_2\T2\right) \nn\\
    &= \int_S\left(\frac{u^2}{2}D^2\tq_0-uD\tQ_1+\tQ_2-3\tq_0\pui C\right)\T2,
\end{align}
\begin{align}
    (4\pi \GN)\cq^u_{\T3} &=\int_S\left(\tq_0\left(\frac{u^3}{3!}D^3\T3-4D\big(\T3\pui[2]C\big)-3D\T3\pui(uC)\right)\right. \nn\\
    &\qquad+\left.\tQ_1\left(\frac{u^2}{2}D^2\T3-4\T3\pui C\right)+\tQ_2 uD\T3+\tQ_3\T3\right) \nn\\
    &=\int_S\left(-\frac{u^3}{3!}D^3\tq_0+4D\tq_0\pui[2]C+3D\big(\tq_0\pui(uC)\big)\right. \\
    &\,\hspace{3cm}\left.+\frac{u^2}{2}D^2\tQ_1-4\tQ_1\pui C-uD\tQ_2+\tQ_3\right)\T3, \nn
\end{align}
\es
and similarly for $\cq^u_{\T4}$.
We can thus identify the charge aspects
\bs
\label{renormcharge}
\begin{align}
\tq_0 &\equiv q_0-\bN C, \\
\tq_1 &\equiv q_1+uD(\bN C), \\
\tq_2 &\equiv q_2-\frac{u^2}{2}D^2(\bN C)-3\tq_0\pui C, \\
\tq_3 &\equiv q_3+\frac{u^3}{3!}D^3(\bN C)+4D\tq_0\pui[2]C+3D\big(\tq_0\pui(uC)\big)-4\tQ_1\pui C, \\
\tq_4 &\equiv q_4-\frac{u^4}{4!}D^4(\bN C)-5D^2\tq_0\pui[3]C-4D\big(D\tq_0\pui[2](uC)\big)-\frac32 D^2\big(\tq_0\pui(u^2C)\big) \nn\\
& +5D\tQ_1\pui[2]C+ 4D\big(\tQ_1\pui(uC)\big)-5\tQ_2\pui C +15\tq_0\pui\big(C\pui C\big), \label{tq4}
\end{align}
\es
where 
\be 
q_s =\sum_{n=0}^s \frac{(-u)^n}{n!} D^n \tQ_{s-n}. \label{defqs}
\ee 
The reader can check that the $\tq_s$ defined by \eqref{renormcharge} are conserved when no radiation is present.
More precisely, $\pa_u\tq_s$ involves only terms that contain $\bN$---see App.\,\ref{App:tq4} for the explicit computation of $\pa_u\tq_4$.
We dub $\tq_s$ the \emph{renormalized charge} aspect of helicity $s$.\footnote{Such a renormalization procedure depends on the choice of cut $S(U)$.
For simplicity, as we did in section \ref{Sec:SolDualEOM}, we present the discussion associated with the cut $S\equiv S(0)$.}
Up to the term\footnote{This term is also just the consequence of swapping $\tQ_0$ for $\tq_0$ in the definition \eqref{defqs} of $q_s$.} $(-1)^{s+1}\frac{u^s}{s!}D^s(\bN C)$ and to the fact that the Bondi mass $\tq_0$ enters in our expressions rather than the covariant mass $\tQ_0$, the expressions for the spins $0,1,2$ were first given in \cite{Freidel:2021dfs} while the form of $\tq_3$ was first given in \cite{Geiller:2024bgf}.
The fact that the renormalized charges $\tq_s$ only involve the combination $\tQ_0-\bN C$, and never the charge $\tQ_0$ alone is perfectly natural in our construction, since this is a direct consequence of the initial constraint \eqref{tauintial}, cf. \eqref{bondimass}.
The expression $\tq_4$ was also partially given  in \cite{Geiller:2024bgf}: the part linear in $(C,\tQ)$ is correctly reproduced by their equation (6.18); however the term of higher order, namely
\begin{equation}
    15\tq_0\pui\big(C\pui C\big), \label{cubicUs}
\end{equation}
is  incorrectly written as
\begin{equation}
    \frac{15}{2}\tQ_0(\pui C)(\pui C) \label{cubicMarc}
\end{equation}
in \cite{Geiller:2024bgf}.
Notice however that the time derivative of \eqref{cubicMarc}, namely
\begin{equation}
    \pa_u\left(\frac{15}{2}\tQ_0(\pui C)(\pui C)\right)= 15\tQ_0C\pui C+\frac{15}{2}\pa_u\tQ_0(\pui C)(\pui C),
\end{equation}
equals the time derivative of \eqref{cubicUs}, namely
\begin{equation}
    \pa_u\big(15\tq_0\pui(C\pui C)\big)=15\tq_0C\pui C+15\pa_u\tq_0\pui(C\pui C),
\end{equation}
up to terms that vanish when there is no radiation.
We understand that fact as an indication of how one can easily be misled in the construction of the renormalized charges.
Two terms can have the same behavior upon time derivation, but actually lead to a different action on the phase space.
Without a general procedure, it is incredibly cumbersome to work out the correct notion of $\tq_s$ for arbitrary $s$ and at arbitrary order in $\gN$.
The fact is, our construction precisely realizes this for free.
It guarantees that the charges form an algebra and is non-ambiguous when it comes to defining the renormalized charge.
Besides, as we mentioned already, our algorithm (understand the systematic usage of the dual EOM) is blind to the peculiar value of helicity one wishes to consider or to the order in $\gN$ one wishes to work at.
\medskip

Summarizing the results so far, we showed that 
\begin{equation} \label{defcqTsu}
    \cq^u_{\T{s}}\equiv Q^u_{\bft(\T{s})}=\frac{1}{4\pi \GN}\sum_{n=-1}^s \int_S\tQ_n\t{n}(\T{s})=\frac{1}{4\pi \GN}\int_S\tq_s\T{s},\qquad s=0,\ldots,4.
\end{equation}
We are thus ready to state the following theorem.

\begin{tcolorbox}[colback=beige, colframe=argile]
\textbf{Theorem [Noether charge of spin $s$]}\\
The Noether charge $\cq_{\T{s}}$ associated to the symmetry parameter $\T{s}$ of helicity $s$ is written as the following corner integral:
\begin{equation}
    \cq_{\T{s}}\equiv \cq^{-\infty}_{\T{s}}=\frac{1}{4\pi \GN}\int_S\hq_s\T{s}, \qquad s\geqslant 0, \label{Noetherhatqs}
\end{equation}
with $ \cq_{\T{s}}^u$ given by \eqref{defcqTsu} for all $s\in\N$ and
\begin{equation}
    \hq_s(z,\bz)=\lim_{u\to -\infty}\tq_s(u,z,\bz).
\end{equation}
The renormalized charge aspect $\tq_s$ satisfies $\pa_u\tq_s=0$ in the absence of radiation $\bN=0$.
\end{tcolorbox}

\paragraph{Proof:}
We present the way to construct $\tq_s$ systematically in the appendix \ref{App:renormcharge}, cf. \eqref{deftqs}, where we also show that the renormalized charge $\th_s$ associated to the covariant charge $\Ht$ matches with the one recently proposed in \cite{Kmec:2024nmu}.
To show that $\pa_u\tq_s=0$ in the absence of left-handed radiation, simply notice on the one hand that, cf. \eqref{Qtdot},
\begin{equation}
    \pa_u \cq^u_{\T{s}} =-\frac{1}{4\pi \GN}\int_S\bN\delta_{\tau(\T{s})}C=0 \qquad\textrm{if}\qquad\bN=0,
\end{equation}
while on the other hand, we also have that
\begin{equation}
    \pa_u \cq^u_{\T{s}} =\frac{1}{4\pi \GN}\pa_u\left(\int_S\tq_s\T{s}\right)= \frac{1}{4\pi \GN}\int_S\pa_u\tq_s\T{s}.
\end{equation}
This concludes the proof.

\paragraph{Remark:}
In a non-radiative strip of $\scri$, where $N=0$ for $u\in[0,u_0]$, we get that
\begin{equation}\label{deftqs0rad}
    \boxed{\tq_s=\sum_{k=0}^s\frac{(-u)^k}{k!}\big(\Dcal^{\ast k}\tQcal\big)_s}.
\end{equation}
where $\tQcal_s\equiv\tQ_s$ for $s >0$; $\tQcal_0\equiv\tq_0=\tQ_0-\bN C$ and $\tQcal_s\equiv 0$ for $s< 0$; while $\big(\Dcal^*\tQ\big)_s=D\tQ_{s-1}+(s+1)C \tQ_{s-2}$. 
We present the proof of \eqref{deftqs0rad} at the end of App.\,\ref{App:renormcharge}.
We can easily check that this renormalized charge aspect is conserved in the strip if we also assume that $\bN=0$ since
\begin{equation}
    \pa_u\tq_s=-\sum_{k=1}^s\frac{(-u)^{k-1}}{(k-1)!}\big(\Dcal^{\ast k}\tQcal\big)_s +\sum_{k=0}^s\frac{(-u)^k}{k!}\big(\Dcal^{\ast k+1}\tQcal\big)_s=\frac{(-u)^s}{s!}\big(\Dcal^{\ast s+1}\tQcal\big)_s=0.
\end{equation}
In the last step, we used that $\big(\Dcal^{\ast s+1}\tQcal\big)_s$ only contains $\tQcal_s$ for $s<0$.

\subsection{Soft and Quadratic actions for arbitrary spin $s$ \label{SecComparison}} 

We now study the action $\dt C$ in details at quadratic order and show that we recover the result from \cite{Freidel:2021ytz} and \cite{Geiller:2024bgf}.

The Noether charge $\cq_{\T{s}}$, or equivalently the charge aspect $\hq_s$, depends linearly on $\bN$ and polynomially\footnote{In a generalized sense since the coefficients of the polynomial can be differential operators.} on $C$. 
More precisely we can decompose $\hq_s$ for $s\geqslant -2$  as
\be \label{hqsk}
\hq_s =\sum_{k=0}^{\lfloor s/2\rfloor+1} \q{s}{k},
\ee 
where $\q{s}{k}$ is homogeneous of degree $k$ in $C$ and linear in $\bN$.\footnote{As a reminder, this amounts to the expansion in terms of the coupling constant $\gN$ (while all charges are at leading order in $\bgN$).}
The charge aspect $\q{s}{0}\equiv\soft{\hq_s}$ is the soft charge while $\q{s}{1}\equiv\hard{\hq_s}$ is the hard charge and $\sum_{k=2}^{\lfloor s/2\rfloor+1} \q{s}{k}$ is the super-hard contribution.

In this section, we compute the soft and hard action for any spin. 
Actually \eqref{deltaC} allows us to predict much more than just the quadratic action, but for now we shall focus mostly on the latter and show how the action \eqref{deltaC} combined with the dual EOM \eqref{tausevolb} contain all the information previously obtained after tedious computations. 
Nevertheless, even if we let for further work the study of the super-hard action, we emphasize that the closure of the algebra generated by $\Qt$---in full generality, i.e. without relying on any sort of soft or hard truncation---is a huge achievement and a highly non-trivial consistency check that \eqref{deltaC} bears a fundamental status.

In order to classify the action as a function of the helicity, we define 
\begin{equation}
    \d{s}{\T{s}}C :=\big(\dt C\big)\big|_{\tau=\bft(\T{s})}= \poisson{\cq_{\T{s}},C}, \label{defdeltaTs}
\end{equation}
where $\bft(\T{s})$ is the solution \eqref{soltaupTs}.
We introduce a superscript $\d{s}{}$, which refers to the helicity of the charge the transformation is associated with.
We also use it to emphasize that $\d{s}{\T{s}}\neq\dT$ for $T=(0,\ldots,0,\T{s},0,\ldots)$.

\paragraph{Warm-up: Spin 0, 1 and 2:}

Before giving a general proof, we start with a warm-up and focus on the action of super-translations (spin 0), sphere diffeomorphisms (spin 1) and the helicity 2 charge. 
This was one of the results of \cite{Freidel:2021dfs}.
Using the solutions \eqref{soltau43210},
\begin{subequations}
\label{deltaC012}
\begin{align}
\d{0}{\T0} C&=-\big(D^2\T0\big)+\big(N\T0\big), \\
\d{1}{\T1} C&=-\big(uD^3\T1\big)+\big(uND\T1+3CD\T1+2DC\T1\big), \label{deltaYC}\\
\d{2}{\T2} C&= -\left(\frac{u^2}{2}D^4\T2\right)+ \left(\frac{u^2}{2}N D^2\T2+3D^2\big(\T2\pui C\big)+2 uDCD\T2+3uCD^2\T2\right) \nn\\
&\quad -\big(3N\T2\pui C +3 C^2 \T2\big),
\end{align}
\end{subequations}
where we have highlighted the soft, hard and super hard (in the spin 2 case) actions with the parentheses.
The reader can directly compare the 3 lines of \eqref{deltaC012} with the equations (59), (75) and (89) from \cite{Freidel:2021dfs}: they are identical upon changing $\T0\to T$, $\T1\to Y/2$, $\T2\to Z/3$ and $C\to C/2$.
These transformations were confirmed in \cite{Geiller:2024bgf}, cf. formula (6.14) upon changing $\T0\to T$, $\T1\to \overbar{\mathcal Y}$, $\T2\to Z$ and $C\to -C$.

In order to facilitate the comparison with the expressions of \cite{Freidel:2021ytz} this time, we rewrite \eqref{deltaC012} in this equivalent way:
\begin{subequations}
\label{deltaC012bis}
\begin{align}
\d{0}{\T0} C&=-\big(D^2\T0\big)+\big(\T0\big)\pa_u C, \\
\d{1}{\T1} C&=-\big(uD^3\T1\big)+\big(D\T1(u\pa_u+3)+2\T1 D\big)C, \\
\d{2}{\T2} C&= -\left(\frac{u^2}{2}D^4\T2\right)+ \left(D^2\T2\Big(\frac{u^2}{2}\pa_u^2+3u\pa_u+3\Big) +2 D\T2 D \big(u\pa_u+3\big)+3\T2 D^2\right)\pui C \nn\\
&\quad -\left(3\T2\pa_u(C\pui C)\right).
\end{align}
\end{subequations}
We emphasized that the hard action $\Hd{p}{\,\cdot}C$ can be viewed as a differential operator acting on $\pa_u^{1-p}C$. We will see that this property holds for any $p$ and not only for $p=0,1,2$ as we showed explicitly in this warp-up.

The transformation $\dt C$ \eqref{deltaC}, combined with the dual EOM \eqref{tausevolb}, thus reproduces the correct action of the charges of spin 0, 1 and 2 on the gravitational phase space. 

\paragraph{General case:}

We then focus on the soft and quadratic actions for arbitrary values of $s$, namely\footnote{$\soft{\cq_{\T{s}}} =\frac{1}{4\pi \GN}\int_S\soft{\hq_s}\T{s}$ and similarly for $\hard{\cq_{\T{s}}}$.}
\begin{equation}
    \boxed{\poisson{\soft{\cq_{\T{s}}},C} =\Sd{s}{\T{s}} C=-D^2\soft{\t0}(\T{s})} \label{deltatauSoft}
\end{equation}
and
\begin{equation}
    \boxed{\poisson{\hard{\cq_{\T{s}}},C}=  \Hd{s}{\T{s}}C= -D^2\hard{\t0}(\T{s})+N\soft{\t0}(\T{s})+ 2DC\soft{\t1}(\T{s})+3CD\soft{\t1}(\T{s})}. \label{deltatauHard}
\end{equation}
Before stating the lemmas, we define yet another piece of notation, namely 
\begin{equation}
    \td{s}{\T{s}}C\equiv\deg^u_0\big(\d{s}{\T{s}}C\big), \label{deftildedeltaTs}
\end{equation}
where by $\deg^u_0(O)$, we mean the coefficient of the term $u^0$ in the expression $O$ (when the latter is a polynomial in $u$).

\begin{tcolorbox}[colback=beige, colframe=argile]
\textbf{Lemma [Soft action]}\\
The soft action for arbitrary spin $s\in\N$ is given by
\begin{equation}
    \Sd{s}{\T{s}}C=-\frac{u^s} {s!} D^{s+2}\T{s}.
\end{equation}
\end{tcolorbox}

\ni Notice that $\Sd{s}{\T{s}}C=\frac{u^s}{s!}\Std{0}{D^s\T{s}}C$, where $\Std{0}{\T0}C=-D^2\T0$.
This means that the soft transformation of spin $s$, parametrized by the tensor $\T{s}$, is nothing more than the soft part of a super-translation parametrized by the tensor $D^s\T{s}$ (times the $u$ dependence $\frac{u^s}{s!}$).

\begin{tcolorbox}[colback=beige, colframe=argile]
\textbf{Lemma [Hard action][Part 1]}\\
The hard action for arbitrary spin $s\in\N$ takes the form
\begin{subequations}
\label{deltatildeHard}
\begin{equation}\label{deltatildeHardbb}
    \Hd{s}{\T{s}}C=\sum_{p=0}^s\frac{u^{s-p}}{(s-p)!}\Htd{p}{D^{s-p}\T{s}}C,
\end{equation}
where
\begin{equation}
    \Htd{p}{\T{p}}C=\sum_{k=0}^{\alpha_p}(-1)^k\frac{(p-2)_k}{k!}(p-k+1) D^{p-k}\big(D^k\T{p}\,\pa_u^{1-p}C\big), \label{deltatildeHardb}
\end{equation}
\end{subequations}
with $(x)_k=x(x-1)\ldots(x-k+1)$ the falling factorial, $\alpha_p=\max[\mathrm{mod}_2(p),p-2]$ where $p$ modulo 2 dominates if $p=0,1$.
\end{tcolorbox}

\ni Finally, the formula \eqref{deltatildeHardb} can also be written in another useful form:

\begin{tcolorbox}[colback=beige, colframe=argile]
\textbf{Lemma [Hard action][Part 2]}\\
\begin{equation}
    \Htd{p}{\T{p}}C=\sum_{k=0}^{\min[3,p]}\binom{3}{k}(p+1-k)D^k\T{p}\,D^{p-k}\pa_u^{1-p}C. \label{deltatildeHardbis}
\end{equation}
\end{tcolorbox}

\ni We present the proofs of these lemmas in Appendix \ref{AppSoftHardAction}.
The formulae \eqref{deltatildeHardbb} and \eqref{deltatildeHardbis} match with the equations (73) and (74) of \cite{Freidel:2021ytz}.

\paragraph{Remark:}
In a non-radiative strip of $\scri$, where $N=0$ for $u\in[0,u_0]$, we have that \eqref{soltaunTs} reduces to
\begin{equation}\label{deftn0rad}
    \t{n}(\T{s})=\sum_{k=\lfloor\frac{s+1-n}{2}\rfloor}^{s-n}\frac{u^k}{k!}\big(\Dcal^kT\big)_n,
\end{equation}
which implies that
\begin{equation}
    \d{s}{\T{s}}C=-\sum_{k=\lfloor\frac{s-1}{2}\rfloor}^{s}\frac{u^k}{k!}\big(\Dcal^{k+2}T\big)_{-2}.
\end{equation}
Since $\d{s}{\T{s}}C=\poisson{\cq_{\T{s}}, C}$, this expression parallels \eqref{deftqs0rad}.

\section{Algebroid section of the $\Wcal_\sigma$-algebra \label{Sec:algebroidTs}}

We now study the map $\bft$ in more details.
In particular, we prove that it is a Lie algebroid isomorphism.
Let us start with some preliminary calculations which follow from the previous sections.

Given $\tau,\tau'\in\sfTh$, we have seen in section \ref{sec:TimeEvol} that $\lbr\tau,\tau'\rbr\in\sfTh$ then satisfies the dual EOM.
It is therefore in the image of $\bft$ and we can write that
\begin{equation}
    \big\lbr\bfm{\tau}(T),\bfm{\tau} (T')\big\rbr\equiv\bfm{\tau}\big(\Tpp{}(T,T') \big)\in\sfTh. \label{tauppTTp}
\end{equation}
Indeed, since the LHS is fully determined by $T$ and $T'$, so should be the RHS.
We are thus looking for the symmetry parameter $T''$ determined in terms of $T$ and $T'$ such that $\bfm{\tau}(T'')$ matches with $\big\lbr\bfm{\tau}(T),\bfm{\tau} (T')\big\rbr\in\sfTh$.
By evaluating \eqref{tauppTTp} at $u=0$, we are then able to deduce this associated symmetry parameter $\Tpp{}$.
By construction of the map $\bft$ as a solution of the dual EOM, we have that $\bft(T)\big|_{u=0}=T$.
It is then clear that
\begin{equation}
    \big[\bfm{\tau}(T),\bfm{\tau} (T')\big]^C\Big|_{u=0}=[T,T']^\sigma,\qquad \sigma:=C\big|_{u=0}. \label{CsigmaBracket}
\end{equation}
Besides, using the concise form \eqref{solEOMcompact} plus the fact that $\pui=\int_0^u$ so that $\big(\pui(\ldots)\big)\big|_{u=0}=0$, we have that\footnote{Had we considered $\bft(T)\in\sfT$, then $\bft_{-1}(T)$ would not be defined (where $\bft_s\equiv\pi_s \circ\bft$).}
\begin{equation}
    \Big(\delta_{\bft(T)}\bft(T')\Big) \Big|_{u=0}=\Big(\pui\big( \delta_{\bft(T)}\big(\Dcal\bft(T')\big)\big)+\delta_{\bft(T)}T'\Big) \Big|_{u=0}=\dT T', \label{dtauTtauTp}
\end{equation}
with $\dT \sigma = \nu\T0  + \hdT \sigma$, cf.\,\eqref{dTCv1}, and $\nu:=N|_{u=0}$,
so that $\dT\sigma$ is indeed $\big(\delta_{\bft(T)} C\big)\big|_{u=0}$.
\smallskip

We thus define the Lie algebroid bracket onto the sphere, denoted by $\lbr \cdot\,,\cdot \rbr^\sigma$:
\begin{equation}
    \boxed{~\big\lbr T,T'\big\rbr^\sigma= \big[T,T'\big]^\sigma + \dTp T-\dT T'~}. \label{TbracketSphere}
\end{equation}
The space $\big(\V(S),\lbr \cdot\,,\cdot\rbr^\sigma, \delta\big)$ is clearly an algebroid over $S$ since it results from the projection of the $\Ah$-algebroid at the cut $u=0$ of $\scri$.
Since we know that $\Ao$ is also an algebroid, we can similarly consider the projection coming from $\Ao$, for which it is guaranteed that $D\lbr T, T'\rbr^\sigma_{-1} =\sigma\lbr T, T' \rbr^\sigma_0$.\footnote{In that case, the map $\bft_{-1}(T)$ is implicitly defined via the constraint $D\bft_{-1}(T)=C\bft_{0}(T)$. We never need more than this relation.}
The previous results (\ref{tauppTTp}-\ref{dtauTtauTp}) show that the map $\bft$ is an isomorphism of Lie algebroids, i.e.
\be  \label{IsoAlgebroids}
\boxed{~\big\lbr \bfm{\tau}(T), \bft(T') \big\rbr =  \bft\big( \lbr T,T'\rbr^\sigma  \big)~}.
\ee 
Besides, $\bft$ maps the base space $S$ of the sphere algebroid, to the base space $\scri$ of $\Ah$ (or $\Ao$), via the inclusion map at a constant $u$ cut $\iota_S:S\to S(0)\subset\scri$.
We let for future work the careful discussion of an arbitrary cut $S(U)$; cf. also section \ref{sec:ArbitraryCut}.

\paragraph{Remark:} Why did we not name $\big(\V(S),\lbr \cdot\,,\cdot\rbr^\sigma, \delta\big)$ as $\A_\sigma$?
The subtlety comes from the transformation property of $\nu$.
Knowing about the $\A$-algebroid structure over $\scri$ (here we use $\A$ to refer irrespectively to $\Ah$ or $\Ao$), we have by construction that 
\begin{align}
    \dT\nu:=\big(\delta_{\bft(T)}N\big) \big|_{u=0} &=\Big(\delta_{\dot{\bft}(T)}C+ \bft_0(T) \dot N+2DN\bft_1(T)+3ND \bft_1(T)-6CN\bft_2(T)\Big)\Big|_{u=0} \nn\\
    &=\delta_{\Dcal T}\sigma+\Cc2\T0+2D\nu\T1+3\nu D\T1-6\sigma\nu\T2, \label{dTnu}
\end{align}
where $\Cc2:=\big(\pa_u N\big)\big|_{u=0}$ and we use the fact that $\pa_u\bft(T) =\Dcal\bft(T)$.
The important feature to notice here is that while the $\A$-algebroid is determined by only one parameter, namely the shear $C$, we expect its projection at a cut to depend on $\bfm{\sigma}:= \{\Cc{n}\}_{n=0}^\infty$, where $\Cc{n} \equiv \big(\pa_u^n C\big)\big|_{u=0}$.
We know that onto $\scri$ the action of the anchor map on $\pa_u^n C$ is given by $\pa_u^n(\dt C)$. 
This defines recursively $\dT\Cc{n}$  from the sphere viewpoint.
Repeating the computation \eqref{algebroidRepinterm} for the bracket $\lbr\cdot\,,\cdot\rbr^\sigma$, we find that $\dT\nu\equiv\dT\Cc1$ needs to be equal to \eqref{dTnu} in order for $[\dT,\dTp]\sigma+\delta_{\lbr T,T'\rbr^\sigma}\sigma=0$ to hold (where $\Cc0\equiv\sigma$).
Similarly we expect that $[\dT,\dTp]\Cc{n} +\delta_{\lbr T,T'\rbr^\sigma}\Cc{n}=0$ to follow from the transformation property of $\dT\Cc{n+1}$.
The systematic study (at the intrinsic corner level) of the potential $\A_{\bfCc}$-algebroid structure goes beyond the scope of the present manuscript and we let it for future investigation.
Notice however that $\A_{\bfCc}$ can still be defined as the restriction of $\A$ to an arbitrary cut.\\

A great simplification arises when the chosen cut at which $T=\tau|_{u=S{(U)}}$ is non-radiative. 
In this case $\nu=0$ so that $\dT\sigma\to\hdT\sigma$ and we can restrict $T$ to belong to the generalized wedge algebra $\Wcal_\sigma(S)$. 
The latter is characterized by $\hdT\sigma\equiv 0$ so that the algebroid bracket $\lbr\cdot\,,\cdot\rbr^\sigma$ reduces to the Lie algebra $\sigma$-bracket $[\cdot\,,\cdot]^\sigma$ that we studied in chapter \ref{secGRI}.
We also know that $\hdT\sigma\equiv 0$ amounts to the condition $(\Dcal^{s+2}T)_{-2}=0$ on the graded vector $T$.\footnote{Note that $s=-1$ is the initial constraint $\sI_\tau\big|_{u=0}=0$.}
To consistently satisfy the non-radiation condition $\nu=0$, we impose that the transformation $\dT\nu$ vanishes as well.
Evaluating \eqref{dTnu} at $\nu=0$, we see that we have to take $\Cc2=0$.
Iterating this reasoning, we find that a non-radiative cut is characterized by $\Cc{n}=0,\,n>0$, so that $\Wcal_\sigma$ is indeed parametrized by the sole parameter $\sigma$.
In this case, we get that\footnote{We similarly get that $\dT\Cc{n}\big|_{\{\Cc{p}\}_{p=1}^n=0}=\hat \delta_{\Dcal^nT}\sigma=-(\Dcal^{n+2}T)_{-2}=0$ for $T\in\Wcal_\sigma$.}

\begin{equation}
    \dT\nu\big|_{\nu=0}=\hat\delta_{\Dcal T}\sigma=-(\Dcal^3T)_{-2}=0\quad\mathrm{for}\quad T\in\Wcal_\sigma.
\end{equation}

\paragraph{Remark:}
The discussion of this section is alike the one for YM in chapter \ref{secNAGTII} where $\Wcal_\sigma(S)$ here corresponds to $\Wcal_{\bfAc}(S)$ there.
In both cases, the covariant wedge comes about as the non-radiative restriction of the projection of $\Wcal_C(\scri)$ (or $\Wcal_A(\scri)$ in YM) at a cut.
In the gravitational study, there is however another layer where the covariant wedge appears intrinsically from the sphere (i.e. not from the projection $\scri\to S$).
The reason being that the deformed $\sigma$-bracket is a Lie bracket only \textit{inside} $\Wcal_\sigma(S)$.
The Jacobi identity is otherwise violated, with the violation proportional to $\dT\sigma$, where $\dT\sigma$ is the gravitational analog of $\dA \Ac0$.
The bracket $[\cdot\,,\cdot]^\g$ (equivalently $[\cdot\,,\cdot]_\g$) does not undergo any deformation in the gauge theory case and is thus a genuine Lie bracket irrespectively of any conditions on $\Aa{s}$.
Remarkably, in the gravitational case, $\hdT\sigma=0$ \textit{amounts} to the covariant wedge onto the sphere; namely it leads to conditions on all $\T{s},\,s\geq 0$.
For YM, we saw that $\Dcal\Aa0=0$ does not impose any constraint on $\Aa1$.\footnote{This behavior can be traced back to the fact that $[\cdot\,,\cdot]^\g$ is of degree 0 while its gravitational counterpart, namely the $\sigma$-bracket $[\cdot\,,\cdot]^\sigma$ is of degree $-1$.} 
In that sense, $\Wcal_{\bfAc}(S)$ appears \textit{solely} as the projection of $\Wcal_A(\scri)$ at a non-radiative cut.\\

Then, as a consequence of equation \eqref{IsoAlgebroids}, we obtain the following theorem.

\begin{tcolorbox}[colback=beige, colframe=argile, breakable]
\textbf{Theorem [Representation of $\Wcal_\sigma$ on $\scri$]}\\
At any non-radiative cut of $\scri$, the Lie algebra $\Wcal_\sigma(S)$ admits a Lie algebroid section in $\Ao$, realized via the map $\bft$,
\begin{align}
    \bft :\,&\Wcal_\sigma(S) \to \Ao \nn\\
     &\, T\mapsto\bft(T),
\end{align}
\vspace{-0.7cm}

\ni which satisfies
\vspace{-0.3cm}
\begin{align}
    \bft\big([T,T']^\sigma\big)= \big\lbr\bft(T),\bft (T')\big\rbr. \label{tautoTbracket}
\end{align}
\end{tcolorbox}

When we study a gravitational system in between two non-radiative cuts $u_0$ and $u_0'$, with definite shears $\sigma,\sigma'$ and symmetry parameters $T,T'$ respectively, the covariant wedge conditions $\hdT\sigma=0=\hdTp\sigma'$ come from the requirement that the symmetry action leaves these shears unchanged.
In other words, the covariant wedge algebra is the Lie algebra that preserves the boundary conditions $C|_{u=u_0}=\sigma$ and $C|_{u=u'_0}=\sigma'$.
In the limit $u_0\to\infty$, this is the algebra that preserves the late time fall-off condition, cf. footnote \ref{foot1}.
Notice that imposing the normal wedge condition $D^{s+2}\T{s}=0$ is not sufficient (if $\sigma\neq 0$ at the cut in question) since it only kills the inhomogeneous, i.e. soft, part of the shear transformation.
Nevertheless, recall that one can build a dressed symmetry parameter $\mTG(T)$ which does satisfy $D^{s+2}\mTG_s(T)=0$, see \ref{sec:CovWedgeSol}.
In the next section, we come back to this fact and give it an interpretation in terms of Newman's good cut equation.

We let for further investigation the proper description of the transition between $\Wcal_\sigma(S(u_0))$ and $\Wcal_{\sigma'}(S(u'_0))$. 
This is an interesting question that requires a careful treatment of radiation.

We conclude this chapter by making contact with the twistor description of higher spin symmetries, especially the work \cite{Kmec:2024nmu}.

\section{Good cut and relation to twistor theory \label{sec:twistor}}

While finishing \cite{Cresto:2024mne}, the Oxford group published a remarkable work \cite{Kmec:2024nmu} which overlapped our work, although their starting point was totally different from ours. 
They study self-dual GR in twistor space and show that via a gauge fixing adapted to the asymptotic twistor space, the residual gauge transformations form the $Lw_{1+\infty}$ algebra on twistor space.
From there, they build Noether charges on $\scri$ that realize the algebra to all order in $\gN$.
The construction of the Noether charges is non-perturbative because the renormalization scheme necessary to form a charge conserved in the absence of radiation to all order in $\gN$ only involves the knowledge of the dual EOM \eqref{tausevolb}. 
In their perspective, the latter come as a result of a gauge fixing, which is totally independent from our standpoint.

They also show that the charge conservation laws \eqref{Qevolve} can naturally be derived from a generalized Gauss's law in twistor space that follows from the twistor equation of motion on the Lagrange multiplier that imposes self-duality. 
Connecting these equations non perturbatively directly to the Bianchi identities for the Weyl tensor is still a challenge for us.
What is remarkable is that the twistor space construction is based on a twistor Poisson bracket which is independent of the field while we have seen that the canonical analysis done in this work involves a shear deformation of the original $\W$-bracket.
In the next subsection, we show that the twistor  Poisson bracket equates our $C$-bracket on-shell of the dual EOM, cf. \eqref{TCbracket} and \eqref{TwistorPoissonBracket}.

\subsection{Trading the grading for an extra dimension}

The key element needed to understand the connection between our derivation of the $\Ah$-algebroid and the twistor derivation is the introduction of a spin $1$ variable $q \in \Ccar{0,1}$. This variable can be used to promote the series $\tau=(\tau_s)_{s+1\in \N}$ to a holomorphic function $\ft$ of $q$\footnote{We keep the Bondi coordinates dependence hidden. 
The full expression reads $\ft(q, u, z, \bz)=\sum_{s=-1}^{\infty} \tau_s( u, z, \bz)  q^{s+1}$. } 
valued onto vector fields on $\scri$,
\be
\ft(q)=\sum_{s=-1}^{\infty} \tau_s  q^{s+1} \in \tV_{-1}. \label{ftauDef}
\ee 
To avoid misunderstanding and shorten the notation, we denote this function by $\ft$.
The graded vector $\tau$ and the function $\ft$ can be viewed as two different representations of the same abstract vector in $\tV(\scri)$.
The covariant derivative operator is simply represented on these functions as the  functional
\be 
(\Dcal \ft)(q) = \sum_{s=-1}^{\infty} (\Dcal \tau)_s q^{s+1} =
 \pa_u \ft(q) - \sE_{\ft}(q), \label{Dcaltauq}
\ee 
where we introduced 
$
\sE_{\ft}(q):= \sum_{s=-1}^\infty \E_s(\tau) q^{s+1}.
$
Using this we can write our bracket in terms of the spin functionals. 
Quite remarkably we find that it is related to
the Poisson bracket in the $(q,u)$ plane, namely we introduce 
\be \label{bracketquPlane}
\{\ft,\ft'\}:= \pa_q \ft \pa_u \ft' - \pa_q \ft' \pa_u \ft.
\ee
Denoting $[\ft, \ft']^{C}(q) := \sum_s     [ \tau, \tau']_s^{C}q^{s+1}$, we find that 
\begin{align}
    \sum_{s=-1}^\infty [\tau, \tau']_s^{C}q^{s+1}&=\left(\sum_{s=-1}^\infty\sum_{n=0}^{s+1}(n+1)\t{n}q^n(\Dcal\tau')_{s-n}q^{s-n+1}\right)-\tau\leftrightarrow\tau'\cr
    &=\left(\sum_{n=0}^\infty\sum_{s=n-1}^\infty\pa_q(\t{n}q^{n+1})(\Dcal\tau')_{s-n}q^{s-n+1}\right)-\tau\leftrightarrow\tau'\\
    &=\pa_q\left(\sum_{n=0}^\infty \t{n}q^{n+1}\right)\left(\sum_{s=-1}^\infty(\Dcal\tau')_{s}q^{s+1}\right)-\tau\leftrightarrow\tau'. \nn
\end{align}
Using \eqref{Dcaltauq}, we conclude that
\begin{align}
\boxed{ [ \ft, \ft']^{C}=   
 \{ \ft, \ft'\} - \pa_q \ft \E_{\ft'} - \pa_q \ft' \E_{\ft} }.\label{TCbracket}
\end{align}
This shows that the $C$-bracket is given by the Poisson bracket, once we assume the dual equations of motion, i.e. for $\tau,\tau' \in \sfTh$.
The Poisson bracket is the canonical twistor Poisson bracket under the parametrization of the twistor fiber coordinates $\bar\mu^\alpha = u n^\alpha + q \lambda^\alpha$ where $\lambda^\alpha$ is the spinor parametrizing holomorphic homogeneous coordinates on the sphere and $n^\alpha$ is a spinor such that $\epsilon_{\alpha \beta}= \lambda_\alpha n_\beta-\lambda_\beta n_\alpha$.\footnote{It is customary to take 
$ n_\alpha = \frac{\hat{\lambda}_\alpha }{\lambda^\beta \hat\lambda_\beta}$ where $ \hat{\lambda}_\alpha=T_{\alpha \dot\alpha}\bar\lambda^{\dot\alpha} $ with $T_{\alpha \dot\alpha}$ a timelike vector.}${}^,$\footnote{If we consider the perturbation around flat space due to a single graviton, i.e. a plane wave propagating in the direction $\omega^\alpha$ on the celestial sphere, then $q=-\bar\mu^\alpha n_\alpha=r\lambda^\alpha\omega_\alpha$, cf. \cite{Donnay:2024qwq}. In that case, we also have $q\pa_u\to\pa_z$ in the (on-shell) Poisson bracket \eqref{bracketquPlane}.}
This directly follows from $\pa_u = n^\alpha \frac{\pa}{\pa \bar\mu^\alpha}$ and $\pa_q = \lambda^\alpha \frac{\pa}{\pa \bar\mu^\alpha}$ so that 
\vspace{-0.1cm}
\be \label{TwistorPoissonBracket}
\{\ft,\ft'\} 
= (\lambda^\alpha n^\beta-\lambda^\beta  n^\alpha )\frac{\pa\ft}{\pa \bar\mu^\alpha}\frac{\pa\ft'}{\pa \bar\mu^\beta} = \epsilon^{\alpha\beta} \frac{\pa \ft }{\pa \bar\mu^\alpha} \frac{\pa \ft'}{\pa \bar\mu^\beta}.
\ee 
Finally we can establish that the transformation $\tdt h $ of the twistor potential given in \eqref{dtauh} is equivalent to a twistor gauge transformation.
One first introduces the covariant derivative 
\be 
\nabla := q\pa_u -D + C \pa_q,
\ee
and we obtain  that 
\begin{align}
    \nabla \ft(q) 
    &= \sum_{s=-1}^\infty \big(  \pa_u\tau_s - 
    D \tau_{s+1} 
     + (s+3) C \tau_{s+2}\big) q^{s+2}
    - D \tau_{-1} + C \tau_0 \cr
    &=  \sI_\tau + q \E_{\ft}(q) .\label{bNI}
\end{align}
The condition $\tau \in \sfTh$, which imposes $\E_{\ft}=0$, simply  reads, in twistor variables, $\pa_q \nabla \ft=0$.
This shows that when $\tau \in \Ah$, then $\nabla\ft$ is independent of $q$\footnote{This is natural since the asymptotic potential $h$ is only a function of $(u,z,\bz)$.} and
\be 
\tdt h= \nabla \ft = \sI_\tau.
\ee
We recognize \eqref{dtauh},
which establishes the equivalence between the twistor and phase space transformations for $\tau\in \Ah$.

Although we agree with most of the results of \cite{Kmec:2024nmu}, there seems to be a discrepancy with their equation (4.16) whose RHS misses the shear dependence of the term  $\sigma\paren{\cdot\,,\cdot}$ (cf.\,\eqref{DaliBracket}) in the $\sigma$-bracket \eqref{Wbracket2}.
We suspect that the reason of the difference is that there bracket is evaluated at $\scri^+_-$, where according to the Schwartz boundary conditions, the shear vanishes.
This shear dependence of the $\sigma$-bracket when written on an arbitrary non-radiative cut of $\scri$ is the origin of the many subtleties of the present work.

\subsection{Good cut \label{sec:GoodCut}}

From \cite{Adamo:2021lrv,Kmec:2024nmu} we know that $q$ is a fiber coordinate associated with the projection $p:\mathbb{PT}\to \scri_{\mathbb{C}}$ given by $(\bar{\mu}^\alpha, \bar\lambda_{\dot\alpha}) \to (u =\bar{\mu}^\alpha \lambda_\alpha, \lambda_\alpha, \bar\lambda_{\dot\alpha})$.\footnote{The coordinate we call $(u,q)$ here are denoted $(\bar{u}, \bar{q})$ in \cite{Kmec:2024nmu}.} 
We now want to show that the spin coordinate $q$ possesses a natural geometrical interpretation from the point of view of  $\scri$ and its spacetime embedding.
First we know that $\scri$, being a null surface, is equipped with a Carrollian structure $(\ell^a, q_{ab})$ where $\ell$ is the vector tangent to the null congruence and the degeneracy vector for $q_{ab}$, i.e. $\ell^a q_{ab}=0$ \cite{Ashtekar:1981bq, Riello:2024uvs}. 
Choosing a complex structure on the metric amounts to the choice of a frame field $m_a$ such that $q_{ab}= m_a\bm_b +\bm_a m_b$. This frame is such that $\ell^a m_a=\ell^a \bm_a=0$. 
In order to equip $\scri$ with a connection we need to choose a ruling, that is a one form $k_a$ such that $k_a\ell^a=1$ \cite{Freidel:2022vjq,Freidel:2024emv}. 
In the Bondi analysis, one selects the ruling form to be exact and given by the differential of the Bondi time $k=\rd u$.
The carrollian vector is then simply $\ell=\pa_u$.
However, it is useful to allow more general choices such that $k$ carrying vorticity, i.e. $\rd k \neq 0$. 
For instance we can  choose $k_a$ to be associated to a null rigging structure \cite{Mars:1993mj}. 
This means that we can see the ruling form as deriving from a spacetime \emph{null} vector $k^a$ transverse to $\scri$. 
Such a transverse vector is understood as labeling a congruence of null geodesics transverse to $\scri$. 
Such general ruling vector, like the ones associated with geodesics null congruence, can be parametrized by a pair of spin variables $(q,\bar{q})$  and given by 
 \be
 k_{(q,\bar{q})} = k + q \bm +\bar{q} m - q\bar{q} \ell.
 \ee 
The complex structure vectors normal to $k_q$ are then given by
 \be
 m_q= m- q\ell.
 \ee
The transformation $(k,m, \ell)\to(k_q,m_q,\ell_q)$, which fixes $\ell=\ell_q$ corresponds to a null boost with angle $q$.
This shows, as explained in the work of Adamo and Newman \cite{Adamo:2009vu, Adamo:2010ey}, that the value of $q$ at any point of $\scri$ is the stereographic angle which describes the null direction of each geodesic intersecting $\scri$.

The supertranslations $T=T(z,\bz)$ acts non trivially on the pair $(u,q)$\footnote{For the special case of a supertranslation, note that $\delta_{\tau(\T0)}\cdot=\d{0}{\T0}\cdot=\delta_{\T0}\cdot$, and here we just rename $\T0\to T$ since the context is clear.}
 \be 
 \delta_T u= T, \qquad \delta_T q = -D T.
 \ee 
The transformation for $u$ is the supertranslation definition. The transformation  for $q$ follows from the fact that while $m$ is a vector tangent to the cut $u=\mathrm{cst}$, the vector  $m_{q=-DT}=m + DT \ell$ is the vector tangent to the cut $u-T=\mathrm{cst}$.

In order to use the power of the null rotations we  consider a Goldstone field $G(u,z,\bz)$, which transforms linearly under supertranslation $\dT G= -T$.
This field defines a diffeomorphism of $\scri$ denoted  $\hG: \scri \to \scri$ and given by $\hG(u,z,\bz):= (G(u,z,\bz),z, \bz)$. 
Under this map the Bondi cuts $u=u_0$ are mapped onto supertranslated cuts $u= G(u_0,z,\bz)$. 
Moreover, the holomorphic derivative is shifted  by a spin $1$ connection $L$
\be \label{DefConnectionL}
D (F\circ \hG)= (D F + L\pa_u F)\circ \hG, \quad\mathrm{with}\quad L= DG\circ \hG^{-1}. 
\ee 
The Goldstone is defined to satisfy the good cut equation\footnote{For the equivalence, apply \eqref{DefConnectionL} to $ L\circ\hG $.}$^,$\footnote{Notice how a super-translation acts on $C\circ\hG$:
\vspace{-0.1cm}
\be
\dT\big(C\circ \hG\big)
 &= (\dT C)\circ \hG + \dT G (\pa_uC \circ \hG) \nn\\
&= (T\pa_u C - D^2 T )\circ \hG - T \pa_uC \circ \hG 
= -D^2T \circ \hG = -D^2T=\dT (D^2G).
\ee
\vspace{-0.4cm}
} \cite{newman_heaven_1976}
\be 
D^2 G = C \circ \hG\quad \Leftrightarrow\quad   DL + L \pa_u L=C, \label{GoodCutEq}
\ee 
which determines $G$ from the shear $C$ up to a 4 dimensional freedom interpreted as complex spacetime \cite{Adamo:2010ey, newman_heaven_1976}.

Given $\ft\in \sfT $,\footnote{Recall that the function $\ft$ is also a representation of the vector space $\sfT$.} we define the map $\mTG: \sfT \to \sfT $  with image $ \mTG[\ft]\in \sfT$ given by 
\be 
\boxed{\mTG[\ft](q):= \ft(q -L) \circ \hG}.
\ee
$\mTG$ is a map  which combines a shift of $q$ by the spin $1$ connection $L$ and a redefinition of time, both determined by the supertranslation Goldstone field. 
Explicitly, this means that  
\be 
\mTG[\ft](q,u,z, \bz)
= \ft\big(q - DG(u,z,\bz), G(u,z,\bz),z,\bz\big). \label{defmTGtwistor}
\ee
This transformation is such that
\begin{equation}
    \pa_u \mTG[\ft] =\dot G\mTG[\pa_u\ft]-D\dot G\mTG[\pa_q\ft].
\end{equation}
Similarly for $D$, we can write compactly\footnote{$D\dot G= {\dot G} (\pa_u L \circ\hG) $.}
\bs\label{DuTGq}
\begin{align}
\pa_u \mTG[\ft] &= \dot{G} \mTG\big[\pa_u \ft - \pa_u L \pa_q\ft\big], \\
  D \mTG[\ft] &= \mTG\big[D \ft + L \pa_u \ft - C\pa_q\ft \big].
\end{align}
\es
These relations can be inverted once  we introduce the frames 
\be
\ell^G:= \dot{G}^{-1}(\pa_u + D\dot{G} \pa_q),\qquad  
D^G:= D - DG \ell^G. 
\ee
We see that $DG$ and $D\dot{G}$ play the role of rotation coefficients respectively deforming the usual derivatives $(D, \pa_u)$ along $(\ell^G,\pa_q)$.
We recast \eqref{DuTGq} as\footnote{$\pa_q\mTG[\ft]=\mTG[\pa_q\ft]$.}
\begin{equation}\label{lGintertwine}
    \ell^G\mTG[\ft]=\mTG\big[\pa_u \ft\big], \qquad D^G\mTG[\ft]=\mTG\big[(D-C\pa_q)\ft\big].
\end{equation}
Notice also that
\begin{equation}
    \mTG[q\ft]=(q-DG)\mTG[\ft]\equiv q^G\mTG[\ft],
\end{equation}
where $q^G:=q-DG$ is the new $q$ variable after the change of frame.
The dual equation of motions for $\tau$ are then mapped into simpler equations where the shear has been removed.
Indeed, by combining the last two equations, we readily get that
\be \label{IntertT1}
\boxed{\big(q^G\ell^G-D^G\big)\mTG[\ft]= \mTG[\nabla\ft]}.
\ee 
For $\ft\in\sfTh$, we can use \eqref{bNI} to deduce the following form of the dual EOM:
\begin{equation}
    \pa_q \big(q^G\ell^G-D^G\big)\mTG[\ft]=0.
\end{equation}
It is interesting to formalize the previous findings. 
First we introduce the \emph{Newman space}\footnote{It is related to twistor space via the relations around \eqref{TwistorPoissonBracket}.} $\TN\equiv\C\times\scri_{\C} $ which is a one dimensional fibration of $\scri$ with fiber coordinate $q$ and we extend the map $\hG$ into a diffeomorphism $\hhG:\TN\to\TN$ given by 
\begin{align}
    \hhG:(q,u,z)\mapsto(q^G,G,z). \label{hhGdiff}
\end{align}
This map extends as an inverse pushforward to the vector fields $(\pa_q,\ell,m)$ according to\footnote{If we treat $\hhG_*$ as a holomorphic map, it leaves $\bm$ invariant.} 
\be 
\hhG_*(\pa_q,\ell,m) = \big(\pa_q,\ell^G,m^G\big).
\ee 
The map $\hhG$ is such that $\big(\pa_{q^G},\ell^G,m^G\big)(F \circ \hhG) = \big[(\pa_{q},\ell,m-C\pa_q) F\big]\! \circ \hhG $.
It is designed such that the shear variable $C$ is mapped onto 0!
Hence, the result \eqref{IntertT1} simply means that $\mTG$ implements the pullback $\hhG{}^*$ of $\ft$ from a radiative $\TN$ to a non radiative one.
In other words, the evolution operator $\nabla=(q\ell-m+C\pa_q)$ is mapped onto its non radiative version $\nabla^G:=(q^G\ell^G-m^G)$ such that
\begin{equation}
    \hhG{}^*(\nabla\ft)=\nabla^G\big(\hhG{}^*\ft\big).
\end{equation}

\subsection{Relation to the dressing map \label{sec:RelationDressingMap}}

We show that the map $\mTG$ defined here is the same as the dressing map $\mTG$ introduced in \ref{sec:CovWedgeSol}.
In a way this follows from the previous calculations by noticing that\footnote{We can readily check that $(D-C\pa_q)\ft=-\sI_\tau+q\Dcal\ft$.}
\begin{equation}
    D^G\mTG[\ft]=\mTG \big[q\Dcal\ft\big]-\sI_\tau\circ\hG.
\end{equation}
Therefore $\mTG$ intertwines the action of $D^G$ and $q\Dcal$ when $\ft\in\sfTo$, which was the definition of the intertwining map in section \ref{sec:CovWedgeSol}.
It is also instructive to prove the isomorphism explicitly.
First notice that by definition \eqref{defmTGtwistor},\footnote{This equation defines the map $\mTG$ on the graded vector $\tau$.}
\begin{align}
    \mTG[\ft](q) =\sum_{s=-1}^\infty\mTG_s[\tau]q^{s+1} =\sum_{s=-1}^\infty\t{s}(G,z,\bz)\sum_{k=0}^{s+1}\binom{s+1}{k}q^{s+1-k}(-DG)^k.
\end{align}
Rearranging the summations we can write this expression in terms of the map 
$T^G$ \eqref{TGdef}. Namely if we define
\begin{equation}
T^G_s[\tau]:=\sum_{k=0}^\infty \frac{(s+k+1)!}{k!(s+1)!}(-DG)^k\t{s+k}, \qquad s\geqslant -1,
\end{equation}
we have that 
\begin{align}
    \mTG[\ft](q) =\sum_{k=0}^\infty \sum_{s=-1}^\infty\frac{(s+k+1)!}{k!(s+1)!}(-DG)^k q^{s+1}\big(\t{s+k}\circ\hG\big)= \sum_{s=-1}^\infty T^G_s \big[\tau\circ\hG\big]q^{s+1}.
\end{align}
This implies that while $T^G[\ft](q)= \ft(q-DG)$, we get that $\mTG$ is the composition of $T^G$ with the pullback of $\hG$:
\begin{equation}
   \mTG[\ft](q)=\ft (q-L)\circ \hG =(\ft\circ \hG) (q-DG) = T^G[\ft\circ \hG](q).
\end{equation}

Notice finally that in the particular case where $\hG$ is a supertranslation, i.e. when $G=G(z,\bz)$ is time independent, then
\begin{equation}
    \tau\circ\hG=\big(e^{G\pa_u}\tau\big) \big|_{u=0}=e^{G\Dcal}T.
\end{equation}
Therefore,
\begin{equation}
    \mTG[\tau]=T^G\big[\tau\circ\hG\big]= T^G\big[e^{G\Dcal}T\big]= \sum_{n=0}^\infty\frac{G^n}{n!}T^G\big[\Dcal^n T\big],
\end{equation}
as it was defined in \eqref{mTGdef}.
In the previous chapter, we had denoted this map $\mTG[T]$ to emphasize its expression intrinsic to the sphere $S$.
Here we see explicitly that the peculiar dependence in $G$ and $\Dcal$ comes from $e^{G\Dcal}$ and reflects the evaluation of the Carrollian symmetry parameter $\tau$ at the super-translated cut defined by $G$.

\chapter{Einstein-Yang-Mills Theory \label{secEYM}}

This chapter gathers the material of \cite{Cresto:2025ubl}.

\section{Introduction}

In this chapter, we piece together our analysis on asymptotic higher spin symmetries in General Relativity and Yang-Mills theory, and study the symmetry algebra and symmetry algebroid resulting from the minimal coupling of gravity and non-abelian gauge theory, namely Einstein-Yang-Mills (EYM) theory.
The main input of this part, i.e. the EYM evolution equations for the charge aspects, comes from \cite{Agrawal:2024sju}.

Notice that in this chapter we change notation for few objects, especially the name of some vector spaces, mostly because we will not repeat certain parts of the analysis of the previous chapters and instead present yet another viewpoint on the brackets, putting the emphasis on their off-shell versus on-shell versions. 
The new notation shall thus reflect this approach and the reader may refer to the new glossary in section \ref{secEYM:glossary} if needed. 

We derive in section \ref{secEYM:DualEOM+Noether} the set of dual EOM that the Carrollian symmetry parameters $\t{}$ and $\a{}$ have to satisfy in order for the higher spin symmetries to be canonically realized via Noether charges.
We deduce the infinitesimal symmetry transformation onto the shear $C$ and the transverse gauge field $A$ in section \ref{secEYM:NoetherChargeBondi}.
This gives us the opportunity to shed a new light on the gravitational analysis using a complementary standpoint compare to the previous chapters \ref{secGRI} and \ref{secGRII}.
We show that these transformations are generated by a Noether charge $\Qta$, which is given by the sum of the Noether charges $\Qt$ and $\Qa$ introduced in sections \ref{Sec:CanAction} and \ref{secYM:Noether} plus cross terms.

Next, in subsection \ref{secEYM:T-bracket}, we define the $\sfST$-algebroid-bracket $\lbr\cdot\,,\cdot\rbr$, which generalizes the pure GR and YM ones worked out in the previous chapters to incorporate the cross terms specific to the EYM coupling.
We then introduce a kinematical Lie algebra bracket defined \emph{off-shell} of the dual EOM in subsection \ref{secEYM:YMBracket}. 
We prove that the $\sfST$-algebroid bracket is obtained from this kinematical Lie algebra bracket after imposition of the dual EOM. 
The off-shell algebra has the structure of a semi-direct product. 
Imposing the dual EOM however destroys this structure. 
We prove in section \ref{secEYM:BicrossedJacobi} that the $\cST$-algebroid is instead realized through a combination of a bicrossed product and a semi-direct action of $\cT$ on $\cS$. 
This means that
$\cST\simeq\cTo\Bprod\big(\bcT \Wprod\cS\big)$, where $\cTo$ is the sub-algebroid of super-translations with the center, while $\bcT$ represents the sub-algebroid of sphere diffeomorphisms and higher-spin parameters. 
$\cS$ is the YM algebroid and an ideal.
This structure allows us to prove in subsec.\,\ref{secEYM:JacobiIdentity} that the $\sfST$-bracket satisfies the Jacobi identity.

Besides, we study in section \ref{secEYM:CovWedge} the covariant wedge algebra $\Wcal_{CA}(\scri)$ associated to the $\cST$-algebroid and how it represents a deformation of the celestial $sw_{1+\infty}$ bracket, cf.\,\ref{secEYM:celestialBracket}, while on the other hand it can also be recast as a shifted Schouten-Nijenhuis bracket.
Interestingly, we find that the 3 actions forming the bicrossed/semi-direct products of Lie algebroids recombine into a single action $\blackwhitetriangle$ such that at a non-radiative cut $S$ of $\scri$, the EYM wedge algebra takes the form of the semi-direct product of the gravitational wedge algebra with the Yang-Mills one. 

Finally we point out in section \ref{secEYM:twistor} the relationship with twistor theory and its Poisson bracket.
On top of the main text, a large collection of appendices \ref{AppEYM} details most demonstrations.

\section{Dual EOM, infinitesimal action and Noether charge \label{secEYM:DualEOM+Noether}}

From the paper of Agrawal,  Charalambous and Donnay \cite{Agrawal:2024sju}, we know that the EYM equations of motion for the higher spin charge aspects are given by\footnote{We use an anti-hermitian basis of generators for $\g$. The pairing $\Pairg{Y,Z}=-\Tr(YZ),~Y,Z\in\g$,  is positive definite. }
\bs\label{EOMEYM}
\begin{align}
    \pa_u\tQ_s^\ym &=D\tQ_{s-1}^\ym+\big[A,\tQ_{s-1}^\ym\big]_\g +sC\tQ_{s-2}^\ym, && s\geqslant 0,\\
    \pa_u\tQ_s^\gr &=D\tQ_{s-1}^\gr+(s+1)C\tQ_{s-2}^\gr+(s+1)\Pairg{F,\tQ_{s-1}^\ym}, & & s\geqslant -1,
\end{align}
\es
with $C$ and $N=\pa_u C$ the shear and curvature news, and $A$ and $F=\pa_u A$ the asymptotic gauge field and Yang-Mills curvature. 
The lowest charge aspects encode the radiative data\footnote{Our conventions are such that $\tQ_{-3}^\gr=\tQ_{-2}^\ym=0$.}
\be
\tQ_{-2}^\gr=\frac{\dot{\bN}}{4\pi \GN}, \qquad\tQ_{-1}^\gr=\frac{D\bN}{4\pi \GN}, \qquad 
\tQ_{-1}^\ym=- \frac{ 2\bnews}{\gYM^2}.
\ee
The gravitational charge aspects  $\tQ_{s}^\gr$  have a Carrollian weight equal to $3$ and a Carrollian spin equal to $s$ which we denote as $\tQ_s^\gr\in\Ccar{3,s}$. Similarly, the charge aspects  $\tQ_{s}^\ym$ are Lie algebra valued and of Carrollian weight equal to $2$ and spin $s$ which we denote as $\tQ_s^\ym\in\Ccarg{2,s}$.

To define the integrated Noether charge we  introduce the dual transformation parameters
\be 
\t{s}\in\Ccar{-1,-s}, \qquad \a{s}\in\Ccarg{0,-s},
\ee
which can be used to form the integrated  master charges 
\bs
\label{defMasterChargeEYM}
\begin{align}
    \Qt^u &:=\sum_{s=-1}^{\infty}\QGR{s}[\t{s}]\equiv\sum_{s=-1}^{\infty}\int_{S_u}\tQ_s^\gr\t{s}, \\
    \Qa^u &:=\sum_{s=0}^{\infty}\QYM{s}[\a{s}]\equiv\sum_{s=0}^{\infty}\int_{S_u}\Pairg{\tQ_s^\ym,\a{s}}.
\end{align}
\es
Here and in the following we denote the series $\tau=(\t{-1},\t0,\t1,\ldots)$ and $\a{}=(\a0,\a1,\a2,\ldots)$ simply by the letters without the subscript.
The integrals are taken over the sphere $S_u$ at the cut $u=\mathrm{cst}$. We naturally define the EYM master charge $\Qta^u$ as the sum of the respective master charges $\Qt^u$ and $\Qa^u$:
\begin{equation} \label{defMasterChargeEYM2}
    \boxed{\QPta^u\equiv\Qta^u:=\Qt^u+\Qa^u},
\end{equation}
where $\Pta$ stands for the pair of elements $(\tau,\a{})$.
We denote $\Pta_s=(\t{s},\a{s}),\, s\geq 0$ and $\Pta_{-1}=(\t{-1},0)$.
We assume that there is no massive matter so that the master charge at $i^+$ vanishes: $\QPta^{+\infty}=0$.

\subsection{Dual equations of motion}

As in chapter \ref{secNAGTII}, we find the dual set of EOM for $\gamma$ in order for the  master charge to be  conserved in the absence of left-handed radiation.\footnote{See section \ref{SecGR:Noether} (and what follows) for an extensive discussion about the subtleties of defining the non-radiative condition either as $\bN\equiv 0$ or as $\dot{\bN}\equiv 0$. \label{footNNdot}}

\begin{tcolorbox}[colback=beige, colframe=argile]
\textbf{Theorem [Master charge conservation]}\\
In the absence of left-handed radiation, $\dot\bN=0=\bnews$, the master charge is conserved 
\be 
\pa_u \QPta^u =0,
\ee 
if and only if the dual EOM  $\E(\Pta)=0$ are satisfied,  where $\E(\Pta)=\big( \E^\gr(\t{}), \E^\ym(\Pta)\big)$ and
\vspace{-0.7cm}

\bs \label{DualEOMEYM}
\begin{align} 
    \E_s^\gr(\t{})&:=\pa_u\t{s}-\Big(D\t{s+1}-(s+3)C\t{s+2}\Big), & s\geq -1, \\
    \E_s^\ym(\Pta)&:=\pa_u\a{s}-\Big(\DYM\a{s+1}-(s+2)C\a{s+2}-(s+2)F\t{s+1}\Big), & s\geq 0.
\end{align}
\es
\end{tcolorbox}

\ni We have denoted the asymptotic gauge covariant derivative onto the sphere by\footnote{What we were denoting by $\Dcal$ in chapters \ref{secNAGTI} and \ref{secNAGTII}.}
\begin{equation}
    \DYM:=D+\adA=D+[A,\cdot\,]_\g.
\end{equation}

\paragraph{Remark:} In the following it will be useful to note that the equations of motion can be simply written as 
\begin{equation}\label{dEOM}
    \boxed{\E(\Pta)=\pa_u\Pta-\Dcal\Pta},
\end{equation}
where we define a covariant derivative $ \Dcal\Pta :=\big(\DGR\tau,\DTOT\Pta\big)$ with\footnote{$\DGR$ is what we called $\Dcal$ in chapters \ref{secGRI} and \ref{secGRII}.}
\bs \label{covd}
\begin{align}
    \big(\DGR\t{}\big)_s &:=D\t{s+1}-(s+3)C\t{s+2}, \\
    \big(\DTOT\Pta\big)_s &:=\DYM\a{s+1}-(s+2)\big(C\a{s+2}+F\t{s+1}\big).
\end{align}
\es

Notice that the gravitational evolution equations for the Carrollian symmetry parameters $\tau$ are unchanged compare to \eqref{tausevolb}.
However, the dual EOM for $\a{}$ get corrected by interaction terms of the type $C\a{}$ and $F\tau$.
It is worth emphasizing the structure of the couplings by rewriting $\E(\Pta)$ in matrix form as
\begin{equation}
    \pa_u\Pta_s=(D+\cA_s)\Pta_{s+1} -\mathcal{C}_s\Pta_{s+2},
\end{equation}
where
\begin{equation}
    \Pta_s=\binom{\t{s}}{\a{s}},\qquad \cA_s=\mat{0}{0}{-(s+2)F}{\adA},\qquad \mathcal{C}_s=\mat{(s+3)C}{0}{0}{(s+2)C}.
\end{equation}
We see that the gravitational field acts diagonally as $-(\delta+ s+2) C\phi_{s+2}$,  on  a symmetry parameter $\phi_s\in\Ccar{-\delta,-s}$. Note that $\dCar+s+2$ is the homogeneous holomorphic weight (see \cite{gelfand1966generalized} and \eqref{defCarrollHomogeneous}) of the field $\phi_{s+2}$.\\

We define $\dot\bN=\bnews=0$ as the self-dual non-radiative conditions. 
They are implied by but not equivalent to the self-duality of the gravitational and Yang-Mills fields.\footnote{Self-duality imposes also that $\psi_2=\psi_1=\psi_0=\phi_1=\phi_0=0$, in the Newman-Penrose notation \cite{Newman:1968uj}.}
If one relaxes these conditions, we find that the master charge satisfies a simple flux equation.

\begin{tcolorbox}[colback=beige, colframe=argile]
\textbf{Theorem [Charge flux]}\\
Assuming that the dual EOM $\E(\gamma)=0$ are satisfied, in the presence of left-handed radiation the master charge flux is linear in $(\dot\bN,\bnews)$ and given by 
\be \label{Qflux}
\pa_u \QPta^u =
\frac{1}{4\pi\GN}\dot\bN\big[\sI_\tau\big]-\frac{2}{\gYM^2}\bnews\big[\sI_\Pta^\ym\big],
\ee 
where we define the \emph{initial constraint} $\sI_\Pta$ as
\begin{equation}
    \label{defI} \sI_\Pta\equiv\big(\sI_\tau,\sI_\Pta^\ym \big):= \Big(C\t0-D\t{-1}\,,-\DYM\a0+C\a1+F\t0\Big).
\end{equation}
\end{tcolorbox}

\paragraph{Proof of Theorems:}
The proof of both theorems follows from the same calculation. 
Taking the time derivative of \eqref{defMasterChargeEYM2}, we find, using the charge conservation laws \eqref{EOMEYM} and the convention $\QYM{-2}=\QGR{-3}=0$, that
\begin{align}
    \pa_u\QPta^u 
&=\sum_{s=0}^\infty \left(\QYM{s}\big[\pa_u\a{s}\big]-\QYM{s-1}\big[\DYM\a{s}\big] + \QYM{s-2}\big[sC\a{s}\big]  \right) \nn\\  
&+\sum_{s=-1}^\infty \left( \QGR{s}\big[ \pa_u\t{s}\big]- \QGR{s-1} \big[D\t{s}\big] +\QGR{s-2}\big[(s+1)C\t{s}\big]+ \QYM{s-1}\big[(s+1)F\t{s}\big]\right) \nn\\
&=\sum_{s=0}^\infty \QYM{s}\big[\pa_u\a{s} - \DYM\a{s+1} + (s+2) (C\a{s+2} + F \t{s+1})\big] \nn\\*
&+\sum_{s=-1}^\infty \QGR{s}\big[ \pa_u\t{s}- D\t{s+1} + (s+3)C\t{s+2}\big] \\*
&+ \QYM{-1}\big[-\DYM\a0+C\a1+F\t0\big]
+\QGR{-2}\big[C\t0-D\t{-1}\big]. \nn
\end{align}
Therefore we get that
\begin{equation} \label{QuPtaDot}
 \boxed{  \pa_u\QPta^u =\frac1{4\pi \GN}\dot\bN\big[\sI_\tau\big]-\frac2{\gYM^2}\bnews\big[\sI_\Pta^\ym\big] 
    + \sum_{s=-1}^\infty \QGR{s}\big[\E_{s}^\gr\big]
+\sum_{s=0}^\infty\QYM{s}\big[\E_{s}^\ym\big] }
\end{equation}
where we used the definitions (\ref{DualEOMEYM}-\ref{defI}).
Imposing the dual evolution equations 
$\E_s^\gr(\tau)=0$, $\forall\,s\geq -1$ and $\E_s^\ym(\Pta)=0$, $\forall\,s\geq 0$, implies \eqref{Qflux}.
Setting $\dot\bN=\bnews=0$ in addition, we see that $\pa_u\QPta^u=0$, which concludes the proofs.

\subsection{Infinitesimal variations and Noether charge \label{secEYM:NoetherChargeBondi}}

Given the central relevance of the dual EOM we introduce the following vector spaces:

\begin{tcolorbox}[colback=beige, colframe=argile]
\textbf{Definition [$\sfST$-space]} 
\bs \label{DefSTspace}
\begin{align}
\sfST^+ :=&\Big\{\,\Pta\equiv(\tau,\a{})\in\sfTK^+ \oplus\sfSK~\big|~\E_s(\Pta)=0,~  \forall\,s\geqslant 0~ \Big\},\\
\sfST :=& \Big\{\,\Pta\equiv(\tau,\a{})\in\sfTK\oplus\sfSK~\big|~\E_{s-1}^\gr(\tau )=\E_{s}^\ym(\Pta)=0,~  \forall\,s\geqslant 0~ \Big\}.
\end{align}
\es
\end{tcolorbox}

\ni $\sfTK$ and $\sfSK$ are short-hand notations for the spaces 
\be \label{defSTKspace}
\sfTK :=\bigoplus_{s=-1}^\infty\Ccar{-1,-s}, \qquad 
\sfSK:=\bigoplus_{s= 0}^\infty\Ccarg{0,-s}.
\ee 
The subscript $K$ stands for \emph{Kinematical}, i.e. without any dual EOM imposed.
Similarly we denote $\sfTK^+ :=\bigoplus_{s=0}^\infty\Ccar{-1,-s}$ as the space similar to $\sfTK$ with the $\tau_{-1}$ parameter excluded. 
For shortness, we write $\sfSTK\equiv\sfTK\oplus\sfSK$ while $\sfST$ is the space where the dual EOM are imposed.
The difference between $\sfST$ and $\sfST^+\subset \sfST$ is in the range of parameters and of  dual EOM imposed. 
In $\sfST^+$ only dual EOM for positive $s$ hold.
$\Pta\equiv(\tau,\a{})$ denotes an element of $\sfST$ (or $\sfSTK$ according to the context).
The $\sfS$-space and the $\sfT$-space are just the restrictions of $\sfST$ to the YM, respectively GR sub-spaces.

Another master charge plays a key role in the analysis of gravitational symmetry.
It is defined as\footnote{This perspective is equivalent to the initial constraint $\sI_\tau=0$ and the $\Ao$ discussion of chapters \ref{secGRI} and \ref{secGRII}.}
\be 
\boxed{\widehat{Q}^u_\Pta:= Q^u_\Pta - \frac{1}{4\pi \GN} \bN[\sI_\tau]}. 
\ee 
By construction $\widehat{Q}^u_\Pta $ is independent of $\t{-1}$.
While $\tQ^\gr_0$ is the covariant mass aspect introduced in \cite{Freidel:2021qpz}, $\widehat{Q}^\gr_0 = \tQ^\gr_0-\bN C/{4\pi\GN}$ corresponds to the Bondi mass aspect. 
The other (higher spin) charge aspects are unmodified.

If we assume that $\gamma\in \sfST^+$  we obtain that
this master charge satisfies the conservation law  
\be \label{cons2}
\boxed{\pa_u  \widehat Q^u_\Pta  = 
- \frac{1}{4\pi \GN} \bN\big[\dPta C\big]
- \frac2{\gYM^2} \bnews\big[\dPta A\big]},
\ee 
where we have defined the infinitesimal variations of the shear and gauge potential as

\begin{tcolorbox}[colback=beige, colframe=argile]
\textbf{Definition [Infinitesimal variations]}\\
For $\Pta\in \sfST^+$,
\vspace{-0.6cm}

\bs \label{deltaPtaCA}
\begin{align}
    \dPta C &=N\t0-D^2\t0+2DC\t1+3CD\t1-3C^2\t2, \\
    \dPta A &=F\t0-\DYM\a0+C\a1.
\end{align}
\es
\end{tcolorbox}

\ni The proof of \eqref{cons2} follows from \eqref{QuPtaDot} from which we infer that the modified master charge satisfies the conservation law
\be 
\pa_u  \widehat Q^u_\Pta  = 
- \frac{1}{4\pi \GN} \bN\big[\pa_u \sI_\tau + D \E_{-1}^\gr(\tau)\big]
- \frac2{\gYM^2} \bnews\big[\sI_\Pta^\ym\big].
\ee
Moreover we have that 
\begin{align}
\dPta A  =\sI_\Pta^\ym  \qquad\textrm{and}\qquad
\dPta C\equiv \dt C =\pa_u\sI_\tau+D\E_{-1}^\gr(\t{}). \label{dtCoffShell}
\end{align}

According to \eqref{deltaPtaCA}, note that the action of $\t0$ is the supertranslation action $\t0\pa_u$ on $C$ and $A$ while $\a0$ acts as a gauge transformation on $A$. 
The subleading parameter $\t1$ acts as a holomorphic diffeomorphism on $C$ while $\a1$ and $\t2$ respectively act on $A$ and $C$ via a term proportional to the gravitational shear.
The action on $C$ is unmodified compare to the purely gravitational case while the action on $A$  can be written as 
\be 
\delta_{(\tau,\alpha)}A= \da A + F\t0 + C\a1.
\ee 
The term $\da A = -\DYM\a0$ is the usual gauge action, the extra two terms represent the coupling of Yang-Mills to gravity.
\medskip

The symmetry charge $\widehat{Q}_\Pta:= \widehat{Q}_\Pta^{-\infty}$ is defined as the limit of $\widehat Q^u_\Pta$ to the asymptotic corner $\scri^+_-$, i.e. when $u\to -\infty$.  
In the absence of matter fields, the boundary condition at timelike infinity is such that $\lim_{u\to +\infty} \widehat{Q}^u_\Pta =0$. This means that for $\gamma \in\sfST^+$, we can express the charge as an integral over $\scri$ 
\begin{equation} \label{NoetherChargeEYM}
    \boxed{\widehat{Q}_{(\tau,\alpha)}=\frac{1}{4\pi\GN}\int_\scri\bN\dt C+\frac{2}{\gYM^2}\int_\scri\Pairg{\bnews,\dta A}}.
\end{equation}
Besides, the holomorphic Ashtekar-Streubel symplectic potential \cite{ashtekar1981symplectic} for EYM is given by  
\begin{equation} \label{HolAshtekarStreubel}
    \Theta^{\mathsf{HAS}}=\frac1{4\pi \GN} \int_{\scri}\bN \delta C+\frac{2}{\gYM^2}\int_{\scri}\Pairg{\bnews,\delta A}.
\end{equation}
We see that the charge is obtained as the contraction of the symplectic potential with the field space variation:
$ \widehat{Q}_{\Pta}=I_{\dPta} \Theta^{\mathsf{HAS}}$ and therefore, following  the Noether analysis of \ref{secNAGTII} and \ref{secGRII}, we know that the limit to the corner $\scri^+_-$ of the master charge is the Noether charge.\footnote{Assuming the fields satisfy Schwartzian fall-off conditions \cite{schwartz2008mathematics}. Going beyond is an interesting challenge which is necessary to include hyperbolic scattering data, see \cite{Choi:2024ajz, Choi:2024mac}.}
By construction, $\widehat Q_\Pta$ generates the symmetry transformation, namely
\begin{equation}
    \poisson{\widehat Q_{(\tau,\a{})},C}=\dt C\qquad\textrm{and} \qquad\poisson{\widehat Q_{(\tau,\a{})},A}=\dta A.
\end{equation}

\paragraph{Remark:}
The Noether charge splits as follows
\be \label{NoetherChargeEYM3}
\widehat{Q}_\Pta= \widehat{Q}^\gr_\tau + \Qa^\ym + \frac{2}{\gYM^2}
\int_{\scri} \Pairg{\bnews, F} \t0  +\frac{2}{\gYM^2}\int_{\scri} \Pairg{\bnews C,\a1},
\ee 
where the gravitational and Yang-Mills charges were already constructed in \ref{SecGR:Noether} and \ref{secYM:Noether}:
\bs \label{NoetherChargeEYM4}
\begin{align}
\widehat{Q}^\gr_\tau &\equiv 
\frac1{4\pi \GN}\left(
\int_{\scri} \left(\bN N - D^2\bN\right)\t0
-\int_{\scri} \left(3D\bN C + \bN DC\right)\t1
-\int_{\scri} 3\bN C^2\t2
\right),\\
\Qa^\ym &= -\frac2{\gYM^2} \int_{\scri} \Pairg{\bnews, \DYM\a0}.
\end{align}
\es
In particular, the Noether charge associated to a global time translation $\t0=1$, i.e. the total energy, takes the Bondi form (with $\hat\Ham:=(1,0,0,\ldots)\in\sfT^+$)
\begin{equation}
    \widehat{Q}_{\hat\Ham}=\frac{1}{4\pi\GN}\int_\scri\bN N+\frac{2}{\gYM^2}\int_\scri\Pairg{\bnews, F}.
\end{equation}
Although we have a complete Noether understanding for all spin $s$ charges, our work does not provide any spacetime interpretation for the higher spin symmetries.
Connecting our results to the analysis of  \cite{Campiglia:2021oqz, Nagy:2022xxs, Nagy:2024dme, Nagy:2024jua}, which interprets the spin 2 symmetry as over-leading gauge transformations is left for future work.
Note however than the cubic term for the spin 2 charge has recently been understood by \cite{Jorstad:2025qxu} as part of the collinear corrections to the soft theorem in the $S$-matrix analysis.

\subsection{Explicit action of the low spin symmetry}

In order to write the Noether charge for the spin 0, 1 and 2 transformations, we solve the dual EOM \eqref{DualEOMEYM} for some initial conditions given at the $u=0$ cut for simplicity.
These are the usual celestial symmetry parameters.
We first focus on the gravity side.
The spin 0 symmetry is labeled by a supertranslation parameter $\T0$.
The spin 1 is labeled by the time independent vector $\T1$ while the parameter $\T2\in\Ccel{-1,-2}$ labels the spin 2 transformation.
On the gauge theory side, $\Aa0$ is a super-phase rotation while $\Aa1$ labels the subleading spin 1 transformation.
To proceed, it is convenient to introduce the potentials $(h,a)$, $h\in \Ccar{0,2}$ and $a\in \Ccar{0,1}$, which are such that 
\be 
\pa_u h=C, \qquad \pa_u a= A.
\ee 
We write the solutions of the dual EOM for parameters that vanish at high spin: $\t{s}=0$, $s\geqslant 3$ and $\a{s}=0$ for $s\geqslant 2$.
These solutions are expressed in terms of functions on the sphere $(T_0,T_1,T_2)$ for the gravitational sector and $(\Aa0,\Aa1)$ for YM. 
These encode the first subleading symmetries.
\bs \label{solEOMEYM}
\begin{align}
    \t2 &=\T2, \\*
    \t1 &=uD\T2+\T1, \\*
    \t0 &=\frac{u^2}{2}D^2\T2+uD\T1+\T0-3h\T2.
\end{align}
Similarly,
\begin{align}
    \a1 &=\Aa1 -3 \pa_ua \T2, \\
    \a0 &=uD\Aa1+\Aa0+[a,\Aa1]_\g-2 \pa_ua \T1-2u\pa_ua  D\T2-aD\T2-3Da\T2,
\end{align}
\es
The validity of the dual EOM can be checked by direct inspection.
The various Noether charges for $\T0,\T1,\T2,\Aa0,\Aa1$ are then obtained by plugging the solutions \eqref{solEOMEYM} into \eqref{NoetherChargeEYM}.
After some algebra we get\footnote{Since the context is clear, we write $\widehat{Q}_{\T{s}}\equiv\widehat{Q}_{(\tau(\T{s}),\a{}(\T{s}))}$ and $\widehat{Q}_{\text{\scalebox{1.24}{$\Aa{s}$}}}\equiv\widehat{Q}_{(0,\a{}(\Aa{s}))}$ for conciseness.}
\bs
\begin{align}
    \widehat{Q}_{\T0} &= \frac1{4\pi \GN}
    \int_{\scri} \bN\Big(\!- D^2\T0+\Ll_{\T0} h\Big)+\frac{2}{\gYM^2}\int_{\scri} \Pairg{\bnews,\Ll_{\T0} a}, \\
    \widehat{Q}_{\T1} &=\frac1{4\pi \GN}    \int_{\scri} \bN\Big(\!-u D^3\T1+\Ll_{\T1}h\Big)+\frac{2}{\gYM^2}\int_{\scri} \Pairg{\bnews,\Ll_{\T1}a}, \\
    \widehat{Q}_{\T2} &=\frac1{4\pi \GN}    \int_{\scri} \bN\Big(\!-\frac{u^2}{2} D^4\T2+\Ll_{\T2}h\Big)+\frac{2}{\gYM^2}\int_{\scri} \Pairg{\bnews,\Ll_{\T2}a}\\
    &+\frac{2}{\gYM^2}\int_{\scri} \Big\langle\bnews,\big[A,(3\T2 D+2D\T2)a\big]_\g\Big\rangle_\g, \nn\\
    \widehat{Q}_{\text{\scalebox{1.24}{$\Aa0$}}} &= \frac{2}{\gYM^2}\int_{\scri}\Big\langle\bnews,-D\Aa0+[\Aa0,\pa_u a]_\g\Big\rangle_\g, \\
    \widehat{Q}_{\text{\scalebox{1.24}{$\Aa1$}}} \!&= \frac{2}{\gYM^2}\int_{\scri}\!\Big\langle\bnews, -uD^2\Aa1+[\Aa1,Da]_\g-\big[[\Aa1,a]_\g,\pa_u a\big]_\g +\big[D\Aa1(1+u\pa_u),a\big]_\g +C\Aa1\Big\rangle_\g,
\end{align}
\es
where the differential operators $\Ll$ are defined as
\bs
\begin{align}
    \Ll_{\T0} &:=\T0\pa_u^2, \\
    \Ll_{\T1} &:=\big(2\T1 D+(s+1+u\pa_u)D\T1 \big)\pa_u, \\
    \Ll_{\T2} &:=3\T2\big(D^2-C\pa_u-h\pa_u^2\big)+2D\T2 (s+1+u\pa_u)D+\frac12 D^2\T2(s+u\pa_u)(s+1+u\pa_u).\!\!
\end{align}
\es
Here $s$ is the spin-weight of the field $\Ll$ acts on, i.e. 1 for YM and 2 for GR.
$\widehat{Q}_{\T1}$ is the Dray-Streubel charge, which is covariant \cite{Dray:1984rfa, Flanagan:2015pxa,Chen:2022fbu, Freidel:2024jyf}. The charges $ \widehat{Q}_{\text{\scalebox{1.24}{$\Aa1$}}}$ and $\widehat{Q}_{\T2}$ contain terms which are cubic in the field variables $(\bN, \bnews, h, a)$.\footnote{The master charge, respectively the Noether charge, is either written as a flux integral over a portion of $\scri$, respectively over full $\scri$, or as a corner integral \eqref{defMasterChargeEYM}.
We refer the reader to \ref{secYM:RenormCharge}-\ref{sec:RenormCharge} for similar corner expressions involving the renormalized charge aspects and the initial condition $\Pta\big|_{u=0}\equiv(\T{},\Aa{})$.}

\subsection{Self-dual phase space}

The self dual phase space is defined in terms of the potential $h$ which is related to the shear via 
\be 
C=\pa_u h.
\ee 
The asymptotic self-dual symplectic potential is then given by 
\begin{equation}
    \Theta^{\mathsf{SD}}= -\frac1{4\pi \GN} \int_{\scri} \dot \bN \delta h +\frac{2}{\gYM^2}\int_{\scri}\Pairg{\bnews,\delta A}.
\end{equation} 
It is related to the holomorphic Ashtekar-Streubel potential \eqref{HolAshtekarStreubel} by a corner term which vanishes under the Schwartzian fall-off conditions.
An analysis similar to the subsection \ref{secEYM:NoetherChargeBondi} lets us conclude that when $\Pta\in\sfST$ the charge 
\be \label{NoetherChargeEYM2}
\boxed{\QPta := \QPta^{-\infty}=-\frac1{4\pi \GN} \int_{\scri} \dot \bN \dt h +\frac{2}{\gYM^2}\int_{\scri}\Pairg{\bnews,\dPta A}}
\ee 
is the Noether charge for the action $\dPta$,
where $\dPta A$ is the same as before while 
\be \label{dtahEYM}
\boxed{\dt h = \sI_\tau= C\t0 -D\t{-1}}.
\ee  
Note that thanks to \eqref{dtCoffShell} and the fact that $E_{-1}^\gr(\tau)=0$ for $\Pta\in \sfST$ we have that 
\be 
\pa_u \dt h = \dt C, \label{dthdtC}
\ee 
so the action of $\sfST$ on $h$ extends the action of $\sfST^+$ on $C=\pa_u h$.
Notice that the analog of the Bondi mass aspect involves now the covariant mass introduced in \cite{Freidel:2021qpz}.
Given $\Ham= (V, 1, 0,0,\ldots)\in\sfT$,
\be 
Q_{\Ham}= 
\frac{1}{4\pi\GN} \int_\scri  \dot \bN \big(DV-C\big) +\frac{2}{\gYM^2}\int_\scri\Pairg{\bnews, F},
\ee 
where $V=\tau_{-1}$ such that $\pa_u V=0$ labels the value of the central element.
We see that the central element evaluates to a vanishing boundary term under the boundary conditions we work with.

\section{$\cST$-algebroid \label{secEYM:algebroid}}

In the previous section we have defined the action $\dPta$ for $\Pta \in \sfST^+ $ or $\Pta \in \sfST$.
We intend to show in this section that this action closes under the commutator, namely
\begin{equation}
\big[\dPta,\dPtap\big]\subset\mathrm{Im}(\delta).
\end{equation}
The fact that this action closes means that there exists a bracket 
on $\sfST$, denoted $\lbr\cdot\,,\cdot\rbr$, which is such that 
\be \label{morphismEYM1}
\big[\dPta,\dPtap\big]=-\delta_{\lbr\Pta,\Ptap\rbr}
\ee 
when acting on functionals of $C$ and $A$, for  $\Pta,\Ptap\in\sfST$. 
In this section we first define the bracket, then prove that it preserves the dual EOM and that it satisfies \eqref{morphismEYM1} and the Jacobi identity.

\subsection{$C$-, YM- and $\sfST$-brackets}\label{secEYM:T-bracket}

For the reader's convenience, we recall the expressions of $[\t{},\tp{}]^C$ and $\paren{\t{},\tp{}}^\gr$, respectively called the $C$-bracket and GR Dali-bracket, both defined in section \ref{sec:celestialToScri}.

\begin{tcolorbox}[colback=beige, colframe=argile]
\textbf{Definition [$C$-bracket]}\\
The $C$-bracket $[\cdot\,,\cdot]^C:\sfTK \times\sfTK\to\sfTK$ is given by 
\bs \label{TbracketEYM}
\begin{align}
    [\tau,\tau']_s^C &:=\sum_{n=0}^{s+1}(n+1)\big(\t{n}D\tp{s+1-n}-\tp{n}D\t{s+1-n}\big)-C\paren{\t{},\tp{}}_s^\gr, \\
    \paren{\t{},\tp{}}_s^\gr &:=(s+3)\big(\t0\tp{s+2}-\tp0\t{s+2}\big).
\end{align}
\es
\end{tcolorbox}

\ni We then introduce the YM-bracket. 
It contains a piece  $[\a{},\ap{}]^\g$ which denotes the algebra part of the $\sfS$-bracket \eqref{Sbracket},
\begin{equation}
\big[\a{},\ap{}\big]^\g_s:=\sum_{n=0}^s\big[\a{n} ,\ap{s-n}\big]_\g. \label{SbracketEYM}
\end{equation}

\begin{tcolorbox}[colback=beige, colframe=argile]
\textbf{Definition [YM-bracket]}\\
The YM-bracket $[\cdot\,,\cdot]^\ym:\sfSTK\times\sfSTK\to\sfSK$  between $\Pta,\Ptap\in\sfSTK$ is defined as
\begin{align}
    \big[\Pta,\Ptap\big]^\ym_s &:= \big[\a{},\ap{}\big]^\g_s+ \bigg\{ \sum_{n=0}^s(n+1) \Big(\t{n}\DYM\ap{s+1-n}-\ap{n+1}D\t{s-n}\Big)-\Pta\leftrightarrow\Ptap\bigg\} \label{YMbracketCFdef}\\
    &\quad -(s+2)C \Big(\t0\ap{s+2}-\tp0\a{s+2}+\a1\tp{s+1}-\ap1\t{s+1}\Big)-F\paren{\t{},\tp{}}^\ym_s, \nn
\end{align}
where the YM Dali-bracket takes the form
\begin{equation}
\paren{\t{},\tp{}}^\ym_s :=(s+2)\big(\t0\tp{s+1}-\tp0\t{s+1}\big) \quad \in~\Ccar{-2,-s-1}.
\end{equation}
\end{tcolorbox}

These brackets are antisymmetric pairings defined off-shell. 
In order to connect with the Noether charge construction and to prove the Jacobi identity we need to impose the dual EOM. 
The on-shell condition for $(\tau,\alpha)$ means that they are field dependent and therefore the brackets need to be promoted to an algebroid bracket with the anchor map $\delta$. 
See \ref{Sec:Liealg} for a reminder about algebroids \cite{algebroid, Ciambelli:2021ujl}. 

\begin{tcolorbox}[colback=beige, colframe=argile, breakable]
\textbf{Definition [$\sfST$-bracket]}\\
The $\sfST$-bracket $\lbr\cdot\,,\cdot\rbr:\sfSTK\times\sfSTK\to\sfSTK$ between $\Pta,\Ptap\in\sfSTK$ is given by
\begin{align} \label{STbracket}
    \big\lbr\Pta,\Ptap\big\rbr=\big[\Pta, \Ptap\big]+\dPtap\Pta-\dPta\Ptap=
    \Big( \big\lbr\Pta,\Ptap\big\rbr^\gr ,\big\lbr\Pta,\Ptap\big\rbr^\ym\Big),
\end{align}
with the projection onto the GR subspace given by the $\sfT$-bracket \eqref{Tbracket}, namely the sum of the $C$-bracket \eqref{TbracketEYM} with the anchor map,
\begin{equation}
    \big\lbr\Pta,\Ptap \big\rbr^\gr:=\lbr\tau,\tau'\rbr =[\tau,\tau']^C+\dtp\tau-\dt\tau',
\end{equation}
and the projection onto the YM subspace given by the sum of the YM-bracket \eqref{YMbracketCFdef} with the EYM anchor map,
\begin{align}
    \big\lbr\Pta,\Ptap\big\rbr^\ym :=\big[\Pta,\Ptap\big]^\ym+\dPtap\a{}-\dPta\ap{}.
\end{align}
\end{tcolorbox}

\paragraph{Remark:}
From \eqref{YMbracketCFdef} and the results of chapter \ref{secGRI}, it is clear that $\t{-1}$ spans the center of the $\sfST$-bracket.\\

The first major property of this bracket is its closure in the $\sfST$-space.

\begin{tcolorbox}[colback=beige, colframe=argile] \label{LemmaClosureEYM}
\textbf{Lemma [$\sfST$-bracket closure]}\\
The $\sfST$-bracket closes, i.e. satisfies the dual EOM \eqref{DualEOMEYM}.
If $\Pta,\Ptap\in\sfST$ then $\E\big(\lbr\Pta,\Ptap\rbr\big)=0$. 
Hence $\lbr\Pta,\Ptap\rbr \in \sfST$.
\end{tcolorbox}

\paragraph{Proof of Lemma [$\sfST$-bracket closure]:}
We showed already in \hyperref[LemmaClosure]{Lemma [$\sfT$-bracket closure]} that if $\tau,\tau'\in\sfT$, then
\begin{equation} \label{dualEOMGRbracket}
    \pa_u\lbr\tau,\tau'\rbr=\DGR \lbr\tau,\tau'\rbr.
\end{equation}
We report the brute force demonstration of
\begin{equation} \label{dualEOMEYMbracket}
    \pa_u\big\lbr\Pta,\Ptap\big\rbr^\ym= \DTOT\big\lbr\Pta,\Ptap\big\rbr,\qquad\Pta,\Ptap \in\sfST
\end{equation}
in App.\,\ref{App:dualEOMEYM}. 
The covariant derivatives are introduced in \eqref{covd}.
The lengthy part is to compute
\begin{align}
    \Big(\big(\pa_u-\Dcal\big)\big[\Pta,\Ptap\big] \Big)^\ym_s=\Big\{\big[\dPta A,\ap{s+1}\big]_\g-(s+2)\ap{s+2}\dt C-(s+2)\tp{s+1}\dPta F\Big\}-\Pta\leftrightarrow\Ptap.
\end{align}
It is then easy to show that 
\begin{equation}
    \Big(\big(\pa_u-\Dcal\big)\big(\dPtap\Pta-\dPta\Ptap\big)\Big)^\ym=-\Big(\big(\pa_u-\Dcal\big)\big[\Pta,\Ptap\big]\Big)^\ym,
\end{equation}
so that \eqref{dualEOMEYMbracket} holds.

\paragraph{Remark:}
The time evolution of $\Pta$ amounts to requiring $\lbr\Ham,\Pta\rbr=0$, for $\Ham$ the Hamiltonian, see also the remark around \eqref{HamiltonianConstraint}.
If we had started by the proof of the Jacobi identity for the $\sfST$-bracket, the previous \hyperref[LemmaClosureEYM]{lemma} would be a corollary.\\

One of the main results of this chapter is then the following theorem.

\begin{tcolorbox}[colback=beige, colframe=argile, breakable] \label{theoremSTAlgebroid}
\textbf{Theorem [$\cST$-algebroid]}\\
The space $\cST\equiv\big(\sfST,\lbr\cdot\,,\cdot\rbr,\delta\big)$ equipped with the $\sfST$-bracket \eqref{STbracket} and the anchor map $\delta$,
\vspace{-0.3cm}
\begin{align}
    \delta : \,\,&\cST\to \X(\PS) \nn\\
     &(\tau,\alpha)\mapsto\dta, \label{defdeltaEYM}
\end{align}
is a Lie algebroid over the Einstein-Yang-Mills phase space $\PS$. 
This means in particular that the anchor map is an anti-homomorphism of Lie algebroids, i.e.
\begin{equation}
    \big[\dPta,\dPtap\big]\cdot=-\delta_{\lbr\Pta,\Ptap\rbr}\cdot
\end{equation}
and that the $\sfST$-bracket satisfies the Jacobi identity,\footnote{$\cyc$ indicates an equality valid upon cyclic permutation of both the sides.}
\begin{equation}
    \big\lbr\Pta,\lbr\Ptap,\Ptapp\rbr\big\rbr\cyc 0.
\end{equation}
\end{tcolorbox}

\ni The demonstration of this theorem is split into three lemmas.
We already showed the first one, namely \hyperref[LemmaClosureEYM]{[$\sfST$-bracket closure]}.
The second lemma goes as follows:

\begin{tcolorbox}[colback=beige, colframe=argile] \label{lemmaAlgebroidAction}
\textbf{Lemma [Algebroid action]}\\
The action $\delta$ defined in \eqref{deltaPtaCA} and \eqref{defdeltaEYM} is a realization of the $\sfST$-bracket \eqref{STbracket} onto the phase space $\PS$:
\begin{equation} \label{AlgebroidActionEYM}
    \big[\dPta,\dPtap\big]\cdot=-\delta_{\lbr\Pta,\Ptap\rbr}\cdot.
\end{equation}
\end{tcolorbox}

\paragraph{Proof of Lemma [Algebroid action]:}
The fact that the anchor map restricted to the GR subspace is an anti-homomorphism of Lie algebroids, i.e.
\begin{equation}
    \big[\dPta,\dPtap\big]C=\big[\dt,\dtp\big]C =-\delta_{\lbr\tau,\tau'\rbr}C,
\end{equation}
is already one of the main result of chapter \ref{secGRI}.
Concerning the YM subspace, we report in appendix \ref{AppEYM:AlgAct} the demonstration of the morphism property 
\begin{equation} \label{AlgebroidActionEYM1}
    \big[\dPta,\dPtap\big]A=-\delta_{\lbr\Pta,\Ptap\rbr}A.
\end{equation}
for $\Pta,\Ptap\in \sfST$.

The third lemma \hyperref[LemmaJacEYM]{[Jacobi identity]} is given is section \ref{secEYM:JacobiIdentity} and necessitates a careful analysis of the algebraic structure of the $\sfST$-bracket.
We dedicate the entire section \ref{secEYM:BicrossedJacobi} to the latter.
\medskip 

We now come back to the YM-bracket \eqref{YMbracketCFdef} and detail one way to construct it, highlighting the complementarity of the Carrollian and celestial perspectives.
Doing so, we also give a complementary proof of the aforementioned lemma \hyperref[LemmaJacEYM]{[Jacobi identity]}.

\subsection{Kinematical algebra \label{secEYM:YMBracket}}

In this subsection, we provide a simplified construction of  the YM-bracket \eqref{YMbracketCFdef}, as the on-shell restriction of a simple \emph{kinematical} bracket.

Let us start with the pure gravitational analysis.
In \eqref{CFbrackettauDcal}, we showed that the $C$-bracket \eqref{TbracketEYM} can equivalently be written as
\begin{equation}\label{CbracketDGRdefEYM}
    [\tau,\tau']_s^C =\sum_{n=0}^{s+1}(n+1)\big(\t{n}(\DGR\tp{})_{s-n}-\tp{n}(\DGR\tau)_{s-n}\big)
\end{equation}
with the covariant derivative defined in \eqref{covd}.
This was a consequence of the identity\footnote{This is obvious upon changing $ n \to s+2-n$.}
\begin{align}
    &\sum_{n=0}^{s+2}(n+1)(s+3-n)\big(\t{n}\tp{s+2-n}-\tp{n}\t{s+2-n}\big)=0 \nn\\*
    \Leftrightarrow \quad & \sum_{n=0}^{s+1}(n+1)(s+3-n)\big(\t{n}\tp{s+2-n}-\tp{n}\t{s+2-n}\big)=\paren{\t{},\tp{}}_s^\gr.
\end{align}
The remarkable point is that by working on $\scri$ and using the dual EOM $\pa_u\tau=\DGR\tau$, one can replace the derivative onto the sphere by $\pa_u$ so that the $C$-bracket takes the simple form
$[\t{},\tp{}]^C_s=\{\t{},\tp{}\}_s$ with 
\begin{equation}
    \boxed{\big\{\t{},\tp{}\}_s:=\sum_{n=0}^{s+1}(n+1)\big(\t{n}\pa_u\tp{s-n} -\tp{n}\pa_u\t{s-n}\big) }.\label{CbracketScridef}
\end{equation}
In appendix \ref{AppEYM:JacobiScri} we show that $\{\tau,\tp{}\}$ is a Lie bracket, namely
\begin{equation} \label{JacobiKGRbracket}
    \big\{\tau,\{\tp{},\tpp{}\}\big\}\cyc 0.
\end{equation}
This means that the kinematical algebra 
$\cTK \equiv\big(\sfTK,\{\cdot\,,\cdot\}\big)$ is a Lie algebra which admits a natural action on $\cSK\equiv\big(\sfSK,[\cdot\,,\cdot]^\g\big)$.

\begin{tcolorbox}[colback=beige, colframe=argile]
\textbf{Definition [$\barWhiteTriangle$ action]}\\
We define the action\! $\barWhiteTriangle$ of GR onto YM by
\begin{align}
    \barWhiteTriangle :\sfTK &\to\mathrm{End}\big( \sfSK\big)\\
    \tau&\mapsto\bWt,
\end{align}
such that
\begin{equation} \label{KinematicalAction}
    \big(\bWt\ap{}\big)_s :=\sum_{n=0}^s(n+1) \Big(\t{n}\pa_u\ap{s-n}-\ap{n+1}\pa_u\t{s-n-1}\Big).
\end{equation}
\end{tcolorbox}

\ni We prove in App.\,\ref{AppEYM:tauBarAction} and \ref{AppEYM:HomtauBarAction} that the action $\!\barWhiteTriangle$ is a derivative and a homomorphism of Lie algebras, i.e.
\begin{equation}
    \barWhiteTriangle :\cTK\to\mathrm{Der}\big(\cSK\big),
\end{equation}
which amounts to the two properties
\bs \label{barWhiteProperties}
\begin{align}
    \bWt\big[\ap{},\app{}\big]^\g &=\big[\bWt\ap{},\app{}\big]^\g+\big[\ap{}, \bWt\app{}\big]^\g,\\
    \big\{\t{},\tp{}\big\} \barWhiteTriangle\app{} &=\big[\bWt,\bWtp\!\big]\app{}.
\end{align}
\es
This shows that $\barWhiteTriangle$ is a semi-direct action, which in turns guarantees that the Jacobi identity holds for $\{\cdot\,,\cdot\}^\ym$ defined as follows (so that $\cSTK=\cTK\bWprod\cSK$ is a Lie algebra).

\begin{tcolorbox}[colback=beige, colframe=argile]
\textbf{Theorem [$\cSTK$-algebra]}\\
$\cTK\equiv\big(\sfTK,\{\cdot\,,\cdot\}\big)$ and $\cSK\equiv\big(\sfSK,[\cdot\,,\cdot]^\g\big)$, respectively equipped with the brackets \eqref{CbracketScridef} and \eqref{SbracketEYM} are Lie algebras.
Moreover, $\cSTK\equiv\big(\sfSTK,\{\cdot\,,\cdot\}\big)$, equipped with the  kinematical $\sfK$-bracket $\{\cdot\, ,\cdot\}:\sfSTK\times\sfSTK\to\sfSTK$  defined as
\begin{equation}
    \{\Pta,\Ptap\}:=\big(\{\t{},\tp{}\},\{\Pta,\Ptap \}^\ym\big), \label{Kbracket}
\end{equation}
where the GR projection is given by \eqref{CbracketScridef} and the YM projection takes the form 
\begin{align}
    \{\Pta,\Ptap \}^\ym:=[\a{},\ap{}]^\g+\bWt\ap{}-\bWtp\a{}, \label{YMbracketScridef}
\end{align}
is the  semi-direct product Lie algebra 
\begin{equation}
    \cSTK=\cTK\bWprod\cSK. \label{Kalgebra}
\end{equation}
\end{tcolorbox}
\medskip

We now establish that the Carrollian kinematical bracket projected on-shell is the same as the celestial bracket\footnote{Celestial in the sense that it involves differential operators onto the sphere. The relation with the usual celestial $sw_{1+\infty}$ bracket is given in Sec.\,\ref{secEYM:celestialBracket}.} defined earlier.

\begin{tcolorbox}[colback=beige, colframe=argile]
\textbf{Theorem [on-shell $\sfK$-bracket]} \label{KbracketTheorem} \\
The on-shell projection of the $\sfK$-bracket \eqref{Kbracket}, i.e. its restriction to the $\sfST$-space, is the $\sfST$-bracket \eqref{STbracket}:
\begin{equation}
   \big[\Pta, \Ptap\big]= \big\{\Pta,\Ptap\big\}\qquad \textrm{if}\quad\Pta,\Ptap\in \sfST,
\end{equation} and therefore the algebroid brackets coincide on-shell
\begin{equation}
   \big\lbr\Pta, \Ptap\big\rbr= \big\{\Pta,\Ptap\big\} +\dPtap\Pta-\dPta\Ptap \qquad \textrm{if}\quad\Pta,\Ptap\in \sfST.
\end{equation}
\end{tcolorbox}

\paragraph{Proof of Theorem [on-shell $\sfK$-bracket]:}
We start by showing how to get \eqref{YMbracketCFdef} from \eqref{YMbracketScridef} after the  on-shell projection.
For this we use the dual EOM in the form \eqref{dEOM} and the definition \eqref{covd} of the covariant derivatives in order to recast the $\sfK$-bracket as a bracket involving derivative operators over the sphere.
Hence, when $\gamma,\gamma'\in \sfST$ we have 
\begin{equation}
    \boxed{\big\{\Pta,\Ptap\big\}^\ym_s\overset{\sfST}{=} \big[\a{},\ap{}\big]^\g_s+\bigg\{\sum_{n=0}^s(n+1) \Big(\t{n}(\DTOT\Ptap)_{s-n}-\ap{n+1}(\DGR\t{})_{s-n-1}\Big)-\Pta\leftrightarrow\Ptap\bigg\}}. \label{YMbracketDTOTdef}
\end{equation} 
Next we expand $\Dcal\Pta$ and leverage two identities.
The first one is that
\begin{equation}
    \sum_{n=0}^{s+1}(n+1)(s+2-n)\big(\t{n}\tp{s+1-n}-\tp{n}\t{s+1-n}\big)=0,
\end{equation}
or equivalently 
\begin{equation}
    \sum_{n=0}^{s}(n+1)(s+2-n)\big(\t{n}\tp{s+1-n}-\tp{n}\t{s+1-n}\big)=(s+2)\big(\t0\tp{s+1}-\tp0\t{s+1}\big)=\paren{\tau,\tau'}^\ym_s.
\end{equation}
The second one is
\begin{equation}
    \sum_{n=0}^{s+1}(n+1)(s+2-n)\big(\t{n}\ap{s+2-n}-\ap{n+1}\t{s+1-n}\big)=0,
\end{equation}
which amounts to
\begin{equation}
    C\sum_{n=0}^s(n+1)(s+2-n)\big(\t{n}\ap{s+2-n}-\ap{n+1}\t{s+1-n}\big)=(s+2)C \big(\t0\ap{s+2}-\ap1\t{s+1}\big).
\end{equation}
From there we deduce that
\begin{align}
    \big\{\Pta,\Ptap\big\}^\ym_s &\overset{\sfST}{=} \big[\a{},\ap{}\big]_s+ \bigg\{ \sum_{n=0}^s(n+1) \Big(\t{n}\DYM\ap{s+1-n}-\ap{n+1}D\t{s-n}\Big)-\Pta\leftrightarrow\Ptap\bigg\} \\
    &\qquad-(s+2)C \Big(\t0\ap{s+2}-\tp0\a{s+2}+\a1\tp{s+1}-\ap1\t{s+1}\Big)-F\paren{\t{},\tp{}}^\ym_s, \nn
\end{align}
which is indeed the YM-bracket \eqref{YMbracketCFdef}.
Similarly, we already showed in the discussion at the beginning of this subsection that $\{\tau,\tp{}\} \overset{\sfST}{=}[\tau,\tp{}]^C$.
The general structure of Lie algebroid ensures that there is a unique choice of anchor map $\delta_\gamma$ on $(C,A)$ which preserves the dual EOM. 
This phenomenon is generic in the context of gauge fixing \cite{Barnich:2009se, Barnich:2011mi, Freidel:2021fxf, Riello:2024uvs} where an algebra is promoted after gauge fixing to an algebroid which preserves the gauge fixing in question.
Once this choice is made the bracket is an algebroid bracket by construction.

Let us illustrate how this works  on the gravitational sector.
It is useful to introduce the concept of Leibniz rule anomaly of a bracket $[\cdot\,,\cdot]$ with respect to a differential operator $\mathscr{D}$, encoded in the expression 
\begin{equation} \label{LeibnizRuleNotation}
    \cA\big([\cdot\,,\cdot],\mathscr{D} \big):=\mathscr{D}[\cdot\,,\cdot]-[\mathscr{D}\cdot\,,\cdot]-[\cdot\,,\mathscr{D}\cdot].
\end{equation}
We find that
\begin{equation}
    \cA\Big(\big\{\tau,\tp{}\big\},\pa_u-\DGR\Big)=-\cA\Big(\big\{\tau,\tp{}\big\},\DGR\Big),
\end{equation}
where
\begin{equation}\label{Leibnizgr}
    \!\!\cA_s\Big(\big\{\tau,\tp{}\big\},\DGR\Big) =\left((s+3) \t{s+2} \pa_u\big(\DGR\tp{} \big)_{-2}+ \sum_{n=-1}^{s+1} \big(\DGR\tau\big)_n\big(\pa_u\tau'-\DGR\tau'\big)_{s-n}\right)-\tau\leftrightarrow\tau'.
\end{equation}
This clearly shows that even when $\tau,\tp{}\in\sfT$, such that $\pa_u\tau=\DGR\tau$, we are still left with
\begin{equation}
    \Big(\big(\pa_u-\DGR\big)\big\{\tau,\tp{}\big\} \Big)_{\!s} \overset{\sfT}=(s+3) \left[\tp{s+2} \pa_u\big(\DGR\t{} \big)_{-2}-\t{s+2} \pa_u\big(\DGR\tp{} \big)_{-2}\right],
\end{equation}
where $\big(\DGR\t{} \big)_{-2}=D\t{-1}-C\t{0}$.
The action of $\pa_u-\DGR$ is thus anomalous on the \emph{Lie bracket}.
The algebroid correction provides a counter-term. 
Indeed it is direct to check that 
\begin{equation}
    \Big(\big(\pa_u-\DGR\big)\big(\dtp\tau-\dt\tp{}\big)\Big)_{\!s}\overset{\sfT}=(s+3)\left(\tp{s+2}\dt C-\t{s+2}\dtp C\right).
\end{equation}
This means that for $\tau,\tau'\in\sfT$,\footnote{Albeit presented differently, this computation was done already in Sec.\,\ref{sec:TimeEvol}.}
\begin{equation}
    \Big(\big(\pa_u-\DGR\big)\big\lbr\tau, \tp{}\big\rbr\Big)_{\!s}=(s+3)\tp{s+2}\Big(\dt C +\pa_u\big(\DGR\tau\big)_{-2}\Big)-\tau\leftrightarrow\tp{}.
\end{equation}
Requiring that $\dt C=-\pa_u\big(\DGR\tau\big)_{-2}$ in order for the RHS of this last equation to vanish can be viewed as a \emph{definition} of $\dt C$. 
This definition matches the one given in \eqref{deltaPtaCA} since  
\be 
\pa_u\big(\DGR\tau\big)_{-2} &=  
D\pa_u\t{-1}-C\pa_u\t0 - N\t0 \cr
&\overset{\sfT}= D^2\t0 - 2D(C\t1)- C D\t1 +3C^2 \t2 -N\t0\equiv -\dt C,
\ee
where we use the dual EOM $\E_{-1}(\tau)= \E_0(\tau)=0$ and the definition of the action \eqref{deltaPtaCA} in the last equality (see also \eqref{dthdtC}). 
The story is similar for the YM projection of the $\sfK$-bracket.
This thus concludes the proof that 
\begin{equation}
    \big\{\Pta,\Ptap\big\}+ \dPtap\Pta-\dPta\Ptap \overset{\sfST}=\big\lbr\Pta, \Ptap\big\rbr.
\end{equation}

\paragraph{Remark:}
The various ways \eqref{TbracketEYM} or  \eqref{CbracketScridef} and \eqref{YMbracketCFdef} or \eqref{YMbracketScridef} of writing the $C$- and YM-brackets each have pros and cons. 
On one hand \eqref{TbracketEYM}, \eqref{YMbracketCFdef} and \eqref{CbracketDGRdefEYM}, \eqref{YMbracketDTOTdef} emphasize the \emph{corner algebra} structure, since only derivative operators onto the sphere appear.
Once projected at a cut of $\scri$, the bracket keeps the same functional form with the Carrollian parameters $(\t{},\a{})$ replaced by the celestial ones $(T,\Aa{})$.
Moreover, as we will see in section \ref{secEYM:celestialBracket}, once we impose the wedge condition, the resulting algebra is a deformation of the $sw_{1+\infty}$ algebra studied in \cite{Agrawal:2024sju}, where the radiative data play the role of deformation parameters.
We expect  the deformed algebra to be isomorphic to the undeformed one, as this was the case in gravity, cf.\,\ref{sec:CovWedgeSol}.\footnote{Note that the brackets \eqref{TbracketEYM}, \eqref{YMbracketCFdef} do not satisfy the Jacobi identity outside of the wedge, see \eqref{JacobiEYMCbracket} and \eqref{JacobiEYMYMbracket} below (and chapter \ref{secGRI}).}
On the other hand, \eqref{CbracketScridef} and \eqref{YMbracketScridef} emphasize the simplicity of the brackets once expressed on $\scri$, fully committing to a Carrollian picture, where this is now the derivative along the null generator which appears explicitly. 
These brackets satisfy the Jacobi identity off-shell and exhibit the semi-direct product structure of the off-shell algebra. 
The algebroid arises naturally by imposing the dual EOM, which mimics a gauge fixing procedure.

\section{Bicrossed product and Jacobi identity \label{secEYM:BicrossedJacobi}}

In the previous section, we showed that the kinematical $\sfK$-bracket $\{\cdot\,,\cdot\}$ satisfies the Jacobi identity and projects to the $\sfST$-bracket on shell of the dual EOM.
We also showed that the $\cSTK$-algebra possesses a semi-direct product structure \eqref{Kalgebra}.
Unfortunately once we impose the dual EOM, the semi-direct product structure no longer holds on the $\sfST$-space. 

In this section, we show that the $\cST$-algebroid is instead built as the combination of a bicrossed product \cite{Majid_2002, michor1992knitproductsgradedlie} and a semi-direct product of Lie sub-algebroids.
Unraveling this algebraic structure also provides a direct proof that the Jacobi identity holds for the $\sfST$-bracket.
If the reader wishes to have a glance at the results, the main theorem is \hyperref[STisomorphismBis]{[Bicrossed product structure]}.

\subsection{Bicrossed product structure \label{secEYM:Bicrossed}}

We already mentioned the two vector subspaces $\sfS=\sfST\cap\sfSK$ and $\sfT=\sfST\cap\sfTK$.
Let us introduce the $\sfS$-bracket, which generalizes \eqref{Sbracket} when the anchor map depends on the shear.

\begin{tcolorbox}[colback=beige, colframe=argile]
\textbf{Definition [$\sfS$-bracket]}\\
The $\sfS$-bracket $\lbr\a{},\ap{}\rbr^\g =[\a{},\ap{}]^\g+\hdap\a{}-\hda\ap{}$ is simply the $\sfST$-bracket restricted to $\sfS$. 
It naturally generalizes the one introduced in \ref{secNAGTII}, where the only difference appears in the usage of the anchor $\hda$, satisfying 
\vspace{-0.6cm}

\be \label{hatdaA}
\hda A=-\DYM\a0+C\a1.
\ee 
\vspace{-0.5cm}

$\cS\equiv\big(\sfS,\lbr\cdot\,,\cdot\rbr^\g,\hat\delta\big)$ is the $\cS$-algebroid in the presence of a non-zero shear. 
\end{tcolorbox}

\ni We can readily check that the $\cS$-algebroid is an ideal. 
Indeed, taking $\Ptap=(\tp{},\ap{})$, we have that
\begin{equation}
    \big\lbr(0,\a{}),\Ptap\big\rbr=\Big(0,\big\lbr (0,\a{}),\Ptap\big\rbr^\ym\Big).
\end{equation}
However the $\cT$-algebroid is \emph{not} a sub-algebroid! 
This is clear since 
\begin{equation} \label{anomalyTsubAlgebroid}
    \big[(\t{},0),(\tp{},0)\big]= \Big([\tau,\tau']^C,-F\paren{\tau,\tau'}^\ym\Big).
\end{equation}
In other words, the Dali-bracket prevents us from having a semi-direct action of $\cT$ onto $\cS$.
The situation is thus subtler.
To address it, we define $\osfT$ and $\bsfT$ as follows.

\begin{tcolorbox}[colback=beige, colframe=argile, breakable]
\textbf{Definition [$\osfT$- \& $\bsfT$-subspaces]}

%\vspace{-0.5cm}
\bs \label{defSpaceobar}
\begin{align}
    \osfT &:=\sfST\cap\big( \Ccar{-1,1}\oplus\Ccar{-1,0}\big), \\
    \bsfT &:=\sfST\cap\bigg(\bigoplus_{s=1}^\infty \Ccar{-1,-s}\bigg).
\end{align}
\es

$\osfT$ is the subspace of ``Carrollian super-translations'' together with the center. 
We denote its elements as $\ot\equiv(\t{-1},\t0,0,0,\ldots)$.
$\bsfT$ is the subspace of ``Carrollian sphere diffeomorphisms'' and higher-spin parameters. 
We denote its elements by $\bt\equiv(0,0,\t1,\t2,\t3,\ldots)$.
\end{tcolorbox}

\ni For convenience we denote an element of $\bsfT\oplus\sfS$ as $\bPta=(\bt,\a{})$ and an element of $\sfT$ as $\t{}=(\ot;\bt)$.

\begin{tcolorbox}[colback=beige, colframe=argile]
\textbf{Definition [$\circ$- \& bar-brackets]}\\
We define the $\circ$-bracket $\big\lbr\ot,\otp\big\rbr^\circ=\big[\ot,\otp \big]^\circ+\odtp\ot-\odt\otp$ and the bar-bracket $\overline{\lbr\bt,\btp\rbr}=\overline{[\bt,\btp]} +\dbtp\bt-\dbt\btp$, respectively as the restriction of the $\sfST$-bracket to $\osfT$ and $\bsfT$ :
\vspace{-0.5cm}

\begin{adjustwidth}{-0.4cm}{-0.1cm}
\bs \label{defBracketobar}
\begin{align}
    \lbr\cdot\,,\cdot\rbr^\circ&:\osfT\times \osfT\to\osfT \!\!\qquad\textrm{with}\qquad\! \big[\ot,\otp\big]^\circ= \big(\t0D\tp0-\tp0D\t0\,,0,0,\ldots\big), \\
    \overline{\lbr\cdot\,,\cdot\rbr} \,&:\bsfT\times \bsfT\to \bsfT\!\!\qquad\!\textrm{with}\mkern6mu\qquad\! \overline{[\bt,\btp]}_s=\sum_{n=1}^s(n+1)\big(\t{n}D\tp{s+1-n}-\tp{n}D\t{s+1-n}\big).
\end{align}
\es
\end{adjustwidth}
\end{tcolorbox}

\ni Since $\paren{\bt,\btp}^\ym=0$, we infer that $\big\lbr(\bt,0),(\btp,0)\big\rbr= \big(\overline{\lbr\bt,\btp\rbr},0\big)$, so that $\bcT\equiv\big(\bsfT,\overline{\lbr\cdot\,,\cdot \rbr},\delta\big)$ is a sub-algebroid.
Similarly, since $\paren{\ot,\otp}^\ym=0$, we deduce that $\big\lbr(\ot,0),(\otp,0)\big\rbr= \big(\lbr\ot,\otp\rbr^\circ,0\big)$, so that $\ocT\equiv\big(\osfT,\lbr\cdot\,,\cdot\rbr^\circ, \delta\big)$ is a sub-algebroid of $\cST$ as well.
Notice that $\big\lbr(\bt,0),\Ptap\big\rbr^\ym\neq 0$ and $\big\lbr(\ot,0),\Ptap\big\rbr^\ym\neq 0$, so that neither $\bcT$ nor $\ocT$ are ideals.
Finally, we see that $\big\lbr\bPta,\bPtap \big\rbr^\gr=\big(0;\overline{\lbr\bt,\btp\rbr}\big)\in\bsfT$, which proves that $\big(\bsfT\oplus\sfS,\lbr\cdot\,,\cdot\rbr,\delta \big)$\footnote{Here it is understood that $\lbr\cdot\,,\cdot\rbr$ is the restriction of the $\sfST$-bracket to the subspace $\bsfT\oplus\sfS$.
We do not introduce any new notation for this bracket. 
The context should always make it clear which one we consider.} is itself a sub-algebroid of $\cST$ (though it is not an ideal).

We now define several actions and a co-reaction to describe the algebroid structure. The white action $\vartriangleright$ allows to construct the semi-direct product of algebroids $\bcT\Wprod\cS$.
We also introduce the black action $\blacktriangleright$ and the black back-reaction $\blacktriangleleft$ to form the bicrossed product \cite{michor1992knitproductsgradedlie} $\ocT\Bprod\big(\bcT\Wprod\cS\big)$.

\begin{tcolorbox}[colback=beige, colframe=argile, breakable]
\textbf{Definition [$\vartriangleright,\blacktriangleright, \blacktriangleleft$ Actions]}\\
We define 
\begin{align} \label{defAction2}
    \vartriangleright\,&:\bsfT_{\mkern-3mu K}\to\textrm{End} \big( \sfSK\big)&   \blacktriangleright\,&: \osfT_{\mkern-3mu K}\to\textrm{End} \big(\bsfT_{\mkern-3mu K} \oplus\sfSK \big)&  \blacktriangleleft\,&:\bsfT_{\mkern-3mu K}\oplus \sfSK\to\textrm{End}(\osfT_{\mkern-3mu K}) \\
    &:\bt\mapsto\Wt, &  &:\ot\mapsto\Bt, &  &: \bPta\mapsto\,\,\blacktriangleleft\bPta, \nn
\end{align}
such that\footnote{The first 3 lines are valid for $s\geqslant 0, s\geqslant 1$ and $s\geqslant 0$ respectively.}
\vspace{-0.7cm}

\begin{adjustwidth}{-0.3cm}{-0.1cm}
\bs \label{defAction}
\begin{align} 
    \big(\Wt\ap{}\big)_s \,&:=\sum_{n=1}^s\Big( (n+1)\t{n}\DYM\ap{s+1-n}-n\ap{n}D\t{s+1-n}\Big)+(s+2)C\ap1\t{s+1}, \label{defWhiteAction} \\
    \big(\Bt\bPtap\big)_s^\gr &:=\t0 D\tp{s+1}-(s+2)\tp{s+1}D\t0-(s+3)C\t0\tp{s+2}, \label{defBlackActionGR} \\
    \big(\Bt\bPtap\big)_s^\ym &:=\t0\DYM\ap{s+1}-(s+1)\ap{s+1}D\t0-(s+2)C\t0\ap{s+2}-(s+2)F\t0\tp{s+1}, \label{defBlackActionYM} \\
    \big(\rBt\bPtap\big)_{-1} &:=-2C\t0\tp1, \label{defReverseBlackAction-1} \\
    \big(\rBt\bPtap\big)_0 \,&:=\t0D\tp1-2\tp1D\t0-3C\t0\tp2. \label{defReverseBlackAction0}
\end{align}
\es
\end{adjustwidth}
\end{tcolorbox}

\paragraph{Remark:} 
Notice that if we restrict ourselves to $\gbms$ transformations, i.e. spin 0 and 1, then $\Bt\bPtap\equiv 0$ and the entire $\gbms$ bracket comes from $\big(\rBt\bPtap\big)_0$ after setting the center, namely the degree $-1$ element, to 0.\\

One of the main result of this chapter is the following theorem, see also \eqref{STisomorphismBis}.\footnote{We could also have written the algebroid as a different cross-product structure, of the type $\bcT\,'\!\Bprod\!'\big(\ocT\,'\!\Wprod\cS\big)$ for some primed actions related to the unprimed ones.}

\begin{tcolorbox}[colback=beige, colframe=argile]
\textbf{Theorem [Bicrossed product structure]}\\
\vspace{-0.6cm}

\begin{equation} \label{STisomorphism}
\cST\simeq\ocT\Bprod\big(\bcT\Wprod\cS\big),
\end{equation}
\vspace{-0.6cm}

such that in particular
\vspace{-0.6cm}

\begin{align}
    \big[\Pta,\Ptap\big] &=\Big(\big[ \ot,\otp\big]^\circ+\rBt\bPtap-\rBtp\bPta\,;\overline{\big[\bt,\btp\big]}+ \big(\Bt\bPtap-\Btp\bPta\big)^\gr, \label{bicrossedBracket}\\
    &\qquad\big[\a{},\ap{}\big]^\g+\Wt\ap{}-\Wtp\a{}+ \big(\Bt\bPtap-\Btp\bPta\big)^\ym\Big).\nn
\end{align}
\end{tcolorbox}

\paragraph{Proof:} 
By definition of a bicrossed product bracket \cite{michor1992knitproductsgradedlie}, we have that
\begin{align}
    \big[(\ot,\bPta),(\otp,\bPtap)\big]=\Big(\big[ \ot,\otp\big]^\circ+\rBt\bPtap-\rBtp\bPta\,, \big[\bPta,\bPtap\big]+\Bt\bPtap-\Btp\bPta\Big). 
\end{align}
Using the semi-direct structure $\bcT\Wprod\cS$, we can further expand the bracket as
\begin{align}
    \big[(\ot,\bPta),(\otp,\bPtap)\big] &= \big[(\ot;\bt,\a{}),(\otp;\btp,\ap{})\big] \nn\\
    &=\Big(\big[ \ot,\otp\big]^\circ+\rBt\bPtap-\rBtp\bPta\,;\overline{\big[\bt,\btp\big]}+ \big(\Bt\bPtap-\Btp\bPta\big)^\gr, \label{bicrossedBracket1}\\
    &\qquad\big[\a{},\ap{}\big]^\g+\Wt\ap{}-\Wtp\a{}+ \big(\Bt\bPtap-\Btp\bPta\big)^\ym\Big). \nn
\end{align}
From the definitions \eqref{defBracketobar} and \eqref{defAction} of the various brackets and actions, we find by inspection that \eqref{bicrossedBracket1} is indeed equal to $\big[\Pta,\Ptap\big]$ as defined in \eqref{STbracket}.

\paragraph{Remark:}
At this stage, \eqref{STisomorphism} is technically valid only if the anchor map $\delta$ is identically vanishing.
Indeed, we only expressed $[\Pta,\Ptap]$, and not $\lbr\Pta,\Ptap\rbr$, in terms of the bicrossed product.
We made that choice to highlight a feature of the proof of the Jacobi identity of the $\sfST$-bracket, that we address in the next subsection. 
For completeness, we report the algebroid version of this discussion in section \ref{secEYM:BicrossedMathy}.
On top of that, all the computations of the next subsection are necessary for \ref{secEYM:BicrossedMathy} anyway.

\subsection{Jacobi identity \label{secEYM:JacobiIdentity}}

We can now come back to the proof of the Jacobi identity for the $\sfST$-bracket.
It is a well-known fact that the Jacobi identity for a bracket constructed via bicrossed (or semi-direct) product amounts to a list of properties that the actions have to satisfy (we recall the proof in what follows).
For instance, the semi-direct action $\vartriangleright$ has to be a derivative operator on the product over the $\sfS$-space, namely the bracket $[\cdot\,,\cdot]^\g$.
$\vartriangleright$ must also be a Lie algebra homomorphism.
As we just mentioned in the previous remark, at this stage we do not include the anchor map part of the brackets. 
We thus proceed as follows: we check all the properties that the actions must satisfy. 
Most of them are ``anomalous'', where the anomaly is the difference between what the property is and what the property should be in a traditional Lie algebra setting \cite{Majid_2002, michor1992knitproductsgradedlie}.
We show that these anomalies can always be expressed in terms of the action of the anchor map onto the deformation parameters, i.e. the infinitesimal variations of $C,A$ and $F$.
We can then easily compute the Jacobi identity for the algebra part of the $\sfST$-bracket, namely $\big[\Pta,\Ptap\big]$ and see how the various anomalies recombine to define a \emph{Jacobi identity anomaly}.
Next, we show that the latter is equal to the \emph{Leibniz anomaly} of the anchor onto $\big[\Pta,\Ptap\big]$. 
This fact then allows us to conclude that the Jacobi identity for the algebroid $\sfST$-bracket $\big\lbr\Pta,\Ptap\big\rbr$ holds.
Recall that the Leibniz anomaly is defined in \eqref{LeibnizRuleNotation}.

\paragraph{Properties of $\vartriangleright$:}
We can show that $\Wt$ is a derivative operator up to an anomaly (see appendix \ref{AppEYM:tauAction}):
\begin{equation} \label{LeibnizAnomalyTauAction}
    \boxed{\cA_s\big([\ap{},\app{}]^\g,\Wt\!\big)
    =-\sum_{n=1}^s (n+1)\t{n}\big[\hdap A,\app{s+1-n}\big]_\g-(\ap{}\leftrightarrow\app{})}.
\end{equation}
Recall that $\hda A=-\DYM\a0+C\a1$.
The action $\vartriangleright$ is also a homomorphism of Lie algebras up to an anomaly (see appendix \ref{AppEYM:tauHom})
\begin{equation}
    \boxed{\Big(\overline{[\bt{},\btp{}]} \vartriangleright \app{}-\big[\Wt,\Wtp\!\big]\app{} \Big)_s=(s+2)\app1\Big(\t{s+1}\dbtp C-\tp{s+1}\dbt C\Big)}, \label{HomAnomalyTauAction}
\end{equation}
where $\dbt C$ is the variation $\dt C$ restricted to the $\bsfT$-subspace, i.e.
\begin{equation}
    \dbt C=2DC\t1+3CD\t1-3C^2\t2.
\end{equation}

\paragraph{Properties of $\blacktriangleright$:}
Similarly for the black action, we get (see App.\,\ref{AppEYM:BlackTauActionLeibnizGR})
\begin{equation} \label{LeibnizAnomalyBlackTauActionGR}
    \boxed{\cA_s^\gr\Big(\big[\bPtap,\bPtapp \big], \Bt\!\Big)=\Big\{ \big((\rBt\bPtap)\blacktriangleright \bPtapp\big)^\gr_s+(s+3)\t0\tpp{s+2} \dbtp C\Big\}-\bPtap\leftrightarrow\bPtapp}
\end{equation}
and (see App.\,\ref{AppEYM:BlackTauActionLeibnizYM})
\vspace{-0.4cm}

\begin{adjustwidth}{-0.3cm}{-0.1cm}
\begin{empheq}[box=\fbox]{align}
    \cA_s^\ym\Big(\big[\bPtap,\bPtapp \big], \Bt\!\Big) &=\bigg\{\big((\rBt\bPtap)\blacktriangleright \bPtapp\big)^\ym_s+(s+2)\t0\app{s+2}\dbtp C-(s+2)\ap1\tpp{s+1}\odt C \nn\\
    &\quad -\t0 \big[\hdap A,\app{s+1}\big]_\g+(s+2)\t0\tpp{s+1}\dbPtap F \label{LeibnizAnomalyBlackTauActionYM}\\
    &\quad +\sum_{n=1}^{s}(n+1)\tp{n}\big[\odt A, \app{s+1-n}\big]_\g\bigg\}-\bPtap\leftrightarrow\bPtapp, \nn
\end{empheq}
\end{adjustwidth}
where $\odt C=-D^2\t0+N\t0$ and (using the dual EOM \eqref{DualEOMEYM})
\begin{align}
    \dPta F=\pa_u(\dPta A)&=-\DYM^2\a1+3C\DYM\a2+ 2DC\a2-3C^2\a3-[F,\a0]_\g+2\DYM F\t1 \nn\\
    &+3FD\t1+N\a1-6CF\t2+\dot F\t0.
\end{align}
$\dbPta F$ is just $\dPta F$ with $\t0=0$.
These Leibniz rule anomalies contain 2 different parts.
The term $(\rBt\bPtap)\blacktriangleright \bPtapp$ mixing the black action and co-reaction is a classic feature of a bicrossed product \cite{michor1992knitproductsgradedlie}.
The other part of the anomalies involves the infinitesimal variation of the asymptotic data $C,A$ and $F$.
Concerning the homomorphism anomaly, we show in appendix \ref{AppEYM:BlackTauHom} that
\begin{equation} \label{HomAnomalyBlackTauActionYM}
    \boxed{\Big(\big[\ot,\otp\big]^\circ \blacktriangleright\bPtapp-\big[\Bt,\Btp\!\big]\bPtapp\Big)^\ym_s=(s+2)\app{s+2}\Big(\tp0\odt C-\t0\odtp C\Big)}.
\end{equation}
Similarly, the projection onto the $\bsfT$-subspace is
\begin{equation} \label{HomAnomalyBlackTauActionGR}
    \boxed{\Big(\big[\ot,\otp\big]^\circ \blacktriangleright\bPtapp-\big[\Bt,\Btp\!\big]\bPtapp\Big)^\gr_s=(s+3)\tpp{s+2}\Big(\tp0\odt C-\t0\odtp C\Big)}.
\end{equation}

\paragraph{Properties of $\blacktriangleleft$:}
The Leibniz rule anomaly at degree $-1$ is (proof in App.\,\ref{AppEYM:BackBlackTauLeibniz})
\bs \label{LeibnizAnomalyrBlackTauAction}
\begin{equation}
    \boxed{\cA_{-1}\Big(\big[\otp,\otpp \big]^\circ, \blacktriangleleft\bPta \Big)=\Big\{ \big(\rBtp(\Btpp\bPta)\big)_{-1}+2\t1\tpp0 \odtp C\Big\}-\otp\leftrightarrow\otpp},
\end{equation}
while the degree 0 is anomaly free,
\begin{equation}
     \boxed{\cA_0\Big(\big[\otp,\otpp \big]^\circ, \blacktriangleleft\bPta \Big)=0}.
\end{equation}
\es
Finally, the homomorphism anomaly takes the form (proof in App.\,\ref{AppEYM:BackBlackTauHom})
\begin{equation} \label{HomAnomalyrBlackTauAction}
    \boxed{\Big(\rBtpp[\bPta,\bPtap]-\otpp\big[\!\blacktriangleleft\bPta, \blacktriangleleft\bPtap\big] \Big)_{s}= (s+3)\tpp0\big(\tp{s+2}\dbt C-\t{s+2}\dbtp C\big)},\qquad s=-1,0.
\end{equation}

\paragraph{Jacobi identity anomaly:}
As we have proven in \eqref{JacCFC}, the GR projection of the Jacobi anomaly of the bracket $[\Pta,\Ptap]$ is given by
\begin{equation} \label{JacobiEYMCbracket}
    \big[\Pta,[\Ptap,\Ptapp]\big]^\gr_s=\big[\tau,[\tau',\tau'']^C\big]^C_s \cyc-\dt C\paren{\tp{},\tpp{}}^\gr_s.
\end{equation}
Besides, the Jacobi anomaly of the YM-bracket takes the form
\vspace{-0.5cm}

\begin{adjustwidth}{-0.2cm}{0cm}
\begin{empheq}[box=\fbox]{align}
    \big[\Pta,[\Ptap,\Ptapp]\big]^\ym_s &\cyc \sum_{n=0}^s(n+1)\Big(\tp{n}\big[\dPta A,\app{s+1-n}\big]_\g-\tpp{n}\big[\dPta A,\ap{s+1-n}\big]_\g\Big) \label{JacobiEYMYMbracket}\\
    &-(s+2)\dt C\Big(\tp0\app{s+2}-\tpp0\ap{s+2}+ \ap1\tpp{s+1}- \app1\tp{s+1}\Big)-\dPta F\paren{\tp{}, \tpp{}}^\ym_s.  \nn
\end{empheq}
\end{adjustwidth}
We present the proof of this result in appendix \ref{AppEYM:Jacobi} (which at the same time serves as a reminder of the equivalence between demanding the Jacobi identity and requiring the actions $\vartriangleright,\blacktriangleright, \blacktriangleleft$ to satisfy the derivation and homomorphism properties).
\medskip

Next we prove that the Jacobi identity for the $\sfST$-bracket holds if and only if $\delta$ is a morphism, namely a realization of the $\sfST$-bracket onto the phase space $\PS$. 
This is the usual result when one deals with an algebroid bracket of the type ``$\lbr\cdot\,,\cdot\rbr=[\cdot\,,\cdot]+\delta_\cdot\cdot-\,\delta_\cdot\cdot$''.
We had an illustration of that fact in chapter \ref{secNAGTII} when treating the pure YM $\sfS$-bracket.
The subtlety here (and the reason why the proof is non-trivial) is that the $C$- and YM-brackets are not Lie brackets \emph{per se}, in the sense that they have the aforementioned \emph{Jacobi identity anomaly}.
Besides the anchor map $\delta$ is not a derivative operator onto the YM-bracket (nor the $C$-bracket) as it would be in the usual symmetry algebroid setup.
Instead it satisfies a \emph{Leibniz rule anomaly} precisely equal to the Jacobi identity anomaly, so that the two compensate one another.
The essence of the demonstration is thus to establish this cancellation.
We encountered the same mechanism when treating the pure GR case and we refer the reader to remark \ref{remarkBB} and to the upcoming proof for details.

\begin{tcolorbox}[colback=beige, colframe=argile]
\textbf{Lemma [Jacobi identity]} \label{LemmaJacEYM}\\
The $\sfST$-bracket $\lbr\cdot\,,\cdot\rbr$ satisfies Jacobi if and only if  $\delta$ is an algebroid action:
\begin{equation}
    \big\lbr\Pta,\lbr\Ptap,\Ptapp\rbr\big\rbr\cyc 0 \qquad\Leftrightarrow \qquad \big[\dPta,\dPtap\big]+\delta_{\lbr\Pta,\Ptap \rbr}=0. \label{JacobiMorphismEYM}
\end{equation}
\end{tcolorbox}

\paragraph{Proof:}
For any bracket $\lbr\Pta,\Ptap\rbr=[\Pta,\Ptap]+\dPtap\Pta-\dPta\Ptap$, we have that
\begin{align}
    \big\lbr\Pta,\lbr\Ptap,\Ptapp\rbr\big\rbr &=\big[\Pta,\lbr\Ptap,\Ptapp\rbr\big]+ \delta_{\lbr\Ptap,\Ptapp\rbr}\Pta-\dPta\lbr\Ptap,\Ptapp\rbr \nn\\
    &=\big[\Pta,[\Ptap,\Ptapp]\big]+ \big[\Pta,\dPtapp\Ptap\big]-\big[\Pta,\dPtap\Ptapp\big]+ \delta_{\lbr\Ptap,\Ptapp\rbr}\Pta-\dPta\big[\Ptap,\Ptapp\big]-\dPta\dPtapp\Ptap+\dPta\dPtap\Ptapp \nn\\
    &\cyc \big[\Pta,[\Ptap,\Ptapp]\big]-\cA\big([\Ptap,\Ptapp],\dPta\big)+ \delta_{\lbr\Pta,\Ptap\rbr}\Ptapp+\big[\dPta, \dPtap\big]\Ptapp. \label{computationJacobiEYM}
\end{align}
Hence if the Jacobi identity anomaly of the bracket $[\cdot\,,\cdot]$ equates the Leibniz rule anomaly of the anchor onto this bracket, we get the result \eqref{JacobiMorphismEYM}.
In our specific case where $[\Pta,\Ptap]$ reduces to the $C$- and YM-brackets \eqref{TbracketEYM} and \eqref{YMbracketCFdef}, it is clear that 
\begin{equation}\label{LeibnizAnchorEYM}
\cA^\gr\big([\Ptap,\Ptapp],\dPta\big)= \eqref{JacobiEYMCbracket}\qquad\textrm{and}\qquad \cA^\ym\big([\Ptap,\Ptapp],\dPta\big)= \eqref{JacobiEYMYMbracket}.
\end{equation}

The combination of the 3 lemmas \hyperref[LemmaClosureEYM]{[$\sfST$-bracket closure]}, \hyperref[lemmaAlgebroidAction]{[Algebroid action]} and \hyperref[LemmaJacEYM]{[Jacobi identity]} proves the theorem \hyperref[theoremSTAlgebroid]{[$\cST$-algebroid]}.

\subsection{Bicrossed product of Lie algebroids \label{secEYM:BicrossedMathy}}

In this subsection, we come back to the generalization of the bicrossed product to Lie algebroids.
In \eqref{bicrossedBracket} we emphasized the bicrossed product structure of the single staff bracket $[\Pta,\Ptap]$, to which we added by hand the contribution $\dPtap\Pta-\dPta\Ptap$ of the anchor map in order to form the double staff bracket $\lbr\Pta,\Ptap\rbr$. 
Doing so, the Jacobi identity for the $\sfST$-bracket split into 3 easily identifiable contributions, cf.\,\eqref{computationJacobiEYM}.
However, the equality of the Jacobi identity anomaly \eqref{JacobiEYMCbracket}-\eqref{JacobiEYMYMbracket} with the Leibniz rule anomaly $\cA\big(\lbr\Ptap,\Ptapp\rbr,\dPta\big)$ is revealing of the fact that one should be able to define algebroid actions $\dotvartriangleright, \dotblacktriangleright$ and $\dotblacktriangleleft$, denoted with a dot, which are no longer anomalous.

\begin{tcolorbox}[colback=beige, colframe=argile]
\textbf{Definition [$\dotvartriangleright,\dotblacktriangleright, \dotblacktriangleleft$ Actions]}\\
We define\footnote{For simplicity we do not add dots on the products appearing in \eqref{defActionDot} or \eqref{STisomorphismBis}.}
\begin{align} \label{defActionDot}
    \dotvartriangleright\,&:\bcT\to\textrm{Der}(\cS)&   \dotblacktriangleright\,&: \ocT\to\textrm{End}\big(\bcT\Wprod\cS\big) &  \dotblacktriangleleft\,&:\bcT\Wprod\cS\to\textrm{End}(\ocT) \\
    &:\bt\mapsto\dWt, &  &:\ot\mapsto\dBt, &  &: \bPta\mapsto\,\dotblacktriangleleft\,\bPta. \nn
\end{align}
such that
\bs \label{defActionDot2}
\begin{align} 
    \dWt\ap{} &:=\Wt\ap{}-\dbt\ap{}, \\
    \dBt\bPtap &:=\Bt\bPtap-\odt\bPtap, \\
    \drBt\bPtap &:=\rBt\bPtap+\dbtp\ot.
\end{align}
\es
\end{tcolorbox}

\begin{tcolorbox}[colback=beige, colframe=argile]
\textbf{Lemma [Properties of $\dotvartriangleright,\dotblacktriangleright, \dotblacktriangleleft$]}\\
The dotted white action is a semi-direct action and thus satisfies the following derivative and homomorphism properties:
\bs \label{dotWhiteProperties}
\begin{align}
    \dWt\big\lbr\ap{},\app{}\big\rbr^\g &=\big\lbr\dWt\ap{},\app{}\big\rbr^\g+\big\lbr\ap{}, \dWt\app{}\big\rbr^\g,\\
    \overline{\big\lbr\bt,\btp\big\rbr} \mkern5mu\dotvartriangleright\mkern5mu\app{} &=\big[\dWt,\dWtp\!\big]\app{}.
\end{align}
\es
The dotted black action and back-reaction are bicrossed product actions and thus satisfy the following derivative and homomorphism properties characteristic of bicrossed products \cite{michor1992knitproductsgradedlie} (where we use the notation \eqref{LeibnizRuleNotation}):
\bs \label{dotBlackProperties}
\begin{align}
    \cA\Big(\big\lbr\bPtap,\bPtapp \big\rbr, \dBt\!\Big) &=(\drBt\bPtap)\,\dotblacktriangleright\, \bPtapp-(\drBt\bPtapp)\,\dotblacktriangleright\, \bPtap, \\
    \big\lbr\ot,\otp\big\rbr^\circ \,\dotblacktriangleright\,\bPtapp &=\big[\dBt,\dBtp\!\big]\bPtapp, \\
    \cA\Big(\big\lbr\otp,\otpp \big\rbr^\circ, \dotblacktriangleleft\, \bPta\Big)  &=\drBtp\big(\dBtpp\bPta\big)-\drBtpp\big(\dBtp\bPta\big), \\
    \drBtpp\big\lbr\bPta,\bPta'\big\rbr &=\otpp\big[ \dotblacktriangleleft\,\bPta\,,\dotblacktriangleleft\,\bPta'\,\big].
\end{align}
\es
\end{tcolorbox}

\ni We give the proofs for the actions $\dotvartriangleright$ and $\dotblacktriangleleft$ in App.\,\ref{AppEYM:DotWhiteLeibniz} and \ref{AppEYM:DottauHom}.
The other demonstrations go along the same lines.
We can then state again the theorem \hyperref[STisomorphism]{[Bicrossed product structure]}, this time in a full algebroid fashion.

\begin{tcolorbox}[colback=beige, colframe=argile, breakable]
\textbf{Theorem [Bicrossed product structure]}\\
\begin{equation} \label{STisomorphismBis}
\cST\simeq\ocT\Bprod\big(\bcT\Wprod\cS\big).
\end{equation}
Hence, using the definitions \eqref{defBracketobar} and \eqref{defActionDot2},
\begin{align}
    \big\lbr\Pta,\Ptap\big\rbr &=\Big(\big\lbr \ot,\otp\big\rbr^\circ+\drBt\bPtap-\drBtp\bPta\,;\overline{\big\lbr\bt,\btp\big\rbr}+ \big(\dBt\bPtap-\dBtp\bPta\big)^\gr, \label{bicrossedBracketBis}\\
    &\qquad\big\lbr\a{},\ap{}\big\rbr^\g+\dWt\ap{}-\dWtp\a{}+ \big(\dBt\bPtap-\dBtp\bPta\big)^\ym\Big).\nn
\end{align}
\end{tcolorbox}

\section{Covariant wedge algebras \label{secEYM:CovWedge}}

From the analysis done in the previous section we see that the Jacobi and Leibniz anomalies  associated with the bracket $[\cdot\,,\cdot]=\big([\cdot\,,\cdot]^C,[\cdot\,,\cdot]^\ym\big)$ defined in (\ref{TbracketEYM}-\ref{YMbracketCFdef})
vanish when the symmetry parameters are such that $\dPta \Ptap=0$. 
This imposes constraints on $\Pta$ which define a wedge algebra.
In other words, picking $\Pta$ in the kernel of the anchor map in the $\cST$-algebroid is the definition of the associated algebra $\Wcal_{CA}(\scri)$.

\begin{tcolorbox}[colback=beige, colframe=argile]
\textbf{Definition/Corollary 
[Covariant wedge on $\scri$]}\\
The covariant wedge space $\W_{CA}(\scri)$ and algebra $\Wcal_{CA}(\scri)$ are given by
\bs \label{CovWedgeScri}
\begin{align}
    \W_{CA}(\scri) &:=\Big\{\,\Pta\in\sfST~ \big|~\dt h=0=\dPta A\,\Big\}, \\
    \Wcal_{CA}(\scri) &:=\big(\W_{CA}(\scri),[\cdot\,,\cdot]\big),
\end{align}
\es
where the $\sfST$-bracket \eqref{STbracket} reduces to (\ref{CbracketCompact}-\ref{YMbracketCompact}). $\Wcal_{CA}(\scri) $ is a \emph{Lie} algebra.
\end{tcolorbox}

\paragraph{Remark:}
The fact that $\Wcal_{CA}(\scri)$ is a Lie algebra follows from the fact that $\cST$ is a Lie algebroid.
Indeed, the closure of the bracket $[\cdot\,,\cdot]$, namely that $[\Pta,\Ptap]\in \W_{CA}(\scri)$ when $\Pta,\Ptap \in \W_{CA}(\scri)$, is obvious from \eqref{LeibnizAnchorEYM}.\footnote{Using that $\dPta\Ptap\propto\dPta (h,A)$.}
Moreover the dual EOM are respected by construction.
The important point to emphasize is that since $\dPta(C,A)=0$ we can view the radiative data $(C,A)$ as \emph{background data} which are not varied and label the wedge algebra.
On top of that, we show in (\ref{CbracketCompact}-\ref{YMbracketCompact}) that in the wedge, the Lie algebra brackets $[\cdot\,,\cdot]^C$ and $[\cdot\,,\cdot]^\ym$ can be recast in a form which makes manifest that their structure constants are indeed \emph{constant}, i.e. independent of the radiative data.
The dependence in these background fields only remains as a constraint among the symmetry generators.\footnote{For instance that $D\t{-1}=C\t0$, but also that $D^2\t0=N\t0+2DC\t1+3CD\t1-3C^2\t2$, which follows from the time derivative of the previous condition, and so on so forth.}
\medskip

From \eqref{Leibnizgr} and a similar condition for $[\cdot\,,\cdot]^\ym$, one establishes that $\Dcal$ is a covariant derivative in the covariant wedge and thus satisfies the Leibniz rule.
In essence the corollary simply follows from the well-known fact that an algebroid reduces to an algebra when the anchor map is trivial.  
The Leibniz property is a bonus result which is a consequence of the fact that the algebroid satisfies the dual EOM as constraints.
To see it, just notice that\footnote{To be accurate, one should think of this equation as being evaluated at a non-radiative cut.}
\begin{equation} \label{HamConstraintVSLeibniz}
    \big\lbr\Ham,\lbr\Pta,\Ptap\rbr\big\rbr= \big\lbr\lbr\Ham,\Pta\rbr,\Ptap\big\rbr+ \big\lbr\Pta,\lbr\Ham,\Ptap\rbr\big\rbr \quad\ensurestackMath{\stackon[4pt]{\Longrightarrow}{\scriptstyle{\Wcal_{CA}}}}\quad \Dcal[\Pta,\Ptap]=\big[\Dcal\Pta,\Ptap\big]+ \big[\Pta,\Dcal\Ptap\big].
\end{equation}

We are especially interested in the projection of $\Wcal_{CA}(\scri)$ onto a non-radiative cut $S$ of $\scri$.
For simplicity, let us consider the cut $u=0$.
The celestial symmetry parameters $T$ and $\Aa{}$ are then defined as the initial condition of the dual EOM \eqref{DualEOMEYM}, i.e. $\Pta\big|_{u=0}=(T,\Aa{})$.
On $\scri$, $\dPta(h,A)=0$ trivially imposes $\pa_u^n\big(\dPta(h,A)\big)\big|_{u=0}=0,\,\forall \,n\in\N$.
At a cut all conditions labeled by $n$ are independent.
In order to write them we need to encode the radiative data in terms of a collection of fields $\Cc{n} \in \Ccel{n+1,2}$ and $\Ac{n} \in \Ccel{n+1,1}$. 
These background fields are defined as
\be
 \Cc{n}\equiv(\pa_u^n C)\big|_{u=0}\qquad \textrm{and}\qquad\Ac{n}\equiv(\pa_u^n A)\big|_{u=0}.
\ee 
Besides we denote the collections as $\bfCc=(\Cc0,\Cc1,\Cc2,\ldots)$ and $\bfAc=(\Ac0,\Ac1,\Ac2,\ldots)$. 
Evaluating $\dPta(h,A)=0$ and $\dPta(C,F)=0$ at $u=0$, we get\footnote{It is understood that $\DYMc\Aa{s}=D\Aa{s}+\big[\Ac0,\Aa{s}\big]_\g$ and  $(\DGR T)_s=D\T{s+1}-(s+3)\Cc0\T{s+2}$.}
\bs
\begin{align}
    \DYMc\Aa0 &=\Cc0\Aa1+\Ac1\T0, \\
    \DYMc^2\Aa1 &=3\Cc0D_{\beta_0}\Aa2+ 2D\Cc0\Aa2-3\Cc0^2\Aa3-[\Ac1,\Aa0]_\g+2D_{\beta_0}\Ac1\T1 \\
    &+3\Ac1 D\T1+\Cc1\Aa1-6\Cc0\Ac1\T2+\Ac2\T0, \nn\\
    \DYMc^3\Aa2 &= \cdots, \nn
\end{align}
together with
\begin{align}
    D\T{-1} &=\Cc0\T0, \\
    D^2\T0 &=\Cc1\T0+2D\Cc0\T1+3\Cc0 D\T1-3\Cc0^2\T2, \\
    D^3\T1 &=\Cc2\T0+2D\Cc1\T1+4\Cc1 D\T1+8D\Cc0 D\T2+6\Cc0 D^2\T2+3D^2\Cc0\T2+\cdots,
\end{align}
\es
where $\cdots$ contains higher order terms in $\bfCc$.
This set of relations thus takes the form of a constraint on the derivatives $\DYMc^{s+1}\Aa{s}$ and $D^{n+2}\T{n}$ for $s\geq 0$ and $n\geq -1$.
If we focus on a non-radiative (NR) cut, for which $\Cc{n}=0=\Ac{n}, \forall\,n\geq 1$, then the constraint takes a very compact form.
Indeed, simply notice that for $n\in\N$,
\bs \label{CovWedgeEYMdef}
\begin{align}
    \big(\pa_u^n(\dPta C)\big)\big|_{u=0} &\equiv\dT \Cc{n} \overset{\textsc{nr}}{=}\big(\DGR{}^{n+2}T \big)_{-2}=0, \\
    \big(\pa_u^n(\dPta A)\big)\big|_{u=0}&\equiv \delta_{(T,\Aa{})}\Ac{n} \overset{\textsc{nr}}{=}\big(\DEYM{}^{n+1}\Aa{} \big)_{-1}=0,
\end{align}
\es
where $\DEYM\a{}$ is the non-radiative part of $\DTOT\Pta$, namely\footnote{In \eqref{CovWedgeEYMdef}, we use its projection onto $S$.}
\begin{equation} \label{defDEYM}
    \big(\DEYM\a{}\big)_s:= \big(\DTOT\Pta \big)_s\big|_{F=0}=\DYM\a{s+1}-(s+2)C\a{s+2}.
\end{equation}
Therefore, in order to preserve the non-radiative conditions $\Cc{n}=0=\Ac{n}, n\geq 1$ of the cut, together with the initial values of the shear $\Cc0$ and of the gauge potential $\Ac0$, we must impose the RHS of \eqref{CovWedgeEYMdef} to be 0.

\paragraph{Remark:}
Another way to phrase the result \eqref{CovWedgeEYMdef} is to say that the no radiation condition of a cut $S$ of $\scri$ is an integrability requirement, in the sense that
\begin{equation} \label{NonRadIntegrability}
    \boxed{\big[\pa_u^n,\DGR\big]\Big|_S \overset{\textsc{nr}}{=}0 \qquad\textrm{and}\qquad \big[\pa_u^n,\DTOT\big]\Big|_S \overset{\textsc{nr}}{=}0},\quad \forall\,n\geq 1.
\end{equation}
\smallskip

This discussion then motivates the definition of the covariant wedge $\Wcal_{\bfCc\bfAc}(S)$ onto the sphere, or more precisely a non-radiative section of $\scri$ for which $\bfCc=(\Cc0,0,0,\ldots)$ and $\bfAc=(\Ac0,0,0,\ldots)$.
In the following $\sfSTK(S)$ is the same as $\sfSTK$, cf.\,\eqref{defSTKspace}, but for celestial fields.

\begin{tcolorbox}[colback=beige, colframe=argile]
\textbf{Definition [Covariant wedge space \& algebra on $S$]}\\
The covariant wedge space $\W_{\bfCc\bfAc}(S)$ at a non-radiative cut $S$ of $\scri$ characterized by the condition $\Cc{n}=0=\Ac{n}, \forall\,n\geqslant 1$ is defined as
\begin{equation} \label{CovWedgeEYMdefBis}
    \W_{\bfCc\bfAc}(S):=\Big\{\,(T,\Aa{})\in\sfSTK(S)~ \big|~\big(\DGR{}^{s+1}T \big)_{-2}=0=\big(\DEYM{}^{s+1}\Aa{} \big)_{-1},\,\forall\,s\in\N\,\Big\}.
\end{equation}
$\Wcal_{\bfCc\bfAc}(S)\equiv\big(\W_{\bfCc\bfAc}(S),[\cdot\,,\cdot]\big)$ is the associated Lie algebra with the bracket given by (the projection on $S$) of (\ref{CbracketCompact}-\ref{YMbracketCompact}).
\end{tcolorbox}

\ni $\Wcal_{\bfCc\bfAc}(S)$ is a Lie algebra since it corresponds to a special case of the projection of $\Wcal_{CA}(\scri)$ at a cut.

\paragraph{Remark:}
We leave for future investigation a more refined study of the non-radiative structure of the covariant wedge.
Indeed, it is natural to wonder what happens if we relax the non-radiation condition $\Cc{n}=\Ac{n}=0,\,\forall\, n\geq 1$ into, for instance, $\Cc{n}=\Ac{n}=0,\,\forall\, n\geq 2$ while keeping $\Cc0,\Cc1,\Ac0$ and $\Ac1$ different than 0. 
Note that the parameters $(\Cc{n},\Ac{n})$ are the higher goldstone fields introduced in \cite{Freidel:2022skz} which allowed the construction of the discrete basis of states.
\medskip

We now give a compact form for the $\sfST$-bracket when restricted to the covariant wedge.
At the same time, we point out how the 3 actions $\vartriangleright, \blacktriangleright$ and $\blacktriangleleft$ recombine into a single semi-direct action $\blackwhitetriangle$ when the symmetry parameters pertain to $\Wcal_{\bfCc\bfAc}(S)$.

\subsection{Semi-direct structure \label{secEYM:SemiDirectWedge}}

We have seen already in \eqref{anomalyTsubAlgebroid} that the gravitational $\cT$-algebroid is not a sub-algebroid of $\cST$.
However, the culprit term in \eqref{anomalyTsubAlgebroid} involves $F$.
This suggests that in the non-radiative sector, where $F=0$, a semi-direct structure should hold.
Let us show that indeed, the symmetry algebra $\Wcal_{\bfCc\bfAc}(S)$, where $S$ is non-radiative as in the previous subsection, can be recast in the form $\Wcal_{\bfCc}^\gr(S)\BWprod \Wcal_{\bfCc\bfAc}^\ym(S)$ for a certain action $\blackwhitetriangle$.
Here $\Wcal_{\bfCc}^\gr(S)$ denotes the GR covariant wedge algebra from sections \ref{sec:Wedge} and \ref{Sec:algebroidTs} and $\Wcal_{\bfCc\bfAc}^\ym(S)$ is the YM covariant wedge in an asymptotically flat background.\footnote{Namely characterized by $\big(\DEYM{}^{s+1}\Aa{}\big)_{-1}=0$.
We show in App.\,\ref{AppEYM:LeibnizDEYM} that $\cA\big([\Aa{},\Aap{}]^\g,\DEYM\big)=0$ if $\Aa{},\Aap{}\in\Wcal_{\bfCc\bfAc}^\ym(S)$ such that $\big(\DEYM{}^{s+1}[\Aa{},\Aap{}]^\g\big)_{-1}=0$ if $\Aa{},\Aap{}\in\Wcal_{\bfCc\bfAc}^\ym(S)$.
We proved already in \hyperref[LemmaLeibniz]{Lemma [Leibniz rule]} that $\cA\big([T,T']^{\sigma_0},\DGR\big)=0$ when $T,T'\in \Wcal^\gr_{\bfCc}(S)$.
Overall, this is also a consequence of \eqref{HamConstraintVSLeibniz}.} 
Their brackets are respectively the $C$-bracket and $[\cdot\,,\cdot]^\g$.

To avoid too many changes of notation, but also because the first part of the computation is valid on $\scri$, we keep using the Carrollian parameter $\Pta$ in the following.
The projection onto $S$ can be applied at any point.
For $\tau\in\Wcal_{C}^\gr(\scri)$, we know that $D\t{-1}=C\t0$, which allows us to recast the deformation along the shear of the $C$-bracket as a constraint on $\t{-1}$.
Therefore, for $\tau,\tau'\in\Wcal^\gr_C(\scri)$, the $C$-bracket \eqref{TbracketEYM} takes the form
\begin{equation} \label{CbracketCompact}
    \big[\tau,\tau'\big]^C_s=\sum_{n=0}^{s+2}(n+1)\big(\t{n}D\tp{s+1-n}-\tp{n}D\t{s+1-n}\big).
\end{equation}
Note that the sum now runs till $s+2$ and the term $n=s+2$ precisely reproduces the GR Dali-bracket.

The same phenomenon occurs with the YM-bracket \eqref{YMbracketCFdef} when imposing $\dPta A=0$, i.e. $\DYM\a0=C\a1+F\t0$, together with $\dt h=C\t0-D\t{-1}=0$.
Indeed, for $\Pta,\Ptap\in\Wcal_{CA}(\scri)$, we have that\footnote{We correctly recover the action of a diffeomorphism onto a gauge transformation $\a0$ studied in \cite{Barnich:2013sxa}.}
\begin{align} \label{YMbracketCompact}
   \big[\Pta,\Ptap\big]^\ym_s= \big[\a{},\ap{}\big]^\g_s +\bigg\{\sum_{n=0}^{s+1}(n+1)\Big(\t{n}\DYM\ap{s+1-n}-\ap{n+1}D\t{s-n}\Big)-\Pta\leftrightarrow\Ptap\bigg\},
\end{align}
where the sum now runs till $s+1$.
The equations \eqref{CbracketCompact} and \eqref{YMbracketCompact} represent a compact way of writing the \emph{algebra part} of the $\sfST$-bracket.
It also emphasizes the fact that the brackets associated to the covariant wedge algebras have field space independent structure constants, as they should.
We had noticed that feature in Sec.\,\ref{sec:CovFormBracket} already.

Defining the action $\blackwhitetriangle$ as 
\begin{align}
     \blackwhitetriangle&:\Wcal_C^\gr(\scri)\to \textrm{End}\big(\Wcal^\ym_{CA}(\scri)\big) \nn\\
     &\quad \tau\mapsto\BWt, \label{BlackWhiteAction}\\
     \big(\BWt\ap{}\big)_s &:=\sum_{n=0}^{s+1}(n+1)\Big(\t{n}\DYM\ap{s+1-n}-\ap{n+1}D\t{s-n}\Big), \nn
\end{align}
we get that 
\begin{equation}
    \boxed{\big[\Pta,\Ptap\big]^\ym= [\a{},\ap{}]^\g+\BWt\ap{}-\BWtp\a{}}.
\end{equation}
The reader can check\footnote{We do not report these demonstrations since they are similar to the ones presented in appendices \ref{AppEYM:tauAction}-\ref{AppEYM:tauHom}.} that in the wedge (supplemented by the non-radiative condition $F\equiv 0$), the action $\blackwhitetriangle$ is a derivative and a homomorphism of Lie algebras, i.e.
\bs
\begin{align}
    \cA_s\big([\ap{},\app{}]^\g,\BWt\big) &=\t0\big[\ap1\dt h-\hdap A,\app{s+1}\big]_\g-\ap{}\leftrightarrow\app{} \overset{\Wcal_{\bfCc\bfAc}}{=}0, \\
    \Big([\t{},\tp{}]^C \blackwhitetriangle \app{}-\big[\BWt,\BWtp\!\big]\app{} \Big)_s\! &=\Big\{ (s+3)\app{s+2}D\big(\tp0\dt h\big)+(s+3)\t0\tp{s+2}\DYM\big(\hdapp A\big) \nn\\
    &\hspace{-3.3cm}\!+\dt h\big(\tp0\DYM\app{s+3}-(s+3)\tp{s+2}\DYM\app1-\app1 D\tp{s+2}\big)-(s+3)\app1\tp{s+2}D\big(\dt h\big) \nn\\
    &\hspace{-3.3cm}\!+\hdapp A \big(\t0D\tp{s+2}+(s+3)\tp{s+2}D\t0\big)\Big\}-\t{}\leftrightarrow\tp{}\overset{ \Wcal_{\bfCc\bfAc}}{=}0.
\end{align}
\es
with $\dt h$ and $\hda A$ given by \eqref{dtahEYM} and \eqref{hatdaA}.
This means in particular that the covariant wedge algebra associated to the $\cST$-algebroid at a non-radiative cut $S$ of $\scri$ is given by the semi-direct product\footnote{This structure was pointed out in \cite{Peraza:2022glu} for the restriction to $\gbms$.}
\begin{equation} \label{WedgeAlgebraEYMsemidirect}
    \boxed{\Wcal_{\bfCc\bfAc}(S)\simeq \Wcal^\gr_{\bfCc}(S)\BWprod \Wcal^\ym_{\bfCc\bfAc}(S)}.
\end{equation}

\subsection{Relation with the celestial bracket \label{secEYM:celestialBracket}}

Since we considered Schwartz fall-off conditions for $C$ and $A$, the deformation parameters are zero on the celestial sphere $\scri^+_-$.
From the Carrollian YM-bracket \eqref{YMbracketCFdef}, one can define its projection onto the sphere $S=\scri^+_-$, to obtain the celestial bracket which reduces to\footnote{For the upcoming argument, we just focus on the part of the bracket that mixes $T$ and $\Aa{}$.}
\begin{equation} \label{CelestialBracket}
    \big[(T,\Aa{}),(\Tp{},\Aap{})\big]^{\mathsf{cel}}_s=\bigg\{ \sum_{n=0}^s(n+1)\big(\T{n}D\Aap{s+1-n}-\Aap{n+1}D\T{s-n}\big)\bigg\} -(T,\Aa{})\leftrightarrow(\Tp{},\Aap{}),
\end{equation}
where $T$ and $\Aa{}$ are formally defined at the cut $\scri^+_-\equiv u\to-\infty$.

\paragraph{Relation with the celestial modes $w^p_m$ and $S^{q,b}_n$:}
The celestial bracket \eqref{CelestialBracket} corresponds to a symmetry \textit{algebra} bracket when $\Aa{}\in\Wcal^\ym_{\bfCc=\bfAc=0}(S)$ and $T\in\Wcal^\gr_{\bfCc=0}(S)$, which means that $D^{s+1}\Aa{s}=0$ and $D^{s+2}\T{s}=0$, see \eqref{CovWedgeEYMdefBis}.
Notice that if we work onto the twice punctured sphere $S=\C^*$, then we can expand $\Aa{s}$ and $\T{s}$ as\footnote{$\{Z_b\}$ is the set of generators of $\g$.}
\bs
\begin{align}
    \Aa{s} &=\sum_{m=0}^s\Aa{(s,m)}^b z^m Z_b,\qquad \textrm{with}\quad \Aa{(s,m)}^b=\Aa{(s,m)}^b(\bz), \\
    \T{s} &=\sum_{m=0}^{s+1}\T{(s,m)}z^m ,\qquad\quad\! \textrm{with}\quad \T{(s,m)}=\T{(s,m)}(\bz).
\end{align}
\es
Setting $s=2p-3$ for GR and $s=2p-2$ for YM,\footnote{This leads in both cases to $p=1,\frac32,2,\ldots$.} we equivalently get
\bs
\begin{align}
    \Aa{s} &=\sum_{m=1-p}^{p-1}\Aa{(2p-2,p-1+m)}^b z^{p-1+m} Z_b, \\
    \T{s} &=\sum_{m=1-p}^{p-1}\T{(2p-3,p-1+m)} z^{p-1+m}.
\end{align}
\es
Besides, notice that if we take $\Aa{}=0=T'$ and $T=\iota(\T{s})$, $\Aap{}=\iota(\Aap{s'})$,\footnote{$\iota$ is the inclusion map, cf. section \ref{sec:Vspace}.} then the only non-zero term in \eqref{CelestialBracket} reduces to
\begin{equation}
    \big[\big(\iota(\T{s}),0\big),\big(0, \iota(\Aap{s'})\big) \big]^{\mathsf{cel}}_{s+s'-1}=(s+1)\T{s}D\Aap{s'}-s'\Aap{s'}D\T{s}.
\end{equation}
In particular, the bracket of the modes 
\begin{equation}
    w^p_m=-\iota\big(\tfrac12 z^{p-1+m}\big)\qquad\textrm{and}\qquad S^{q,b}_n=-\iota\big(\tfrac12 z^{q-1+n}Z_b\big)
\end{equation}
is given by
\begin{equation}
    \big[w^p_m,S^{q,b}_n\big]^{\mathsf{cel} }_{2(p+q-2)-2}=\big(m(q-1)-n(p-1)\big)S^{p+q-2,b}_{m+n}.
\end{equation}
This is the usual $sw_{1+\infty}$ bracket introduced in \cite{Strominger:2021mtt}.

\subsection{The return of the Schouten-Nijenhuis bracket}

In chapter \ref{secGRI}, we started the analysis of the higher spin symmetry bracket by pointing out the relation to the Schouten-Nijenhuis (SN) bracket for symmetric tensors \cite{Schouten, SNbracket} on the 2d manifold $S$.
In our notation, as explained in \eqref{SNbracketDefbis}, the SN-bracket reads
\begin{equation}
    [\tau,\tau']^\sn_s :=\sum_{n=0}^{s+1} n\big(\t{n}D\tp{s+1-n}-\tp{n}D\t{s+1-n}\big)=\sum_{n=0}^{s} (n+1)\big(\t{n+1}D\tp{s-n}-\tp{n+1}D\t{s-n}\big). \label{SNbracketEYM}
\end{equation}
If we define the tilde variable $\tt{s+1}\equiv\t{s}$, then (still assuming $\Pta,\Ptap\in\Wcal_{CA}(\scri)$ so that we can use \eqref{CbracketCompact} and \eqref{YMbracketCompact})
\begin{align}
    [\tau,\tau']^C_s=\sum_{n=0}^{s+2}(n+1)\big(\tt{n+1}D\ttp{s+2-n}-\ttp{n+1}D\tt{s+2-n}\big).
\end{align}
Furthermore,
\begin{align}
    \big[\Pta,\Ptap\big]^\ym_s =\big[\a{},\ap{}\big]^\g_s+\sum_{n=0}^{s+1}(n+1)\Big(\tt{n+1} \DYM\ap{s+1-n} &-\ap{n+1}D\tt{s+1-n} \\
    &-\ttp{n+1}\DYM\a{s+1-n}+\a{n+1}D\ttp{s+1-n}\Big). \nn
\end{align}
Comparing with \eqref{SNbracketEYM}, we obtain the following lemma.

\begin{tcolorbox}[colback=beige, colframe=argile] \label{lemmaSNbracket}
\textbf{Lemma [SN-bracket]}\\
In the covariant wedge algebra $\Wcal_{CA}(\scri)$, the $C$-bracket and the action $\blackwhitetriangle$ (cf.\,\eqref{BlackWhiteAction}) at degree $s$ are SN-brackets at degree $s+2$ and $s+1$ respectively, over the shifted variables $\tt{}$ and $\ttp{}$,
\bs
\begin{align}
    [\tau,\tau']^C_s &=\big[\tt{},\ttp{}\big]^\sn_{s+2}, \\
    \big(\BWt\ap{}\big)_s &=\big[\tt{},\ap{}\big]^\sn_{s+1}, \label{CFasSN}
\end{align}
\es
where $D\to\DYM$ when the argument is Lie algebra valued.
\end{tcolorbox}

\ni Since the SN-bracket is known to be a Lie bracket, this is a complementary way to confirm that $\Wcal_{CA}(\scri)$ is indeed a Lie algebra\footnote{This lemma implies that the $C$- and YM-brackets when $\dPta\equiv 0$ are Lie \textit{algebra} brackets, hence Jacobi holds. 
This however does \textit{not} imply that Jacobi is satisfied for the Lie \textit{algebroid} bracket.
See \hyperref[LemmaJacEYM]{Lemma [Jacobi identity]} for details.}---the dual EOM being preserved by the brackets is a particular case of \eqref{dualEOMGRbracket}-\eqref{dualEOMEYMbracket}.

\section{Summary}

We now recapitulate the main features of the phase space action $\dPta$.
First of all, it acts on the radiative variables as follows, cf.~\eqref{deltaPtaCA},
\bs \label{deltaPtaCAbis}
\begin{align}
    \dPta C &=N\t0-D^2\t0+2DC\t1+3CD\t1-3C^2\t2, \\
    \dPta A &=F\t0-\DYM\a0+C\a1.
\end{align}
\es
Next, as proven in \eqref{AlgebroidActionEYM}, this action is the realization of the algebroid $\sfST$-bracket \eqref{STbracket}, namely
\begin{equation}
    \big[\dPta,\dPtap\big]\cdot=-\delta_{\lbr\Pta,\Ptap\rbr}\cdot.
\end{equation}

Importantly, the time dependency of the Carrollian symmetry parameters $\Pta_s$ is determined by a set of dual EOM \eqref{DualEOMEYM} and the $\sfST$-bracket satisfies the latter as well (see the \hyperref[LemmaClosureEYM]{Lemma [$\sfST$-bracket closure]}).
Since the dual evolution equations involve the radiative data $(C,A,F)$, a major consequence is the field dependency of $\Pta$, which explains the necessity of working with an algebroid framework.
On top of that, the phase space variables appear explicitly in the $\sfST$-bracket itself.
Although the action $\dPta$ is not an algebroid action in the sense that it is not a derivative on the $\sfST$-bracket (cf.~\eqref{LeibnizAnchorEYM}), we still show that the Jacobi identity \eqref{JacobiMorphismEYM} holds for this bracket.
This comes from the remarkable property that the Leibniz rule anomaly of $\dPta$ precisely equals and compensates for the Jacobi identity anomaly of the $C$- and YM-brackets.

Besides, we showed that one can start from a kinematical $\sfK$-bracket \eqref{Kbracket}, for which the time evolution of $\Pta$ is arbitrary, and then construct the $\sfST$-bracket and its associated anchor map $\delta$ as the unique bracket and action which preserve the dual time evolution \eqref{DualEOMEYM}, cf.~\hyperref[KbracketTheorem]{Theorem [on-shell $\sfK$-bracket]}.
The phase space action is then the same as the one deduced from the conservation of the master charge, that we recalled in \eqref{deltaPtaCAbis}.

While the $\cSTK$-\textit{algebra} associated to the $\sfK$-bracket admits a semi-direct structure between its GR and YM parts---see \hyperref[KbracketTheorem]{Theorem [on-shell $\sfK$-bracket]} for details---the $\cST$-\textit{algebroid} splits into a more refined bicrossed product structure analyzed in details in section \ref{secEYM:BicrossedJacobi}, cf.~in particular \eqref{STisomorphismBis}:
\begin{equation}
\cST\simeq\ocT\Bprod\big(\bcT\Wprod\cS\big).
\end{equation}

Moreover, by restricting the anchor map to its kernel, namely
\begin{equation}
    \Big\{\Pta\in\sfST~ \big|~\dt h=0=\dPta A\Big\},
\end{equation}
the $\cST$-algebroid reduces to the covariant wedge Lie algebra $\Wcal_{CA}(\scri)$.
Of particular relevance, its projection onto a non-radiative cut $S$ of null infinity is characterized by the following constraints on the celestial parameters $(\T{},\Aa{})$:
\begin{equation}
    \big(\DGR{}^{s+1}T \big)_{-2}=0=\big(\DEYM{}^{s+1}\Aa{} \big)_{-1},\,\forall\,s\in\N.
\end{equation}
One can prove, cf.~\eqref{WedgeAlgebraEYMsemidirect}, that the EYM covariant wedge algebra is the semi-direct product of its GR sub-algebra onto the YM sub-algebra (in an asymptotically flat background), where the 3 actions $\vartriangleright, \blacktriangleright,\blacktriangleleft$ recombine to form the action $\blackwhitetriangle$.

Finally, let us mention that the algebroid symmetry acts on the dynamical phase space while the algebra symmetry obtained after imposing the wedge condition preserves the no-radiation boundary condition at $i^0$.

\section{Twistorial standpoint \label{secEYM:twistor}}

We now give one last perspective on the $\sfST$-bracket, by recasting it as a Poisson bracket in the $(q,u)$ plane, where $q$ is a spin 1 coordinate in the Newman space.
The latter is defined as the homogeneous bundle\footnote{Recall that a Carrollian field of weights $(\delta,s)$ can be matched to a homogeneous section of the bundle
$\Ocal\big(-(\delta+s),-(\delta-s)\big)$ once one introduces homogeneous coordinates onto the sphere.} $\TN=\Ocal(-1,1)\to\scri_{\C}$ over complexified $\scri$. Since $\scri_{\C}$ is a $\Ocal(1,1)$ bundle over  $\CP1$  \cite{Eastwood_Tod_1982} we have  equivalently that
$\TN=\Ocal(-1,1)\oplus \Ocal(1,1) \to\CP1$ when fibered over the Riemann sphere.
One can supplement the holomorphic coordinates $(q,u)$ of the fibers over $\CP1$ by their anti-holomorphic counterpart $(\bar q,\bar u)$.
The variables $q,\bar q$ have a remarkable geometric interpretation since they parametrize the notion of horizontality on $\scri_\C$.
In other words, they define an Ehresmann connection on $\scri_\C$.
To see this, recall that on top of the Carrollian structure $(\ell^a,q_{ab})$ on $\scri$, where $\ell=\pa_u$ is the null generator and $q_{ab}$ the degenerate metric, we need to specify a one-form $k_a$ to get a notion of connection \cite{Ashtekar:1981bq, Riello:2024uvs,Freidel:2022vjq,Freidel:2024emv}.
It satisfies $k_a\ell^a=1$ by definition.
Using a congruence of null geodesics transverse to $\scri$, labeled by a null vector $k^a$, the Ehresmann one-form can arise from a so-called null rigging structure \cite{Mars:1993mj}.
Upon a choice of reference one-form,\footnote{This means that $\TN$ is an affine bundle, since a reference one-form defines an origin for the coordinate $q$.} for instance $k_a\rd x^a=\rd u$, a generic Ehresmann connection, or equivalently the associated ruling vector, is parametrized as follows (see also Sec.\,\ref{sec:GoodCut}):
\begin{equation}
    k_{(q,\bar q)}=k+q\bm+\bar q m-q\bar q\ell.
\end{equation}
The sphere dyad normal to $k_{(q,\bar q)}$ is given by 
\begin{equation}
    m_q=m-q\ell,\qquad\bm_{\bar q}=\bm-\bar q\ell.
\end{equation}
As such, $\big(\ell,k_{(q,\bar q)},m_q,\bm_{\bar q} \big)$ forms a null tetrad.
The coordinate $q$ appeared in \cite{Adamo:2009vu, Adamo:2010ey} and recently in \cite{Kmec:2024nmu} as a natural way to parametrize the fibers of twistor space over $\CP1$.
\medskip

We can straightforwardly convert the expressions written in terms of $\tau$ and $\a{}$ as formulas involving the functions $\ft$ and $\fa$, cf.\,\ref{secYM:twistor} and \ref{sec:twistor}.
For this we define the vector\footnote{The
graded vector $\Pta$ and the function $\fPta$ are two different representations of the
same abstract vector in $\sfSTK$.} 
\begin{equation}
\fPta(q,u,z,\bz)\equiv(\ft,\fa):=\left(\sum_{s=-1}^\infty\t{s}q^{s+1},\sum_{s=0}^\infty\a{s}q^{s}\right),
\end{equation}
where $q\in\Ccar{0,1}$, $\ft\in\Ccar{-1,1}$, and $\fa\in\Ccarg{0,0}$.
Besides we define the differential operators
\bs \label{defNabla}
\begin{align}
\nablaGR &:=q\pa_u-D+C\pa_q, \\
\nablaYM &:=q\pa_u-D-\adA, \\
\nablaEYM &:=\nablaYM+C\pa_q.
\end{align}
\es
Denoting also $\E_{\ft}(q)=\sum_{s=-1}^\infty\E^\gr_s(\tau)q^{s+1}$ and $\E^\ym_{\fPta}(q)=\sum_{s=0}^\infty\E^\ym_s(\Pta)q^{s}$, we deduce that
\begin{equation}
\nabla\left(\begin{array}{l}
        \ft \\
        \fa
        \end{array}\right):=
        \left( \begin{array}{lr}
        \nablaGR & 0 \\
        F\pa_q & \nablaEYM
        \end{array}\right)
        \left(\begin{array}{c}
        \ft \\
        \fa
        \end{array}\right)=
        \left(\begin{array}{c}
        \dt h+q\E_{\ft} \\
        \dPta A+q\E^\ym_{\fPta}
        \end{array}\right),
\end{equation}
which shows that $\nabla(\ft,\fa)$ is an infinitesimal gauge transformation if $(\ft,\fa)\in\sfST$.

Finally, as in the gravitational case \ref{sec:twistor}, the YM-bracket corresponds to the Poisson bracket in the $(q,u)$ plane (on-shell of the dual EOM).
Indeed, after some algebra, we get that
\vspace{-0.7cm}

\begin{align} \label{CFasPoisson}
    \big[\fPta,\fPta'\big]^\ym(q) &:=\sum_{s=0}^\infty \big[\Pta,\Ptap\big]^\ym_sq^s \nn\\
    &=\big[\fa,\fa'\big]^\g(q)+ \Big(\Big(\poisson{\ft,\fa'}-\pa_q\ft \E^\ym_{\fPta'}+\pa_q\fa'\E_{\ft}\Big)(q)-\fPta\leftrightarrow\fPta'\Big),
\end{align}
where
\begin{equation}
    \poisson{\ft,\fa'}:=\pa_q\ft\pa_u\fa'-\pa_u\ft\pa_q\fa'.
\end{equation}
In other words,
\begin{align}
    \poisson{\fPta,\fPta'}=\left(\sum_{s=-1}^\infty \poisson{\tau,\tau'}_s q^{s+1},\sum_{s=0}^\infty \poisson{\Pta,\Ptap}^\ym_s q^s\right),
\end{align}
where we used the kinematical $\sfK$-brackets \eqref{CbracketScridef}-\eqref{YMbracketScridef}.
The Poisson bracket on the LHS (up to the pure $\g$ part $\big[\fa,\fa'\big]^\g$) is the canonical Poisson bracket on the twistor fibers, expressed using the coordinates $(q,u)$.
Indeed, taking $\bar\mu^\alpha = u n^\alpha + q \lambda^\alpha$ where $\lambda^\alpha$ parametrizes holomorphic homogeneous coordinates on $\CP1$ and $n^\alpha$ is a spinor such that $\epsilon_{\alpha \beta}= \lambda_\alpha n_\beta-\lambda_\beta n_\alpha$, we see that
\begin{equation}
    \{\cdot\,,\cdot\}=\epsilon^{\alpha\beta} \frac{\pa\,\cdot }{\pa \bar\mu^\alpha} \frac{\pa\, \cdot}{\pa \bar\mu^\beta}=\pa_q\cdot\pa_u\cdot-\,\pa_u\cdot\pa_q\cdot.
\end{equation}

Notice finally that if we define the potential $\hat h \in \Ccar{0,2}$ which depends on $q\in \Ccar{0,1}$ as
\begin{equation}
    \hat h(q,u,z,\bz):=-\frac{q^2}{2}+h(u,z,\bz),
\end{equation}
then we can understand the covariant derivatives as 
\begin{equation}
-\big(\nabla\fPta\big)^\gr=-\nablaGR\ft=D\ft+\poisson{\hat h,\ft},\qquad 
    -\big(\nabla\fPta\big)^\ym=\DYM\fa +\poisson{\hat h,\fa}+\poisson{A,\ft},
\end{equation}
while $-\nablaEYM \fa=\DYM\fa+\poisson{\hat h,\fa}$.
These operators thus represent a deformation by the asymptotic data of the original complex structure $D$ (or $\DYM$) \cite{Adamo:2021lrv, Kmec:2024nmu}.

\newpage
\section{Glossary \label{secEYM:glossary}}

\begin{multicols}{2}

\ni\textbf{Vector spaces}
\begin{itemize}
    \item $\sfSK,\sfTK,\sfTK^+,\sfSTK$: Kinematical spaces \eqref{DefSTspace}.
    \item $\sfSK=\bigoplus_{s=0}^\infty\Ccar{0,-s}$.
    \item $\sfTK=\bigoplus_{s=-1}^\infty\Ccar{-1,-s}$.
    \item $\sfTK^+=\bigoplus_{s=0}^\infty\Ccar{-1,-s}$.
    \item $\sfSTK=\sfSK\oplus\sfTK$.
    \item $\sfS,\sfT,\sfST,\sfST^+$: On-shell spaces \eqref{DefSTspace}. Same vector space structure as their kinematical pendant, but with their respective dual EOM imposed. 
    \item $\osfT,\bsfT$: sub-spaces of $\sfST$ \eqref{defSpaceobar}.
    \item $\W_{CA}(\scri), \W_{\bfCc\bfAc}(S)$: Wedge spaces (\ref{CovWedgeScri}-\ref{CovWedgeEYMdefBis}).
\end{itemize}

\ni\textbf{Algebras and algebroids}
\begin{itemize}
    \item $\cS$: YM algebroid \ref{secNAGTII} and \eqref{hatdaA}.
    \item $\cT$: GR algebroid \ref{secGRI} and \eqref{defdeltaEYM}.
    \item $\cST$: EYM algebroid \eqref{defdeltaEYM}.
    \item $\ocT, \bcT$: sub-algebroids of $\cST$ \eqref{defBracketobar}.
    \item $\cSK,\cTK,\cSTK$: kinematical algebras \eqref{Kbracket}.
    \item $\Wcal_{CA}(\scri), \Wcal_{\bfCc\bfAc}(S), \Wcal^\gr_{\bfCc}(S), \Wcal^\ym_{\bfCc\bfAc}(S)$: covariant wedges (\ref{CovWedgeScri}-\ref{CovWedgeEYMdefBis}-\ref{secEYM:SemiDirectWedge}).
\end{itemize}

\ni\textbf{Brackets}
\begin{itemize}
    \item $[\cdot\,,\cdot]_\g$: YM Lie algebra bracket.
    \item $[\a{},\ap{}]^\g, \lbr\a{},\ap{}\rbr^\g$: $\sfS$-bracket \eqref{SbracketEYM} and \eqref{hatdaA}.
    \item $\{\tau,\tau'\}$: GR kinematical bracket \eqref{CbracketScridef}.
    \item $\{\Pta,\Ptap\}^\ym$: YM kinematical bracket \eqref{Kbracket}.
    \item $\{\Pta,\Ptap\}$: EYM kinematical bracket \eqref{Kbracket}.
    \item $[\tau,\tau']^C$: GR $C$-bracket \eqref{TbracketEYM}.
    \item $[\Pta,\Ptap]^\ym$: YM-bracket \eqref{YMbracketCFdef} or \eqref{YMbracketDTOTdef}.
    \item $\lbr\tau,\tp{}\rbr$: GR $\sfT$-bracket \eqref{STbracket}.
    \item $\lbr\Pta,\Ptap\rbr$: EYM $\sfST$-bracket \eqref{STbracket}.
    \item $\big\lbr\ot,\otp\big\rbr^\circ$: $\circ$-bracket \eqref{defBracketobar}.
    \item $\overline{\lbr\bt,\btp\rbr}$: bar-bracket \eqref{defBracketobar}.
\end{itemize}
\bigskip

\ni\textbf{Covariant derivatives}
\begin{itemize}
    \item $\Dcal,\DGR,\DYM,\DTOT$: \eqref{covd}.
    \item $\DEYM$: \eqref{defDEYM}.
    \item $\nabla,\nablaGR,\nablaYM,\nablaEYM$: \eqref{defNabla}.
\end{itemize}

\ni\textbf{Charges}
\begin{itemize}
    \item $\Qt^u,\Qa^u,\QPta^u$: master charges (\ref{defMasterChargeEYM}-\ref{defMasterChargeEYM2}).
    \item $\widehat{Q}_\Pta,\QPta$: Noether charges (\ref{NoetherChargeEYM}-\ref{NoetherChargeEYM3}) and \eqref{NoetherChargeEYM2}.
\end{itemize}

\ni\textbf{Actions}
\begin{itemize}
    \item $\barWhiteTriangle$: \eqref{KinematicalAction}.
    \item $\vartriangleright,\blacktriangleright, \blacktriangleleft$: \eqref{defAction}.
    \item $\dotvartriangleright, \dotblacktriangleright,\dotblacktriangleleft$: \eqref{defActionDot2}. 
    \item $\blackwhitetriangle$: \eqref{BlackWhiteAction}.
\end{itemize}
\end{multicols}

\chapter{Conclusion}

In this thesis, we constructed a symmetry algebroid $\cST^+$ onto $\scri$, which is realized on the holomorphic Ashtekar-Streubel asymptotic Einstein-Yang-Mills phase space non linearly. 
We also built the algebroid $\cST$, which admits a linear realization on a self-dual phase space that naturally appears from the reduction of twistor space to $\scri$. 
They differ on the way one deals with the gravitational symmetry transformation of degree $-1$. 
Both algebroids have the same action on the shear $C$ and asymptotic transverse gauge field $A$ and both are realized canonically via Noether charges. 
Interestingly, the charge associated with the time translation is the Bondi mass for $\cST^+$ and the covariant mass \cite{Freidel:2021qpz} for $\cST$.
These charges satisfy a Hamiltonian flow equation valid for every spin and at all order in the holomorphic coupling constants $g$ and $\gN$ and are conserved in the absence of anti-holomorphic radiation $\bN=\bnews=0$ (or $\dot\bN=\bnews=0$).
The algebroid bracket represented on phase space is built out of the $C$-bracket \eqref{CFbrackettau} and YM-bracket \eqref{YMbracketCFdef}, which are a deformation of the celebrated $sw_{1+\infty}$ bracket.
Deforming the $sw_{1+\infty}$ algebra is essential to go beyond the wedge and realize the whole symmetry algebra canonically.

All the aforementioned results are derived for pure YM, pure GR, and then build up to deal with EYM.
We give a strong emphasis on the algebraic structure, which in practice comes by studying the many facets of these brackets, from their Carrollian to their celestial realizations, in their algebroid or their algebra forms, cf.\,\eqref{CFbrackettau}-\eqref{CFbrackettauDcal}-\eqref{CbracketCompact}-\eqref{CbracketScridef}-\eqref{CelestialBracket}-\eqref{YMbracketCFdef}-\eqref{YMbracketDTOTdef}-\eqref{YMbracketCompact}-\eqref{YMbracketScridef} to cite only a few.
This analysis leads to a rich internal structure of the algebroid $\cST \simeq \ocT\Bprod\big( \bcT\Wprod\cS\big)$, which is made of a bicrossed product of 3 sub-algebroids, the time algebroid $\ocT$, the diffeomorphism plus higher spin algebroid $\bcT$ and the Yang-Mills algebroid $\cS$.
By restricting $\cST$ to its algebra part $\Wcal_{\bfCc\bfAc}(S)$ at a non-radiative cut $S$ of $\scri$, the bicrossed product boils down to the semi-direct structure $\Wcal_{\bfCc\bfAc}(S)\simeq \Wcal^\gr_{\bfCc}(S)\BWprod \Wcal^\ym_{\bfCc\bfAc}(S)$.

The key for our classically non-perturbative treatment is the introduction of time and field dependent symmetry parameters $\t{s}$ \& $\a{s}$ constrained by a set of evolution equations \eqref{DualEOMEYM} dual to (a truncation of) the asymptotic Einstein-Yang-Mills equations \eqref{EOMEYM}.
The infinitesimal shear transformation under the algebroid is then parametrized entirely by $\t0$, $\t1$ and $\t2$ while the infinitesimal variation of $A$ is given in terms of $\a0,\a1$ and $\t0$, see \eqref{deltaPtaCA}.
The usual celestial symmetry parameters $\T{s},\Aa{s}$ that generalize super-translations $\T0$, sphere diffeomorphisms $\T1$ and super-phase rotation $\Aa0$ are understood as initial conditions for the dual EOM.
In other words, they correspond to the value of $(\t{s},\a{s})$ at a certain cut of $\scri$.
When evaluating $\t0,\t1,\t2,\a0,\a1$ on such a solution parametrized by $\T{s}$ \& $\Aa{s}$, the dual EOM encode all the non-linearity and non-locality of the radiative data variations.
Similarly, the Noether charge (defined at $\scri^+_-$) associated to the spin $s$ parameter $\T{s}$ or $\Aa{s}$ is the result of a renormalization procedure fully determined by the dual EOM.

We also find that the symmetry transformations that preserve the shear and gauge potential at any non-radiative cut of $\scri$ define an algebra---called the covariant wedge algebra $\Wcal_{\bfCc\bfAc}(S)$. 
This is the latter which represents the true symmetry transformations preserving the late time fall-off condition of the asymptotic data. 
We explain how these algebras (one of for each shear and gauge field) are embedded inside the algebroid. 
Our formalism guarantees that $\Wcal_{\bfCc\bfAc}(S)$ is canonically represented on $\scri$ through the $\cST$-algebroid.

Moreover, we prove that the $C$-bracket (and YM-bracket) amounts to the twistor Poisson bracket discussed in \cite{Kmec:2024nmu}.
The correspondence relies heavily on two essential features: 
The higher spin parameters can all be recast in terms of a generating functional which depends on a spin variable $q$. 
This variable was first introduced by Newman as the parameter that describes the choice of Carrollian ruling of $\scri$ \cite{Freidel:2024emv, Riello:2024uvs}. 
It also can be understood from the twistor perspective as the fiber coordinate of the twistor fibration over $\scri$. 
The second ingredient is the dual equations of motion that allow to recast the $C$-bracket defined on the celestial sphere, which explicitly depends on the shear, into a linear Poisson bracket. 
One therefore sees that the twistor description amounts to a linearization of the gravitational symmetry representation (again, only possible on-shell of the dual EOM). 
\bigskip

Our work opens up several questions. 
First the analysis of \cite{Kmec:2024nmu} and recently \cite{Kmec:2025ftx} clearly shows that the  asymptotic EOM \eqref{Qevolve} can be obtained as charge conservation in self-dual gravity. 
What is missing, from our perspective, is to understand clearly the relationship between these equations and the asymptotic Bianchi identities. 
It was conjectured in \cite{Geiller:2024bgf} that the conservation laws can be extracted from the Weyl Bianchi identities associated with self-dual gravity after a proper choice of asymptotic frame, but that still needs to be demonstrated.
Equivalently in non-abelian gauge theory, the self-dual condition which consists in setting the Newman-Penrose scalar $\bar\phi_{\bm},\bar\phi$ and $\bar\phi_m$ to 0 (and thus $\tbQ_s=0$) does not reduce \eqref{phi0n} (or similarly \eqref{Qnevol}) into the fundamental pattern \eqref{fundamentalPattern}.

On another front, while the symmetry algebras are identified as generalized wedge algebras, the algebroid, which goes beyond the wedge, has to be the appropriate algebraic entity to describe radiation.
Some questions are still to resolve, in particular it would be interesting to understand more deeply how the algebroid structure allows us to interpolate between two non radiative cuts. 
A better understanding of the algebra $\Wcal_C(\scri)$ and potential algebroid structure $\A_{\bfCc}$ will probably be necessary to address this question.

Also, while we constructed explicitly the renormalized symmetry charges parametrized in terms of $\T{s}$ or $\Aa{s}$ defined at a cut $u=u_0$, we did not study directly the flux laws for these renormalized charges. 
This should allow us to define directly the charges in terms of data defined as limits to $\scri^+_-$ and $\scri^+_+$, which would be interesting to construct.

Moreover, while we have mostly described the renormalized charges at constant $u=u_0$ cuts (see however section \ref{sec:ArbitraryCut}), we can more generally define them at arbitrary cuts $u=U(z,\bar{z})$.
It would be desirable to understand better the covariance properties of the charges, and how does that relate to the notion of dressing map that enables to reabsorb the shear into a change of frame, as described in sections \ref{sec:CovWedgeSol} and \ref{sec:GoodCut}. 

It would also be interesting to understand the (anti-)holomorphicity structure determined by the equation \eqref{dualEOM}, which does not involve the operator $\bD$. 
See \cite{Chen:2023rxf, Garner:2023zqn} for 3d generalization of the notion of chirality, relevant for extending the celestial modes expansion in $(z,\bz)$ to include the $u$ direction. 
Working solely with a cut of $\scri$ for instance does not capture the higher degree negative modes (i.e. the raviolo polynomials \cite{Garner:2023zqn}) which could be present when working globally on null infinity. 
The relevance, if any, of these modes in celestial holography seems worth investigating.

Besides, an important question concerns whether we could generalize the canonical analysis and relax the Schwartzian fall-off condition for the time dependency of the shear/gauge potential and the charge aspects. 
Recently the work \cite{Nagy:2024jua} investigated the possibility of constructing an extended phase space for Yang-Mills---generalizing the dressing field ideas of \cite{Donnelly:2016auv} to higher spin (see also \cite{Campiglia:2024uqq} for an application of the phase space extension to include GBMS)---in order to allow for a relaxation of the asymptotic boundary conditions. 
A similar construction should be available in gravity and it seems important to understand the connections between \cite{Nagy:2024jua} and the current work.

Another important aspect left to understand is the link with OPEs and soft theorems. 
A natural question is that if we assume that the $S$-matrix is invariant under the action of the Noether charge $\Qt$ (including the super-hard contributions), what does it imply for the scattering of soft gravitons (see the recent work \cite{Jorstad:2025qxu} for super-hard corrections to the sub-subleading soft graviton theorem)? 
Is there a quantum anomaly in the charge algebra as suggested by \cite{Costello:2022upu, Bittleston:2022jeq} that can be re-derived from the canonical perspective?
See \cite{Baulieu:2024oql, Baulieu:2025itt} for a BRST approach on that question for super-rotations.

After having treated theories in vacuum, we naturally expect our formalism to hold in the presence of matter.
In particular, including massive matter and getting a proper understanding of the matching conditions between future timelike infinity $i^+$ and null infinity at $\scrip_+$ is desirable.
More generally, we seek a unified picture from past timelike infinity $i^-$ to $i^+$, passing through spatial infinity $i^0$.
See \cite{Campiglia:2015lxa, Compere:2023qoa, Ashtekar:2023zul, Fiorucci:2024ndw} for works in the context of the BMS group.
We expect the conservation laws at $i^0$ (where there is no flux of radiation) to determine a set of antipodal matching conditions for the higher spin charge aspects, as originally proposed for the Bondi mass \cite{Strominger:2013jfa}.

Besides, being able to relate the charge aspects to the more common language of multipoles expansion would shed a new light on the way bulk information can be collected by a distant observer \cite{Compere:2022zdz}. 
In conjunction with the previous paragraph, let us mention the recent work \cite{Compere:2025tzr}, which studies electromagnetic multipoles around $i^0$ and indeed deduces an infinite set of antipodal matching relations between them.
Upon inclusion of long-range Coulombic interactions, the antipodal matching conditions become subtler due to the appearance of tails leading in particular to logarithmic corrections to soft theorems \cite{He:2014bga, Sahoo:2018lxl, Campiglia:2019wxe, AtulBhatkar:2020hqz, Sahoo:2020ryf}.
How to treat these logarithmic corrections from an amplitude point of view, both their classical and quantum parts, and the extent to which they are determined by asymptotic symmetries is an active topic of research \cite{Choi:2024ygx, Choi:2024ajz, Choi:2024mac}.
The question is also related to the inclusion of Goldstone modes in the phase space \cite{Donnay:2022hkf, Agrawal:2023zea}, which could admit a higher spin generalization.

Finally, from \cite{Kmec:2024nmu} it is clear that the higher spin symmetries described here are symmetries of self-dual gravity. 
However we also know that they are exact symmetry of full Einstein gravity up to spin 3 (and up to spin 1 in YM) and that spin 2 (resp. spin 1 for YM) generates the entire tower of higher spin charges. 
It therefore begs the question to understand what can be the use of the higher spin charges in full GR.

We can hope that the understanding of the quantization of the higher spin charges serves as a solid basis to describe asymptotic dressed states and develop a new type of interacting picture for full gravity. 
Whether or not part of the algebraic structure (or a deformation of it) will survive the inclusion of the corrections to the fundamental pattern \eqref{EOMEYM} necessary to reproduce the full asymptotic Einstein-Yang-Mills equations is a major open question.

\newpage
\appendix
\phantomsection
\addcontentsline{toc}{chapter}{Appendices}
\chapter{Non-Abelian Gauge Theory I \label{AppNAGTI}}

\section{Covariant derivative onto the sphere \label{AppCNCD}}

The formulas (\ref{intrinsic1}) can be obtained starting from the definition of the intrinsic derivative \cite{Chandrasekhar}:
\begin{align}
O_{m\ldots m|m}&:=m^{A_1}\ldots m^{A_s}m^AD_AO_{A_1\ldots A_s} \nonumber\\
&=\pa_m O_s-s\big(m^AD_Am^B\big)\bm_B O_s \nonumber\\
&=P\partial_z  O_s+s\frac{P}{\bP}\partial_z\bP O_s \nonumber\\
&\equiv D O_s \label{intrinsic2}\\
\textrm{and similarly}\qquad  O_{m\ldots m|\bm}&\equiv\bD O_s. \nn
\end{align}
Moreover, as it is clear from the definition (\ref{intrinsic2}), $D O_s$ has helicity $s+1$ and $\bD O_s$ has helicity $s-1$. 
Besides, $D$ and $\bD$ satisfy the commutation relation
\begin{equation}
\boxed{[\bD,D] \Ocal_s=\cR s\Ocal_s}.
\end{equation}
In the following, we will need another useful relation,
\begin{align}
D\bD^n O_{n+p}&=\big(\bD D-\cR(p+1)\big)\bD^{n-1} O_{n+p}\nn\\
&=\bD\big(\bD D-\cR(p+2)\big)\bD^{n-2} O_{n+p}-\cR(p+1)\bD^{n-1} O_{n+p}\nn\\
&=\bD^2 D\bD^{n-2} O_{n+p}-\cR\big((p+1)+(p+2)\big)\bD^{n-1} O_{n+p}\nn\\
&=\bD^n D O_{n+p}-\cR\sum_{k=1}^{n}(p+k)\,\bD^{n-1} O_{n+p}.
\end{align}
Therefore,
\begin{equation}
\boxed{D\bD^n O_{n+p}=\bD^nD O_{n+p}-\cR\frac{n(n+1+2p)}{2}\bD^{n-1} O_{n+p}}. \label{ethcommutn}
\end{equation}

Using the last relation recursively, we can even generalize it further and get $D^m\bD^n\Ocal_{n+p}$. 
After a lengthier but similar in spirit computation, we can show that
\begin{equation}
    \boxed{D^m\bD^n\Ocal_{n+p}=\sum_{k=0}^{\min[m,n]} (-\cR)^k\cF_{k,m} \bD^{n-k}D^{m-k}\Ocal_{n+p}},
\end{equation}
where
\begin{align}
    \cF_{0,m} &=1, \nn\\
    \cF_{1,m} &=\sum_{q=1}^m\frac{n(n+2q-1+2p)}{2}, \nn\\
    \cF_{m,m} &=\prod_{q=1}^m\frac{(n+1-q)(n+q+2p)}{2}, \\
    \cF_{k,m} &=\cF_{k,k}+\sum_{q=k}^{m-1}\cF_{k-1,q}\frac{(n+1-k)(n+2q-k+2+2p)}{2},\qquad k\neq\{0,1,m\}. \nn
\end{align}

\subsection{For a gauge covariant derivative}

When dealing with non-abelian gauge theory, the gauge field becomes charged and we need to replace the covariant derivative onto the sphere $D$ by the gauge covariant derivative $\Dcal$ (\ref{gaugecovd}).

First of all, (\ref{ethcommut}) now includes the usual curvature term:
\begin{equation}
\boxed{[\bDcal,\Dcal]  O_s=\cR s O_s+\big[F_{\bm m}^{(0)}, O_s\big]_\g}. \label{ethcommutYM}
\end{equation}
When one has to rearrange terms, one can use\footnote{In our case, $ O_s$ always transforms under the adjoint representation.}
\begin{equation}
\bD D O_s =\bDcal\Dcal O_s-[\bA,D O_s]_\g-\bD[A, O_s]_\g-\big[\bA,[A, O_s]_\g\big]_\g, \label{switchDcovd}
\end{equation}
or its complex conjugate. The equation (\ref{ethcommutn}) also admits a generalization, namely
\begin{align*}
\Dcal\bDcal^n O_{n+s} &=\bDcal^n \Dcal O_{n+s}-\cR\frac{n(n+1+2s)}{2}\bDcal^{n-1} O_{n+s}-\sum_{k=0}^{n-1}\bDcal^k\big[F_{\bm m}^{(0)},\bDcal^{n-1-k} O_{n+s}\big]_\g \\
&= \bDcal^n \Dcal O_{n+s}-\cR\frac{n(n+1+2s)}{2}\bDcal^{n-1} O_{n+s}-\sum_{k=0}^{n-1}\sum_{p=0}^k\binom{k}{p}\big[\bDcal^p F_{\bm m}^{(0)},\bDcal^{n-1-p} O_{n+s}\big]_\g \\
&=\bDcal^n \Dcal O_{n+s}-\cR\frac{n(n+1+2s)}{2}\bDcal^{n-1} O_{n+s}-\sum_{p=0}^{n-1}\sum_{k=p}^{n-1}\binom{k}{p}\big[\bDcal^p F_{\bm m}^{(0)},\bDcal^{n-1-p} O_{n+s}\big]_\g.
\end{align*}
Finally,
\begin{empheq}[box=\fbox]{align}
\Dcal\bDcal^n O_{n+s} &=\bDcal^n \Dcal O_{n+s}-\cR\frac{n(n+1+2s)}{2}\bDcal^{n-1} O_{n+s} \nn\\
&-\sum_{p=0}^{n-1}\frac{n!}{(p+1)!(n-p-1)!}\big[\bDcal^p F_{\bm m}^{(0)},\bDcal^{n-1-p} O_{n+s}\big]_\g. \label{ethcommutnYM}
\end{empheq}
One also easily deduce that 
\begin{equation} \label{pauDcalnCommut}
    \pa_u\bDcal^nO_s=\bDcal^n\pa_u O_s+\sum_{p=0}^{n-1}\bDcal^p\big[\bnews,\bDcal^{n-p-1}O_s\big]_\g.
\end{equation}

\section{Evolution equation for $\np{m}{n}$ \label{APPphi0}}

The starting point is the closed form (\ref{phiclose0}) for $\phi_m$ in the radial gauge. 
We reproduce it here for convenience:
\begin{equation}
\begin{split}
\underbrace{\left(\cR+\cR\frac{r\partial_r(3+r\partial_r)}{2}-r\partial_u(3+r\partial_r)\right)\phi_m+D\bD\phi_m}_{\dn{1}}  + \underbrace{{D[A_{\bm},\phi_m]_\g}_{\vphantom{\big(}}}_{\dn{2}}\\
+\underbrace{(2+r\partial_r)\big(r[\phi_m,A_u]_\g \big)_{\vphantom{\big(}}}_{\dn{3}} +\underbrace{{(2+r\pa_r)[A_m,\phi]_\g}_{\vphantom{\big(}}}_{\dn{4}}=0.
\end{split}
\end{equation}
Using the radial expansions (\ref{radialexp}) and for now only shifting the sums when necessary, we get\footnote{In the following, we will always write expressions valid for $n\geqslant 0$ unless stated otherwise. 
If a sum does not contain a term $n=0$, it is assumed to be 0. Moreover, notice that $\sum_{p=0}^{n-1}[A_{\bm}^{(p+1)},\np{m}{n-p-1}]_\g=\sum_{p=0}^{n-1}[A_{\bm}^{(n-p)},\np{m}{p}]_\g$. Both ways can be convenient sometimes.}
\begin{align}
    \dn{1} &=\aunderbrace[l1r]{\cR\np{m}{n}}+\cR\frac{n(n+3)}{2}\np{m}{n}+(n+1)\pa_u\np{m}{n+1}+\aunderbrace[l1r]{D\bD\np{m}{n}}, \nn\\
    \dn{2} &= \sum_{p=0}^n D\big[A_{\bm}^{(n-p)},\np{m}{p}\big]_\g,\\
    \dn{3} &=(n+1)\sum_{p=0}^{n}\big[A_u^{(n+1-p)},\np{m}{p}\big]_\g, \nn\\
    \dn{4} &= -(n+1)\sum_{p=0}^{n+1}\big[A_m^{(n+1-p)},\np{}{p}\big]_\g. \nn
\end{align}

First of all, we get rid of the field $\phi$ using \eqref{phi10}, reproduced here,
\begin{equation}
    \np{}{p+1}=-\frac{1}{p+1}\left(\bDcal\np{m}{p}+\sum_{k=0}^{p-1}\big[A_{\bm}^{(p-k)},\np{m}{k}\big]_\g\right),\qquad n\geqslant 1. \nn
\end{equation}
We also start to extract the leading charge $\tQ_{0}$ and the phase space variables $A,\bA$ since they have a different status than the sub-leading terms (e.g. $A,\bA$ will later recombine into the gauge covariant derivative $\Dcal$). 
Second of all, taking advantage of the radial gauge $A_r=0=A_u^{(0)}$, we rewrite every appearance of $A_\mu$'s in terms of $\phi$'s. Indeed, $F_{rm}=r\phi_m=\pa_r A_{m}$ which amounts to
\begin{equation}
\boxed{-(n+1)A_{m}^{(n+1)}=\np{m}{n}},\qquad n\geqslant 0.
\end{equation}
Moreover $F_{ru}=\phi+\bar\phi=\pa_r A_u$ meaning that\footnote{For completeness, notice also that $r\phi_{\bm}=F_{\bm u}-\frac12 F_{\bm r}=F_{\bm u}+\frac12 \pa_r A_{\bm}$ so that that $\tQ_{-1}=F_{\bm u}^{(0)}$.}
\begin{equation}
\boxed{-(n+1)A_u^{(n+1)}=\np{}{n}+\bnp{}{n}},\qquad n\geqslant 0. \label{Auphiphibar}
\end{equation} 
Hence the fourth term is recast as
\begin{align}
\dn{4} &=-(n+1)\big[A_m^{(n+1)},\np{}{0}\big]_\g-(n+1)\sum_{p=0}^n\big[A_{m}^{(n-p)},\np{}{p+1}\big]_\g \nn\\
&=\big[\np{m}{n},\tQ_{0}\big]_\g+(n+1)\sum_{p=0}^n\frac{1}{p+1}\big[A_{m}^{(n-p)},\bDcal\np{m}{p}\big]_\g+(n+1)\sum_{p=0}^n\sum_{k=0}^{p-1}\frac{1}{p+1}\big[A_{m}^{(n-p)},[A_{\bm}^{(p-k)},\np{m}{k}]_\g\big]_\g \nn\\
&=\big[\np{m}{n},\tQ_{0}\big]_\g+\aunderbrace[l1r]{\big[A,\bDcal\np{m}{n}\big]_\g}-\sum_{p=0}^{n-1}\frac{n+1}{(p+1)(n-p)}\overbrace{\big[\np{m}{n-p-1},\bDcal\np{m}{p}\big]_\g} \label{term4interm}\\
&-\aoverbrace[L1R]{\sum_{k=0}^{n-1}\frac{1}{k+1}\big[A,[\bnp{m}{k},\np{m}{n-1-k}]_\g\big]_\g}+\sum_{p=0}^{n-1}\sum_{k=0}^{p-1}\frac{n+1}{(p+1)(n-p)(k+1)}\big[\np{m}{n-1-p},[\bnp{m}{k},\np{m}{p-1-k}]_\g\big]_\g. \nn
\end{align}
Without further ado the second term is
\begin{equation}
    \dn{2}=\aunderbrace[l1r]{D\big[\bA,\np{m}{n}\big]_\g}-\aoverbrace[L1R]{\sum_{p=0}^{n-1}\frac{1}{p+1}D\big[\bnp{m}{p},\np{m}{n-1-p}\big]_\g}. 
\end{equation}

The 2 terms denoted with a $\aoverbrace[L1R]{\cdot\,\cdot\,\cdot}$ form a total gauge covariant derivative. On the other hand, the four terms marked with a $\aunderbrace[l1r]{\cdot\,\cdot\,\cdot}$ recombine as follows
\begin{align}
\left(\np{m}{n}+D\bD\np{m}{n}+D\big[\bA,\np{m}{n}\big]_\g+\big[A,\bDcal\np{m}{n}\big]_\g\right) &=\np{m}{n}+\Dcal\bDcal\np{m}{n} \nn\\
&=\bDcal\Dcal\np{m}{n}-\big[F^{(0)}_{\bm m},\np{m}{n}\big]_\g \nn\\
&=\bDcal\Dcal\np{m}{n}-\big[\tQ_{0}-\tbQ_{0},\np{m}{n}\big]_\g,
\end{align}
where in the last step we used the fact that $F_{\bm m}=r^2(\phi-\bar\phi)$, meaning that
\begin{equation}
\boxed{F^{(0)}_{\bm m}=\tQ_{0}-\tbQ_{0}}.
\end{equation}

We now address the third term, using \eqref{Auphiphibar}:
\begin{align}
    \dn{3} &=(n+1)\sum_{p=0}^{n}\big[A_u^{(p+1)},\np{m}{n-p}\big]_\g=-\sum_{p=0}^{n}\frac{n+1}{p+1}\big[\np{}{p}+\bnp{}{p},\np{m}{n-p}\big]_\g \nn\\
    &=-(n+1)\big[\tQ_{0}+\tbQ_{0},\np{m}{n}\big]_\g-\sum_{p=0}^{n-1}\frac{n+1}{p+2}\big[\np{}{p+1}+\bnp{}{p+1},\np{m}{n-1-p}\big]_\g \nn\\
    &=-(n+1)\big[\tQ_{0}+\tbQ_{0},\np{m}{n}\big]_\g+\sum_{p=0}^{n-1}\frac{n+1}{(p+1)(p+2)}\big[\Dcal\bnp{m}{p}+\overbrace{\bDcal\np{m}{p},\np{m}{n-1-p}\big]_\g} \nn\\
    &-\sum_{p=0}^{n-1}\sum_{k=0}^{p-1}\frac{n+1}{(p+1)(p+2)(p-k)}\big[[\bnp{m}{p-1-k},\np{m}{k}]_\g+[\np{m}{p-1-k},\bnp{m}{k}]_\g,\np{m}{n-1-p}\big]_\g \label{term3interm}
\end{align}
If we add the terms involving a double commutator in \eqref{term4interm} and \eqref{term3interm}, we get
\begin{equation}
    -\sum_{p=0}^{n-1}\sum_{k=0}^{p-1}\frac{(n+1)\big[(n+2)(p-k)-(k+1)(n-p)\big]}{(p+1)(p+2)(p-k)(k+1)(n-p)}\big[[\bnp{m}{k},\np{m}{p-1-k}]_\g,\np{m}{n-1-p}\big]_\g.
\end{equation}
Furthermore, the 2 terms topped by a $\overbrace{\cdot\,\cdot\,\cdot}$ sum up to
\begin{equation}
    +\sum_{p=0}^{n-1}\frac{(n+1)(n+2)}{(p+1)(p+2)(n-p)}\big[\bDcal\np{m}{p},\np{m}{n-p-1}\big]_\g.
\end{equation}

Collecting everything, we find the evolution equation \eqref{phi0n} for $\np{m}{n}$:
\begin{align}
    &\qquad (n+1)\pa_u\np{m}{n+1}+\cR\frac{n(n+3)}{2}\np{m}{n}+\bDcal\Dcal\np{m}{n} \nn\\
    &\quad -(n+3)\big[\tQ_{0},\np{m}{n}\big]_\g+\sum_{p=0}^{n-1}\frac{(n+1)(n+2)}{(p+1)(p+2)(n-p)}\big[\bDcal\np{m}{p},\np{m}{n-p-1}\big]_\g \nn\\
    &=n\big[\tbQ_{0},\np{m}{n}\big]_\g+\sum_{p=0}^{n-1}\frac{1}{p+1}\Dcal\big[\bnp{m}{p},\np{m}{n-1-p}\big]_\g-\sum_{p=0}^{n-1}\frac{n+1}{(p+1)(p+2)}\big[\Dcal\bnp{m}{p},\np{m}{n-1-p}\big]_\g \nn\\
    &\quad+\sum_{p=0}^{n-1}\sum_{k=0}^{p-1}\frac{(n+1)\big[(n+2)(p-k)-(k+1)(n-p)\big]}{(p+1)(p+2)(p-k)(k+1)(n-p)}\big[[\bnp{m}{k},\np{m}{p-1-k}]_\g,\np{m}{n-1-p}\big]_\g.
 \end{align}
In the next section, we shall refer to each line by \dn{31}, \dn{32}, \dn{33} and \dn{34} respectively.

\section{Evolution equation for $\tQ_{n}$ \label{APPphi0Q}}

We have computed the evolution equation of the charges $\tQ_{0}$ and $\tQ_{1}$, cf. equations \eqref{Q0evol} and \eqref{Q1evol}, let us do the same for $\tQ_{n}$, $n>1$. 
Starting from \eqref{phi0n} and using the local ansatz \eqref{ansatzcharge}, i.e. $\np{m}{n}=\frac{(-1)^n}{n!}\bDcal^n\tQ_{n+1}$, we obtain\footnote{We have multiplied each line by $(-1)^{n+1}n!$ and used \eqref{ethcommutnYM} to massage \dn{31}.}
\begin{align}
    \dn{31}&=\bDcal^{n+1}(\pa_u\tQ_{n+2}-\Dcal\tQ_{n+1})+\sum_{p=0}^n\bDcal^p\big[\bnews, \bDcal^{n-p}\tQ_{n+2}\big]_\g \nn\\
    &\quad +\aunderbrace[l1r]{\sum_{p=0}^{n-1}\frac{n!}{(p+1)!(n-p-1)!}\bDcal \big[\bDcal^p(\tQ_{0}-\tbQ_{0}),\bDcal^{n-1-p}\tQ_{n+1}\big]_\g}, \nn\\
    \dn{32}&=\aunderbrace[l1r]{(n+3)\big[\tQ_{0},\bDcal^n\tQ_{n+1}\big]_\g}+\sum_{p=0}^{n-1}\frac{(n+2)!}{(p+2)!(n-p)!}\big[\bDcal^{p+1}\tQ_{p+1},\bDcal^{n-1-p}\tQ_{n-p}\big]_\g, \nn\\
    \dn{33}&=\aunderbrace[l1r]{-n\big[\tbQ_{0},\bDcal^n\tQ_{n+1}\big]_\g}+\sum_{p=0}^{n-1}\frac{n!}{(p+1)!(n-1-p)!}\Dcal\big[\Dcal^p\tbQ_{p+1},\bDcal^{n-1-p}\tQ_{n-p}\big]_\g\nn\\
    & \quad -\sum_{p=0}^{n-1}\frac{(n+1)!}{(p+2)!(n-1-p)!}\big[\Dcal^{p+1}\tbQ_{p+1},\bDcal^{n-1-p}\tQ_{n-p}\big]_\g, \\
    \dn{34}&=-\sum_{p=0}^{n-1}\sum_{k=0}^{p-1}\frac{(n+1)!\big[(n+2)(p-k)-(k+1)(n-p)\big]}{(p+1)(p+2)(p-k)!(k+1)!(n-p)!}\big[[\Dcal^k\tbQ_{k+1},\bDcal^{p-1-k}\tQ_{p-k}]_\g,\bDcal^{n-1-p}\tQ_{n-p}\big]_\g \nn
\end{align}
The 3 terms denoted with $\aunderbrace[l1r]{\cdot\,\cdot\,\cdot}$ reduce to
\begin{equation}
    +\sum_{p=0}^{n-1}\frac{(n+1)!}{(p+2)!(n-p-1)!}\big[\bDcal^{p+1}(\tQ_{0}-\tbQ_{0}),\bDcal^{n-p-1}\tQ_{n+1}\big]_\g+(2n+3)\big[\tQ_{0},\bDcal^n\tQ_{n+1}\big]_\g
\end{equation}
The evolution equation for the charge $\tQ_{n}$ is then \eqref{Qnevol}
\vspace{-0.5cm}

\begin{adjustwidth}{-0.6cm}{-0.6cm}
\begin{align}
    &\quad \bDcal^{n+1}(\pa_u\tQ_{n+2}-\Dcal\tQ_{n+1})-\sum_{p=0}^n\bDcal^p\big[ \tQ_{-1},\bDcal^{n-p}\tQ_{n+2}\big]_\g +(2n+3)\big[ \tQ_{0},\bDcal^n\tQ_{n+1}\big]_\g \nn\\
    &+\sum_{p=0}^{n-1}\frac{(n+1)!}{(p+2)!(n-p-1)!}\big[\bDcal^{p+1}\tQ_{0},\bDcal^{n-p-1}\tQ_{n+1}\big]_\g +\sum_{p=0}^{n-1}\frac{(n+2)!}{(p+2)!(n-p)!}\big[\bDcal^{p+1}\tQ_{p+1},\bDcal^{n-1-p}\tQ_{n-p}\big]_\g \nn\\
    &=\sum_{p=0}^{n-1}\frac{(n+1)!}{(p+2)!(n-p-1)!}\big[\bDcal^{p+1}\tbQ_{0},\bDcal^{n-p-1}\tQ_{n+1}\big]_\g \nn\\
    &-\sum_{p=0}^{n-1}\frac{(n+1)!}{(p+2)!(n-1-p)!}\big[\Dcal^{p+1}\tbQ_{p+1},\bDcal^{n-1-p}\tQ_{n-p}\big]_\g \nn\\
    &+\sum_{p=0}^{n-1}\frac{n!}{(p+1)!(n-1-p)!}\Dcal\big[\Dcal^p\tbQ_{p+1},\bDcal^{n-1-p}\tQ_{n-p}\big]_\g\nn\\
    &+ \sum_{p=0}^{n-1}\sum_{k=0}^{p-1}\frac{(n+1)!\big((n+2)(p-k)-(k+1)(n-p)\big)}{(p+1)(p+2)(p-k)!(k+1)!(n-p)!}\big[\bDcal^{n-1-p}\tQ_{n-p},[\Dcal^k\tbQ_{k+1},\bDcal^{p-1-k}\tQ_{p-k}]_\g\big]_\g. \nn
\end{align}
\end{adjustwidth}
It is also possible to expand the total derivative so that the fourth and the fifth line of the latter equation recombine into
\begin{equation}
    \sum_{p=0}^{n-2}\frac{n!}{(n-p)!p!}\big[\bDcal^{p+1}\tQ_{p+2},\Dcal^{n-p-1}\tbQ_{n-p-1}\big]_\g+\sum_{p=0}^{n-1}\frac{n!}{(n-p)!p!}\big[\Dcal^{n-p-1}\tbQ_{n-p},\Dcal\bDcal^p\tQ_{p+1}\big]_\g.
\end{equation}

\chapter{Non-Abelian Gauge Theory II \label{AppNAGTII}}

\section{Proof of the Jacobi identity \label{App:JacobiYM}}

To prove that the $\sfS$-bracket $\lbr\cdot,\cdot\rbr$ is indeed a Lie (algebroid) bracket, we check that it satisfies the Jacobi identity.
First notice that $\big\lbr\a{},\lbr\ap{},\app{}\rbr\big\rbr_s$ splits into 3 contributions:
\begin{subequations}
\begin{align}
    \!\!\!\!\big\lbr\a{},\lbr\ap{},\app{}\rbr \big\rbr_s &=-\da\lbr\ap{},\app{}\rbr_s+ \delta_{\lbr\ap{},\app{}\rbr}\a{s} +\sum_{n=0}^s\big[\a{n},\lbr\ap{},\app{}\rbr_{s-n}\big]_\g \nn\\
    &=\da\dap\app{s}-\da\dapp\ap{s}+ \delta_{\lbr\ap{},\app{}\rbr}\a{s}\label{line1}\\
    &-\sum_{n=0}^s\Big(\big[\da\ap{n},\app{s-n}\big]_\g+\big[\ap{n},\da\app{s-n}\big]_\g\Big)-\sum_{n=0}^s\big[\a{n},\dap\app{s-n}-\dapp\ap{s-n}\big]_\g \label{line2}\\
    &+\sum_{n=0}^s\sum_{k=0}^{s-n}\big[\a{n},[\ap{k},\app{s-n-k}]_\g\big]_\g. \label{line3}
\end{align}
\end{subequations}
Taking the cyclic permutation,\footnote{We use the notation $\cyc$ to denote an equality valid upon adding the cyclic permutation of the terms on the left and right hand sides.} we get that the first line vanishes,
\begin{equation}
    \eqref{line1}\cyc \da\dap\app{s}-\dap\da\app{s}+ \delta_{\lbr\a{},\ap{}\rbr}\app{s}= 0,
\end{equation}
where we used the morphism property \eqref{SalgebraMorphism}.
Similarly, all terms of \eqref{line2} cancel once we consider its cyclic permutation.
Indeed,
\begin{align}
    -\eqref{line2}&\cyc \sum_{n=0}^s\Big(\big[\da\ap{n},\app{s-n}\big]_\g+\big[\ap{n},\da\app{s-n}\big]_\g\Big)+\sum_{n=0}^s\big[\app{n},\da\ap{s-n}\big]_\g -\sum_{n=0}^s\big[\ap{n},\da\app{s-n}\big]_\g \nn\\
    &\cyc \sum_{n=0}^s\big[\da\ap{n},\app{s-n}\big]_\g+\sum_{n=0}^s\big[\app{s-n},\da\ap{n}\big]_\g = 0,
\end{align}
where to get the second equality, we changed $n\to s-n$ in the third term of the first line.
The contribution \eqref{line3} vanishes thanks to the Jacobi identity of the Lie algebra bracket $[\cdot\,,\cdot]_\g$.\footnote{Simply notice that $\eqref{line3}=\sum_{a+b+c=s}\big[\a{a},[\ap{b},\app{c}]_\g\big]_\g$ and that the sum is now invariant under permutation of $a,b,c$. \label{footJacobi}}
We thus infer that $\big\lbr\a{},\lbr\ap{},\app{}\rbr \big\rbr\cyc 0$.

\section{Closure of the $\sfS$-bracket and Leibniz anomalies \label{App:dualEOMYM}}

In this section we use that the variational derivative $\da$ commutes with the differentials 
\be 
[\pa_u, \da]=0, \qquad [D, \da]=0.
\ee 
This is a consequence of the fact that the variational Cartan calculus commutes with the differential Cartan calculus \cite{Freidel:2020xyx}.\footnote{In general we have $[\delta, \LL_V]= \LL_{\delta V}$ for an arbitrary vector field.} 

The action $\da$ can be written explicitly as a vector field on the jet bundle  \cite{saunders1989geometry} $J^\infty\bfm p$ over $\scri$, where $\bfm p:\PS\to\scri$ is the line bundle of Carrollian fields of weight $(1,1)$:
\begin{align}
    \da O   = \sum_{n,m=0}^\infty \int_{\scri}  \pa_u^nD^m\big(\da A\big)
    \frac{\delta O}{\delta \big(\pa_u^nD^m A\big)}. \label{actiondelta}
\end{align}
\medskip

Recall that we define the Leibniz rule anomaly for any bracket $[\cdot\,,\cdot]$ and differential operator $\mathscr{D}$ as
\begin{equation}
    \cA\big([\cdot\,,\cdot],\mathscr{D} \big)=\mathscr{D}[\cdot\,,\cdot]-[\mathscr{D}\cdot\,,\cdot]-[\cdot\,,\mathscr{D}\cdot].
\end{equation}
We already proved in \eqref{DcalLeibniz} that
\begin{equation}
    \cA\big([\cdot\,,\cdot]_\g,\Dcal\big)=0.
\end{equation}
However, notice that
\begin{align}
    \big(\Dcal[\a{},\ap{}]^\g\big)_{s} &= \Dcal[\a{},\ap{}]^\g_{s+1}=\sum_{n=0}^{s+1}\Big(\big[(\Dcal\a{})_{n-1},\ap{s+1-n} \big]_\g+\big[\a{n},(\Dcal\ap{})_{s-n}\big]_\g\Big) \\
    &=\big[(\Dcal\a{})_{-1},\ap{s+1}\big]_\g \!+\!\sum_{n=0}^{s}\big[(\Dcal\a{})_n,\ap{s-n} \big]_\g\!+\big[\a{s+1},(\Dcal\ap{})_{-1}\big]_\g\!+\!\sum_{n=0}^{s} \big[\a{n},(\Dcal\ap{})_{s-n}\big]_\g, \nn
\end{align}
while
\vspace{-0.2cm}
\begin{equation}
    \big[\Dcal\a{},\ap{}\big]^\g_s+\big[\a{}, \Dcal\ap{}\big]^\g_s= \sum_{n=0}^{s}\big[(\Dcal\a{})_n,\ap{s-n} \big]_\g+\sum_{n=0}^{s} \big[\a{n},(\Dcal\ap{})_{s-n}\big]_\g.
\end{equation}
Consequently, using $(\Dcal\a{})_{-1}=\Dcal\a0=-\da A$, we get that\footnote{Note that this anomaly vanishes inside $\Wcal_A(\scri)$ (or $\Wcal_{\bfAc}(S)$ if we trade $\a{}$ for $\Aa{}$), see Sec.\,\ref{secYM:wedge}.}
\begin{equation} \label{DcalLeibnizApp}
    \boxed{\cA_s\big([\a{},\ap{}]^\g,\Dcal\big)= \big[\dap A,\a{s+1}\big]_\g-\big[\da A,\ap{s+1}\big]_\g}.
\end{equation}
Next, using that
\begin{align}
    \Dcal\big(\dap\a{s+1}-\da\ap{s+1}\big) &=\dap(\Dcal\a{})_s-\da(\Dcal\ap{})_s-\big[\dap A,\a{s+1}\big]_\g+\big[\da A,\ap{s+1}\big]_\g \nn\\
    &=\dap(\Dcal\a{})_s-\da(\Dcal\ap{})_s-\cA_s\big([\a{},\ap{}],\Dcal\big),
\end{align}
we infer that
\begin{align}
    \big(\Dcal\lbr\a{},\ap{}\rbr\big)_s &=\big(\Dcal[\a{},\ap{}]^\g\big)_s+\Dcal\big(\dap\a{s+1}-\da\ap{s+1}\big) \cr
    &=\Big(\big[\Dcal \a{}, \ap{}\big]^\g_s+\dap(\Dcal\a{})_s\Big)-\a{}\leftrightarrow\ap{}.
\end{align}
Since $\big\lbr\Dcal\a{},\ap{}\big\rbr_s= \big[\Dcal \a{}, \ap{}\big]^\g_s+\dap (\Dcal\a{})_s-\delta_{\Dcal \a{}}\ap{s}$, we finally obtain that
\begin{equation}
    \cA\big(\lbr\a{},\ap{}\rbr, \Dcal\big)=\delta_{\Dcal \a{}}\ap{}-\delta_{\Dcal \ap{}}\a{}.
\end{equation}
We can also compute similar anomalies for $\pa_u$:
\begin{equation}
     \cA\big([\cdot\,,\cdot]_\g,\pa_u\big) =0\qquad\textrm{and}\qquad \cA\big([\cdot\,,\cdot]^\g,\pa_u\big)=0,
\end{equation}
while
\begin{equation}
    \cA\big(\lbr\a{},\ap{}\rbr, \pa_u\big)=\delta_{\pa_u\a{}}\ap{}-\delta_{\pa_u\ap{}}\a{}.
\end{equation}

\section{General expression for $(\pui)^*$ \label{App:pui}}

If we keep $\beta$ arbitrary in the definition of $\pui$, namely $\pui=\int_\beta^u$, then
\begin{align}
    \int_{-\infty}^\infty\rd u\,A(u)[\pui B](u) &=\int_{-\infty}^\infty\rd u\, A(u)\left[\int_\infty^u \rd u'B(u')\right]+\int_{-\infty}^\infty\rd u\,A(u)\left[\int_\beta^\infty \rd u'B(u')\right] \nn\\
    &=\int_{-\infty}^\infty\!\!\rd u\left[\int^{-\infty}_u\!\!\!\rd u' A(u')\right]\!B(u)+\int_{-\infty}^\infty\!\!\rd u\left[\int_{-\infty}^\infty\!\!\rd u'\,\theta(u-\beta)A(u')\right]\!B(u)\nn\\
    &\equiv\int_{-\infty}^\infty\rd u\big[(\pui)^*A\big](u)B(u).
\end{align}
We thus get that in general, 
\begin{equation}
    (\pui O)(u)=\int_\beta^u O(u')\rd u',\quad\big((\pui)^*O\big)(u)=\int_u^{-\infty}\!\!O(u')\rd u'+\theta(u-\beta)\int_{-\infty}^\infty\!O(u')\rd u'.
\end{equation}
Equivalently,
\begin{equation}
    \big((\pui)^*O\big)(u)=\left\{
    \begin{array}{lr}
        \int_u^{-\infty}\rd u'O(u') & \rm{if}~u<\beta, 
        \vspace{0.2cm}\\
        \int_u^{\infty}\rd u'O(u') & \rm{if}~u\geqslant\beta.
    \end{array}\right.
\end{equation}
Notice that if we require $\big((\pui)^*O\big)(u)$ to be continuous, then 
\begin{equation}
    \int_\beta^\infty\rd u'O(u')=\int_\beta^{-\infty}\rd u'O(u')\qquad\Leftrightarrow\qquad\int_{-\infty}^\infty\rd u'O(u')=0,
\end{equation}
so that any memory effect between $\scri^+_-$ and $\scri^+_+$ is excluded, which is expected from imposing Schwartz falloffs.

\section{Proof of Theorem \hyperref[NoetherhatqsYM]{[Noether charge at spin $s$]} \label{AppYM:renormcharge}}

The goal of this demonstration is to recast the Noether charge $\Qa$, when $\a{}=\bfa(\Aa{s})$, as the integral over the (possibly punctured) sphere $S$ of a certain renormalized charge aspect $\tq_s(u,z,\bz)$ smeared against the symmetry parameter $\Aa{s}$ that defines the $\bfa(\Aa{s})$.

Let us start with an example for the lowest spin-weights.
Take $\alpha\in\cS^2$, which implies that $\Aa{}=(\Aa0,\Aa1,\Aa2,0,\ldots)$, then
\bs \label{sola210}
\begin{align}
    \a2(\Aa{}) &=\Aa2, \\
    \a1(\Aa{}) &=uD\Aa2+\Aa1+\pui[1][A,\Aa2]_\g, \\
    \a0(\Aa{}) &=\frac{u^2}{2}D^2\Aa2+uD\Aa1+ \Aa0+\pui[2]D[A,\Aa2]_\g \\
    &+\pui\big[A,uD\Aa2+ \Aa1\big]_\g+\pui\big[A,\pui[1][A,\Aa2]_\g\big]_\g. \nn
\end{align}
\es
Consider successively $\a{}=\bfa(\Aa{s})$ for $s=0,1,2$.
Using the explicit computation \eqref{sola210}, we deduce that\footnote{We use the trace for convenience of writing, which introduces a sign, cf. footnote \ref{footPairg}.}${}^,$\footnote{Notice that some steps need care. 
For instance, $\Tr\Big(\tQ_0\pui\big[A,uD \Aa2\big]_\g\Big)=\Tr\Big(\tQ_0\big[\pui(uA),D \Aa2\big]_\g\Big)=\Tr\Big(\big[\tQ_0, \pui(uA)\big]_\g D\Aa2\Big)=\Tr\Big(-D\big[\tQ_0, \pui(uA)\big]_\g\Aa2\Big)$ up to a total derivative that drops upon integration.}
\bs
\begin{align}
    \frac{\gYM^2}{-2}\cq^u_{\text{\scalebox{1.24}{$\Aa0$}}} &=\int_S\Tr\big(\tQ_0\Aa0\big), \\
    \frac{\gYM^2}{-2}\cq^u_{\text{\scalebox{1.24}{$\Aa1$}}} &=\int_S\Tr\left(\tQ_0 \big(uD\Aa1+\pui[1][A,\Aa1]_\g\big)+\tQ_1\Aa1\right) \nn\\
    &=\int_S\Tr\bigg(\Big(-uD\tQ_0+\big[\tQ_0,\pui A\big]_\g+\tQ_1\Big)\Aa1\bigg), \\
    \frac{\gYM^2}{-2}\cq^u_{\text{\scalebox{1.24}{$\Aa2$}}} &=\int_S\Tr\bigg(\tQ_0\left(\frac{u^2}{2}D^2\Aa2+\pui[2]D[A,\Aa2]_\g+\pui\big[A,uD\Aa2\big]_\g+\pui\big[A,\pui[1][A,\Aa2]_\g\big]_\g\right) \nn\\
    &\qquad\quad\, +\tQ_1\big(uD\Aa2+\pui[1][A,\Aa2]_\g\big)+\tQ_2\Aa2\bigg) \nn\\
    &= \int_S\Tr\bigg(\bigg(\frac{u^2}{2}D^2\tQ_0-\big[D\tQ_0,\pui[2]A\big]_\g -D\big[\tQ_0,\pui(uA)\big]_\g+\Big[\big[\tQ_0, \pui \big(A\big]_\g,\pui A\big)\Big]_\g \nn\\
    &\qquad\qquad -uD\tQ_1+\big[\tQ_1,\pui A\big]_\g+\tQ_2\bigg)\Aa2\bigg).
\end{align}
\es
We can thus identify the renormalized charge aspects\footnote{Be mindful of the peculiar set of parenthesis when one starts to have nested commutators in $\tq_2$.}
\vspace{-0.4cm}
\begin{adjustwidth}{-0.25cm}{-0.1cm}
\bs
\label{renormchargeYM}
\begin{align}
\tq_0 &\equiv q_0, \\
\tq_1 &\equiv q_1+\big[\tQ_0,\pui A\big]_\g, \\
\tq_2 &\equiv q_2-\!\big[D\tQ_0,\pui[2]A\big]_\g -D\big[\tQ_0,\pui(uA)\big]_\g\! +\Big[\big[\tQ_0, \pui \big(A\big]_\g,\pui A\big)\Big]_\g\!+\big[\tQ_1,\pui A\big]_\g, 
\end{align}
\es
\end{adjustwidth}
such that $\cqAas^u=\cq_s^u[\Aa{s}]$, $s=0,1,2$, where 
\be 
q_s =\sum_{n=0}^s \frac{(-u)^n}{n!} D^n \tQ_{s-n}. \label{defqsYM}
\ee 
The reader can check that $\pa_u\tq_s$ only contains terms that involve $\bnews$ so that $\tq_s$ is conserved when no left-handed radiation is present.

Let us mention that the expression for $\cq^u_{\text{\scalebox{1.24}{$\Aa1$}}}$ in the abelian case first appeared in \cite{Lysov:2014csa}.
In that case we have $\cq^u_{\text{\scalebox{1.24}{$\Aa1$}}}=\frac{2}{e^2}\int_S\big(\tQ_0(u\pa_z\Aa1)+\tQ_1\Aa1\big)$, with $e$ the electric charge, which matches with (4.9) of \cite{Lysov:2014csa} since $\lim_{r\to\infty}\big(r^2\mathcal{F}_{ru}\gamma_{z\bz}+\mathcal{F}_{\bz z}\big)=2\tQ_0$ and $\lim_{r\to\infty}\big(r^2\mathcal{F }_{rz}\big)=\tQ_1$, while $\Aa1=Y^z=Y_{\bz}$.\footnote{Just for simplicity, we wrote the expressions on the celestial plane where $D\to\pa_z$.}

We now construct $\tq_s$ for a general $s\geqslant 0$.
For this we use the solution \eqref{solAlphanAs} and the general formula \eqref{puiDcalkYM}.
We obtain\footnote{Recall that $P=\sum_{i=0}^\l p_i$.}
\begin{align}
    \frac{\gYM^2}{-2} \cqAas^u &=\sum_{k=0}^s\int_S\Tr\Big(\tQ_{s-k} \big(\pui\Dcal\big)^k \Aa{s}\Big) \nn\\
    &=\sum_{k=0}^s\int_S\Tr\Bigg( \tQ_{s-k} \sum_{\l=0}^k\sum_{P=k-\l}\pui[p_0]D^{p_0}\bigg(\pui\Big[A,\pui[p_1]D^{p_1}\Big(\pui\Big[A,\pui[p_2]D^{p_2}\Big(\ldots \nn\\
    &\qquad\quad\qquad\ldots \pui\Big[A,\pui[p_{\l-1}]D^{p_{\l-1}}\Big(\pui \Big[A,\pui[p_\l]D^{p_\l}\Aa{s}\Big]_\g\Big)\Big]_\g\ldots \Big)\Big]_\g\Big)\Big]_\g\bigg)\Bigg). \label{intermrenormYM}
\end{align}
Each term in the sum over $k$ behaves similarly.
We thus get (the strategy consists in alternatively integrating by parts over $S$ and using the trace invariance over the adjoint action)\footnote{We use that $\pui[p] \Aa{s} = \frac{u^p}{p!} \Aa{s}$.}${}^,$\footnote{Remember that if there is no opening parenthesis/bracket right after $\pui$, then it acts on everything on its right till the first closing parenthesis/bracket.}
\begin{align}
    \eqref{intermrenormYM} &=\sum_{k=0}^s\int_S\Tr\Bigg(\tQ_{s-k}\bigg(\frac{u^k}{k!}D^k\Aa{s}+\!\!\sum_{P=k-1}\!\!\pui[p_0]D^{p_0}\Big(\pui\Big[A,\pui[p_1]D^{p_1}\Aa{s}\Big]_\g\Big) \nn\\
    &\qquad\qquad+\!\!\sum_{P=k-2} \!\!\pui[p_0]D^{p_0}\Big(\pui\Big[A,\pui[p_1]D^{p_1}\Big(\pui\Big[A,\pui[p_2]D^{p_2}\Aa{s}\Big]_\g\Big)\Big]_\g\Big) \nn\\
    &\qquad\qquad+\ldots+\pui\Big[A,\pui \Big[A,\ldots \pui\Big[A,\Aa{s}\Big]_\g\ldots\Big]_\g\Big]_\g\bigg)\Bigg) \nn\\
    &=\sum_{k=0}^s\int_S\Tr\Bigg((-1)^k \frac{u^k}{k!}D^k\tQ_{s-k}\Aa{s}+\!\!\sum_{P=k-1}\!(-1)^{p_0}D^{p_0}\tQ_{s-k}\Big[\pui[p_0-1]A,\pui[p_1]D^{p_1}\Aa{s}\Big]_\g \nn\\
    &\qquad\qquad+\!\!\sum_{P=k-2}\!(-1)^{p_0}D^{p_0}\tQ_{s-k}\Big[\pui[p_0-1]A,\pui[p_1]D^{p_1}\Big( \pui\Big[A,\pui[p_2]D^{p_2}\Aa{s}\Big]_\g\Big)\Big]_\g \nn\\
    &\qquad\qquad+\ldots+\Big[\tQ_{s-k},\pui\Big(A\Big]_\g\pui\Big[A,\ldots \pui\Big[A,\Aa{s}\Big]_\g\ldots\Big]_\g\Big)\Bigg)\\
    &=\sum_{k=0}^s\int_S\Tr\Bigg((-1)^k \frac{u^k}{k!}D^k\tQ_{s-k}\Aa{s}+\!\!\sum_{P=k-1}\!(-1)^{p_0}\Big[D^{p_0}\tQ_{s-k},\pui[p_0-1]\Big(A\Big]_\g \pui[p_1](1)\Big)D^{p_1}\Aa{s}\nn\\
    &\qquad\qquad+\!\!\!\sum_{P=k-2}\!(-1)^{p_0}\Big[D^{p_0}\tQ_{s-k},\pui[p_0-1]\Big(A\Big]_\g\pui[p_1]D^{p_1}\Big( \pui\Big[A,\pui[p_2]D^{p_2}\Aa{s}\Big]_\g\Big)\Big) \nn\\
    &\qquad\qquad+\ldots+\Big[\Big[\tQ_{s-k},\pui\Big(A\Big]_\g,\pui\Big( A\Big]_\g\ldots \pui\Big[A,\Aa{s}\Big]_\g\ldots\Big)\Big)\Bigg) \nn\\
    &=\sum_{k=0}^s\int_S\Tr\Bigg((-1)^k \frac{u^k}{k!}D^k\tQ_{s-k}\Aa{s}+\!\!\sum_{P=k-1}\!(-1)^{k-1}D^{p_1}\Big[D^{p_0}\tQ_{s-k},\pui[p_0-1]\Big(A\Big]_\g \pui[p_1](1)\Big)\Aa{s}\nn\\
    &\qquad\qquad+\!\!\!\sum_{P=k-2}\!(-1)^{p_0+p_1}D^{p_1}\Big[D^{p_0}\tQ_{s-k},\pui[p_0-1]\Big(A\Big]_\g \pui[p_1-1]\Big[A,\pui[p_2]D^{p_2}\Aa{s}\Big]_\g\Big) \nn\\
    &\qquad\qquad+\ldots+\Big[\Big[\Big[ \tQ_{s-k},\pui\Big(A\Big]_\g,\pui\Big( A\Big]_\g,\pui\Big(A\Big]_\g\ldots \pui\Big[A,\Aa{s}\Big]_\g\ldots\Big)\Big)\Big)\Bigg) \nn\\
    &=\sum_{k=0}^s\int_S\Tr\bigg((-1)^k \sum_{\l=0}^k(-1)^\l\!\!\sum_{P=k-\l}\!D^{p_0}\Big[D^{p_1}\Big[D^{p_2}\Big[\ldots \Big[D^{p_\l}\tQ_{s-k},\pui[p_\l-1]\Big(A\Big]_\g,\ldots\Big]_\g,\,\cdot \nn\\
    &\hspace{6cm}\cdot\pui[p_2-1]\Big(A\Big]_\g,\pui[p_1-1]\Big(A\Big]_\g\pui[p_0](1)\Big)\Big)\ldots\Big)\bigg)\Aa{s}. \nn
\end{align}
Hence the renormalized charge aspect for arbitrary spin-weight $s$ takes the form
\begin{empheq}[box=\fbox]{align}
    \tq_s &=\sum_{k=0}^s\sum_{\l=0}^k (-1)^{k+\l}\!\!\sum_{P=k-\l}\!D^{p_0}\Big[D^{p_1}\Big[D^{p_2}\Big[\ldots \Big[D^{p_\l}\tQ_{s-k},\pui[p_\l-1]\Big(A\Big]_\g,\ldots\Big]_\g,\,\cdot \nn\\
    &\hspace{5cm}\cdot\pui[p_2-1]\Big(A\Big]_\g,\pui[p_1-1]\Big(A\Big]_\g\pui[p_0](1)\Big)\Big)\ldots\Big). \label{deftqsYM}
\end{empheq}

\section{Proof of Lemma \hyperref[YMHardAction]{[Hard action]} \label{App:HardActionYM}}

From the generalized Leibniz rule for pseudo-differential calculus \cite{PseudoDiffBakas},
\begin{equation}
    \pui[\beta](fg)=\sum_{n=0}^\infty\frac{(-\beta)_n}{n!}(\pa_u^nf)\pui[(n+\beta)]g, \qquad \beta\in\R, 
\end{equation}
where $(\beta)_n=\beta(\beta-1)\cdots(\beta-n+1)$ is the falling factorial, we get in particular that
\begin{equation}
    \pui[p]\left(\frac{u^{s-p}}{(s-p)!}A\right)=\sum_{n=0}^{s-p}\frac{(-p)_n}{n!} \frac{u^{s-p-n}}{(s-p-n)!}\pui[(n+p)]A.
\end{equation}
From there we can compute \eqref{puiDcalHard} using \eqref{puiDcalkYM}:
\begin{align}
    \hard{\Big(\pa_u\big(\pui\Dcal\big)^{s+1} \Aa{s}\Big)} &=\sum_{p_0+p_1=s} \pui[p_0+1]D^{p_0}\Big(\pui\Big[A,\pui[p_1]D^{p_1}\Aa{s}\Big]_\g\Big) \nn\\*
    &=\sum_{p=0}^sD^{p}\Big(\pui[p] \Big[A,\frac{u^{s-p}}{(s-p)!}D^{s-p}\Aa{s}\Big]_\g\Big) \nn\\*
    &=\sum_{p=0}^s\sum_{n=0}^{s-p}\frac{(-p)_n}{n!}\frac{u^{s-p-n}}{(s-p-n)!} D^p\big[\pui[(n+p)]A,D^{s-p}\Aa{s}\big]_\g \\
    &=\sum_{k=0}^s\sum_{n=0}^k\frac{(-(k-n))_n}{n!}\frac{u^{s-k}}{(s-k)!} D^{k-n}\big[\pui[k]A,D^{n+s-k}\Aa{s}\big]_\g, \nn
\end{align}
where in the last step we changed variable $n+p\to k$.
Therefore,
\begin{equation}
    \Hd{s}{\Aa{s}}A=\sum_{p=0}^s \frac{u^{s-p}}{(s-p)!}\Htd{p}{D^{s-p}\Aa{s}}A,
\end{equation}
with\footnote{$(-(p-n))_n=(-1)^n(p-1)_n$.}
\begin{equation}
    \Htd{p}{\Aa{p}}A=\sum_{n=0}^p(-1)^{n+1} \frac{(p-1)_n}{n!}D^{p-n}\big[ \pui[p]A,D^n\Aa{p}\big]_\g. \label{tdpHYM}
\end{equation}
When $p=0$, then $\Htd{0}{\Aa0}A=-[A,\Aa0]_\g$, but if $p>0$ then we can simplify \eqref{tdpHYM} much further. 
Indeed, using that $(p-1)_p=0$, we have that
\begin{align}
    \Htd{p}{\Aa{p}}A &=-\sum_{n=0}^{p-1}(-1)^n \frac{(p-1)_n}{n!}\sum_{k=0}^{p-n} \frac{(p-n)_k}{k!}\big[ \pui[p]D^k A,D^{p-k}\Aa{p}\big]_\g \nn\\
    &=-\sum_{k=1}^p\big[\pui[p]D^kA,D^{p-k}\Aa{p}\big]_\g \sum_{n=0}^{p-k}(-)^{n} \frac{(p-1)_n}{n!}\frac{(p-n)_k}{k!}-\sum_{n=0}^{p-1}(-)^n \frac{(p-1)_n}{n!}\big[\pui[p]A,D^p\Aa{p}\big]_\g \nn\\
    &=-\sum_{k=1}^p\big[\pui[p]D^kA,D^{p-k}\Aa{p}\big]_\g\binom{p}{k}{}_2F_1\big(\!-(p-k),1-p;-p;1\big). \label{intermYMHardAction}
\end{align}
To get the first line, we just used the Leibniz rule on $D^{p-n}$.
In the second equality, we extracted the term $k=0$ in order to then be able to commute the sums over $n$ and $k$.
The third line is a rewriting of the binomial coefficients in terms of hypergeometric functions, which will allow us to use the identity (cf. formula 7.3.5.4 in \cite{HyperGmathbook})
\begin{equation}\label{HyperID}
    {}_2F_1(-m,-b;-c;1)=\sum_{n=0}^m(-1)^n \binom{m}{n}\frac{(b)_n}{(c)_n}=\frac{(c-b)_m}{(c)_m}.
\end{equation}
So to get the third line of \eqref{intermYMHardAction}, we used that
\begin{align}
    \sum_{n=0}^{p-k}(-1)^n \frac{(p-1)_n}{n!}\frac{(p-n)_k}{k!} &=\sum_{n=0}^{p-k}(-1)^n \binom{p-k}{n}\frac{(p-1)_n(p-n)!}{(p-k)!k!} \nn\\*
    &=\binom{p}{k}\sum_{n=0}^{p-k}(-1)^n \binom{p-k}{n}\frac{(p-1)_n}{(p)_n} \\
    &=\binom{p}{k}{}_2F_1\big(\!-(p-k),1-p;-p;1\big), \nn
\end{align}
together with the binomial identity
\begin{align}
    \sum_{n=0}^{p-1}(-1)^n \binom{p-1}{n}=0.
\end{align}
By \eqref{HyperID}, we infer that for $p\geq 1$,\footnote{In terms of Kronecker $\delta$, $(1)_{p-k}=\delta_{p,k}+\delta_{p-1,k}$.}
\begin{align}
    \Htd{p}{\Aa{p}}A &=-\sum_{k=1}^p\big[\pui[p]D^kA,D^{p-k}\Aa{p}\big]_\g\binom{p}{k}\frac{(1)_{p-k}}{(p)_{p-k}} \nn\\
    &=-\big[\pui[p]D^pA,\Aa{p}\big]_\g- \big[\pui[p]D^{p-1}A,D\Aa{p}\big]_\g=-D\big[\pui[p]D^{p-1}A,\Aa{p}\big]_\g,
\end{align}
which corresponds to the formula \eqref{YMHardAction}.

\section{Hard action: comparison with literature \label{AppYM:comparison}}

Here we show that the equation (50) in \cite{Freidel:2023gue} amounts to \eqref{YMHardAction}.
Indeed, on one hand we get from \cite{Freidel:2023gue} that
\begin{equation} \label{conformalAction3}
    \Hd{s}{\Aa{s}}\hA(\Delta)=i^s\sum_{p=0}^s \frac{(\Delta)_p}{p!}\big[D^p\Aa{s}, 
    D^{s-p}\hA(\Delta-s)\big]_\g,
\end{equation}
where $\hA(\Delta,z,\bz)$\footnote{We keep the dependence in $(z,\bz)$ implicit henceforth.} is the gauge potential in the conformal basis (i.e. the conformal gluon primary operator\footnote{\eqref{ConformalA} is also equal to the Mellin transform in energy of the Fourier transform $\tilde A(\omega)$ of $A(u)$, i.e. 
\begin{equation}
    \hA(\Delta)=\int_0^\infty\rd \omega\,\omega^{\Delta-1}\tilde A(\omega).
\end{equation}}):
\begin{equation} \label{ConformalA}
    \hA(\Delta):=i^\Delta\Gamma(\Delta) \int_\R\rd u\,(u+i\epsilon)^{-\Delta}A(u).
\end{equation}
$\Gamma$ denotes the Gamma function.
On the other hand, we can apply the change to the conformal basis to \eqref{YMHardAction}.
For this we use the following properties---that can be proven by induction:\footnote{The operator $u^{-n} \pui[n]=(\hD+n-1)_n^{-1}$ is the formal inversion of $\pa_u^nu^n=(\hD+n-1)_n$. We prove the latter by induction using that $\pa_u^{n+1}u^{n+1}=\pa_u\big((\pa_u^nu^n)u\big)=(\hD+n)_n\pa_u u=(\hD+n)_{n+1}$, which itself follows from $\pa_u(\hD+\alpha)_n=(\hD+\alpha+1)_n\pa_u$.}
\begin{align} \label{conformalProperties}
    u^\alpha(\hD+n-1)_n^{-1}=(\hD+n-1-\alpha)_n^{-1}u^\alpha,\qquad
    u^{-n} \pui[n] =(\hD+n-1)_n^{-1},
\end{align}
valid for $n\in\N$ and $\alpha\in\R$ and where $\hD:=\pa_u u=1+u\pa_u$.
From \eqref{YMHardAction}, we thus obtain that\footnote{We keep the $i\epsilon$ prescription implicit for shortness. It is understood that it appears all along the calculation, as in \eqref{ConformalA}.
Besides, we also write $\int_\R\equiv \int_\R\rd u$ since the context is clear.}
\begin{align} \label{ConformalAction1}
    \Hd{s}{\Aa{s}}\hA(\Delta)&=-i^\Delta \Gamma(\Delta)\int_\R u^{-\Delta}\left(\frac{u^s}{s!}[A,D^s\Aa{s}]_\g+\sum_{p=1}^s\frac{u^{s-p}}{(s-p)!}D\big[\pui[p]D^{p-1}A,D^{s-p}\Aa{s}\big]_\g\right).
\end{align}
\vspace{-0.5cm}

\ni We then use \eqref{conformalProperties} to infer that
\begin{align}
    \int_\R u^{s-\Delta-p}\pui[p]A &=\int_\R u^{s-\Delta}(\hD+p-1)^{-1}_pA \nn\\
    &=\int_\R(\hD+\Delta-s+p-1)^{-1}_p u^{s-\Delta}A \nn\\
    &=\int_\R\frac{\Gamma(\Delta-s)}{\Gamma(\Delta-s+p)}u^{s-\Delta}A \\
    &=\frac{i^{s-\Delta}}{\Gamma(\Delta-s+p)}\hA(\Delta-s), \nn
\end{align}
where in the penultimate line, we used that $\hD$ (and any analytic function of it) integrates to 0 for fields in the Schwartz space.
Therefore, \eqref{ConformalAction1} simplifies to
\begin{align} \label{ConformalAction2}
    \Hd{s}{\Aa{s}}\hA(\Delta)&=\frac{i^s \Gamma(\Delta)}{s!\Gamma(\Delta-s)}\big[D^s\Aa{s}, \hA(\Delta-s)\big]_\g \nn\\
    &+\sum_{p=1}^s\frac{i^s\Gamma(\Delta)}{(s-p)!\Gamma(\Delta-s+p)}D\big[D^{s-p}\Aa{s},D^{p-1}\hA(\Delta-s)\big]_\g.
\end{align}
Notice that we can recast the sum as follows: We first act with the total $D$ derivative, change $p\to p-1$ in the second sum and then recombine both sums. This gives
\begin{align}
    &\sum_{p=1}^s\frac{i^s\Gamma(\Delta)}{(s-p)!\Gamma(\Delta-s+p)}D\big[D^{s-p}\Aa{s},D^{p-1}\hA(\Delta-s)\big]_\g \nn\\
    =&\sum_{p=1}^s\frac{i^s\Gamma(\Delta)}{(s-p)!\Gamma(\Delta-s+p)}\big[D^{s-p}\Aa{s},D^{p}\hA(\Delta-s)\big]_\g \nn\\*
    +&\sum_{p=0}^{s-1}\frac{i^s\Gamma(\Delta)}{(s-p-1)!\Gamma(\Delta-s+p+1)}\big[D^{s-p}\Aa{s},D^p\hA(\Delta-s)\big]_\g \nn\\
    =&\,\frac{i^s\Gamma(\Delta)}{(s-1)!\Gamma(\Delta-s+1)}\big[D^s\Aa{s},\hA(\Delta-s)\big]_\g+i^s\big[\Aa{s},D^s\hA(\Delta-s)\big]_\g \\
    +&\sum_{p=1}^{s-1}\frac{i^s\Gamma (\Delta+1)}{(s-p)!\Gamma(\Delta-s+p+1)}\big[D^{s-p}\Aa{s},D^p\hA(\Delta-s)\big]_\g. \nn
\end{align}
Hence \eqref{ConformalAction2} reduces to
\begin{align}
    \Hd{s}{\Aa{s}}\hA(\Delta)&=\sum_{p=0}^s \frac{i^s\Gamma (\Delta+1)}{(s-p)!\Gamma(\Delta-s+p+1)}\big[D^{s-p}\Aa{s},D^p\hA(\Delta-s)\big]_\g \nn\\
    &=i^s\sum_{p=0}^s\frac{\Gamma(\Delta+1)}{p!\Gamma(\Delta-p+1)}\big[D^p\Aa{s},D^{s-p}\hA(\Delta-s)\big]_\g,
\end{align}
which is indeed \eqref{conformalAction3}.

\chapter{General Relativity I \label{appGRI}}

\section{Solution of the wedge condition on $S_0$ \label{AppWedgeOnSphere}}

We study the solution of the constraint $D^{s+2}\T{s}=0$ on the sphere $S_0$.
Expanding $\T{s}$ in spin-spherical harmonics \eqref{sshExpansion}, we find
\begin{equation}
    \T{-1}=\sum_{\l\geqslant 1}\sum_m T_{-1}^{\l,m}\, Y^1_{\l,m}\qquad\Rightarrow\qquad D\T{-1}=\sum_{\l\geqslant 2} \sqrt{\frac{(\l-1)(\l+2)}{2}}\sum_m T_{-1}^{\l,m}\, Y^2_{\l,m},
\end{equation}
for some coefficients $T_{-1}^{\l,m}\in\C$.
Hence, 
\begin{equation}
    D\T{-1}=0\qquad\Rightarrow\qquad \T{-1}=\sum_{m=-1}^1 T_{-1}^{1,m}\,Y^1_{1,m}. \label{solT-1}
\end{equation}
Similarly,
\begin{equation}
    \T0=\sum_{\l\geqslant 0}\sum_m T_{0}^{\l,m}\, Y^0_{\l,m}\qquad\Rightarrow\qquad D^2\T0=\sum_{\l\geqslant 2}\frac{\sqrt{(\l+2)!}}{2\sqrt{(\l-2)!}}\sum_m T_{0}^{\l,m}\, Y^2_{\l,m}.
\end{equation}
Therefore, 
\begin{equation}
    D^2\T0=0\qquad\Rightarrow\qquad \T0=T^{0,0}_0\,Y^0_{0,0}+\sum_{m=-1}^1 T_0^{1,m}\,Y^0_{1,m}. \label{solT0}
\end{equation}
Note that $Y^0_{0,0}\propto 1$ while  the basis of functions 
$Y^{0}_{1,m}$ is equivalent to the normalized sphere vector $n^i$, with $n^2=1$. Therefore this means that 
\be 
T_0= x^0 + x^i n_i= x^a q_a,
\ee 
for $q_a$ defined by \eqref{T0T}.
In the same vein, we find that
\begin{equation}
    \T1=\sum_{m=-1}^1 T_1^{1,m}Y^{-1}_{1,m}. \label{solT1}
\end{equation}
Then we use \eqref{raisingD} and \eqref{loweringbD} to show that 
\begin{align}
    D Y^{0}_{1,m}  =Y^{1}_{1,m},\qquad 
    \bD Y^{0}_{1,m} =- Y^{-1}_{1,m}.
\end{align}
This establishes \eqref{TDT}.

For $s\geqslant 2$,  one uses that 
\begin{equation}
    \T{s}=\sum_{\l\geqslant s\geqslant 2}\sum_m T_s^{\l,m}\, Y^{-s}_{\l,m}\quad\Rightarrow\quad D^{s+2}\T{s}=\sum_{\l\geqslant s\geqslant 2}\sqrt{\frac{(\l+s)!(\l+2)!}{2^{s+2}(\l-s)!(\l-2)!}}\sum_m T_s^{\l,m}\, Y^2_{\l,m},
\end{equation}
Therefore 
\begin{equation}
    D^{s+2}\T{s}=0\qquad\Rightarrow\qquad \T{s}=0\qquad \textrm{for } s\geqslant 2. \label{solT2}
\end{equation}

\section{Proof of Lemma [Wedge algebra]}

\subsection{Closure \label{AppProofWedge}}

The closure of the $\W$-bracket \eqref{Wbracket} follows from the Leibniz rule which implies that 
\be
D^{s+2} (\T{n} D \Tp{s+1-n})
&=\sum_{k=0}^{s+2}\binom{s+2}{k} D^k \T{n} D^{s+3-k} \Tp{s+1-n} \nn\\
&= \binom{s+2}{n+1}D^{n+1}\T{n}D^{s+2-n}\Tp{s+1-n},
\ee 
To get the second equality, we used that $D^k \T{n}$ vanishes when $k\geq  n+2$ while $D^{s+3-k} \Tp{s+1-n}$ vanishes when $s+3-k \geq s+3 -n$, i.e. when $k\leq n$.
Therefore,
\begin{align}
    D^{s+2}[T,T']^\W_s &= \sum_{n=0}^{s+1}\frac{(s+2)!}{n!(s+1-n)!}\left( D^{n+1}\T{n} D^{s+2-n} \Tp{s+1-n}-D^{n+1}\Tp{n} D^{s+2-n}\T{s+1-n}\right) \nn\\
    &=0, \label{checkWedgeCond}
\end{align}
since by changing $s+1-n\to n$, we see that the sum is equal to its opposite.
This proof holds for $s\geqslant -1$.
This bracket thus acts in the wedge. 

\subsection{Jacobi identity \label{AppJacobiVbracket}}

Here we show that the $\W$-bracket satisfies the Jacobi identity when the wedge condition holds.
\begin{align}
    \big[T,[T',T'']^\W\big]_s^\W &= \sum_{n=0}^{s+1}(n+1)\Big(\T{n}D[T',T'']^\W_{s+1-n}-[T',T'']^\W_n D\T{s+1-n}\Big) \\
    &=\sum_{n=0}^{s+1}\sum_{k=0}^{s+2-n}(n+1)(k+1)\Big(\T{n}D\Tp{k}D\Tpp{s+2-n-k}-\T{n}D\Tpp{k}D\Tp{s+2-n-k} \nn\\*
    &\hspace{5cm}+\T{n}\Tp{k}D^2\Tpp{s+2-n-k}-\T{n}\Tpp{k}D^2\Tp{s+2-n-k}\Big) \nn\\
    &-\sum_{n=0}^{s+1}\sum_{k=0}^{n+1}(n+1)(k+1)\Big(\Tp{k}D\Tpp{n+1-k}-\Tpp{k}D\Tp{n+1-k}\Big)D\T{s+1-n}. \nn
\end{align}
It turns out to be very convenient to rewrite the sums as follows
\begin{align}
  \big[T,[T',T'']^\W\big]_s^\W 
    &=\sum_{\overset{a+b+c=s+2}{\scriptscriptstyle{b+c>0}}} \Big( (a+1)(b+1)\T{a}D\Tp{b}D\Tpp{c}-
    (a+1)(c+1)\T{a}D\Tpp{c}D\Tp{b} \Big)\nn\\
    & - \sum_{\overset{a+b+c=s+2}{\scriptscriptstyle{b+c>0}}} \Big( (b+c)(b+1)\Tp{b}D\Tpp{c}D\T{a}-
    (b+c)(c+1)\Tpp{c}D\Tp{b}D\T{a} \Big)\cr
    &+\sum_{\overset{a+b+c=s+2}{\scriptscriptstyle{b+c>0}}}
    \big( (a+1)(b+1)\T{a}\Tp{b}D^2\Tpp{c} -
    (a+1)(c+1)\T{a}\Tpp{c}D^2\Tp{b}\Big).
\end{align} 
In the first two lines we relax the restriction $b+c>0$ as the terms $b=c=0$ and $a =s+2$ vanish. 
In the last line the restriction $b+c>0$ can be relaxed if we  subtract the term\footnote{The latter vanishes inside the $\W$-space. 
For later convenience, we keep it explicit.}
\be 
(s+3)\T{s+2} \big(\Tp0 D^2\Tpp0 - \Tpp0 D^2\Tp0\big).
\ee 
When the summations are simply over $a+b+c= s+2$ we can freely exchange indices. For instance the last line becomes $\sum_{a+b+c=s+2}
     (a+1)(b+1)\big(  \T{a}\Tp{b}D^2\Tpp{c} -\T{a}\Tpp{b}D^2\Tp{c}\Big)$ and therefore  after cyclic permutation we have that 
\begin{align}
  \big[T,[T',T'']^\W\big]_s^\W
    &\cyc\sum_{a+b+c=s+2} \Big( (a+1)(b+1)-
    (a+1)(c+1) \Big) \T{a}D\Tp{b}D\Tpp{c} \nn\\
    & - \sum_{a+b+c=s+2 } \Big( (a+b)(a+1)-
    (a+c)(a+1) \Big) \T{a}D\Tp{b}D\Tpp{c} \nn\\ 
    &+ (s+3) D^2\T0\big(\Tp0 \Tpp{s+2} - \Tp{s+2} \Tpp0\big).
\end{align} 
The first two lines are easily seen to cancel and we are left with \eqref{Jac1}:
\be
\big[T,[T',T'']^\W\big]_s^\W \cyc (s+3) D^2\T0\big(\Tp0 \Tpp{s+2} - \Tp{s+2} \Tpp0\big).
\ee

\section{Poincaré algebra \label{app:Poinc}} 

In this section we establish the results needed for the proof of \eqref{PoincB} and \eqref{nids}. 
One starts with the expression for the unit vector and its derivative 
$Dn = P \pa_z n$, with $P=(1+|z|^2)/\sqrt{2}$:
\be
n&=\frac1{1+|z|^2} \big(z + \bz , -i(z-\bz), 1-|z|^2\big), \cr
Dn&=\frac1{\sqrt{2}} \frac1{1+|z|^2} \big(1- \bz^2 ,-i(1+ \bz^2), -2\bz\big).
\ee 
We then evaluate (notice that $\bD n=\overline{(Dn)}$)
\be
2 D n_1 \bD n_1 + n_1n_1 &= \frac{1}{(1+|z|^2)^2}\left(  (1- \bz^2)(1- z^2) + (z+\bz)^2 \right) = 1,\cr
2 D n_1 \bD n_2 + n_1n_2 &= 
\frac{i}{(1+|z|^2)^2}\left( (1- \bz^2)(1+ z^2) - (z+\bz)(z-\bz) \right) \cr
& = \frac{i}{(1+|z|^2)^2}\left( 1 -|z|^4  \right)  = i n_3.
\ee 
We can evaluate similarly the other components and get
\be 
2 D n_i \bD n_j + n_in_j&= \delta_{ij} + i \epsilon_{ij}{}^k n_k. \label{PoinAlgInterm}
\ee
Besides $D^2 n = \pa_z (P D n)=0$.
From \eqref{casimirDD} we know that $D\bD Y^0_{\ell,m}= -\frac{\ell}{2}(\ell+1) Y^0_{\ell,m}$. Specifying for $\ell=1$ gives $D\bD n_i=-n_i$.
Taking the $D$ derivative of \eqref{PoinAlgInterm}, we thus infer that 
\be 
 n_iDn_j - n_j Dn_i&=  i \epsilon_{ij}{}^k D n_k,
\ee
which establishes \eqref{nids}.

We can now straightforwardly compute the different commutators.
We start with the rotation generators, for which\footnote{Since the notation in terms of the Poincaré generators is clear, we denote for shortness $[\bJ_i,\bJ_j]\equiv \big[\iota(\bJ_i),\iota(\bJ_j)\big]^\W_1$, $[P_a,P_b]\equiv \big[\iota(P_a),\iota(P_b)\big]^\W_{-1}$ and $[\bJ_i,P_b]\equiv \big[\iota(\bJ_i),\iota(P_b)\big]^\W_0$.}
\be 
[\bJ_i,\bJ_j] &= 
2 \big(\bD n_i D\bD n_j - \bD n_j D\bD n_i\big)\cr
&= 2 (n_i \bD n_j - n_j \bD n_i)\cr
&= -2i \epsilon_{ij}{}^k \bD n_k = -2i \epsilon_{ij}{}^k \bJ_k.
\ee 
The translation generators bracket is given by
\be 
[P_a, P_b] &= 
q_a D q_b - q_b D q_a,\cr 
[P_0, P_j] &= 
 q_0 D q_j - q_j D q_0=  D n_j =  C_j,\cr
[P_i, P_j] &= 
q_i D q_j - q_j D q_i= n_iD n_j- n_j Dn_i=
i \epsilon_{ij}{}^k Dn_k
= i \epsilon_{ij}{}^k C_k,
\ee 
where 
we used that $q_0=1$ and $q_i=n_i$.
For the mixed commutators we get 
\be 
[\bJ_i, P_b] &= 
2\bD n_i D q_b - q_b D \bD n_i=  2\bD n_i D q_b + n_i q_b, \cr 
[\bJ_i, P_0 ] &=  n_i = P_i, \cr
[\bJ_i, P_j] &= 
2\bD n_i D n_j + n_i n_j = \delta_{ij} + i \epsilon_{ji}{}^k n_k
= \delta_{ij} P_0 - i\epsilon_{ij}{}^k P_k.
\ee
Using the definition of the 't Hooft symbol \eqref{tHooftSymbol}, we recover \eqref{PoincB}.

\section{Proofs related to the Jacobi identity \label{AppJacobi}}

\subsection{SN bracket \label{JacobiSN}}

As well summarized by \cite{SNbracket}, the fact that the SN bracket satisfies the Jacobi identity is obvious once we know two things. 
Firstly, (and keeping the notation of the \hyperref[SNreminder]{reminder}), that each $S_n\in\X^n(M)$ is uniquely related to a function $f$ on $T^*M$ by 
\begin{equation}
    f(x,p):=S_n(\underbrace{p, \ldots,p}_{n\textrm{ times}})\big|_x = S_n^{k_1\cdots k_n}\big|_x\, p_{k_1} \cdots p_{k_n}\label{HPOdef}
\end{equation}
for all $x\in M$ and co-vector $p\in T^*_x(M)$. 
Such a function is called a \textit{homogeneous polynomial observable of degree n} on $T^*M$.
Given such a function, it determines $S$ uniquely, i.e. \eqref{HPOdef} can be read from right to left too. 
We dub $S_n(f)$ the tensor resulting from this mapping.
Secondly, that the SN bracket \eqref{SNbracketDef} (and by extension \eqref{SNbracketDefbis}) between $S_n(f)$ and $S_q(g)$, for $f$ and $g$ homogeneous polynomial observables of degree $n$ and $q$ respectively, can be defined as 
\begin{equation}
    [S_n(f),S_q(g)]^\sn_{n+q-1}:=S_{n+q-1}\big(\{f,g\}\big),
\end{equation}
where $\{f,g\}=\sum_{i=1}^d\frac{\pa f}{\pa p_i}\frac{\pa g}{\pa x^i}-\frac{\pa g}{\pa p_i}\frac{\pa f}{\pa x^i}$ is the canonical Poisson bracket on $T^*M$ such that $\{p_i,x^j\}=\delta^j_i$.
Jacobi for $[\cdot\,,\cdot]^\sn$ thus follows from Jacobi for $\{\cdot\,,\cdot\}$.

\subsection{Proofs of formulae \eqref{Jacobi1} \& \eqref{Jacobi2} \label{AppJacobiCF}}

We first deal with the two terms proportional to $\sigma$:
\begin{align}
    -\big[T,\sigma\paren{T',T''}\big]_s^\V &- \sigma\paren{T,[T',T'']^\V}_s = \nn\\
    &-\sum_{n=0}^{s+1}(n+1)(s+4-n)\T{n}D\Big(\sigma\big(\Tp0\Tpp{s+3-n}-\Tpp0\Tp{s+3-n}\big)\Big) \nn\\
    & +\sigma\sum_{n=0}^{s+1}(n+1)(n+3)\big(\Tp0\Tpp{n+2}-\Tpp0\Tp{n+2}\big)D\T{s+1-n} \nn\\
    &-(s+3)\sigma\T0\left(\sum_{n=0}^{s+3}(n+1)\big(\Tp{n}D\Tpp{s+3-n}-\Tpp{n}D\Tp{s+3-n}\big)\right) \label{intermJacobi}\\
    &+(s+3)\sigma\T{s+2}\left(\sum_{n=0}^1(n+1) \big(\Tp{n}D\Tpp{1-n}-\Tpp{n}D\Tp{1-n}\big)\right). \nn
\end{align}
In the following, we refer to each line as \circled 1 to \circled 4 respectively. 
First of all,\footnote{Since in this first step we extract the $n=0$ and $n=1$ terms, a whole part of \eqref{eq204} drops if we consider the special cases $s=-1$ and $s=0$.
The final result of the demonstration is however unchanged.}
\begin{align}
    \circled 1 &= -\sum_{n=2}^{s+1}(n+1)(s+4-n)D\sigma\T{n}\big(\Tp0\Tpp{s+3-n}-\Tpp0\Tp{s+3-n}\big) \nn\\*
    &\quad -\sigma\sum_{n=2}^{s+1}(n+1)(s+4-n)\T{n}D\big(\Tp0\Tpp{s+3-n}-\Tpp0\Tp{s+3-n}\big) \nn\\*
    &\quad -(s+4)\T0 D\big(\sigma(\Tp0\Tpp{s+3}-\Tpp0\Tp{s+3})\big)-2(s+3)\T1 D\big(\sigma(\Tp0\Tpp{s+2}-\Tpp0\Tp{s+2})\big) \nn\\
    &\cyc -\sum_{n=2}^{s+1}(n+1)(s+4-n)D\sigma\Tpp0\big(\cancel{\Tp{n}\T{s+3-n}}-\cancel{\T{n}\Tp{s+3-n}}\big) \label{eq204}\\
    &\quad -\sigma\sum_{n=2}^{s+1}(n+1)(s+4-n)D\T0 \big(\cancel{\Tpp{n}\Tp{s+3-n}}-\cancel{\Tp{n}\Tpp{s+3-n}}\big) \nn\\
    &\quad -\sigma\sum_{n=2}^{s+1}(n+1)(s+4-n)\T0 \big(\Tpp{n}D\Tp{s+3-n}-\Tp{n}D\Tpp{s+3-n}\big) \nn\\
    &\quad +\sigma(s+4)D\T0 \big(\Tp0\Tpp{s+3}-\Tpp0\Tp{s+3}\big)-2(s+3)\T1 D\sigma\big(\Tp0\Tpp{s+2}-\Tpp0\Tp{s+2}\big) \nn\\
    &\quad -2\sigma(s+3)\big(\Tp1 D\Tpp0\T{s+2}-\Tpp1 D\Tp0\T{s+2}\big)-2\sigma(s+3)\big(\Tp1\Tpp0 D\T{s+2}-\Tpp1\Tp0 D\T{s+2}\big), \nn
\end{align}
where we used that
\begin{equation}
    D\big(\sigma\T0\Tp0\Tpp{s+3}-\sigma\T0\Tpp0\Tp{s+3}\big)\cyc 0
\end{equation}
and changed few prime indices for later convenience. 
The sums cancel since equal to their opposite (change variable $n\to s+3-n$ to see it).
Second of all,
\begin{align}
    \circled 2 &\cyc \sum_{n=0}^{s-1}(n+1)(n+3)\sigma\T0\big(\Tp{n+2}D\Tpp{s+1-n}-\Tpp{n+2}D\Tp{s+1-n}\big) \\
    &+(s+2)(s+4)\sigma D\T0\big(\Tp0\Tpp{s+3}-\Tpp0\Tp{s+3}\big)+ (s+1)(s+3)\sigma D\T1\big(\Tp0\Tpp{s+2}-\Tpp0\Tp{s+2}\big). \nn
\end{align}
Third of all,
\begin{align}
    \circled 3 &\cyc -(s+3)\sigma\T0\left(\sum_{n=2}^{s+1}(n+1)\big(\Tp{n}D\Tpp{s+3-n}-\Tpp{n}D\Tp{s+3-n}\big)\right) \\
    &\quad -(s+3)(s+4)\sigma\big(\Tp0\Tpp{s+3}D\T0-\Tpp0\Tp{s+3}D\T0\big)-(s+3)^2\sigma\big(\Tp0\Tpp{s+2}D\T1-\Tpp0\Tp{s+2}D\T1\big) \nn\\
    &\quad -(s+3)\sigma\big(\cancel{\T0\Tp0 D\Tpp{s+3}}-\cancel{\Tp0\T0 D\Tpp{s+3}}\big)-2(s+3)\sigma\big(\Tp0\Tpp1 D\T{s+2}-\Tpp0\Tp1 D\T{s+2}\big) \nn
\end{align}
Fourth of all,
\begin{equation}
    \circled 4\cyc(s+3)\sigma\big(\Tp{s+2}\Tpp0 D\T1-\Tpp{s+2}\Tp0 D\T1+2\T{s+2}\Tp1 D\Tpp0 -2\T{s+2}\Tpp1 D\Tp0\big).
\end{equation}
Summing all of them,
\begin{align}
    \eqref{intermJacobi} &\cyc\left(-(s+3)+(s+1)(s+3)-(s+3)^2\right)\sigma D\T1\big(\Tp0\Tpp{s+2}-\Tpp0\Tp{s+2}\big) \nn\\*
    &+2\big((s+3)-(s+3)\big)\sigma\T{s+2}\big(\Tp1 D\Tpp0 -\Tpp1 D\Tp0\big) \nn\\*
    &+\left(-(s+3)(s+4)+(s+4)+(s+2)(s+4)\right) \sigma D\T0\big(\Tp0\Tpp{s+3}-\Tpp0\Tp{s+3}\big) \\*
    &+2\big((s+3)-(s+3)\big)\sigma D\T{s+2}\big(\Tp0\Tpp1 -\Tpp0\Tp1\big)-2(s+3)\T1 D\sigma\big(\Tp0\Tpp{s+2}-\Tpp0\Tp{s+2}\big) \nn\\
    &-\sigma\T0\left(\sum_{n=2}^{s+1}\big((n+1)(s+4-n)-(s+3)(n+1)\big)\big(\Tpp{n}D\Tp{s+3-n}-\Tp{n}D\Tpp{s+3-n}\big)\right) \nn\\
    &+\sigma\T0\left(\sum_{n=0}^{s-1}(n+1)(n+3)\big(\Tp{n+2}D\Tpp{s+1-n}-\Tpp{n+2}D\Tp{s+1-n}\big)\right) \nn\\
    &\cyc -3(s+3)\sigma D\T1\big(\Tp0\Tpp{s+2}-\Tpp0\Tp{s+2}\big)-2(s+3)\T1 D\sigma\big(\Tp0\Tpp{s+2}-\Tpp0\Tp{s+2}\big) \nn\\
    &=-\big(3\sigma D\T1+2D\sigma\T1\big)\paren{T',T''}_s. \nn
\end{align}

Concerning the term proportional to $\sigma^2$, we get straightforwardly
\begin{align}
    \sigma \paren{T, \sigma\paren{T',T''}}_s &=(s+3)\sigma^2\left((s+5)\big(\T0\Tp0\Tpp{s+4}-\T0\Tpp0\Tp{s+4}\big)-3(\Tp0\Tpp2-\Tpp0\Tp2)\T{s+2}\right) \nn\\
    &\cyc 3(s+3)\sigma^2\T2\big(\Tp0\Tpp{s+2}-\Tpp0\Tp{s+2}\big) \\
    &\cyc 3 \sigma^2\T2\paren{T',T''}_s. \nn
\end{align}

\section{Proof of the morphism formula \eqref{algebraRepinterm}\label{AppAlgebraAction}}

For convenience, here is the equation \eqref{dTCv1} again:
\begin{equation}
    \hdT \sigma =-D^2\T0+2D\sigma \T1+3\sigma D\T1-3 \sigma ^2\T2.
\end{equation}
This implies that\footnote{Here $T$ does not depend on $\sigma$, hence $\dTp \T{s}=0,\,s\geqslant 0$.}
\begin{align}
    \big[\hdTp,\hdT\big]\sigma  &=\bigg\{2\T1 D\left(-D^2\Tp0+2D\sigma \Tp1+3\sigma D\Tp1-3 \sigma ^2\Tp2\right) \nn\\
    &\qquad +3D\T1\left(-D^2\Tp0+2D\sigma \Tp1+3\sigma D\Tp1-3 \sigma ^2\Tp2\right) \\
    &\qquad -6\sigma \T2\left(-D^2\Tp0+2D\sigma \Tp1+3\sigma D\Tp1-3 \sigma ^2\Tp2\right)\bigg\} -T\leftrightarrow T'. \nn
\end{align}
On the other hand we know that
\begin{align}
    \hat\delta_{[T,T']^\sigma}\sigma  &=\bigg\{-D^2\left(\T0D\Tp1+2\T1D\Tp0-3 \sigma \T0\Tp2\right) \nn\\
    &\qquad +2D\sigma \big(\T0D\Tp2+2\T1D\Tp1+3\T2D\Tp0-4\sigma \T0\Tp3\big) \\
    &\qquad +3\sigma D\big(\T0D\Tp2+2\T1D\Tp1+3\T2D\Tp0-4\sigma \T0\Tp3\big) \nn\\*
    &\qquad -3 \sigma ^2\big(\T0D\Tp3+2\T1D\Tp2+3\T2D\Tp1+4\T3D\Tp0-5 \sigma \T0\Tp4\big)\bigg\}-T\leftrightarrow T'. \nn
\end{align}
Hence, by expanding every term, we compute
\begin{align}
    &\quad ~\big[\hdTp,\hdT\big]\sigma -\hat\delta_{[T,T']^\sigma}\sigma = \nn\\
    &=\Big\{-\cancel{2}\big(\T1D^3\Tp0-\Tp1D^3\T0\big)-\cancel{3}\big(D\T1D^2\Tp0-D\Tp1D^2\T0\big)+\cancel{1}\big(D^2\T0D\Tp1-D^2\Tp0D\T1\big) \nn\\
    &\qquad+\cancel{2}\big(D\T0D^2\Tp1-D\Tp0D^2\T1\big)+\big(\T0D^3\Tp1-\Tp0D^3\T1\big) +\cancel{2}\big(D^2\T1D\Tp0-D^2\Tp1D\T0\big) \nn\\
    &\qquad +\cancel{4}\big(D\T1D^2\Tp0-D\Tp1D^2\T0\big)+\cancel{2}\big(\T1D^3\Tp0-\Tp1D^3\T0\big)\Big\} \nn\\
    &+\Big\{\cancel{4D\sigma }\big(\T1D\Tp1-\Tp1D\T1\big)+\cancel{6\sigma }\big(\T1D^2\Tp1-\Tp1D^2\T1\big)+6\sigma \big(\T2D^2\Tp0-\Tp2D^2\T0\big) \nn\\
    &\quad +\cancel{6D\sigma }\big(\T1D\Tp1-\Tp1D\T1\big)-\cancel{6D\sigma }\big(\T1D\Tp1-\Tp1D\T1\big)-9\sigma \big(\T2D^2\Tp0-\Tp2D^2\T0\big) \nn\\
    &\quad -3D^2\sigma \big(\T0\Tp2-\Tp0\T2\big)-6D\sigma D\big(\T0\Tp2-\Tp0\T2\big)-3\sigma D^2\big(\T0\Tp2-\Tp0\T2\big) \nn\\
    &\quad -2D\sigma \big(\T0D\Tp2-\Tp0D\T2\big)-\cancel{4D\sigma }\big(\T1D\Tp1-\Tp1D\T1\big)-6D\sigma (\T2D\Tp0-\Tp2D\T0\big) \nn\\
    &\quad +6\sigma \big(D\T0D\Tp2-D\Tp0D\T2\big)-3\sigma \big(\T0D^2\Tp2-\Tp0D^2\T2\big)-\cancel{6\sigma }\big(\T1D^2\Tp1-\Tp1D^2\T1\big)\Big\} \nn\\
    &+2\sigma \Big\{-\cancel{6D\sigma }\big(\T1\Tp2-\Tp1\T2\big)-\cancel{3\sigma }\big(\T1D\Tp2-\Tp1D\T2\big)+\cancel{\frac92 \sigma }\big(\T2D\Tp1-\Tp2D\T1\big) \nn\\
    &\qquad~~ -\cancel{6D\sigma }\big(\T2\Tp1-\Tp2\T1\big)-\cancel{9\sigma }\big(\T2D\Tp1-\Tp2D\T1\big)+4D\sigma \big(\T0\Tp3-\Tp0\T3\big) \nn\\
    &\qquad~~ +6D\sigma \big(\T0\Tp3-\Tp0\T3\big)+6\sigma \big(\T0D\Tp3-\Tp0D\T3\big)-\cancel{6\sigma }\big(\T3D\Tp0-\Tp3D\T0\big) \nn\\
    &\qquad~~+\frac32 \sigma \big(\T0D\Tp3-\Tp0D\T3\big)+\cancel{3\sigma }\big(\T1D\Tp2-\Tp1D\T2\big)+\cancel{\frac92 \sigma }\big(\T2D\Tp1-\Tp2D\T1\big) \nn\\
    &\qquad~~+\cancel{6\sigma }\big(\T3D\Tp0-\Tp3D\T0\big)-\frac{15}{2}\sigma ^2\big(\T0\Tp4-\Tp0\T4\big)\Big\}.\nn
\end{align}
We emphasized the obvious simplifications.
A few more occur between the terms involving one $\sigma $.
We notice that we end up only with terms proportional to $\T0$ (and $\Tp0$).
Therefore,
\begin{align}
    &\quad~\big[\hdTp,\hdT\big]\sigma -\hat\delta_{[T,T']^\sigma}\sigma = \nn\\
    &=\T0\Big\{D^3\Tp1 \underbrace{-3\sigma D^2\Tp2-3D^2\sigma \Tp2-6D\sigma D\Tp2}_{-3D^2(\sigma \Tp2)}-2D\sigma D\Tp2-3\sigma D^2\Tp2+8\sigma D\sigma \Tp3 \nn\\
    &\qquad +12\sigma D\sigma \Tp3+12\sigma ^2D\Tp3+3 \sigma ^2D\Tp3-15\sigma ^3\Tp4\Big\}-T\leftrightarrow T' \nn\\
    &=\T0\Big\{D^2\big(D\Tp1-3\sigma \Tp2\big)-2D\sigma \big(D\Tp2-4\sigma \Tp3\big) \nn\\
    &\qquad -3\sigma D\big(D\Tp2-4\sigma \Tp3\big)+3 \sigma ^2\big(D\Tp3-5 \sigma \Tp4\big)\Big\} -T\leftrightarrow T' \\
    &=\T0\bigg\{D^2(\Dcal T')_0-2D\sigma (\Dcal T')_1-3\sigma D(\Dcal T')_1+3 \sigma^2(\Dcal T')_2\bigg\} -T\leftrightarrow T'\nn\\
    &= \T0\Big\{ -\hat\delta_{[\hHam,T']^\sigma} \sigma\Big\}-T\leftrightarrow T',\nn
\end{align}
which is indeed \eqref{algebraRepinterm}.\footnote{Again, strictly speaking, $\Ham_0\big(\big[\hdTp,\hdT\big]\sigma -\hat\delta_{[T,T']^\sigma}\sigma\big)=\Tp0 \hat\delta_{[\hHam,T]^\sigma} \sigma-\T0 \hat\delta_{[\hHam,T']^\sigma} \sigma$.}

\section{Covariant wedge initial condition \label{AppCovWed1}}

To prove the formula \eqref{com1}, notice that 
\begin{align}
    D[T,T']^\sigma_{-1} &=D\big(\T0 D\Tp0-2\sigma\T0\Tp1\big)-T\leftrightarrow T' \nn\\
    &=\big(\T0 D^2\Tp0-2D\sigma\T0\Tp1-2\sigma\Tp1 D\T0-2\sigma\T0 D\Tp1\big) -T\leftrightarrow T' \\
    &=\Big(T_0\big( D^2\Tp0-2D\sigma\Tp1-3\sigma D\Tp1\big) +2\sigma\T1 D\Tp0+\sigma\T0 D\Tp1\Big) -T\leftrightarrow T', \nn
\end{align}
while
\begin{align}
    \sigma[T,T']^\sigma_0 =\big(\sigma\T0 D\Tp1+2\sigma\T1 D\Tp0-3\sigma^2\T0\Tp2\big)-T\leftrightarrow T'.
\end{align}
Therefore, using the definition \eqref{dTCv1} of the variation $\hdT\sigma$, we see that 
\begin{equation}
    D[T,T']^\sigma_{-1}- \sigma[T,T']^\sigma_0 = \Tp0\hdT\sigma-\T0\hdTp\sigma,
\label{IniConstInterm}
\end{equation}
which matches with \eqref{com1} up to the usual factor of $\Ham_0$ that we keep implicit.

\section{Proof of the Jacobi identity \eqref{Jacobi-1Version} \label{AppJacobi-1Version}}

Here we study the condition that the bracket \eqref{CFbracketSphereVersion} has to satisfy in order to respect the Jacobi identity, if we were not aware that $D\T{-1}$ has to be equal to $\sigma\T0$.
To avoid confusion, we remove the superscript $\sigma$ to denote this preliminary version of the $\sigma$-bracket.
\begin{align}
    \big[T,[T',T'']\big]_s &=\sum_{n=0}^{s+2}(n+1)\Big(\T{n}D[T',T'']_{s+1-n}-[T',T'']_n D\T{s+1-n}\Big) \label{Jac-1interm}\\
    &= \sum_{n=0}^{s+2}\sum_{k=0}^{s+3-n}(n+1)(k+1)\Big(\T{n}D\Tp{k}D\Tpp{s+2-n-k}-\T{n}D\Tpp{k}D\Tp{s+2-n-k} \nn\\
    &\hspace{5cm}+\T{n}\Tp{k}D^2\Tpp{s+2-n-k}-\T{n}\Tpp{k}D^2\Tp{s+2-n-k}\Big) \nn\\
    &-\sum_{n=0}^{s+2}\sum_{k=0}^{n+2}(n+1)(k+1)\Big(\Tp{k}D\Tpp{n+1-k}-\Tpp{k}D\Tp{n+1-k}\Big)D\T{s+1-n}. \nn
\end{align}
The goal is to extract the boundary terms of the sums, namely those that involve the element $\T{-1}$.
Therefore
\begin{align}
    \big[T,[T',T'']\big]_s &=\sum_{a+b+c=s+2}(a+1)(b+1)\Big(\T{a}D\Tp{b}D\Tpp{c}-\T{a}D\Tpp{b}D\Tp{c}\Big) \nn\\
    &+\sum_{n=0}^{s+2}(n+1)(s+4-n)\Big(\T{n}D\Tp{s+3-n}D\Tpp{-1}-\T{n}D\Tpp{s+3-n}D\Tp{-1}\Big) \nn\\
    &+\sum_{a+b+c=s+2}(a+1)(b+1)\Big(\T{a}\Tp{b}D^2\Tpp{c}-\T{a}\Tpp{b}D^2\Tp{c}\Big) \nn\\
    &+\sum_{n=0}^{s+2}(n+1)(s+4-n)\Big(\T{n}\Tp{s+3-n}D^2\Tpp{-1}-\T{n}\Tpp{s+3-n}D^2\Tp{-1}\Big) \nn\\
    \qquad\qquad\quad &-\sum_{a+b+c=s+2}(b+c)(b+1)\Big(\Tp{b}D\Tpp{c}D\T{a}-\Tpp{b}D\Tp{c}D\T{a}\Big) \nn\\
    &-\sum_{k=0}^{s+4}(s+3)(k+1)\Big(\Tp{k}D\Tpp{s+3-k}D\T{-1}-\Tpp{k}D\Tp{s+3-k}D\T{-1}\Big) \\
    &-\sum_{n=0}^{s+1}(n+1)(n+3)\Big(\Tp{n+2}D\Tpp{-1}D\T{s+1-n}-\Tpp{n+2}D\Tp{-1}D\T{s+1-n}\Big). \nn
\end{align}
According to \eqref{Jac-1interm}, the first sum that comes with a minus sign should be restricted to $b+c>0$, but of course we can include the term $b+c=0$ for free since that contribution vanishes. 
Upon cyclic permutations, we get
\begin{align}
    \big[T,[T',T'']\big]_s &\cyc \!\!\!\!\!\sum_{a+b+c=s+2}\!\!\!\!\!\T{a}D\Tp{b}D\Tpp{c} \big((a+1)(b+1)-(a+1)(c+1)-(a+b)(a+1)+(a+c)(a+1)\big) \nn\\
    &+\sum_{a+b+c=s+2}\T{a}\Tp{b}D^2\Tpp{c}\big((a+1)(b+1)-(b+1)(a+1)\big) \nn\\
    &+D\T{-1}\left\{\sum_{n=0}^{s+2}(n+1)(s+4-n)\Big(\Tp{n}D\Tpp{s+3-n}-\Tpp{n}D\Tp{s+3-n}\Big)\right. \nn\\
    &\qquad\quad~ -\sum_{n=0}^{s+4}(s+3)(n+1)\Big(\Tp{n}D\Tpp{s+3-n}-\Tpp{n}D\Tp{s+3-n}\Big) \\
    &\qquad\quad~ -\left.\sum_{n=0}^{s+1}(n+1)(n+3)\Big(\Tpp{n+2}D\Tp{s+1-n}-\Tp{n+2}D\Tpp{s+1-n}\Big)\right\} \nn\\
    &-D^2\T{-1}(s+4)\Big(\Tp{s+3}\Tpp0-\Tpp{s+3}\Tp0\Big). \nn
\end{align}
The first two lines vanish. Hence
\begin{align}
    \big[T,[T',T'']\big]_s &\cyc D\T{-1}\left\{\sum_{n=0}^{s+2}(n+1)(1-n)\Big(\Tp{n}D\Tpp{s+3-n}-\Tpp{n}D\Tp{s+3-n}\Big)\right. \nn\\
    & \qquad\quad~ -(s+3)(s+4)\Big(\Tp{s+3}D\Tpp0-\Tpp{s+3}D\Tp0\Big) \\
    &\quad\qquad~ -\left.\sum_{n=0}^{s+1}(n+1)(n+3)\Big(\Tpp{n+2}D\Tp{s+1-n}-\Tp{n+2}D\Tpp{s+1-n}\Big)\right\} \nn\\
    &-D^2\T{-1}(s+4)\Big(\Tp{s+3}\Tpp0-\Tpp{s+3}\Tp0\Big), \nn
\end{align}
where we got rid of
\begin{equation}
    -D\T{-1}(s+3)(s+5)\big(\Tp{s+4}D\Tpp{-1}-\Tpp{s+4}D\Tp{-1}\Big)\cyc 0.
\end{equation}
Massaging the sums (as usual by extracting some boundary terms), we have
\begin{align}
    \big[T,[T',T'']\big]_s &\cyc  D\T{-1}\bigg\{\Tp0 D\Tpp{s+3}-\Tpp0 D\Tp{s+3}-\sum_{n=2}^{s+2}(n+1)(n-1)\Big(\Tp{n}D\Tpp{s+3-n}-\Tpp{n}D\Tp{s+3-n}\Big) \nn\\
    &\qquad\quad~ -\sum_{n=2}^{s+2}(n-1)(n+1)\Big(\Tpp{n}D\Tp{s+3-n}-\Tp{n}D\Tpp{s+3-n}\Big) \\
    & \qquad\quad~ -(s+4)\Big(\Tp{s+3}D\Tpp0-\Tpp{s+3}D\Tp0\Big)\bigg\} -D^2\T{-1}(s+4)\Big(\Tp{s+3}\Tpp0-\Tpp{s+3}\Tp0\Big), \nn
\end{align}
The two sums cancel one another and we can also get rid of the $D^2\T{-1}$ term using the Leibniz's rule (over the last term in $\{\cdots\}$), so that one ends up with
\begin{align}
    \big[T,[T',T'']\big]_s &\cyc -(s+4)D\Big(D\T{-1}\Tpp0\Tp{s+3}-D\T{-1}\Tp0\Tpp{s+3}\Big)+(s+3)D\T{-1}\Big(\Tpp0 D\Tp{s+3}-\Tp0 D\Tpp{s+3}\Big) \nn\\
    &\cyc -(s+4)D\Big(\T{s+3}\big(\Tp0 D\Tpp{-1}-\Tpp0 D\Tp{-1}\big)\Big)+(s+3)D\T{s+3}\Big(\Tp0 D\Tpp{-1}-\Tpp0 D\Tp{-1}\Big) \nn\\
    &\cyc -D\T{s+3}\big[T',T''\big]_{-2}-(s+4)\T{s+3}D\big[T',T''\big]_{-2}.
\end{align}
In the second line we used the cyclic permutation one more time to highlight the combination
\begin{equation}
    \big[T',T''\big]_{-2}=\Tp0 D\Tpp{-1}-\Tpp0 D\Tp{-1}.
\end{equation}
This concludes the proof of \eqref{Jacobi-1Version}.

\section{Proof of Lemma [Leibniz] \label{AppLeibniz}}

We proved that $\big(\Dcal[T,T']^\sigma\big)_{-2}=0$ when $(\Dcal^2 T)_{-2}=0$ in App.\,\ref{AppCovWed1}.
Moreover, the original version of the proof of the \hyperref[LemmaLeibniz]{Lemma} at any degree was presented in \cite{Cresto:2024fhd}. 
We then had to repeat it in \cite{Cresto:2024mne}.
Since the latter is slightly more general than the former, we refer the reader to appendix \ref{App:dualEOM} for the newest version, where one simply needs to make the change of notation $(C,\tau)\to(\sigma,T)$ to recover the version which concerns us here.
Both proofs are otherwise identical.

\section{Proof of Lemma [Intertwining properties]}
\subsection{Adjoint action}
\label{app:Adj} 

We evaluate the derivative
\begin{align}
    \pa_\alpha \big(\mathcal{T}^{ G_\alpha}_s (T_\alpha)\big)
    &= 
    \sum_{n,k =0}^{\infty} 
    \frac{(s+k+1)!}{n! (s+1)! k!} G_\alpha^{n}(-DG_\alpha)^k \Big( (n+k)  
    \big(\Dcal_{\sigma_\alpha}^nT_\alpha\big)_{s+ k} + \pa_\alpha \big(\Dcal_{\sigma_\alpha}^n T_\alpha\big)_{s+k}\Big) \cr
    &=\sum_{n=1}^\infty\sum_{k=0}^{\infty} \left(
    \frac{(s+k+1)!}{(n-1)! (s+1)! k!} G_\alpha^{n}(-DG_\alpha)^k\right)\big(\Dcal_{\sigma_\alpha }^nT_\alpha\big)_{s+k} \nn\\*
    &+\sum_{n=0}^\infty\sum_{k=1}^{\infty} \left( 
    \frac{(s+k+1)!}{n! (s+1)! (k-1)!} G_\alpha^n(-DG_\alpha)^{k} \right)\big(\Dcal_{\sigma_\alpha}^nT_\alpha\big)_{s+k} \nn\\*
    &+ \sum_{n,k =0}^{\infty} 
    \frac{(s+k+1)!}{n! (s+1)! k!} G_\alpha^{n}(-DG_\alpha)^k \left( \pa_\alpha \big(\Dcal_{\sigma_\alpha}^nT_\alpha\big)_{s+k}\right)\nn\\
    &= \sum_{n,k =0}^{\infty} 
    \frac{(s+k+1)!}{n! (s+1)! k!} G_\alpha^{n}(-DG_\alpha)^k
    \Big(\!
    \big(G_\alpha \Dcal_{\sigma_\alpha}^{n+1}T_\alpha\big)_{s+k}+\pa_\alpha \big(\Dcal_{\sigma_\alpha}^nT_\alpha \big)_{s+k} \nn\\
    &\hspace{6cm}-(s+k+2) DG_\alpha \big(\Dcal^n_{\sigma_\alpha} T_\alpha\big)_{s+k+1}\Big) \nn\\
    &= \sum_{n,k =0}^{\infty} 
    \frac{(s+k+1)!}{n! (s+1)! k!} G_\alpha^{n}(-DG_\alpha)^k
    \left(\big(\pa_\alpha+\adaG\big) \big(\Dcal_{\sigma_\alpha}^n T_\alpha\big)
    \right)_{s+k},
\end{align} 
which corresponds to \eqref{palT}.

\subsection{Intertwining relation}
\label{app:Int}
Here we prove \eqref{DTrecursion}.
Starting from 
\be 
T_{s}^G(T):=
\sum_{k=0}^\infty 
\frac{(s+k+1)!}{(s+1)! k!} (-DG)^k T_{s+k},\qquad s\geqslant -1,
\ee 
we easily see that we have for $T\in\V^p$,\footnote{If $T\in \V^p$ then the series truncates to a finite sum where $k\leq p-s$ since  $T_{s+k}=0$ for $ k > p-s$.}
\begin{align} 
DT_{s}^G(T)& =
\sum_{k=0}^{p-s}
\frac{(s+k+1)!}{(s+1)! k!} \Big((-DG)^k D T_{s+k} - k (-DG)^{k-1} D^2G \,T_{s+k}\Big) \cr
& =
\sum_{k=0}^{p-s} 
\frac{(s+k+1)!}{(s+1)! k!} (-DG)^k D T_{s+k}
- \sum_{k=1}^{p-s} 
\frac{(s+k)!}{(s+1)! (k-1)!}  (-DG)^{k-1}  (s+k+1) \sigma  T_{s+k}\cr
& =
\sum_{k=0}^{p-s} 
\frac{(s+k+1)!}{(s+1)! k!} (-DG)^k D T_{s+k}-\sum_{k=0}^{p-s-1} 
\frac{(s+k+1)!}{(s+1)! k!} (-DG)^k  (s+k+2) \sigma  T_{s+k+1} \cr
& =
\sum_{k=0}^{p-s} 
\frac{(s+k+1)!}{(s+1)! k!} (-DG)^k  (\Dcal T)_{s+k-1}.
\end{align}
Next we evaluate ($\Dcal$ shifts the filtration since it comes from a bracket, i.e. $\Dcal:\V^p\to\V^{p-1}$)
\begin{align}
    &\quad\big(D T^G(T)\big)_{s-1} + \big(DG T^G\big([\hHam,T]^\sigma\big)\big)_{s-1} =
    D T_{s}^G(T) + D G T^G_{s}\big([\hHam,T]^\sigma\big) \cr
    &= \sum_{k=0}^{p-s} 
\frac{(s+k+1)!}{(s+1)! k!}(-DG)^k  (\Dcal T)_{s+k-1} - \sum_{k=0}^{p-s-1} 
\frac{(s+k+1)!}{(s+1)! k!} (-DG)^{k+1}  (\Dcal T)_{s+k} \cr
&= (\Dcal T)_{s-1} + \sum_{k=1}^{p-s} 
\left(\frac{(s+k+1)!}{(s+1)! k!}- 
\frac{(s+k)!}{(s+1)! (k-1)!} \right)  (-DG)^k  (\Dcal T)_{s+k-1} \label{diffs}.
\end{align}
At this stage there are two cases to consider: 
If $s=-1$ then we see that the two coefficients appearing in the difference inside the sum are equal to $1$. 
Their contributions cancel and we are left with
\be \label{TG-2interm}
D T_{-1}^G(T) + DG T_{-1}^G\big([\hHam,T]^\sigma\big)= (\Dcal T)_{-2}=T^G_{-2}\big([\hHam, T]^\sigma \big).
\ee
When $s>-1$ we can evaluate the difference and get that \eqref{diffs} reduces to
\be 
(\Dcal T)_{s-1} + \sum_{k=1}^{p-s} 
\frac{(s+k)!}{s! k!}(-DG)^k  (\Dcal T)_{s+k-1}. \label{DTG}
\ee
On the other hand we have that 
\be \label{Teta}
   T_{s-1}^G\big([\hHam, T]^\sigma \big) =
    \sum_{k=0}^{p-s} 
\frac{(s+k)!}{s! k!} (-DG)^k  (\Dcal T)_{s+k-1}.
\ee
Equation \eqref{TG-2interm} and the equality of \eqref{DTG} with \eqref{Teta} proves \eqref{DTrecursion}.

\subsection{Inverse intertwining relation}
\label{app:IInt}

Here we prove  that $(\mTG)^{-1}$ defined in \eqref{mTinv} satisfies the intertwining relation  $\Dcal (\mTG)^{-1}(T) = (\mTG)^{-1}(DT)$.
Using that
\be 
T_{s}^{-G}(T)=
\sum_{k=0}^\infty 
\frac{(s+k+1)!}{(s+1)! k!} (DG)^k T_{s+k},\qquad s\geqslant -1,
\ee
one evaluates 
\begin{align}
    \big(\Dcal T^{-G}(T)\big)_{s-1} &= 
    DT_{s}^{-G}(T) - (s+2) \sigma T_{s+1}^{-G}(T)\cr
    &=\sum_{k=1}^\infty\frac{(s+1+k)!}{(s+1)!(k-1)!}(DG)^{k-1}D^2G\,\T{s+k}+\sum_{k=0}^\infty\frac{(s+1+k)!}{(s+1)!k!}(DG)^k D\T{s+k} \cr
& - \sigma 
\sum_{k=0}^{\infty} 
\frac{(s+k+2)!}{(s+1)! k!}
(DG)^k T_{s+k+1}
\cr
&=\sum_{k=0}^{\infty} 
\frac{(s+k+1)!}{(s+1)! k!} (DG)^k  (D T)_{s+k-1},
\end{align}
where in the second line, the first and third terms cancel out since $D^2G=\sigma$.
Next we evaluate\footnote{For $s=-1$, $\big(\Dcal T^{-G}(T)- D G T^{-G}(DT)\big)_{-2}=(DT)_{-2}=T^{-G}_{-2}(DT)$.}
\begin{align}
    &\quad\big(\Dcal T^{-G}(T)\big)_{s-1} - \big(DG T^{-G}(D T)\big)_{s-1} =
    \big(\Dcal T^{-G}(T)\big)_{s-1} - D G T^{-G}_{s}(DT) \cr
    &= \sum_{k=0}^{\infty} 
\frac{(s+k+1)!}{(s+1)! k!}(DG)^k  (D T)_{s+k-1} - \sum_{k=0}^{\infty} 
\frac{(s+k+1)!}{(s+1)! k!} (DG)^{k+1}  (D T)_{s+k} \cr
&= (DT)_{s-1} + \sum_{k=1}^\infty 
\left(\frac{(s+k+1)!}{(s+1)! k!}- 
\frac{(s+k)!}{(s+1)! (k-1)!} \right)  (DG)^k  (D T)_{s+k-1} \cr
& =
    \sum_{k=0}^\infty 
\frac{(s+k)!}{s! k!} (DG)^k  (D T)_{s+k-1}
=  T_{s-1}^{-G}(DT),\qquad\qquad s\geqslant 0.
\end{align}
This in turn implies that 
\begin{align}
 \Dcal (\mTG)^{-1}(T) 
 &= \sum_{n=0}^{\infty} \frac{(-G)^n}{n!}\Big( \Dcal T^{-G}\big(D^n T \big)- DG T^{-G}\big(D^{n+1} T \big)\Big) \cr
 &=\sum_{n=0}^{\infty} \frac{(-G)^n}{n!} T^{-G}\big(D^{n+1}T\big) = (\mTG)^{-1}(DT).
\end{align}

\section{Proof of the Jacobi anomaly of the $C$-bracket \label{AppJacobiDcal}}

Here we tackle the computation of the cyclic permutation of $\big[\tau,[\tau',\tau'']^C\big]^C$, making use of the Leibniz anomaly \eqref{AnomalyLeibniz1}, that we prove in App.\,\ref{App:dualEOM}.
We use that $\hdt C= -(\Dcal^2\tau)_{-2}$.
\begin{align}
    \big[\tau,[\tau',\tau'']^C\big]^C_s &=\sum_{n=0}^{s+1}(n+1)\Big(\t{n}\big(\Dcal[\tau',\tau'']^C\big)_{s-n}-[\tau',\tau'']^C_n(\Dcal\tau)_{s-n}\Big) \cr
    &=\sum_{n=0}^{s+1}(n+1)\bigg(\t{n}[\Dcal\tau',\tau'']^C_{s-n}+\t{n}[\tau',\Dcal\tau'']^C_{s-n} \cr
    &\hspace{3cm} +(s-n+3)\t{n}\Big(\tp{s-n+2}(\Dcal^2\tau'')_{-2}-\tpp{s-n+2}(\Dcal^2\tau')_{-2}\Big)\bigg) \cr
    &-\sum_{n=0}^{s+1}(n+1)[\tau',\tau'']^C_n(\Dcal\tau)_{s-n} \cr
    &=\left\{\sum_{n=0}^{s+1}\sum_{k=0}^{s-n+1}(n+1)(k+1)\Big(\t{n}(\Dcal\tau')_k(\Dcal\tau'')_{s-n-k}-\t{n}\tpp{k}(\Dcal^2\tau')_{s-n-k}\Big)\right. \cr
    &\quad -\sum_{n=0}^{s+1}\sum_{k=0}^{n+1}(n+1)(k+1)\tp{k}(\Dcal\tpp{})_{n-k}(\Dcal\tau)_{s-n} \label{jacobiDcalinterm}\\
    &\quad \left.+\sum_{n=0}^{s+1}(n+1)(s-n+3)\t{n}\tp{s-n+2}(\Dcal^2\tau'')_{-2}\right\}-(\tau'\leftrightarrow\tau''). \nn
\end{align}
The last equation then splits into three types of contributions, that we shall refer to as \circled{1}, \circled{2} and \circled{3}.
Let us start with the third line of \eqref{jacobiDcalinterm}, that we dub \circled{3}. 
Extracting the $n=0$ term and then taking appropriate cyclic permutations, we get

\begin{align}
    \circled{3}&\cyc (s+3)\t0\tp{s+2}(\Dcal\tau'')_{-2}-(s+3)\tp0\t{s+2}(\Dcal\tau'')_{-2} \\*
    &+\sum_{n=1}^{s+1}(n+1)(s-n+3)\t{n}\tp{s-n+2}(\Dcal^2\tau'')_{-2}-\sum_{n=1}^{s+1}(n+1)(s-n+3)\tp{n}\t{s-n+2}(\Dcal^2\tau'')_{-2}. \nn
\end{align}
Changing $n\to s-n+2$ in the last sum, the two sums cancel and \circled{3} reduces to
\begin{equation} \label{jacobiDcalinterm3}
    \circled{3}\cyc (\Dcal\tau'')_{-2}\paren{\tau,\tau'}_s \cyc -\hdt C\paren{\tau',\tau''}_s, 
\end{equation}
which is precisely \eqref{JacCFC}. 
We thus have to show that all the other terms in \eqref{jacobiDcalinterm} cancel.

The trick first consists in isolating from the summation the boundary terms that contain degree $-1$ element such as $(\Dcal \tau)_{-1}$ or $(\Dcal^2 \tau)_{-1}$.
We denote all the latter by $\circled{1}$ and the rest by \circled{2}.
For instance, in the first line of \eqref{jacobiDcalinterm},
\begin{equation}
    \sum_{n=0}^{s+1}\sum_{k=0}^{s-n+1}= \sum_{n=0}^{s+1}(k=s-n+1)+\sum_{n=0}^s\sum_{k=0}^{s-n}
\end{equation}
and we collect the first  term in \circled{1} and the second  term in \circled{2}.
We proceed similarly with the second line of \eqref{jacobiDcalinterm} and obtain that
\begin{align}
    \circled{1} &=\left\{\sum_{n=0}^{s+1}(n+1)(s-n+2)\t{n}(\Dcal\tau')_{s-n+1}(\Dcal\tau'')_{-1}\right. \cr
    &\quad - \sum_{n=0}^{s+1}(n+1)(s-n+2)\t{n}\tpp{s-n+1}(\Dcal^2\tau')_{-1} \label{jacobiDcalinterm2}\\
    &\hspace{-1cm}\left.-\sum_{n=0}^s(n+1)(n+2)\tp{n+1}(\Dcal\tpp{})_{-1}(\Dcal\tau)_{s-n}-\sum_{k=0}^{s+2}(s+2)(k+1)\tp{k}(\Dcal\tau'')_{s+1-k}(\Dcal\tau)_{-1}\right\} -(\tau'\leftrightarrow\tau''). \nn
\end{align}
The second term of this equation evaluates  after skew-symmetrization to
\begin{equation}
    -(\Dcal^2\tau')_{-1} \sum_{n=0}^{s+1}(n+1)(s-n+2)\t{n}\tpp{s-n+1}+(\Dcal^2\tau'')_{-1} \sum_{n=0}^{s+1}(n+1)(s-n+2)\t{n}\tp{s-n+1}\cyc 0,
\end{equation}
where we used the cyclic permutation and the change of variable $n\to s-n+1$ in one of the sums. Hence, by extracting the terms $n=s$, $k=s+1$ and $k=s+2$ in the last line of \eqref{jacobiDcalinterm2} (and renaming $k\to n$), \circled{1} is equal to
\begin{align}
    \circled{1}
    &= \left\{\sum_{n=0}^{s+1}(n+1)(s-n+2)\t{n}(\Dcal\tau')_{s-n+1}(\Dcal\tau'')_{-1}-\sum_{n=0}^{s+2}(s+2)(n+1)\tp{n}(\Dcal\tau'')_{s+1-n}(\Dcal\tau)_{-1} \right. \nn\\
    &\quad \left. -\sum_{n=0}^{s}(n+1)(n+2)\tp{n+1}(\Dcal\tpp{})_{-1}(\Dcal\tau)_{s-n}\right\}-(\tau'\leftrightarrow\tau'') \cyc 0. 
\end{align}
These terms  vanish upon cyclic permutation, which becomes clear as soon as one puts $(\Dcal\tau)_{-1}$ as a prefactor of every term.

We still have to deal with \circled{2}, which can now very conveniently be written as
\begin{align}
    \circled{2}&=-\sum_{a+b+c=s}(a+1)(b+1)\Big(\t{a}\tpp{b}(\Dcal^2\tau')_c-\t{a}\tp{b}(\Dcal^2\tau'')_c\Big) \cr
    &\quad +\sum_{a+b+c=s}(a+1)(b+1)\Big(\t{a}(\Dcal\tau')_b(\Dcal\tau'')_c-\t{a}(\Dcal\tau'')_b(\Dcal\tau')_c\Big) \cr
    &\quad -\sum_{a+b+c=s}(a+b+1)(a+1)\Big(\tp{a}(\Dcal\tau'')_b(\Dcal\tau)_c-\tpp{a}(\Dcal\tau')_b(\Dcal\tau)_c\Big).
\end{align}
Notice that $a,b$ and $c$ run from 0 to $s$. 
Without extracting the degree $-1$ elements in \circled{1}, the ranges for $a,b$ and $c$ would have been different, preventing us from treating them on par. 
Using cyclic permutation, we have that
\begin{align}
    \circled{2} &\cyc -\sum_{a+b+c=s}(a+1)(b+1)\Big(\t{a}\tpp{b}(\Dcal^2\tau')_c-\tpp{a}\t{b}(\Dcal^2\tau')_c\Big) \cr
    &\quad +\sum_{a+b+c=s}(a+1)(b+1)\Big(\t{a}(\Dcal\tau')_b(\Dcal\tau'')_c-\t{a}(\Dcal\tau'')_b(\Dcal\tau')_c\Big) \cr
    &\quad -\sum_{a+b+c=s}(a+b+1)(a+1)\Big(\t{a}(\Dcal\tau')_b(\Dcal\tau'')_c-\t{a}(\Dcal\tau'')_b(\Dcal\tau')_c\Big) \\
    &\cyc -\sum_{a+b+c=s}a(a+1)\Big(\t{a}(\Dcal\tau')_b(\Dcal\tau'')_c-\t{a}(\Dcal\tau'')_b(\Dcal\tau')_c\Big) \cyc 0. \nn
\end{align}
Therefore $\circled{1}=0=\circled{2}$ and $\big[\tau,[\tau',\tau'']^C\big]^C_s=\circled{3}= \eqref{jacobiDcalinterm3}$.

\section{Proofs of closure for the $\sfT$-bracket \label{Apptppsevol}}

\subsection{Leibniz anomalies \label{App:dualEOM}}

In this section we provide the proof of \eqref{AnomalyLeibniz}.
We start by computing explicitly $\big(\Dcal[\tau,\tau']^C\big)_{s-1}$, for $s\geq 0$, using the formula \eqref{CFbrackettauDcal} for the $C$-bracket. 
\begin{align}
    \big(\Dcal[\tau,\tau']^C\big)_{s-1} &=D[\tau,\tau']^C_s-(s+2)C [\tau,\tau']_{s+1}^C \nn\\
    &=\left\{\sum_{n=0}^{s+1}(n+1)\Big(D\t{n}(\Dcal \tau')_{s-n}+\t{n}D(\Dcal \tau')_{s-n}\Big)\right. \\
    & \left.-(s+2)C\sum_{n=0}^{s+2}(n+1)\t{n}(\Dcal \tau')_{s+1-n}\right\}-(\tau\leftrightarrow \tau'). \nn
\end{align}
We now add and subtract the necessary terms to transform the $D$ into $\Dcal$ in the first sum:
\begin{align}
    \big(\Dcal[\tau,\tau']^C\big)_{s-1}  &= \left\{\sum_{n=0}^{s+1}(n+1)\Big((\Dcal \tau)_{n-1}(\Dcal \tau')_{s-n}+\t{n}(\Dcal^2 \tau')_{s-n-1}\Big)\right. \nn\\
    &+\sum_{n=0}^{s+1}(n+1)(n+2)C\t{n+1}(\Dcal \tau')_{s-n}+\sum_{n=0}^{s+1}(n+1)(s-n+2)C\t{n}(\Dcal \tau')_{s-n+1} \nn\\
    &-\left.(s+2)C\sum_{n=0}^{s+1}(n+1)\t{n}(\Dcal \tau')_{s+1-n}-(s+2)(s+3)C\t{s+2}(\Dcal \tau')_{-1}\right\} -(\tau\leftrightarrow \tau') \nn\\
    &=\left\{\sum_{n=0}^{s+1}(n+1)\Big((\Dcal \tau)_{n-1}(\Dcal \tau')_{s-n}+\t{n}(\Dcal^2 \tau')_{s-n-1}\Big)\right. \nn\\
    &\quad +\sum_{n=1}^{s+2}n(n+1)C\t{n}(\Dcal \tau')_{s+1-n}-\sum_{n=1}^{s+1}n(n+1)C\t{n}(\Dcal \tau')_{s+1-n} \nn\\
    &\quad -(s+2)(s+3)C\t{s+2}(\Dcal \tau')_{-1}\Bigg\} -(\tau\leftrightarrow \tau')\\
    &=\left\{\sum_{n=0}^{s+1}(n+1)\Big((\Dcal \tau)_{n-1}(\Dcal \tau')_{s-n}+\t{n}(\Dcal^2 \tau')_{s-n-1}\Big)\right\} -(\tau\leftrightarrow \tau'). \nn
\end{align}
We now evaluate the second term in the anomaly,
\begin{align}
  & \big[\tau,\Dcal \tau'\big]^C_{s-1}+\big[\Dcal \tau,\tau'\big]^C_{s-1} \cr
 = &  
 \sum_{n=0}^{s}(n+1)\Big((\Dcal\tau)_{n}(\Dcal \tau')_{s-1-n}+ \t{n}(\Dcal^2 \tau')_{s-1-n}\Big) 
  -(\tau\leftrightarrow \tau').
  \cr
 = &  
 \left\{\sum_{n=1}^{s+1} n(\Dcal\tau)_{n-1}(\Dcal \tau')_{s-n}+  \sum_{n=0}^{s} (n+1)\t{n}(\Dcal^2 \tau')_{s-1-n} \right\}
  -(\tau\leftrightarrow \tau').
\end{align}
Therefore the difference gives 
\begin{align}
\cA_{s-1}\big([\tau,\tau']^C,\Dcal \big) &=
 \left\{\sum_{n=0}^{s+1} (\Dcal \tau)_{n-1}(\Dcal \tau')_{s-n}
 +(s+2) \t{s+1}(\Dcal^2 \tau')_{-2}\right\} -(\tau\leftrightarrow \tau') \cr
 &=(s+2) \left( \t{s+1}(\Dcal^2 \tau')_{-2} -\tp{s+1}(\Dcal^2 \tau)_{-2} \right).
\end{align}
In the last equality we use that the first term is symmetric in $\tau\leftrightarrow \tau'$ which follows from the relabelling $n\leftrightarrow s+1-n$.

Moreover, we can also consider the special case $\big(\Dcal[\tau,\tau']^C\big)_{-2}$ for which
\begin{align}
    \big(\Dcal[\tau,\tau']^C\big)_{-2} &= D[\tau,\tau']^C_{-1}-C[\tau,\tau']^C_0 \cr
    &=\Big(D\big(\t0(\Dcal\tau')_{-1}\big)-C\big(\t0(\Dcal\tau')_0+2\t1(\Dcal\tau')_{-1}\big)\Big)-\tau\leftrightarrow\tau' \cr
    &=\Big(D\t0(\Dcal\tau')_{-1}+\t0( \Dcal^2\tau')_{-2}-2C\t1(\Dcal\tau')_{-1}\Big)-\tau\leftrightarrow\tau' \cr
    &=\t0( \Dcal^2\tau')_{-2}-\tp0( \Dcal^2\tau)_{-2}. 
    \label{AnomalyLeibniz-2}
\end{align}
In the third line, we gathered $D(\Dcal\tau')_{-1}-C(\Dcal\tau')_0=(\Dcal^2\tau')_{-2}$ while in the last step, we expanded the rest of the terms and used the anti-symmetry between $\tau$ and $\tau'$.
Now notice that the $C$-bracket at degree $-2$ vanishes identically, $[\cdot\,,\cdot]^C_{-2}\equiv 0$. 
This is a natural extension of our definition for the $C$-bracket, that we discuss at length in section \ref{sec:DeformingWedge}.
It implies that the Leibniz anomaly of $\Dcal$ on $[\cdot\,,\cdot]^C$ takes the general form
\begin{equation}
    \boxed{\cA_{s}\big([\tau,\tau']^C,\Dcal \big) =(s+3) \left( \t{s+2}(\Dcal^2 \tau')_{-2} -\tp{s+2}(\Dcal^2 \tau)_{-2} \right)},\qquad s\geqslant-2. \label{AnomalyLeibniz1bis}
\end{equation}
This establishes \eqref{AnomalyLeibniz1} since $(\Dcal^2 \tau)_{-2}=-\hdt C$.

We then consider the quantity
\begin{align}
    \big(\Dcal(\dtp\tau)\big)_s &=D(\dtp\t{s+1})-(s+3)C\dtp\t{s+2} \cr
    &=\dtp D\t{s+1}-(s+3)\dtp(C\t{s+2})+(s+3)\t{s+2}\dtp C \nn\\
    &=\dtp (\Dcal\tau)_s+ (s+3)\t{s+2}\dtp C.
\end{align}
This allows us to compute the anomaly to the Leibniz rule of $\Dcal$ onto the algebroid bracket,
\begin{align}
    \cA_s\big(\lbr\tau,\tau'\rbr,\Dcal \big) &= 
    \cA_s\big([\tau,\tau']^C,\Dcal \big) + \cA_s\big((\delta_{\tau'}\tau-\delta_\tau\tau'),\Dcal \big)\cr
    &= - \delta_{\Dcal\tau'}\tau + \delta_{\Dcal\tau}\tau' + (s+3) \left( \t{s+2}(\dtp C-\hdtp C)  -\tp{s+2}(\dt C-\hdt C)
    \right)\cr
    & =  - \delta_{\Dcal\tau'}\tau + \delta_{\Dcal\tau}\tau' + (s+3)N  \big( \t{s+2}\tp{0}  -\tp{s+2}\t{0}
    \big),
\end{align}
which corresponds to \eqref{AnomalyLeibniz2}.
Finally, we compute similarly the time derivative of $\lbr\cdot\,,\cdot\rbr$ and find that
\begin{align}
    \pa_u\lbr\tau,\tau'\rbr_s &=\Big([\tau,\pa_u\tau']_s^C-(s+3)N\t0\tp{s+2}+\dtp\pa_u\t{s}\Big) -(\tau\leftrightarrow\tau') \cr
    &=\Big(\lbr\tau,\pa_u\tau'\rbr_s -\delta_{\pa_u\tau'}\t{s}-(s+3)N\t0\tp{s+2}\Big) -(\tau\leftrightarrow\tau'),
\end{align}
which matches with \eqref{AnomalyLeibniz3}.

\subsection{Proof of \eqref{tt1dot} $\Rightarrow$ \eqref{t3dot} \label{App:1implies3}}

Here we assume only $\pa_u\t{s}=(\Dcal\tau)_s,\,s=0,1,2$ and compute $\pa_u \lbr\tau,\tau'\rbr_1-\big(\Dcal \lbr\tau,\tau'\rbr\big)_1$. 
We just need to use the Leibniz anomaly \eqref{AnomalyLeibniz}, which at degree 1 gives
\begin{align}
    \big((\pa_u-\Dcal)\lbr\tau,\tau'\rbr\big)_1 &= \big\lbr(\pa_u-\Dcal)\tau,\tau'\big\rbr_1 +\big\lbr\tau,(\pa_u-\Dcal)\tau'\big\rbr_1+\delta_{(\pa_u-\Dcal)\tau}\tp1-\delta_{(\pa_u-\Dcal)\tau'}\t1 \nn\\
    &=\big\lbr(\pa_u-\Dcal)\tau,\tau'\big\rbr_1 -\tau\leftrightarrow\tau',
\end{align}
where we use that $\delta_{(\pa_u-\Dcal)\tau}\tp1\propto \delta_{(\pa_u-\Dcal)\tau} C=0$ by hypothesis. 
Hence,
\begin{align}
     \big\lbr(\pa_u-\Dcal)\tau,\tau'\big\rbr_1 &=\sum_{n=0}^2(n+1)\Big(\cancel{\big(\pa_u\t{n}-(\Dcal\tau)_n\big)}(\Dcal\tau')_{1-n}-\tp{n}\big(\Dcal(\pa_u-\Dcal)\tau\big)_{1-n}\Big) \cr
     &=-\sum_{n=0}^2(n+1)\tp{n}D \cancel{\big(\pa_u\tau-\Dcal\tau\big)}_{2-n}+\sum_{n=0}^2(n+1)\tp{n}(4-n)C\big(\pa_u\tau-\Dcal\tau\big)_{3-n} \cr
     &=4C\tp0\big(\pa_u\t3-(\Dcal\tau)_3\big).
\end{align}
This means that  $\pa_u\lbr\tau,\tau'\rbr_1=\big(\Dcal\lbr\tau,\tau'\rbr\big)_1$ precisely if $\dot\tau_3-(\Dcal\tau)_3 = \t0 M_3$, where $M_3 \in \Ccar{1,-3}$. 
No such local functional can be constructed using only $C$ and its (holomorphic) derivatives and we therefore conclude that $\dot\tau_3=(\Dcal\tau)_3$ which is the dual equation of motion for $s=3$, namely \eqref{t3dot}.

\subsection{Initial condition \label{App:InitialConstraint}}

If we assume that $\tau,\tau'\in\sfTo$  
we show that the $\sfT$-bracket respects the initial condition \eqref{initialCondition}, i.e.
\begin{equation}
    \boxed{D\lbr\tau,\tau'\rbr_{-1}= C\lbr\tau,\tau'\rbr_0}.
\end{equation}
Indeed, notice that thanks to the computation \eqref{AnomalyLeibniz-2}, we have that
\begin{align}
    \big(\Dcal\lbr\tau,\tau'\rbr\big)_{-2}&= \Big(\big(\Dcal[\tau,\tau']^C\big)_{-2}+\big(\Dcal(\dtp\tau)\big)_{-2}\Big)-\tau\leftrightarrow\tau' \cr
    &=\Big(\t0(\Dcal^2\tau')_{-2} +\dtp (D\t{-1})-C\dtp\t0\Big)-\tau\leftrightarrow\tau' \cr
    &=\Big(\t0(\Dcal^2\tau')_{-2} +\t0\dtp C\Big)-\tau\leftrightarrow\tau \\
    &=0, \nn
\end{align}
since $\dtp C=-(\Dcal^2\tau')_{-2}+N\tp0$ and we leveraged the anti-symmetry in $\tau,\tau'$.

\section{Proof of formula \eqref{hatdTT} \label{AppAlgebraAction2}}

We compute the following quantity:
\begin{align}
    \Big(\Big\{2\tp1 D\big(\hdt C\big)+3D\tp1\big(\hdt C\big)-6C\tp2\big(\hdt C\big)\Big\}-\tau\leftrightarrow\tau'\Big)+\hat\delta_{[\tau,\tau']^C}C, \label{morphismInterm}
\end{align}
using $\hdt C=-(\Dcal^2\tau)_{-2}$ and the Leibniz anomaly \eqref{AnomalyLeibniz1bis}.
This is a complementary demonstration to the one already performed in App.\,\ref{AppAlgebraAction}.
Indeed,
\begin{align}
    \hat\delta_{[\tau,\tau']^C}C &=-\big(\Dcal^2[\tau,\tau']^C\big)_{-2} =
    -\Big(\Dcal[\Dcal\tau,\tau']^C+ \Dcal[\tau,\Dcal\tau']^C+\Dcal\cA\big([\tau,\tau']^C,\Dcal\big)\Big)_{-2} \cr
    &=-\big[\Dcal^2\tau,\tau' \big]^C_{-2}-2 \big[\Dcal\tau, \Dcal\tau'\big]^C_{-2} -\big[\tau,\Dcal^2\tau'\big]^C_{-2} \\
    &\quad \,-\cA_{-2}\big([\Dcal\tau,\tau']^C, \Dcal\big)-\cA_{-2}\big([\tau, \Dcal\tau']^C, \Dcal\big)-D\cA_{-1}\big([\tau,\tau']^C, \Dcal\big)+C \cA_0\big([\tau,\tau']^C, \Dcal\big). \nn
\end{align}
We can now exploit the fact that the $C$-bracket at degree $-2$ vanishes identically, $[\cdot\,,\cdot]^C_{-2}\equiv 0$. 
Therefore, we just have to evaluate the various anomalies using \eqref{AnomalyLeibniz1bis}. 
We get
\begin{align}
    \hat\delta_{[\tau,\tau']^C}C &=\Big((\Dcal\tau)_0\hdtp C-\tp0\hat\delta_{\Dcal\tau}C+D \big(2\t1\hdtp C\big)-3C\t2\hdtp C\Big)-\tau\leftrightarrow\tau'.
\end{align}
Hence,
\begin{align}
    \eqref{morphismInterm} &=\Big(D\tp1\big( \hdt C\big)-3C\tp2\hdt C -(\Dcal\tau')_0\hdt C+\t0\hat \delta_{\Dcal\tau'}C\Big)-\tau\leftrightarrow\tau' \cr
    &=\t0\hat \delta_{\Dcal\tau'}C-\tp0\hat \delta_{\Dcal\tau}C,
\end{align}
which concludes the proof.

\chapter{General Relativity II \label{appGRII}}

\section{Symmetry action on $h$ \label{App:hE-1}}

On the one hand,
\begin{equation}
    \big[\tdt,\tdtp\big]h=\tdt\big(C\tp0-D\tp{-1}\big)-\tau\leftrightarrow\tau'=\big(\tp0\tdt C+\tilde\delta_{\tdt\tau'}h\big)-\tau\leftrightarrow\tau'.
\end{equation}
On the other hand (using \eqref{AnomalyLeibniz-2}),
\begin{align}
    \tilde\delta_{\lbr\tau,\tau'\rbr}h=-\big(\Dcal[\tau,\tau']^C\big)_{-2}+ \tilde\delta_{\tdtp\tau-\tdt\tau'}h &=\big(\tp0(\Dcal^2\tau)_{-2}-\tilde\delta_{\tdt\tau'}h\big)-\tau\leftrightarrow\tau' \cr
    &=-\big(\tp0\dt C+\tilde\delta_{\tdt\tau'}h\big)-\tau\leftrightarrow\tau',
\end{align}
where in the last step, we used that $(\Dcal^2\tau)_{-2}=-\hdt C$ and we then replaced $\hdt C$ by $\dt C$ since the two differ by a term that vanishes upon anti-symmetrization in $\tau,\tau'$.
Finally, using the fact that $\tdt C=\pa_u(\tdt h)=\dt C$, we readily see that $\big[\tdt,\tdtp\big]h+\tilde\delta_{\lbr\tau,\tau'\rbr}h=0$.

\section{Time derivative of $\tq_4$ \label{App:tq4}}

We compute the time derivative of 
\begin{align}
    \tq_4 &\equiv q_4-\frac{u^4}{4!}D^4(\bN C)-5D^2\tq_0\pui[3]C-4D\big(D\tq_0\pui[2](uC)\big)-\frac32 D^2\big(\tq_0\pui(u^2C)\big) \nn\\
    & +5D\tQ_1\pui[2]C+ 4D\big(\tQ_1\pui(uC)\big)-5\tQ_2\pui C +15\tq_0\pui\big(C\pui C\big).
\end{align}
We get
\begin{align}
    \pa_u\tq_4 &=-\sum_{n=1}^4\frac{(-u)^{n-1}}{(n-1)!}D^n\tQ_{4-n}+\sum_{n=0}^4\frac{(-u)^n}{n!}D^n\big(D\tQ_{3-n}+(5-n)C\tQ_{2-n}\big) \nn\\
    &-\frac{u^3}{3!}D^4(\bN C)-\frac{u^4}{4!}D^4\pa_u(\bN C)-5D^2\pa_u\tq_0\pui[3]C-5D^2\tq_0\pui[2]C-4D\big(D\pa_u\tq_0\pui[2](uC)\big) \nn\\
    &-4D\big(D\tq_0\pui(uC)\big)-\frac32 D^2\big(\pa_u\tq_0\pui(u^2C)\big)-\frac32 D^2\big(\tq_0 u^2C\big)+5D\tQ_1\pui C \\
    &+5D\big(D\tQ_0+2C\tQ_{-1}\big) \pui[2]C + 4D\big((D\tQ_0+2C\tQ_{-1}) \pui(uC)\big) + 4D\big(\tQ_1 uC\big) \nn\\
    &-5\big(D\tQ_1+3C\tQ_0\big)\pui C-5\tQ_2 C+15\pa_u\tq_0\pui\big(C\pui C\big)+15\tq_0\big(C\pui C\big). \nn
\end{align}
Let us reorganize the terms to emphasize the simplifications:
\begin{align}
    \pa_u\tq_4 &=-\sum_{n=0}^3\frac{(-u)^n}{n!}D^{n+1}\tQ_{3-n}+\sum_{n=0}^3\frac{(-u)^n}{n!}D^n\big(D\tQ_{3-n}\big) +\frac{u^4}{4!}D^4\pa_u\tQ_0-\frac{u^4}{4!}D^4\pa_u(\bN C) \nn\\
    &+\sum_{n=0}^3\frac{(-u)^n}{n!}D^n\big((5-n)C\tQ_{2-n}\big)-\frac{u^3}{3!}D^4(\bN C)-\frac32 D^2\big(\tq_0 u^2C\big)+ 4D\big(\tQ_1 uC\big) -5\tQ_2 C\nn\\
    &-5D^2\tq_0\pui[2]C -4D\big(D\tq_0\pui(uC)\big)+5D^2\tQ_0 \pui[2]C + 4D\big(D\tQ_0 \pui(uC)\big)\nn\\
    &+5D\tQ_1\pui C-5D\tQ_1\pui C-15C\tQ_0\pui C+15\tq_0\big(C\pui C\big) \\
    &-5D^2\pa_u\tq_0\pui[3]C-4D\big(D\pa_u\tq_0\pui[2](uC)\big)-\frac32 D^2\big(\pa_u\tq_0\pui(u^2C)\big)+15\pa_u\tq_0\pui\big(C\pui C\big) \nn\\
    &+10D\big(C\tQ_{-1}\big) \pui[2]C+ 8D\big(C\tQ_{-1} \pui(uC)\big). \nn
\end{align}
The first line reduces to
\bs 
\label{leftover}
\begin{equation}
    \frac{u^4}{4!}D^4\pa_u\tq_0,
\end{equation}
the second line to
\begin{equation}
    -\frac{u^3}{3!}D^4(\bN C)+\frac{3u^2}{2}D^2(\bN C^2)-\frac{2u^3}{3!}D^3\big(C\tQ_{-1}\big),
\end{equation}
where we used that $\tQ_{-1} = D \bN$.
The third line reduces to
\begin{equation}
    5D^2(\bN C)\pui[2]C +4D\big(D(\bN C)\pui(uC)\big),
\end{equation}
and the fourth to
\begin{equation}
    -15\bN C^2\pui C,
\end{equation}
\es
while the last and penultimate lines always involve $\pa_u\tq_0$ or $\tQ_{-1}= D\bN$, which are 0 if $\bN\equiv 0$.
Similarly, the leftover \eqref{leftover} annihilates when $\bN\equiv 0$, so that $\pa_u\tq_4=0$ when no left-handed radiation is present.

\section{Proof of Theorem \hyperref[Noetherhatqs]{[Noether charge at spin $s$]} \label{App:renormcharge}}

We construct $\tq_s$ for a general $s\geqslant 0$.
For this we have to leverage the computation \eqref{soltaunTs} and \eqref{puiDcalk}.
We define $\shs$ and $\shS$ the dual spin and shift\footnote{$\shS$ and $\hS$ are actually equal.} operators according to
\begin{equation}
    \shs=-\hs,\qquad \shs\tQ_s=s\tQ_s\qquad\textrm{and}\qquad \shS\tQ_s=\tQ_{s-1},
\end{equation}
with commutation relations given by
\begin{equation}
    \shS D=D\shS,\quad\shS C=C\shS,\quad D\shs=(\shs-1)D,\quad\shs C=C(\shs+2),\quad\shs\shS=\shS(\shs-1).
\end{equation}
By definition, and using \eqref{soltaunTs},
\begin{equation}
    4\pi G\,\cq^u_{\T{s}}=\int_S\tQ_{-1} \t{-1}+ \sum_{n=0}^s \sum_{k=\lfloor\frac{s+1-n}{2}\rfloor}^{s-n}\int_S\tQ_n\left(\pui D-\pui C(\hs+1)\hS\right)^k\T{n+k}.
\end{equation}
Then, since
\begin{equation}
    \sum_{n=0}^s\sum_{k=\lfloor\frac{s+1-n}{2}\rfloor}^{s-n}=\sum_{k=0}^s\sum_{0\leqslant n=s-2k}^{s-k}, \label{changesum}
\end{equation}
we need to compute, using \eqref{puiDcalk},\footnote{Recall that $P=\sum_{i=0}^\l p_i$.}
\begin{align}
    &\sum_{n=s-2k}^{s-k}\int_S\tQ_n \left(\pui D-\pui C(\hs+1)\hS\right)^k\T{n+k} = \nn\\
    =&\sum_{n=s-2k}^{s-k}\int_S\tQ_n\Bigg( \sum_{\l=0}^k(-1)^\l\sum_{P=k-\l}\pui[p_0]D^{p_0}\bigg(\pui\Big\{C(\hs+1)\pui[p_1]D^{p_1}\Big(\ldots \label{intermrenorm}\\
    &\hspace{7cm}\ldots\pui\Big\{C(\hs+1) \pui[p_\l]D^{p_\l}\hS^\l\Big\}\ldots \Big)\Big\}\bigg)\Bigg)\T{n+k}. \nn
\end{align}
Recall that we study $\tau\in\sfT_s$, so that only $\T{s}\neq 0$.
Hence, for each term in the sum over $n$, there is only one term in the sum over $\l$ that survives, namely the one for which $n+k+\l=s$.
We thus get\footnote{We use that $\pui[p] T_s = \frac{u^p}{p!} T_s.$}
\begin{align}
    \!\eqref{intermrenorm} &=\!\int_S\bigg(\tQ_{s-k}(-1)^0\frac{u^k}{k!}D^k\T{s}+\tQ_{s-k-1}(-1)^1\!\!\!\sum_{P=k-1}\!\!\pui[p_0]D^{p_0}\Big(\pui\Big\{C(\hs+1)D^{p_1}\T{s}\pui[p_1](1)\Big\}\Big) \nn\\
    &+\tQ_{s-k-2}(-1)^2\!\!\!\sum_{P=k-2} \!\!\pui[p_0]D^{p_0}\Big(\pui\Big\{C(\hs+1)\pui[p_1]D^{p_1}\Big(\pui\Big\{C(\hs+1)D^{p_2}\T{s}\pui[p_2](1)\Big\}\Big)\Big\}\Big) \nn\\
    &+\ldots+\tQ_{s-2k}(-1)^k\pui\Big\{C(\hs+1)\pui\Big\{C(\hs+1)\ldots \pui\Big\{C(\hs+1)\T{s}\Big\}\ldots\!\Big\}\Big\}\!\bigg)\! \label{intermrenorm2}\\
    &=\int_S\bigg((-1)^k\frac{u^k}{k!}D^k\tQ_{s-k}\T{s}+\!\!\sum_{P=k-1}\!(-1)^{p_0+1}D^{p_0}\tQ_{s-k-1}\pui[p_0-1]\Big(C(\hs+1)\pui[p_1](1)\Big)D^{p_1}\T{s} \nn\\
    &+\!\!\sum_{P=k-2}\!(-1)^{p_0+2}D^{p_0}\tQ_{s-k-2}\pui[p_0-1]\Big\{C(\hs+1)\pui[p_1]D^{p_1}\Big( \pui\Big\{C(\hs+1)D^{p_2}\T{s}\pui[p_2](1)\Big\}\Big)\Big\} \nn\\
    &+\ldots+(-1)^k\Big((\shs+3)\tQ_{s-2k}\Big)\pui\Big\{C\pui\Big\{C(\hs+1)\ldots \pui\Big\{C(\hs+1)\T{s}\Big\}\ldots\Big\}\Big\}\bigg), \nn
\end{align}
where in the last line we trade the action of $\hs$ for its dual.
Indeed, $(\hs+1)$ acts on the object $C^{k-1}\T{s}$ (the $\pui$ are irrelevant for counting the spin), so that
\begin{equation}
    (\hs+1)(C^{k-1}\T{s})=(s-2k+3)C^{k-1}\T{s}.
\end{equation}
On the other hand,
\begin{equation}
    (\shs+3)\tQ_{s-2k}=(s-2k+3)\tQ_{s-2k},
\end{equation}
so that the prefactors do match.
We proceed further,\footnote{When trading $\hs$ for $\shs$ in the first line for instance, we use that $(\hs+1)D^{p_1}\T{s}=(s-p_1+1)D^{p_1}\T{s}=(s+p_0-k+2)D^{p_1}\T{s}$, while $(\shs+3)D^{p_0}\tQ_{s-k-1}=(s+p_0-k+2)D^{p_0}\tQ_{s-k-1}$.}
\begin{align}
    \!\!\eqref{intermrenorm2} &=\int_S\bigg((-1)^k\frac{u^k}{k!}D^k\tQ_{s-k}\T{s}+(-1)^k\!\!\sum_{P=k-1}\!\!D^{p_1}\Big((\shs+3)D^{p_0}\tQ_{s-k-1}\pui[p_0-1]\Big\{C\pui[p_1](1)\Big\}\Big)\T{s} \nn\\
    &+\!\!\!\sum_{P=k-2}\!(-1)^{2+p_0+p_1}D^{p_1}\Big((\shs+3)D^{p_0}\tQ_{s-k-2}\pui[p_0-1]\Big\{C\Big)\pui[p_1-1]\Big\{C(\hs+1)D^{p_2}\T{s}\pui[p_2](1)\Big\}\Big\} \nn\\
    &+\ldots+(-1)^k(\shs+3)\Big((\shs+3)\tQ_{s-2k}\pui\Big\{C\Big)\pui\Big\{C\ldots\pui\Big\{C(\hs+1)\T{s}\Big\}\ldots\Big\}\Big\}\Bigg) \nn\\
    &=\int_S(-1)^k\bigg(\frac{u^k}{k!}D^k\tQ_{s-k}+\!\!\sum_{P=k-1}\!\!D^{p_1}\Big((\shs+3)D^{p_0}\tQ_{s-k-1}\pui[p_0-1]\Big\{C\Big)\pui[p_1](1)\Big\} \\
    &+\!\!\!\sum_{P=k-2}\!D^{p_2}\Big((\shs+3)D^{p_1}\Big((\shs+3)D^{p_0}\tQ_{s-k-2}\pui[p_0-1]\Big\{C\Big)\pui[p_1-1]\Big\{C\Big)\pui[p_2](1)\Big\}\Big\} +\ldots\nn\\
    &+(\shs+3)\Big(\ldots\Big((\shs+3) \Big((\shs+3) \tQ_{s-2k}\pui\Big\{C\Big)\pui\Big\{C\Big)\ldots C\Big)\pui\Big\{C\Big\}\ldots\Big\}\Big\}\Bigg)\T{s} \nn\\
    &=\int_S\bigg((-1)^k \sum_{\l=0}^k\sum_{P=k-\l}D^{p_0}\Big((\shs+3)D^{p_1}\Big((\shs+3)D^{p_2}\Big(\ldots \Big((\shs+3)\hS^{*\l}D^{p_\l}\tQ_{s-k}\cdot \nn\\
    &\hspace{4cm}\cdot\pui[p_\l-1]\Big\{C\Big)\ldots\Big)\pui[p_2-1]\Big\{C\Big)\pui[p_1-1]\Big\{C\Big)\pui[p_0](1)\Big\}\Big\}\ldots\Big\}\bigg)\T{s}. \nn
\end{align}
Most of the work is now done.
To relate this expression to $\cq^u_{\T{s}}$, we need two slight modifications.
Firstly, as we discussed already in several occasions, adding the term $Q_{-1}^u[\t{-1}]$ amounts to replacing every appearance of $\tQ_0$ by $\tQ_0-\bN C$.
For that purpose, we introduce $\tQcal_s\equiv \tQ_s$, for $s>0$ and $\tQcal_0=\tQ_0-\bN C$.
Secondly, according to \eqref{changesum}, the variable $n$ had to be positive.
The easiest way to implement this restriction is to take $\tQcal_s\equiv 0$, for $s<0$.
Finally, taking the sum $\sum_{k=0}^s$, we get the expression for $\tq_s$, namely
\vspace{-0.6cm}

\begin{adjustwidth}{-0.1cm}{0cm}
\begin{empheq}[box=\fbox]{align}
    \tq_s=\sum_{k=0}^s(-1)^k\sum_{\l=0}^k &\sum_{P=k-\l} \!D^{p_0}\Big((\shs+3) D^{p_1}\Big((\shs+3)D^{p_2}\Big(\!\ldots\Big((\shs+3)\hS^{*\l} D^{p_\l}\widetilde\Qcal_{s-k}\cdot \nn\\
    &\cdot\pui[p_\l-1]\Big\{C\Big)\ldots\Big)\pui[p_2-1]\Big\{C\Big)\pui[p_1-1]\Big\{C\Big)\pui[p_0](1)\Big\}\Big\}\ldots\Big\}. \label{deftqs}
\end{empheq}
\end{adjustwidth}

This proof is straightforwardly amended if we wish to work with the covariant charge $\Ht$ instead.
Indeed, since the element $\T{-1}$ now behaves on par with $\T0,\T1,\ldots$, we just have to consider
\begin{equation}
    \cH^{\,u}_{\T{s}}= \sum_{n=-1}^s\sum_{k=\lfloor\frac{s+1-n}{2}\rfloor}^{s-n}\int_S\tQ_n\left(\pui D-\pui C(\hs+1)\hS\right)^k\T{n+k},
\end{equation}
and then notice that
\begin{equation}
    \sum_{n=-1}^s\sum_{k=\lfloor\frac{s+1-n}{2}\rfloor}^{s-n}=\sum_{k=0}^{s+1}\sum_{-1\leqslant n=s-2k}^{s-k}. \label{changesumH}
\end{equation}
We thus find that 
\begin{equation} \label{RenormChargeCov}
    \cH^{\,u}_{\T{s}}=\frac{1}{4\pi G}\int_S\th_s\T{s},\qquad s\geqslant -1,
\end{equation}
for $\th_s$ given by the formula \eqref{deftqs} where $\tQcal_s$ is now defined as $\tQcal_s\equiv\tQ_s, \,s\geqslant -1$ and $\tQcal_s=0,s<-1$.
As an example,
\begin{align}
    \th_{-1} &=\tQ_{-1},\nn\\
    \th_0 &=q_0-uD\tQ_{-1}, \nn\\
    \th_1 &=q_1+\frac{u^2}{2}D^2\tQ_{-1}-2\tQ_{-1}\pui C, \\
    \th_2 &=q_2-\frac{u^3}{3!}D^3\tQ_{-1}+3D\tQ_{-1}\pui[2]C+2D\big(\tQ_{-1}\pui(uC)\big)-3\tQ_0\pui C, \nn
\end{align}
where $q_s$ is given in \eqref{defqs}.
The reader can check that $\pa_u\th_s=0$ when $\dot\bN=0$, as expected.
Of course, the associated Noether charge is defined as 
\begin{equation}
    \cH_{\T{s}}=\frac{1}{4\pi G}\int_S\hh_s\T{s},\qquad s\geqslant -1,
\end{equation}
with $\hh_s= \lim_{u\to -\infty}\th_s$.
\medskip

Furthermore, it is interesting to consider the special case of \eqref{deftqs} where $N=0$ in a strip $u\in[0,u_0]$, namely a non-radiative strip.
The renormalized charge aspect then takes a very compact form.
Indeed, first notice that in this strip, $C=C\big|_{u=0}=\sigma$ and the shear goes through the inverse $u$ derivatives\footnote{Where $\pui=\int_0^u$, with $u\in[0,u_0]$.} of \eqref{deftqs}.
The latter then reduces to 
\begin{equation}
    \tq_s=\!\sum_{k=0}^s\frac{(-u)^k}{k!}\sum_{\l=0}^k \sum_{P=k-\l} D^{p_0}\Big(\sigma(\shs+3)D^{p_1}\Big(\sigma(\shs+3)D^{p_2}\Big(\!\ldots\!\Big(\sigma(\shs+3)\hS^{*\l} D^{p_\l}\widetilde\Qcal_{s-k}\Big)\ldots\Big)\!\Big)\!\Big),
\end{equation}
where we used that
\begin{equation}
    \pui[p_\l-1]\cdots\pui[p_1-1]\pui[p_0](1)=\pui[(P+\l)](1)=\pui[k](1)=\frac{u^k}{k!}.
\end{equation}
Defining the dual operator $\Dcal^*$ acting on $\tQ$ as
\begin{equation}
    \big(\Dcal^*\tQ\big)_s=D\tQ_{s-1}+(s+1)C\tQ_{s-2}=\big(D+C(\shs+3)\shS\big)\tQ_{s-1},
\end{equation}
we conclude using formula \eqref{puiDcalk} that when $N=0$ in a strip of $\scri$, then the renormalized charge aspect inside this interval is given by \eqref{deftqs0rad}
\begin{equation}
    \boxed{\tq_s=\sum_{k=0}^s\frac{(-u)^k}{k!}\big(\Dcal^{\ast k}\widetilde\Qcal\big)_s}.
\end{equation}

\section{Proofs of Lemmas \hyperref[deltatildeHard]{[Soft and Hard actions]} \label{AppSoftHardAction}}

We start with few preliminary computations. 
From \eqref{soltaupTs}, written again here,
\begin{equation}
    \t{p}(\T{s}) =\frac{u^{s-p}}{(s-p)!}D^{s-p}\T{s}-\sum_{n=p+2}^s(n+1)\big(\pui D\big)^{n-p-2}\pui (C\t{n}),
\end{equation}
we find in particular that
\begin{equation}
    \soft{\t{p}}(\T{s})= \frac{u^{s-p}}{(s-p)!}D^{s-p}\T{s},\qquad 0\leqslant p\leqslant s.
\end{equation}
Hence
\begin{equation}
    \soft{\t0}(\T{s})=\frac{u^s}{s!}D^s\T{s}\qquad\textrm{and}\qquad \soft{\t1}(\T{s})=\frac{u^{s-1}}{(s-1)!}D^{s-1}\T{s}. \label{soltau01Soft}
\end{equation}
Moreover
\begin{equation}
    \hard{\t0}(\T{s})=-\sum_{n=2}^s(n+1)\big(\pui D\big)^{n-2}\pui \left(C\soft{\t{n}}\right),\qquad s\geqslant 2.
\end{equation}
Using the generalized Leibniz rule for pseudo-differential calculus \cite{PseudoDiffBakas},
\begin{equation}
    \pui[\alpha](fg)=\sum_{n=0}^\infty\frac{(-\alpha)_n}{n!}(\pa_u^nf)\pui[(n+\alpha)]g, \qquad \alpha\in\R, \label{LeibnizRuleGeneral}
\end{equation}
which implies in particular that
\begin{equation}
    \pui[(n-1)]\left(\frac{u^{s-n}}{(s-n)!}C\right)=\sum_{k=0}^{s-n}(-1)^k\binom{n+k-2}{k} \frac{u^{s-n-k}}{(s-n-k)!}\pui[(k+n-1)]C,
\end{equation}
we obtain
\begin{align}
     \hard{\t0}(\T{s}) &=-\sum_{n=2}^s\sum_{k=0}^{s-n}(-1)^k\binom{n+k-2}{k}(n+1)\frac{u^{s-n-k}}{(s-n-k)!}D^{n-2}\Big(D^{s-n}\T{s}\,\pui[(k+n-1)]C\Big) \nn\\
     &=-\sum_{p=2}^s\sum_{k=0}^{p-2}(-1)^k\binom{p-2}{k}(p-k+1)\frac{u^{s-p}}{(s-p)!}D^{p-k-2}\Big(D^{s-p+k}\T{s} \,\pa_u^{1-p}C\Big), \label{tau0HardTs}
\end{align}
where we changed variable $p=n+k$ in the last step.

\subsection{Proof of Lemma [Soft action]}

Using \eqref{deltatauSoft}, combined with \eqref{soltau01Soft} and the definition \eqref{defdeltaTs}, we infer that
\begin{equation}
    \Sd{s}{\T{s}}C=-D^2\soft{\t0}(\T{s})=-\frac{u^s}{s!}D^{s+2}\T{s}
\end{equation}
and in particular
\begin{equation}
    \Sd{0}{\T0}C=-D^2\T0\equiv\Std{0}{\T0}C.
\end{equation}
Therefore
\begin{equation}
    \boxed{\Sd{s}{\T{s}}C=\frac{u^s}{s!}\Std{0}{D^s\T{s}}C}.
\end{equation} 

\subsection{Proof of Lemma [Hard action][Part 1]}

The simplest way to prove \eqref{deltatildeHard} is to expand the result and check that it matches with \eqref{deltatauHard} (combined with \eqref{defdeltaTs}).
For $\alpha=\max[\mathrm{mod}_2(p),p-2]$ we have that\footnote{Recall that the falling factorial satisfies $(x)_0=1$.}
\begin{align}
    \Hd{s}{\T{s}}C &=\sum_{p=0}^s\frac{u^{s-p}}{(s-p)!}\sum_{k=0}^{\alpha}(-1)^k\frac{(p-2)_k}{k!}(p-k+1) D^{p-k}\big(D^{s-p+k}\T{s}\,\pa_u^{1-p}C\big) \nn\\
    &=\sum_{p=2}^s\frac{u^{s-p}}{(s-p)!}\sum_{k=0}^{p-2}(-1)^k\binom{p-2}{k}(p-k+1) D^{p-k}\big(D^{s-p+k}\T{s}\,\pa_u^{1-p}C\big) \nn\\
    &+\frac{u^s}{s!}D^s\T{s}\,\pa_u C+\frac{u^{s-1}}{(s-1)!}\Big(2D\big(D^{s-1}\T{s}C\big)-(-1)_1D^s\T{s}C\Big) \\
    &=-D^2\hard{\t0}(\T{s})+N\soft{\t0}(\T{s}) +2DC\frac{u^{s-1}}{(s-1)!} D^{s-1}\T{s} +3C\frac{u^{s-1}}{(s-1)!}D^s\T{s}  \nn\\
    &=-D^2\hard{\t0}(\T{s})+N\soft{\t0}(\T{s})+2DC\soft{\t1}(\T{s})+3CD\soft{\t1}(\T{s}), \nn
\end{align}
as desired.

\subsection{Proof of Lemma [Hard action][Part 2]}

We have to distinguish the special cases $s=0$ and 1.
\medskip

$\bullet$ \textbf{Spin 0:} 
\begin{equation}
    \boxed{\Hd{0}{\T0}C=\T0\pa_u C\equiv\Htd{0}{\T0}C}. \label{deltatildeHard0}
\end{equation}

$\bullet$ \textbf{Spin 1:} 
\begin{equation}
    \Hd{1}{\T1}C= D\T1\big(u\pa_u+3\big)C+2\T1DC\equiv u\,\Htd{0}{D\T1}C+\Htd{1}{\T1}C,
\end{equation}
where we identify
\begin{equation}
    \boxed{\Htd{1}{\T1}C=2\T1 DC+3CD\T1}. \label{deltatildeHard1}
\end{equation}
The equations \eqref{deltatildeHard0} and \eqref{deltatildeHard1} indeed agree with the general formula \eqref{deltatildeHardbis}.
\medskip

$\bullet$ \textbf{Spin $p\geqslant 2$:} 
\medskip

Next let us massage \eqref{deltatildeHardb} for $p\geqslant 2$:
\begin{align}
    \Htd{p}{\T{p}}C &=\sum_{k=0}^{p-2}(-1)^k\frac{(p-2)_k}{k!}(p-k+1) D^{p-k}\big(D^k\T{p}\,\pa_u^{1-p}C\big) \nn\\
    &=\sum_{k=0}^{p-2}\sum_{n=0}^{p-k}(-1)^k\frac{(p-2)_k}{k!}\frac{(p-k)_n}{n!}(p-k+1) D^{k+n}\T{p}\,D^{p-k-n}\pa_u^{1-p}C. \label{intermdeltatilde}
\end{align}
We can change variable $k+n\to m$ such that the sums schematically become
\begin{equation}
    \sum_{k=0}^{p-2}\sum_{n=0}^{p-k}= \sum_{m=2}^p\sum_{n=2}^m+\sum_{m=1}^{p-2}\sum_{n=0}^1+\big[k=0;n=0\big] +\big[k=p-2;n=1\big].
\end{equation}
Explicitly, \eqref{intermdeltatilde} turns into
\begin{align}
    \Htd{p}{\T{p}}C &= \sum_{m=2}^p D^m\T{p}\,D^{p-m}\pa_u^{1-p}C\sum_{n=2}^m (-1)^{m-n}\frac{(p-2)_{m-n}}{(m-n)!}\frac{(p+n-m)_n}{n!}(p+n-m+1) \nn\\
    &+ \sum_{m=1}^{p-2}D^m\T{p}\,D^{p-m}\pa_u^{1-p}C\sum_{n=0}^1 (-1)^{m-n}\frac{(p-2)_{m-n}}{(m-n)!}\frac{(p+n-m)_n}{n!}(p+n-m+1) \nn\\
    &+(p+1)\T{p}\,D^p\pa_u^{1-p}C+6(-1)^p D^{p-1}\T{p}\,D\pa_u^{1-p}C.
\end{align}
We then split the first line into three parts: $[m=p-1]+[m=p]+\sum_{m=2}^{p-2}$. 
Similarly we split the second line into two terms: $[m=1]+\sum_{m=2}^{p-2}$. 
Therefore,
\begin{align}
    \Htd{p}{\T{p}}C &=\sum_{m=2}^{p-2} D^m\T{p}\,D^{p-m}\pa_u^{1-p}C\sum_{n=0}^m (-1)^{m-n}\frac{(p-2)_{m-n}}{(m-n)!}\frac{(p+n-m)_n}{n!}(p+n-m+1) \nn\\
    &-D^{p-1}\T{p}\,D\pa_u^{1-p}C\sum_{n=2}^{p-1}(-1)^{p-n}\frac{(p-2)_{p-1-n}}{(p-1-n)!}(n+1)(n+2) \nn\\
    &+D^p\T{p}\,\pa_u^{1-p}C\sum_{n=2}^p (-1)^{p-n}\frac{(p-2)_{p-n}}{(p-n)!}(n+1) \label{intermdeltatilde2}\\
    &-D\T{p}\,D^{p-1}\pa_u^{1-p}C\sum_{n=0}^1(-1)^n(p-2)_{1-n}(p+n-1)_n(p+n) \nn\\
    &+(p+1)\T{p}\,D^p\pa_u^{1-p}C+6(-1)^p D^{p-1}\T{p}\,D\pa_u^{1-p}C, \nn
\end{align}
where for now we only simplified trivial factorial factors.
In the first line, we have to deal with the expression
\begin{align}
    &\sum_{n=0}^m (-1)^{m-n}\frac{(p-2)_{m-n}}{(m-n)!}\frac{(p+n-m)_n}{n!}(p+n-m+1) \nn\\
    =&\sum_{n=0}^m (-1)^n\binom{m}{n}\frac{(p-2)!(p-n+1)!}{m!(p-2-n)!(p-m)!} \nn\\
    =& ~(p+1)\binom{p}{m}\sum_{n=0}^m (-1)^n\binom{m}{n}\frac{(p-2)_n}{(p+1)_n} \nn\\
    =& ~(p+1)\binom{p}{m} {}_2F_1(-m,2-p;-(p+1);1) \nn\\
    =& ~(p+1)\binom{p}{m}\frac{(3)_m}{(p+1)_m} \nn\\
    =& ~(p+1-m)\frac{(3)_m}{m!} =\left\{
    \begin{array}{lc}
       0 & \textrm{if } m>3, \\
       (p+1-m)\binom{3}{m}  & \textrm{if } m\leqslant 3,
    \end{array}\right.
\end{align}
where we used the fact that the hyper-geometric function (written in falling factorial convention using $(n)^{(k)}=(-1)^k(-n)_k$) simplifies to---cf. formula 7.3.5.4 in \cite{HyperGmathbook}---
\begin{equation}
    {}_2F_1(-m,-b;-c;1)=\frac{(c-b)_m}{(c)_m}.
\end{equation}
Then notice that in the penultimate line of \eqref{intermdeltatilde2},
\begin{equation}
    -\sum_{n=0}^1(-1)^n(p-2)_{1-n}(p+n-1)_n(p+n)=3p.
\end{equation}
Moreover, the second line of \eqref{intermdeltatilde2} recombines with the term proportional to 6 in the last line of the same equation:
\begin{align}
    \hspace{-0.3cm}-\sum_{n=2}^{p-1}(-1)^{p-n}\frac{(p-2)_{p-1-n}}{(p-1-n)!}(n+1)(n+2)+6(-1)^p &=\sum_{n=0}^{p-2} (-1)^{p-n}\binom{p-2}{n}(n+2)(n+3) \nn\\
    &=6(-1)^p{}_2F_1\big(-(p-2),4;2;1\big) \nn\\
    &=6(-1)^p\frac{(2)_{p-2}}{(-2)_{p-2}} \nn\\
    &=6\frac{(2)_{p-2}}{(p-1)!}=\left\{
    \begin{array}{ll}
       \!6 & \textrm{if ~} p=2,3, \\
       \!2 & \textrm{if ~} p=4, \\
       \!0 & \textrm{if ~} p\geqslant 5.
    \end{array}\right.\!\!\!\!
\end{align}
Concerning the third line of \eqref{intermdeltatilde2},
\begin{align}
    \sum_{n=2}^p (-1)^{p-n}\frac{(p-2)_{p-n}}{(p-n)!}(n+1) &=\sum_{n=0}^{p-2}(-1)^{p-n}\binom{p-2}{n}(n+3) \nn\\
    &=3(-1)^p{}_2F_1\big(-(p-2),4;3;1\big) \nn\\
    &=3(-1)^p\frac{(1)_{p-2}}{(-3)_{p-2}} \nn\\
    &=6\frac{(1)_{p-2}}{p!}=\left\{
    \begin{array}{ll}
       3 & \textrm{if ~} p=2, \\
       1 & \textrm{if ~} p=3, \\
       0 & \textrm{if ~} p\geqslant 4.
    \end{array}\right.
\end{align}
Gathering those simplifications, \eqref{intermdeltatilde2} reduces to\footnote{$\delta_{p,n}$ is just the Kronecker symbol.}
\begin{align}
    \Htd{p}{\T{p}}C &=\sum_{m=2}^{\min[3,p-2]} \binom{3}{m}(p+1-m)D^m\T{p}\,D^{p-m}\pa_u^{1-p}C \nn\\
    &+3p\,D\T{p}\,D^{p-1}\pa_u^{1-p}C+(p+1)\T{p}\,D^p\pa_u^{1-p}C +\delta_{p,4}\big(2D^3\T4 D\pui[3]C\big)\nn\\
    &+\delta_{p,2}\big(6D\T2 D\pui C+3D^2\T2\pui C\big)+\delta_{p,3}\big(6D^2\T3 D\pui[2]C+D^3\T3\pui[2]C\big) \nn\\
    &= \sum_{m=0}^{\min[3,p]} \binom{3}{m}(p+1-m)D^m\T{p}\,D^{p-m}\pa_u^{1-p}C,
\end{align}
which is indeed \eqref{deltatildeHardbis}.

\chapter{Einstein-Yang-Mills Theory \label{AppEYM}}

\section{Closure of the $\sfST$-bracket \label{App:dualEOMEYM}}

In this appendix, we prove that
\begin{align} \label{force1}
    \pa_u\big\lbr\Pta,\Ptap\big\rbr^\ym =\DTOT\big\lbr\Pta,\Ptap\big\rbr.
\end{align}
For this, we shall take the time derivative of the algebroid YM-bracket.
This is a brute force computation, so we simply expand most of the terms to then reconstruct the RHS of \eqref{force1}, i.e.
\begin{align}
    \DTOT\big\lbr\Pta,\Ptap\big\rbr
    =\DTOT\big[\Pta,\Ptap \big] +\DTOT\big(\dPtap\Pta-\dPta\Ptap\big),
\end{align}
where
\begin{align}
    \Big(\DTOT\big[\Pta,\Ptap\big]\Big)_s &=\DYM\big[ \Pta,\Ptap\big]^\ym_{s+1} -(s+2)C\big[ \Pta,\Ptap\big]^\ym_{s+2}-(s+2)F[\tau,\tau']^C_{s+1}.
\end{align}
Let us proceed. 
In the following, for shortness, it is understood that equalities are valid upon anti-symmetrization of the RHS. 
Using the form \eqref{YMbracketCFdef} of the YM-bracket, we get that
\begin{align}
    &\pa_u \bigg\{\bigg\{ \sum_{n=0}^s(n+1) \Big(\t{n}\DYM\ap{s+1-n}+\a{n+1}D\tp{s-n}\Big)-(s+2)C \big(\t0\ap{s+2}+\a1\tp{s+1}\big)\bigg\}-\Pta\leftrightarrow\Ptap\bigg\}= \nn\\
    &=\sum_{n=0}^s(n+1)\Big\{\big(\DGR\tau\big)_n\DYM\ap{s+1-n}+\t{n}\DYM\big(\DTOT\Ptap\big)_{s+1-n}+\t{n}\big[F,\ap{s+1-n}\big]_\g \nn\\
    &\qquad\qquad\qquad+\big(\DTOT\Pta \big)_{n+1} D\tp{s-n}+\a{n+1}D\big(\DGR \tp{}\big)_{s-n}\Big\} -(s+2)N\big(\t0 \ap{s+2}+\a1\tp{s+1}\big)\nn\\
    &-(s+2)C\Big\{\big(\DGR\tau\big)_0\ap{s+2} +\t0\big(\DTOT\Ptap\big)_{s+2}+\big(\DTOT \Pta\big)_1\tp{s+1}+\a1\big(\DGR\tp{}\big)_{s+1}\Big\} \nn\\
    &=\sum_{n=0}^s(n+1)\Big\{D\t{n+1}\DYM \ap{s+1-n}+\t{n}\DYM^2\ap{s+2-n} +\DYM\a{n+2}D\tp{s-n}+\a{n+1}D^2\tp{s+1-n} \nn\\
    &\qquad\qquad\qquad -(n+3)C\t{n+2}\DYM\ap{s+1-n}-(s+3-n)DC\t{n}\ap{s+3-n}-(s+3-n)C\t{n}\DYM\ap{s+3-n} \nn\\
    &\qquad\qquad\qquad -(n+3)C\a{n+3} D\tp{s-n}-(s+3-n)DC\a{n+1}\tp{s+2-n}-(s+3-n)C \a{n+1}D\tp{s+2-n} \nn\\*
    &\qquad\qquad\qquad -(s+3-n)\t{n} \DYM\big(F\tp{s+2-n}\big) +\t{n}\big[F,\ap{s+1-n}\big]_\g -(n+3)F\t{n+2}D\tp{s-n}\Big\} \nn\\*
    &-(s+2)C\Big\{\big(\DGR\tau\big)_0 \ap{s+2}+\t0\DYM\ap{s+3}-(s+4)C\t0\ap{s+4}-(s+4)F\t0\tp{s+3} \nn\\
    &\qquad\qquad\qquad +\DYM\a2\tp{s+1}-3C\a3\tp{s+1}-3F\t2\tp{s+1}+\a1D\tp{s+2}-(s+4)C\a1\tp{s+3}\Big\} \nn\\
    &-(s+2)N\big(\t0\ap{s+2}+\a1\tp{s+1}\big) \nn\\
    &=\DYM\left(\sum_{n=0}^{s+1}(n+1)\t{n}\DYM\ap{s+2-n}\right)-\sum_{n=0}^{s+1}(n+1)\underline{ D\t{n}\DYM\ap{s+2-n}}-(s+2)\dashuline{\t{s+1}\DYM^2\ap1} \nn\\
    &+\DYM\left(\sum_{n=0}^{s+1}(n+1)\a{n+1}D\tp{s+1-n}\right)-\sum_{n=0}^{s+1}(n+1)\underline{ \DYM\a{n+1}D\tp{s+1-n}}-(s+2)\uwave{\a{s+2}D^2\tp0} \nn\\
    &+\sum_{n=0}^s(n+1)\underline{ D\t{n+1}\DYM\ap{s+1-n}}+\sum_{n=0}^s(n+1)\underline{\DYM\a{n+2}D\tp{s-n}} \nn\\
    &-(s+3)\DYM\big(C\t0\ap{s+3}\big) +\aunderbrace[l1r]{C\t0\DYM\ap{s+3}}+(s+3)\aunderbrace[l1r]{DC\t0\ap{s+3}}+(s+3)\aunderbrace[l1r]{CD\t0\ap{s+3}}\nn\\
    &-(s+3)\DYM\big(C\a1\tp{s+2}\big) +\aunderbrace[l1r]{C\a1D\tp{s+2}} +(s+3)\aunderbrace[l1r]{C\tp{s+2}\DYM\a1}+(s+3)\aunderbrace[l1r]{DC\a1\tp{s+2}} \nn\\
    &-(s+3)\DYM\big(F\t0\tp{s+2}\big)+(s+3) \aunderbrace[l3*{4}{01}03r]{F\t0 D\tp{s+2}}+(s+3)\aunderbrace[l3*{5}{01}03r]{\DYM (F\t0)\tp{s+2}} \nn\\
    &-(s+2)C\sum_{n=0}^{s+2}(n+1)\t{n}\DYM\ap{s+3-n}+(s+2)C\sum_{n=0}^{s+2}(n+1)\uuline{\t{n}\DYM\ap{s+3-n}} \nn\\
    &-C\sum_{n=0}^s(n+1)(s+3-n)\uuline{\t{n}\DYM\ap{s+3-n}}-C\sum_{n=0}^s(n+1)(n+3)\uuline{\t{n+2}\DYM\ap{s+1-n}} \nn\\
    &-(s+2)C\sum_{n=0}^{s+2}(n+1)\a{n+1}D\tp{s+2-n}+(s+2)C\sum_{n=0}^{s+2}(n+1) \dotuline{\a{n+1}D\tp{s+2-n}} \nn\\
    &-C\sum_{n=0}^s(n+1)(n+3)\dotuline{\a{n+3}D\tp{s-n}}-C\sum_{n=0}^s(n+1)(s+3-n)\dotuline{\a{n+1}D\tp{s+2-n}} \nn\\
    &-\sum_{n=0}^s(n+1)(s+3-n)\underbrace{DC\t{n}\ap{s+3-n}}-\sum_{n=0}^s(n+1)(s+3-n)\underbrace{DC\a{n+1}\tp{s+2-n}} \nn\\
    &+(s+2)(s+4)C^2\big(\t0\ap{s+4}+\a1 \tp{s+3}\big)+(s+2)(s+4)CF\t0\tp{s+3} \nn\\
    &-(s+2)\uwave{C\ap{s+2}\big(\DGR\t{}} \big)_0-(s+2)\uwave{N\ap{s+2}\t0} \nn\\
    &-(s+2)C\dashuline{\tp{s+1}\DYM\a2}+ 3(s+2)\dashuline{C^2\a3\tp{s+1}}+3(s+2)\dashuline{CF\t2\tp{s+1}} -(s+2)\dashuline{N\a1\tp{s+1}} \nn\\
    &+\sum_{n=0}^s(n+1)\Big\{ \t{n}\big[F,\ap{s+1-n}\big]_\g-(s+3-n)\aunderbrace[l3*{5}{01}03r]{\t{n} \DYM\big(F\tp{s+2-n}}\big) -(n+3)\aunderbrace[l3*{5}{01}03r]{F\t{n+2}D\tp{s-n}}\Big\}. \nn
\end{align}

\ni Moreover, the time derivative of the YM Dali-bracket is simply
\begin{align}
    -\pa_u\big(F\paren{\t{},\tp{}}^\ym_s\big)
    &=-(s+2)\dashuline{\dot F\t0\tp{s+1}} -(s+2)\dashuline{F\tp{s+1}\big(\DGR\tau}\big)_0-(s+2)\aunderbrace[l3*{5}{01}03r]{F\t0 D\tp{s+2}} \nn\\*
    &+(s+2)(s+4)CF\t0\tp{s+3}.
\end{align}
Next, it is not hard to check that the \underline{underlined} sums add up to 0 upon anti-symmetrization.
Moreover, the wavy underlined terms sum up to
\begin{equation}
    \uwave{~\cdots~}=-(s+2)\ap{s+2}\dt C+2(s+2)\aunderbrace[l1r]{DC\ap{s+2}\t1}+2(s+2)\aunderbrace[l1r]{C\ap{s+2}D\t1},
\end{equation}
while for the dashed underlined terms, we get
\begin{equation}
    \dashuline{~\,\cdots\,~}=-(s+2)\tp{s+1}\dPta F-(s+2)\tp{s+1}[F,\a0]_\g+2(s+2)\tp{s+1} \DYM\big(\aunderbrace[l1r]{C\a2}+\aunderbrace[l*{2}{01}0r]{F\t1}\big),
\end{equation}
where
\begin{align}
    \dPta F=\pa_u\big(\dPta A\big) &=-\DYM^2\a1+3C\DYM\a2+ 2DC\a2-3C^2\a3-[F,\a0]_\g+2\DYM F\t1 \nn\\
    &+3FD\t1+N\a1-6CF\t2+\dot F\t0.
\end{align}
Besides, the double underlined sums simplify to
\begin{equation}
    \uuline{~\cdots~}=(s+3)\aunderbrace[l1r]{C\t{s+2}\DYM\ap1}+ 2(s+2)\aunderbrace[l1r]{C\t{s+1}\DYM\ap2}-\aunderbrace[l1r]{C\t0\DYM\ap{s+3}},
\end{equation}
while the dotted ones reduce to
\begin{equation}
    \dotuline{~\cdots~}=2(s+2)\aunderbrace[l1r]{C\a{s+2}D\tp1}+(s+3)\aunderbrace[l1r]{C\a{s+3}D\tp0}-\aunderbrace[l1r]{C\a1 D\tp{s+2}}.
\end{equation}
Furthermore, the under-braced sums contribute as
\begin{equation}
    \underbrace{~\cdots~}=(s+3)DC\big(\aunderbrace[l1r]{\a{s+3}\tp0}-\aunderbrace[l1r]{\a1\tp{s+2}}\big)+2(s+2)DC\big( \aunderbrace[l1r]{\a{s+2} \tp1}-\aunderbrace[l1r]{\a2\tp{s+1}}\big).
\end{equation}
Finally, all the terms $\aunderbrace[l1r]{\,\cdots\,}$ cancel one another (upon anti-symmetrization) and the dashed under-braced contributions take the form
\begin{equation}
    \aunderbrace[l1*{2}{01}01r]{~\cdots~}=-(s+2)F\sum_{n=0}^{s+2}(n+1)\t{n}D\tp{s+2-n}.
\end{equation}
Therefore, gathering the factors that form the YM- and $C$-brackets, we obtain that
\begin{align} \label{force2}
    &\quad \,\pa_u\Big([\Pta,\Ptap]^\ym-[\a{},\ap{}]^\g\Big)_s \\
    &-\Big\{\DYM\Big([\Pta,\Ptap]^\ym-[\a{},\ap{}]^\g\Big)_{s+1} -(s+2)C\Big([\Pta,\Ptap]^\ym-[\a{},\ap{}]^\g\Big)_{s+2}-(s+2)F[\tau,\tau']^C_{s+1}\Big\} \nn\\
    &=\sum_{n=0}^s(n+1)\dashuline{\t{n} \big[F,\ap{s+1-n}}\big]_\g \!-(s+2)\dashuline{\tp{s+1}\mkern-2mu \big[F,\a0}\big]_\g\!-(s+2)\ap{s+2}\dt C\mkern-2mu-(s+2)\tp{s+1}\dPta F. \nn
\end{align}
On the other hand the time derivative of $[\a{},\ap{}]^\g$ is given by (using \eqref{LeibnizAnomalyDYM})
\begin{align} \label{force3}
    \pa_u[\a{},\ap{}]^\g_s -\DYM[\a{},\ap{}]^\g_{s+1} &=-\big[\dotuline{\DYM\a0,\ap{s+1}}\big]_\g-C\sum_{n=0}^s(n+2)\underline{ \big[\a{n+2},\ap{s-n}}\big]_\g \\
    &-\sum_{n=0}^s(n+2)\dashuline{\t{n+1} \big[F,\ap{s-n}}\big]_\g. \nn
\end{align}
The dashed underlined terms in \eqref{force2} and \eqref{force3} recombine into
\begin{equation} \label{force4}
    \dashuline{~\,\cdots\,~}=\t0 \big[\dotuline{F,\ap{s+1}}\big]_\g.
\end{equation}
Furthermore, writing the anti-symmetrization of the \underline{sum} proportional to $C$ explicitly, we notice that
\begin{align} 
    \underline{~\cdots~} &=-C\sum_{n=0}^s(n+2) \big[\a{n+2},\ap{s-n}\big]_\g+C \sum_{n=0}^s(n+2)\big[\ap{n+2},\a{s-n}\big]_\g \nn\\
    &=C\left\{-\!\sum_{n=-1}^s(n+2) \big[\a{n+2},\ap{s-n}\big]_\g+ \!\sum_{n=-1}^s(n+2)\big[\ap{n+2},\a{s-n}\big]_\g +\big[\a1,\ap{s+1}\big]_\g-\big[\ap1,\a{s+1}\big]_\g\right\} \nn\\
    &=C\left\{\sum_{n=-2}^{s-1}(s-n)\big[\ap{n+2},\a{s-n}\big]_\g+ \!\sum_{n=-1}^s(n+2)\big[\ap{n+2},\a{s-n}\big]_\g +\big[\a1,\ap{s+1}\big]_\g-\big[\ap1,\a{s+1}\big]_\g\right\} \nn\\
    &=C\left\{(s+2)\sum_{n=-2}^s \big[\ap{n+2},\a{s-n}\big]_\g +\big[\a1,\ap{s+1}\big]_\g-\big[\ap1,\a{s+1}\big]_\g\right\} \nn\\
    &=C\left\{-(s+2)[\a{},\ap{}]^\g_{s+2} +\big[\dotuline{\a1,\ap{s+1}}\big]_\g-\big[\dotuline{\ap1,\a{s+1}}\big]_\g \right\}. \label{force5}
\end{align}
The dotted underlined factors in \eqref{force3}, \eqref{force4} and \eqref{force5} simply add up to
\begin{equation}
    \dotuline{~\cdots~}=\big[\dPta A,\ap{s+1}\big]_\g.
\end{equation}
Hence
\begin{align}
    &\quad~\pa_u \big[\Pta,\Ptap\big]^\ym_s-\DYM \big[\Pta,\Ptap\big]^\ym_{s+1}+(s+2)C\big[\Pta,\Ptap\big]^\ym_{s+2}+(s+2)F[\t{},\tp{}]^C_{s+1} \nn\\
    &=\big[\dPta A,\ap{s+1}\big]_\g-(s+2)\ap{s+2}\dt C-(s+2)\tp{s+1}\dPta F. \label{force6}
\end{align}
Finally, the last contribution comes from the anchor map, for which we get
\begin{align} \label{force7}
    \pa_u\big(\dPtap\a{s}-\dPta\ap{s}\big)&=-\dPta\big(\DTOT\Ptap\big)_s \\
    &=-\big(\DTOT(\dPta\Ptap)\big)_s-\big[\dPta A,\ap{s+1}\big]_\g+(s+2)\ap{s+2}\dt C+(s+2)\tp{s+1}\dPta F. \nn
\end{align}
Summing \eqref{force6} and \eqref{force7}, we conclude that, cf. \eqref{dualEOMEYMbracket},
\begin{equation}
    \pa_u \big\lbr\Pta,\Ptap \big\rbr^\ym_s=\Big(\DTOT\big\lbr\Pta,\Ptap \big\rbr\Big)_s.
\end{equation}

\newpage
\section{Proof of the algebroid action \eqref{AlgebroidActionEYM1} \label{AppEYM:AlgAct}}

Firstly, using the variation \eqref{deltaPtaCA} and the dual EOM \eqref{DualEOMEYM}, we compute
\bs
\begin{align}
    \dPta\dPtap A &=\dPta\Big(-\DYM\ap0+C\ap1+F\tp0\Big) \nn\\
    &=-\DYM\big(\dPta\ap0\big)-\big[-\DYM\a0+C\a1+F\t0,\ap0\big]_\g+C\dPta\ap1+F\dt\tp0 \nn\\
    &+\ap1\Big(N\t0-(\DGR{}^2\t{})_{-2}\Big) +\tp0\pa_u\Big(-\DYM\a0+C\a1+F\t0\Big) \nn\\
    &=-\DYM\big(\dPta\ap0\big)+\big[\ap0,-\DYM\a0\big]_\g +C\big(\dPta\ap1+[\ap0,\a1]_\g+\tp0(\DTOT\Pta)_1\big) \label{EYMline1}\\
    &-\t0[F,\ap0]_\g-\tp0[F,\a0]_\g+N\ap1\t0+N\a1\tp0+\dot F\t0\tp0 \label{EYMline2}\\
    &-\ap1\Big(D(\DGR\tau)_{-1}- C(\DGR\tau)_0\Big)-\tp0\DYM\big(\DTOT\Pta\big)_0 \label{EYMline3}\\
    &+F\big(\dt\tp0 +\tp0(\DGR\tau)_0\big). \label{EYMline4}
\end{align}
\es
Upon anti-symmetrization between $\Pta$ and $\Ptap$, the line \eqref{EYMline2} cancels out.
Moreover, we can easily identify the appearance of $[\ap{},\a{}]^\g$ and $\lbr\tp{},\t{}\rbr$.
Indeed, the anti-symmetrization of \eqref{EYMline1} reduces to
\begin{align}
    -\DYM\big([\ap{},\a{}]^\g_0+\dPta\ap0-\dPtap\a0\big)+C\Big( [\ap{},\a{}]^\g_1+\dPta\ap1-\dPtap\a1+\tp0(\DTOT\Pta)_1-\t0(\DTOT\Ptap)_1\Big), \label{EYMline5}
\end{align}
while the anti-symmetrization of \eqref{EYMline4} is simply
\begin{equation}
    F\Big(\lbr\tp{},\t{}\rbr_0-2\tp1(\DGR\t{})_{-1}+2\t1(\DGR\tp{})_{-1}\Big). \label{EYMline6}
\end{equation}
Finally, \eqref{EYMline3} can be recast as
\begin{align} \label{EYMline7}
    -\DYM\big(\ap1(\DGR\tau)_{-1}\big) +(\DYM\ap{})_0(\DGR\tau)_{-1}-\DYM\big(\tp0 (\DTOT\Pta)_0\big) +D\tp0(\DTOT\Pta)_0+C\ap1(\DGR\t{})_0.
\end{align}
Taking the anti-symmetrization of \eqref{EYMline7} and summing with \eqref{EYMline5} and \eqref{EYMline6}, we get that
\begin{align} \label{EYMline8}
    \big[\dPta,\dPtap\big]A &=-\DYM\lbr\Ptap,\Pta\rbr^\ym_0 +C[\ap{},\a{}]^\g_1+F\lbr\tp{},\t{}\rbr_0 +\bigg\{\bigg\{C\Big(\tp0( \DTOT\Pta)_1+\ap1(\DGR\t{})_0+\dPta\ap1 \Big) \nn\\
    &+2F\t1(\DGR\tp{})_{-1}+(\DYM\ap{})_0(\DGR\tau)_{-1} +D\tp0(\DTOT\Pta)_0\bigg\}-\Pta\leftrightarrow\Ptap\bigg\},
\end{align}
where we used the definition \eqref{YMbracketDTOTdef} of the YM-bracket.
We still have to massage the last line, which upon expanding each derivative turns out to be proportional to C:
\begin{align}
    &\quad \bigg\{2F\t1(\DGR\tp{})_{-1}+(\DYM\ap{})_0(\DGR\tau)_{-1} +D\tp0(\DTOT\Pta)_0\bigg\}-\Pta\leftrightarrow\Ptap \label{EYMline9}\\
    &= C\Big(2\tp1\DYM\a1+2\ap2 D\t0\Big)-\Pta\leftrightarrow\Ptap=C\Big(2\tp1(\DTOT\Pta )_0+2\ap2 (\DGR\tau)_{-1}\Big)-\Pta\leftrightarrow\Ptap. \nn
\end{align}
\vspace{-0.3cm}

\ni In the last step, we just added and subtracted the necessary terms to form $\DTOT$ and $\DGR$.
Using \eqref{EYMline9} in \eqref{EYMline8} we see that the terms proportional to the shear recombine to give the YM-bracket at degree 1.
We thus conclude that
\begin{align}
    \hspace{-0.2cm}\big[\dPta,\dPtap\big]A &=-\DYM\big\lbr\Ptap,\Pta\big\rbr^\ym_0+C \big\lbr\Ptap,\Pta\big\rbr^\ym_1+F \big\lbr\Ptap,\Pta\big\rbr^\gr_0=-\Big(\DTOT \big\lbr\Ptap,\Pta \big\rbr\Big)_{-1}\!=-\delta_{\lbr \Pta,\Ptap\rbr}A.\!
\end{align}

\section{Jacobi identity for the gravitational $\sfK$-bracket \label{AppEYM:JacobiScri}}

One uses that 
\be 
\left\{\t{},\tp{}\right\}_s=\sum_{a+b=s}(a+1)\big(\t{a}\pa_u\tp{b} -\tp{a}\pa_u\t{b}\big).
\ee 
The sum is over integers $-1\leq a,b\leq s+1$ such that $a+b=s$.
\begin{align}
    \big\{\t{},\{\tp{},\tpp{}\}\big\}_s &= \sum_{a+b=s}(a+1)\Big(\t{a}\pa_u\{\tp{},\tpp{}\}_b-\{\tp{}, \tpp{}\}_a\pa_u\t{b}\Big) \nn\\
    &=\sum_{a+b=s}\sum_{c+d=b}(a+1)(c+1)\t{a}\pa_u\big( \tp{c}\pa_u\tpp{d}-\tpp{c}\pa_u\tp{d}\big) \nn\\
    &-\sum_{a+b=s}\sum_{c+d=a}(a+1)(c+1)\pa_u\t{b}\big( \tp{c}\pa_u\tpp{d}-\tpp{c}\pa_u\tp{d}\big) \nn\\
    &=\sum_{a+b+c=s}(a+1)(c+1)\Big(\t{a} \pa_u\tp{c}\pa_u\tpp{b}+\t{a}\tp{c}\pa_u^2 \tpp{b}-\t{a}\pa_u\tpp{c}\pa_u\tp{b}-\t{a}\tpp{c} \pa_u^2\tp{b}\Big) \nn\\
    &-\sum_{a+b+c=s}(a+c+1)(c+1)\Big(\tp{c}\pa_u\t{b} \pa_u\tpp{a}-\tpp{c}\pa_u\t{b}\pa_u\tp{a}\Big) \\
    &\cyc \sum_{a+b+c=s}\t{a}\pa_u\tpp{b}\pa_u\tp{c} (a+1)\Big((c+1)-(b+1)-(a+c+1)+(a+b+1)\Big) \nn\\
    &+\sum_{a+b+c=s}\t{a}\tp{c}\pa_u^2\tpp{b} \big((a+1)(c+1)-(c+1)(a+1)\big) \nn\\
    &\cyc 0. \nn
\end{align}
Therefore we have proven that the Jacobi identity holds, cf.\,\eqref{JacobiKGRbracket}.

\section{Properties of the action  \includegraphics[width=0.40cm]{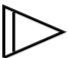}}

\subsection{Leibniz rule \label{AppEYM:tauBarAction}}

We first evaluate 
\begin{align}
\big(\bWt[\ap{},\app{}]^\g\big)_s &=\sum_{n=0}^s(n+1)\Big(\t{n}\pa_u\big[\ap{},\app{} \big]^\g_{s-n}-\big[\ap{},\app{}\big]^\g_{n+1} \pa_u\t{s-n-1} \Big) \\
&=\sum_{n=0}^s(n+1)\Big(\t{n}\big[\pa_u\ap{},\app{} \big]^\g_{s-n}+\t{n}\big[\ap{},\pa_u\app{} \big]^\g_{s-n}-\big[\ap{},\app{}\big]^\g_{n+1} \pa_u\t{s-n-1} \Big).\nn
\end{align}
We then focus on 
\begin{align}
    \big[\bWt\ap{},\app{}\big]^\g_s &=\sum_{k=0}^s\big[ (\bWt\ap{})_k,\app{s-k}\big]_\g \\
    &=\sum_{k=0}^s\sum_{n=0}^k(n+1)\t{n}\big[\pa_u \ap{k-n},\app{s-k}\big]_\g-\sum_{k=0}^s \sum_{n=0}^k(n+1)\pa_u\t{k-n-1}\big[\ap{n+1}, \app{s-k}\big]_\g \nn\\
    &=\sum_{n=0}^s\sum_{k=0}^{s-n}(n+1)\t{n} \big[\pa_u\ap{k},\app{s-k-n}\big]_\g-\sum_{k=0}^s \sum_{n=0}^{s-k}(n+1)\pa_u\t{s-n-k-1}\big[\ap{n+1},\app{k}\big]_\g, \nn
\end{align}
where we commuted the first double sum, namely $\sum_{k=0}^s\sum_{n=0}^k=\sum_{n=0}^s\sum_{k=n}^s$, and then changed $k-n\to n$; while we also changed $s-k\to k$ in the second double sum.
Taking $p=n+k$, we have that $\sum_{k=0}^s\sum_{n=0}^{s-k}=\sum_{p=0}^s \sum_{k=0}^p$.
Simply renaming $p\to n$, we then get
\begin{equation}
    \big[\bWt\ap{},\app{}\big]^\g_s =\sum_{n=0}^s(n+1) \t{n}\big[\pa_u\ap{},\app{}\big]^\g_{s-n}-\sum_{n=0}^s\sum_{k=0}^n(n-k+1)\pa_u\t{s-n-1} \big[\ap{n-k+1},\app{k}\big]_\g.
\end{equation}
The anti-symmetrization of this last double sum reduces to
\begin{align}
    &\quad -\sum_{n=0}^s\sum_{k=0}^n(n-k+1)\pa_u\t{s-n-1} \big[\ap{n-k+1},\app{k}\big]_\g\!-\ap{}\leftrightarrow\app{} \\
    &=- \sum_{n=0}^s\sum_{k=0}^n(n+1)\pa_u\t{s-n-1} \big[\ap{n-k+1},\app{k}\big]_\g =-\sum_{n=0}^s(n+1)\big[\ap{},\app{}\big]^\g_{n+1} \pa_u\t{s-n-1}. \nn
\end{align}
Therefore, the action $\!\barWhiteTriangle$ is a derivative, namely satisfies the Leibniz rule, cf.\,\eqref{barWhiteProperties}:
\begin{equation}
    \boxed{\bWt\big[\ap{},\app{}\big]^\g= \big[\bWt\ap{},\app{}\big]^\g+ \big[\ap{},\bWt\app{}\big]^\g}.
\end{equation}

\subsection{Homomorphism \label{AppEYM:HomtauBarAction}}

In this appendix, unless otherwise stated, we take the convention that equalities are valid upon anti-symmetrization of the RHS.
Firstly, we evaluate the action of the $\sfK$-bracket:
\begin{align}
    \big(\{\tau,\tp{}\}\barWhiteTriangle\app{}\big)_s &=\sum_{n=0}^s(n+1)\sum_{k=0}^{n+1}(k+1)\t{k} \pa_u\tp{n-k}\pa_u\app{s-n} \nn\\
    &-\sum_{n=0}^s(n+1) \app{n+1}\pa_u\left(\sum_{k=0}^{s-n}(k+1)\t{k}\pa_u\tp{s-n-k-1}\right). \label{intermHomBarAction}
\end{align}
Secondly, the commutator of the actions is given by
\begin{align}
    \Big(\big[\bWt,\bWtp\!\big]\app{}\Big)_s &=\sum_{ n=0}^s(n+1)\Big(\t{n}\pa_u\big(\bWtp\app{}\big)_{s-n}-\big(\bWtp\app{}\big)_{n+1}\pa_u\t{s-n-1}\Big) \nn\\
    &=\sum_{n=0}^s(n+1)\t{n}\sum_{k=0}^{s-n}(k+1) \Big(\dashuline{\pa_u\tp{k}\pa_u\app{s-n-k}}+\uline{\tp{k}\pa_u^2\app{s-n-k}} \\
    &\hspace{4.6cm}-\dashuline{\pa_u\app{k+1}\pa_u\tp{s-n-k-1}}-\uwave{\app{k+1}\pa_u^2\tp{s-n-k-1}}\Big) \nn\\
    &-\sum_{n=0}^s(n+1)\pa_u\t{s-n-1}\sum_{k=0}^{n+1}(k+1)\Big(\uuline{\tp{k}\pa_u\app{n+1-k}}-\dotuline{\app{k+1}\pa_u \tp{n-k}}\Big).\nn
\end{align}
The explicit anti-symmetrization of the underlined term vanishes,\footnote{Using that $\sum_{n=0}^s\sum_{k=0}^{s-n}A_nB_k =\sum_{n=0}^s\sum_{k=0}^{s-n}A_kB_n$.}
\begin{equation}
    \uline{~\cdots~}=\sum_{n=0}^s\sum_{k=0}^{s-n}(n+1)(k+1)\big(\t{n}\tp{k}-\tp{n}\t{k}\big)\pa_u^2 \app{s-n-k}=0.
\end{equation}
The wavy underlined term gives us
\begin{align}
    \uwave{~\cdots~} &=-\sum_{n=0}^s\sum_{k=0}^{s-n}(n+1)(k+1)\Big(\app{n+1}\pa_u\big(\t{k}\pa_u\tp{s-n-k-1}\big)-\dotuline{\app{k+1}\pa_u\t{n}\pa_u\tp{s-n-k-1}}\Big).
\end{align}
The dotted underlined terms reduce to\footnote{Changing $s-n-1\to n$ in the first sum.}
\begin{align}
    \dotuline{~\cdots~}&\!=\!\sum_{n=-1}^s \sum_{k=0}^{s-n}(s-n)(k+1)\app{k+1}\pa_u\t{n}\pa_u\tp{s-n-k-1}+\sum_{n=0}^s\sum_{k=0}^{s-n}(n+1)(k+1)\app{k+1}\pa_u\t{n}\pa_u\tp{s-n-k-1} \nn\\
    &=\sum_{n=-1}^s \sum_{k=0}^{s-n}(s+1)(k+1)\app{k+1}\pa_u\t{n}\pa_u\tp{s-n-k-1}=0,
\end{align}
which vanishes due to the anti-symmetrization.
To see it, make use of
\begin{equation}
    \sum_{n=-1}^s \sum_{k=0}^{s-n} A_k B_n C_{s-n-k-1} =\sum_{n=-1}^s \sum_{k=0}^{s-n} A_k B_{s-n-k-1} C_n.
\end{equation}
Next the double underlined contribution is (in the first step we already permuted the sums and used the anti-symmetry)
\begin{align}
    \uuline{~\cdots~}&=\sum_{k=0}^{s+1}\sum_{n=k-1}^s (n+1)(k+1)\t{k}\pa_u\tp{s-n-1}\pa_u\app{n+1-k} \nn\\
    &=\sum_{k=0}^{s+1}\sum_{n=k-1}^s(s+k-n)(k+1)\t{k} \pa_u\tp{n-k}\pa_u\app{s-n} \\
    &=\sum_{n=-1}^s\sum_{k=0}^{n+1}(s+k-n)(k+1)\t{k} \pa_u\tp{n-k}\pa_u\app{s-n}. \nn
\end{align}
Finally the dashed underlined factors gather as (where we already swapped $n\leftrightarrow k$ in the first step)
\begin{align}
     \dashuline{~\,\cdots\,~}&=\sum_{k=0}^s \sum_{n=-1}^{s-k}(n+1)(k+1)\t{k}\pa_u\tp{n}\pa_u\app{s-n-k}-\sum_{k=0}^s\sum_{n=-1}^{s-k}(n+1)(k+1)\t{k} \pa_u\tp{s-n-k-1}\pa_u\app{n+1} \nn\\
     &=\sum_{k=0}^s \sum_{n=-1}^{s-k}(s-2n-k-1)(k+1)\t{k} \pa_u\tp{s-n-k-1}\pa_u\app{n+1}\\*
     &=\sum_{k=0}^s\sum_{n=k-1}^s(2n-k-s+1)(k+1)\t{k} \pa_u\tp{n-k}\pa_u\app{s-n}, \nn
\end{align}
where we changed $s-n-1\to n$ to get the last line. 
Moreover we can add the term $k=s+1$ for free since its contribution vanishes, which allows us to commute the sums.
Thus, we obtain that
\begin{equation}
    \uuline{~\cdots~}+\dashuline{~\,\cdots\,~}= \sum_{n=-1}^s\sum_{k=0}^{n+1}(n+1)(k+1)\t{k} \pa_u\tp{n-k}\pa_u\app{s-n}.
\end{equation}
Combining the latter with $\uwave{\,\cdots\,}$, we get \eqref{intermHomBarAction}.
Hence, cf.\,\eqref{barWhiteProperties}
\begin{equation}
    \boxed{\big\{\tau,\tp{}\big\}\barWhiteTriangle \app{}= \big[\bWt,\bWtp\!\big]\app{}}.
\end{equation}

\section{Properties of the action $\vartriangleright$}

\subsection{Leibniz rule anomaly \label{AppEYM:tauAction}}

We start by computing the action of $\bt$ onto the bracket $[\ap{},\app{}]$.
For this we use \eqref{DcalLeibnizApp}, namely
\begin{equation}
    \cA_s\big([\ap{},\app{}]^\g,\DYM\big)=\big[ \DYM\ap0,\app{s+1}\big]_\g-\big[\DYM\app0, \ap{s+1}\big]_\g. \label{LeibnizAnomalyDYM}
\end{equation}
Hence
\begin{align}
    \!\!\big(\Wt[\ap{},\app{}]^\g\big)_s\! &=\sum_{n=1}^s\Big((n+1)\t{n}\DYM\big[\ap{},\app{}\big]^\g_{s+1-n}-n\big[\ap{},\app{}\big]^\g_n D\t{s+1-n}\Big)+(s+2)C\t{s+1}\big[\ap{},\app{}\big]^\g_1 \nn\\
    &=\sum_{n=1}^s\!\Big((n+1)\Big(\t{n}\big[ \DYM\ap{},\app{}\big]^\g_{s-n}\!+\t{n}\big[ \ap{},\DYM\app{}\big]^\g_{s-n}\Big)-n\big[\ap{},\app{}\big]^\g_{n}D\t{s+1-n}\Big) \label{tauAction2} \\
    &+(s+2)C\t{s+1}\big[\ap{},\app{}\big]^\g_1 +\sum_{n=1}^s(n+1)\Big(\t{n}\big[ \DYM\ap0,\app{s+1-n}\big]_\g-\t{n} \big[\DYM\app0, \ap{s+1-n}\big]_\g\Big).\nn
\end{align}
Next we calculate
\begin{align}
    \big[\Wt\ap{},\app{}\big]^\g_s &=\sum_{k=0}^s\big[(\Wt\ap{})_k,\app{s-k}\big]_\g \nn\\
    &=\sum_{k=1}^s\sum_{n=1}^k(n+1)\t{n} \big[\DYM\ap{k+1-n},\app{s-k}\big]_\g-\sum_{k=1}^s\sum_{n=1}^k n\big[\ap{n}, \app{s-k}\big]_\g D\t{k+1-n} \nn\\
    &+C\sum_{k=0}^s(k+2)\Big(\t{k+1}[\ap1,\app{s-k}]_\g\Big) \nn\\
    &=\sum_{n=1}^s\sum_{k=0}^{s-n}(n+1)\t{n} \big[\DYM\ap{k+1},\app{s-k-n}\big]_\g-\sum_{n=1}^{s}\sum_{k=1}^n k[\ap{k},\app{n-k}]_\g D\t{s+1-n} \label{tauAction1}\\
    &+C\sum_{k=0}^s(k+2)\t{k+1}[\ap1,\app{s-k}]_\g, \nn
\end{align}
where in the last step, for the term containing $\DYM\ap{}$, we commuted the sums and then changed $k-n\to k$; while for the term containing $D\t{}$, we first change $k-n\to n$, then commuted the sums, changed $k-n\to k$ and finally $s-n\to n$.
Still in that same double sum, we can add the term $k=0$ since its contribution is 0.
Anti-symmetrizing in $\ap{},\app{}$, we get in particular that the term involving $D\t{}$ reduces to
\begin{align}
    -\sum_{n=1}^s\sum_{k=0}^{n} D\t{s+1-n}\Big(k\big[\ap{k},\app{n-k}\big]_\g+k\big[\ap{n-k},\app{k}\big]_\g\Big) &=-\sum_{n=1}^s\sum_{k=0}^{n}n D\t{s+1-n}\big[\ap{k},\app{n-k}\big]_\g \nn\\
    &=-\sum_{n=1}^s n\big[\ap{},\app{}\big]^\g_nD\t{s+1-n}.
\end{align}
Besides, the anti-symmetrization of the term proportional to $C\ap1$ gives\footnote{The first step consists in extracting the term $n=s$ from the sum.}
\begin{align}
    C\sum_{n=0}^s(n+2)\t{n+1}&\Big(\big[\ap1, \app{s-n}\big]_\g+\big[\ap{s-n},\app1\big]_\g\Big)= \\
    &=(s+2)C\t{s+1}\big[\ap{},\app{}\big]^\g_1+C \sum_{n=1}^{s}(n+1)\t{n}\Big(\big[\ap1, \app{s+1-n}\big]_\g+\big[\ap{s+1-n},\app1\big]_\g\Big). \nn
\end{align}
Therefore, combining the last two equations with \eqref{tauAction1}, we infer that
\begin{align}
    &\quad \big[\Wt\ap{},\app{}\big]^\g_s+\big[ \ap{}, \Wt\app{}\big]^\g_s= -\sum_{n=1}^s n\big[\ap{},\app{}\big]^\g_n D\t{s+1-n}+(s+2)C\t{s+1}\big[\ap{},\app{}\big]^\g_1 \label{tauAction3}\\
    &\!+\!\sum_{n=1}^s (n+1)\t{n}\Big(\big[\DYM \ap{},\app{} \big]^\g_{s-n}\!+\big[\ap{},\DYM \app{}\big]^\g_{s-n}\Big)\!+ C\sum_{n=1}^{s}(n+1)\t{n}\Big(\big[\ap1, \app{s+1-n}\big]_\g\!+\big[\ap{s+1-n},\app1\big]_\g\Big).\nn
\end{align}
Upon comparing \eqref{tauAction3} with \eqref{tauAction2}, we deduce that the Leibniz anomaly of the $\bt$-action takes the form \eqref{LeibnizAnomalyTauAction}
\begin{align}
    \boxed{\cA_s\big([\ap{},\app{}]^\g,\Wt\!\big) = \sum_{n=1}^s(n+1)\t{n}\big[ \DYM\ap0-C\ap1,\app{s+1-n}\big]_\g-\ap{}\leftrightarrow\app{}}.
\end{align}

\subsection{Homomorphism anomaly \label{AppEYM:tauHom}}

We start by computing the commutator of the actions of $\bt$ and $\btp$.
In the following, for shortness, it is understood that equalities are valid upon anti-symmetrization of the RHS.
\begin{align}
    &\big(\Wt(\Wtp\app{})\big)_s-\big(\Wtp(\Wt\app{})\big)_s =\nn\\
    &=\sum_{n=1}^s \Big((n+1)\t{n}\DYM(\Wtp\app{})_{s+1-n}-n(\Wtp\app{})_{n}D\t{s+1-n}\Big)+(s+2)C\t{s+1}(\Wtp\app{})_1 \nn\\
    &=\sum_{n=1}^s(n+1)\t{n}\DYM\bigg\{ \sum_{k=1}^{s+1-n}\Big((k+1)\tp{k}\DYM\app{s+2-n-k}-k\app{k}D\tp{s+2-n-k}\Big)+(s+3-n)C\app1\tp{s+2-n}\bigg\} \nn\\
    &-\sum_{n=1}^s nD\t{s+1-n}\bigg\{ \sum_{k=1}^n\Big((k+1)\tp{k}\DYM\app{n+1-k}-k\app{k}D\tp{n+1-k}\Big) +(n+2)C \app1\tp{n+1}\bigg\} \nn\\
    &+(s+2)C\t{s+1}\bigg\{2\tp1 \DYM\app1-\app1 D\tp1+3C\app1\tp2\bigg\} \nn\\
    &=\sum_{n=1}^s\sum_{k=1}^{s+1-n}(n+1)(k+1)\uline{\t{n}\tp{k}\DYM^2\app{s+2-n-k}}+\sum_{n=1}^s\sum_{k=1}^{s+1-n}(n+1)(k+1)\dotuline{\t{n}D\tp{k}\DYM\app{s+2-n-k}} \nn\\
    &-\sum_{n=1}^s\sum_{k=1}^{s+1-n}(n+1)k \dotuline{\t{n}D\tp{s+2-n-k}\DYM\app{k}}-\sum_{n=1}^s\sum_{k=1}^{s+1-n}(n+1)k \dashuline{\t{n}D^2\tp{s+2-n-k}\app{k}} \nn\\
    &+\sum_{n=1}^s(n+1)(s+3-n)\uwave{\t{n}D(C\tp{s+2-n}) \app1}+ \,C\sum_{n=1}^s(n+1)(s+3-n)\underline{\t{n}\tp{s+2-n}\DYM\app1}\nn\\
    &-\sum_{n=1}^s\sum_{k=1}^{n}n(k+1) \dotuline{\tp{k}D\t{s+1-n}\DYM\app{n+1-k}}+\sum_{n=1}^s \sum_{k=1}^{n}nk\dashuline{D\t{s+1-n}D\tp{n+1-k}\app{k}} \\
    &-C\sum_{n=1}^s n(n+2)D\t{s+1-n}\uwave{\app1\tp{n+1}}+(s+2)C\t{s+1} \bigg\{\underline{2\tp1 \DYM\app1}-\uwave{\app1 D\tp1}+3C\app1\tp2\bigg\}. \nn
\end{align}
Writing the anti-symmetrization of the underlined term explicitly, we immediately notice that\footnote{In the following, we often make use of the fact that $\sum_{n=1}^s\sum_{k=1}^{s+1-n}A_nB_k=\sum_{n=1}^s\sum_{k=1}^{s+1-n}A_kB_n$. \label{footEYM:sum}}
\begin{align}
    \underline{~\cdots~}=0.
\end{align}
We then treat the dotted underlined terms. 
In the first step, we do the appropriate changes of variables to get $\app{s+2-n-k}$ to appear in each sum and we leverage the fact that the RHS is understood to be anti-symmetrized to swap $\tau$ and $\tau'$ in the second double sum.
\begin{align}
    \dotuline{~\cdots~}\!&=\sum_{n=1}^s \sum_{k=1}^{s+1-n}(n+1)(k+1)\t{n}D\tp{k}\DYM\app{s+2-n-k}+\sum_{n=1}^s\sum_{k=1}^{s-n+1}(s-n+1)(k+1) \t{k}D\tp{n}\DYM\app{s-n+2-k} \nn\\
    &-\sum_{n=1}^s\sum_{k=1}^{s+1-n}(n+1)(s+2-n-k) \t{n}D\tp{k}\DYM\app{s+2-n-k}. \label{dottedUnderline}
\end{align}
Next, still in the second double sum, we use the property emphasized in the footnote \ref{footEYM:sum}, such that each term in \eqref{dottedUnderline} now involves $\t{n}D\tp{k}\DYM\app{s+2-n-k}$.
We thus obtain
\begin{align}
    \dotuline{~\cdots~}=\sum_{n=1}^s \sum_{k=1}^{s+1-n}(n+1)(n+k)\t{n}D\tp{k}\DYM\app{s+2-n-k}.
\end{align}
For later purposes, let us define the variable $p=n+k$ such that 
\begin{align}
    \dotuline{~\cdots~}=\sum_{p=2}^{s+1} \sum_{n=1}^{p-1} p(n+1)\t{n}D\tp{p-n}\DYM\app{s+2-p}=\sum_{n=1}^s \sum_{k=1}^n(n+1)(k+1)\aunderbrace[l1r]{\t{k}D\tp{n+1-k}\DYM\app{s+1-n}},
\end{align}
where in the last step, we renamed $p-1\to n$ and $n\to k$.
Concerning the dashed underlined contributions, we get
\vspace{-0.4cm}

\begin{adjustwidth}{-0.2cm}{0cm}
\begin{align}
    &\dashuline{\,~\cdots~\,}\!=-\sum_{n=1}^s \sum_{k=1}^{s+1-n}(n+1)kD\big(\t{n}D \tp{s+2-n-k}\big)\app{k}+\sum_{n=1}^s \sum_{k=1}^{s+1-n}(s+2)kD\t{n}D\tp{s+2-n-k}\app{k} \\
    &=-\sum_{n=1}^s \sum_{k=1}^{s+1-n}(k+1)nD\big(\t{k}D\tp{s+2-n-k}\big)\app{n}+\sum_{n=1}^s \sum_{k=1}^{s+1-n}(s+2)(s+2-n-k)D\t{n}D\tp{k}\app{s+2-n-k}. \nn
\end{align}
\end{adjustwidth}
The last double sum vanishes because of the anti-symmetrization. Hence,
\begin{align}
    \dashuline{\,~\cdots~\,}\!&=
    -\sum_{n=1}^s n\app{n}D\bigg\{ \sum_{k=1}^{s+1-n}(k+1)\aunderbrace[l1r]{\t{k}D\tp{s+2-n-k}}\bigg\}.\nn
\end{align}
We also get that the wavy underlined terms take the form
\begin{align}
    \uwave{~\cdots~}&=\app1\sum_{n=1}^s (n+1)(s+3-n)DC\t{n}\tp{s+2-n}+C\app1\sum_{n=1}^s(n+1)(s+3-n)\t{n}D\tp{s+2-n} \nn\\
    &+C\app1\sum_{n=2}^{s+1}(n-1)(n+1)\t{n} D\tp{s+2-n}-(s+2)C\app1\t{s+1} D\tp1 \\
    &=-2(s+2)DC\app1\t{s+1}\tp1-3(s+2)C\app1 \t{s+1}D\tp1+(s+2)C\app1\sum_{n=1}^{s+1}(n+1)\aunderbrace[l1r]{\t{n} D\tp{s+2-n}}. \nn
\end{align}
It is now straightforward to notice that the terms emphasized with the under-brace recombine to form the projection of the $C$-bracket onto the $\bcT$-sub-algebra as follows:\footnote{Of course here the anti-symmetrization of the RHS is explicit.}
\begin{equation}
    \aunderbrace[l1r]{~\cdots~}=\sum_{n=1}^s \Big((n+1)\overline{[\bt{},\btp{}]}_n\DYM\app{s+1-n}-n\app{n}D\overline{[\bt{} ,\btp{}]}_{s+1-n}\Big)+(s+2)C\app1\overline{[\bt{},\btp{}]}_{s+1}.
\end{equation}
We recognize the latter as the action of the bar-bracket onto $\app{}$, namely
\begin{equation}
    \aunderbrace[l1r]{~\cdots~}=\big(\overline{[\bt{},\btp{}]} \vartriangleright \app{}\big)_s.
\end{equation}
Therefore, just gathering the various contributions, we get that
\begin{align}
    &\big(\Wt(\Wtp\app{})\big)_s-\big(\Wtp(\Wt\app{})\big)_s -\big(\overline{[\bt{},\btp{}]} \vartriangleright \app{}\big)_s= \\
    &=-2(s+2)DC\app1\tp1\t{s+1} -3(s+2)C\app1 \t{s+1}D\tp1+3(s+2)C^2\app1\t{s+1}\tp2. \nn
\end{align}
The final result takes the form (cf. \eqref{HomAnomalyTauAction})
\begin{equation}
    \boxed{\Big(\overline{[\bt{},\btp{}]} \vartriangleright \app{}-\big[\Wt,\Wtp\!\big]\app{} \Big)_s=(s+2)\app1\Big(\t{s+1}\dbtp C-\tp{s+1}\dbt C\Big)},
\end{equation}
where $\dbt C$ is the variation $\dt C$ restricted to the $\bsfT$-subspace, i.e.
\begin{equation}
    \dbt C=2DC\t1+3CD\t1-3C^2\t2.
\end{equation}

\section{Properties of the action $\blacktriangleright$ \label{AppEYM:BlackTauAction}}

\subsection{Leibniz rule anomaly: GR projection \label{AppEYM:BlackTauActionLeibnizGR}}

We first compute the GR projection of the action of $\ot$ onto the bracket $\big[\bPtap,\bPtapp\big]$.
\begin{align}
    \big(\Bt\big[\bPtap,\bPtapp\big] \big)^\gr_s &=\t0 D\overline{\big[\btp,\btpp\big]}_{s+1}-(s+2)\overline{\big[\btp,\btpp\big]}_{s+1} D\t0-(s+3)C\t0\overline{\big[\btp, \btpp\big]}_{s+2} \nn\\
    &=\bigg\{\t0\sum_{n=1}^{s+1}(n+1)\big(D\tp{n}D\tpp{s+2-n}+\tp{n}D^2\tpp{s+2-n}\big)-(s+2)D\t0\sum_{n=1}^{s+1}(n+1)\tp{n}D\tpp{s+2-n} \nn\\
    &-(s+3)C\t0\sum_{n=1}^{s+2}(n+1)\tp{n}D\tpp{s+3-n}\bigg\}-\btp\leftrightarrow\btpp. \label{eq101}
\end{align}
Next,
\begin{align}
    \big[\Bt\bPtap,\bPtapp\big]^\gr_s &=\overline{\big[(\Bt\bPtap)^\gr,\btpp \big]}_s \nn\\
    &=\sum_{n=1}^s(n+1)\Big(\big(\Bt\bPtap \big)^\gr_n D\tpp{s+1-n}-\tpp{n}D\big(\Bt \bPtap\big)^\gr_{s+1-n}\Big) \nn\\
    &=\sum_{n=1}^s(n+1)D\tpp{s+1-n}\Big\{\t0 D\tp{n+1}-(n+2)\tp{n+1}D\t0-(n+3)C\t0\tp{n+2}\Big\} \nn\\
    &-\sum_{n=1}^s(n+1)\tpp{n}D\Big\{\t0 D\tp{s+2-n}-(s+3-n)\tp{s+2-n}D\t0-(s+4-n)C\t0\tp{s+3-n}\Big\} \nn\\
    \qquad\qquad &=\t0\sum_{n=1}^s(n+1)\uuline{D\tp{n+1} D\tpp{s+1-n}}-D\t0\sum_{n=1}^s(n+1)(n+2)\underline{\tp{n+1}D\tpp{s+1-n}} \nn\\
    &-C\t0\sum_{n=1}^s(n+1)(n+3)\uwave{\tp{n+2}D\tpp{s+1-n}} -D\t0\sum_{n=1}^s(n+1)\underline{\tpp{n}D\tp{s+2-n}} \nn\\
    &-\t0\sum_{n=1}^{s+1}(n+1)\aunderbrace[l1r]{\tpp{n}D^2\tp{s+2-n}}+(s+2)\t0\tpp{s+1}D^2\tp1  \\
    &+D\t0\sum_{n=1}^s(n+1)(s+3-n)\underline{\tpp{n}D\tp{s+2-n}}+D^2\t0\sum_{n=1}^s(n+1)(s+3-n)\dotuline{\tp{s+2-n}\tpp{n}} \nn\\
    &+D(C\t0)\!\sum_{n=1}^s(n+1)(s+4-n)\dashuline{\tp{s+3-n}\tpp{n}}+C\t0\!\sum_{n=1}^s(n+1)(s+4-n)\uwave{\tpp{n}D\tp{s+3-n}}. \nn
\end{align}
Anti-symmetrizing the straight underlined terms leads to
\begin{equation}
    \underline{~\cdots~}=D\t0\bigg\{(s+2)\sum_{n=1}^{s+1}(n+1)\aunderbrace[l1r]{\tpp{n}D\tp{s+2-n}}-2\tpp1D\tp{s+1}-(s+2)\tpp{s+1}D\tp1\bigg\}-\btp\leftrightarrow\btpp.
\end{equation}

\ni Similarly for the wavy underlined terms, we get
\begin{align}
    \uwave{~\cdots~} &=C\t0\bigg\{(s+3) \sum_{n=1}^{s+2}(n+1)\aunderbrace[l1r]{\tpp{n}D\tp{s+3-n}}-3(s+2)\tpp{s+1}D\tp2-2(s+3)\tpp{s+2}D\tp1 \nn\\*
    &\qquad\quad -3\tpp2D\tp{s+1}\bigg\}-\btp\leftrightarrow\btpp.
\end{align}
Also notice that the anti-symmetrization of the dashed underlined term reduces to
\begin{equation}
    \dashuline{\,~\cdots~\,}\!=-D(C\t0)\Big\{ 3(s+2)\tp2\tpp{s+1}+2(s+3)\tp1\tpp{s+2}\Big\}-\btp\leftrightarrow\btpp,
\end{equation}
and in a similar way
\begin{equation}
    \dotuline{~\cdots~} =-D^2\t0\Big\{2(s+2)\tp1\tpp{s+1}\Big\}-\btp\leftrightarrow\btpp.
\end{equation}
In order to identify the double underlined term with its counterpart in \eqref{eq101}, we conveniently write that
\begin{equation}
    \uuline{~\cdots~} =\t0\bigg\{\sum_{n=1}^{s+1}(n+1)\aunderbrace[l1r]{D\tp{n}D\tpp{s+2-n}}-D\tp1D\tpp{s+1}\bigg\}-\btp\leftrightarrow\btpp.
\end{equation}
Therefore, using the fact that the under-braced factors equate \eqref{eq101}, we find that
\begin{align}
    &\quad \big[\Bt\bPtap,\bPtapp\big]^\gr_s +\big[\bPtap,\Bt\bPtapp\big]^\gr_s =\big(\Bt\big[\bPtap,\bPtapp\big] \big)^\gr_s+ \nn\\
    &+\bigg\{\bigg\{(s+2)\t0\tpp{s+1}D^2\tp1 +D\t0\Big( 2\tp1D\tpp{s+1}-(s+2)\tpp{s+1}D\tp1\Big) -2(s+2)D^2\t0\tp1\tpp{s+1} \nn\\
    &-C\t0\Big(3(s+2)\tpp{s+1}D\tp2+ 2(s+3)\tpp{s+2}D\tp1-3\tp2D\tpp{s+1}\Big) \label{BlackLeibnizInterm1}\\
    &-D(C\t0)\Big( 3(s+2)\tp2\tpp{s+1}+2(s+3) \tp1\tpp{s+2}\Big)-\t0 D\tp1D\tpp{s+1}\bigg\}-\btp\leftrightarrow\btpp\bigg\}. \nn
\end{align}
Let us now compare the terms that did not recombine into $\big(\Bt\big[\bPtap,\bPtapp\big] \big)^\gr$ with
\begin{align}
    \big((\rBt\bPtap)\blacktriangleright \bPtapp\big)^\gr_s &=\big(\rBt\bPtap \big)_0D\tpp{s+1}-(s+2)\tpp{s+1}D\big(\rBt\bPtap \big)_0 -(s+3)C\tpp{s+2}\big(\rBt\bPtap \big)_0 \nn\\
    &=D\tpp{s+1}\Big(\t0D\tp1-2\tp1D\t0-3C\tp2\t0\Big) \label{BlackLeibnizInterm2}\\
    &-(s+2)\tpp{s+1}\Big(-D\t0D\tp1+\t0D^2\tp1 -2\tp1D^2\t0-3D(C\t0)\tp2-3C\t0D\tp2\Big) \nn\\
    &-(s+3)C\tpp{s+2}\Big(\t0D\tp1-2\tp1D\t0-3C\tp2\t0\Big). \nn
\end{align}
All the factors proportional to $D\tpp{s+1}$ and to $\tpp{s+1}$ (i.e. the first two lines of \eqref{BlackLeibnizInterm2}) appear as such in \eqref{BlackLeibnizInterm1}, up to a sign.
It is then not hard to see that the terms proportional to $\tpp{s+2}$ add up to form the variation $\dbtp C$.
Thence the Leibniz rule anomaly for the action $\blacktriangleright$ projected onto $\bsfT$ is given by \eqref{LeibnizAnomalyBlackTauActionGR}
\begin{equation}
    \boxed{\cA_s^\gr\Big(\big[\bPtap,\bPtapp \big], \Bt\!\Big)=\Big\{ \big((\rBt\bPtap)\blacktriangleright \bPtapp\big)^\gr_s+(s+3)\t0\tpp{s+2} \dbtp C\Big\}-\bPtap\leftrightarrow\bPtapp}.
\end{equation}

\subsection{Leibniz rule anomaly: YM projection \label{AppEYM:BlackTauActionLeibnizYM}}

The YM projection of the action of $\ot$ onto $\big[\bPtap,\bPtapp\big]$ takes the form
\begin{align}
    \big(\Bt\big[\bPtap,\bPtapp\big] \big)^\ym_s &=\t0\DYM\big[ \bPtap,\bPtapp\big]^\ym_{s+1}-(s+1)\big[\bPtap,\bPtapp\big]^\ym_{s+1}D\t0-(s+2)C\t0\big[\bPtap,\bPtapp \big]^\ym_{s+2} \nn\\
    &-(s+2)F\t0\overline{\big[\btp,\btpp\big] }_{s+1} \nn\\
    &=\t0\DYM\big[\ap{},\app{}\big]^\g_{s+1}-(s+1)\big[\ap{},\app{}\big]^\g_{s+1}D\t0-(s+2)C\t0\big[\ap{},\app{}\big]^\g_{s+2} \nn\\
    &+\bigg\{\bigg\{\t0\DYM\big(\Wtp\app{} \big)_{s+1}-(s+1)\big(\Wtp\app{} \big)_{s+1}D\t0-(s+2)C\t0\big(\Wtp\app{} \big)_{s+2} \nn\\
    &-(s+2)F\t0\sum_{n=1}^{s+1}(n+1)\tp{n}D\tpp{s+2-n}\bigg\}-\bPtap\leftrightarrow\bPtapp\bigg\}. \label{LeibnizYMinterm}
\end{align}
On the other hand,
\begin{align}
&~\quad \big[\Bt\bPtap,\bPtapp\big]^\ym_s= \nn\\ &=\big[(\Bt\bPtap)^\ym,\app{}\big]^\g_s+\big((\Bt \bPtap)^\gr\vartriangleright\app{}\big)_s-\big( \Wtpp(\Bt\bPtap)^\ym\big)_s \nn\\
&=\sum_{n=0}^s\big[(\Bt\bPtap)^\ym_n,\app{s-n}\big]_\g+\sum_{n=1}^s\Big((n+1)\big(\Bt\bPtap\big)^\gr_n\,\DYM\app{s+1-n}-n\app{n}D\big(\Bt\bPtap\big)^\gr_{s+1-n}\Big) \nn\\
&+(s+2)C\app1\big(\Bt\bPtap\big)^\gr_{s+1}-\sum_{n=1}^s\Big((n+1)\tpp{n}\DYM\big(\Bt\bPtap\big)^\ym_{s+1-n}-n\big(\Bt\bPtap\big)^\ym_n D\tpp{s+1-n}\Big) \nn\\
&-(s+2)C\big(\Bt\bPtap\big)^\ym_1\tpp{s+1}\nn\\
&=\sum_{n=0}^s\Big[\underline{\t0\DYM\ap{n+1}}-(n+1)\underline{\ap{n+1}D\t0}-(n+2)\underline{C\t0\ap{n+2}}-(n+2)F\t0\tp{n+1},\app{s-n}\Big]_\g \nn\\
&+\sum_{n=1}^s(n+1)\DYM\app{s+1-n}\Big\{\uuline{\t0 D\tp{n+1}}-(n+2)\dashuline{\tp{n+1}D\t0}-(n+3)\uwave{C\t0\tp{n+2}}\Big\} \nn\\
&-\sum_{n=1}^sn\app{n}\Big\{\dashuline{D \t0D\tp{s+2-n}}+\uuline{\t0D^2\tp{s+2-n}}-(s+3-n)\dashuline{D\tp{s+2-n}D\t0}-\aunderbrace[l2*{8}{10}3r]{(s+3-n)\tp{s+2-n}D^2\t0} \nn\\
&\quad\qquad\qquad -(s+4-n)\aunderbrace[l1r]{D(C\t0)\tp{s+3-n}}-(s+4-n)\uwave{C\t0D\tp{s+3-n}}\Big\} \nn\\
&+(s+2)C\app1\Big\{\underbrace{\t0D\tp{s+2}}-(s+3)\underbrace{\tp{s+2}D\t0}-(s+4)\underbrace{C\t0\tp{s+3}}\Big\} \nn\\
&-\sum_{n=1}^s(n+1)\tpp{n}\Big\{ \dashuline{D\t0\DYM\ap{s+2-n}}+\uuline{\t0\DYM^2\ap{s+2-n}}-(s+2-n)\dashuline{\DYM\ap{s+2-n}D\t0} \nn\\
&\qquad\qquad -\aunderbrace[l2*{8}{10}3r]{(s+2-n)\ap{s+2-n}D^2\t0}-(s+3-n)\aunderbrace[l1r]{D(C\t0)\ap{s+3-n}}-(s+3-n)\uwave{C\t0\DYM\ap{s+3-n}} \nn\\
&\qquad\qquad -(s+3-n)\DYM(F\t0)\tp{s+2-n}-(s+3-n)\aunderbrace[l1*{3}{01}01D10*{3}{10}1r]{F\t0D\tp{s+2-n}}\Big\} \\
&+\sum_{n=1}^snD\tpp{s+1-n}\Big\{\uuline{\t0\DYM\ap{n+1}}-(n+1)\dashuline{\ap{n+1}D\t0}-(n+2)\uwave{C\t0\ap{n+2}}-\aunderbrace[l1*{3}{01}01D10*{3}{10}1r]{(n+2)F\t0\tp{n+1}}\Big\} \nn\\*
&-(s+2)C\tpp{s+1}\Big\{\t0\DYM\ap2-\underbrace{2\ap2D\t0}-3C\t0\ap3-3F\t0\tp2\Big\}. \nn
\end{align}
After anti-symmetrization in $\bPtap,\bPtapp$, the underlined terms contribute as\footnote{We use that
\vspace{-0.2cm}
\begin{equation*}
    \sum_{n=0}^s(n+1)\big[\ap{n+1},\app{s-n}\big]_\g-\ap{}\leftrightarrow\app{}= \sum_{n=0}^{s+1}\Big(n\big[\ap{n},\app{s+1-n}\big]_\g+n\big[\ap{s+1-n},\app{n}\big]_\g \Big)=\sum_{n=0}^{s+1}(s+1)\big[\ap{n},\app{s+1-n}\big]_\g.
\end{equation*}
\vspace{-0.3cm}}
\begin{align}
    \underline{~\cdots~} &=\t0\big[\DYM\ap{}, \app{}\big]^\g_s+\t0\big[\ap{},\DYM\app{}\big]^\g_s-(s+1)D\t0\big[\ap{},\app{}\big]^\g_{s+1}-(s+2)C\t0\big[\ap{},\app{}\big]^\g_{s+2} \nn\\
    &+C\t0\Big(\big[\ap1,\app{s+1}\big]_\g +\big[\ap{s+1},\app1\big]_\g \Big) \nn\\
    &=\t0\DYM\big[\ap{},\app{}\big]^\g_{s+1}+\t0 \big[\hdap A,\app{s+1}\big]_\g-\t0 \big[\hdapp A,\ap{s+1}\big]_\g \\
    &-(s+1)D\t0\big[\ap{},\app{}\big]^\g_{s+1}-(s+2)C\t0\big[\ap{},\app{}\big]^\g_{s+2}, \nn
\end{align}
where in the last step, we used the property \eqref{LeibnizAnomalyDYM}.
Besides, the double underlined terms partially recombine (leveraging the anti-symmetrization) to give $\t0\DYM(\Wtp\app{})_{s+1}$:
\begin{align}
    \uuline{~\cdots~} &=\t0\bigg\{\sum_{n=2}^{s+1}n D\tp{n}\DYM\app{s+2-n}+\sum_{n=1}^s(n+1)\tp{n}\DYM^2\app{s+2-n}-\sum_{n=2}^{s+1}(n-1)\DYM\app{n}D\tp{s+2-n} \nn\\
    &\qquad -\sum_{n=1}^s n\app{n}D^2\tp{s+2-n}\bigg\}-\bPtap\leftrightarrow\bPtapp \nn\\
    &=\t0\Big\{\DYM\big(\Wtp\app{} \big)_{s+1}-D\tp1\DYM\app{s+1}-(s+2) \tp{s+1}\DYM^2\app1+(s+1)\app{s+1}D^2\tp1 \\
    &\qquad -(s+3)\underbrace{\DYM\big(C\app1\tp{s+2} \big)}\Big\}-\bPtap\leftrightarrow\bPtapp. \nn
\end{align}
Concerning the dash-underlined terms, they partially recombine into $D\t0(\Wtp\app{})_{s+1}$:
\begin{align}
    \dashuline{~\,\cdots~\,}\! &=D\t0\bigg\{ \sum_{n=1}^sn(s+2-n)\app{n}D\tp{s+2-n}+\sum_{n=2}^{s+1}n(n-1)\app{n}D\tp{s+2-n} \nn\\
    &\qquad~~-\sum_{n=1}^s(n+1)(s+1-n)\tp{n}\DYM\app{s+2-n}-\sum_{n=2}^{s+1}n(n+1)\tp{n}\DYM\app{s+2-n}\bigg\}-\bPtap\leftrightarrow\bPtapp \nn\\
    &=-D\t0\Big\{ (s+1)\big(\Wtp\app{}\big)_{s+1}+(s+1)\app{s+1}D\tp1-2\tp1\DYM\app{s+1} \\
    &\qquad\quad~ -(s+1)(s+3)\underbrace{C\app1\tp{s+2}}\Big\}-\bPtap\leftrightarrow\bPtapp. \nn
\end{align}
We can rearrange the wavy underlined factors as
\begin{align}
    \uwave{~\cdots~} &=-C\t0\bigg\{ \sum_{n=3}^{s+2}(n-1)(n+1)\tp{n}\DYM\app{s+3-n}-\sum_{n=1}^sn(s+4-n) \app{n}D\tp{s+3-n} \\*
    &\qquad\quad~ +\sum_{n=1}^s(n+1)(s+3-n)\tp{n}\DYM\app{s+3-n}-\sum_{n=3}^{s+2}n(n-2)\app{n}D\tp{s+3-n} \bigg\}-\bPtap\leftrightarrow\bPtapp \nn\\
    &=-C\t0\Big\{(s+2)\big(\Wtp\app{}\big)_{s+2}-3\tp2\DYM\app{s+1}-2(s+2)\tp{s+1}\DYM \app2-(s+3)\underbrace{\tp{s+2}\DYM\app1}\nn\\
    &-\underbrace{\app1D\tp{s+2}}+3(s+1)\app{s+1} D\tp2+2(s+2)\app{s+2}D\tp1-(s+2)(s+4)\underbrace{C\app1\tp{s+3}}\Big\}-\bPtap\leftrightarrow\bPtapp. \nn
\end{align}
The straight under-braced contributions sum up to
\begin{align}
    \aunderbrace[l1r]{~\cdots~}&=D(C\t0)\bigg\{ \sum_{n=1}^sn(s+4-n)\app{n}\tp{s+3-n}-\sum_{n=3}^{s+2}n(s+4-n)\tp{s+3-n}\app{n} \bigg\}-\bPtap\leftrightarrow\bPtapp\nn\\
    &=CD\t0\Big\{(s+3)\underbrace{\app1\tp{s+2}}+2(s+2) \underbrace{\app2\tp{s+1}}- 3(s+1)\app{s+1}\tp2-2(s+2)\app{s+2}\tp1\Big\} \\
    &+DC\t0\Big\{(s+3)\underbrace{\app1\tp{s+2}}+2(s+2) \app2\tp{s+1}- 3(s+1)\app{s+1}\tp2-2(s+2)\app{s+2}\tp1 \Big\}-\bPtap\leftrightarrow\bPtapp.\nn
\end{align}
The curly under-braced terms simply add up to 0,
\begin{equation}
    \underbrace{~\cdots~}=0,
\end{equation}
while the dash-under-braced factors reduce to
\begin{equation}
    \aunderbrace[l1*{4}{10}1r]{~\,\cdots\,~}= D^2\t0\Big\{-2(s+1)\app{s+1}\tp1+(s+2)\tp{s+1}\app{1}\Big\}-\bPtap\leftrightarrow\bPtapp.
\end{equation}
Finally the curly dash-under-braced terms simplify to
\begin{align}
    \aunderbrace[l0*{2}{02}00D00*{2}{20}0r]{~\,\cdots\cdots\,~}=-(s+2)F\t0\bigg\{\sum_{n=1}^{s+1}(n+1)\tp{n}D\tpp{s+2-n}-2\tp{s+1}D\tpp1\bigg\}-\bPtap\leftrightarrow\bPtapp.
\end{align}
Using the fact that
\begin{equation}
    \bigg\{\DYM(F\t0)\sum_{n=1}^s(n+1)(s+3-n)\tpp{n}\tp{s+2-n}\bigg\}-\btp\leftrightarrow\btpp=2(s+2)\DYM(F\t0) \tp{s+1}\tpp1 -\btp\leftrightarrow\btpp,
\end{equation}
we gather all these contributions and get that
\begin{align}
    &~\quad \Big(\big[\Bt\bPtap,\bPtapp\big]+\big[\bPtap,\Bt\bPtapp\big]\Big)^\ym_s= \nn\\ 
    &=\t0\DYM\big[\ap{},\app{}]_{s+1}-(s+1)D\t0\big[\ap{},\app{}\big]_{s+1}-(s+2)C\t0\big[\ap{},\app{}\big]_{s+2}-(s+2)F\t0\overline{\big[\btp, \btpp\big]}_{s+1} \nn\\
    &+\bigg(\!\bigg(\!\t0 \big[\hdap A,\app{s+1}\big]_\g+\t0 \DYM\big(\Wtp\app{}\big)_{s+1}-(s+1)D\t0\big(\Wtp\app{}\big)_{s+1}-(s+2)C\t0\big(\Wtp\app{}\big)_{s+2} \nn\\
    &+\t0\Big\{-D\tp1\DYM\app{s+1}-(s+2)\dotuline{\tp{s+1}\DYM^2\app1}+(s+1)\app{s+1}D^2\tp1\Big\} \nn\\
    &-C\t0\Big\{-3\tp2\DYM\app{s+1}-2(s+2)\dotuline{\tp{s+1}\DYM \app2}+3(s+1)\app{s+1} D\tp2+2(s+2)\app{s+2}D\tp1\Big\} \nn\\
    &-D\t0\Big\{(s+1)\app{s+1}D\tp1-2\tp1\DYM\app{s+1}\Big\}+CD\t0\Big\{- 3(s+1)\app{s+1}\tp2-2(s+2)\app{s+2}\tp1 \Big\} \nn\\
    &+DC\t0\Big\{2(s+2)\dotuline{\app2\tp{s+1}}- 3(s+1)\app{s+1}\tp2-2(s+2)\app{s+2}\tp1 \Big\} \label{LeibnizYMBig}\\
    &+D^2\t0\Big\{-2(s+1)\app{s+1}\tp1+(s+2)\tp{s+1}\app{1}\Big\}+(s+2)C\tp{s+1}\Big\{\dotuline{ \t0\DYM\app2}-3\dotuline{C\t0\app3}-\dotuline{3F\t0 \tpp2}\Big\} \nn\\
    &+2(s+2)\dotuline{F\t0\tp{s+1}D\tpp1} +2(s+2)\dotuline{\DYM F\t0 \tp{s+1}\tpp1}+2(s+2)FD\t0 \tp{s+1}\tpp1 \nn\\
    &-(s+2)\dotuline{\t0\tp{s+1}[F,\app0}]_\g-\sum_{n=0}^{s-1}(n+2)\t0\tp{n+1}\big[F, \app{s-n}\big]_\g\bigg)-\bPtap\leftrightarrow\bPtapp\bigg). \nn
\end{align}
In order to simplify this last expression, we compute (using the dual EOM \eqref{DualEOMEYM})
\begin{align}
    \dPta F=\pa_u\big(\dPta A\big) &=-\DYM^2\a1+3C\DYM\a2+ 2DC\a2-3C^2\a3-[F,\a0]_\g+2\DYM F\t1 \nn\\
    &+3FD\t1+N\a1-6CF\t2+\dot F\t0.
\end{align}
The dot-underlined terms then form the following combination:
\begin{equation}
    \dotuline{~\cdots~}=(s+2)\t0\tp{s+1}\Big(\dbPtapp F-N\app1-FD\tpp1+3CF\tpp2\Big).
\end{equation}
Hence, comparing with \eqref{LeibnizYMinterm} and grouping terms proportional respectively to $\app{s+1}, \DYM\app{s+1}$, $\app{s+2}$ and $\tpp{s+1}$, we deduce that
\begin{align}
    &\quad\Big(\big[\Bt\bPtap,\bPtapp\big]+ \big[\bPtap,\Bt\bPtapp\big]\Big)^\ym_s=\big(\Bt\big[\bPtap,\bPtapp\big]\big)^\ym_s+ \nn\\
    &+\!\bigg(\!\bigg(\!\!-\!\Big(\underbrace{\t0D\tp1-2\tp1D\t0-3C\t0\tp2}_{\textrm{$\big(\rBt\bPtap\big)_0$}}\Big)\DYM\app{s+1}+(s+2)\app{s+2}\Big(\!\underbrace{-2C\t0D \tp1-2C\tp1D\t0-2DC\t0\tp1}_{\textrm{$-\t0\big(3CD\tp1+2DC\tp1-3C^2\tp2\big)+C\big(\rBt\bPtap\big)_0$}}\Big) \nn\\
    &+(s+1)\app{s+1}\Big\{\underbrace{\t0D^2\tp1-2\tp1D^2\t0-D\t0D\tp1-3C\t0D\tp2-3D(C\t0)\tp2}_{\textrm{$D\big(\rBt\bPtap\big)_0$}}\Big\} \nn\\
    &+(s+2)\tpp{s+1}\Big\{\underbrace{N\t0\ap1 +F\t0D\tp1-3CF\t0\tp2- \ap1D^2\t0-2F\tp1 D\t0}_{\textrm{$\ap1\big(N\t0-D^2\t0\big)+F\big(\rBt\bPtap\big)_0$}} \Big\} \nn\\
    &-\sum_{n=0}^{s-1}(n+2)\t0\tp{n+1}\big[F, \app{s-n}\big]_\g+\t0 \big[\hdap A,\app{s+1}\big]_\g+(s+2)\t0\tp{s+1}\dbPtapp F\bigg)-\bPtap\leftrightarrow\bPtapp\bigg). \nn
\end{align}
As expected, we recognize the combination of actions
\begin{align}
    \big((\rBt\bPtap)\blacktriangleright \bPtapp\big)^\ym_s &=\big(\rBt\bPtap\big)_0 \DYM\app{s+1}-(s+1)\app{s+1}D\big(\rBt \bPtap\big)_0 \nn\\
    &-(s+2)C\big(\rBt\bPtap\big)_0 \app{s+2}-(s+2)F\big(\rBt\bPtap\big)_0 \tpp{s+1}.
\end{align}
Therefore the Leibniz rule anomaly for the action $\blacktriangleright$ projected onto $\sfS$ is given by \eqref{LeibnizAnomalyBlackTauActionYM}
\begin{adjustwidth}{-0.3cm}{-0.1cm}
\begin{empheq}[box=\fbox]{align}
    \cA_s^\ym\Big(\big[\bPtap,\bPtapp \big], \Bt\!\Big) &=\bigg\{\!\big((\rBt\bPtap)\blacktriangleright \bPtapp\big)^\ym_s+(s+2)\t0\app{s+2}\dbtp C-(s+2)\ap1\tpp{s+1}\odt C \nn\\
    &\quad -\t0 \big[\hdap A,\app{s+1}\big]_\g+(s+2)\t0\tpp{s+1}\dbPtap F \\
    &\quad +\sum_{n=1}^{s}(n+1)\tp{n}\big[\odt A, \app{s+1-n}\big]_\g\bigg\}-\bPtap\leftrightarrow\bPtapp, \nn
\end{empheq}
\end{adjustwidth}
where we used that $\odt A=F\t0$.

\subsection{Homomorphism anomaly \label{AppEYM:BlackTauHom}}

From the definitions \eqref{defBlackActionGR}-\eqref{defBlackActionYM} of the action $\blacktriangleright$, we immediately get that the action of the commutator $\big[\ot,\otp\big]^\circ$ vanishes,
\begin{equation}
    \big[\ot,\otp\big]^\circ \blacktriangleright \bPtapp=(0,0).
\end{equation}
We then compute the commutator of the actions.
We give all the details for the YM sub-space part, since the GR part is analogous.
\begin{align}
    \big(\Bt\, &(\Btp\bPtapp)\big)^\ym_s =\nn\\
    &=\t0\DYM \big(\Btp\bPtapp\big)^\ym_{s+1}-(s+1)\big(\Btp\bPtapp\big)^\ym_{s+1}D\t0-(s+2)C\t0\big(\Btp\bPtapp\big)^\ym_{s+2} \nn\\
    &-(s+2)F\t0 \big(\Btp\bPtapp\big)^\gr_{s+1} \nn\\
    &=\t0\DYM\Big(\tp0\DYM\app{s+2}-(s+2)\app{s+2}D\tp0-(s+3)C\tp0\app{s+3}-(s+3)F\tp0\tpp{s+2}\Big) \nn\\
    &-(s+1)D\t0\Big(\tp0\DYM\app{s+2}-(s+2)\app{s+2}D\tp0-(s+3)C\tp0\app{s+3}-(s+3)F\tp0\tpp{s+2}\Big) \nn\\
    &-(s+2)C\t0\Big(\tp0\DYM\app{s+3}-(s+3)\app{s+3}D\tp0-(s+4)C\tp0\app{s+4}-(s+4)F\tp0\tpp{s+3}\Big) \nn\\
    &-(s+2)F\t0\Big(\tp0D\tpp{s+2}-(s+3)\tpp{s+2}D\tp0-(s+4)C\tp0\tpp{s+3}\Big).
\end{align}
After anti-symmetrization in $\ot,\otp$, we obtain that
\begin{align}
    &\quad\Big(\Bt\big(\Btp\bPtapp\big)-\Btp\big(\Bt\bPtapp\big)\Big)^\ym_s =-\big(\ot\leftrightarrow\otp\big)\,+\nn\\
    &+\Big\{\underline{\t0D\tp0\DYM\app{s+2}}-(s+2)\underline{\t0D\tp0\DYM\app{s+2}}-(s+2)\app{s+2}\t0D^2\tp0-(s+3)C\uwave{\app{s+3}\t0D\tp0} \nn\\
    &-(s+3)F\dashuline{\tpp{s+2}\t0D\tp0}-(s+1)\underline{\tp0D\t0\DYM\app{s+2}}+(s+1)(s+3)C\uwave{\tp0 D\t0\app{s+3}} \\
    &+(s+1)(s+3)F\dashuline{\tpp{s+2}\tp0D\t0}+(s+2)(s+3)C\uwave{\app{s+3}\t0D\tp0}+(s+2)(s+3)F\dashuline{\tpp{s+2}\t0D\tp0}\Big\}. \nn
\end{align}
The various underlined terms cancel each other, so that we are left with\footnote{The term $N\t0$ in $\odt C$ does not contribute because of the anti-symmetrization. We can thus include it for free.}
\begin{equation}
    \boxed{\Big(\big[\ot,\otp\big]^\circ \blacktriangleright\bPtapp-\big[\Bt,\Btp\!\big]\bPtapp\Big)^\ym_s=(s+2)\app{s+2}\Big(\tp0\odt C-\t0\odtp C\Big)}.
\end{equation}
Similarly, the projection onto the $\bsfT$-subspace results in \eqref{HomAnomalyBlackTauActionGR}
\begin{equation}
    \boxed{\Big(\big[\ot,\otp\big]^\circ \blacktriangleright\bPtapp-\big[\Bt,\Btp\!\big]\bPtapp\Big)^\gr_s=(s+3)\tpp{s+2}\Big(\tp0\odt C-\t0\odtp C\Big)}.
\end{equation}

\section{Properties of the action $\blacktriangleleft$}

\subsection{Leibniz rule anomaly \label{AppEYM:BackBlackTauLeibniz}}

Since $\big[\otp,\otpp\big]^\circ$ is in the center of $\cST$, i.e. is non-zero only at degree $-1$, we readily see from the definitions \eqref{defReverseBlackAction-1}-\eqref{defReverseBlackAction0} that
\begin{equation}
    \big[\otp,\otpp\big]^\circ \blacktriangleleft\bPta =(0,0).
\end{equation}
Moreover, since $\big[\rBtp\bPta,\otpp\big]_0^\circ=0$, we trivially get that the Leibniz rule anomaly at degree 0 vanishes:
\begin{equation}
    \boxed{\cA_0\Big(\big[\otp,\otpp \big]^\circ, \blacktriangleleft\bPta \Big)=0}.
\end{equation}
In the following, we will also need to calculate
\begin{align}
    \big(\rBtp(\Btpp\bPta)\big)_0 &=\tp0D\big(\Btpp\bPta\big)^\gr_1-2\big(\Btpp\bPta\big)^\gr_1 D\tp0-3C\tp0 \big(\Btpp\bPta\big)^\gr_2 \nn\\
    &=\tp0D\big(\tpp0D\t2-3\t2D\tpp0-4C\tpp0\t3\big)-2D\tp0\big(\tpp0D\t2-3\t2D\tpp0-4C\tpp0\t3\big) \nn\\
    &-3C\tp0\big(\tpp0D\t3-4\t3D\tpp0-5C\tpp0\t4\big),
\end{align}
which upon anti-symmetrization in $\otp,\otpp$ simplifies to\footnote{In the last step we added $N\tp0\tpp0-N\tpp0\tp0=0$.}
\begin{align}
    \big(\rBtp(\Btpp\bPta)\big)_0-\big(\rBtpp(\Btp\bPta)\big)_0 &=3\t2\tpp0D^2\tp0-3\t2\tp0D^2\tpp0 \nn\\
    &=3\t2\big(\tp0\odtpp C-\tpp0\odtp C\big). \label{IntermImportant}
\end{align}

Concerning the degree $-1$, we compute
\begin{align}
    &\quad~\big[\rBtp\bPta,\otpp\big]_{- 1}^\circ+ \big[\otp,\rBtpp\bPta\big]_{- 1}^\circ =\Big(\big(\rBtp\bPta\big)_0 D\tpp0-\tpp0 D\big(\rBtp\bPta\big)_0\Big)-\otp\leftrightarrow\otpp \nn\\
    &=\Big(\big(\tp0D\t1-2\t1D\tp0-3C\tp0\t2\big)D\tpp0+\tp0D\big(\tpp0D\t1-2\t1D\tpp0-3C\tpp0\t2\big)\Big)-\otp\leftrightarrow\otpp \nn\\
    &=\Big(-2\tp0\t1D^2\tpp0-6C\tp0\t2D\tpp0\Big)-\otp\leftrightarrow\otpp.
\end{align}
Now notice that
\begin{align}
    \big(\rBtp(\Btpp\bPta)\big)_{-1}=-2C\tp0 \big(\Btpp\bPta\big)_1^\gr=-2C\tp0\big( \tpp0D\t2-3\t2D\tpp0-4C\tpp0\t3\big),
\end{align}
which upon anti-symmetrization reduces to
\begin{equation}
    \big(\rBtp(\Btpp\bPta)\big)_{-1}-\big(\rBtpp(\Btp\bPta)\big)_{-1}=6C\tp0\t2 D\tpp0-6C\tpp0\t2 D\tp0.
\end{equation}
Consequently,
\begin{equation}
    \big[\rBtp\bPta,\otpp\big]_{-1}^\circ+ \big[\otp,\rBtpp\bPta\big]_{-1}^\circ =\Big(2\tp0\t1\odtpp C-\big(\rBtp(\Btpp\bPta)\big)_{-1}\Big)-\otp\leftrightarrow\otpp,
\end{equation}
where $\odt C=-D^2\t0+N\t0$.
The Leibniz rule anomaly at degree $-1$ is thus \eqref{LeibnizAnomalyrBlackTauAction}
\begin{equation}
    \boxed{\cA_{-1}\Big(\big[\otp,\otpp \big]^\circ, \blacktriangleleft\bPta \Big)=\Big\{ \big(\rBtp(\Btpp\bPta)\big)_{-1}+2\t1\tpp0 \odtp C\Big\}-\otp\leftrightarrow\otpp}.
\end{equation}

\subsection{Homomorphism anomaly \label{AppEYM:BackBlackTauHom}}

To check that the back reaction $\blacktriangleleft$ is a homomorphism of Lie algebras, we first calculate
\begin{align}
    \big((\rBtpp\bPta)\blacktriangleleft\bPtap \big)_{-1}-\big((\rBtpp\bPtap) \blacktriangleleft\bPta \big)_{-1} &=-2C\tp1\big(\rBtpp\bPta\big)_0-\bt\leftrightarrow\btp \nn\\
    &=-2C\tp1\big(\tpp0D\t1-2\t1D\tpp0-3C\tpp0\t2\big)-\bt\leftrightarrow\btp \nn\\
    &=-2\tpp0\tp1\big(CD\t1-3C^2\t2\big)-\bt\leftrightarrow\btp.
\end{align}
On the other hand,
\begin{equation}
    \big(\rBtpp[\bPta,\bPtap]\big)_{-1}= -2C\tpp0\overline{[\bt,\btp]}_1-\bt\leftrightarrow\btp=-4C\tpp0\t1D\tp1-\bt\leftrightarrow\btp.
\end{equation}
Therefore,\footnote{We can include the term $2DC\t1$ appearing in $\dbt C$ since the anti-symmetrization cancels it.}
\begin{equation} \label{HomAnomaly-1interm}
    \boxed{\Big(\rBtpp[\bPta,\bPtap]-\otpp\big[\!\blacktriangleleft\bPta, \blacktriangleleft\bPtap\big] \Big)_{-1}= 2\tpp0\big(\tp1\dbt C-\t1\dbtp C\big)}.
\end{equation}

At degree 0, we have that
\begin{align}
    \big((\rBtpp\bPta)\blacktriangleleft\bPtap \big)_0=\big(\rBtpp\bPta\big)_0D\tp1-2\tp1 D\big(\rBtpp\bPta\big)_0-3C\tp2\big(\rBtpp\bPta\big)_0.
\end{align}
Anti-symmetrizing between $\bPta$ and $\bPtap$ and using that
\begin{equation}
    \big(\rBtpp\bPta\big)_0=\tpp0D\t1-2\t1D\tpp0-3C\t2\tpp0,
\end{equation}
we get after some simple algebra that
\begin{equation}
    \big(\otpp\big[\!\blacktriangleleft\bPta, \blacktriangleleft\bPtap\big]\big)_0=\Big(-2\tpp0\tp1D^2\t1+4\tp1D\t1D\tpp0+6DC\tp1\t2\tpp0+6C\tp1D\t2\tpp0\Big)-\bt\leftrightarrow\btp.
\end{equation}
Since
\begin{align}
    \big(\rBtpp[\bPta,\bPtap]\big)_0 &=\tpp0 D\overline{[\bt,\btp]}_1-2\overline{[\bt,\btp]}_1 D\tpp0-3C\tpp0\overline{[\bt,\btp]}_2 \\
    &=\Big(2\tpp0\t1D^2\tp1-4\t1D\tp1D\tpp0-3C\tpp0\big(2\t1D\tp2+3\t2D\tp1\big)\Big)-\bt\leftrightarrow\btp, \nn
\end{align}
we infer that
\begin{equation}
    \Big(\rBtpp[\bPta,\bPtap]-\otpp\big[\!\blacktriangleleft\bPta, \blacktriangleleft\bPtap\big] \Big)_0= -3\tpp0\t2\big(2DC\tp1+3CD\tp1\big)-\bt\leftrightarrow \btp.
\end{equation}
The latter recombines with \eqref{HomAnomaly-1interm} such that for $s=-1,0$, one gets \eqref{HomAnomalyrBlackTauAction}
\begin{equation} 
    \boxed{\Big(\rBtpp[\bPta,\bPtap]-\otpp\big[\!\blacktriangleleft\bPta, \blacktriangleleft\bPtap\big] \Big)_{s}= (s+3)\tpp0\big(\tp{s+2}\dbt C-\t{s+2}\dbtp C\big)}.
\end{equation}

\section{Proof of the Jacobi identity anomaly \label{AppEYM:Jacobi}}

We compute the Jacobi identity anomaly for the YM-bracket. 
We first evaluate $\big[\Pta,[\Ptap,\Ptapp]\big]$ by splitting $\Pta$ as $(\ot,\bPta)$:
\begin{align}
    \big[\Pta,[\Ptap,\Ptapp]\big] &=\Big[ (\ot,\bPta),\big[(\otp,\bPtap),(\otpp,\bPtapp)\big]\Big] \nn\\
    &=\bigg(\Big[\ot,\big[\otp,\otpp \big]^\circ+\rBtp\bPtapp-\rBtpp\bPtap\Big]^\circ +\rBt\Big(\big[ \bPtap,\bPtapp\big]+\Btp\bPtapp-\Btpp\bPtap\Big) \nn\\
    &\qquad -\Big(\big[\otp,\otpp \big]^\circ+\rBtp\bPtapp-\rBtpp\bPtap\Big) \blacktriangleleft\bPta\,, \\
    &\qquad\Big[\bPta,\big[\bPtap,\bPtapp \big]+ \Btp\bPtapp-\Btpp\bPtap\Big]+\Bt \Big(\big[\bPtap,\bPtapp \big]+ \Btp\bPtapp-\Btpp\bPtap\Big) \nn\\
    &\qquad -\Big(\big[\otp,\otpp \big]^\circ+\rBtp\bPtapp-\rBtpp\bPtap\Big) \blacktriangleright\bPta\bigg). \nn
\end{align}
Using the symbol $\cyc$ to denote an equality valid upon cyclic permutation of both the LHS and the RHS, we conveniently rearrange the terms as follows:
\begin{align}
    \big[\Pta,[\Ptap,\Ptapp]\big] &\cyc \bigg(\cancel{\big[\ot,[\otp,\otpp]^\circ\big]^\circ}+\big[\otp, \rBtpp\bPta\big]^\circ+\big[ \rBtp\bPta,\otpp\big]^\circ+\rBtpp\big[\bPta,\bPtap\big]+\rBtp\big(\Btpp\bPta\big) \nn\\
    &\qquad -\rBtpp\big(\Btp\bPta\big)-[\otp,\otpp]^\circ\blacktriangleleft\bPta-\big(\rBtpp\bPta\big)\blacktriangleleft \bPtap+\big(\rBtpp\bPtap\big)\blacktriangleleft \bPta\,, \nn\\
    &\qquad\big[\bPta,[\bPtap,\bPtapp]\big]-\big[\Bt\bPtap,\bPtapp\big]-\big[\bPtap,\Bt\bPtapp\big]+\Bt\big[\bPtap,\bPtapp\big]+\Bt\big(\Btp\bPtapp\big) \nn\\
    &\qquad -\Btp\big(\Bt\bPtapp\big)-\big[\ot,\otp\big]^\circ\blacktriangleright\bPtapp-\big(\rBt\bPtap\big) \blacktriangleright \bPtapp+ \big(\rBt\bPtapp\big)\blacktriangleright \bPtap\bigg) \\
    &\hspace{-2cm}\cyc\bigg(\!-\cA\Big(\big[\otp,\otpp \big]^\circ, \blacktriangleleft\bPta\Big) +\rBtp\big(\Btpp\bPta\big)-\rBtpp\big(\Btp\bPta\big)+\otpp\Big(\! \blacktriangleleft\big[\bPta,\bPtap\big]-\big[\!\blacktriangleleft\bPta, \blacktriangleleft \bPtap\big]\Big)\,,\nn\\
    &\hspace{-2cm} \big[\bPta,[\bPtap,\bPtapp]\big]+\cA\Big(\big[\bPtap,\bPtapp\big],\Bt\!\!\Big)-\big(\rBt\bPtap\big) \blacktriangleright \bPtapp+ \big(\rBt\bPtapp\big)\blacktriangleright \bPtap-\Big(\big[\ot,\otp\big]^\circ\! \blacktriangleright-\big[\Bt,\Btp\!\big]\Big)\bPtapp\bigg). \nn
\end{align}
Since the cyclic permutation of $\big[\bPta,[\bPtap,\bPtapp]\big]$ is similarly given by
\begin{align}
    \big[\bPta,[\bPtap,\bPtapp]\big] &=\Big[(\bt,\a{}),\big[(\btp,\ap{}),(\btpp,\app{})\big]\Big] \\
    &\cyc \bigg(\overline{\big[\bt,\overline{[\btp,\btpp]}\big]},\big[\a{},[\ap{},\app{}]^\g\big]^\g+ \cA\Big(\big[\ap{},\app{}\big]^\g,\Wt\!\Big)-\Big(\overline{[\bt,\btp]} \vartriangleright-\big[\Wt,\Wtp\!\big]\Big)\app{}\bigg),\nn
\end{align}
we can gather all the properties of the actions $\vartriangleright, \blacktriangleright$ and $\blacktriangleleft$ to deduce the Jacobi identity anomaly of the $\sfST$-bracket.
More precisely, using (\ref{LeibnizAnomalyBlackTauActionGR}-\ref{LeibnizAnomalyrBlackTauAction}-\ref{HomAnomalyBlackTauActionGR}-\ref{HomAnomalyrBlackTauAction}) and \eqref{IntermImportant}, we get that\footnote{The reader can check that $\overline{\big[\bt,\overline{[\btp,\btpp]}\big]}\cyc 0$. 
Of course, since we already computed the Jacobi identity anomaly of the $C$-bracket in \eqref{JacCF} and also in \eqref{JacCFC}, we know that its restriction to the $\bsfT$-sub-space vanishes (the anomaly involving the degree 0 elements $\t0,\tp0,\tpp0$, cf. \eqref{JacobiAnomalyCbracketEYMinterm}).}
\begin{align}
    \big[\Pta,[\Ptap,\Ptapp]\big]^\gr_{-1} &\cyc 2\tpp0\big(\tp1\dbt C-\t1\dbtp C\big)-2\t1\big(\tpp0\odtp C-\tp0\odtpp C\big), \nn\\
    \big[\Pta,[\Ptap,\Ptapp]\big]^\gr_0 &\cyc 3\tpp0\big(\tp2\dbt C-\t2\dbtp C\big)+3\t2\big(\tp0\odtpp C-\tpp0\odtp C\big),\label{JacobiAnomalyCbracketEYMinterm}\\
    \big[\Pta,[\Ptap,\Ptapp]\big]^\gr_s &\cyc (s+3)\t0\big(\tpp{s+2}\dbtp C-\tp{s+2}\dbtpp C\big)-(s+3)\tpp{s+2}\big(\tp0\odt C-\t0\odtp C\big), \qquad  s\geqslant 1, \nn
\end{align}
which thanks to the cyclic permutation is more compactly written for all $s\geq -1$ as
\begin{equation}
    \boxed{\big[\Pta,[\Ptap,\Ptapp]\big]^\gr_s \cyc-\dt C\paren{\tp{},\tpp{}}^\gr_s}.
\end{equation}
Besides, collecting (\ref{LeibnizAnomalyTauAction}-\ref{LeibnizAnomalyBlackTauActionYM}-\ref{HomAnomalyTauAction}-\ref{HomAnomalyBlackTauActionYM}) we obtain that the YM projection of the Jacobi anomaly of the $\sfST$-bracket takes the form\footnote{Using that $\big[\a{},[\ap{},\app{}]^\g\big]^\g\cyc 0$, cf. footnote \ref{footJacobi}.}
\begin{align}
    \big[\Pta,[\Ptap,\Ptapp]\big]^\ym_s&\cyc\bigg\{\!\bigg\{ (s+2)\t0\app{s+2}\dbtp C-(s+2)\ap1\tpp{s+1}\odt C -\t0 \big[\hdap A,\app{s+1}\big]_\g\!+(s+2)\t0\tpp{s+1}\dbPtap F \nn\\
    &\quad +\sum_{n=1}^{s}(n+1)\tp{n}\big[\odt A, \app{s+1-n}\big]_\g-\sum_{n=1}^s(n+1)\t{n}\big[\hdap A,\app{s+1-n}\big]_\g\bigg\}-\bPtap\leftrightarrow\bPtapp\bigg\} \nn\\
    &\quad -(s+2)\app{s+2}\Big(\tp0\odt C-\t0\odtp C\Big)-(s+2)\app1\Big(\t{s+1} \dbtp C-\tp{s+1}\dbt C\Big).
\end{align}
Simplifying and permuting cyclically some of the terms, we get
\vspace{-0.5cm}

\begin{adjustwidth}{-0.1cm}{0cm}
\begin{empheq}[box=\fbox]{align}
    \big[\Pta,[\Ptap,\Ptapp]\big]^\ym_s &\cyc \sum_{n=0}^s(n+1)\Big(\tp{n}\big[\dPta A,\app{s+1-n}\big]_\g-\tpp{n}\big[\dPta A,\ap{s+1-n}\big]_\g\Big) \\
    &-(s+2)\dt C\Big(\tp0\app{s+2}-\tpp0\ap{s+2}+ \ap1\tpp{s+1}- \app1\tp{s+1}\Big)-\dbPta F\paren{\tp{}, \tpp{}}^\ym_s,  \nn
\end{empheq}
\end{adjustwidth}
where we used the fact that
\begin{align}
    &\bigg\{\!-\sum_{n=1}^{s}(n+1)\tpp{n}\big[\odt A, \ap{s+1-n}\big]_\g-\sum_{n=1}^s(n+1)\t{n}\big[\hdap A,\app{s+1-n}\big]_\g-\t0 \big[\hdap A,\app{s+1}\big]_\g\bigg\}-\bPtap\leftrightarrow\bPtapp \nn\\
    &\cyc -\sum_{n=0}^s(n+1) \t{n}\big[\dPtap A,\app{s+1-n}\big]_\g-\Ptap\leftrightarrow\Ptapp.
\end{align}
Finally, notice that we can swap $\dbPta F$ for $\dPta F$ since $\t0\paren{\tp{},\tpp{}}^\ym\cyc 0$, thence getting \eqref{JacobiEYMYMbracket}.

\section{Properties of the algebroid actions}

\subsection{Leibniz rule \label{AppEYM:DotWhiteLeibniz}}

\paragraph{Action} \!\!$\dotvartriangleright\,$:
First notice that
\begin{equation}
    \big(\Wt(\hdapp\ap{})\big)_s=\hdapp\big( \Wt\ap{}\big)_s-\sum_{n=1}^s (n+1)\t{n}\big[\hdapp A,\ap{s+1-n}\big]_\g.
\end{equation}
Moreover we have that
\begin{equation}
    \big\lbr\Wt\ap{},\app{}\big\rbr^\g =\big[\Wt\ap{} ,\app{}\big]^\g +\hdapp\big(\Wt\ap{}\big)-\hat\delta_{\bt\mkern1mu \vartriangleright\mkern1mu\ap{}}\app{}.
\end{equation}
Using \eqref{LeibnizAnomalyTauAction}, this directly leads to
\begin{equation}
    \cA\big(\lbr\ap{},\app{}\rbr^\g ,\Wt\!\big)=\cA\big([ \ap{},\app{}]^\g ,\Wt\!\big)+\cA\big(\hdapp\ap{}-\hdap\app{},\Wt\!\big)= \hat\delta_{\bt\mkern1mu\vartriangleright \mkern1mu\ap{}}\app{}-\hat\delta_{\bt\mkern1mu \vartriangleright\mkern1mu\app{}}\ap{}.
\end{equation}
From \eqref{defActionDot2}, we also know that
\begin{equation}
    \cA\big(\lbr\ap{},\app{}\rbr^\g,\dWt\!\big)= \cA\big(\lbr\ap{},\app{}\rbr^\g,\Wt\!\big)- \cA\big(\lbr\ap{},\app{}\rbr^\g,\dbt\big),
\end{equation}
where
\begin{align}
    \cA\big(\lbr\ap{},\app{}\rbr^\g,\dbt\big) &=\Big(\dbt\hdapp\ap{}-\hdapp\dbt\ap{}+\hat\delta_{\dbt\ap{}}\app{}\Big)-\ap{}\leftrightarrow\app{} \nn\\
    &=\Big(-\hat\delta_{\bt\mkern1mu\vartriangleright \mkern1mu\app{}-\dbt\app{}}\ap{} +\hat\delta_{\dbt\ap{}}\app{}\Big)-\ap{}\leftrightarrow\app{} \\
    &=\hat\delta_{\bt\mkern1mu\vartriangleright \mkern1mu\ap{}}\app{}-\hat\delta_{\bt\mkern1mu\vartriangleright \mkern1mu\app{}}\ap{}. \nn
\end{align}
In the second line, we used the morphism property \eqref{AlgebroidActionEYM} of the anchor map, i.e.
\begin{equation}
    \big[\dbt,\hdapp\big]\ap{}=-\delta_{\lbr(0;\bt,0),(0;0,\app{})\rbr}\ap{},
\end{equation}
where the only non-vanishing part of $\lbr(0;\bt,0),(0;0,\app{})\rbr$ is in the YM subspace and equals
\begin{equation}
    \lbr(0;\bt,0),(0;0,\app{})\rbr^\ym=\dWt\app{}= \Wt\app{}-\dbt\app{}.
\end{equation}
Therefore, cf.\,\eqref{dotWhiteProperties},
\begin{equation}
    \boxed{\cA\big(\lbr\ap{},\app{}\rbr^\g ,\dWt\!\big) =0}.
\end{equation}

\paragraph{Action} \!\!$\dotblacktriangleleft\,$:
On the one hand,
\begin{align}
    \big\lbr\otp,\otpp\big\rbr^\circ\, \dotblacktriangleleft\,\bPta &=\big\lbr\otp,\otpp\big\rbr^\circ\!\blacktriangleleft \bPta+\dbt\big\lbr\otp,\otpp\big\rbr^\circ \\
    &=\big[\otp,\otpp\big]^\circ\! \blacktriangleleft\bPta +\big(\odtpp\otp-\odtp\otpp\big)\blacktriangleleft\bPta+ \dbt\big[\otp,\otpp\big]^\circ+\dbt\odtpp\otp-\dbt\odtp\otpp. \nn
\end{align}
On the other hand, 
\begin{align}
    \big\lbr\drBtp\bPta,\otpp\big\rbr^\circ&=\big[ \rBtp\bPta,\otpp\big]^\circ+\big[\dbt\otp,\otpp\big]^\circ+\odtpp\big(\rBtp\bPta\big)+\odtpp\dbt \otp-\delta_{\text{\scalebox{1.1}{$\ensurestackMath{\stackon[-1.9pt]{\tau'}{\mkern-4mu\scriptscriptstyle{\circ}}}$}} \scriptdotblacktriangleleft\,\bPta}\otpp.
\end{align}
We can now use the fact that, for $s=-1,0$,
\begin{equation}
    \odtpp\big(\rBtp\bPta\big)_s=\big((\odtpp\otp) \blacktriangleleft\bPta\big)_s+\big(\rBtp(\odtpp\bPta) \big)_s-(s+3)\tp0\t{s+2}\odtpp C,
\end{equation}
so that
\begin{align}
    \cA_s\Big(\big\lbr\otp,\otpp\big\rbr^\circ, \dotblacktriangleleft\,\bPta\Big) &=\cA_s\Big(\big[\otp,\otpp\big]^\circ, \blacktriangleleft\bPta\Big)+\cA_s\Big(\big[\otp,\otpp\big]^\circ, \dbt\Big)+\Big\{\Big\{\big[ \dbt, \odtpp\big]\otp_{\!\!s} \\
    & -\delta_{\text{\scalebox{1.1}{$\ensurestackMath{\stackon[-2pt]{\tau''}{\mkern-9mu\scriptscriptstyle{\circ}}}$}} \scriptdotblacktriangleleft\,\bPta}\otp_{\!\!s} -\big(\rBtp(\odtpp\bPta) \big)_s+(s+3)\tp0\t{s+2}\odtpp C\Big\}-\otp\leftrightarrow\otpp\Big\}. \nn
\end{align}
Note that for the $\circ$-bracket, $\cA\big([\otp,\otpp]^\circ,\dbt\big)=0$.
We then have that the $\sfST$-bracket resulting from $[\dbt,\odtpp]$ is
\begin{equation}
    \big\lbr(0,\bPta),(\otpp,0)\big\rbr=-\Big( \drBtpp\bPta,\dBtpp\bPta\Big),
\end{equation}
such that
\begin{equation}
    \big[\dbt,\odtpp\big]\otp= \delta_{\text{\scalebox{1.1}{$\ensurestackMath{\stackon[-2pt]{\tau''}{\mkern-9mu\scriptscriptstyle{\circ}}}$}} \scriptdotblacktriangleleft\,\bPta}\otp+ \delta_{(\text{\scalebox{1.1}{$\ensurestackMath{\stackon[-2pt]{\tau''}{\mkern-9mu\scriptscriptstyle{\circ}}}$}} \scriptdotblacktriangleright\,\bPta)^\gr}\otp.
\end{equation}
Next recall that $\cA_0\big([\otp,\otpp]^\circ, \blacktriangleleft\bPta\big)=0$.
Hence,
\begin{align}
    \cA_0\Big(\big\lbr\otp,\otpp\big\rbr^\circ, \dotblacktriangleleft\,\bPta\Big) &=\Big\{ \delta_{(\text{\scalebox{1.1}{$\ensurestackMath{\stackon[-2pt]{\tau''}{\mkern-9mu\scriptscriptstyle{\circ}}}$}} \scriptdotblacktriangleright\,\bPta)^\gr}\otp_{\!\!0}-\big(\rBtp(\odtpp\bPta) \big)_0+3\tp0\t{2}\odtpp C\Big\}-\otp\leftrightarrow\otpp.
\end{align}
Comparing with
\begin{equation}
    \drBtp\big(\dBtpp\bPta\big)=\rBtp\big(\Btpp \bPta-\odtpp\bPta\big)+ \delta_{(\text{\scalebox{1.1}{$\ensurestackMath{\stackon[-2pt]{\tau''}{\mkern-9mu\scriptscriptstyle{\circ}}}$}} \scriptdotblacktriangleright \,\bPta)^\gr}\otp
\end{equation}
and using \eqref{IntermImportant}, we see that
\begin{equation}
    \cA_0\Big(\big\lbr\otp,\otpp \big\rbr^\circ, \dotblacktriangleleft\, \bPta\Big)=\Big(\drBtp\big(\dBtpp\bPta\big)-\drBtpp\big(\dBtp\bPta\big) \Big)_0.
\end{equation}
Concerning the degree $-1$, we use the Leibniz rule anomaly \eqref{LeibnizAnomalyrBlackTauAction} and get
\begin{align}
    \cA_{-1}\Big(\big\lbr\otp,\otpp \big\rbr^\circ, \dotblacktriangleleft\, \bPta\Big) &=\Big\{ \big(\rBtp(\Btpp\bPta)\big)_{-1} -\cancel{2\tp0\t1\odtpp C} \\
    & +\delta_{(\text{\scalebox{1.1}{$\ensurestackMath{\stackon[-2pt]{\tau''}{\mkern-9mu\scriptscriptstyle{\circ}}}$}} \scriptdotblacktriangleright\,\bPta)^\gr}\otp_{\!\!-1}-\big(\rBtp(\odtpp\bPta) \big)_{-1}+\cancel{2\tp0\t1\odtpp C}\Big\}-\otp\leftrightarrow\otpp. \nn
\end{align}
We are therefore left with the same result both at degree $-1$ and 0, namely \eqref{dotBlackProperties}
\begin{equation}
    \boxed{\cA\Big(\big\lbr\otp,\otpp \big\rbr^\circ, \dotblacktriangleleft\, \bPta\Big)=\drBtp\big(\dBtpp\bPta\big)-\drBtpp\big(\dBtp\bPta\big)}.
\end{equation}

\subsection{Homomorphism \label{AppEYM:DottauHom}}

\paragraph{Action} \!\!$\dotvartriangleright\,$:
We first compute
\begin{equation}
    \overline{\lbr\bt,\btp\rbr} \,\dotvartriangleright\, \app{}=\overline{[\bt,\btp]} \vartriangleright \app{}-\delta_{\mkern3mu\overline{\lbr\bt,\btp\rbr} \mkern1mu} \app{}+\big(\dbtp\bt\big) \vartriangleright \app{}- \big(\dbt\btp\big)\vartriangleright \app{}.
\end{equation}
Next
\begin{align}
    \big[\dWt,\dWtp\!\big]\app{} =\Big\{\Wt \big(\Wtp\app{}\big)-\Wt\big(\dbtp\app{}\big) -\dbt\big(\Wtp\app{}\big)+\dbt\dbtp\app{}\Big\} -\bt\leftrightarrow\btp.
\end{align}
Using that\footnote{Notice that $\dbtp A=0$.}
\begin{equation}
    \big(\dbtp(\Wt\app{})\big)_s=\big((\dbtp\bt) \vartriangleright\app{}\big)_s+\big(\Wt(\dbtp \app{})\big)_s+(s+2)\dbtp C\t{s+1}\app1,
\end{equation}
and the fact that
\begin{equation}
    \big[\dbt,\dbtp\big]\app{}=-\delta_{\mkern3mu\overline{\lbr\bt,\btp\rbr} \mkern1mu}\app{},
\end{equation}
we infer that the commutator of the dotted action reduces to
\begin{align}
    \Big(\big[\dWt,\dWtp\!\big]\app{}\Big)_s &=\Big(\big[\Wt,\Wtp\!\big]\app{}\Big)_s-\delta_{\mkern3mu\overline{\lbr\bt,\btp\rbr} \mkern1mu}\app{s} \\
    &+\Big\{\Big\{\big((\dbtp\bt) \vartriangleright\app{}\big)_s+(s+2)\dbtp C\t{s+1}\app1\Big\}-\bt\leftrightarrow\btp\Big\}. \nn
\end{align}
From the result \eqref{HomAnomalyTauAction}, we conclude that
\begin{equation}
    \boxed{\overline{\lbr\bt,\btp\rbr} \,\dotvartriangleright \,\app{}-\big[\dWt,\dWtp\!\big]\app{}=0}.
\end{equation}

\paragraph{Action} \!\!$\dotblacktriangleleft\,$:
Firstly,
\begin{equation}
    \drBtpp\big\lbr\bPta,\bPta'\big\rbr=\rBtpp\Big( \big[\bPta,\bPta'\big]+\dbPtap\bPta-\dbPta\bPtap\Big)+ \delta_{\lbr\bt,\btp\rbr}\otpp.
\end{equation}
Secondly,
\begin{equation}
    \otpp\big[\dotblacktriangleleft\,\bPta\,, \dotblacktriangleleft\,\bPtap\,\big]=\Big\{ \big(\rBtpp\bPta\big)\blacktriangleleft\bPtap +\big(\dbt\otpp\big)\blacktriangleleft\bPtap +\dbtp\big(\rBtpp\bPta\big)+\dbtp\dbt\otpp\Big\} -\bPta\leftrightarrow\bPtap.
\end{equation}
Hence, for $s=-1,0$,
\begin{align}
    \Big(\otpp\big[\dotblacktriangleleft\,\bPta\,, \dotblacktriangleleft\,\bPtap\,\big]\Big)_s &=\Big(\otpp\big[\!\blacktriangleleft\bPta\,, \blacktriangleleft\bPtap\,\big]\Big)_s+ \delta_{\lbr\bt,\btp\rbr}\otpp_{\!\!\!s}+\Big\{\Big\{\cancel{\big((\dbt\otpp) \blacktriangleleft \bPtap}\big)_s \\
    &-\cancel{\big((\dbt\otpp)\blacktriangleleft \bPtap}\big)_s-\big(\rBtpp(\dbt\bPtap)\big)_s +(s+3)\tpp0\tp{s+2}\dbt C\Big\}-\bPta\leftrightarrow\bPtap\Big\}. \nn
\end{align}
Since $\rBtpp(\dbPta\bPtap)=\rBtpp(\dbt\bPtap)$, we just have to use \eqref{HomAnomalyrBlackTauAction} to infer that, cf.\,\eqref{dotBlackProperties}
\begin{equation}
    \boxed{\drBtpp\big\lbr\bPta,\bPta'\big\rbr -\otpp\big[ \dotblacktriangleleft\,\bPta\,, \dotblacktriangleleft\,\bPta'\,\big]=0}.
\end{equation}

\section{Covariant derivative in the wedge \label{AppEYM:LeibnizDEYM}}

Starting from $\big(\DEYM\a{}\big)_s=\DYM\a{s+1}-(s+2)C\a{s+2}$, we get that
\begin{align}
    \big(\DEYM[\a{},\ap{}]^\g\big)_s &=\DYM \big[\a{},\ap{}\big]^\g_{s+1}-(s+2)C\big[\a{},\ap{}\big]^\g_{s+2} \nn\\
    &=\big[\DYM\a{},\ap{}\big]^\g_s+ \big[\a{},\DYM\ap{}\big]^\g_s+\big[\DYM\a0,\ap{s+1}\big]_\g-\big[\DYM\ap0,\a{s+1}\big]_\g-(s+2)C\big[\a{},\ap{}\big]^\g_{s+2} \nn\\
    &=\sum_{n=0}^s\big[(\DEYM\a{})_n,\ap{s-n}\big]_\g+C\sum_{n=0}^s(n+2)\uline{\big[\a{n+2},\ap{s-n}\big]_\g}+\big[\DYM\a0, \ap{s+1}\big]_\g \nn\\
    &+\sum_{n=0}^s\big[\a{n},(\DEYM\ap{})_{s-n}\big]_\g+C\sum_{n=0}^s(s+2-n)\uline{\big[\a{n},\ap{s+2-n}\big]_\g}-\big[\DYM\ap0,\a{s+1}\big]_\g \nn\\
    &-(s+2)C\sum_{n=0}^{s+2} \uline{\big[\a{n},\ap{s+2-n}\big]_\g}.
\end{align}
The underlined terms simply give
\begin{equation}
    \uline{~\cdots~}=-C\big([\a1,\ap{s+1}]_\g- [\ap1,\a{s+1}]_\g\big).
\end{equation}
Hence
\begin{equation}
    \big(\DEYM[\a{},\ap{}]^\g\big)_s=\Big( \big[\DEYM \a{},\ap{}\big]^\g_s-\big[\hda A,\ap{s+1}\big]_\g\Big) -\a{}\leftrightarrow\ap{},
\end{equation}
where $\hda A=-\DYM\a0+C\a1=-\big(\DEYM \a{}\big)_{-1}$.
Projecting this equation onto a non-radiative cut, we infer that $\DEYM$ satisfies the Leibniz rule:
\begin{equation}
    \boxed{\cA\big([\Aa{},\Aap{}]^\g,\DEYM\big)=0,} \qquad \textrm{if}\quad\Aa{},\Aap{} \in\Wcal_{\bfCc\bfAc}^\ym(S).
\end{equation}

\newpage
\addcontentsline{toc}{chapter}{Bibliography}

\bibliographystyle{JHEP}
\bibliography{source_NoetherYM}

\end{document}